\documentclass[11pt]{article}
\usepackage{booktabs}
\usepackage{fullpage}
\usepackage{geometry}
\usepackage{titlesec}
\usepackage{changepage}

\usepackage{amsmath,amsfonts,amsthm,mathrsfs,xspace,graphicx}
\usepackage[backref,colorlinks,citecolor=blue,bookmarks=true]{hyperref}
\usepackage{rotating}
\usepackage{mathpazo}
\usepackage{endnotes}
\usepackage{float}
\usepackage{mdframed}
\usepackage{bbm}
\usepackage{suffix} 
\usepackage{times}
\usepackage{tabularx}
\usepackage{makecell}
\usepackage{amssymb,latexsym}
\usepackage[capitalize]{cleveref}
\usepackage{enumitem}
\usepackage[dvipsnames]{xcolor}
\usepackage{tikz}
\usetikzlibrary{cd}
\usepackage{mathdots}
\usepackage[sans]{dsfont}
\usepackage{multirow}
\usepackage[section]{placeins}
\usepackage[affil-it]{authblk}
\usepackage{nicefrac,xfrac}
\usepackage{comment}
\usepackage{caption}

\mdfdefinestyle{figstyle}{ %
  linecolor=black!7, %
  backgroundcolor=black!7, %
  innertopmargin=10pt, %
  innerleftmargin=25pt, %
  innerrightmargin=25pt, %
  innerbottommargin=10pt %
}

\newtheorem{theorem}{Theorem}[section]
\newtheorem{proposition}[theorem]{Proposition}

\newtheorem{lemma}[theorem]{Lemma}
\newtheorem{claim}[theorem]{Claim}
\newtheorem{fact}[theorem]{Fact}
\newtheorem{corollary}[theorem]{Corollary}

\theoremstyle{definition}
\newtheorem{definition}[theorem]{Definition}
\newtheorem{example}[theorem]{Example}
\newtheorem{remark}[theorem]{Remark}
\newtheorem{observation}[theorem]{Observation}
\newcommand{\beq}{\begin{eqnarray}}
\newcommand{\eeq}{\end{eqnarray}}

\newcommand{\code}{\mathcal{C}}

\newcommand{\Tr}{\mbox{\rm Tr}}
\newcommand{\Id}{\ensuremath{{\rm Id}}}

\DeclareMathOperator*{\Expectation}{\mathbb{E}}
\newcommand{\Es}[1]{\Expectation_{#1}}

\DeclareMathOperator*{\Probability}{\mathbb{P}}
\newcommand{\Pro}[1]{\Probability_{#1}}

\newcommand{\ol}[1]{\overline{#1}}

\newcommand{\N}{\ensuremath{\mathbb{N}}}

\newcommand{\complex}{\ensuremath{\mathbb{C}}}

\newcommand{\F}{\ensuremath{\mathbb{F}}}

\newcommand{\cM}{\ensuremath{\mathcal{M}}}

\newcommand{\mX}{\ensuremath{\mathcal{X}}}

\newcommand{\ind}{\ensuremath{\mathrm{ind}}}

\DeclareMathOperator{\poly}{poly}

\newcommand{\val}{\ensuremath{\mathrm{val}}}

\newcommand{\desc}[1]{\overline{\cal{#1}}}

\newcommand{\eps}{\varepsilon}

\newcommand{\mZ}{\mathcal{Z}}

\newcommand{\LevelConstant}{5}

\newcommand{\UnaryTimeHalt}{{\rm UnaryTimeHalt}}
\newcommand{\BinaryTimeHalt}{{\rm BinaryTimeHalt}}

\DeclareMathOperator{\polylog}{polylog}

\newcommand{\abs}[1]{\left\vert {#1} \right\vert}

\DeclareMathOperator{\tr}{Tr}

\newcommand{\E}{\mathop{\mathbb{E}}\displaylimits} 

\newcommand{\tvora}{\hat{\verifier}^\ora}
\newcommand{\tsora}{\hat{\sampler}^\ora}
\newcommand{\taora}{\hat{\length}^\ora}
\newcommand{\tlora}{\hat{\linproc}^\ora}
\newcommand{\tdora}{\hat{\decider}^\ora}

\newcommand{\game}{\mathfrak{G}}
\newcommand{\sampler}{\mathcal{S}}
\newcommand{\decider}{\mathcal{D}}
\newcommand{\verifier}{\mathcal{V}}
\newcommand{\linproc}{\mathcal{L}}
\newcommand{\length}{\mathcal{A}}
\newcommand{\strategy}{\mathscr{S}}

\newcommand{\Sym}{{\rm Sym}}

\newcommand{\IP}{{\rm IP}}

\newcommand{\cF}{\mathcal{F}}
\newcommand{\FF}{\mathbb{F}}

\newcommand{\WH}{{\rm P}}
\newcommand{\Pauli}{{\rm P}}
\newcommand{\Img}{{\rm Im}}

\newcommand{\PauliBasis}{{\mathfrak{Pauli\ Basis}}}
\newcommand{\PXm}{{\mathds{X}}}
\newcommand{\PZm}{{\mathds{Z}}}
\newcommand{\PYm}{{\mathds{Y}}}

\newcommand{\frL}{\frak{L}}
\newcommand{\frR}{\frak{R}}
\newcommand{\frS}{\frak{s}}
\newcommand{\frSS}{\frak{S}}
\newcommand{\frM}{\frak{M}}
\newcommand{\frC}{\frak{C}}

\newcommand{\sJ}{\mathsf{J}}
\newcommand{\sX}{\mathsf{X}}
\newcommand{\sO}{\mathsf{O}}
\newcommand{\sY}{\mathsf{Y}}
\newcommand{\sZ}{\mathsf{Z}}
\newcommand{\sU}{\mathsf{U}}
\newcommand{\sP}{\mathsf{P}}
\newcommand{\sQ}{\mathsf{Q}}

\newcommand{\sAns}{\mathsf{Ans}}
\newcommand{\sQue}{\mathsf{Que}}

\newcommand{\sPoint}{\mathsf{Point}}

\newcommand{\sDLine}{\mathsf{DLine}}
\newcommand{\sALine}{\mathsf{ALine}}

\newcommand{\QueRed}{\mathsf{QuestionReduction}}

\newcommand{\mttx}{\mathtt{x}}
\newcommand{\mtty}{\mathtt{y}}
\newcommand{\mttz}{\mathtt{z}}
\newcommand{\mttw}{\mathtt{w}}
\newcommand{\mttv}{\mathtt{v}}
\newcommand{\mttA}{\mathtt{A}}
\newcommand{\mttB}{\mathtt{B}}
\newcommand{\mttO}{\mathtt{O}}

\newcommand{\mttX}{\mathtt{X}}

\newcommand{\mttZ}{\mathtt{Z}}
\newcommand{\mttSpace}{\mathtt{Space}}
\newcommand{\Ex}{\ensuremath{\mathbb{E}}}

\newcommand{\type}{\mathcal{T}}

\newcommand{\gamestyle}[1]{\ensuremath{\textsc{#1}}\xspace}

\newcommand{\ora}{\gamestyle{Orac}}
\newcommand{\pcp}{\gamestyle{PCP}}
\newcommand{\ar}{\gamestyle{AR}}
\newcommand{\ld}{\gamestyle{LD}}
\newcommand{\qr}{\gamestyle{QR}}

\newcommand{\labelstyle}[1]{\ensuremath{\textsc{#1}}\xspace}

\newcommand{\xpt}{\labelstyle{X}}
\newcommand{\ypt}{\labelstyle{Y}}

\newcommand{\dir}[1]{\labelstyle{V#1}}
\newcommand{\coord}{\labelstyle{I}}

\newcommand{\lnf}{\labelstyle{Ln}}

\newcommand{\tvarstyle}[1]{\mathsf{#1}}
\newcommand{\tvar}{\ensuremath{\tvarstyle{t}}}
\newcommand{\lvar}{\ensuremath{\tvarstyle{u}}}
\newcommand{\rvar}{\ensuremath{\tvarstyle{v}}}

\newcommand{\alice}{\labelstyle{A}}
\newcommand{\bob}{\labelstyle{B}}

\newcommand{\oracle}{\mathtt{Oracle}}
\newcommand{\ab}{\{\alice, \bob\}}

\newcommand{\typestyle}[1]{\ensuremath{\mathtt{#1}}\xspace}

\renewcommand{\line}{\mathbf{\ell}}

\newcommand{\Point}{\typestyle{Point}}

\newcommand{\Line}{\typestyle{Line}}
\newcommand{\ALine}{\typestyle{ALine}}
\newcommand{\DLine}{\typestyle{DLine}}

\renewcommand{\Pauli}{\typestyle{Pauli}}
\newcommand{\Sample}{\typestyle{Sample}}
\newcommand{\Read}{\typestyle{Read}}

\newcommand{\Hide}[1]{\typestyle{Hide}_{#1}}

\newcommand{\Intro}{\typestyle{Intro}}

\newcommand{\Anchor}{\typestyle{Anchor}}

\newcommand{\Introspect}{\mathfrak{Intro}}

\newcommand{\ldc}{k} 

\newcommand{\class}[1]{\ensuremath{\mathsf{#1}}\xspace}
\newcommand{\NP}{\class{NP}} %
\renewcommand{\IP}{\class{IP}} %
\newcommand{\NEXP}{\class{NEXP}} %
\newcommand{\PSPACE}{\class{PSPACE}} %
\newcommand{\PCP}{\class{PCP}} %
\newcommand{\MIP}{\class{MIP}} %
\newcommand{\NTIME}{\class{NTIME}} %
\newcommand{\MIPSTIME}{\class{MIP}^*\class{TIME}} %
\newcommand{\TMIPSTIME}{\class{TailoredMIP}^*\class{TIME}} %
\newcommand{\MIPEXP}{\class{MIPEXP}} %
\newcommand{\MIPTIME}{\class{MIPTIME}} %
\newcommand{\MIPP}{\class{MIPP}} %
\newcommand{\MIPSP}{\class{MIP}^*\class{P}} %
\newcommand{\MIPSEXP}{\class{MIP}^*\class{EXP}} %
\newcommand{\TMIPSP}{\class{TailoredMIP}^*\class{P}} %
\newcommand{\TMIPSEXP}{\class{TailoredMIP}^*\class{EXP}} %
\newcommand{\TMIP}{\class{TailoredMIP}} %
\newcommand{\RE}{\class{RE}} %

\newcommand{\Ent}{\mathscr{E}}

\newcommand{\machine}{\cal{M}}
\renewcommand{\cal}[1]{\mathcal{#1}}

\mathchardef\mhyphen="2D

\newcommand{\pcpparams}{\mathsf{pcpparams}}

\newcommand{\TIME}{\mathbb{T}}
\newcommand{\answer}{\mathsf{ANS}}

\newcommand{\anch}{\gamestyle{Anch}}

\newcommand{\tmstyle}[1]{\ensuremath{\mathsf{#1}}}
\newcommand{\Compress}{\tmstyle{Compress}}

\newcommand{\OracleVerifier}{\tmstyle{Oracularize}}
\newcommand{\Triangulate}{\tmstyle{Triangulate}}
\newcommand{\Purify}{\tmstyle{Purify}}

\newcommand{\ComputeParrepVerifier}{\tmstyle{ParRep}}

\newcommand{\ZPC}{\tmstyle{ZPC}}

\newenvironment{gamespec}{
  \begin{mdframed}[style=figstyle]}{
  \end{mdframed}}

\newcommand{\zero}{\mathrm{zero}}

\newcommand{\simulpolymeas}[4]{\mathrm{PolyMeas}(#1,#2,#3, #4)}

\newcommand{\eval}{\mathrm{eval}}

\newcommand{\circuit}{\mathcal{C}}

\newcommand{\pcpeval}{\Xi}
\newcommand{\pcpverifier}{\mathcal{M}_\ar}
\newcommand{\qlen}{Q}
\DeclareMathOperator{\ev}{eval}

\newcommand{\enc}{\textrm{enc}}
\newcommand{\dec}{{\rm dec}}
\newcommand{\coded}{\mathrm{Dec}}

\newcommand{\soundness}{\mathrm{sound}}

\newcommand{\rep}{\gamestyle{Rep}}

\newcommand{\Res}{{\rm Res}}
\newcommand{\Ind}{{\rm Ind}}

\newcommand{\binary}[1]{\mathrm{binary}_{#1}}

\newcommand{\verteq}[0]{\begin{turn}{90} $=$\end{turn}}

\renewcommand{\rvar}{\frak{R}}
\renewcommand{\lvar}{\frak{L}}
\newcommand{\canonical}{\mathrm{can}}
\newcommand{\decouple}{\mathsf{Decouple}}
\newcommand{\paddedsuccinctdecider}{\mathsf{SuccinctTOI}}
\newcommand{\TMAnchoring}{\mathsf{Anchor}}
\newcommand{\TMPR}{\mathsf{PartialParRep}}

\newcommand{\PartialAnsRed}{\mathsf{PartialAnsRed}}
\newcommand{\AnsRed}{\mathsf{AnswerReduction}}
\newcommand{\triples}{\mathrm{triples}}

\def\dsize{D}
\def\on{\overline{n}}
\newcommand{\valns}{\val^{\rm non-sync}}
\newcommand{\Entns}{\Ent^{\rm non-sync}}

\begin{document}

\title{The Aldous--Lyons Conjecture II: Undecidability}

\author{Lewis Bowen\\
\url{lpbowen@math.utexas.edu}
\and
Michael Chapman\\
\url{mc9578@nyu.edu}
\and
Thomas Vidick\\
\url{thomas.vidick@weizmann.ac.il}}


\maketitle

\begin{abstract}
 This paper, and its companion \cite{BCLV_subgroup_tests}, are devoted to  a negative resolution of the Aldous--Lyons Conjecture \cite{Aldous_Lyons_Conj,Aldous--Lyons_conj_blogpost}.
 \\

    In this part we study \emph{tailored non-local games}. This is a   subclass of \emph{non-local games} ---  combinatorial objects which model certain experiments in quantum mechanics, as well  as interactive proofs in complexity theory. Our main result is that, given a tailored non-local game $\game$, it is undecidable to distinguish between the case where  $\game$ has a special kind of perfect strategy, and the case where every strategy for $\game$ is far from being perfect. Using a reduction introduced in the companion paper~\cite{BCLV_subgroup_tests}, this undecidability result implies a negative answer to the Aldous--Lyons conjecture. Namely, it implies the existence of unimodular networks that are non-sofic.
    
    To prove our result, we use  a variant of the \emph{compression} technique developed in $\MIP^*=\RE$ \cite{MIPRE}. Our main technical contribution is to adapt this technique to the class of tailored non-local games. The main difficulty is in establishing \emph{answer reduction}, which requires a very careful adaptation of existing techniques in the construction of probabilistically checkable proofs. As a byproduct, we are reproving the negation of Connes' embedding problem \cite{connes1976classification} --- i.e., the existence of a   $\textrm{II}_1$-factor which cannot be embedded in an ultrapower of the hyperfinite $\textrm{II}_1$-factor --- first proved in \cite{MIPRE}, using an arguably more streamlined proof. In particular, we incorporate recent simplifications from the literature~\cite{de_la_Salle_spectral_gap,vidick2022almost} due to de la Salle and the third author.
\end{abstract}
\newpage
\setcounter{tocdepth}{2}
\tableofcontents
\newpage

\section{Introduction}

In Part I \cite{BCLV_subgroup_tests} we proved that if the following theorem is true then there are non-sofic unimodular networks,  resolving the Aldous--Lyons conjecture \cite{Aldous_Lyons_Conj} in the negative: 

\begin{theorem}[Main Theorem. See Theorem \ref{thm:tailored_MIP*=RE} for a formal version. Compare to Theorem 7.4 in \cite{BCLV_subgroup_tests}]\label{thm:main-informal}
  There exists a polynomial time algorithm that takes as input  a Turing machine $\cal{M}$ and outputs a \textbf{tailored} non-local game $\game_\cal{M}$ such that:
  \begin{enumerate}
      \item \emph{Completeness}: If $\cal{M}$ halts then there exists a perfect \textbf{$Z$-aligned permutation strategy that commutes along edges} for $\game_\cal{M}$. 
      \item \emph{Soundness}: If $\cal{M}$ never halts then the \textbf{synchronous} quantum value of $\game_\cal{M}$ is bounded from above by $\nicefrac{1}{2}$.
  \end{enumerate}
  \end{theorem}
The reader who is unfamiliar with the study of non-local games should not be discouraged, as all definitions regarding the above theorem are explained within this introduction. The reader familiar with the work $\MIP^*=\RE$ by Ji--Natarajan--Vidick--Wright--Yuen \cite{MIPRE} notices that the above theorem is very similar to  their main theorem. Actually, it is a {strengthening} of their result, namely, this paper reproves that the class of multi-prover interactive proofs with entangled provers contains the Halting problem, which implies a negative solution to Connes' embedding problem \cite{connes1976classification} (see also \cite[Proposition 6.3.5]{brown2006invariant}) as well as to Tsirelson's problem \cite{Tsi06}. 
In the statement above we emphasized the main differences in bold. Elaborating on these differences: 
\begin{itemize}
    \item The game $\game_\cal{M}$ must  belong to 
  the class of  \emph{tailored non-local games}, which is a strict subclass of the synchronous games  used in  \cite{MIPRE}. Tailored games are a  generalization of an important class of games considered in the literature, called \emph{linear constraint system games} (LCS, see~\cite{cleve2014characterization,kim2018synchronous}).
    \item The provers are only allowed to use \emph{synchronous} \cite{paulsen2016estimating} quantum strategies.
    \item  The allowed perfect strategies for $\game_\cal{M}$ in the complete case, \emph{$Z$-aligned permutation strategies that commute along edges} ($\ZPC$ strategies), is a stricter subfamily of the $\mathsf{PCC}$ (projective, consistent and commuting) strategies used in the complete case in \cite{MIPRE}.
\end{itemize} 

Because the class of games considered is more restricted, and because the class of strategies available to show the completeness property is more limited, Theorem~\ref{thm:tailored_MIP*=RE} is more difficult to show than the corresponding reduction from~\cite{MIPRE}. (The restriction to synchronous strategies in the soundness case does play in our favor; however, as we shall see later, this restriction has relatively mild and well-understood consequences.)

The following few subsections of the introduction recall various notions from the theory of non-local games  and then introduce the class of tailored   games. In the process, tailored games are suggested as a ``middle ground'' between synchronous games (which were used in \cite{MIPRE}) and  linear constraint system games; we specifically address  a folklore effort to ``linearize'' $\MIP^*=\RE$ (cf. \cite{paddock2023satisfiability}) --- which would have resulted in the existence of non-hyperlinear groups, and thus refute the Aldous--Lyons conjecture  --- and offer our approach as ``semi-linearization''. Finally, our proof method is discussed, and in particular the similarities and differences between this work and \cite{MIPRE}.

We do not motivate or survey the Aldous--Lyons conjecture nor Connes' embedding problem; even the complexity theoretic aspects of our strengthened version of $\MIP^*=\RE$ are discussed only briefly. Such motivational introductions are already provided both in our companion paper \cite{BCLV_subgroup_tests}, for the Aldous--Lyons conjecture, and in \cite{MIPRE}, for  $\MIP^*=\RE$ and Connes' embedding problem.
\paragraph{Non-local games.}
A non-local game  consists of two finite sets $X,A$, a probability distribution $\mu$ over $X\times X$, and a decision predicate $D\colon X\times X\times A\times A\to \{0,1\}$. The set $X$ is commonly called the \emph{question set} and $A$ the \emph{answer set}.\footnote{The answer set may  depend on the specific question $\mttx\in X$, namely, when $\mttx$ is asked, the allowed answers are from $A_\mttx\subseteq A$. For simplicity, in the introduction, we ignore such dependence.} The game  is  called \emph{synchronous} if $D(\mttx,\mttx,a,b)=0$ for every  $\mttx\in X$ and $a\neq b\in A$; this condition will always be satisfied for us.

The data $\game=(X,A,\mu,D)$ is called a ``game'' because of the following interpretation. We may imagine a referee challenging two players, colloquially referred to as ``Alice'' and ``Bob'', by sending them a pair $(\mttx,\mtty)$ that was sampled according to the distribution $\mu$, such that Alice receives $\mttx$ and Bob receives $\mtty$. Alice then has to respond with some $a\in A$, while Bob has to respond with some $b\in A$. The players are said to win if and only if $D(\mttx,\mtty,a,b)=1$. 

Given a game $\game$, its \emph{value} is defined as the maximum probability, over the referee's choice of a pair of questions and the players' choice of an answer, that the players win the game. To make this formal, one needs to specify how the players may determine their answers, i.e.,\  to define the class of allowed strategies for them. This is where things get interesting, as there are several natural choices. 
The most restricted choice is to require the players to choose a function $f:X\to A$ and return $a=f(\mttx)$ and $b=f(\mtty)$. 
For any game $\game$, maximizing the players' success probability over all such   functions leads to what is known as the (synchronous) \emph{classical value} $\val(\game)$ of the game. Concretely, 
\[\val(\game)=\max_{f\colon X\to A}\ \Big( \sum_{\mttx,\mtty\in X} \mu(\mttx,\mtty) D(\mttx,\mtty,f(\mttx),f(\mtty))\Big)\;.\]
It is not hard to see that allowing ``randomized'' functions, namely letting the players choose $f$ according to some distribution, does not change the value.

In full generality, a strategy for the players is specified by a \emph{correlation}, which is a family of distributions $p(\cdot,\cdot|\mttx,\mtty)$ on $A\times A$, for every pair $(\mttx,\mtty)\in X\times X$. The restriction considered in the previous paragraph leads to the  family of (synchronous) classical strategies. Let us give two other examples of families of strategies. The first example is known as \emph{synchronous quantum strategies}. To define these, first recall the notion of a \emph{projective valued measure} (PVM). 
A PVM is a collection of operators $\{\cal{P}_a\}_{a\in A}$ acting on a Hilbert space $\mathcal{H}$, where $A$ is any finite set,   the operators $\cal{P}_a$ are orthogonal projections ($\cal{P}_a^*=\cal{P}_a=\cal{P}_a^2$),  and $\sum_a \cal{P}_a=\Id_\cal{H}$. A synchronous quantum strategy is then specified by a \emph{finite-dimensional} Hilbert space $\mathcal{H}$ and PVMs $\{\cal{P}^\mttx_a\}_{a\in A}$ for each $\mttx\in X$.  Such a strategy is said to be \emph{commuting along edges}\footnote{The reason for the name commuting along edges, is that the support of $\mu$ induces a graph structure on $X$, and the condition is indeed that PVMs that are associated with neighboring vertices must commute.} (or just commuting in \cite{MIPRE}), if for every pair of questions $(\mttx,\mtty)$ that can be sampled in the game (namely, in the support of $\mu$), the projections $\cal{P}^\mttx_a$ and $\cal{P}^\mtty_b$ commute for every $a,b\in A$.
The correlation that the  strategy $\cal{P}$ induces is 
\begin{equation}\label{eq:correlation_induced_by_synch_quantum_strat}
    p(a,b|\mttx,\mtty)=\tau(\cal{P}^\mttx_a \cal{P}^\mtty_b)\ ,
\end{equation} 
where $\tau$ is the dimension-normalized trace on $\mathcal{H}$.
The resulting maximum success probability is called the (synchronous) \emph{quantum value} of the game and is denoted by $\val^*(\game)$. Concretely,
\begin{equation}\label{eq:qval-intro}
  \val^*(\game)=\sup_{\mathcal{H}, \{\cal{P}^\mttx_a\}} \ \Big(\sum_{\mttx,\mtty} \mu(\mttx,\mtty) \sum_{a,b}\tau(\cal{P}^\mttx_a \cal{P}^\mtty_b) D(\mttx,\mtty,a,b)\Big)\;.
\end{equation}
In general, $\val(\game)\leq \val^*(\game)$ always holds, and furthermore the inequality can be strict.\footnote{The first to provide  an example with a strict inequality was John Bell in \cite{bell1964einstein}.} 
This is demonstrated, for example, by the \emph{magic square game} described in Example~\ref{example:magic-square}. The fact that $\val(\game)<\val^*(\game)$ is interpreted as a witness of the \emph{non-locality} of quantum mechanics. It has led to experiments (e.g.~\cite{hensen2015loophole}) which verify that the quantum mechanical prediction for $\val^*(\game)$ is indeed achievable using a physical system (such as a pair of photons). Such experiments demonstrate that non-classical aspects of quantum mechanics are necessary to explain the physical world. 

In this paper, we force the perfect strategies in the complete case to be  \emph{$Z$-aligned permutation strategies that commute along edges}, or $\ZPC$ strategies for short. Let us define this subfamily of synchronous quantum strategies. Assume  that the answer set of the game is $A=\FF_2^\Lambda$ for some fixed integer $\Lambda$.\footnote{In general we allow $\Lambda$ to depend on the question $\mttx$.} In a \emph{permutation strategy}, a finite set  $\Omega$  is chosen, and we let $\Omega_\pm=\{\pm\}\times \Omega$ be the signed version of $\Omega$, and $\sigma_\sJ\in \Sym(\Omega_\pm)$ be the sign flip; namely $\sigma_\sJ(\pm,\star)=(\mp,\star)$ for every $\star\in \Omega$.\footnote{We later denote this sign flip by $-\Id$ instead of $\sigma_\sJ$, but for the sake of clarity decided on this different notation in the introduction.} Then, to each $\mttx\in X$,  a family of $\Lambda$ pairwise commuting, involutive permutations that commute with the sign flip $\{\sigma_{\mttx,i}\}_{i=1}^\Lambda\subseteq \Sym(\Omega_\pm)$ are associated --- this is the same as choosing for every vertex $\mttx$, a \emph{signed permutations} representation of $\FF_2^{\Lambda}$ (acting on $\Omega_\pm$).  Using the natural  embedding of permutations acting on $\Omega_\pm$ in the unitary matrices acting on $\complex^{\Omega_\pm}$, and as all the $\sigma_{\mttx,i}$'s commute with the sign flip permutation, it can be checked that 
\begin{equation}\label{eq:strategy_associated_with_perm_intorduction}
\forall \mttx\in X\ ,\ a\in \FF_2^{\Lambda}\ \colon\ \ \cal{P}^\mttx_a=\frac{\Id-\sigma_\sJ}{2}\cdot \prod_{i=1}^\Lambda \Big(\frac{\Id+(-1)^{a_i}\sigma_{x,i}}{2}\Big)
\end{equation}
induces PVMs on the $|\Omega|$-dimensional Hilbert space $\cal{H}=\frac{\Id-\sigma_\sJ}{2}\complex^{\Omega_\pm}$, which is the space of anti-symmetric functions from $\Omega_\pm$ to $\complex$, namely functions satisfying $f(-,\star)=-f(+,\star)$ for every $\star\in \Omega$. These PVMs form a quantum strategy  $\cal{P}$ called the \emph{quantum strategy associated with the permutation strategy} $\sigma$.  The permutation strategy  $\sigma$  is said be \emph{commuting along edges} if the associated $\cal{P}$ is commuting along edges.
The correlation  $p(\cdot,\cdot|\cdot,\cdot)$ induced by the PVMs $\{\cal{P}^\mttx_a\}$, as in \eqref{eq:correlation_induced_by_synch_quantum_strat}, is said to be \emph{induced} by the permutation strategy $\sigma$.
In words, $p(a,b|\mttx,\mtty)$ is the relative dimension in $\cal{H}$ of the joint eigenspace of each $\sigma_{\mttx,i}$ associated with eigenvalue $(-1)^{a_i}$, and of each $\sigma_{\mtty,j}$ associated with eigenvalue $(-1)^{b_j}$. The notion of $\sigma$ being $Z$-aligned can be described only after we  introduce the class of tailored non-local games. 

\paragraph{Tailored games.}

A tailored game is a non-local game that has the following structure. First, the answer set is $\FF_2^{\Lambda_\frR}\times \FF_2^{\Lambda_\frL}$, where $\Lambda_\frR$ and $\Lambda_\frL$ are integers, and let $\Lambda=\Lambda_\frR+\Lambda_\frL$.\footnote{In the formal definition, $\Lambda_\frR$ and $\Lambda_\frL$ may vary depending on the question $\mttx$.}  So, the answer $a=(a^\rvar,a^\lvar)$ to a question $\mttx$ consists of two parts: a \emph{readable} part $a^{\rvar}$ and an \emph{unreadable}  (or \emph{linear}) part $a^{\lvar}$. Furthermore, the decision procedure of a tailored game is required to be \emph{controlled-linear}: Given a pair of questions $(\mttx,\mtty)$ and answers $(a,b)=((a^\rvar,a^\lvar),(b^\rvar,b^\lvar))$, it first reads only the pair $(a^{\rvar},b^{\rvar})$, and depending on it returns a system of linear equations with $\F_2$-coefficients  $L=L_{\mttx\mtty}(a^\frR,b^\frR)$  over $2\Lambda$ variables. Then, the pair $(a,b)\in \FF_2^{2\Lambda}$ is accepted by the decision procedure, namely $D(\mttx,\mtty,a,b)=1$, if and only if   $L$ is satisfied by the assignment $(a,b)$. 

Of course, a tailored game such that the entire answer is marked as readable, i.e. $\Lambda_\rvar=\Lambda$ for all questions $\mttx$, is nothing but a general non-local game. For more restricted choices of $\Lambda_\rvar<\Lambda$ to be useful,  we need to describe the kinds of strategies which we consider for tailored games. A permutation strategy 
\[
\big\{\sigma_{\mttx,\frR,i},\sigma_{\mttx,\frL,j}\mid \mttx\in X\ ,\ i\in [\Lambda_\frR]\ ,\ j\in [\Lambda_\frL]\big\}
\]
for a tailored non-local game, acting on $\Omega_\pm$, is said to be \emph{$Z$-aligned} if the readable permutations act as  controlled sign flips. I.e., for every   $\star\in \Omega$, $i\in [\Lambda_\frR]$ and $\mttx\in X$, the permutation $\sigma_{\mttx,\frR,i}$ maps the set $\{(+,\star), (-,\star)\}$
 to itself.
 This means, in particular, that the readable permutations are mutually diagonalizable in the standard basis of $\cal{H}=\frac{\Id-\sigma_\sJ}{2}\complex^{\Omega_\pm}$, which consists of the functions ${\bf 1}_{(+,\star)}-{\bf 1}_{(-,\star)}$ (with ${\bf 1}_\cdot$ being the indicator function). A $\ZPC$ strategy for a tailored non-local game is a permutation strategy that commutes along edges and is $Z$-aligned, and the $\ZPC$ value of a game will be the maximum success probability of a $\ZPC$ strategy in the game. \textbf{The reader can now parse our main theorem}.

One can now see why the tailoring of $\game$ may  affect the value of a game if we restrict it to use only $\ZPC$ strategies: the more answer bits are marked as readable, the more restricted the class of strategies that is allowed; thus an ``aggressive'' tailoring (e.g.\ marking all answer bits as readable) may lead to a smaller $\ZPC$-value, while a more ``relaxed'' tailoring of the same game would have higher $\ZPC$-value. In fact, one can easily verify that, for any game $\game$ such that $\Lambda_\rvar=\Lambda$, the $\ZPC$-value agrees with the classical one. Naturally, ``fully relaxed'' tailoring of a given game (e.g.\ marking all answer bits as unreadable) is not always possible, because the decision function $D$ may simply not be linear. But, when such a relaxation is possible, the resulting game is said to be a \emph{linear constraint system game} (LCS,~\cite{cleve2014characterization,kim2018synchronous}). 
So, tailored games are a natural generalization of LCS games. LCS games  are widely studied, and their  values are  related to   approximation properties --- such as hyperlinearity and soficity --- of a certain finitely presented group associated with the LCS game. Let us say more about this subclass.

\paragraph{Linear constraint system games.}

 LCS games are a restricted class of non-local games such that the function $D(\mttx,\mtty,a,b)$ is a conjunction of \emph{linear} functions of its input $(a,b)$, seen as an element of $\F_2^{2\Lambda}$. Namely, for every $\mttx,\mtty$ that can be sampled by $\mu$, there is a system of linear equations $\mathscr{A}_{\mttx\mtty}\vec x=\vec c_{\mttx\mtty}$ with $\FF_2$-coefficients and with $\vec x\in \FF_2^{2\Lambda}$, and $D(\mttx,\mtty,a,b)=1$ if and only if $\vec x=(a,b)$ is a solution to this system of equations.\footnote{The formal definition of an LCS is slightly more restricted, see Example~\ref{example:LCSs}, but this generalized setup is essentially equivalent to the standard definition.}

A natural $C^*$-algebra $\mathcal{A}(\game)$, known as the \emph{game algebra}, can be associated to every synchronous game. In case $\game$ is an LCS, $\mathcal{A}(\game)$ happens to be a {group} von Neumann algebra. Namely, there is a finitely presented group $\Gamma(\game)$, often referred to as the \emph{solution group} (cf. \cite{slofstra2019tsirelson}), such that $\mathcal{A}(\game)$ is the von Neumann closure of (a quotient of) the group ring $\complex[\Gamma(\game)]$. 

There is an additional game value $\val_{qc}$, known as the (synchronous) \emph{quantum commuting value}, defined by taking the supremum as in~\eqref{eq:qval-intro} over all tracial von Neumann algebras $(\mathcal{M},\tau)$ (instead of only finite dimensional ones). Using known connections between the existence of perfect strategies and $*$-homomorphisms of $\mathcal{A}(\game)$~\cite{kim2018synchronous}, it is a folklore result that the existence of an LCS game such that $\val_{qc}(\game)=1>\val^*(\game)$
 implies the existence of a non-hyperlinear group, which is thus  non-sofic, and in turn refutes the Aldous--Lyons conjecture. 
 
In $\MIP^*=\RE$~\cite{MIPRE}, synchronous games that satisfy  
 $\val_{qc}(\game)=1>\val^*(\game)$
 are constructed.
 Unfortunately, these games are not LCS. Moreover, it seems essential for some of the key steps of the construction from~\cite{MIPRE} that the game decision function $D$ is allowed to depend non-linearly on the answers $a,b$ --- this is due to the use of techniques from the field of efficient proof verification in computer science; we describe this obstacle in more detail when discussing \emph{answer reduction} in Section~\ref{sec:intro-proof}. In turn, the results of \cite{paddock2023satisfiability} demonstrate that implementing the non-linear OR function cannot be done in a ``naive'' way using LCS only.

 Now, tailored non-local games \emph{are} allowed to have decision functions that depend non-linearly on the answers --- at least, on the readable part of the answers. Crucially however, the form of strategies which we consider is required to be more limited, as a function of the tailoring. Thus, tailored non-local games are a broader class of games than LCS, but ones with a restricted class of strategies. The combination of these two ingredients allow us to carry through the proof approach from~\cite{MIPRE} (because our class of games is sufficiently general) while, to some extent, maintaining the connection with group theory (through the reduction to subgroup tests proved in the companion paper \cite{BCLV_subgroup_tests}). However, we are not able to determine whether there exists a non-sofic group; that remains an open problem.

\paragraph{The complexity theoretic angle.}
Theorem~\ref{thm:main-informal} is formulated as a reduction from the problem of deciding if a Turing machine $\cM$ halts to the problem of deciding if a game $\game_\cM$, that is polynomial-time computable from the description of $\cM$, has $\ZPC$ value $1$ or synchronous quantum value at most $\frac12$. The existence of such a reduction can be reformulated succinctly as the equality of two complexity classes. 

Let $\RE$ be the class of problems that are polynomial-time reducible to the Halting Problem. Here, $\RE$ stands for ``recursively enumerable.'' An equivalent definition of $\RE$ is that it consists of all problems such that there is an algorithm which, given an instance of the problem, always terminates with the answer ``yes'' when indeed the answer should be yes; when the answer should be no, the algorithm can either say ``no'', or it is also allowed to never terminate.\footnote{We provide an  overview of complexity classes in Section \ref{sec:prelude_decision_problems_PCPs}, which includes a formal definition of $\RE$ as well.} Turing showed that the Halting Problem is a complete problem for this class. 

Let $\TMIP^*$ be the class of languages that are polynomial-time reducible to the problem of deciding if the $\ZPC$ value of a tailored game provided as input is $1$, or if its synchronous value is at most $\frac12$ (given that one of these is promised to be the case).\footnote{To make this definition precise, one needs to clarify how a tailored game is represented; this is discussed in Section~\ref{sec:normal-form}.} Then Theorem \ref{thm:main-informal} can be formulated succinctly as
  \[ {\TMIP^*}={\RE}\;.\]
Reformulated in this way, our result bears a clear analogy with the result $\MIP^*=\RE$. It is also clear that it is a strengthening of the latter, as $\TMIP^* \subseteq \MIP^*$ (and the inclusion $\MIP^*\subseteq \RE$ is not hard; it is the reverse inclusion that requires work). Such characterization inscribes itself in a long tradition of complexity theory, where equalities such as $\IP=\PSPACE$~\cite{lund1990algebraic,shamir1990ip} or $\MIP=\NEXP$~\cite{BFL91} are taken as fundamental statements about the nature of computation, which tend to have important consequences in areas ranging from cryptography to hardness of approximation. In a different direction, extending the class of strategies allowed for the provers has led to analogues of $\MIP^*=\RE$ for higher classes of the arithmetical hierarchy~\cite{mousavi2022nonlocal}.

\subsection{Proof ideas}
\label{sec:intro-proof}
While our proof follows the same template as~\cite{MIPRE}, and indeed re-uses the most important ideas therein, it is arguably more streamlined. In particular we are able to take advantage of some simplifications that were discovered after the publication of~\cite{MIPRE}. Most notably, we take advantage of the fact that synchronous games can without loss of generality be analyzed by considering their synchronous value only \cite{kim2018synchronous,vidick2022almost},\footnote{This simplification is already taken into account in the expression~\eqref{eq:qval-intro}, which technically represents the synchronous value.} and the simplification of~\cite{de_la_Salle_spectral_gap} for the step of question reduction (further discussed below). 

At the heart of our work  is a result about \emph{compression} of non-local games. Informally, compression reduces the size of a game (measured by the number of questions and answers) while preserving its quantum value. The fact that a form of compression implies undecidability as in Theorem~\ref{thm:main-informal} is very general, as shown in~\cite{marks24recursive}. For this to be possible, of course, one must introduce certain computational considerations; in particular the procedure which achieves compression must be computable.
For clarity of this introduction,  we, for the most part,  set computational aspects aside, and focus on compression as a \emph{combinatorial} transformation. In this respect, the following is what needs to be done. 

\paragraph{Compression.}
Let $N=2^n$ for some integer $n$. Our starting point is a tailored game $\game$, that has questions and answers of length $N$, i.e.\ the sets $X,A$ each have cardinality $2^N$. The goal of compression is to construct a new tailored game $\frak{Compr}(\game)$ with the following properties:
\begin{enumerate}
  \item Questions and answers in $\frak{Compr}(\game)$ have length $\poly(n)$.\footnote{We use the $O$ and $\poly$ notations although we have not yet specified the asymptotics. For now, it can be assumed that there is a universal constant $C$ such that $\poly(n)$ is bounded from above by $Cn^C$ and $O(N)$ is bounded by $CN$ (cf.\ Remark \ref{rem:asymptotic_notation}) This is a bit misleading, as the length of the encoding of $\game$ plays a role as well, but we are trying to postpone complexity theoretic considerations for now. Note that this guarantees a genuine compression only for large enough values of $n$, which is enough for the undecidability result to hold.} Namely, an exponential reduction in the length of questions and answers.
  \item $\frak{Compr}(\game)$ \emph{simulates} $\game$, as follows: 
  \begin{enumerate}
    \item \emph{Completeness}: If there exists a perfect $\ZPC$ strategy for $\game$, then there is also such a strategy for $\frak{Compr}(\game)$. 
    \item \emph{Soundness}: If $\val^*(\game)\leq \frac12$, then $\val^*(\frak{Compr}(\game))\leq \frac12$. 
  \end{enumerate}
\end{enumerate}
Compression is composed of three main steps:
\begin{enumerate}
  \item In the first step the length of questions is reduced through a technique referred to as ``introspection'': informally, each player is instructed to generate its own question by itself; shorter questions are used to enforce that the player samples according to the right question distribution $\mu$. This step produces a game $\game'=\frak{QueRed}(\game)$, whose questions have length $\poly\log(N)=\poly(n)$ and answers have length $O(N)$.
  \item In the second step the length of the answers is reduced. This is achieved using techniques from probabilistic proof checking. Loosely speaking, the players encode their answers in $\game'$ using an error-correcting code that allows probabilistic checking of computational statements (such as ``this answer is a valid answer to that question'') by reading only a small number of bits of the encoding --- which constitute the player's new answer. This step results in a game $\game''=\frak{AnsRed}(\game')$ whose questions and answers have length $\poly\log(N)=\poly(n)$.\footnote{As opposed to the previous step, this step depends heavily on the  way $\game$ is encoded. More specifically, $\game$ needs to be encoded \textbf{succinctly}. In our case, we encode an infinite family of games in a uniform manner, and compress them all at once, which means in particular that the games are succinctly encoded as needed.}
  \item The combination of the two preceding transformations does not quite satisfy item 2(b) above. Instead, whenever $\val^*(\game)\leq \frac12$ we only have $\val^*(\game'')\leq 1-\nicefrac{1}{\poly(n)}$. To remedy this, the game is repeated in parallel $\poly(n)$ times to yield $\game'''=\frak{ParRep}(\game'')$, which still has $\poly(n)$-question and answer length, and moreover satisfies item 2(b). 
\end{enumerate}
 All in all, 
  \[
\frak{Compr}(\game)=\game'''=\frak{ParRep}(\frak{AnsRed}(\frak{QueRed}(\game)))\ .
  \]
Each of the three transformations satisfies item 2(a), and so at the end both 2(a) and 2(b) are satisfied. We now discuss each step in more detail. 

\paragraph{Question reduction.}

The introspection technique goes back to the work of Natarajan and Wright~\cite{NW19}. Intuitively, the idea is to ``force the players to sample their own questions''. Let $\mu$ be the question distribution in $\game$. In the game $\frak{QueRed}(\game)$, there is a special pair of questions $(\Intro_A,\Intro_B)$ such that answers to this pair of questions are expected to take the form $((\mttx,a),(\mtty,b))$ (each answer thus has length $2N$). We would like that three conditions hold. Firstly, it should be that, whenever this pair of questions is asked, the marginal distribution of the players' answers on $(\mttx,\mtty)$ is exactly $\mu$. Secondly, it should be that the $a$  part of the answer is determined using only  the question $\Intro_A$  and $\mttx$ part of the answer, namely without ``peeking'' into the other players' question $\mtty$ (and similarly for $b$ with all roles reversed).  Finally, it should be that $(a,b)$ are valid answers to $(\mttx,\mtty)$ in $\game$. 

The last condition is easy to verify, as the referee in $\frak{QueRed}(\game)$ can check it by themselves. The first two conditions require work. In particular, one may not expect to enforce a condition on the \emph{distribution} of an answer from a test that depends on that answer only; as this could only restrain the support of the answer, but not its distribution. To achieve the first requirement one must thus consider a more complicated test that involves additional questions in the game. The method for forcing the distribution also allows  to limit peeking, by leveraging the Heisenberg uncertainty principle --- which states that information stored in a quantum state can be destroyed by performing a measurement in the complementary basis. Let us describe now the underlying ideas. 

The main tool for forcing $(\mttx,\mtty)$ to be distributed according to a specific distribution $\mu$, is to  verify that the PVMs associated with the bits of $\mttx$ and $\mtty$  come from (the Fourier transform of) a non-commutative representation of the Pauli group. To explain this a little more, let us focus on the case where $\mu$ is uniform on pairs $(\mttx,\mtty)\in\F_2^N\times\F_2^N$. The technique we describe generalizes to more complex, although far from arbitrary, distributions --- it is known to apply to the class of \emph{conditionally linear} distributions introduced in~\cite{MIPRE} (see Section \ref{sec:CLMs}). 

Let $\PXm=\begin{pmatrix} 0 & 1 \\ 1 & 0 \end{pmatrix}$ and $\PZm=\begin{pmatrix} 1 & 0 \\ 0 & -1 \end{pmatrix}$. These are commonly referred to as the $\PXm$ and $\PZm$ Pauli matrices.
The Pauli group acting on $k$ qubits, sometimes called the Weyl--Heisenberg group or the $k$-dimensional Heisenberg group over $\FF_2$, is the subgroup of unitaries acting on $(\complex^{2})^{\otimes k}$ generated by length $k$ Kronecker tensor products of $\PXm$ and $\PZm$ Pauli matrices.  Namely, if for every $\alpha,\beta\in \FF_2^k$ we define $\PXm^{\otimes \alpha}=\bigotimes_{i=1}^k\PXm^{\alpha_i}$ and $\PZm^{\otimes \beta}=\bigotimes_{j=1}^k\PZm^{\otimes \beta}$, then $\WH_k=\{\pm \PXm^{\otimes\alpha}\PZm^{\otimes \beta}\mid \alpha,\beta\in \FF_2^k\}$.
What is relevant for us is that this group contains two copies of $\FF_2^k$ as subgroups, the $\PXm$-subgroup $\{\PXm^{\otimes \alpha}\mid \alpha\in \FF_2^k\}$ and the $\PZm$-subgroup $\{\PZm^{\otimes \beta}\mid \beta\in \FF_2^k\}$, and that the above representation is its unique non-abelian irreducible representation.
By taking the Fourier transform of the $\PZm$-subgroup, we get a PVM $\{\mathscr{F}^\PZm_a\}_{a\in \FF_2^k}$, and measuring according to it provides  a string $a$ of length $k$ which, following~\eqref{eq:correlation_induced_by_synch_quantum_strat}, is uniformly distributed. So, by choosing $k=2N$, if we are able to force the provers to use a non-commutative representation of $\WH_{2N}$, we are able to force them to sample a uniform $(\mttx,\mtty)\in \FF_2^N\times \FF_2^N$ as required.

A key property that is used to force the provers to measure according to the {representation} of $\WH_k$ described above  is that there exists a specific {presentation} of it (with generators and relations) that is very \emph{stable} (cf.~\cite{GowersHatami,HadwinShulman,CL_part1,GlebskyRivera}). The  chosen presentation is due to de la Salle \cite{de_la_Salle_spectral_gap}, who showed that in our setup the stability result is quite easily deduced by combining a  spectral gap argument with a technique that translates anti-commutation to commutation due to Natarajan--Vidick \cite{natarajan2018low}. Another important property of $\WH_k$ is that $\PXm^{\otimes \alpha}$ and $\PZm^{\otimes \beta}$ anti-commute whenever $\langle\alpha,\beta\rangle=1$; this is commonly referred to as mutual non-measurability, or the Heisenberg uncertainty principle. This property is used to guarantee the ``no peeking'' requirement mentioned above, by comparing the answers of the players in a clever way to certain $\PXm$-measurement outcomes.

All in all, the method we use for question reduction is a combination of~\cite{de_la_Salle_spectral_gap} and~\cite{MIPRE}. The main new observation is that the irreducible representation introduced above can be described as a permutation strategy, which is essential to showing that the introspected game has a perfect $\ZPC$ strategy (when the original game does). 


\paragraph{Answer reduction.}

To reduce the length of answers in the game we use techniques from the area of probabilistically checkable proofs (PCPs) in computer science. At a high level, the idea is that instead of directly providing an answer that the referee checks, the player will first encode its answer in a suitable error-correcting code. The referee then requests a small number of bits from the encoded answer, and this will suffice for him to verify that these symbols are taken from a well-encoded answer (or one that is sufficiently close to such) that would have satisfied the original checks. 

This step crucially relies on complexity theoretic  assumptions on the way the original game $\game$ was encoded. Specifically, it should be possible to represent the decision procedure from $\game$ using a circuit of size $\poly(n)$ --- this is possible, in our case, due to the scaled up Cook--Levin theorem together with the fact that the decision procedure runs in exponential time. 
The reason why translating the decision procedure to a circuit is crucial, is that the provers are not asked to encode their original answer, but instead to encode an assignment to all wires in the verifier's verification circuit,\footnote{Circuits are a standard model of computation, alongside Turing machines. In this paper we need very little about them; whatever we need is recalled in detail in Section~\ref{sec:circuits_and_S3SAT}.} with the question inputs  $(\mttx,\mtty)$ being hard coded. The assignment to all the wires must include the answers $(a,b)$ themselves, but also a lot of additional information that is relevant for verifying that the answer would have satisfied the original checks, without actually reading the entire answer and performing the entire original computation. 

Because of this new answer, the encoded assignment to all wires in the verification circuit depends on \emph{both} questions $\mttx,\mtty$ and both answers $a,b$, and only a player that has access to this entire information may compute it. To make this possible, before answer reduction is performed the game $\game'$ is \emph{oracularized}. The referee in the oracularized game sends $(\mttx,\mtty)$ to one player and $\mttx$ or $\mtty$ to the other. It checks the first player's answers according to the game's decision function, and the second player's answer for consistency with the first.\footnote{It is to guarantee that the oracularized game has a perfect strategy whenever the original game does that we need to restrict to strategies that are commuting along edges.}

Because the verification circuit is in general non-linear, some of the bits of the assignment described in the previous paragraph are obtained as e.g.\ the ``AND'' of some of the bits of the original answers. This creates a difficulty: if two answer bits $a_i$ and $a_j$ are computed by permutations $\sigma_i$, $\sigma_j$ in a perfect strategy for the original game (in the sense of the associated quantum strategy~\eqref{eq:strategy_associated_with_perm_intorduction} and the measurement rule~\eqref{eq:correlation_induced_by_synch_quantum_strat}), the bit $a_i \wedge a_j$ may not have a permutation that measures to it. Indeed the natural way to define a permutation that ``computes'' the AND bit is to  take the minus of the projection on the joint $(-1,-1)$ eigenspace of $\sigma_i$ and $\sigma_j$, plus the projection on all other eigenspaces;\footnote{Here we use the association ${\rm True}\mapsto-1=(-1)^{1}$ and ${\rm False}\mapsto 1=(-1)^0$.} while this operation is an involution it is easy to see that it may not be a permutation ---   cf.\ \eqref{eq:example_of_permutations_with_non_perm_AND}.

A possible attempt to overcome this obstacle would be to require the decision function to be linear, as the parity of two answer bits computed by commuting permutations $\sigma_i$ and $\sigma_j$ is naturally computed by the permutation $\sigma_i\sigma_j$. Unfortunately, there are known obstacles to implementing answer reduction using error-correcting codes in a way that requires only a linear decision function. In particular, it is well-known in the classical literature on probabilistic proof checking that linear constraints can only lead to probabilistic checkers that do not have perfect completeness, i.e.\ even in the ideal case one has to abandon the requirement that the value of the game equals $1$ --- intuitively this is  because linear systems of equations can be solved in polynomial time (using Gaussian elimination);  so, deciding if there is a perfect strategy (in the classical model) for a linear verifier can be done efficiently; hence,  there is no advantage to considering a nonlocal game in the first place. One may hope that the situation changes when one considers quantum  (or permutation)  strategies. However, even in that case there are strong obstacles to performing answer reduction  in a liniar manner. In particular, it was shown in~\cite{paddock2023satisfiability} that, in general, games that involve an AND verification predicate cannot be embedded in LCS games; this is shown using the fact that the algebra of an LCS game has a more structured collection of representations than a general game algebra---we refer to~\cite{paddock2023satisfiability} for further discussion.  

It is necessary to overcome this issue, and the generalized setup of tailored games (compared to  LCS games) enables us to resolve it. Loosely speaking, in a tailored game, the checks performed on the unreadable part of the answer are linear, and hence amenable to linear checking. The checks performed on the readable part require the full power of non-linear proof checking techniques; but because perfect $\ZPC$ strategies are required to be $Z$-aligned, it is possible to define a permutation that computes the AND of two (or more) $Z$-aligned permutations --- this fact is captured in Corollary~\ref{cor:encodings}.

To perform answer reduction while preserving the category of tailored games we need to design a bespoke probabilistic proof checker (PCP). This is one of the contributions of our paper. Indeed at first it is not obvious that the computation performed by a tailored verifier can be encoded and verified in a way that preserves its structure. In particular, any bit of the encoded answer that depends on an unreadable bit of the original answer must do so in a linear manner only; that is, the bit cannot be multiplied by any other unreadable bit. We carefully design the required PCP using standard techniques in probabilistic proof checking, including the use of multivariate polynomials and the Reed--Muller code; combined with observations specific to the linear case from~\cite{ben2004robust}.

\paragraph{Parallel repetition}

The two preceding transformations have successfully reduced the size of the game; however, the soundness parameter has degraded. To restore this we perform parallel repetition. This consists in executing $k$ instances of the game in parallel, and accepting only if all tuples of answers are valid with respect to the corresponding tuple of questions. Similar to~\cite{MIPRE}, we apply ``anchored'' repetition, which is known (by \cite{bavarian2017hardness}) to reduce the game value at an exponential rate whenever it was initially strictly smaller than $1$. (Here there is a small subtlety, as we are in the synchronous soundness setup, and parallel repetition assumes a stronger soundness assumption, but this is resolved using the results from \cite{vidick2022almost}.)  It is not hard to verify that this transformation preserves perfect completeness with $\ZPC$ strategies and we do not describe it any further here.

\subsection{Organization of the paper}
In Section \ref{sec:tailoredmip*=re}, we provide minimal  preliminaries so to be able to formulate our main theorem    $\TMIP^*=\RE$ (Theorem \ref{thm:tailored_MIP*=RE}) and the Compression theorem (Theorem \ref{thm:compression}), and to deduce the former assuming the latter.  In Section \ref{sec:compress_toolbox} we proivde a wide range of preliminaries which are needed for the proof of Compression, some are quite standard and some are very particular to this paper.  The next three sections are devoted to the three transformations, described in this introduction, which are the components of the Compression transformation: Section \ref{sec:quered} describes the Question Reduction transformation, Section \ref{sec:answer_reduction} describes the Answer Reduction transformation, and Section \ref{sec:parallel_rep} describes the Parallel Repetition transformation. Lastly, in Section \ref{sec:proof_of_compression} we prove  Compression  by composing these three transformations.

\subsection{Notations, naming conventions, and some general remarks}\label{sec:notations}

\begin{remark}[Asymptotic notation]\label{rem:asymptotic_notation}
    We often use the asymptotic notation  $O,\Theta,\Omega$; namely, for two functions $f,g\colon \mathbb{N}\to\mathbb{N}$,  $f(n)=O(g(n))$ if there is a universal constant $C>0$ such that $f(n)\leq C\cdot g(n)$ for every $n\in \mathbb{N}$ (similarly, $f(n)=\Omega(g(n))$ if $f(n)\geq C\cdot g(n)$, and $\Theta$ is the combination of the two). In addition, we use the following somewhat less conventional notation. For positive integers $a,b,c$ and so on, that we treat as growing to infinity, we write $\poly(a,b,c,...)$ to denote a function bounded by $C (a^C + b^C + c^C+...)$ for some universal constant $C \geq 1$. For non-negative real numbers $0 \leq \alpha,\beta,\gamma,... < 1$, that we treat as going to $0$, we write $\poly(\alpha,\beta,\gamma,...)$ to denote a function bounded by $C(\alpha^{1/C} + \beta^{1/C} + \gamma^{1/C}+...)$ for some universal constant $C \geq 1$. In either case, the universal constant $C$ can vary each time the $\poly(\cdot)$ notation is used. We write $\polylog(a,b,c,...)$ for $\poly(\log a,\log b,\log c,...)$ and $\exp(a,b,c,...)$ for $2^{\poly(a,b,c,...)}$. Finally, we may use $O_h,\Theta_h,\Omega_h,\poly_h,\polylog_h,\exp_h$ and so on, which means that the constant $C$ involved in the bound is some function of $h$.
\end{remark}

\begin{remark}[Additional conventions throughout the paper]
\ 
    \begin{itemize}[noitemsep,topsep=0pt,parsep=0pt,partopsep=0pt] 
    \item Questions in our games  (namely, vertices in the underlying graphs of the games)  are denoted using the typewriter style (mathtt):  
$\mathtt{x},\mathtt{y},\mathtt{Intro},\Hide{}, \mathtt{var},\mathtt{row}$
    and so on. 
    \item Games, and combinatorial transformations on games, are denoted using the gothic style  (mathfrak): $\game$, $\mathfrak{PauliBasis}$,  $\mathfrak{M}$, $\mathfrak{Lcs}$, $\frak{QueRed}$
    and so on. 
    \item Formal variables that control the bits of the player's answers in our games are denoted using the sans serif style (mathsf): $\sX$, $\sY$, $\sAns$, $\sQue$, $\mathsf{Var}$, $\mathsf{ReadQue}$, 
    and so on.
    This style is also used for certain transformations applied by Turing machines (algorithms) such as $\Compress,\decouple$ and so on. It is also used for certain acronyms such as $\ZPC$ and $\TMIP^*$.
    \item PVMs and observables are denoted using the caligraphic style (mathcal): $\cal{P},\cal{Q},\cal{U}$. Some Turing machines also use this style, usually with the letter $\cal{M}$, as well as the components of a tailored normal form verifier $\verifier$ which are $\sampler,\length,\linproc$ and $\decider$.
    \item 
$\mathds{X,Z}$ are the Pauli matrices.
\item  For a positive integer $n$, $[n]$  is the set $\{1,...,n\}$.
    \item We often use
$\sigma$ for permutations, and for permutation strategies. The elements in the sets $\Omega$ on which our permutations act are usually denoted by $\star$ and $\diamond$. The elements of the signed set $\Omega_\pm$ are often denoted by $\spadesuit,\diamondsuit$, by which we mean $\spadesuit$ is $+\star \  \textrm{or}\  -\star$ for some $\star\in \Omega$. 
\item We use $\cdot$ (and less frequently $\star$ and $\circ$) as an input which is not specified. It should be understood from context what are the possible inputs for $\cdot$ (respectively $\star,\circ$). Sometimes this notation actually means ``for all possible inputs'' in this position, and again this should be understood from context.
\item We use $|\cdot|$ to denote the length of a word in some finite alphabet $\Sigma$. Usually, this word is over bits $\FF_2=\{0,1\}$, but occasionally, it is over larger alphabets. The set $\Sigma^*$ is the free monoid over the alphabet $\Sigma$, namely all strings with letters from $\Sigma$. Namely, $\{0,1\}^*$ it is the set of all bit strings with the concatenation $*$ of words as the product. Similar to other products, we often write $ww'$ instead of $w*w'$ for the concatenation of two words $w,w'$.
\item The sign $\eps$ is used in various contexts in the text: as a small positive real number; as an $\FF_2$-exponent of order-$2$ elements in a group; or as the empty word in the mononid $\{0,1\}^*$. These use cases should be understood from context.
\item We usually write $\Id$ for the identity element $\Id_\Gamma$ of a group $\Gamma$. The specific group should be understood from context. Most commonly, $\Id$ is used instead of $\Id_k$ for the $k\times k$ matrix, where $k$ should be understood from context.
\item Bits with $2$-modular arithmetic $\{0,1\}$ and the field with two elements $\FF_2$  are used interchangeably throughout the paper. Given a vector space $\{0,1\}^S$ where $S$ is a finite set, we use ${\bf 1}_\sX\colon S\to \{0,1\}$ for the indicator function of $\sX\in S$, and  $\langle u,v\rangle=\sum_{\sX\in S} u(\sX)v(\sX)$ for the \emph{standard bilinear form} on it. If $S=[k]$, we often denote by $e_i$ the indicator of $i\in [k]$, instead of ${\bf 1}_i$, and $\{e_1,...,e_k\}$ is commonly called the \emph{standard basis}. Given an (ordered) set $S$, we commonly think of $u\colon S\to \FF_2$ both as a \textbf{function} and as a \textbf{string of bits} parameterized by the set $S$. 
\end{itemize}
\end{remark}

\subsection*{Acknowledgements}
We would like to thank Mikael de la Salle for his ongoing effort to simplify the proof of  $\MIP^*=\RE$ \cite{MIPRE}. The formalism he developed, as well as  various simplifications he suggested, made writing this paper much simpler, and hopefully clearer to the readers. 
We would also like to thank Alon Dogon for reading an early version of this paper and suggesting many useful improvements to the presentation.

Lewis Bowen is supported by NSF grant DMS-2154680.
Michael Chapman acknowledges with gratitude the Simons Society of Fellows and is supported by a grant from the Simons Foundation (N. 965535).
Thomas Vidick is supported by AFOSR Grant No. FA9550-22-1-0391 and
ERC Consolidator Grant VerNisQDevS (101086733).

\section{Tailored games and deducing $\TMIP^*=\RE$ from Compression}\label{sec:tailoredmip*=re}
The goal of this section is to provide the minimal preliminaries so that our main theorem (Theorem \ref{thm:tailored_MIP*=RE}) can be rigorously formulated, and then show how Compression of tailored normal form verifiers (Theorem \ref{thm:compression}) implies it. The rest of the paper is devoted to the proof of Compression.

Throughout this section we use $\{0,1\}$ and $\FF_2$ interchangeably to describe the field with two elements, namely bits with $2$-modular arithmetic. Given a vector space $\{0,1\}^S$ where $S$ is a finite set, let ${\bf 1}_\sX\colon S\to \{0,1\}$ be the indicator function of $\sX\in S$, and let $\langle u,v\rangle=\sum_{\sX\in S} u(\sX)v(\sX)$ be the \emph{standard bilinear form} on it (referred to also as the \emph{dot product} of $u$ and $v$). If $S=[k]$, we often denote by $e_i$ the indicator of $i\in [k]$, instead of ${\bf 1}_i$. Given an (ordered) set $S$, we commonly think of $u\colon S\to \FF_2$ both as a \textbf{function} and as a \textbf{string of bits} parameterized by the set $S$. 

\subsection{Measurements}
The concept of \emph{measurement} plays a key role in quantum mechanics. Measurements are modeled using \emph{positive operator valued measures}, which can be viewed as non-commutative counterparts of standard probability measures.  A  finite probability measure is, on the one hand, just a tuple of non-negative real numbers that adds up to $1$, and on the other hand  a \emph{sampling scheme} with finitely many results. In a similar way:

\begin{definition}[POVMs and PVMs]\label{defn:PVM}
	A \emph{positive operator valued measure} (POVM) of dimension $n$ with outcomes in a (finite) set $A$  is a mapping $\cal{P}\colon A\to M_{n\times n}(\complex)$ such that  for every $a\in A$, $\cal{P}_a$ is a positive matrix --- i.e., $\cal{P}_a=C^*C$ for some matrix $C$, where $*$ is the conjugate transpose operation ---  and $\sum_{a\in A}\cal{P}_a=\Id_n$, where $\Id_n$ is the $n\times n$ identity matrix. It is called a projective valued measure (PVM) if in addition $\cal{P}_a$ is an orthogonal projection for every $a\in A$, namely $(\cal{P}_a)^2=\cal{P}_a=(\cal{P}_a)^*$. 
	
	As its name suggests, every POVM $\cal{P}$ defines a probability distribution over its outcome set $A$ as follows:\footnote{The reader who is familiar with quantum measurements may notice that this is not the most general setup of finite dimensional measurements, as the normalized trace occurs when measuring a system in  a specific mixed state. In the vast majority of this paper, this special case is all we need. But, for the soundness analysis of the parallel repetition theorem, we need the more general theory, which is discussed in Section \ref{sec:non-synch_setup}.}
	\begin{equation}\label{eq:sampling_according_POVM}
	\Pro{}[a\ {\rm is\ sampled}]:=\tau(\cal{P}_a)\ ,
	\end{equation}
	where $\tau=\frac{1}{n}\tr$ is the dimension normalized trace on $n\times n$ matrices. Such an answer is said to be \emph{sampled according} to $\cal{P}$ and we denote it by $a\sim \cal{P}$.
\end{definition}
This definition alone seems like a complicated way of generating probability distributions on finitely many outcomes. The following definition is what makes this model interesting:
\begin{definition}[Joint measurements]\label{defn:joint_measurement}
	Given two POVMs $\cal{P}^\mttx$ and $\cal{P}^\mtty$ of the same dimension $n$, where $\cal{P}^\mttx$ is with outcomes in $A$ and $\cal{P}^\mtty$ is with outcomes in $B$, we define their \emph{joint measurement} to be the following $n$-dimensional POVM with outcomes in $A\times B$: $$\cal{P}^{\mttx\mtty}_{a,b}=(\cal{P}^\mttx_a)^{\nicefrac{1}{2}}\cal{P}^\mtty_b(\cal{P}^\mttx_a)^{\nicefrac{1}{2}}\ ,$$
	where $A^{\nicefrac{1}{2}}$ is a well defined (positive) matrix given $A$ is a positive matrix.
	The joint measurement of $\cal{P}^\mttx$ and $\cal{P}^\mtty$ defines a probability distribution over $\gamma=(a,b)\in A\times B$, which we refer to as their \emph{joint sampling}:
	\begin{equation}\label{eq:joint_sampling_according_PVM}
	\Pro{}[a,b\ {\rm are\ sampled}]:=\tau(\cal{P}^{\mttx\mtty}_{a,b})=\tau(\cal{P}^\mttx_a\cal{P}^\mtty_b)\ .
	\end{equation}
\end{definition}

\begin{remark}\label{rem:procedural_joint_sampling}
	In case $\cal{P}^\mttx$ and $\cal{P}^\mtty$ are projective measurements, namely PVMs, there is a ``procedural'' viewpoint of jointly sampling according to them. Let $\mathscr{B}$ be an orthonormal basis of eigenvectors for all of the matrices $\{\cal{P}^\mttx_a\}_{a\in A}$ and $\mathscr{C}$ an orthonormal basis of eigenvectors for $\{\cal{P}^\mtty_b\}_{b\in B}$. Sample $\vec v\in \mathscr{B}$ uniformly at random. Sample $\vec w\in \mathscr{C}$ with probability $|{\langle \vec v|\vec w\rangle}|^2$, where $\langle \cdot|\cdot\rangle$ is the standard inner product on $\complex^n$. As $\cal{P}^\mttx$ and $\cal{P}^\mtty$ are PVMs, there is only one $a\in A$ and $b\in B$ such that $\vec v\in \Img(\cal{P}^\mttx_a)$ and $\vec w\in \Img(\cal{P}^\mtty_b)$. Output $(a,b)$.
\end{remark}
In case $A=\FF_2^S$ for some finite set $S$, there is a close connection between unitary representations of (the group) $\FF_2^S$ and PVMs with outcomes in $\FF_2^S$. As the images of a unitary representation of $\FF_2^S$ are commuting, they have  mutual eigenspaces, and there is an algebraic way of extracting the orthogonal projections onto them.

\begin{definition}[The Fourier transform of a representation]\label{defn:Fourier_transform_reps}
	Let $\cal{U}\colon \FF_2^S\to U(n)$ be a unitary representation. The \emph{Fourier transform} of $\cal{U}$ is a  PVM $\cal{P}\colon \FF_2^S\to M_{n\times n}(\complex)$ defined as follows
	\[
	\forall a\in \FF_2^S\ \colon \ \ \cal{P}_a=\Es{\alpha\in \FF_2^S}\left[(-1)^{\langle a,\alpha\rangle}\cal{U}(\alpha)\right]\;.
	\]
	Indeed for every $\vec v\in \Img (\cal{P}_a)$ and $\alpha\in \FF_2^S$ we have $\cal{U}(\alpha)\vec v=(-1)^{\langle a,\alpha\rangle}\vec v$.
	The inverse Fourier transform in this case is 
	\[
	\forall \alpha\in \FF_2^S\ \colon \ \ U(\alpha)=\sum_{a\in \FF_2^S} (-1)^{\langle a,\alpha\rangle}\cal{P}_a\;.
	\]
\end{definition}

\begin{definition}[Projective, Representation and Observable form of a PVM]\label{defn:projective_representation_and_observable_form_PVM}
	Let $S$ be a finite set. The following three objects contain the same data:
	\begin{itemize}
		\item \emph{Projective form}: A map $\cal{P}\colon \FF_2^S\to M_{n\times n}(\complex)$ whose images are orthogonal projections that sum up to the identity.
		\item \emph{Representation form}: A unitary representation $\cal{U}\colon \FF_2^S\to U(n)$.
		\item \emph{Observable form}: A map $\cal{U}\colon S\to U(n)$ whose images are  commuting  involutions (i.e., square to the identity).
	\end{itemize}
	So, a PVM can be given in any of these forms, and we refer to them as the projective, representation, and observable form of the PVM respectively.  Furthermore, if we have a PVM $\cal{U}$ in representation (or observable) form, we still denote by $a\sim \cal{U}$ an outcome sampled according to the PVM (and similarly $(a,b)\sim (\cal{U}^\mttx,\cal{U}^\mtty)$ for the joint measurement).
\end{definition}
\begin{remark}
	Note that we use the same notation  $\cal{U}$ for the representation and observable form of a PVM. This may be a bit confusing, as for $\sX\in S$, $\cal{U}({\bf 1}_\sX)$   in representation form is the same as $\cal{U}(\sX)$ in  observable form. But it is in fact a natural choice, as we use the ``universal property'' of $\FF_2^S$, which says that any map $\cal{U}\colon S\to U(n)$ whose images are commuting involutions can be extended to a unitary representation of $\FF_2^S$ through the embedding of $S$ in $\FF_2^S$ through the map $\sX\mapsto {\bf 1}_\sX$.
\end{remark}
\begin{definition}[Diagonal PVM]\label{defn:diagonal_PVM}
	A PVM $\cal{P}\colon A\to M_{n\times n}(\complex)$ is \emph{diagonal} if all its images $\cal{P}_a$ are diagonal matrices. Namely, the projections are on spaces spanned by subsets of  the standard basis. 
	In case $A=\FF_2^S$, this property is preserved under the Fourier transform. Namely, it is equivalent to the representation (and thus observable) form $\cal{U}$ of the PVM to consist of only diagonal unitaries.
\end{definition}
\begin{definition}[Readably $Z$-aligned PVM]\label{defn:readably_Z-aligned_PVM}
	Let $S^\frR$ and $S^\frL$ be disjoint finite sets.\footnote{This notation is $\frR$ for \textbf{readable} variables and $\frL$ for \textbf{linear} or \textbf{unreadable} variables.} A PVM in observable form  $\cal{U}\colon {S^\frR\sqcup S^\frL}\to U(n)$ is said to be  \emph{readably $Z$-aligned} if its restriction to $S^\frR$ is diagonal (Definition \ref{defn:diagonal_PVM}). 
\end{definition}
\begin{remark}
	The standard basis in quantum information theory is commonly called the $Z$-basis, as it is the mutual eigenbasis of the  $\PZm$-matrices in the Pauli group  (more on that in Section \ref{sec:Pauli_gp}). Hence the term  ``readably $Z$-aligned''  for  one whose readable observables are diagonal with respect to the standard basis.
\end{remark}
\subsection{Permutations and Signed permutations}

As described in the introduction, the perfect strategies in our category should be induced by permutation representations --- actually, by signed permutation representations. To that end we give the following definition.

\begin{definition}[Permutation matrices and representations]
	Let $\Omega$ be a finite set. As $\Sym(\Omega)$ acts naturally on $\Omega$, its action  extends to $\complex^\Omega$ as follows: Given $f\colon \Omega\to \complex$ and $\sigma\in \Sym(\Omega)$, let $\sigma.f(\star):=f(\sigma^{-1}.\star)$. The standard basis of $\complex^\Omega$ consists of the indicators ${\bf 1}_\star$ for every $\star\in \Omega$, and we have
	\[
	\forall \diamond\in \Omega\ \colon\ \   \sigma.{\bf 1}_\star(\diamond)={\bf 1}_\star(\sigma^{-1}.\diamond)=\begin{cases} 1 & \sigma^{-1}.\diamond=\star\ ,\\
	0 & \sigma^{-1}.\diamond\neq \star\ , 
	\end{cases}
	\]
	namely $\sigma.{\bf 1}_\star={\bf 1}_{\sigma.\star}$.
	Representing $\Sym(\Omega)$ via this action as $\Omega\times \Omega$ matrices gives rise to the subset of $U(\complex^{\Omega})$ consisting of all $0/1$ matrices with exactly one $1$ in every row and column. Unsurprisingly, these matrices are called \emph{permutation matrices}. An action is a homomorphism from a group to $\Sym(\Omega)$, and by using the above embedding of permutations  into $U(\complex^\Omega)$, we get a unitary representation of the group. Such representations are called \emph{permutation representations}.
\end{definition}
\begin{definition}[Signed sets]\label{defn:signed_set}
	Given a finite set $\Omega$, we define its signed version $\Omega_\pm$ to be 
	$\{\pm\}\times \Omega$; we commonly denote $+\star$ and $-\star$ instead of $(+,\star)$ and $(-,\star)$. We  commonly use $\star,\diamond$ for elements of $\Omega$ and $\spadesuit,\diamondsuit$ for elements of $\Omega_{\pm}$.
\end{definition}

\begin{definition}[The sign flip]
	The \emph{sign flip} $-\Id$ is a permutation on $\Omega_\pm$ that, as its name suggests, flips the sign of every vertex. Namely,
	\[
	\forall \star\in \Omega\ \colon \ \ -\Id.\pm\star=\mp\star\ .
	\]
	A function $f\colon \Omega_\pm\to \mathbb{C}$ is said to be \emph{symmetric} if $f(+\star)=f(-\star)$ for every $\star\in \Omega$ and \emph{anti-symmetric} if $f(+\star)=-f(-\star)$. The symmetric functions are the $(+1)$-eigenspace of $-\Id$ and we denote them by $W^+\subseteq \complex^{\Omega_\pm}$, while the anti-symmetric functions are its $(-1)$-eigenspace and are denoted by $W^{-}\subseteq \complex^{\Omega_\pm}$.
	Let $\Xi\colon \mathbb{C}^{\Omega_{\pm}}\to W^{-}$ be the orthogonal  projection on $W^{-}$.  We fix  
	\begin{equation}\label{eq:bases_for_symmetric_and_anti-symmetric_functions}
	B^+=\left\{\frac{{\bf 1}_{+\star}+{\bf 1}_{-\star}}{\sqrt 2}\right\}_{\star\in \Omega}\quad,\quad  B^-=\left\{\frac{{\bf 1}_{+\star}-{\bf 1}_{-\star}}{\sqrt 2}\right\}_{\star\in \Omega}
	\end{equation}
	to be the  standard orthonormal bases for $W^+$ and $W^{-}$ respectively. Note that  these bases  are indeed the images (up to a sign in case of $B^{-}$) of the standard basis of $\complex^{\Omega_\pm}$ via its orthogonal projection onto the symmetric and anti-symmetric functions (i.e., $\Xi$) respectively. 
\end{definition}

\begin{definition}[Signed permutations and representations]\label{defn:signed_permutations_and_reps}
	
	A \emph{signed permutation} is a permutation $\sigma\in \Sym(\Omega_\pm)$ that commutes with the sign flip, and we denote by $\Sym_\pm(\Omega)$ the subgroup of all signed permutations. 
	The action of the signed permutations on $\complex^{\Omega_\pm}$ preserves the spaces of anti-symmetric functions $W^-$, which induces an embedding $\Sym_\pm(\Omega)\hookrightarrow U(W^-)$. 
	The image of this embedding is called the group of \emph{signed permutations}. By representing the matrices in  ${\rm End}(W^-)$ with respect to the basis $B^-$ from \eqref{eq:bases_for_symmetric_and_anti-symmetric_functions}, the image of $\Sym_\pm(\Omega)$ consists of  all matrices with coefficients in  $\{0,+1,-1\}$, such that in each row and column there is a single non-zero entry (which must be either $+1$ or $-1$).  A \emph{signed action} is a homomorphism of a group into $\Sym_\pm(\Omega)$, and by composing it with the above embedding into $U(W^-)$ we get a \emph{signed permutation representation}.
\end{definition}
\begin{remark}[Signed permutations as a semidirect product]
	Every signed permutation matrix $\mathscr{A}\in U(n)$ can be written (uniquely) as a product $\mathscr{B}\cdot \mathscr{D}$, where $\mathscr{B}$ is a (non-signed) permutation matrix, and $\mathscr{D}$ is a diagonal matrix with $\pm 1$ on the diagonal. As the subgroup   of diagonal matrices with $\pm 1$ on the diagonal is normal in the signed permutations, and is isomorphic to $\FF_2^n$, we deduce that
	\[
	\Sym_\pm(\Omega)\cong  \Sym(\Omega)\ltimes \FF_2^\Omega \ .
	\]
\end{remark}

\begin{definition}[Signed permutation PVM]\label{defn:perm_PVM}
	Let $S$ be a finite set. A \emph{signed permutation PVM} (in representation form and with outcomes in $\FF_2^S$) is a  signed permutation representation  of $\FF_2^S$, namely a homomorphism $\cal{U}\colon \FF_2^S\to \Sym_\pm(\Omega)\subseteq U(W^-)$. We seldomly extend $\cal{U}$  (in observable form) to be defined on an additional element  $\sJ\notin S$ such that $\cal{U}(\sJ)=-\Id$ --- this yields a representation of $\FF_2^{S\cup \{\sJ\}}$. 
\end{definition}

\subsection{Non-local Games}

\begin{definition}[Games]\label{defn:non-local_game}
	A ($2$-player, $1$-round, synchronous non-local) game $\game$ 
	consists of a finite (oriented)  graph $G=(V,E)$, a length function $\ell\colon V\to \mathbb{N}$, (distinct\footnote{Namely, there are no formal generators that belong to $S_\mttx$ and to $S_\mtty$ for any $\mttx\neq \mtty$.}) formal sets of generators $S_\mathtt{x}$ of size $\ell(\mathtt{x})$ for every vertex $\mathtt{x}\in V$, a distribution $\mu$ over the edge set $E$, and decision functions $D_{\mathtt{xy}}\colon \{0,1\}^{S_{
			\mathtt{xy}}}\to \{0,1\}$ for every edge $\mathtt{xy}\in E$, where $S_\mathtt{xy}=S_\mathtt{x}\cup S_\mathtt{y}$.\footnote{In case $\mttx\neq\mtty$, $S_{\mttx\mtty}$ is the disjoint union $S_\mttx\sqcup S_\mtty$, but in case  $\mttx=\mtty$ then $S_{\mttx\mtty}=S_\mttx=S_\mtty$.} We denote by $S$ the set $\bigcup_{\mathtt{x}\in V} S_\mathtt{x}$ consisting of all formal variables used in the game. 
\end{definition}
\begin{remark}[Standard definition of a  game]\label{rem:subgp_test_defn_of_game}
	It is common to define a game with less data, as follows: It consists of two finite sets $X,A$, a probability distribution $\mu$ over $X\times X$, and a decision predicate $D\colon X\times X\times A\times A\to \{0,1\}$. The set $X$ is commonly called the \emph{question set} and $A$ the \emph{answer set}. Such a game  is  called \emph{synchronous} if $D(\mttx,\mttx,a,b)=0$ for every $a\neq b\in A$ and $\mttx\in X$. 
	
	One can extract the data of Definition \ref{defn:non-local_game} from the above as follows:
	Let $\ell$ be the constant function $\Lambda=\lceil \log |A|\rceil$, and fix an embedding of $A$ into $\{0,1\}^\Lambda$. The vertices $V$ of the underlying graph $G$ will be $X$, and the support of $\mu$ will be the edge set $E\subseteq X\times X$. 
	There is a unique formal generator in $S_\mathtt{x}$ that corresponds to each bit of the answer $a$ when $\mathtt{x}\in X$ is asked as a question --- this is the case as all the $S_\mttx$ are disjoint. Then, given that $\mathtt{x}\neq\mathtt{y}$ were asked, a pair of answers $a,b$ can be encoded as a map $\gamma \colon S_\mathtt{x}\cup S_\mathtt{y}\to \{0,1\}$, where $\gamma|_{S_\mathtt{x}}=a$ and $\gamma|_{S_\mathtt{y}}=b$. Lastly, $D_\mathtt{xy}(\gamma)=D(\mathtt{x},\mathtt{y},a,b)$, where $\gamma$ is the aforementioned encoding.  Note that under this formulation, if $\mttx=\mtty$, then $S_\mttx=S_\mtty$, which implies that $D_\mathtt{xx}$ is a function only of  $a=\gamma|_{S_\mttx}=\gamma|_{S_\mtty}=b$. As our strategies are (almost) always synchronous (Definition \ref{defn:quantum_strategy}), this will mostly not be an issue --- see Section \ref{sec:non-synch_setup} for the non-synchronous setup, which is used only in the soundness argument of the parallel repetition theorem. 
\end{remark}

\begin{definition}[Strategies]\label{defn:quantum_strategy}\label{defn:permutation_strategy}
	A (synchronous, quantum) $n$-dimensional strategy $\strategy$ for a game $\game$ (Definition \ref{defn:non-local_game}) is a map  that associates to every vertex $\mttx\in V$ a $n$-dimensional PVM (Definition \ref{defn:PVM}) with outcomes in $\FF_2^{S_\mttx}$. I.e.,
	\begin{itemize}
		\item \emph{Projective form}: A function $\cal{P}$ that takes as input a vertex $\mathtt{x}\in V$ and a bit string $a\colon S_\mathtt{x}\to \{0,1\}$ and outputs a $n\times n$ matrix with complex coefficients, where for every $\mathtt{x}\in V$ the restriction $\cal{P}^\mathtt{x}=\cal{P}(\mathtt{x},\cdot)\colon \{0,1\}^{S_\mathtt{x}} \to M_{n\times n}(\complex)$ is a PVM in projective form. In such a case, we  denote 
		\[
		\strategy=\{\cal{P}\}=\{\cal{P}^\mathtt{x}_a\mid \mttx\in V, a\colon S_\mttx\to \FF_2\}\ .
		\]
		\item \emph{Representation form}: A function $\cal{U}$ that takes as input a vertex $\mathtt{x}\in V$ and a vector $\alpha\in \FF_2^{S_\mttx}$ and outputs an $n\times n$ unitary $\cal{U}^\mttx(\alpha)$, where for every $\mttx$ the map $\cal{U}^\mttx(\cdot)\colon \FF_2^S\to U(n)$ is a unitary representation.  In such a case we denote
		\[
		\strategy=\{\cal{U}\}=\{\cal{U}^\mttx(\alpha)\mid \mttx\in V, \alpha\in \FF_2^{S_\mttx} \}\ .
		\]
		\item \emph{Observable form}: A function $\cal{U}$ that takes as input a formal variable $\sX\in S$ and outputs an $n\times n$ unitary $\cal{U}(\sX)$, such that its restriction to $S_\mttx$   consists of commuting unitary involutions for every fixed $\mathtt{x}\in V$. In such a case we denote
		\[
		\strategy=\{\cal{U}\}=\{\cal{U}(\sX)\mid \sX\in S \}\ .
		\]
	\end{itemize}
	
	We say that the strategy $\strategy$ \emph{commutes along edges} if for every $\mathtt{xy}\in E$,  the images of $\cal{P}^\mathtt{x}$ and $\cal{P}^\mathtt{y}$ commute (equivalently, the images of $\cal{U}^\mathtt{x}$ and $\cal{U}^\mathtt{y}$ commute, or for every $\sX\in S_\mttx$ and $\sY\in S_\mtty$ the matrices $\cal{U}(\sX)$ and $\cal{U}(\sY)$ commute). We say that $\strategy$ is a (signed) \emph{permutation strategy} if it associates to each vertex a signed permutation PVM (Definition \ref{defn:perm_PVM}).
	\\
	The game distribution $\mu$ specifies a way to sample edges $\mathtt{xy}\in E$. After an edge $\mathtt{xy}$ is sampled, one can jointly measure according to the PVMs at the vertices $\mttx$ and $\mtty$, which gives an outcome $(a,b)$ where $a\colon S_\mttx\to  \FF_2$ and $b\colon S_\mtty\to  \FF_2$.  
	Namely, 
	\begin{equation}\label{eq:sampling_according_to_strategy}
	\Pro{}[(a,b)\in \FF_2^{S_{\mttx}}\times \FF_2^{S_{\mtty}}\ {\rm is\ sampled}\mid \mathtt{x},\mathtt{y} {\rm\  were \ sampled}]=\tau(\cal{P}^\mathtt{x}_a\cal{P}^\mathtt{y}_b)\;.
	\end{equation}
	We often denote the concatenation of  $a$ and $b$ as $\gamma=ab\colon S_\mttx\cup S_\mtty\to \FF_2$.
	A function $\gamma=ab$ sampled as in \eqref{eq:sampling_according_to_strategy} is said to be \emph{sampled according} to the strategy $\strategy$, and we denote it by $\gamma\sim \strategy$ (with the dependence on $\mttx,\mtty$ usually left implicit).  
\end{definition}
\begin{remark}
	In \cite{BCLV_subgroup_tests} permutation strategies were defined slightly differently. There, we distinguished between the signed permutation PVMs (in observable form) associated to each vertex, which have images in $\Sym_\pm(\Omega)$ --- the collection of them was called the \emph{permutation strategy} (Definition 6.11 therein, where the image of $\sJ$ plays the role of the sign flip $-\Id$) ---  and the quantum strategy induced by embedding $\Sym_\pm(\Omega)$ in $U(W^{-})$ --- which is called \emph{the quantum strategy induced by a permutation strategy} (Definition 6.14 therein). As these obejcts provide the same information, here we decided to drop the distinction between them and just think of $\Sym_\pm(\Omega)$  as embedded in the natural way in the unitaries on anti-symmetric functions  $U(W^{-})$.
\end{remark}

\begin{example}[Classical  strategies]\label{example:classical_perm_strategies}
	The subgroup $\{\pm 1\}\subseteq U(1)$ is the collection of $1\times 1$ signed permutation matrices (Definition \ref{defn:signed_permutations_and_reps}). Let $\game$ be a game and $S$ its formal set of generators. For every fixed $f\colon S\to \{0,1\}$, we can define a  strategy $\cal{U}\colon S\to \{\pm 1\}$ as follows
	\[
	\cal{U}(\sX)=\begin{cases}
	+1 & f(\sX)=0\ ,\\
	-1 & f(\sX)=1\ .
	\end{cases}
	\]
	Given that we have sampled an edge $\mttx\mtty$ according to $\mu$, it is straightforward that $\gamma\colon S_\mathtt{xy}\to \{0,1\}$ which is sampled according to $\strategy=\{\cal{U}\}$ (Definition \ref{defn:quantum_strategy}) is deterministically $f|_{S_{\mathtt{xy}}}$. Such strategies are usually called \emph{deterministic}. By taking direct sums of such deterministic strategies (for potentially different $f$'s)  --- which is the same as requiring that the strategy  associates to every vertex a  diagonal PVM (Definition \ref{defn:diagonal_PVM}) --- we can get any (rational) distribution over deterministic strategies. Such strategies are usually called \emph{classical}. Hence, every (rational) classical strategy can be obtained as a permutation strategy. 
\end{example}

\begin{definition}[Value]\label{defn:value_of_a_game}
	We can ``run'' the strategy $\strategy$ against the game $\game$:
	sample $\mathtt{xy}\in E$ according to $\mu$; sample $\gamma\colon S_\mathtt{xy}\to \{0,1\}$  according to $\strategy$; \emph{Accept} if $D_\mathtt{xy}(\gamma)=1$, and otherwise \emph{Reject}.
	The \emph{value} of $\strategy$ against $\game$ is its acceptance probability in the above procedure, namely
	\[
	\begin{split}
	\val(\game,\strategy)&=\Es{\mathtt{xy}\sim\mu }\Es{\gamma\sim \strategy}[D_\mathtt{xy}(\gamma)]\\
	&=\sum_{\mathtt{xy}\in E}\sum_{\substack{{a\colon{S_\mathtt{x}}\to \FF_2}\\ b\colon {S_\mathtt{y}}\to \FF_2}} \mu(\mathtt{xy})\tau\left(\cal{P}^\mathtt{x}_a\cal{P}^\mathtt{y}_b\right)D_{\mathtt{xy}}(ab)\ .
	\end{split}
	\]
	We say that a strategy $\strategy$ is \emph{perfect} (for $\game$) if $\val(\game,\strategy)=1$.
	The (synchronous quantum) value $\val^*(\game)$ of $\game$ is the supremum of its value against every quantum strategy $\strategy$. 
	
\end{definition}

\begin{remark}[Correlations]\label{rem:induced_correlations}
	Usually, the collection of conditional distributions
	$$\Pro{}[\gamma=(a,b)\ {\rm is\ sampled\ by\ }\strategy\mid \mathtt{x},\mathtt{y} {\rm\  were \ sampled}]$$  for every pair $\mathtt{x},\mathtt{y}\in V$ is called the \emph{correlation} induced by the quantum strategy, and is denoted by $p(a,b|\mathtt{x},\mathtt{y})$ (or $p_{\strategy}$ when wanting to emphasize the dependence on $\strategy$). 
\end{remark}
\begin{remark}[Dramatization of a game]\label{rem:dramatization_non_local_game}
	The reason for the name ``game'' for the data described in Definition \ref{defn:non-local_game}, and for the name ``strategy'' for the collection of PVMs described in Definition \ref{defn:quantum_strategy} is the following:
	
	Two players, that can share a maximally entangled state of any dimension $n$, are separated spatially --- e.g., they are seated in far away rooms. 
	A referee  samples a pair of questions --- i.e., an edge $\mttx\mtty\in E$ --- and sends one question to each player --- namely, $\mttx$ to player $A$ and $\mtty$ to player $B$. 
	The players agreed beforehand, for every possible question in the game, how  they will measure their part of the state --- namely, they chose a map from $V$ to PVMs acting on $\complex^n$. After receiving their questions, each player measures their part of the state as agreed beforehand,  comes up with answers ---  $a$ for player $A$ and $b$ for player $B$ --- according to what they have measured, and send them back to the referee. 
	The referee then decides, using the decision predicate $D_{\mttx\mtty}(ab)$, whether the players \emph{won} or \emph{lost}. 
	The decision predicate $D$ as well as the distribution $\mu$ over possible questions are assumed to be known to the players before they choose their \emph{strategy}, namely the dimension of their maximally entangled state and the projective measurements associated to each vertex.

\end{remark}

\subsection{Tailored games}\label{sec:Tailored_Games}

\begin{definition}[Tailored games]\label{defn:tailored_games}
	Colloquially,  a \emph{tailored} game is one where $D_{\mathtt{xy}}$ reads \textbf{part} of (the answer pair) $\gamma=ab$, and decides according to this partial view which \textbf{parity checks} to apply on \textbf{the whole} of $\gamma$.\footnote{We considered calling such games \emph{controlled linear}, since it is more informative. But, since conditionally linear is a term we use in this paper, and terms containing linear are generally overused, we decided to use a less informative notion.}
	
	Formally,
	a tailored (non-local) game $\game$ is equipped with extra structure, described shortly, and its decision functions $D_\mathtt{xy}$ behave \textbf{canonically} with respect to this extra data.
	Instead of a single length function $\ell$, $\game$ has two length functions $\ell^\frR\colon V\to \mathbb{N}$ and $\ell^{\frL}\colon V\to \mathbb{N}$, and $\ell=\ell^\frR+\ell^\frL$. Before, the length function described the size of the formal  set of generators at each vertex. 
	Now, the formal set of generators $S_\mathtt{x}$ at $\mathtt{x}\in V$ will be a disjoint union of the sets $S_\mathtt{x}^\frR$ and $S_\mathtt{x}^\frL$, where $S_\mathtt{x}^\frR$ is of size $\ell^{\frR}(\mathtt{x})$ and $S_\mathtt{x}^\frL$ is of size $\ell^\frL(\mathtt{x})$. The elements of  $S_\mathtt{x}^\frR$ are  called the \emph{readable} variables at $\mathtt{x}\in V$ and the elements of $S_\mathtt{x}^{\frL}$ the \emph{linear} or \emph{unreadable} variables at $\mathtt{x}$. 
	In addition, $\game$ is equipped with  a collection of \emph{controlled linear constraints} functions $L_\mathtt{xy}$ that take as input a function $\gamma^\frR \colon S_\mathtt{x}^\frR\sqcup S_\mathtt{y}^\frR\to \FF_2$, and outputs a sequence of subsets of $S_\mathtt{xy}\sqcup \{\sJ\}$, where $\sJ$ is a new formal variable not in any other set. Namely,
	\[
	L_{\mathtt{xy}}\colon \FF_2^{S_{\mathtt{x}}^\frR \cup S_\mathtt{y}^\frR}\to \FF_2^{\FF_2^{S_{\mathtt{xy}}\cup \{\sJ\}}}\;.
	\]
	The image of $L_{\mathtt{xy}}$ is interpreted as a collection of linear constraints that will be verified by the decision function.   
	The decision function $D_{\mathtt{xy}}(\gamma)$ behaves as follows: It restricts $\gamma$ to the readable variables, namely looks at $\gamma^\frR=\gamma|_{S_\mathtt{x}^\frR\cup S_\mathtt{y}^\frR}\colon S_\mathtt{x}^\frR\cup S_\mathtt{y}^\frR\to \FF_2$, and calculates $L_{\mathtt{xy}}(\gamma^\frR)$. Then, it extends $\gamma$ such that $\gamma(\sJ)=1$. 
	Finally, for every  $c\in L_{\mathtt{xy}}(\gamma^\frR)$, we have $c\colon S_{\mathtt{xy}}\cup \{\sJ\}\to \FF_2$,  and $D_{\mathtt{xy}}$ verifies that 
	$$\langle c,\gamma\rangle=\sum_{\sX\in S_{\mathtt{xy}}\cup \{\sJ\}}c(\sX) \cdot \gamma(\sX)=0\;.$$ 
	Namely, $L_{\mathtt{xy}}(\gamma^\frR)$ consists of linear constraints that $\gamma$ needs to satisfy. 
	If all of the above were satisfied, then $D_{\mathtt{xy}}(\gamma)=1$, and otherwise it is $0$.
	In the spirit of Remark \ref{rem:subgp_test_defn_of_game}, we often denote 
	\begin{equation}\label{eq:a_b_to_gamma}
	a^\rvar=\gamma|_{S_\mathtt{x}^\frR}\; ,\quad a^\lvar=\gamma|_{S_\mathtt{x}^\frL}\; ,\quad b^\rvar=\gamma|_{S_\mathtt{y}^\frR}\quad\textrm{and}\quad b^\lvar=\gamma|_{S_\mathtt{y}^\frL}\;,
	\end{equation}
	and conversely $\gamma=a^\rvar a^\lvar b^\rvar b^\lvar$ or  $\gamma= (a^\rvar,a^\lvar,b^\rvar,b^\lvar)$. Hence, it is common, for example, to see $L_{\mttx\mtty}(a^\frR,b^\frR)$  instead of $L_{\mttx\mtty}(\gamma^\frR)$ throughout the paper.
\end{definition}

\begin{definition}[Underlying combinatorial game]\label{defn:combinatorial_game_underlying_tailored}
	Given a tailored (non-local) game, we can refer to its \emph{underlying combinatorial game}. By this, we mean the game with the same graph, a length function that disregards readability $\ell=\ell^\frR+\ell^\frL$, and the same decision predicates. Note that in any operative way, these are the same game, we just \emph{forget} about the tailored structure that governs $D_{\mathtt{xy}}$.
\end{definition}

\begin{remark}[Naive tailoring of any game]\label{rem:naive_tailoring}
	Being tailored may at first seem to be quite a restrictive form for a non-local game. Indeed, while the dependence of the decision function on the readable variables is allowed to be arbitrary, the dependence on the linear variables is restrictive --- as not every boolean function can be expressed as a conjunction of affine-linear functions --- for example, consider the OR function. However, observe that because the definition allows one to ``tailor'' according to any partition of the variables in ``readable'' and ``unreadable'' variables, every game can be tailored in a trivial manner, as follows. First, all variables are declared readable, namely $\ell^\frR=\ell$ and $\ell^\frL=0$. Then, if the decision function $D_{\mathtt{xy}}$ decided to accept $\gamma$ according to the original game, then it lets $L_{\mathtt{xy}}(\gamma^\frR)$ be empty (and thus all linear conditions will be satisfied regardless of what $\gamma$ is). And, if $D_{\mathtt{xy}}$ decided to reject $\gamma$ according to the original game, 
	then it chooses $L_{\mathtt{xy}}(\gamma^\frR)$ to contain the singleton $\{\sJ\}$ as the single subset appearing in  $L_{\mathtt{xy}}$. Note that $\{\sJ\}$ represents the linear equation $1\cdot \gamma(\sJ)=0$, which is $1=0$, and thus cannot be satisfied by any $\gamma$. 
\end{remark}

\textbf{This raises the question}: What have we gained by defining tailored non-local games, if any game can be tailored in a straightforward manner? 

\begin{definition}[$Z$-aligned permutation strategies]\label{defn:Z-aligned_strategy}
	A strategy $\cal{U}$ for a tailored non-local game $\game$ is said to be a $Z$-\emph{aligned permutation strategy} if it associates to each vertex $\mttx\in V$ a readably $Z$-aligned (Definition \ref{defn:readably_Z-aligned_PVM}) signed permutation PVM (Definition \ref{defn:perm_PVM}). Namely, in observable form, for every $\sX\in S$ we have $\cal{U}(\sX)\in \Sym_\pm(\Omega)\subseteq U(W^-)$ (which is the permutation strategy condition) and for every readable variable $\sX$ the observable $\cal{U}(\sX)$ is diagonal (which is the readably $Z$-aligned condition). This is equivalent to having a permutation strategy such that each  readable variable acts on each point in the signed set  $\Omega_\pm$ either like the identity  or like the sign flip $-\Id$. 
	
	We use the acronym $\ZPC$  to describe a  $Z$-aligned permutation strategy that commutes along edges.
\end{definition}

\begin{remark}
	The classical strategies described in Example \ref{example:classical_perm_strategies} are $Z$-aligned permutation strategies. But, one can construct permutation strategies that induce a classical strategy in the standard sense (namely, one whose all outputs are commuting) without it being $Z$-aligned. 
\end{remark}

It  is clearer now why the way one tailors a non-local game matters: The existence of a perfect $\ZPC$ strategy for the game depends on it. Let us demonstrate this with binary linear constraint system (LCS) games, and specifically the Mermin--Peres magic square game. For a thorough introduction to  the magic square game we refer to~\cite{aravind2002simple}, and more generally for an introduction to LCS games see~\cite{cleve2017perfect}. 

\begin{example}[Linear constraint system games]\label{example:LCSs}
	Let $\mathscr{A}$ be an $m\times n$ matrix with $\FF_2$-coefficients, and let $\vec b$ be a column vector in $\FF_2^m$. Classically, such a pair defines a system of linear equations $\mathscr{A}\vec x=\vec b$ over $\FF_2$.  It also defines a certain non-local game $\mathfrak{Lcs}(\mathscr{A},\vec b)$ which is the quantum counterpart of this classical system of equations. In this game,  an assignment to a random linear constraint in $\mathscr{A}\vec x=\vec b$ (i.e., a row) is asked for, and is crossed checked against some ``global'' assignment to the variables (i.e., columns) for consistency. 
	
	The vertices in the underlying graph of $\mathfrak{Lcs}(\mathscr{A},\vec b)$ will be indexed by the rows  (i.e., linear constraints) and columns (i.e., variables) of the matrix $\mathscr{A}$, namely $\{\mathtt{const}_i\mid i\in [m]\}$ and $\{\mathtt{var}_j\mid j\in [n]\}$. There is an edge between $\mathtt{const}_i$ and $\mathtt{var}_j$ if and only if $\mathscr{A}_{ij}=1$ --- which is saying, the $j^{\rm th}$ variable appears in the $i^{\rm th}$ constraint. The length of every column vertex is $1$, and we denote by $\mathsf{Var}_j$ the formal variable associated with the $j^{\rm th}$ column $\mathtt{var}_j$. The length of each row vertex is the number of $1$'s in the row, and we associate formal variables $S_{\mathtt{const}_i}=\{\mathsf{Const}_{ij'}\mid \mathscr{A}_{ij'}=1\}$ to  $\mathtt{const}_i$. The decision function $D_{\mathtt{const}_i\ \mathtt{var}_j}$ gets as input an assignment $\gamma$ to $\mathsf{Var}_j$ and $\{\mathsf{Const}_{ij'}\mid \mathscr{A}_{ij'}=1\}$, and accepts if and only if 
	\begin{equation}\label{eq:LCS_check}
	\sum_{j'\colon \mathscr{A}_{ij'}=1}\gamma(\mathsf{Const}_{ij'})=b_i \quad\textrm{and}\quad  \gamma(\mathsf{Var}_j)=\gamma(\mathsf{Const}_{ij}),
	\end{equation}
	namely, if the assignment induced by $\gamma$ satisfies the $i^{\rm th}$ constraint, and is consistent with the global assignment to the $j^{\rm th}$ variable.
	Though for our discussion the distribution $\mu$ over edges in this game is not important, one can consider the following standard sampling scheme: 1) Choose a row uniformly at random. 2) Choose a uniform variable out of the support of the chosen row.
	
	Let us describe a \textbf{non-trivial} tailoring of the LCS game $\mathfrak{Lcs}(\mathscr{A},\vec b)$. First, all variables are chosen to be  \emph{unreadable}, namely $\ell^\frR=0$ and $\ell^\frL=\ell$. Given that the edge $\mathtt{const}_i\ \mathtt{var}_j$ was sampled, the controlled linear constraints $L_{\mathtt{const}_i\ \mathtt{var}_j}$ will consist of two checks, which are derived from 
	\eqref{eq:LCS_check}:\footnote{Note that, as there are no readable variables, $L_{\mathtt{const}_i\ \mathtt{var}_j}$ is constant.} 
	\[
	\begin{split}
	c_{ {\rm consistency}}(\sX)&=\begin{cases}
	0 & \sX\neq \mathsf{Var}_j,\mathsf{Const}_{ij}\\
	1 & \textrm{otherwise}
	\end{cases}\\
	c_{ {\rm linear}}(\sX)&=\begin{cases}
	0 & \sX=\mathsf{Var}_j\\
	1 & \sX=\mathsf{Const}_{ij'}\in S_{\mathtt{row}_i}\\
	b_i & \sX=\sJ
	\end{cases}
	\end{split}
	\]
	Then, $c_{{\rm consistency}}$  forces the canonical decision procedure $D_{\mathtt{const}_i\ \mathtt{var}_j}$ to check consistency between the constraint assignment to the $j^{\rm th}$ variable and the global one, i.e. $\gamma(\mathsf{Var}_j)=\gamma(\mathsf{Const}_{ij})$, and $c_{{\rm linear}}$ forces it to check that the $i^{\rm th}$ linear constraint is indeed sarisfied, i.e. $\displaystyle{\sum_{j'\colon \mathscr{A}_{ij'}=1}}\gamma(\mathsf{Const}_{ij'})=b_i$ --- as required by the definition of the LCS game $\frak{Lcs}(\mathscr{A},\vec b)$.

	The difference between the above tailored form of $\frak{Lcs}(\mathscr{A},\vec b)$ and the one suggested in Remark \ref{rem:naive_tailoring} may seem technical. But, here all the variables are unreadable, and in the version of Remark \ref{rem:naive_tailoring}  all variables are readable. 
	If all variables of a tailored game are readable, a $Z$-aligned permutation strategy for it in observable form is just a collection of diagonal matrices with $\pm 1$ on the diagonal. 
	These strategies are exactly the \emph{classical} ones described in Example \ref{example:classical_perm_strategies}, and having a perfect strategy of this kind for an LCS game is the same as for the linear system $\mathscr{A}\vec x=\vec b$ to have a solution. 
	On the other hand, when all the variables are unreadable, there could be a perfect $Z$-aligned permutation strategy without $\mathscr{A}\vec x=\vec b$ having a solution. This is demonstrated in the next example, which is  used in the proof of Compression (Theorem \ref{thm:compression}).
	
\end{example}

\begin{figure}[h]
	\begin{adjustwidth*}{4em}{4em}
		\begin{tikzpicture}[scale=0.8]
		
		\node[draw, color=black, shape=circle] (x00) at (0,0) {\scriptsize $+00$}; 
		\node[draw, color=black, shape=circle] (x01) at (0,5) {\scriptsize $+01$};
		
		\node[draw, color=black, shape=circle] (x10) at (2,2) {\scriptsize $+10$};
		
		\node[draw, color=black, shape=circle] (x11) at (2,7) {\scriptsize $+11$};
		
		\node[draw, color=black, shape=circle] (y00) at (12,0) {\scriptsize $-00$}; 
		\node[draw, color=black, shape=circle] (y01) at (12,5) {\scriptsize $-01$};
		
		\node[draw, color=black, shape=circle] (y10) at (14,2) {\scriptsize $-10$};
		
		\node[draw, color=black, shape=circle] (y11) at (14,7) {\scriptsize $-11$};

		\draw[purple,-, dashed] (x00) to  [out=315,in=225,looseness=0.5] (y00) ; \node[purple] at (5.5,-2) {\scriptsize $-\Id$};
		
		\draw[purple,-, dashed] (x10) to  [out=315,in=225,looseness=0.3] (y10) ;
		
		\draw[purple,-, dashed] (x11) to  [out=45,in=135,looseness=0.5] (y11) ; \node[purple] at (7.5,9) {\scriptsize $-\Id$};
		
		\draw[purple,-, dashed] (x01) to  [out=45,in=135,looseness=0.3] (y01) ;

		\draw[ForestGreen, -, solid] (x00)--(x01) node[midway,left]{\tiny $\mathds{X}^{\otimes 01}$};
		\draw[ForestGreen, -, solid] (x10)--(x11) node[midway,right]{\tiny $\mathds{X}^{\otimes 01}$};

		\draw[ForestGreen, -, solid] (y00)--(y01) node[midway,left]{\tiny $\mathds{X}^{\otimes 01}$};
		\draw[ForestGreen, -, solid] (y10)--(y11) node[midway,right]{\tiny $\mathds{X}^{\otimes 01}$};

		\draw[SeaGreen, -, solid] (x00)--(x10) node[midway,right]{\tiny $\mathds{X}^{\otimes 10}$};
		
		\draw[SeaGreen, -, solid] (x01)--(x11) node[midway,right]{\tiny $\mathds{X}^{\otimes 10}$};

		\draw[SeaGreen, -, solid] (y00)--(y10) node[midway,right]{\tiny $\mathds{X}^{\otimes 10}$};
		
		\draw[SeaGreen, -, solid] (y01)--(y11) node[midway,right]{\tiny $\mathds{X}^{\otimes 10}$};

		\draw[blue,->, densely dotted] (x00) to  [out=115,in=150,looseness=15] (x00) ; 
		\node[blue] at (-1,1.2) {\tiny $\mathds{Z}^{\otimes 10}$};

		\draw[BlueViolet,->, densely dotted] (x00) to  [out=205,in=240,looseness=15] (x00) ; 
		\node[BlueViolet] at (-1.2,-1.2) {\tiny $\mathds{Z}^{\otimes 01}$};
		
		\draw[blue,->, densely dotted] (y00) to  [out=0,in=330,looseness=15] (y00) ; 
		\node[blue] at (13,0) {\tiny $\mathds{Z}^{\otimes 10}$};

		\draw[BlueViolet,->, densely dotted] (y00) to  [out=315,in=285,looseness=15] (y00) ; 
		\node[BlueViolet] at (12.2,-1.2) {\tiny $\mathds{Z}^{\otimes 01}$};

		\draw[blue,-, densely dotted] (x11) to  [out=60,in=120,looseness=0.7] (y11) ; 
		\node[blue] at (7.5,9.7) {\tiny $\mathds{Z}^{\otimes 10}$};

		\draw[blue,-, densely dotted] (x10) to  [out=60,in=120,looseness=0.7] (y10) ; 
		\node[blue] at (7.5,4.7) {\tiny $\mathds{Z}^{\otimes 10}$};

		\draw[BlueViolet,-, densely dotted] (x11) to  [out=75,in=105,looseness=0.7] (y11) ; 
		\node[BlueViolet] at (7.5,10.1) {\tiny $\mathds{Z}^{\otimes 01}$};
		
		\draw[BlueViolet,-, densely dotted] (x01) to  [out=75,in=105,looseness=0.4] (y01) ; 
		\node[BlueViolet] at (5.5,7.1) {\tiny $\mathds{Z}^{\otimes 01}$};

		\draw[BlueViolet,->, densely dotted] (y10) to  [out=315,in=285,looseness=15] (y10) ; 
		\node[BlueViolet] at (14.2,0.8) {\tiny $\mathds{Z}^{\otimes 01}$};

		\draw[BlueViolet,->, densely dotted] (x10) to  [out=315,in=285,looseness=15] (x10) ; 
		\node[BlueViolet] at (2.2,0.8) {\tiny $\mathds{Z}^{\otimes 01}$};
		
		\draw[blue,->, densely dotted] (y01) to  [out=0,in=330,looseness=15] (y01) ; 
		\node[blue] at (13,5) {\tiny $\mathds{Z}^{\otimes 10}$};

		\draw[blue,->, densely dotted] (x01) to  [out=0,in=330,looseness=15] (x01) ; 
		\node[blue] at (1,5) {\tiny $\mathds{Z}^{\otimes 10}$};

		\end{tikzpicture}
	\end{adjustwidth*}
	\caption{In this figure there are $5$ permutations, $-\Id, \mathds{X}^{\otimes 01},\mathds{X}^{\otimes 10},\mathds{Z}^{\otimes 01},\mathds{Z}^{\otimes 10}$, acting on the set $(\FF_2^2)_{\pm}$. The \textcolor{green}{$\mathds{X}$ permutations} act as bit flips. The \textcolor{blue}{$\mathds{Z}$ permutations} are conditional sign changes, namely, they flip the sign depending on whether the associated bit is $0$ or $1$. Finally, \textcolor{purple}{$-\Id$} flips the sign. }
	\label{tikz:Magic_square_strat}
	
\end{figure}
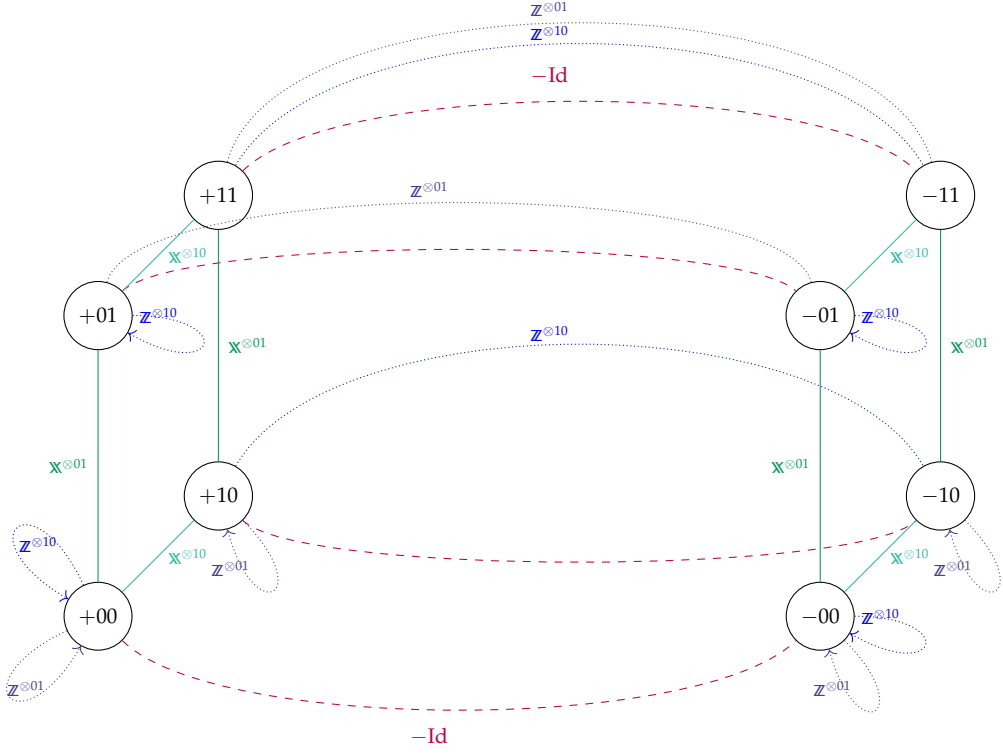

\begin{example}[The Peres--Mermin Magic Square game]\label{example:magic-square}
	The system of linear equations associated with the magic square game has $6$ constraints and $9$ variables, and is defined as follows:
	\[
	\begin{split}
	\mathtt{row_1}\ \colon\ \  \mathsf{Var}_{11}+\mathsf{Var}_{12}+\mathsf{Var}_{13}&=0,\\
	\mathtt{row_2}\ \colon \ \ \mathsf{Var}_{21}+\mathsf{Var}_{22}+\mathsf{Var}_{23}&=0,\\
	\mathtt{row_3}\ \colon \ \ \mathsf{Var}_{31}+\mathsf{Var}_{32}+\mathsf{Var}_{33}&=0,\\
	\mathtt{col_1}\ \colon \ \     \mathsf{Var}_{11}+\mathsf{Var}_{21}+\mathsf{Var}_{31}&=1,\\
	\mathtt{col_2}\ \colon\ \ \mathsf{Var}_{12}+\mathsf{Var}_{22}+\mathsf{Var}_{32}&=1,\\
	\mathtt{col_3}\ \colon \ \ \mathsf{Var}_{13}+\mathsf{Var}_{23}+\mathsf{Var}_{33}&=1.
	\end{split}
	\]
	The choice for the names of the variables and  constraints comes from visualising the variables positioned in a $3\times 3$ grid, and asking for the values in each row to sum up to $0$ while the values in each column should sum up to $1$:
	\[
	\begin{array}{ccccccc}
	
	\mathsf{Var}_{11}& +&\mathsf{Var}_{12}&+&\mathsf{Var}_{13}&=0 \\
	+& & + & & +& &\\ 
	\mathsf{Var}_{21}& +&\mathsf{Var}_{22}&+&\mathsf{Var}_{23}&=0 \\
	+& & + & & +& &\\
	\mathsf{Var}_{31}& +&\mathsf{Var}_{32}&+&\mathsf{Var}_{33}&=0 \\
	\verteq& & \verteq & & \verteq& &\\
	1& & 1 & & 1& &
	\end{array}
	\]
	It is straightforward to see that this system has no solution (e.g., by adding up all the constraints). Therefore, it has no classical perfect strategy, and thus no perfect $Z$-aligned permutation strategy according to the naive tailoring of Remark \ref{rem:naive_tailoring}. But, it \textbf{has} a perfect $Z$-aligned permutation strategy, acting on a signed set of size $8$, with respect to the tailoring described in Example \ref{example:LCSs}. In Figure  \ref{tikz:Magic_square_strat}, $5$ permutations are visualized ---  $-\Id, \mathds{X}^{\otimes 01},\mathds{X}^{\otimes 10},\mathds{Z}^{\otimes 01},\mathds{Z}^{\otimes 10}$.\footnote{Throughout this paper, we use $\mathds{X}$ and $\mathds{Z}$ for the Pauli matrices. As the notation $\mathbb{Z}$ for integers is rarely used in this paper, this should not be confusing for the reader.} To get the perfect permutation strategy for the magic square game, take the mapping
	\[
	\begin{array}{ccc}
	\mathsf{Var}_{11}\mapsto \mathds{X}^{\otimes 10} &   \mathsf{Var}_{12}\mapsto \mathds{X}^{\otimes 01} & \mathsf{Var}_{13}\mapsto \mathds{X}^{\otimes 10}\mathds{X}^{\otimes 01}\\
	\mathsf{Var}_{21}\mapsto \mathds{Z}^{\otimes 01} & \mathsf{Var}_{22}\mapsto \mathds{Z}^{\otimes 10} & \mathsf{Var}_{23}\mapsto \mathds{Z}^{\otimes 10}\mathds{Z}^{\otimes 01}\\
	\mathsf{Var}_{31}\mapsto -\Id\cdot  \mathds{X}^{\otimes 10}\mathds{Z}^{\otimes 01} &   \mathsf{Var}_{32}\mapsto -\Id\cdot \mathds{X}^{\otimes 01}\mathds{Z}^{\otimes 10} & \mathsf{Var}_{33}\mapsto -\Id 
	\cdot \mathds{X}^{\otimes 10}\mathds{X}^{\otimes 01}\mathds{Z}^{\otimes 10}\mathds{Z}^{\otimes 01}
	\end{array}
	\] 
	Note that the $\mathds{Z}$ permutations are $Z$-aligned. This is no coincidence, and it will be helpful when later used. 
	We leave the discussion on where this strategy comes from to Section \ref{sec:Pauli_gp}.
\end{example}

As discussed in the introduction, the main result of \cite{MIPRE} is that approximating the quantum value of a game  (Definition \ref{defn:value_of_a_game}) is as hard as the Halting problem. This is shown by a reduction: one exhibits a computable mapping from (encodings\footnote{See Section~\ref{sec:encodings} for a discussion of the role played by encodings.} of) Turing machines $\cal{M}$ to (encodings of) games $\game_{\mathcal{M}}$ such that, if $\cal{M}$ halts then $\val^*(\game_{\mathcal{M}})=1$, and if $\cal{M}$ does not halt then  $\val^*(\game_{\mathcal{M}})\leq \frac{1}{2}$.\footnote{Here, $\frac{1}{2}$ is an arbitrary constant chosen for convenience.}
The goal of this paper is to reprove this result with two extra conditions: The game $\game_\cal{M}$ needs to be tailored,  and the perfect strategy $\strategy$ (in the complete case, i.e.\ the case where $\mathcal{M}$ halts) needs to be $\ZPC$. Formally:

\begin{theorem}[$\TMIP^*=\RE$]\label{thm:tailored_MIP*=RE}
	There exists a polynomial time algorithm that takes as input (the encoding of) a Turing machine $\cal{M}$ and outputs (the encoding of) a \textbf{tailored} game $\game_\cal{M}$ (see Definition \ref{defn:tailored_games}) such that:
	\begin{enumerate}[label=\textcolor{black}{(\arabic*)}, ref= (\arabic*)]
		\item Sampling $\mathtt{xy}\in E$ according to  $\mu$ and evaluating  $D_{\mathtt{xy}}(\cdot)$ from the encoding of the  game $\game_\cal{M}$  can be done in time ${\rm poly}(|\cal{M}|)$, where $|\cal{M}|$ is the bit-length of the encoding of $\cal{M}$.\label{clause:1_in_tailored_MIP*=RE}
		\item If $\cal{M}$ halts, then there exists a perfect \textbf{$Z$-aligned permutation} strategy $\strategy$ for $\game_\cal{M}$ that commutes along edges (see Definitions \ref{defn:permutation_strategy} and \ref{defn:Z-aligned_strategy}). In particular, $\val^*(\game_\cal{M})=1$.\label{clause:2_in_tailored_MIP*=RE}
		\item If $\cal{M}$ never halts, then $\val^*(\game_\cal{M})< \nicefrac{1}{2}$.\label{clause:3_in_tailored_MIP*=RE}
	\end{enumerate}
\end{theorem}

\subsection{Encoding tailored games}
\label{sec:normal-form}

This section is devoted to an encoding scheme for tailored games, and
following \cite{MIPRE} we use the term \emph{tailored normal form verifiers} (TNFV) for it (Section \ref{sec:Descision_problems_complexity_classes} motivates this term). As seen in Theorem \ref{thm:tailored_MIP*=RE}, {some} encoding mechanism for games is needed to be able to prove our result, and also to be able to phrase the compression theorem rigorously.

\subsubsection{Prelude --- Encodings, Running time and Description length}\label{sec:encodings}

\subsubsection*{Encodings}
By \emph{encoding} we  mean a correspondence (not necessarily single valued or onto) between a  collection of objects --- graphs, games, Turing machined,  functions, etc. --- to the set of (finite) bit strings $\{0,1\}^*$. In computer science, whenever one performs manipulations (e.g.\ an algorithm) on a collection of objects, one ought to have in mind an encoding thereof. This is because, ultimately, each instance of the collection is meant to be represented, and manipulated, on a computer --- which processes strings of bits. Encodings are thus essential as a tool to connect high-level language to concrete implementations. Different choices of encodings can affect, as we will shortly demonstrate, the running time of procedures performed on them. They can also, of course, affect the resilience of stored data to errors, which is the fundamental goal in the theory of error correcting codes. Besides their practical importance, encodings are important theoretical tools --- for example encodings enable \emph{self-reference}, which is the backbone of the classical incompleteness and undecidability results of G\"odel and Turing \cite{godel1931formal,turing1937computable}. 

Before proceeding let us first fix a computational model. For us, an algorithm is always  represented by a \emph{Turing machine}. Informally, a Turing machine (TM) is a finite-state machine that processes data presented on an \emph{input tape} (or maybe several input tapes), using a \emph{memory tape} to store intermediate information and an \emph{output tape} to write its output. The input tape is read-only, the memory tape is read-write and the output tape is write-only. Each tape has infinitely many memory cells, indexed by integers. The Turing machine has one head for each tape, initially positioned at location $0$ (of its appropriate tape). At each \emph{time step}, according to its current \emph{internal state} and the bits each of the heads is reading, the Turing machine may move a head by $\pm 1$ (or leave it in place) along its tape, read or write a symbol, and change its internal state (to one of finitely many possible ones). The Turing machine has a designated ``halt'' state; when it reaches that state its \emph{output} is  the contents of the output tape. 
For a more complete description of this standard model, we refer to~\cite{sipser2012introduction}.  

Now let us consider, e.g., a Turing machine that decides if an input graph is connected or not. In high-level language, this can be performed efficiently by, for example, breadth-first search: First, you need to have a queue and a list. Start from an arbitrary vertex, write its name in the beginning of the list, and put it in the rear of the queue (which is also the front of the queue at this point). Then, repeat the following until the queue is empty: Pop the vertex from the front of the queue (namely, remove it from the queue while reading its name); go over the neighbors of the popped vertex --- if a neighbor appears in the list, do nothing, and if it does not appear, add it to the list and put it at the rear of the queue. At the end of this process, you will have a list of vertices (and an empty queue). Go over these vertices and check whether all vertices of the input graph have been visited.  

While this may be straightforward to understand intuitively, implementing the algorithm as a Turing machine requires one to make a number of choices that affect the running time. An important such choice is the way that the graph passed as input is represented (one also needs to specify the implementation of the queue and list using the memory tape of the TM, but we ignore this intricacy). There are two standard possible encodings. The first represents the graph as an adjacency matrix. If there are $n$ nodes, one will first write the binary representation of the integer $n$, then a separator symbol $\sqcup$,\footnote{This symbol should itself be represented as a string of bits. Namely, an encoding of a larger alphabet which includes $\{0,1,\sqcup\}$ is needed. A way of doing that is suggested in Definition \ref{defn:the_alphabet}.} and then $n$ sequences of $n$ bits representing the $n$ rows of the adjacency matrix. The second encoding is by ``adjacency lists:'' we first write $n$ in binary, then $\sqcup$, and then $n$ sequences of a multiple of $\lceil \log n \rceil$ bits each, such that the $i$-th sequence lists the labels of all vertices connected to the $i$-th vertex. For example, a triangle is represented as 
\[11\sqcup 011 \sqcup 101 \sqcup 110 \]
in the first representation, and as 
\[11\sqcup  0110 \sqcup 0010 \sqcup 0001\]
in the second representation. Here, $0110$, $0010$ and $0001$ are unambiguously interpreted as the neighbors $2$ and $3$ of vertex $1$, $1$ and $3$ of vertex $2$, and $1$ and $2$ of vertex $3$; this is because the number of vertices $3$ given first specifies how many bits each vertex is represented with and we naturally label vertices starting with $00$, $01$, etc. 

There are important differences between these representations. Firstly, they generally do not have the same size: for a graph with $n$ vertices and $m$ edges, the first representation has size $O(n^2)$ while the second has size $O((n+m)\log n)$. Secondly, certain algorithms run faster on one or the other representation --- here, it should be clear that the breadth-first search algorithm will take advantage of the second representation for the case of sparse graphs, as it immediately gives access to all neighbors without having to parse a long row which may contain mostly $0$'s.  

Now, because Theorem \ref{thm:tailored_MIP*=RE} states the existence of an algorithm, with certain properties and in particular a certain runtime, that takes as input a Turing machine, for the theorem to be precise we need to fix some encoding of Turing machines. 
However, in contrast to the encoding of tailored games  described in detail in the next section, the specific encoding of Turing machines that we need is not very strict. 
We henceforth assume that a specific encoding of Turing machines has been fixed, that satisfies the conditions mentioned in  Section 3.1 of \cite{MIPRE}; essentially, we need the following:
\begin{fact}[Cf.~\cite{10.5555/1196416} and \cite{arora2009computational}]\label{fact:properties_of_TMs}
	There is an encoding scheme for Turing machines as bit strings which satisfies:
	\begin{enumerate}[label=\textcolor{black}{(\arabic*)}, ref= (\arabic*)]
		\item \label{clause1:conditions_of_TM_encodings} The length of the  encoding of a TM is reasonably sized (polynomial) as a function of the number of states it can be in.
		\item\label{clause2:conditions_of_TM_encodings}  A Turing machine is able to take the encoding of another Turing machine as input, and execute the latter with some polynomial overhead in the running time;
		\item \label{clause3:conditions_of_TM_encodings} Fixing some of the inputs of  a Turing machine enlarges its description length by at most some polynomial in the lengths of the fixed inputs.
	\end{enumerate}
\end{fact}

These conditions are easily satisfied with standard encodings (for a detailed reference, including the universal simulation theorem, see e.g.~\cite{10.5555/1196416}). 
\begin{remark}\label{rem:high-level}
	In the remainder of the paper we describe Turing machines using high-level language and make statements about their runtime and description length; it will always be clear that a low-level formalization, in terms of states and transition functions, of the high-level description can be obtained which satisfies the claimed runtime and description length bounds.
\end{remark}

The inputs to Turing machines are assumed to be {bit strings}, namely elements of the free monoid $\{0,1\}^*$ which are finite sequences of $0$'s and $1$'s. But, as  mentioned above, some larger alphabets are sometimes needed to be able to describe certain objects. To that end, we use the following:
\begin{definition}[The Alphabet]\label{defn:the_alphabet}
	Let $\{0,1,\sqcup,\frak{error}\}$ be the finite alphabet we wish to encode. To that end, we define an encoding map from the above set as follows:
	\[
	\enc(0)=00\quad \enc(1)=01\quad \enc(\sqcup)=10 \quad \enc(\frak{error})=11.
	\]
	This map  extends naturally to an isomorphism of monoids $$\enc\colon \{0,1,\sqcup,\frak{error}\}^*\to \{00,01,10,11\}^*=\{\rm all\ even\ length\ bit\ strings\}.$$
	Given a bit string $\mttx\in \{0,1\}^*$, we define a decoding map 
	\[
	\dec\colon \{0,1\}^*\to \{0,1,\sqcup\}^*\cup \{\frak{error}\}
	\]
	as follows --- given an input $\mttx\in \{0,1\}^*$:
	\begin{itemize}
		\item First,  $\dec$ checks that the number of bits in $\mttx$, which we denote by $|\mttx|$,  is even. If not, it outputs $\frak{error}$. 
		\item Otherwise, $\mttx$ is of even length, and as we mentioned $\enc$ has a unique inverse on bit strings of even length. Let $\mtty=\enc^{-1}(\mttx)\in \{0,1,\sqcup,\frak{error}\}^*$.
		\item Finally, if $\mtty$ contains an $\frak{error}$ symbol, then $\dec$ will output $\frak{error}$. Otherwise, $\dec$ will output $\mtty$, which  in this case is in  $\{0,1,\sqcup\}^*$.
	\end{itemize}
\end{definition}
\begin{definition}[Encodings of integers]\label{def:unary_encoding_int}
	It is common to assume that certain positive integers $\mathbb{N}=\{1,2,3,...\}$ are provided as inputs to Turing machines. There are two natural ways of achieving that, with substantial differences between them:
	\begin{itemize}
		\item \textbf{Binary}: There is a bijection $\overline{(\cdot)}\colon \mathbb{N}\to \{0,1\}^*$ which writes $n$ in binary and chops its leftmost bit. So, e.g., $\overline{1}$ is the empty bit string $\eps$, $\overline{3}$ is $1$ and $\overline{4}$ is $00$. When we say that a certain input $n$ to a TM $\cal{M}$ is an integer in binary, we mean that $\cal{M}$ receives $\overline{n}$. We often abuse notation and denote $\cal{M}(n)$ instead of $\cal{M}(\overline{n})$, though the explicit input must be a bit string.
		\item \textbf{Unary}: The word length function $|\cdot|\colon \{0,1\}^*\to \mathbb{N}$ translates a bit string into an integer. A TM that ignores the specific input bit string $\mttx$, and only uses its length $|\mttx|$ in its computation, is said to take the integer $n=|\mttx|$ as an input in unary. This is not the standard notion of unary input, which assumes $n$ is encoded as 
		$1^{*n}=\underbrace{1...1}_{n-times}$,
		but {any} length $n$ bit string is an encoding for $n$.
	\end{itemize}
\end{definition}
\subsubsection*{Running time}
Though $\RE$, the class of languages for which the Halting Problem is complete, is defined without any running time constraints --- namely, it is a computational class and not a complexity class --- the class  $\TMIP^*$ does have running time restrictions in its definition (see Section \ref{sec:Descision_problems_complexity_classes} for the formal definition of both). Many of the arguments in this paper rely on efficient running time of certain algorithms (Turing machines), and sometimes even not so common variations on efficient running time are needed (e.g., Definition \ref{def:lambda}). To that end, we need to define running time.

Recall that a Turing machine holds a table (function) that tells it  given the current reads from its heads (on the input tapes, memory tape and output tape) and the current internal state of the machine, to which state to move, what to write on the current position (in the memory and output tape) and to which direction each of the heads needs to move. The computation of the Turing machine progresses by following the table and transforming the state, content of tapes and position of heads accordingly --- each application of the table rules is called a \emph{time step}.\footnote{Intuitively, for real machines,  operating  such a step does take physical time, and that is the reason for the name. }

\begin{definition}[Running Time] \label{def:running_time}
	Let $\cal{M}$ be a $k$-input Turing machine, and $\mttx_1,...,\mttx_k\in \{0,1\}^*$ be $k$ bit strings. The computation of $\cal{M}$ given $\mttx_1,...,\mttx_k$ as inputs may halt or not. If it halted, it took some finite amount of time steps to get there, and we denote this number by $\TIME(\cal{M};\mttx_1,...,\mttx_k)$ --- if the TM did not halt, this function outputs $\infty$. In the case it halts, the output of $\cal{M}$ is what is written in the output tape in the end of the computation,   so $\cal{M}$ defines a partial function $\cal{M}\colon (\{0,1\}^*)^k\to \{0,1\}^*$.
	
	Given a function $f\colon \mathbb{N}^k\to \mathbb{N}$, we say that $\cal{M}$ runs in $f$-time if for every $\mttx_1,...,\mttx_k\in \{0,1\}^*$, we have
	\[
	\TIME(\cal{M};\mttx_1,...,\mttx_k)\leq f(|\mttx_1|,...,|\mttx_k|),
	\]
	where $|\cdot|$ is, again, the word length function (in particular, $\cal{M}$ needs to halt regardless of the input) --- this is denoted by $\TIME(\cal{M})\leq  f$, and if this is true only up to some universal constant $C'>0$, then we denote it by $\TIME(\cal{M})=O(f)$ (see Remark \ref{rem:asymptotic_notation} for our asymptotic notation conventions). As is common, we say that $\cal{M}$ runs in \emph{polynomial time} if there is some constant $C>0$  such that for every $\mttx_1,...,\mttx_k\in \{0,1\}^*$, $\TIME(\cal{M};\mttx_1,...,\mttx_k)\leq C|\mttx_1|^C\cdot...\cdot |\mttx_k|^C+C$ --- this is often denoted by $\TIME(\cal{M})=\poly(|\mttx_1|,...,|\mttx_k|)$. Similarly, we say that it runs in \emph{exponential time} if $\TIME(\cal{M})=2^{\poly(|\mttx_1|,...,|\mttx_k|)}$.
	For $r<k$, we often use the notation $$\TIME(\cal{M};\mttx_1,...,\mttx_r,\cdot,...,\cdot)=\sup \{\TIME(\cal{M};\mttx_1,...,\mttx_r,\mttx_{r+1},...,\mttx_k)\mid \mttx_{r+1},...,\mttx_k\in \{0,1\}^*\}\;,$$
	where we emphasize that the supremum on the right-hand side is taken over inputs $x_{r+1},\ldots,x_k$ of arbitrary (unbounded) length. 
\end{definition}

\subsubsection*{Descriptions and description length}
Often Turing machines are fed as input to other Turing machines, so they  need to be encoded somehow. As we remarked in the encoding part of this Prelude, we do not describe this encoding in detail (only assume it satisfies the condition appearing in Section 3.1 of \cite{MIPRE}). But, as we \textbf{do care} about running times, following the size of these encodings and the way they change with the various transformations applied on them is necessary. To that end,
\begin{definition}[Description length]\label{def:Description_length}
	Given a Turing machine $\cal{M}$, let $\desc{M}$ be its \emph{description}, i.e., a bit string which encodes $\cal{M}$. Let $|\cal{M}|=|\desc{M}|$ be  the \emph{description length} of $\cal{M}$,  which is the number of bits in the encoding of $\cal{M}$. 
\end{definition}

\begin{remark}\label{rem:Python_code}
	Although we do not describe the fixed encoding of TMs which we use, it is helpful to think of it as follows: Take your favorite programming language, say, Python. Then, every Turing machine can be written as a function in Python (with the appropriate number of inputs). The code for this function is just a string of symbols, and using ASCII, can be translated to a string of bits. The \emph{description} of the TM is then the Python code for it (translated to bit strings using ASCII), and the \emph{description length} is the length of this code. 
	
	In various places in this paper, we ask one TM to calculate the description of another TM. By this, we mean retrieve the code for the appropriate algorithm, which is described in a high level fashion along the paper. 
\end{remark}

\subsubsection{Tailored normal form verifiers}
Let us motivate the  definition of normal form verifiers. The goal of the verifier is to encode an infinite sequence of games $\{\game_n\}_{n\in \mathbb{N}}$ using a finite amount of data.  This is a common theme in theoretical computer science, and is usually referred to as \emph{uniform generation}. So, we want a finite object that ``calculates''  a function $n\mapsto \game_n$. A natural choice would be an algorithm (i.e., Turing machine) $\verifier$ which on input $n$ outputs the full description of $\game_n$, according to some predefined encoding of underlying graphs, length functions, distributions over edges (which must be rational to be finitely described), and truth tables of decision functions. 

We use a {different type} of encoding which is focused on the procedural manifestation of the $n^{\rm th}$ game. Procedurally, games require both a sampling mechanism of an edge $\mathtt{xy}\in E$ (also known as a pair of questions), and the calculation of the decision predicate $D_\mathtt{xy}(\cdot)=D(\mathtt{x},\mathtt{y},\cdot,\cdot)$. Hence, our verifiers will consist of algorithms that perform the sampling and the decision process required in the $n^{\rm th}$ game. 
There is a subtlety here, which is that the resulting game needs to be tailored (Definition \ref{defn:tailored_games}) --- a restriction not present in \cite{MIPRE}. 
To model this we introduce  {two} additional Turing machines in the definition of normal form verifiers compared to \cite{MIPRE}. One of them calculates the (answer) length functions $\ell^\frR,\ell^\frL$ of $\game$, and the other calculates the controlled linear constraints function  $L_{\mathtt{xy}}$.

For the sampling procedure, $\sampler$ will be a randomized Turing machine that on input $n$ outputs a pair of bit strings $\mathtt{x},\mathtt{y}$ interpreted as two vertices in $V$. 
A (bounded running time) randomized Turing machine  can be assumed to first read a string of random bits from its randomness (whose length depends on the input $n$), and then apply  a deterministic algorithm on the input that consists of $n$ and the string of random bits, to finally produce $\mathtt{x},\mathtt{y}$. 
Though this is a good benchmark for what a sampler is, for compression (Theorem \ref{thm:compression}) to work we need the sampler to be able to provide us with additional details on its inner workings: For example, the number of random bits it uses in the $n^{\rm th}$ game,  partial  computations of its output, and so on --- this appears in Definitions \ref{defn:h-level_CL_Sampler} and \ref{defn:typed_h-level_CL_Sampler}. At this point, let us stick to the simpler to follow definition. 

For the decision algorithm, since $\game_n$ is tailored, we can assume it is done in two steps. First, there is a Turing machine $\length$ which takes the index of the game $n$ and a vertex $x\in V$ as input, and outputs  $\ell^\frR(\mathtt{x}),\ell^\frL(\mathtt{x})$. Then, another Turing machine $\linproc$ takes $\mathtt{xy}$ and a bit string $\gamma^\frR=a^\frR b^\frR$ and calculates $L_{\mathtt{xy}}(\gamma^\frR)$. 
Finally, a canonical Turing machine $\decider$ takes as input a suggested answer $\gamma=a^\frR a^\frL b^\frR b^\frL$ together with the lengths $\ell^\frR(\mathtt{x}),\ell^\frL(\mathtt{x}),\ell^\frR(\mathtt{y}),\ell^\frL(\mathtt{y})$ and the sequence of linear constraints $L_{\mathtt{xy}}(\gamma^\frR)$. It first  checks that the restrictions $a^\frR,a^\frL,b^\frR,b^\frL$ of $\gamma$ are of the appropriate length, and that the linear constraints in $L_{\mathtt{xy}}(\gamma^\frR)$ are properly formatted. Then, it verifies that  $\gamma$ satisfies the constraints in $L_{\mathtt{x}}(\gamma^\frR)$. We now describe this encoding more rigorously.

\begin{definition}[Sampler]\label{def:sampler}
	A sampler $\sampler$ is a $1$-input randomized Turing machine that gets as input an integer $n$ in binary,\footnote{i.e., $\overline{n}\in \{0,1\}^*$ is the input to $\sampler$, as explained in Definition \ref{def:unary_encoding_int}.} and outputs a pair of bit strings $\mathtt{x},\mathtt{y}$.
\end{definition}

\begin{remark}
	We later restrict the family of samplers that we consider (see Definitions \ref{defn:h-level_CL_Sampler} and \ref{defn:typed_h-level_CL_Sampler}), and assume, given $\sampler$, to have access to certain subroutines of its calculation. 
\end{remark}
\begin{definition}[Answer length calculator]\label{defn:answer_length_calc}
	An \emph{answer length calculator} $\length$ is a  $3$-input Turing machine. The input tuple $(n,\mathtt{x},\kappa)$ consists of an integer $n$ in binary (Definition \ref{def:unary_encoding_int}),  a bit string $\mathtt{x}$, and a symbol $\kappa\in \{\frR,\frL\}$. 
\end{definition}

\begin{remark}\label{rem:length_calculator_interpretation}
	\ 
	\begin{itemize}
		\item We  expect $\mathtt{x}$ in the above definition of $\length$ to be the name of one of the vertices sampled by $\sampler$.
		\item The input  $\kappa$ should actually be a single bit $0$ or $1$, where $0$ is interpreted as $\frR$ (namely, $\frR$ is encoded as $0$)   and $1$ is interpreted as $\frL$ (namely, $\frL$ is encoded as $1$). The reason we use the $\frR$ and $\frL$ symbols is mainly for readability, as is clarified in the next clause.
		\item The \textbf{decoded} (Definition \ref{defn:the_alphabet}) output of $\length$ is interpreted as the 
		\textbf{unary} representation of an integer (Definition \ref{def:unary_encoding_int}), which in turn, indicates the length functions in the $n^{\rm th}$ game. This is done as follows (and is repeated in the description of the canonical decider $\decider$, Definition \ref{def:canonical-decider}): Say that $\mtty\in \{0,1\}^*$ is the output of $\length(n,\mathtt{x},\kappa)$. First, we decode $\mtty$ using $\dec$ from Definition \ref{defn:the_alphabet} --- resulting in an element $\dec(\mtty)=\mttz$  in $\{0,1,\sqcup\}^*\cup\{\frak{error}\}$. If $\mttz=\frak{error}$, the decider rejects. Otherwise, it uses the \textbf{length} of $\mttz$, $|\mttz|$,  as the readable answer length  $\ell^\frR(\mathtt{x})$ if $\kappa=\frR$, and  as the linear answer length $\ell^\frL(\mathtt{x})$ if $\kappa=\frL$, in the $n^{\rm th}$ game.
	\end{itemize}
\end{remark}

\begin{definition}[Linear constraints processor]\label{def:linear_constraints_processor}
	A \emph{Linear constraints processor} $\linproc$ is a $5$-input Turing machine. The input tuple $(n,\mathtt{x},\mathtt{y},a^\rvar,b^\rvar)$ consists of an integer $n$ (in binary, Definition \ref{def:unary_encoding_int}), signifying the index of the game, and four bit strings $\mathtt{x},\mathtt{y},a^\rvar,b^\rvar$.  
\end{definition}

\begin{remark}\label{rem:interpretation_of_lin_proc}
	Note that \textbf{any} $5$-input Turing machine can play the role of a linear constraints processor, in particular, one that does not halt. This is important for the way compression is used to deduce Theorem \ref{thm:tailored_MIP*=RE}.
	In the above definition of $\linproc$, the input bit strings $\mathtt{x},\mathtt{y}$ are expected to be the endpoints of the edge sampled by $\sampler$. The bit strings $a^\rvar$ and $b^\rvar$ are expected to be the restrictions of $\gamma$ to the readable variables at $\mathtt{x}$ and $\mathtt{y}$ respectively (which we denoted by $\gamma^\frR$ beforehand). 
	The output of $\linproc$ is expected to be (the encoding of) a sequence of bit strings $(c^1,...,c^k)$, that will be interpreted by $\decider$ as linear constraints on $\gamma$ --- namely, $c_i$ is the $i^{\rm th}$ row of a binary matrix, representing  a system of linear equations over $\FF_2$. This is done  by encoding first the alphabet $\{0,1,\sqcup\}$ as pairs of bits (as is done in Definition~\ref{defn:the_alphabet}), and then writing $c^1\sqcup c^2\sqcup...\sqcup c^k$ as the encoded version.
\end{remark}

\begin{definition}[Canonical Decider]\label{def:canonical-decider}
	The \emph{canonical decider} is a $9$-input Turing machine $\decider$ that either accepts (i.e., outputs $1$) or rejects (i.e., outputs $0$). The input $9$-tuple of $\decider$ is  
	$$(\ell^\rvar_a,\ell^\lvar_a,\ell^\rvar_b,\ell^\lvar_b,a^\rvar,a^\lvar,b^\rvar,b^\lvar,L)\ ,$$
	and all are bit strings.
	The canonical decider works in several steps. First, it checks that the inputs are properly formatted. This includes checking  that 
	\[
	\dec(\ell^\frR_a)\ ,\ \dec(\ell^\frR_b)\ ,\ \dec(\ell^\frL_a)\ ,\ \dec(\ell^\frL_b)\ ,\ \dec(L)\ \neq\  \frak{error}\ ,
	\]
	where $\dec$ is the decoding function from Definition \ref{defn:the_alphabet}. Then, $\decider$ checks that
	\[
	|a^\rvar|=|\dec(\ell^\frR_a)|\ ,\ |a^\lvar|=|\dec(\ell^\frL_a)|\ ,\ |b^\rvar|=|\dec(\ell^\frR_b)|\ ,\ |b^\lvar|=|\dec(\ell^\frL_b)|\ ,
	\]
	and lets 
	\[
	\Delta=|\dec(\ell^\rvar_a)|+|\dec(\ell^\lvar_a)|+|\dec(\ell^\rvar_b)|+|\dec(\ell^\lvar_b)|+1\ .
	\]
	Then, it checks that 
	\[
	\dec(L)=c^1\sqcup...\sqcup c^k
	\]
	for some sequence of bit strings $c^1,...,c^k\in \{0,1\}^*$  each of which of length $\Delta$, namely
	\[
	\forall 1\leq i\leq k\ \colon \ \ |c^i|=\Delta\ .
	\]
	If $L$ is the empty string, then it is decoded to the empty sequence of constraints, which is assumed to be well formatted (and signifies the \emph{no constraints} situation).
	If the inputs are not properly formatted,  then $\decider$ rejects. Otherwise, let $w$ be the concatenation of the bit strings $a^\rvar,a^\lvar,b^\rvar,b^\lvar$ together with an extra $1$ at the end, namely $w=a^\rvar a^\lvar b^\rvar  b^\lvar  1$. Since the inputs are well formatted,  $w$ and $c^i$ are bit strings of the same length $\Delta$. The canonical decider $\decider$ evaluates the dot product (over $\FF_2$, namely $\pmod 2$) between $c^i$ and $w$; i.e., 
	\[
	\forall 1\leq i\leq k\ \colon \ \ \langle c^i ,w\rangle =\sum_{j=1}^\Delta c^i_jw_j\;.
	\]
	Then, $\decider$ accepts if all of the above dot products are zero, and rejects otherwise.
\end{definition}
\begin{remark}
	In the above definition of $\decider$, the inputs $\ell^\rvar_a$, $\ell^\lvar_a$, $\ell^\rvar_b$ and $\ell^\lvar_b$ are expected to be the outputs of $\length(n,\mathtt{x},\frR)$, $\length(n,\mathtt{x},\frL)$, $\length(n,\mathtt{y},\frR)$ and $\length(n,\mathtt{y},\frL)$ respectively, where $\mathtt{x},\mathtt{y}$ is the pair sampled by $\sampler$. As mentioned in Remark \ref{rem:length_calculator_interpretation}, (the decodings of) these outputs are expected to be the (unary representation of the) readable and linear answer lengths  $$\ell^\frR(\mathtt{x}),\ell^\frL(\mathtt{x}),\ell^\frR(\mathtt{y}),\ell^\frL(\mathtt{y})$$ in the encoded game.
	If $\gamma\colon S_{\mathtt{xy}}\to \FF_2$ is the answer produced by running the  strategy $\strategy$, then we use the notation of \eqref{eq:a_b_to_gamma} to obtain 
	\[
	a^\rvar=\gamma|_{S_\mathtt{x}^\frR}\; ,\quad a^\lvar=\gamma|_{S_\mathtt{x}^\frL}\; ,\quad b^\rvar=\gamma|_{S_\mathtt{y}^\frR}\quad\textrm{and}\quad b^\lvar=\gamma|_{S_\mathtt{y}^\frL}\;.
	\]
	Hence, the bit string $w$ is exactly the extension of $\gamma$ such that $\gamma(\sJ)=1$.
	The last input $L$ is expected to be the output of $\linproc(n,\mathtt{x},\mathtt{y},a^\rvar,b^\rvar)$. We intentionally did not require this output to be formatted in a specific way, and thus $\decider$ needs to check on its own that this bit string is indeed an encoding of a sequence $(c^1,c^2,...,c^k)$, where each $c^i$ is a bit string of length $\Delta$.
\end{remark}

\begin{definition} [Tailored normal form verifiers]
	\label{def:normal-ver}
	A \emph{tailored normal form verifier} (TNFV) is a quadruple of Turing machines $\verifier = (\sampler,\length, \linproc, \decider)$,
	where $\sampler$ is a sampler as in Definition \ref{def:sampler}, $\length$ is an answer length calculator as in Definition \ref{defn:answer_length_calc}, $\linproc$ is a linear constraint processor as in Definition \ref{def:linear_constraints_processor}, and $\decider$ is the canonical decider as in Definition \ref{def:canonical-decider}.
	
	Although $\decider$ is fixed, we keep it in the notation.
\end{definition}

Note that while the quadruple $\verifier$ {seems} to encode an infinite sequence of tailored games, it may not. This is because we did not restrict them enough --- e.g., the sampler, answer length calculator and linear constraints processor may never halt (as opposed to the canonical decider that always halts, and in time which is linear in its input length). 
This leads us to the following.

\begin{definition}[The $n^{\rm th}$ game defined by  a tailored normal form verifier]
	\label{def:normal-game}
	Let $\verifier = (\sampler,\length,\linproc, \decider)$ be a TNFV (Definition \ref{def:normal-ver}), and let $n$ be a positive integer. Assume:
	\begin{itemize}
		\item The sampler $\sampler({n})$, which is a randomized TM, always halts in at most $T\in \mathbb{N}$ time steps. In addition, by Definition \ref{def:sampler}, when $\sampler$ halts,  (the encoding) of its output is a pair of bit strings $\mttx\sqcup \mtty$.
		\item The answer length calculator $\length(n,\mttx,\kappa)$ halts for every bit string $\mttx$  of length at most $T$, and $\kappa\in \{\frR,\frL\}$.
		\item The linear constraints processor $\linproc(n,\mttx,\mtty,a^\frR,b^\frR)$ halts for every pair $\mttx,\mtty$ of bit strings of length at most $T$ and every pair of bit strings $a^\frR,b^\frR$ of lengths $|\dec(\length(n,\mttx,\frR))|$ and $|\dec(\length(n,\mtty,\frR))|$ respectively,  where  $\dec$ is the decoding function (Definition \ref{defn:the_alphabet}) and $\abs{\cdot}$ the \emph{word length} function.
	\end{itemize}
	Then $\verifier_{n}$, the \emph{$n^{\rm th}$ game corresponding to $\verifier$}, is  the following tailored non-local game:
	As $\sampler({n})$ runs for at most $T$ steps, the output pair $\mathtt{x},\mathtt{y}$ consists of bit strings of length at most $T$. Then, the vertex set $V$ of the graph underlying $\verifier_n$ consists of  all bit strings of length at most $T$ ---  indeed, the output of $\sampler({n})$ will always be some ordered pair from $V$. The edge set $E$  will consist of all pairs $\mttx\mtty\in V\times V$ that are possible outputs  of  $\sampler({n})$, and  $\mu(\mathtt{xy})$ is the probability $\mathtt{x}\sqcup\mathtt{y}$ was the output of  $\sampler({n})$.
	
	As $\length({n},\mttx,\kappa)$ halts whenever $|\mttx|\leq T$ and $\kappa\in \{\frR,\frL\}$, the readable length at $\mttx\in V$ can be defined to be $\ell^\frR(\mttx)=|\dec(\length({n},\mttx,\frR))|$ and the unreadable length at $\mttx$ can be defined to be $\ell^\frL(\mttx)=|\dec(\length({n},\mttx,\frL))|$.
	
	For any $\gamma\colon S_\mathtt{xy}\to \FF_2$, let us use the notation in \eqref{eq:a_b_to_gamma}, namely  
	\[
	a^\rvar=\gamma|_{S_\mathtt{x}^\frR}\; ,\quad a^\lvar=\gamma|_{S_\mathtt{x}^\frL}\; ,\quad b^\rvar=\gamma|_{S_\mathtt{y}^\frR}\quad\textrm{and}\quad b^\lvar=\gamma|_{S_\mathtt{y}^\frL}\;.
	\]
	The output of  $\linproc({n},\mathtt{x},\mathtt{y},a^\rvar,b^\rvar)$ is either (an encoding of) a sequence of bit strings $c^1\sqcup ...\sqcup c^k$ of lengths $$\Delta=\ell^\frR(\mathtt{x})+\ell^\frL(\mathtt{x})+\ell^\frR(\mathtt{y})+\ell^\frL(\mathtt{y})+1\;,$$ or not. 
	If not, then we let $L_{\mathtt{xy}}(\gamma^\frR)$ be the singleton $\{\sJ\}$ (which translates to definite rejection). 
	Similarly, if one of  the decodings $\dec(\length(\overline{n},\mttx,\frR)),\dec(\length(\overline{n},\mttx,\frL)),\dec(\length(\overline{n},\mtty,\frR))$ or $\dec(\length(\overline{n},\mtty,\frL))$ is $\frak{error}$, then $L_{\mttx\mtty}(\gamma^\frR)$ will also be the singleton $\{\sJ\}$. 
	If the output is well formatted, then we can interpret each term $c^i$ in the sequence as an indicator function $c^i\colon S_{\mathtt{xy}}\cup \{\sJ\}\to \FF_2$. Then we can add $c^i$ to $L_{\mathtt{xy}}(\gamma^\frR)$. 
	This way we get some controlled linear constraint function $L_{\mathtt{xy}}\colon \FF_2^{S_\mathtt{x}^\frR\cup S_\mathtt{y}^\frR}\to \FF_2^{\FF_2^{S_{\mathtt{xy}}\cup \{\sJ\}}}$. 
	Note that the same indicator may appear more than once in the output of $\linproc({n},\mathtt{x},\mathtt{y},a^\rvar,b^\rvar)$; namely, there may be $i\neq j$ such that $c^i=c^j$.  But, this does not affect the function $L_{\mathtt{xy}}$ nor the decision process of the canonical decider $\decider$. 
	
	All in all, it is straightforward to check that the canonical $D_{\mathtt{xy}}(\gamma)$ from Definition \ref{defn:tailored_games} agrees with  the output of 
	\begin{equation} \label{eq:decision_predicate_induced_by_TNFV}
	\decider(\length({n},\mathtt{x},\frR),\length({n},\mathtt{x},\frL),\length({n},\mathtt{y},\frR),\length({n},\mathtt{y},\frL),a^\rvar,a^\lvar,b^\rvar,b^\lvar,\linproc({n},\mathtt{x},\mathtt{y},a^\rvar,b^\rvar))\;.
	\end{equation}
\end{definition}
\begin{remark}\label{rem:n^th_game_less_restrictive_setup}
	A TNFV $\verifier$ that satisfies the three bullets from Definition \ref{def:normal-ver}, is said to have a well defined corresponding $n^{\rm th}$ game $\verifier_n$.
	We often claim that transformations on TNFVs have a combinatorial effect on  the level of  $\verifier_n$ \emph{when it is defined}, by which we mean the above restrictions apply.
\end{remark}

\begin{definition}[$\lambda$-bounded tailored normal form verifiers]
	\label{def:lambda}
	Let $\lambda \in \N$ be an integer.
	A tailored normal form verifier $\verifier = (\sampler,\length, \linproc,\decider)$ is
	\emph{$\lambda$-bounded} if the following two conditions hold
	\begin{enumerate}
		\item Let $n$ be a positive integer (in binary). The running times of $\sampler({n})$, $\length({n},\cdot,\cdot)$ and $\linproc({n},\cdot,\cdot,\cdot,\cdot)$  are bounded by $n^\lambda$ for $n \ge 2$.\footnote{We do
			not require the bound to hold for $n=1$, as $1^\lambda$ is always $1$ and
			hence it is usually not satisfied. } Namely
		\[
		\forall 2\leq  {n}\in \{0,1\}^*\ 
		,\; \forall \mttx,\kappa,\mtty,a^\frR,b^\frR\in \{0,1\}^*\  \colon \ \ \TIME(\sampler;{n})\ ,\ \TIME(\length; {n},\mttx,\kappa)\ ,\ \TIME(\linproc; {n},\mttx,\mtty,a^\frR,b^\frR)\ \leq\  n^\lambda\;.
		\]
		\item The description length $\abs{\verifier}$ of the verifier $\verifier$  (Definition \ref{def:Description_length}) is bounded by $\lambda$.
	\end{enumerate}
\end{definition}
\begin{remark}\label{rem:properties_of_lambda_bounded_TNFVs}
	A few things to note about $\lambda$-bounded TNFVs:
	\begin{itemize}
		\item First, for every $n$, the $n^{\rm th}$ game of a $\lambda$-bounded TNFV is well defined (Remark \ref{rem:n^th_game_less_restrictive_setup}). Namely, such verifiers do define an infinite sequence of tailored games in a uniform way.
		\item At this point, it is not clear what motivates the above running time restriction on $\sampler, \length$ and $\linproc$. This will be clarified in Section \ref{sec:Descision_problems_complexity_classes}. Note the restriction is both very strict and somewhat weak. Usually,  the running time of a TM is bounded as a function of the total length of all of its inputs, and we expect the TMs to be efficient, namely run in time polynomial in these lengths. Here, we ask the running times to be bounded only in terms of the {first} input, which  means that these Turing machines often need to halt before even reading the entirety of their non-$n$ inputs (as they may be too long).  But, the running time is  {exponential} in this first input --- the length $|\overline{n}|$ is $\Theta(\log n)$ and thus $n^\lambda=2^{\lambda \log n}$ is exponential in it.
		\item An easy observation that is later used in a somewhat subtle manner, is that for $\lambda>\lambda'$, a $\lambda'$-bounded  TNFV is also a $\lambda$-bounded TNFV.
	\end{itemize}
\end{remark}

\subsection{Proving $\TMIP^*=\RE$: A protocol for the Halting Problem}\label{sec:proving_Tailored-MIP*=RE}

\subsubsection{Entanglement bounds}
The Halting Problem (HP) is the following decision problem:\footnote{See Section \ref{sec:Descision_problems_complexity_classes} for more on decision problems.} Given (an encoding of) a Turing machine $\cal{M}$, does it ever halt when run on the empty input?  If it does halt, then this can be certified in finite time --- just run the Turing machine. This shows that HP is recursively enumerable (in $\RE$). The reason HP is undecidable (namely, it is not recursive --- in $\mathsf{R}$) is because there is no  bound given $\cal{M}$ on the needed number of steps for it to halt, or alternatively a method for showing in finite time that it does not halt.\footnote{This is a theorem, first shown by Turing~\cite{turing1937computable}.} 

$\TMIP^*$ is the decision problem which was hinted at in Theorem \ref{thm:tailored_MIP*=RE}: Given (an encoding\footnote{The exact encoding does not matter at this point. As  normal form verifiers were defined by now, we can assume the game $\game$ is encoded as a \textbf{pair} consisting of a $\lambda$-bounded normal form verifier $\verifier$ and an index $n\in \mathbb{N}$, and then $\game=\verifier_n$ as defined in Definition \ref{def:normal-game}.} of) a tailored game $\game$, does it have a perfect $Z$-aligned permutation strategy that commutes along edges ($\ZPC$), or does every (quantum) strategy for it have value at most $\nicefrac{1}{2}$? These types of decision problems are called ``promise languages'', as we are not categorizing all possible tailored games, but promised that the input either has a perfect $\ZPC$ strategy or is far from having good strategies. At first glance, it is not clear what is complicated about deciding this. To see that, let us demonstrate why $\TMIP^*$ is in $\RE$. For every dimension $d$, we can define a finite $\nicefrac{1}{d}$-net of quantum strategies in the set of all $d$-dimensional strategies. This provides a countable sequence of strategies, and it can be proven that the $\limsup$ of the value of $\game$ against these strategies is indeed $\val^*(\game)$. Thus, if $\game$ has a perfect $\ZPC$ strategy, then in particular this sequence tends to $1$, and since this sequence can be calculated it will certify that $\val^*(\game)>\nicefrac{1}{2}$, which implies we are in the complete case. The complexity comes exactly from the fact that there is no clear $d$ that depends on (the encoding of) $\game$ which is the correct dimension we should look up to. This leads to the following definition:
\begin{definition}[Entanglement requirements of a game]
	\label{def:ent}
	Given a game $\game$ and a threshold $\nu \in [0, 1]$, let $\Ent(\game, \nu)$ denote the
	minimum integer $d$ such that there exists a $d$-dimensional (synchronous, quantum)
	strategy $\strategy$ whose value against $\game$ is at least $\nu$. 
	If there is no such strategy, then define $\Ent(\game, \nu)$ to be $\infty$.
\end{definition}

\subsubsection{Compression}

We can now formulate the compression theorem, and deduce using it Theorem \ref{thm:tailored_MIP*=RE}. The idea behind compression is to substitute a $\lambda$-bounded tailored normal form verifier $\verifier$ by a $\lambda$-bounded tailored normal form verifier $\verifier'$ that simulates it with exponential speedup. Namely, perfect $\ZPC$ strategies for $\verifier_{2^n}$ translate to perfect $\ZPC$ strategies for $\verifier'_n$, and almost perfect quantum strategies for $\verifier'_n$ translate to almost perfect quantum strategies for $\verifier_{2^n}$.\footnote{Almost perfect strategies are discussed in Section \ref{sec:compress_toolbox}. Results that  translate almost perfect strategies of one game to another are often deep and technical, and usually use ideas from the theory of Robustness of games (see Definition \ref{defn:robustness}). Such results sit at the heart of compression.} In addition, there is a lower bound on the entanglement requirements of $\verifier'_n$ which is exponential in $n$ and independent of $\verifier$. In the body of the text we  prove a slight variation on the following, see Theorem \ref{thm:h_level_compression}. The proved variation assumes an extra condition on $\verifier$ (i.e., it having a sampler which is $h$-level conditionally linear, as defined in Definition \ref{defn:h-level_CL_Sampler}), but ensures that $\verifier'$ also satisfies the same extra condition. This change does not effect the deduction of Theorem \ref{thm:tailored_MIP*=RE}, as can be verified by the reader --- one uses $\Compress_\LevelConstant$ from Theorem \ref{thm:h_level_compression} instead of $\Compress$ from Theorem \ref{thm:compression}.

\begin{theorem}[Compression theorem for tailored games]
	\label{thm:compression}
	There exists a universal positive integer constant $C$ and a polynomial time $2$-input Turing machine $\Compress$ that takes as input a
	TNFV $\verifier=(\sampler,\length,\linproc,\decider)$ and a positive integer $\lambda$ (in binary), and outputs  a  TNFV $\Compress(\verifier,\lambda)=\verifier'= (\sampler^\lambda, \length^\lambda, \linproc',\decider)$, such that: 
	$\bullet$ $\sampler^\lambda$ and  $\length^\lambda$ depend only on $\lambda$, can be calculated from it in time $\polylog(\lambda)$, and run in time $\poly(n,\lambda)$.\footnote{By saying that their running time is $\poly(n,\lambda)$, we mean that no matter what the other inputs are, the runtime of $\sampler^\lambda(n)$ and $\length^\lambda(n,\mttx,\kappa)$ is bounded only by $c\cdot(n^c+\lambda^c)$ for some universal constant $c$.} In addition,  given that $\mttx$ is a possible output of $\sampler^\lambda(n)$, and that $\kappa\in \{\frR,\frL\}$,  the output of $\length^\lambda(n,\mttx,\kappa)$  never decodes (Definition \ref{defn:the_alphabet}) to an $\frak{error}$ sign.  
	$\bullet$ $\linproc'$ depends on both $\lambda$ and $\verifier$,   it can be calculated from them in time $\poly(\log \lambda, |\verifier|)$, and  runs in $\poly(n,\lambda)$-time.
	$\bullet$  The canonical decider $\decider$ (Definition \ref{def:canonical-decider}) is fixed  and runs in  time  which is linear in its input length.
	$\bullet$ If $\verifier$ is $\lambda$-bounded, then 
	$\verifier'$, the output of $\Compress$,  satisfies for all $n \geq C$,
	\begin{enumerate}
		\item \label{enu:compr-completeness} \textbf{Completeness}: If $\verifier_{2^n}$
		has a perfect $Z$-aligned permutation strategy that commutes along edges ($\ZPC$ strategy), then so does $\verifier'_n$.
		\item \label{enu:compr-soundness} \textbf{Soundness}:
		$\Ent(\verifier_{n}',\frac{1}{2}) \geq \max \big \{
		\Ent(\verifier_{2^n}, \frac{1}{2}), 2^{2^{\lambda n}-1}
		\big \}$.
	\end{enumerate}
\end{theorem}

\begin{remark}\label{rem:poly(n,lambda)}
	It may seem technical, and even unnatural, that the time complexities of the sampler, answer length calculator and linear constraints processor of the compressed verifier are $\poly(n,\lambda)$. One would expect $\verifier'$ to be $\lambda$-bounded, which requires a bound of the form  $n^\lambda$.  Note that for every $c>0$ there is a large enough $\lambda$ such that $n^\lambda$ upper bounds $c(n^c+\lambda^c)$ for $n\geq 2$ (cf.\  Lemma 12.4 in \cite{MIPRE}). This \emph{better than} $\lambda$-bounded condition is crucial for the Halting problem reduction to work out. Specifically, it is  used in Lemma \ref{lem:lambda}, which plays a key role in the reduction.
\end{remark}

\begin{remark}\label{rem:the_method_of_proving_compression}
	The formulation of the compression theorem, Theorem \ref{thm:compression}, is hiding the approach to prove it in some sense. 
	Disregarding the complexity theoretic part (which is critical, but independent of  what we are emphasizing now), the point is that we transform a game $\verifier_{2^n}$ to a game $\verifier'_n$ in a complete and sound way. The completeness and soundness that we prove are actually stronger than what the formulation reveals.
	
	The completeness that is actually proven is that any perfect $\ZPC$ strategy for $\verifier_{2^n}$ can be transformed into a perfect $\ZPC$ strategy for $\verifier'_n$.
	The soundness that we actually prove is that every strategy for $\verifier'_n$ with value $1-\eps$ can be perturbed so that a strategy with value $1-f(\eps)$ for $\verifier_{2^n}$ can be extracted out of it. 
	Following the bounds deduced on the function $f$ throughout the steps of compression, one can show that for $\eps<\nicefrac{1}{2}$, also $f(\eps)<\nicefrac{1}{2}$, namely $\val^*(\verifier_{2^n})<\nicefrac{1}{2}$ implies $\val^*(\verifier'_n)<\nicefrac{1}{2}$.  
	Furthermore, the entanglement needed to win $\verifier'_{n}$ with probability $1-\eps$ is (morally) the product of the entanglement  needed  to win $\verifier_{2^n}$ with probability $1-f(\eps)$ and $(1-f(\eps))\cdot 2^{2 ^{\lambda n}}$, which is substantially larger than  the maximum between them. Namely, we can  deduce something of the form $\Ent(\verifier'_n,1-\eps)\geq \Ent(\verifier_{2^n},1-f(\eps))\cdot (1-f(\eps))\cdot 2^{2^{\lambda n}}$ for all $\eps>0$.
	
	This viewpoint is  better for understanding the structure of  completeness and soundness proofs of the transformations associated with compression. We elaborate on this in Section \ref{sec:compress_toolbox}.
\end{remark}

\subsubsection{A $\TMIP^*$-protocol for the Halting Problem: Proving Theorem \ref{thm:tailored_MIP*=RE} assuming Theorem \ref{thm:compression}}
\label{sec:halting-protocol}

This section is devoted to the proof of our main theorem, Theorem \ref{thm:tailored_MIP*=RE}, assuming Compression, Theorem \ref{thm:compression}.
The idea in the reduction is to transform a Turing machine $\cal{M}$ and an integer $\lambda$ into a tailored normal form verifier $\verifier^{\cal{M},\lambda}$ that is a ``fixed point'' of the algorithm $\Compress$ from Theorem \ref{thm:compression} --- this approach is part of a long tradition of fixed point theorems in computation theory, cf.\ \cite{Rogers1987}; see also \cite{marks24recursive} for a broader perspective on the connection between compression techniques and undecidability. 
Then, we show that there is a constant $\lambda=\lambda(\cal{M})$ --- that is bounded by a polynomial in the description length of $\cal{M}$ --- such that $\verifier^{\cal{M},\lambda}$ is $\lambda$-bounded. 
Finally, using the properties described in Theorem \ref{thm:compression} and the fact that $\verifier^{\cal{M},\lambda}$ is a fixed point, we can choose $\game_{\cal{M}}=\verifier^{\cal{M},\lambda}_C$ --- the $C^{\rm th}$ game defined by the verifier $\verifier^{\cal{M},\lambda}$, as in Definition \ref{def:normal-game}, where $C$ is the constant promised by Theorem \ref{thm:compression} --- and it satisfies the requirements of our main theorem,  Theorem \ref{thm:tailored_MIP*=RE}.

Recall that  we fixed some encoding of Turing machines in Section \ref{sec:encodings}. In Definition \ref{def:Description_length}, we denoted by $\desc{M}$ a description of $\cal{M}$, namely a bit string encoding of $\cal{M}$ according to the aforementioned  encoding scheme of TMs. Furthermore, $|\cal{M}|=|\desc{M}|$, the description length  of $\cal{M}$, was the bit length of the description of $\cal{M}$.\footnote{In Remark \ref{rem:Python_code} we provided a helpful heuristic way of thinking about these objects --- $\desc{M}$ is the code of some function in a programming language that behaves exactly as $\cal{M}$, and $|\desc{M}|$ is the bit-length of this code.}
The following is an adaptation of the Turing machine $\cF$ described  in Section  12.2 of \cite{MIPRE}. Note that in our case $\cF$ plays the role of the linear constraints processor $\linproc$ and not the decider $\decider$, which is fixed in the tailored case to be the canonical one.

\begin{definition}   \label{fig:halt_f}
	Let $\cF$ be an $8$-input Turing machine. 
	Its input is $$(\desc{R}, \desc{M}, \lambda, n, \mathtt{x}, \mathtt{y}, a^\rvar, b^\rvar),$$ where $\cal{R}$ is
	an $8$-input Turing machine, $\cal{M}$ is a single input Turing machine, $\lambda$ and $n$ are integers in binary, and $\mathtt{x},\mathtt{y},a^\frR,b^\frR$ are bit strings. The description of  $\cal{F}$ is as follows:
	
	\begin{enumerate}[label=\textcolor{black}{(\arabic*)}, ref= (\arabic*)]
		\item Run $\cal{M}$ on the blank input for $n$ steps. If it halts, then return an empty tape. Continue otherwise.
		\item \label{clause:linprc_in_F_def} Compute the description $\ol{\linproc^{\cal{R},\cal{M},\lambda}}$ of the $5$-input Turing
		machine $\linproc^{\cal{R},\cal{M},\lambda}$ defined by
		\[
		\linproc^{\cal{R},\cal{M},\lambda}(\cdot,\cdot,\cdot,\cdot,\cdot) = \cal{R}(\desc{R}, \desc{M}, \lambda, \cdot,\cdot,\cdot,\cdot,\cdot),
		\]
		i.e.,  on input $(\cdot,\cdot,\cdot,\cdot,\cdot)$ the TM $\linproc^{\cal{R},\cal{M},\lambda}$ calculates the output of  $\cal{R}$ given
		input $(\desc{R}, \desc{M}, \lambda, \cdot,\cdot,\cdot,\cdot,\cdot)$.\footnote{In the heuristic viewpoint of Remark \ref{rem:Python_code}, this is the same as taking the code of the $8$-input function $\cal{R}$, and hard-coding the first three inputs of it to being $\desc{R},\desc{M}$ and $\lambda$. The resulting function has only $5$ free inputs, and is thus a $5$-input TM which can play the role of a linear constraints processor (Definition \ref{def:linear_constraints_processor}).} 
		\item Compute the descriptions $\ol{\sampler^\lambda}$ and $\ol{\length^\lambda}$ of the TMs $\sampler^\lambda$ and $\length^\lambda$ from Theorem \ref{thm:compression}, which are the sampler and answer length calculator that $\Compress(\cdot, \lambda)$ outputs regardless of which input normal form verifier it got. 
		\item Compute the description $\ol{\decider}$ of the canonical decider $\decider$ from Definition \ref{def:canonical-decider}. 
		\item \label{clause:def_of_VRMl_def_F} Let ${\verifier}^{\cal{R},\cal{M},\lambda}=({\sampler}^\lambda,{\length}^\lambda,{\linproc}^{\cal{R},\cal{M},\lambda},{\decider})$ be a TNFV.
		\item \label{clause:compressed_in_def_F} Compute the description $\ol{(\verifier^{\cal{R},\cal{M},\lambda})'}$ of the compressed verifier $$(\verifier^{\cal{R},\cal{M},\lambda})'=\Compress(\verifier^{\cal{R},\cal{M},\lambda},\lambda)=(\sampler^\lambda,\length^\lambda,(\linproc^{\cal{R},\cal{M},\lambda})',\decider)\ ,$$  where $\Compress$ is the algorithm discussed in Theorem \ref{thm:compression}.\footnote{Note that the sampler   and answer length calculator  of both  $\verifier^{\cal{R},\cal{M},\lambda}$ and $(\verifier^{\cal{R},\cal{M},\lambda})'$ are \textbf{the same}. This is because of the way $\Compress$ operates, and our choice of sampler $\sampler^\lambda$ and answer length calculator $\length^\lambda$ for $\verifier^{\cal{R},\cal{M},\lambda}$.}  
		\item \label{claus:output_of_F_def} Output
		$(\linproc^{{\cal{R},\cal{M},\lambda}})'(n, \mathtt{x}, \mathtt{y}, a^\rvar, b^\rvar)$; namely, simulate the operation of the  compressed linear constraints processor $(\linproc^{{\cal{R},\cal{M},\lambda}})'$  on the $5$-tuple input $(n, \mathtt{x}, \mathtt{y}, a^\rvar, b^\rvar)$, and provide the same output as it did.
	\end{enumerate}

\end{definition}

\begin{definition}[The Halting tailored normal form verifier]\label{defn:halting_norm_form_verifier}
	For every Turing machine $\cal{M}$ and  $\lambda\in\mathbb{N}$, define the linear constraints processor $\linproc^{\cal{M}, \lambda}$ to be the $5$-input Turing machine 
	\[
	\linproc^{\cal{M}, \lambda}(\cdot,\cdot,\cdot,\cdot,\cdot) = \cal{F}(\desc{F}, \desc{M},
	\lambda, \cdot,\cdot,\cdot,\cdot,\cdot)\;,
	\]
	where $\cF$ is the Turing machine from Definition \ref{fig:halt_f} --- note again, that this is just hard-coding the first three inputs of $\cF$ to being $\ol{\cF},\desc{M}$ and $\lambda$ respectively, which makes it into a $5$-input TM, and thus it can play the role of a linear constraints processor (Definition \ref{def:linear_constraints_processor}).
	
	Now, define the halting tailored normal form verifier corresponding to $\cal{M}$ and $\lambda$ to be 
	$$\verifier^{\cal{M},\lambda}=(\sampler^\lambda,\length^\lambda,\linproc^{\cal{M},\lambda},\decider)\;,$$
	where, again, $\sampler^\lambda$ and $\length^\lambda$ are the sampler and answer length calculator that $\Compress(\cdot, \lambda)$ always outputs (from Theorem \ref{thm:compression}).
\end{definition}

\begin{remark}\label{rem:props_of_halting_TNFV}
	Let us note some properties of the linear constraints processor $\linproc^{\cal{M},\lambda}$ from Definition \ref{defn:halting_norm_form_verifier}. Specifically, what is the output of 
	\[
	\linproc^{\cal{M},\lambda}(n,\mathtt{x},\mathtt{y},a^\rvar,b^\rvar)=\cF(\ol{\cF},\ol{M},\lambda ,n,\mathtt{x},\mathtt{y},a^\rvar,b^\rvar)
	\] 
	given that $\cal{M}$ does not halt in $n$ steps. 
	By inspecting  the (high-level) description of the Turing machine $\cal{F}$ from Definition \ref{fig:halt_f},
	one can see that the description  of $\linproc^{\cal{R},\cal{M},\lambda}$ computed by
	$\linproc^{\cal{M},\lambda}$ at Step \labelcref{clause:linprc_in_F_def} is the description of $\linproc^{\cal{M},\lambda}$ itself. 
	So,  the TNFV $\verifier^{\cal{R},\cal{M},\lambda}$ computed in Step \labelcref{clause:def_of_VRMl_def_F} is the Halting TNFV $\verifier^{\cal{M},\lambda}$ (Definition \ref{defn:halting_norm_form_verifier}). 
	Thus, as $\cal{M}$ does not halt in $n$ steps, $\linproc^{\cal{M},\lambda}(n,\mttx,\mtty,a^\frR,b^\frR)$ will get to  Step \labelcref{claus:output_of_F_def} and output the same output as $(\linproc^{\cal{M},\lambda})'(n,\mathtt{x},\mathtt{y},a^\rvar,b^\rvar)$, where $(\linproc^{\cal{M},\lambda})'$ is the linear constriants processor of $\Compress(\verifier^{\cal{M},\lambda},\lambda)$; namely,
	\[
	\linproc^{\cal{M},\lambda}(n,\mathtt{x},\mathtt{y},a^\rvar,b^\rvar)=(\linproc^{\cal{M},\lambda})'(n,\mathtt{x},\mathtt{y},a^\rvar,b^\rvar)
	\]
	whenever $\cal{M}$ does not halt in $n$ steps.
	This is the way in which $\linproc^{\cal{M},\lambda}$, and thus $\verifier^{\cal{M},\lambda}$, is a fixed point of $\Compress(\cdot,\lambda)$.
	Furthermore, for all $\cal{M}$ and $\lambda$ the linear constraints processor $\linproc^{\cal{M},\lambda}$ halts on all inputs, and $\verifier^{\cal{M},\lambda}$ is a tailored normal form verifier, though {not necessarily} $\lambda$-bounded. 
\end{remark}

\begin{lemma}\label{lem:VMLn_and_its_compression}
	Let $\cal{M}$ be a Turing machine, and $\lambda$ and $n$ positive integers. Recall the Halting TNFV $\verifier^{\cal{M},\lambda}$ from Definition \ref{defn:halting_norm_form_verifier}, and let $(\verifier^{\cal{M},\lambda})'=\Compress(\verifier^{\cal{M},\lambda},\lambda)$, where $\Compress$ is the transformation from Theorem \ref{thm:compression}.
	\begin{enumerate}[label=\textcolor{black}{\arabic*.}, ref= \arabic*.]
		\item \label{clause1:VLMn_and_its_compression}  The underlying graph $G^\lambda=(V^\lambda,E^\lambda)$, length functions $\ell^{\frR,\lambda},\ell^{\frL,\lambda}$ and distribution $\mu^\lambda$ over edges in   $\verifier^{\cal{M},\lambda}_n$ and $(\verifier^{\cal{M},\lambda})'_n$ --- the $n^{\rm th}$ games associated with $\verifier^{\cal{M},\lambda}$ and $(\verifier^{\cal{M},\lambda})'$ (Definition \ref{def:normal-game}) --- are {the same}. 
		\item \label{clause2:VLMn_and_its_compression}  If $\cal{M}$ does not halt in $n$ steps, then the games $\verifier^{\cal{M},\lambda}_n$ and $(\verifier^{\cal{M},\lambda})'_n$ are {the same}.
		\item \label{clause3:VLMn_and_its_compression}  If $\cal{M}$ halts in less than $n$ steps, then $\verifier^{\cal{M},\lambda}_n$ is the always accepting game  --- namely, $L_{\mathtt{xy}}(\gamma^\frR)$ is empty  regardless of $\gamma$ and $\mathtt{xy}\in E$, and thus $D_{\mathtt{xy}}$  accepts any $\gamma\colon S_{\mathtt{xy}}\to \{0,1\}$.
	\end{enumerate}
\end{lemma}
\begin{proof}
	Clause \ref{clause1:VLMn_and_its_compression} is immediate from the fact that the underlying graph, length functions and distribution over edges in Definition \ref{def:normal-game} depend only on the sampler and answer length calculator, and both $\verifier^{\cal{M},\lambda}$ and $(\verifier^{\cal{M},\lambda})'$ have the same sampler $\sampler^\lambda$ and same answer length calculator $\length^\lambda$.
	
	Clause \ref{clause2:VLMn_and_its_compression} is deduced from Remark \ref{rem:props_of_halting_TNFV}, which states that $$\linproc^{\cal{M},\lambda}(n,\mathtt{x},\mathtt{y},a^\rvar,b^\rvar)=(\linproc^{\cal{M},\lambda})'(n,\mathtt{x},\mathtt{y},a^\rvar,b^\rvar)$$ in case $\cal{M}$ does not halt in $n$ steps. Thus, as the length functions are the same for both games, $L_{\mathtt{xy}}$ is the same for $\verifier^{\cal{M},\lambda}_n$ and $(\verifier^{\cal{M},\lambda})'_n$. Since the rest of the data is the same as well, they are the exact same tailored game. 
	
	For clause \ref{clause3:VLMn_and_its_compression}, note that in this case $\linproc^{\cal{M},\lambda}(n,\mttx,\mtty,a^\frR,b^\frR)$ outputs an empty tape regardless of what $\mttx,\mtty,a^\frR$ or $b^\frR$ are. 
	By the properties of the answer length calculator in Theorem \ref{thm:compression}, for every vertex $\mttx$ in the underlying graph of $\verifier^{\cal{M},\lambda}_n$ and $\kappa\in \{\frR,\frL\}$ we are guaranteed that $\length^\lambda(n,\mttx,\kappa)$ does not decode to $\frak{error}$. Hence, every quadruple $a^\frR,a^\frL,b^\frR,b^\frL$ of respective lengths $|\dec(\length(n,\mttx,\frR)|,|\dec(\length(n,\mttx,\frL)|,|\dec(\length(n,\mtty,\frR)|,|\dec(\length(n,\mtty,\frL)|$ will make the canonical decider $\decider$ (Definition \ref{def:canonical-decider}) output $1$ on input 
	\[
	(\length(n,\mttx,\frR),\length(n,\mttx,\frL),\length(n,\mtty,\frR),\length(n,\mtty,\frL),a^\frR,a^\frL,b^\frR,b^\frL,\linproc^{\cal{M},\lambda}(n,\mttx,\mtty,a^\frR,b^\frR))\ .
	\]
	As described in Definition \ref{def:normal-game}, the decision function $D_{\mttx\mtty}(a^\frR,a^\frL,b^\frR,b^\frL)$ (using the notation of \eqref{eq:a_b_to_gamma})  of the $n^{\rm th}$ game $\verifier^{\cal{M},\lambda}_n$ agrees with the canonical decider in this setup. Namely,  $D_{\mathtt{xy}}$  accept every possible $\gamma\colon S_{\mathtt{xy}}\to \{0,1\}$, as claimed.
\end{proof}

\begin{corollary}\label{lem:dhalt-values}
	Let $\cal{M}$ be a Turing machine and $\lambda$ an integer.
	Then the halting TNFV $\verifier^{\cal{M},\lambda}$ (Definition \ref{defn:halting_norm_form_verifier})  has the following properties. For all $n \in \N$:
	\begin{enumerate}[label=\textcolor{black}{\arabic*.}, ref= \arabic*.]
		\item \label{clause1:dhalt-values} If $\cal{M}$ halts in $n$ steps, then
		$\verifier^{\cal{M},\lambda}_n$ has a perfect $Z$-aligned permutation strategy that commutes along edges ($\ZPC$).
		\item \label{clause2:dhalt-values} If $\cal{M}$ does not halt in $n$ steps, then
		$\verifier^{\cal{M},\lambda}_n$ has a perfect $\ZPC$ strategy if and only if  
		$(\verifier^{\cal{M},\lambda})'_n$ does, where $(\verifier^{\cal{M},\lambda})'=\Compress(\verifier^{\cal{M},\lambda},\lambda)$.
		Furthermore, under the same assumption (that $\cal{M}$ does not halt in $n$ steps), it holds that 
		\begin{equation*}
		\Ent\big(\verifier^{\cal{M},\lambda}_n, \frac{1}{2}\big) \,=\,
		\Ent\big((\verifier^{\cal{M},\lambda})'_n\ , \frac{1}{2}\big)\;.
		\end{equation*}
	\end{enumerate}
\end{corollary}

\begin{proof}
	For item \labelcref{clause1:dhalt-values}, by clause \ref{clause3:VLMn_and_its_compression} of Lemma \ref{lem:VMLn_and_its_compression}, $\verifier^{\cal{M},\lambda}_n$ always accepts. Thus,  any deterministic strategy (as in Example \ref{example:classical_perm_strategies}) is a perfect $Z$-aligned  permutation strategy for it. Since deterministic strategies are globally commuting, they in particular  commute along edges. 
	
	Item \labelcref{clause2:dhalt-values} follows directly from clause \labelcref{clause2:VLMn_and_its_compression} of Lemma \ref{lem:VMLn_and_its_compression}, since they are {the same tailored game}.
\end{proof}

\begin{lemma}
	\label{lem:lambda}
	There is a polynomial-time computable $\lambda=\lambda(\cal{M})$,  
	scaling as $\poly(|\overline{\machine}|)$, such that the verifier
	$\verifier^{\cal{M},\lambda}$ is
	$\lambda$-bounded. Moreover, the time complexities of $\sampler^\lambda,\length^\lambda$ and $\linproc^{\cal{M},\lambda}$ are $\poly(n, \abs{\overline{\cal{M}}})$.\footnote{Note that this extra condition is indeed a strengthening of being $\lambda$-bounded, since the dependence on the description length of $\cal{M}$ appears in the base and not the exponent --- recall Remark \ref{rem:poly(n,lambda)}.}
\end{lemma}

\begin{proof}[Proof sketch]
	This is a combination of:
	\begin{itemize}
		\item The observation from Remark \ref{rem:poly(n,lambda)}, i.e., that $\poly(n,\lambda)$ is dominated by $n^\lambda$ for any large enough $\lambda$. Similarly, $\polylog(\lambda)$ is dominated by $\lambda$.
		\item An accounting argument of the running time and description length of $\linproc^{\cal{M},\lambda}$ through the definition of $\cal{F}$ (Definition \ref{fig:halt_f}).
		\item The time bounds of the sampler, answer length calculator and linear constraint processor of the compressed verifier in Theorem \ref{thm:compression}.
	\end{itemize}  
	It is probably better for the readers to try and follow these calculations for themselves. In any case, a complete proof of the analogous claim appears in~\cite[Lemma 12.5]{MIPRE}. 
\end{proof}

\begin{proof}[Proof of Theorem \ref{thm:tailored_MIP*=RE}]
	For every Turing machine $\cal{M}$, let $\lambda=\lambda(\cal{M})$ be the parameter promised by Lemma \ref{lem:lambda}.  Let  $\verifier^{\cal{M},\lambda}$ be the tailored normal form verifier from Definition \ref{defn:halting_norm_form_verifier}. Then, let $\game_{\cal{M}}=\verifier^{\cal{M},\lambda}_C$ be the $C^{\rm th}$ game defined by $\verifier^{\cal{M},\lambda}$ (as in Definition \ref{def:normal-game}), where $C$ is the  constant promised in Theorem \ref{thm:compression}.
	
	First, let us show that the calculation of (the description of) $\verifier^{\cal{M},\lambda}_C$ takes at most $\poly(|\desc{M}|)$-time. 
	By Lemma \ref{lem:lambda}, calculating $\lambda$ takes $\poly(|\desc{M}|)$-time. 
	Now, calculating the description of $\sampler^\lambda$ and $\length^\lambda$ takes $\polylog(\lambda)$-time, which in turn is $\poly\log(|\desc{M}|)\leq \poly(|\desc{M}|)$. 
	Furthermore, the decider $\decider$ is fixed. Calculating the description of $\linproc^{\cal{M},\lambda}$ requires  $\poly(|\desc{F}|,|\desc{M}|,\log\lambda)$ which is again $\poly(|\desc{M}|)$ (as $|\desc{F}|$ is a constant). 
	Finally, fixing $n=C$ in all of these Turing machines adds at most a constant to their description. This proves that $\game_\cal{M}$ can be calculated in time polynomial in $|\desc{M}|$.
	
	By Lemma \ref{lem:lambda}, $\sampler^\lambda(C)$ runs in time $\poly(C,\lambda)=\poly(|\desc{M}|)$. Recall  Definition \ref{def:normal-game}.
	For the edge set $E$ and the distribution $\mu$ over it, Definition \ref{def:normal-game} took the pushforward along $\sampler^\lambda(C)$. This means that sampling according to $\mu$ is \textbf{exactly} running $\sampler^\lambda(C)$, and that takes $\poly(\desc{M}|)$-time. 
	By Lemma \ref{lem:lambda}, given $\gamma=(a^\rvar,a^\lvar,b^\rvar,b^\lvar)$,  calculating $\linproc^{\cal{M},\lambda}(C,\mathtt{x},\mathtt{y},a^\rvar,b^\rvar)$ takes at most $\poly(C,\lambda)=\poly(|\desc{M}|)$ time, and in particular its output length is bounded by   $\poly(|\desc{M}|)$. Also by Lemma \ref{lem:lambda}, $\length(C,\mathtt{x},t)$ takes at most $\poly(|\desc{M}|)$-time.
	Since $\decider$ runs in time linear in its input, the value 
	\[
	D_{\mathtt{xy}}(\gamma)=\decider(\length(C,\mathtt{x},\frR),\length(C,\mathtt{x},\frL),\length(C,\mathtt{y},\frR),\length(C,\mathtt{y},\frL),a^\rvar,a^\lvar,b^\rvar,b^\lvar,\linproc^{\cal{M},\lambda}(C,\mathtt{x},\mathtt{y},a^\rvar,b^\rvar)),
	\]
	can be calculated in time at most $\poly(|\desc{M}|)$. This proves \labelcref{clause:1_in_tailored_MIP*=RE} in Theorem \ref{thm:tailored_MIP*=RE}.

	Assume that $\cal{M}$ halts. Let $N$ be the number of time steps it takes $\cal{M}$ to halt. For every $n\geq N$,  by Corollary \ref{lem:dhalt-values}, $\verifier^{\cal{M},\lambda}_n$ has a perfect $\ZPC$ strategy. So, if $C\geq N$, then we are done. Otherwise, let $n$ be such that 
	$$\max(C,\log N)\leq n< N.$$ By Lemma \ref{lem:VMLn_and_its_compression},  $\verifier^{\cal{M},\lambda}_n=(\verifier^{\cal{M},\lambda})'_n$ in this case. By the compression theorem \ref{thm:compression}, since $n\geq C$, $(\verifier^{\cal{M},\lambda})'_n$ has a perfect $\ZPC$ strategy given that $\verifier^{\cal{M},\lambda}_{2^n}$ has one. 
	But $2^n\geq N$, and we already argued that these tailored games have perfect $\ZPC$ strategies. Hence, $\verifier^{\cal{M},\lambda}_n$ has a perfect $\ZPC$ strategy when $n\geq \max(C,\log N)$.  
	If $C\geq \log N$, then we are done.  Otherwise, we can iterate this argument and deduce the same for any $n\geq \max(C,\log\log N).$
	Since there exists some $t$ for which  $C\geq \underbrace{\log\dots\log }_{t-\textrm{times}}N$, we deduce that $\game_{\cal{M}}=\verifier^{\cal{M},\lambda}_C$ has a perfect $\ZPC$ strategy. This proves \labelcref{clause:2_in_tailored_MIP*=RE} in Theorem \ref{thm:tailored_MIP*=RE}.
	
	Assume that $\cal{M}$ does not halt. Then, by Lemma  \ref{lem:VMLn_and_its_compression},  
	\begin{equation}\label{eq:1mainthm}
	\verifier^{\cal{M},\lambda}_n=(\verifier^{\cal{M},\lambda})'_n
	\end{equation} 
	for every $n$. If $n\geq C$, then by the compression theorem \ref{thm:compression}, we have 
	\begin{equation}\label{eq:2mainthm}
	\Ent((\verifier^{\cal{M},\lambda})'_n,\nicefrac{1}{2})\geq \Ent(\verifier^{\cal{M},\lambda}_{2^n},\nicefrac{1}{2})\quad {\rm and}\quad \Ent((\verifier^{\cal{M},\lambda})'_n,\nicefrac{1}{2})\geq \underbrace{2^{2^{\lambda n}-1}}_{\geq 2^n}.
	\end{equation}
	So, for every positive integer $t$ we can deduce that 
	\[
	\begin{split}
	\Ent(\verifier^{\cal{M},\lambda}_C,\nicefrac{1}{2})&=_{\eqref{eq:1mainthm}}\Ent((\verifier^{\cal{M},\lambda})'_C,\nicefrac{1}{2})\\
	\geq_{\eqref{eq:2mainthm}}\Ent(\verifier^{\cal{M},\lambda}_{2^C},\nicefrac{1}{2})&=_{\eqref{eq:1mainthm}} \Ent((\verifier^{\cal{M},\lambda})'_{2^C},\nicefrac{1}{2})\\
	&\ \ \vdots\\
	\geq _{\eqref{eq:2mainthm}}\Ent(\verifier^{\cal{M},\lambda}_{\underbrace{{2^{\iddots^{2^C}}}}_{t-times}},\nicefrac{1}{2})&=_{\eqref{eq:1mainthm}}\Ent((\verifier^{\cal{M},\lambda})'_{\underbrace{{2^{\iddots^{2^C}}}}_{t-times}},\nicefrac{1}{2})\\
	&\geq_{\eqref{eq:2mainthm}} {\underbrace{{2^{\iddots^{2^C}}}}_{(t+1)-times}}.
	\end{split}
	\]
	Since this was true for every $t$, we can deduce that $\Ent(\verifier^{\cal{M},\lambda}_C,\nicefrac{1}{2})=\infty$, which in turn proves that $$\val^*(\game_\cal{M})=\val^*(\verifier^{\cal{M},\lambda}_C)<\nicefrac{1}{2}\;,$$
	proving \labelcref{clause:3_in_tailored_MIP*=RE} in Theorem \ref{thm:tailored_MIP*=RE}.
\end{proof}

The rest of the paper is devoted to the proof of  Compression, Theorem \ref{thm:compression}.

\section{The  compression toolbox}\label{sec:compress_toolbox}
In the previous section we provided the minimal amount of preliminaries so that $\TMIP^*=\RE$ (Theorem \ref{thm:tailored_MIP*=RE}) and Compression (Theorem \ref{thm:compression}) can be phrased, and so that the former can be deduced from the latter. 
This section provides additional preliminaries needed for the proof of Compression. Specifically, we introduce various technical tools that are used in the completeness and soundness analysis of the transformations on games that take part in Compression. Section~\ref{sec:toolbox-isometries} provides useful functional analytic definitions and facts. In Section~\ref{sec:toolbox-dist} we introduce a notion of distance between strategies; this notion takes into account the need to compare strategies in different dimensions through the use of isometries. In Section~\ref{sec:toolbox-dp} we consider a frequent transformation on PVMs, \emph{data processing}, and its effect on the distance measure. Section~\ref{sec:About_perm_strategies} contains useful lemmas for manipulating permutation strategies. Section~\ref{sec:composition_of_games} defines  transformations that can be applied on games, which will be used repeatedly in the paper --- specifically, sums, products and double covers of games. In Section \ref{sec:non-synch_setup}, we review the more general setup of non-synchronous strategies for (synchronous) games, and phrase an important Theorem  (Fact \ref{fact:non-synch_high_implies_synch_high}) due to the third author that allows one to move from value and entanglement bounds in the generalized setup back to ours.
Finally, in Section \ref{sec:Pauli_gp}, we recall the definition of $\WH_k$ the Pauli group acting on $k$-qubits, and the generalized Pauli basis test (originally due to Natarajan--Vidick \cite{natarajan2018two}, but here the version of de la Salle \cite{de_la_Salle_spectral_gap} is used); this is a robust self test (Definition \ref{defn:robustness}) that forces any almost perfect strategy for this game to be close to the unique non-commuting irreducible representation of $\WH_k$.

\subsection{Functional analytic preliminaries}
\label{sec:toolbox-isometries}

Let $\langle \cdot |\cdot\rangle$ be the standard euclidean inner product on $\complex^N$, namely 
\[
\forall \vec v,\vec w\in \complex^N\ \colon \ \ \langle \vec v|\vec w\rangle=\sum_{i=1}^N \ol{v_i}\cdot w_i\ ;
\]
note that when $\vec v$ and $\vec w$ are thought of as column vectors, their inner product is exactly $(\vec v)^*\cdot  \vec w$, with $*$ being the conjugate transposition and $\cdot$ the standard product of matrices.
An $N\times N$ complex matrix $A$ is said to be \emph{positive} (semi-definite) if for every $\vec v\in \complex^n$ we have $\langle \vec v|A\vec v\rangle\geq 0$.
Let $\tau=\frac{1}{N}\Tr$ be the normalized trace on $N\times N$ complex matrices. Every such matrix $A$ has a polar decomposition $UP$ where $U$ is unitary and $P$ is positive; the matrix $P$ is unique and often denoted by $|A|$ or $(A^*A)^{\nicefrac{1}{2}}$. By functional calculus (cf.\  \cite[Section I.4.1]{blackadar2006operator}), the $p^{\rm th}$ power of $|A|$, which we denote by $|A|^p$, is defined for every $p\geq 0$.
\begin{definition}[Normalized $p$-norms]\label{defn:normalized_p_norms}
	For every $p\geq 1$ we define the \emph{normalized $p$-norm} $\|A\|_p$ of a $N\times N$ complex matrix $A$ by $(\tau(|A|^p))^{\nicefrac{1}{p}}$. Specifically for the case of $p=2$, this norm is called the \emph{normalized Hilbert--Schmidt norm}; we denote it by $\|A\|_{hs}$, and we note that it is induced by the inner product $\langle A, B\rangle=\tau(A^*B)$. In addition, the case $p=\infty$ is the operator norm, namely $\|A\|_\infty=\|A\|_{op}=\max\{\|A\vec v\|\mid \vec v\in \complex^{n}, \|\vec v\|=1\}$, where $\|\cdot \|$ is the euclidean norm on $\complex^n$ induced by the inner product $\langle\cdot|\cdot\rangle$; the operator norm is well defined for non-square matrices as well.
\end{definition}
\begin{fact}[Useful equations and inequalities. Cf.\ Proposition 2.1 in \cite{quantum_soundness_tensor_codes} and Lemma 6.1 in \cite{GowersHatami}]\label{fact:useful_equations_and_ineqs}
	Let $A,B\in M_{N\times N}(\complex)$, and $p,q\in [1,\infty]$. Then:
	\begin{enumerate}[label=\textcolor{black}{(\arabic*)}, ref= (\arabic*)]
		\item \label{clause1:useful}\emph{Unitary invariance}: If $A$ is unitary, then $\|AB\|_{p}=\|B\|_{p}$.
		\item  \label{clause2:useful}\emph{Cauchy--Schwarz}: $|\tau(A^*B)|\leq \|A\|_{hs}\|B\|_{hs}$.
		\item  \label{clause3:useful} \emph{$1$-norm upper bound}: $|\tau(A)|\leq \tau(|A|)=\|A\|_1$.
		\item  \label{clause4:useful}\emph{H\"older's inequality}: If  $\nicefrac{1}{p}+\nicefrac{1}{q}=1$, then $\tau(|AB|)=\|AB\|_1\leq \|A\|_p\|B\|_q$.
		\item  \label{clause5:useful}\emph{Triangle inequality} ($\triangle$): $\|A+B\|_p\leq \|A\|_p+\|B\|_p$.
		\item  \label{clause6:useful}\emph{Monotonicity}: If $p\leq q$, then $\|A\|_p\leq \|A\|_q$.\footnote{This direction of monotonicity is due the {normalized trace} $\tau$. Without normalization, the monotonicity property is reversed.}
		\item  \label{clause7:useful} \emph{Sub-multiplicativity with operator norm}: $\|AB\|_{p}\leq \|A\|_{p}\|B\|_{op}$.\footnote{The case $p=1$ is covered by H\"older.}
	\end{enumerate}
\end{fact}
\begin{definition}[Projections, isometries and partial isometries]\label{defn:projections_isometries_partial_isometries}
	An (orthogonal) \emph{projection} is an operator (square complex matrix in the finite dimensional case) $A$ satisfying  $A^2=A=A^*$. An \emph{isometry} is a linear map $A$ between Hilbert spaces that satisfies $A^*A=\Id$. A \emph{partial isometry} $\omega\colon \complex^N\to \complex^M$ is a linear map such that both $\omega^*\omega$ and $\omega\omega^*$ are  projections. Any partial isometry can be written as  $\omega=\iota\circ \kappa^*$, where $\iota\colon \complex^K\to \complex^M$ and $\kappa\colon \complex^K\to \complex^N$ are isometries. Given an $M\times M$ matrix $A$, the $N\times N$ matrix  $\omega^* A\omega$  is often called \emph{a corner} of $A$ (with repsect to $\omega$) --- this naming choice is clearer in the case when $M\geq N$ and  $\omega$ is an isometry embedding $\complex^N$ in $\complex^M$.
\end{definition}

We will sometimes need to compare observables, PVMs, or strategies, that act in different spaces. For example, we may have families of operators $\{A_i\}$, $\{B_i\}$ on $\complex^M$ and $\complex^N$ respectively. To compare them, we may measure their distance as the infimum, over all partial isometries $\omega\colon \complex^N\to \complex^M$, of $\sum_i p_i \| A_i - wB_i w^*\|^2_{hs}$, where $p_i$ are some coefficients (e.g.\ probabilities).
The following definition and claims will be useful technical tools in the manipulation of such distance measures.

\begin{definition}[$\eps$-near bijection]\label{defn:near_bijections}
	A partial isometry  $\omega\colon \complex^N\to \complex^M$ is said to be an \emph{$\eps$-near bijection} if $1-\tau(\omega^*\omega),1-\tau(\omega\omega^*)\leq\eps$.\footnote{Note that one of the $\tau$'s is the normalized trace on $N\times N$ matrices and the other on $M\times M$ matrices.}
\end{definition}
\begin{claim}\label{claim:hs_norm_of_corner_is_close_to_original}
	Let $\omega\colon \complex^N\to \complex^M$ be an $\eps$-near bijection (Definition \ref{defn:near_bijections}). Then, for every contraction $A\in M_{M\times M}(\complex)$, i.e. $\|A\|_{op}\leq 1$, we have 
	\[
	\left|\|A\|_{hs}^2-\|\omega^*A\omega\|_{hs}^2\right|\leq 4\eps.
	\]
\end{claim}

\begin{proof}
	If $\eps>\nicefrac{1}{2}$, then the claim is immediate, using $\|A\|_{hs}\leq \|A\|_{op}\leq 1$ and the triangle inequality. Assume otherwise. 
	Let $\omega=\iota\circ\kappa^*$ be the decomposition of $\omega$ as an isometry $\iota\colon \complex^K\to \complex^M$ and co-isometry  $\kappa^*\colon \complex^N\to \complex^K$. First, it is straightforward to check that $\tau(\omega^*\omega)=\nicefrac{K}{N}$ and $\tau(\omega\omega^*)=\nicefrac{K}{M}$. Now,
	\[
	\begin{split}
	\|\omega^*A\omega\|_{hs}^2&=\frac{1}{N}\tr(\omega^*A^*\omega\omega^*A\omega)\\
	&=_{\kappa^*\kappa=\Id_K}\frac{1}{N}\tr(\iota^* A^*\iota\cdot \iota^*A\iota)\\
	&=\frac{1}{N}\tr(A^*\iota\cdot \iota^*A\iota\cdot\iota^*).
	\end{split}
	\]
	We have
	\begin{equation}\label{eq:argument_showing_trace_of_corner_is_close_to_original}
	\begin{split}
	\tr(A^*\iota\cdot\iota^*A)&=\tr(A^*\iota\cdot\iota^*A\iota\cdot\iota^*)+\tr(A^*\iota\cdot\iota^*A(\Id_M-\iota\cdot\iota^*))\\
	&\leq_{\textrm{H\"older}} \tr(A^*\iota\cdot\iota^*A\iota\cdot\iota^*)+\underbrace{\|A^*\iota\cdot\iota^*A\|_{op}}_{\leq 1}\cdot\underbrace{\tr(\Id_M-\iota\cdot\iota^*)}_{M-K}.
	\end{split}
	\end{equation}
	The same argument shows that $\tr(AA^*)\leq \Tr(AA^*\iota\cdot\iota^*)+(M-K)$. Hence,
	\[
	\begin{split}
	\|A\|_{hs}^2&=\frac{1}{M}\tr(AA^*)\\
	&\leq \frac{1}{M}\left(\tr(A^*\iota\cdot\iota^*A\iota\cdot\iota^*)+2(M-K)\right)\\
	&=\frac{N}{M}\|\omega^*A\omega\|_{hs}^2+2\eps.
	\end{split}
	\]
	Finally, $\frac{N}{M}\leq \frac{N}{K}\leq \frac{1}{1-\eps}\leq 1+2\eps$, and since $\|\omega^*A\omega\|_{hs}\leq \|\omega^*A\omega\|_{op}\leq 1$, we deduce that 
	\[
	\|A\|_{hs}^2-\|\omega^*A\omega\|_{hs}^2\leq 4\eps.
	\]
	On the other hand, as $\iota\cdot \iota^*$ and $\Id_M-\iota\cdot \iota^*$ are both positive, we can deduce that 
	\[\begin{split}
	\tr(\iota^*A^*\iota\cdot\iota^*A\iota)&\leq \tr(\iota^*A^*\iota\cdot\iota^*A\iota)+\tr(\iota^*A^*(\Id_M-\iota\cdot\iota^*)A\iota)\\
	&=\tr(\iota^*A^*A\iota)\\
	&\leq\tr(A\iota\cdot\iota^*A^*)+\tr(A(\Id_M-\iota\cdot\iota^*)A^*)\\
	&= \tr(AA^*).
	\end{split}\]
	Therefore,
	\[
	\|\omega^*A\omega\|_{hs}^2=\frac{1}{N}\tr(\iota^*A^*\iota\cdot\iota^*A\iota)\leq\frac{1}{N}\tr(AA^*)=\frac{M}{N}\|A\|_{hs}^2,
	\]
	and as $\frac{M}{N}\leq \frac{M}{K}\leq \frac{1}{1-\eps}\leq 1+2\eps$ and $\|A\|_{hs}^2\leq \|A\|_{op}^2\leq 1$, we deduce that 
	\[
	\|\omega^*A\omega\|_{hs}^2-\|A\|_{hs}^2\leq 2\eps.
	\]
	Combining the two finishes the proof.
\end{proof}

\subsection{Notions of distance between measurements, correlations and strategies}
\label{sec:toolbox-dist}

As mentioned in Remark \ref{rem:the_method_of_proving_compression}, the proof method of the soundness conditions in  Compression (Theorem \ref{thm:compression}) is as follows. Let $\game$ be a tailored game, and $\frak{T}(\game)$ be some transformation of $\game$ into a new game. Assume you are given a strategy $\strategy$ for $\frak{T}(\game)$ with $\val(\frak{T}(\game),\strategy)\geq 1-\eps$. Then, the goal is to extract from $\strategy$ a strategy $\strategy'$ for the original game $\game$ with value at least $1-f(\eps)$ (controlling this $f$ is a recurring technical hurdle). In the first two transformations applied by $\Compress$, \emph{question reduction} and \emph{answer reduction}, the way $\strategy'$ is extracted out of $\strategy$ is by perturbing it until it passes some of the subroutines of $\frak{T}(\game)$ perfectly. After this perturbation, the value of the resulting strategy is not much worse than the value of the original strategy. Using moreover that the new strategy, by definition, passes some subroutines perfectly, then makes it easier for us to extract $\strategy'$ for the original $\game$.  

\subsubsection{Distance between (partial) measurements}
To make this notion of ``perturbation'' formal, we need appropriate notions of distance between POVMs  and between quantum strategies, which is the topic of this section. Let us begin by extending the notion of a measurement. 
\begin{definition}[Partial and Corner POVMs]\label{defn:partial_POVM_submeasurement}
	An $N$-dimensional partial POVM with outcomes in a finite set $A$ is a tuple of positive  $N\times N$ matrices $\{\cal{P}_a\}_{a\in A}$ such that $\sum \cal{P}_a\leq \Id$. It is a partial PVM if every $\cal{P}_a$ is an orthogonal  projection.
	A partial POVM defines a tuple of non-negative real numbers $p_a=\tau(\cal{P}_a)$ satisfying $\sum p_a\leq 1$, which we keep calling the \emph{distribution} induced by $\cal{P}$. The quantity $1-\sum p_a=1-\sum \tau(\cal{P}_a)$ is often called the \emph{deficiency} of $\cal{P}$.
	
	Given an $M$-dimensional partial POVM $\cal{P}$ and  a partial isometry (Defintion \ref{defn:projections_isometries_partial_isometries}) $\omega\colon \complex^N\to \complex^M$, the tuple of $N\times N$ matrices $\cal{P}'_a=\omega^*\cal{P}_a\omega$ parametrized by $A$ is called \emph{the corner}  POVM of $\cal{P}$ with respect to $\omega$. We often denote the corner POVM by $\omega^*\cal{P}\omega$.
\end{definition}

\begin{remark}
	The above definition of a partial POVM (called a submeasurement in \cite{quantum_soundness_tensor_codes}) clearly extends the notion of a POVM (Definition \ref{defn:PVM}), and the ideas of measuring and jointly measuring extend with it (Definition \ref{defn:joint_measurement}) --- though, we may get  partial distributions when measuring instead of  full ones. When needed, we call a POVM (or PVM), as in Definition \ref{defn:PVM}, a \emph{full} or \emph{complete} POVM.
\end{remark}

\begin{claim}\label{claim:corner_POVMs_are_partial_POVMs}
	Given an $M$-dimensional partial POVM $\cal{P}$ and a partial isometry $\omega\colon \complex^N\to \complex^M$, the corner POVM $\omega^*\cal{P}\omega$ is indeed a partial POVM. In addition, if the deficiency  of $\cal{P}$ is $\delta$, and $\omega$ is an $\eps$-near bijection (Defintion \ref{defn:near_bijections}), then the deficiency of the corner $\omega^*\cal{P}\omega$ is at most $\delta+2\eps$.
\end{claim}
\begin{proof}
	The partial order on matrices $A\leq B$ (i.e., $A-B$ being positive) is preserved by corners, namely: If $A,B$ are ${M\times M}$ complex matrices  and $A\leq B$, then $\omega^*A\omega\leq \omega^*B\omega$. This observation implies immediately that the corner POVM consists of positive matrices, and that $\sum_a \omega^*\cal{P}_a\omega\leq \omega^*\omega$; as $\omega^*\omega$ is a projection (Definition \ref{defn:projections_isometries_partial_isometries}), it satisfies $\omega^*\omega\leq \Id_N$, and the proof is complete.
	
	Now, as $\sum \cal{P}_a\leq \Id$, it is a contraction, and the argument in \eqref{eq:argument_showing_trace_of_corner_is_close_to_original} shows that for every $a\in A$, 
	\[
	\sum_{a\in A}\Tr(\cal{P}_a)\leq \Big(\sum_{a\in A}\Tr(\omega^* \cal{P}_a\omega)\Big)+\Tr(\Id-\omega\omega^*)\ .
	\]
	So, rearranging the above inequality and using the deficiency and near bijection assumptions leads to 
	\[
	\sum_{a\in A}\tau(\omega^*\cal{P}_a\omega)=\frac{1}{N}\sum_{a\in A} \Tr(\omega^*\cal{P}_a\omega)\geq \frac{M}{N}(1-\delta-\eps)\geq (1-\eps)(1-\delta-\eps)\geq 1-\delta-2\eps\ .
	\]
	
\end{proof}

\begin{fact}[Naimark's dilation theorem, see e.g.\ Chapter 4 in~\cite{Paulsen_2003}]\label{fact:Naimark_dilation}
	Every (finite dimensional) POVM is a corner of a (finite dimensional) PVM.
\end{fact}

As quantum strategies (Definition \ref{defn:quantum_strategy}), which are the objects of interest for us, are defined using full PVMs, it seems unnecessary to define POVMs, not to mention partial ones. The reason for these intricacies is that we want to be able to compare strategies acting on Hilbert spaces of {different dimensions}. This will require us to use partial isometries between these spaces, and the conjugation of a PVM by a partial isometry --- namely the corner --- is only guaranteed to be a partial POVM by Claim \ref{claim:corner_POVMs_are_partial_POVMs}.  Similarly, in representation form, the conjugation by a partial isometry of a unitary is no longer a unitary. But, as long as the partial isometry is not {too deforming}, namely it is an $\eps$-near bijection (Definition \ref{defn:near_bijections}), these properties are ``almost'' preserved --- see Fact \ref{fact:orthogonalization} and the above claim.

\begin{definition}[Distance and Inconsistency of POVMs]\label{defn:distance_inconsistency_POVMs}
	Let $\cal{P}$ and $\cal{Q}$ be partial POVMs (Definition \ref{defn:partial_POVM_submeasurement}) of the same dimension with outcomes in the same finite set $A$. We say that $\cal{P}$ and $\cal{Q}$ are \emph{$\eps$-close}, and denote it by $\cal{P}_a\approx_\eps \cal{Q}_a$, if 
	\[
	\sum_{a\in A}\|\cal{P}_a-\cal{Q}_a\|_{hs}^2\leq \eps\ .
	\]
	We say that $\cal{P}$ and $\cal{Q}$ are \emph{$\eps$-inconsistent}, and denote it by $\cal{P}\simeq_\eps \cal{Q}$, if 
	\[
	\sum_{a\neq b\in A}\tau(\cal{P}_a\cal{Q}_b)\leq \eps\ .
	\]
\end{definition}
\begin{remark}\label{rem:inconsistency_and_joint_sampling}
	The name inconsistency is appropriate, as by Definition \ref{defn:joint_measurement}, if $\cal{P}$ and $\cal{Q}$ are full POVMs, and we jointly measure $(a,b)\sim  (\cal{P},\cal{Q})$, then the probability $a\neq b$ is exactly the incosistency of $\cal{P}$ and $\cal{Q}$. In particular, note that, as opposed to distance, the inconsistency of a POVM with itself is not necessarily $0$ --- this is true only when the product of $\cal{P}_a$ and $\cal{P}_b$ is $0$ for every $a\neq b$.
	
	Our (tailored) games contain various comparisons between the answers at the endpoints of the sampled edge, and a strategy passing the game along this edge with high probability implies a small inconsistency between the (data processed, Definition \ref{defn:Data_proccessed_PVM}) PVMs at the endpoints of the edge.
\end{remark}

\begin{proposition}[Properties of distance and inconsistency. Cf.\ \cite{quantum_soundness_tensor_codes,NW19} and \cite{CVY_efficient}]\label{prop:properties_of_distance_and_inconsistency}
	Let $\cal{P},\cal{Q},\cal{R}$ be partial POVMs of dimension $N$ with outcomes in a finite set $A$,  let $p,q,r\in \mathbb{R}^A$ be the distributions associated with them, and let $\|\cdot\|_1$ be the standard $L^1$ norm on $\mathbb{R}^A$. 
	\begin{enumerate}
		\item\emph{Inconsistency and distance are the same for projective measurements}: If $\cal{P}$ and $\cal{Q}$ are full PVMs  then 
		$ \cal{P}_a\simeq_{\eps} \cal{Q}_a $ if and only if $\cal{P}_a\approx_{2\eps}\cal{Q}_a$. 
		
		\item \emph{Semi-triangle inequality}: If 
		$\cal{P}_a\approx_{\eps}\cal{Q}_a$ and $\cal{Q}_a\approx_\delta\cal{R}_a$, then  $\cal{P}_a\approx_{2\eps+2\delta}\cal{R}_a$.  More generally, given $k+1$ many partial POVMs $\cal{P}^1,...,\cal{P}^{k+1}$ such that $\cal{P}^i\approx_{\eps_i}\cal{P}^{i+1}$ for every $1\leq i\leq k$, we have
		\[
		\cal{P}^1\approx_{k(\eps_1+...+\eps_k)} \cal{P}^{k+1}\ .
		\]
		
		\item \emph{Consistent almost full POVMs induce close distributions}: Assume the deficiency of $\cal{P}$ is $\delta_1$ and of $\cal{Q}$ is $\delta_2$ --- i.e., $\|p\|_1=\sum\tau(\cal{P}_a)=1-\delta_1$, $\|q\|_1=\sum\tau(\cal{Q}_a)=1-\delta_2$ ---  and assume they are $\eps$-inconsistent --- i.e.,  $\cal{P}\simeq_\eps\cal{Q}$. Then $\|p-q\|_1\leq 2(\delta_1+\delta_2+\eps)$.
		
		\item \emph{Small inconsistency to closeness in case both are full} : Assume $\cal{P},\cal{Q}$ are full POVMs. Then, $\cal{P}\simeq_\eps\cal{Q}$ implies $\cal{P}\approx_{2\eps}\cal{Q}$.
		\item \emph{Closeness to small inconsistency in case one of them is projective}: Assume $\cal{P}$ is projective. Then, $\cal{P}\approx_\eps\cal{Q}$ implies $\cal{P}\simeq_{\sqrt\eps}\cal{Q}$.
	\end{enumerate}
\end{proposition}

\begin{proof}
	\begin{enumerate}
		\item It follows from \begin{equation}\label{eq:consistency_vs_distance_PVM_case}
		\begin{split}
		\sum_{a\in A}\overbrace{\|\cal{P}_a-\cal{Q}_a\|_{hs}^2}^{\tau((\cal{P}_a-\cal{Q}_a)^*(\cal{P}_a-\cal{Q}_a))}&=\sum_{a\in A}\tau(\cal{P}_a)+\tau(\cal{Q}_a)-2\tau(\cal{P}_a\cal{Q}_a)\\
		&=_{\cal{P},\cal{Q}\ \textrm{full\ PVMs}} 2\left(1-\sum_{a\in A}\tau(\cal{P}_a\cal{Q}_a)\right)\\
		&=2\sum_{a\neq b\in A}\tau(\cal{P}_a\cal{Q}_b)\ ,
		\end{split}
		\end{equation}
		where the last equation is since $\sum_{a,b\in A}\tau(\cal{P}_a\cal{Q}_b)=1$. 
		
		\item The first case is immediate from
		\begin{align*}
		\forall a\in A\ \colon \ \ 	\|\cal{P}_a-\cal{R}_a\|_{hs}^2\leq_{\triangle} (\|\cal{P}_a-\cal{Q}_a\|_{hs}+\|\cal{Q}_a-\cal{R}_a\|_{hs})^2\leq 2\|\cal{P}_a-\cal{Q}_a\|_{hs}^2+2\|\cal{Q}_a-\cal{R}_a\|_{hs}^2\ .
		\end{align*}
		The general case uses the same idea together with the inequality $(\sum_{i=1}^k x_i)^2\leq k\sum_{i=1}^k x_i^2$.
		
		\item Choose a new element $\perp\notin A$, and extend the partial POVMs $\cal{P},\cal{Q}$ to full POVMs $\cal{P}',\cal{Q}'$ on $A'=A\cup\{\perp\}$ by letting 
		\begin{align}
		\forall a\in A\ \colon \cal{P}'_a&:=\cal{P}_a\ ,\notag\\
		\cal{P}'_\perp&:=\Id-\sum_{a\in A}\cal{P}_a\ ,\label{eq:extended_completed_POVM}
		\end{align}
		and similarly for $\cal{Q}'$. Furthermore, let $p'$ and $q'$ be the distributions induced by $\cal{P}',\cal{Q}'$. It is immediate that $\|p-q\|_1\leq \|p'-q'\|_1$.
		In addition, 
		\begin{align}
		\sum_{a'\neq b'\in A'}\tau(\cal{P}'_{a'}\cal{Q}'_{b'})&=\sum_{a\neq b\in A}\tau(\cal{P}_a\cal{Q}_b)+\overbrace{\sum_{a\in A} \tau(\cal{P}_a\cal{Q}_\perp)}^{\tau((\sum\cal{P}_a)\cal{Q}_\perp)}+\overbrace{\sum_{b\in A} \tau(\cal{P}_\perp\cal{Q}_b)}^{\tau(\cal{P}_\perp(\sum\cal{Q}_b))}\notag\\
		&\leq_{\cal{P}\simeq_\eps\cal{Q}\ \textrm{and\ H\"older}} \eps+\underbrace{\|\sum \cal{P}_a\|_{op}}_{\leq 1}\cdot \underbrace{\tau(\cal{Q}_\perp)}_{=\delta_1}+\underbrace{\|\sum \cal{Q}_a\|_{op}}_{\leq 1}\cdot \underbrace{\tau(\cal{P}_\perp)}_{=\delta_2} \label{eq:completed_POVMs_are_still_consistent}\\
		&\leq \eps+\delta_1+\delta_2\ ,\notag
		\end{align}
		which means $\cal{P}'\simeq_{\eps+\delta_1+\delta_2}\cal{Q}'$.
		Now, $\cal{P}'$ and $\cal{Q}'$ are full POVMs, and hence for every $a'\in A'$ we have 
		\begin{equation}\label{eq:consistent_POVMs_provide_close_distributions}
		|\tau(\cal{P}'_{a'})-\tau(\cal{Q}'_{a'})|=\Big|\sum_{b'\in A'}\tau(\cal{P}'_{a'}\cal{Q}'_{b'})-\tau(\cal{Q}'_{a'}\cal{P}'_{b'})\Big|\leq \sum_{b'\in A'\colon b'\neq a'}|\tau(\cal{P}'_{a'}\cal{Q}'_{b'})|+|\tau(\cal{Q}'_{a'}\cal{P}'_{b'})|\ .
		\end{equation}
		Summing up over all $a'\in A$ gives us 
		\[
		\|p'-q'\|_1=\sum_{a'\in A'}|\tau(\cal{P}'_{a'})-\tau(\cal{Q}'_{a'})|\leq 2\sum_{a'\neq b'\in A'}|\tau(\cal{P}'_{a'}\cal{Q}'_{b'})|\leq 2(\eps+\delta_1+\delta_2)\ ,
		\]
		as needed.
		\item This is the same calculation as in \eqref{eq:consistency_vs_distance_PVM_case}, except that we need to use  the inequality $\tau(\cal{P}_a^2)\leq \tau(\cal{P}_a)$ and  $\tau(\cal{Q}_a^2)\leq \tau(\cal{Q}_a)$.
		\item We have
		\begin{align*}
		\sum_{a\in A}\sum_{b\in A\colon b\neq a}\tau(\cal{P}_a\cal{Q}_b)&\leq \sum_{a\in A}\tau(\cal{P}_a(\Id-\cal{Q}_a))\\
		&=_{\cal{P}\ \textrm{is\ projective}}\sum_{a\in A}\overbrace{\tau(\cal{P}_a(\cal{P}_a-\cal{Q}_a))}^{=|\langle \cal{P}_a,\cal{P}_a-\cal{Q}_a\rangle|}\\
		&\leq_{\textrm{Cauchy--Schwarz}}\sum_{a\in A}\|\cal{P}_a\|_{hs}\|\cal{P}_a-\cal{Q}_a\|_{hs}\\
		&\leq_{\textrm{Cauchy--Schwarz}}\sqrt{\sum_{a\in A}\|\cal{P}_a\|^2_{hs}}\sqrt{\sum_{a\in A}\|\cal{P}_a-\cal{Q}_a\|^2_{hs}}\\
		&\leq \sqrt \eps\ ,
		\end{align*}
		where the last inequality uses the fact $\cal{P}\approx_{\eps}\cal{Q}$ and the fact that projections satisfy $\|\cal{P}_a\|_{hs}^2=\tau(\cal{P}_a)$.
	\end{enumerate}
\end{proof}
\begin{remark}
	In the full case, $\eps$-inconsistency implies that the POVMs are $\eps$-close (the above clause $4.$, see also \cite[Proposition 2.5]{quantum_soundness_tensor_codes}), but the reverse  is {not} true in general (see \cite[Remark 4.15]{NW19}). Luckily, we have the above clause $5.$, which states that in case one of them is a PVM, there is a way to infer $\sqrt{\eps}$-inconsistency out of $\eps$-closeness (see also \cite[Proposition 2.6]{quantum_soundness_tensor_codes}).  This will be very helpful in the upcoming analysis, as small inconsistency allows to deduce results that closeness cannot (cf.\ \cite[Propositions 2.4 and 2.9]{quantum_soundness_tensor_codes}).
\end{remark}

The value of a measurement being projective leads to the following definition:
\begin{definition}[Almost projective measurements]\label{defn:almost_projective_measurement}
	A partial POVM $\cal{P}$ with outcomes in $A$ is said to be $\eps$-almost projective if $\sum_{a\in A}\|\cal{P}_a-\cal{P}^2_a\|_1\leq \eps$.
\end{definition}
\begin{remark}
	For full POVMs, being $\eps$-almost projective is the same as having $\eps$-self inconsistency, namely satisfying $\cal{P}\simeq_\eps \cal{P}$. This is because
	\[
	\|\cal{P}_a-\cal{P}^2_a\|_1=\tau(\cal{P}_a(\Id-\cal{P}_a))=\sum_{b\in A\colon b\neq a}\tau(\cal{P}_a\cal{P}_b)\ .
	\]
\end{remark}

\begin{claim}[Corners of PVMs are almost projective]\label{claim:corner_of_PVM_almost_projective}
	Let $\cal{P}$ be an $M$-dimensional PVM with outcomes in $A$ and $\omega\colon \complex^N\to \complex^M$ an $\eps$-near bijection. Then the corner POVM $\omega^*\cal{P}\omega$ is $\eps$-almost projective.
\end{claim}
\begin{proof}
	Let us calculate
	\begin{align*}
	\sum_{a\in A}\|\omega^*\cal{P}_a\omega-(\omega^*\cal{P}_a\omega)^2\|_1&=\sum_{a\in A}\tau(\omega^*\cal{P}_a^2\omega-\omega^*\cal{P}_a\omega\omega^*\cal{P}_a\omega)\\
	&=\sum_{a\in A}\tau(\omega^*\cal{P}_a(\Id_M-\omega\omega^*)\cal{P}_a\omega)\\
	&=\tau\Big((\Id_M-\omega\omega^*)\Big(\sum_{a\in A}\cal{P}_a\omega\omega^*\cal{P}_a\Big)\Big)\\
	&\leq_{\textrm{H\"older}} \tau(\Id_M-\omega\omega^*)\Big\|\sum_{a\in A}\cal{P}_a\omega\omega^*\cal{P}_a\Big\|_{op}\ .
	\end{align*}
	By the $\eps$-near bijection assumption, $\tau(\Id_M-\omega\omega^*)\leq \eps$, and as $\cal{P}_a\omega\omega^*\cal{P}_a\leq \cal{P}_a\omega\omega^*\cal{P}_a+\cal{P}_a(\Id_M-\omega\omega^*)\cal{P}_a=\cal{P}_a^2=\cal{P}_a$ we have $\sum\cal{P}_a\omega\omega^*\cal{P}_a\leq \sum\cal{P}_a\leq \Id_M$; as the operator norm respects the order on positive matrices, we are done.
\end{proof}
\begin{claim}[Corners of POVMs produce similar joint distributions]\label{claim:corner_POVMs_joint_distribution}
	Let $\cal{P}$ and $\cal{Q}$ be $M$-dimensional partial POVMs  with outcomes in finite sets  $A$ and $B$ respectively, and let $\omega\colon \complex^N\to \complex^M$ be an $\eps$-near bijection. Then, jointly measuring (Definition \ref{defn:joint_measurement}) according to $(\cal{P},\cal{Q})$ is $4\sqrt{\eps}$-close in $L^1$-distance to jointly measuring according to the corners $(\omega^* \cal{P}\omega,\omega^* \cal{Q}\omega)$.
\end{claim}

\begin{proof}
	If $\eps\geq \nicefrac{1}{2}$, then the conclusion is immediate (as every two partial probability distributions are at most $2$ apart in the $L^1$-norm). Hence, we can assume $\eps<\nicefrac{1}{2}$, and in particular, as $\omega$ is an $\eps$-near bijection, we have
	\[
	1-\eps\leq \nicefrac{M}{N},\nicefrac{N}{M}\leq \nicefrac{1}{1-\eps}\leq 1+2\eps\ .
	\]
	
	By applying H\"older (\cref{clause4:useful} of Fact \ref{fact:useful_equations_and_ineqs}) twice, and using the fact that a projection is a contration, we get
	\begin{equation}\label{eq:trace_of_conjugate_smaller_in_Claim_corners}
	\forall a\in A, b\in B\ \colon\ \  \Tr(\omega^*\cal{P}_a\omega\omega^*\cal{Q}_b\omega)\leq |\Tr(\cal{P}_a\omega\omega^*\cal{Q}_b)|\|\omega\omega^*\|_{op}\leq \Tr(\cal{P}_a\cal{Q}_b)\|\omega\omega^*\|_{op}^2\leq \Tr(\cal{P}_a\cal{Q}_b)\ .
	\end{equation}
	On the other hand, we have
	\begin{align}
	\Tr(\cal{P}_a\cal{Q}_b)&= \Tr(\cal{P}_a\omega\omega^*\cal{Q}_b)+\Tr(\cal{P}_a(\Id_M-\omega\omega^*)\cal{Q}_b)\notag\\
	&= \Tr(\cal{P}_a\omega\omega^*\cal{Q}_b\omega\omega^*)+\Tr(\cal{P}_a\omega\omega^*\cal{Q}_b(\Id-\omega\omega^*))+ \Tr(\cal{P}_a(\Id_M-\omega\omega^*)\cal{Q}_b)\ .\label{eq:double_conjugate_difference_Claim_corners}
	\end{align}
	Using the $\eps$-near bijectiveness of $\omega$ and the fact that $\Id_M-\omega\omega^*$ is a projection, one gets 
	\[
	\Tr((\Id_M-\omega\omega^*)^*(\Id_M-\omega\omega^*))=\Tr(\Id_M-\omega\omega^*)\leq M\eps\ .
	\]
	Therefore, 
	\begin{align}
	\sum_{a,b}\left|\Tr(\cal{P}_a\cal{Q}_b)-\Tr(\omega^*\cal{P}_a\omega\omega^*\cal{Q}_b\omega)\right|&=_{\eqref{eq:trace_of_conjugate_smaller_in_Claim_corners}} \sum_{a,b}\Tr(\cal{P}_a\cal{Q}_b)-\Tr(\omega^*\cal{P}_a\omega\omega^*\cal{Q}_b\omega)\notag\\
	&=_{\eqref{eq:double_conjugate_difference_Claim_corners}}\sum_{a,b}\Tr(\cal{P}_a\omega\omega^*\cal{Q}_b(\Id-\omega\omega^*))+ \Tr(\cal{Q}_b\cal{P}_a(\Id_M-\omega\omega^*))\label{eq:distance_between_joint_measurements_Claim_corners}\\
	&=\Tr\Big(\Big(\sum_{a,b}\cal{P}_a\omega\omega^*\cal{Q}_b\Big)(\Id-\omega\omega^*)\Big)+\Tr\Big(\Big(\sum_{a,b}\cal{Q}_b\cal{P}_a\Big)(\Id-\omega\omega^*)\Big)\ .\notag
	\end{align}
	By applying Cauchy--Schwartz on the two summands we get
	\[
	\Tr\Big(\Big(\sum_{a,b}\cal{Q}_b\cal{P}_a\Big)(\Id-\omega\omega^*)\Big)\leq \sqrt{\Tr\Big(\Big(\sum_{a,b}\cal{Q}_b\cal{P}_a\Big)^*\Big(\sum_{a,b}\cal{Q}_b\cal{P}_a\Big)\Big)}\sqrt{\Tr(\Id-\omega^*\omega)^2}
	\]
	and 
	\[
	\Tr\Big(\Big(\sum_{a,b}\cal{P}_a\omega\omega^*\cal{Q}_b\Big)(\Id-\omega\omega^*)\Big)\leq \sqrt{\Tr\Big(\Big(\sum_{a,b}\cal{P}_a\omega\omega^*\cal{Q}_b\Big)^*\Big(\sum_{a,b}\cal{P}_a\omega\omega^*\cal{Q}_b\Big)\Big)}\sqrt{\Tr(\Id-\omega^*\omega)^2}\ .
	\]
	Now, $\sum\cal{P}_a,\sum\cal{Q}_b\leq \Id_M$ (as they are partial POVMs) which implies $(\sum\cal{P}_a)^2,(\sum\cal{Q}_b)^2\leq \Id_M$; therefore
	\begin{align*}
	\Tr\Big(\Big(\sum_{a,b}\cal{Q}_b\cal{P}_a\Big)^*\Big(\sum_{a,b}\cal{Q}_b\cal{P}_a\Big)\Big)&=\Tr\Big(\Big(\sum_{a}\cal{P}_a\Big)\Big(\sum_{b}\cal{Q}_b\Big)^2\Big(\sum_{a}\cal{P}_a\Big)\Big)\\
	&\leq \Tr\Big(\Big(\sum_{a}\cal{P}_a\Big)^2\Big)\\
	&\leq M\ .
	\end{align*}
	Similarly,
	\begin{align*}    \Tr\Big(\Big(\sum_{a,b}\cal{P}_a\omega\omega^*\cal{Q}_b\Big)^*\Big(\sum_{a,b}\cal{P}_a\omega\omega^*\cal{Q}_b\Big)\Big)&=\Tr\Big(\Big(\sum_{b}\cal{Q}_b\Big)\omega\omega^*\Big(\sum_{a}\cal{P}_a\Big)^2\omega\omega^*\Big(\sum_{b\in B}\cal{Q}_b\Big)\Big)\\
	&\leq \Tr\Big(\Big(\sum_{b}\cal{Q}_b\Big)\omega\omega^*\Big(\sum_{b\in B}\cal{Q}_b\Big)\Big)\\
	&\leq \Tr\Big(\Big(\sum_{b}\cal{Q}_b\Big)^2\Big)\\
	&\leq M\ .
	\end{align*}
	Plugging  all of these upper bounds to \eqref{eq:distance_between_joint_measurements_Claim_corners}, we get
	\[
	\sum_{a,b}\left|\Tr(\cal{P}_a\cal{Q}_b)-\Tr(\omega^*\cal{P}_a\omega\omega^*\cal{Q}_b\omega)\right|\leq 2M\sqrt\eps\ .
	\]
	If we divide both sides by  $M$, we are almost done; the problem is that $\tau(\omega^*\cal{P}_a\omega\omega^*\cal{Q}_b\omega)=\frac{1}{N}\Tr(\omega^*\cal{P}_a\omega\omega^*\cal{Q}_b\omega)$ and not $\frac{1}{M}\Tr(\omega^*\cal{P}_a\omega\omega^*\cal{Q}_b\omega)$. But, for every $a\in A$ and $b\in B$ we have 
	\[
	\left|\frac{1}{N}\Tr(\omega^*\cal{P}_a\omega\omega^*\cal{Q}_b\omega)-\frac{1}{M}\Tr(\omega^*\cal{P}_a\omega\omega^*\cal{Q}_b\omega)\right|=\left|\frac{1}{N}-\frac{1}{M}\right|\Tr(\omega^*\cal{P}_a\omega\omega^*\cal{Q}_b\omega)\ ,
	\]
	and as $\sum_{a,b}\Tr(\omega^*\cal{P}_a\omega\omega^*\cal{Q}_b\omega)\leq N$, we deduce
	\[
	\sum_{a,b}\left|\frac{1}{N}\Tr(\omega^*\cal{P}_a\omega\omega^*\cal{Q}_b\omega)-\frac{1}{M}\Tr(\omega^*\cal{P}_a\omega\omega^*\cal{Q}_b\omega)\right|\leq \left|1-\frac{N}{M}\right|\leq 2\eps\ .
	\]
	Combining all of the above gives
	\[
	\sum_{a,b}\left|\tau(\cal{P}_a\cal{Q}_b)-\tau(\omega^*\cal{P}_a\omega\omega^*\cal{Q}_b\omega)\right|\leq 2\eps+2\sqrt\eps\leq 4\sqrt\eps\ .
	\]
\end{proof}

\begin{claim}[Close almost projective POVMs produce similar joint distributions]\label{claim:close_almost_prjective_POVMs_similar_joint}
	Let $\cal{P}$ and $\cal{Q}$ be two $N$-dimensional partial POVMs with outcomes in $A$ such that  
	$\cal{P}\approx_{\eps}\cal{Q}$,  and let $\cal{R}$ an $N$-dimensional partial POVM with outcomes in $B$. Assume in addition that $\cal{P}$ is $\delta_1$-almost projective (Definition \ref{defn:almost_projective_measurement}) and that $\cal{Q}$ is $\delta_2$-almost projective. Then, jointly measuring (Definition \ref{defn:joint_measurement}) according to $(\cal{P},\cal{R})$ is $(\delta_1+\delta_2+2\sqrt{\eps})$-close in $L^1$-distance to jointly measuring according to $(\cal{Q},\cal{R})$.
\end{claim}
\begin{proof}

	For every $0\neq z\in \complex$ there is a unique  complex number $\alpha$ (with absolute value $1$) such that $|z|=\alpha z$. Hence, for every $a\in A,b\in B$, there is an $\alpha_{a,b}$ such that 
	\begin{align*}
	|\tau(\cal{P}_a\cal{R}_b)-\tau(\cal{Q}_a\cal{R}_b)|=\alpha_{a,b}\tau((\cal{P}_a-\cal{Q}_a)\cal{R}_b)\ .
	\end{align*}
	Summing up the above over $b\in B$ gives
	\begin{align*}
	\sum_{b\in B}|\tau(\cal{P}_a\cal{R}_b)-\tau(\cal{Q}_a\cal{R}_b)|&=\tau((\cal{P}_a-\cal{Q}_a)\sum_{b\in B}\alpha_{a,b}\cal{R}_b)\\
	&\leq_{\textrm{H\"older}}\|\cal{P}_a-\cal{Q}_a\|_1\Big\|\sum_{b\in B}\alpha_{a,b}\cal{R}_b\Big\|_{op}\ .
	\end{align*}
	If $\cal{R}$ consists of projections (i.e., it is a partial PVM), then under an appropriate choice of basis $\sum \alpha_{a,b}\cal{R}_b$ is a diagonal matrix with $\alpha_{a,b}$ on the diagonal, which immediately shows that $\left\|\sum_{b\in B}\alpha_{a,b}\cal{R}_b\right\|_{op}\leq 1$. The general case follows from Naimark's dilation theorem (Fact \ref{fact:Naimark_dilation}). Hence,
	\[
	\sum_{a\in A,b\in B}|\tau(\cal{P}_a\cal{R}_b)-\tau(\cal{Q}_a\cal{R}_b)|\leq \sum_{a\in A}\|\cal{P}_a-\cal{Q}_a\|_1\ .
	\] 
	We now repeat the argument of~\cite[Lemma 5.4]{CVY_efficient} to bound the latter. For every $a\in A$, by the triangle inequality, 
	\begin{equation}\label{eq:summands_in_close_measurements_imply_close_joint_measurements}
	\|\cal{P}_a-\cal{Q}_a\|_1\leq \|\cal{P}_a-\cal{P}_a^2\|_1+\|\cal{P}_a^2-\cal{P}_a\cal{Q}_a\|_1+\|\cal{P}_a\cal{Q}_a-\cal{Q}^2_a\|_1+\|\cal{Q}^2_a-\cal{Q}_a\|_1\ .    
	\end{equation}
	Summing over $a\in A$, the first and latst summands of \eqref{eq:summands_in_close_measurements_imply_close_joint_measurements} are bounded by $\delta_1$ and $\delta_2$ respectively, as $\cal{P}$ is $\delta_1$-almost projective and $\cal{Q}$ is $\delta_2$-almost projective. For the second summand in \eqref{eq:summands_in_close_measurements_imply_close_joint_measurements},
	\begin{align*}
	\sum_{a\in A}\|\cal{P}_a(\cal{P}_a-\cal{Q}_a)\|_1&\leq_{\textrm{H\"older}}\sum_{a\in A}\|\cal{P}_a\|_{hs}\|\cal{P}_a-\cal{Q}_a\|_{hs}\\
	&\leq_{\textrm{Cauchy--Scwhartz}}\sqrt{\sum_{a\in A}\|\cal{P}_a\|_{hs}^2}\sqrt{\sum_{a\in A}\|\cal{P}_a-\cal{Q}_a\|_{hs}^2}
	\end{align*}
	As $\cal{P}$ and $\cal{Q}$ are $\eps$-close, the second factor is bounded by $\sqrt{\eps}$. The first factor is bounded by $1$ as $\|\cal{P}_a\|_{hs}^2=\tau(\cal{P}_a^2)\leq \tau(\cal{P}_a)$ and $\sum \cal{P}_a\leq \Id_M$. The third summand in \eqref{eq:summands_in_close_measurements_imply_close_joint_measurements} is bounded in the exact same way, which leads to 
	\[
	\sum_{a\in A}\|\cal{P}_a-\cal{Q}_a\|_1\leq \delta_1+2\sqrt{\eps}+\delta_2\ ,
	\]
	finishing the proof.
\end{proof}

\subsubsection{Distance between measurements with outcomes in $\FF_2^S$}
The  properties in the previous subsection were for general measurements (or partial measurements). As in our games the set $A$ is always of the form $\FF_2^S$ for some finite set $S$, and as in this case PVMs are closely related to representations (Definition \ref{defn:projective_representation_and_observable_form_PVM}), there are some facts we need to demonstrate in this situation.
In this case, we often view strategies and PVMs as given in observable or representation form; it is thus natural to ask what is the analogous formulation of distance. The following claim is a straightforward application of the Fourier transform (Definition \ref{defn:Fourier_transform_reps}. See also \ \cite[Lemma 3.4]{de_la_Salle_spectral_gap}):
\begin{claim}\label{claim:dist_in_projection_and_observable_are_the_same}
	Let $\cal{P}$  be an $N$-dimensional PVM with outcomes in $\FF_2^S$, where $S$ is a finite set, and let $\cal{U}$ be its representation form. Similarly, let $\cal{Q}$ be an $M$-dimensional PVM with outcomes in the same set $\FF_2^S$, with $\cal{V}$ being its observable form. Then, for every partial isometry $\omega\colon \complex^N\to \complex^{M}$ we have
	\[
	\Es{
		\alpha\colon S\to \FF_2}\left[\left\|\cal{U}(\alpha)-\omega^* \cal{V}(\alpha) \omega\right\|_{hs}^2\right]=\sum_{
		a\colon S\to \FF_2}\left\|\cal{P}_a-\omega^*\cal{Q}_a\omega\right\|_{hs}^2\ .
	\]
	Hence, when we denote $\cal{U}\approx_\eps \omega^*\cal{V}\omega$ where the two sides are in representation form, we mean that the left hand-side of the above equation is smaller or equal to $\eps$.
\end{claim}

\begin{proof}
	Recall that by Definition \ref{defn:Fourier_transform_reps},
	\[
	\forall \alpha\colon S\to \FF_2\ \colon \ \ \cal{U}(\alpha)=\sum_{a\colon S\to \FF_2}(-1)^{\langle \alpha,a\rangle}\cal{P}_a\ .
	\]
	Thus, for every $\alpha\colon S\to \FF_2$, we have
	\[
	\begin{split}
	\left\|\cal{U}(\alpha)-\omega^*\cal{V}(\alpha)\omega\right\|_{hs}^2&=\Big\|\sum_{a\colon S\to \FF_2}(-1)^{\langle \alpha,a\rangle}(\cal{P}_a-\omega^*\cal{Q}_a\omega)\Big\|_{hs}^2\\
	&=\sum_{a,b\colon S\to \FF_2}(-1)^{\langle \alpha,a+b\rangle}\tau\left((\cal{P}_a-\omega^*\cal{Q}_a\omega)(\cal{P}_b-\omega^*\cal{Q}_b\omega)\right)
	\end{split}
	\]
	But, for every fixed $a\neq b\colon S\to\FF_2$, we have $a+b\neq \vec 0$ and thus $$\Es{\alpha\colon S\to \FF_2}\left[(-1)^{\langle \alpha,a+b\rangle}\right]=0\ .$$ Hence,
	\[
	\begin{split}
	\Es{
		\alpha\colon S\to \FF_2}\left[\left\|\cal{U}(\alpha)-\omega^*\cal{V}(\alpha)\omega\right\|_{hs}^2\right]&=\sum_{a\colon S\to \FF_2}\tau((\cal{P}_a-\omega^*\cal{Q}_a\omega)^2)\\
	&=\sum_{a\colon S\to \FF_2}\left\|\cal{P}_a-\omega^*\cal{Q}_a\omega\right\|_{hs}^2\ .
	\end{split}
	\]
\end{proof}

\begin{remark}
	As remarked in \cite{de_la_Salle_spectral_gap}, this is just the standard orthogonality of characters argument for the group $\FF_2^S$. 
\end{remark}

The following is a very useful fact, that states that the corners (Definition \ref{defn:projections_isometries_partial_isometries}) of a representation of $\FF_2^S$ with respect to a nearly bijective partial isometry are close to a genuine representation in the appropriate dimension.

\begin{fact}[Orthonormalization. See Lemma 2.9 in \cite{CVY_efficient} and \cite{de_la_Salle_orthogonalization}]\label{fact:orthogonalization}
	Let $\cal{P}$ be an $N$-dimensional PVM with outcomes in $\FF_2^S$, and $\omega\colon \complex^N\to \complex^M$ be a partial isometry with $1-\tau(\omega\omega^*),1-\tau(\omega^*\omega)\leq \eps$. Then, there is an $M$-dimensional PVM $\cal{Q}$ such that $\omega \cal{P}_a\omega^* \approx_{56\eps} \cal{Q}_a$, namely
	\[
	\sum_{a\colon S\to \FF_2}\|\cal{Q}_a-\omega \cal{P}_a\omega^*\|_{hs}^2\leq 56\eps.
	\]
	In representation form, if $\cal{U}$ is an $N$-dimensional representation of $\FF_2^S$, then there is an $M$-dimensional representation $\cal{V}$ of $\FF_2^S$ such that $\omega \cal{U}\omega^*\approx_{56\eps} \cal{V}$, namely
	\[
	\Es{\alpha\colon S\to\FF_2}\|\cal{V}(\alpha)-\omega\cal{U}(\alpha)\omega^*\|_{hs}^2\leq 56\eps.
	\]
\end{fact}

The above distance between representations is of $L^1$-type. It is also natural to consider the $L^{\infty}$-distance between representations, as many arguments are easier in this setup. The following allows us to move back and forth between the two notions.
\begin{claim}[$L^1$-closeness of representations implies $L^\infty$-closeness]\label{claim:L1_closeness_of_unirep_implies_Linfty}
	Let $\chi\colon \FF_2^S\to U(N)$ and $\zeta\colon \FF_2^S\to U(M)$ be two representations of $\FF_2^S$. Let $\omega\colon \complex^N\to \complex^M$ be a partial isometry such that $1-\tau(\omega^*\omega),1-\tau(\omega\omega^*)\leq \eps$. Then,  for every $\beta\colon S\to \FF_2$, we have 
	\[
	\left\|\chi(\beta)-\omega^*\zeta(\beta)\omega\right\|_{hs}^2\leq 6\Es{
		\alpha\colon S\to \FF_2}\left[\left\|\chi(\alpha)-\omega^*\zeta(\alpha)\omega\right\|_{hs}^2\right]+15\eps.
	\]
\end{claim}
\begin{proof}
	As $\chi$ (and $\zeta$) is a representation of $\FF_2^S$, we have $\chi(\beta)=\chi(\beta+\alpha)\chi(\alpha)$ for every $\alpha\in \FF_2^S$ (and similarly for $\zeta$), and thus by the triangle  and Jensen's inequalities
	\[
	\begin{split}
	\left\|\chi(\beta)-\omega^*\zeta(\beta)\omega\right\|_{hs}^2&=\|\Es{\alpha\colon S\to \FF_2}\chi(\beta+\alpha)\chi(\alpha)-\omega^*\zeta(\alpha+\beta)\zeta(\alpha)\omega\|_{hs}^2\\
	&\leq \Es{\alpha\colon S\to \FF_2}\|\chi(\beta+\alpha)\chi(\alpha)-\omega^*\zeta(\alpha+\beta)\zeta(\alpha)\omega\|_{hs}^2.
	\end{split}
	\]
	By the triangle inequality, for every $\alpha$,
	\[
	\begin{split}
	\|\chi(\beta+\alpha)\chi(\alpha)-\omega^*\zeta(\alpha+\beta)\zeta(\alpha)\omega\|_{hs}&\leq \|\chi(\beta+\alpha)\chi(\alpha)-\omega^*\zeta(\alpha+\beta)\omega\omega^*\zeta(\alpha)\omega\|_{hs}\\
	&+\|\omega^*\zeta(\alpha+\beta)\zeta(\alpha)\omega-\omega^*\zeta(\alpha+\beta)\omega\omega^*\zeta(\alpha)\omega\|_{hs}
	\end{split}
	\]
	and
	\[
	\begin{split}
	\|\chi(\beta+\alpha)\chi(\alpha)-\omega^*\zeta(\alpha+\beta)\omega\omega^*\zeta(\alpha)\omega\|_{hs}&\leq \|\chi(\beta+\alpha)\chi(\alpha)-\chi(\beta+\alpha)\omega^*\zeta(\alpha)\omega\|_{hs}\\
	&+\|\chi(\beta+\alpha)\omega^*\zeta(\alpha)\omega-\omega^*\zeta(\alpha+\beta)\omega\omega^*\zeta(\alpha)\omega\|_{hs}\\
	&\leq \|\chi(\alpha)-\omega^*\zeta(\alpha)\omega\|_{hs}+\|\chi(\beta+\alpha)-\omega^*\zeta(\alpha+\beta)\omega\|_{hs},
	\end{split}
	\]
	where the last inequality uses the unitary invariance of the Hilbert--Schmidt norm (\cref{clause1:useful}), the inequality $\|AB\|_{hs}\leq \|A\|_{op}\|B\|_{hs}$ (\cref{clause7:useful}), and the fact that $\|\omega^*\zeta(\alpha)\omega\|_{op}\leq 1$ (which can also be deduced by the non-square analogue of \cref{clause7:useful}). Therefore,
	\[
	\begin{split}
	\left\|\chi(\beta)-\omega^*\zeta(\beta)\omega\right\|_{hs}^2
	&\leq 3\Es{\alpha\colon S\to \FF_2}\|\chi(\alpha)-\omega^*\zeta(\alpha)\omega\|_{hs}^2\\
	&+3\Es{\alpha\colon S\to \FF_2}\|\chi(\alpha+\beta)-\omega^*\zeta(\alpha+\beta)\omega\|_{hs}^2\\
	&+3\Es{\alpha\colon S\to \FF_2}\|\omega^*\zeta(\alpha+\beta)\zeta(\alpha)\omega-\omega^*\zeta(\alpha+\beta)\omega\omega^*\zeta(\alpha)\omega\|_{hs}^2.
	\end{split}
	\]
	Note that the first and second summand are equal to one another. For the third summand, using Claim \ref{claim:hs_norm_of_corner_is_close_to_original}, we have
	\[
	\begin{split}
	\|\omega^*\zeta(\alpha+\beta)(\Id-\omega\omega^*)\zeta(\alpha)\omega\|_{hs}^2&\leq \|\zeta(\alpha+\beta)(\Id-\omega\omega^*)\zeta(\alpha)\|_{hs}^2 +4\eps \\ 
	&=\underbrace{\|\Id-\omega\omega^*\|_{hs}^2}_{=1-\tau(\omega\omega^*)} +4\eps \\
	&\leq 5\eps.
	\end{split}
	\]
	Combining all of the inequalities,
	\[
	\left\|\chi(\beta)-\omega^*\zeta(\beta)\omega\right\|_{hs}^2
	\leq 6\Es{\alpha\colon S\to \FF_2}\|\chi(\alpha)-\omega^*\zeta(\alpha)\omega\|_{hs}^2+15\eps.
	\]
	which proves the claim.
\end{proof}

\subsubsection{Distance between correlations and strategies}

\begin{definition}[Distance between correlations]
	Recall from Remark \ref{rem:induced_correlations} that every strategy $\strategy$ to a game $\game$ induces a correlation $p(a,b|\mathtt{x},\mathtt{y})$.
	The  distance between correlations associated with a game $\game$ is the following $L^1$-type 
	\[
	d(p,q)=\Es{\mathtt{xy}\sim \mu}\sum_{\substack{{a\colon {S_{\mathtt{x}}}\to \FF_2}\\ b\colon {S_\mathtt{y}}\to \FF_2}} |p(a,b|\mathtt{x},\mathtt{y})-q(a,b|\mathtt{x},\mathtt{y})|.
	\]
\end{definition}
\begin{remark}\label{rem:dist_notations_and_consistency}
	The aforementioned distance between correlations is natural in the following way: If $\strategy=\{\cal{P}^\mttx_a\}$ induces the correlation $p$, and $\strategy'=\{\cal{Q}^\mttx_a\}$ induces  the correlation $q$, then their values are closer than the distance between the correlations, namely
	\[
	\begin{split}
	|\val(\game,\strategy)-\val(\game,\strategy')|&=\Big| \Es{\mathtt{xy}\sim \mu}\sum_{\substack{{a\colon {S_{\mathtt{x}}}\to \FF_2}\\ b\colon {S_\mathtt{y}}\to \FF_2}} \Big(\underbrace{\tau(\cal{P}^\mathtt{x}_{a}\cal{P}^\mathtt{y}_{b})}_{p(a,b|\mathtt{x},\mathtt{y})}-\underbrace{\tau(\cal{Q}^\mathtt{x}_{a}\cal{Q}^\mathtt{y}_{b})}_{q(a,b|\mathtt{x},\mathtt{y})}\Big)D_{\mathtt{xy}}(a,b)\Big|   \\  
	&\leq \Es{\mathtt{xy}\sim \mu}\sum_{\substack{{a\colon {S_{\mathtt{x}}}\to \FF_2}\\ b\colon {S_\mathtt{y}}\to \FF_2}} \left|p(a,b|\mathtt{x},\mathtt{y})-q(a,b|\mathtt{x},\mathtt{y})\right|\underbrace{D_{\mathtt{xy}}(a,b)}_{\leq 1}\\
	&\leq d(p,q)\ .
	\end{split}
	\]
\end{remark}

Similar to the measurements case, we need a generalized notion of strategies for the rest of the arguments to be clear.
\begin{definition}[Partial and Corner strategies]\label{defn:partial_and_corner_strategies}
	Let $\game$ be a game with vertex (question) set $V$, formal generating sets $S_\mttx$ at each vertex\footnote{These sets are unions of readable and unreadable variables at the vertex, but this is irrelevant to this definition, so is ignored.} $\mttx\in V$, and distribution $\mu$ over edges (pairs of questions) of the underlying graph. An $N$-dimensional partial strategy for $\game$ is a map $\cal{P}$ that for every vertex $\mttx\in V$ associates a partial POVM (Definition \ref{defn:partial_POVM_submeasurement}) $\cal{P}^\mttx\colon \FF_2^{S_\mttx}\to M_{N\times N}(\complex)$.
	
	Given an $M$-dimensional (full) strategy $\strategy=\{\cal{P}\}$ (as in Definition \ref{defn:quantum_strategy}) and a partial isometry $\omega\colon \complex^N\to \complex^M$, the $N$-dimensional partial strategy $\strategy'=\{\cal{P}'\}$, defined by $\cal{P}'^\mttx_a=\omega^* \cal{P}^\mttx_a\omega$, is called the \emph{corner strategy} of $\cal{P}$ with respect to $\omega$. We often denote the corner strategy by $\omega^*\cal{P}\omega$.
\end{definition}

The following is the most straightforward notion of distance between  (partial) strategies {of the same dimension}, which just takes the average distance over distance along the POVMs at each vertex.

\begin{definition}[Strict distance between strategies]\label{defn:strict_distance_strategies}
	Let $\strategy=\{\cal{P}^\mttx_a\}$ and $\strategy'=\{\cal{Q}^\mttx_a\}$ be two $N$-dimensional partial strategies (Definition \ref{defn:partial_and_corner_strategies}).  We say that $\strategy$ is $\eps$-(strictly)-close to $\strategy'$, and denote it by $\cal{P}^\mttx_a\approx_\eps \cal{Q}^\mttx_a$, if
	\begin{equation}\label{eq:stricr_dist_between_strategies}
	\Es{\mathtt{x}\sim \mu}\Big[ \sum_{a\colon S_\mathtt{x}\to \FF_2}\Vert \cal{P}^\mathtt{x}_a -\cal{Q}^\mathtt{x}_a \Vert^2_{hs} \Big]\leq \eps\ ,
	\end{equation}
	where $\mathtt{x}\sim \mu$ is the marginalization of $\mu$ to vertices defined by first sampling an edge and then choosing a uniform endpoint of it --- i.e., $\mu(\mttx)=\frac{\sum_{\mtty\in V} \mu(\mttx\mtty)+\mu(\mtty\mttx)}{2}$. Namely, on average over the vertices the associated POVMs are $\eps$-(strictly)-close.
\end{definition}

As we need to be able to compare  strategies of varying dimensions, we define the following generalized notion of distance.

\begin{definition}[Flexible distance between strategies]\label{defn:distance_strategies}
	Let $\strategy=\{\cal{P}^\mttx_a\}$ be an $N$-dimensional strategy and $\strategy'=\{\cal{Q}^\mttx_a\}$ an $M$-dimensional strategy  for a game $\game$ with distribution $\mu$ over its edges. We say that $\strategy$ is $\eps$-(flexibly)-close to $\strategy'$  if there exists an $\eps$-near bijection (Definition \ref{defn:near_bijections}) $\omega\colon \complex^N\to \complex^M$ such that $\cal{P}$ is $\eps$-(strictly)-close to the corner strategy (Definition \ref{defn:partial_and_corner_strategies}) $\omega^*\cal{Q}\omega$. Namely,
	\begin{equation}\label{eq:dist_between_strategies}
	\max\Big\{\Es{\mathtt{x}\sim \mu}\Big[ \sum_{a\colon S_\mathtt{x}\to \FF_2}\Vert \cal{P}^\mathtt{x}_a -\omega^* \cal{Q}^\mathtt{x}_a \omega\Vert^2_{hs} \Big],1-\tau(\omega^*\omega),1-\tau(\omega\omega^*)\Big\}\,\leq\, \eps\ .
	\end{equation}
	
\end{definition}
\begin{remark}
	Note that  in the flexible notion of distance between strategies we measure the expected distance and inconsistency (Definition \ref{defn:distance_inconsistency_POVMs}) between the PVM $\{\cal{P}^\mathtt{x}_a\}$ and the corner POVM $\{\omega^* \cal{Q}^\mathtt{x}_a\omega \}$ (over vertices $\mttx\in V$). 
	The fact that the second object is non-projective and partial causes  technical issues when proving various facts, as was already seen in previous proofs. But, in most cases in this paper, we perturb the strategies (and PVMs) in the {same} dimension, with respect to the trivial isometry $\omega=\Id_n$ --- namely, we have small {strict} distance between the strategies. In this case, many of the technicalities in the proofs become much simpler (and with better parameters). 
\end{remark}

The following demonstrates that flexibly close by strategies produce close by correlations, and thus that their values against the game are close as well --- which shows why this notion is natural in our context.  The  statement appears in Claim \ref{claim:close_strat_implies_close_correlations}, and slightly generalizes~\cite[Lemma 5.5]{CVY_efficient}.

\begin{claim}[Perturbation of strategies]\label{claim:close_strat_implies_close_correlations}
	Let $\strategy = \{\cal{P}^\mttx_a\}$ be an $N$-dimensional (full, projective) strategy and $\strategy'=\{\cal{Q}^\mttx_a\}$ an $M$-dimensional (full, projective) strategy that are $\eps$-(flexibly)-close. Let $p$ and $q$ be the correlations that they induce (respectively), as in Remark~\ref{rem:induced_correlations}. Then 
	$$ d(p,q)\,\leq\, 10\sqrt \eps\;.$$
	In particular, $|\val(\game,\strategy)-\val(\game,\strategy')|\leq 10\sqrt{\eps}.$
\end{claim}

\begin{proof}
	As $\omega$ is an $\eps$-near bijection, by Claim \ref{claim:corner_POVMs_joint_distribution}, for every edge $\mttx\mtty\in E$ in the game, jointly measuring according to $(\cal{Q}^\mttx,\cal{Q}^\mtty)$ is $4\sqrt{\eps}$-close to jointly measuring according to the corners $(\omega\cal{Q}^\mttx\omega^*,\omega\cal{Q}^\mtty\omega^*)$, namely
	\begin{equation}\label{eq:abcdefg1}
	\sum_{\substack{a\colon S_\mttx\to\FF_2\\b\colon S_\mtty\to \FF_2}}|\tau(\cal{Q}^\mttx_a\cal{Q}^\mtty_b)-\tau(\omega^*\cal{Q}^\mttx_a\omega\omega^*\cal{Q}^\mtty_b\omega)|\leq 4\sqrt{\eps}\ .
	\end{equation}
	In addition, as $\cal{Q}$ was projective, by Claim \ref{claim:corner_of_PVM_almost_projective}, the corner strategy $\omega^*\cal{Q}\omega$ is $\eps$-almost projective. 
	
	For every $\mttx\in V$, let $\eps_\mttx$ be the distance (Definition \ref{defn:distance_inconsistency_POVMs}) between the PVM $\cal{P}^\mttx$ and corner POVM $\omega^*\cal{Q}^\mttx\omega$; by the $\eps$-flexible-closeness of $\strategy$ and $\strategy'$, we have 
	\begin{equation}\label{eq:abcdefg0}
	\Es{\mttx\mtty\sim \mu}\left[\frac{\eps_\mttx+\eps_\mtty}{2}\right]=\Es{\mttx\sim \mu}[\eps_\mttx]\leq \eps\ .
	\end{equation}
	For every edge $\mttx\mtty\in E$, by Claim \ref{claim:close_almost_prjective_POVMs_similar_joint}, jointly measuring according to $(\cal{P}^\mttx,\cal{P}^\mtty)$ is $\eps+2\sqrt{\eps_\mttx}$-close to jointly measuring according to $(\omega^*\cal{Q}^\mttx\omega,\cal{P}^\mtty)$, which in turn is $\eps+2\sqrt{\eps_\mtty}$-close to jointly measuring according to $(\omega^*\cal{Q}^\mttx\omega,\omega^*\cal{Q}^\mtty\omega)$. Hence,
	\begin{equation}\label{eq:abcdefg2}
	\sum_{\substack{a\colon S_\mttx\to\FF_2\\b\colon S_\mtty\to \FF_2}}|\tau(\cal{P}^\mttx_a\cal{P}^\mtty_b)-\tau(\omega^*\cal{Q}^\mttx_a\omega\omega^*\cal{Q}^\mtty_b\omega)|\leq 2\eps+2\sqrt{\eps_\mttx}+2\sqrt{\eps_\mtty}\ .
	\end{equation}
	
	Using Jensen's inequality and the bounds above, we deduce
	\begin{align*}
	d(p,q)&=\Es{\mttx\mtty\sim \mu}\Big[\sum_{\substack{a\colon S_\mttx\to\FF_2\\b\colon S_\mtty\to \FF_2}}|p(a,b|\mttx,\mtty)-q(a,b|\mttx,\mtty)\Big]\\
	&=\Es{\mttx\mtty\sim \mu}\Big[\sum_{\substack{a\colon S_\mttx\to\FF_2\\b\colon S_\mtty\to \FF_2}}|\tau(\cal{P}^\mttx_a\cal{P}^\mtty_b)-\tau(\cal{Q}^\mttx_a\cal{Q}^\mtty_b)|\Big]\\
	&\leq_{\triangle+\eqref{eq:abcdefg1}+\eqref{eq:abcdefg2}}\Es{\mttx\mtty\sim \mu}\Big[4\sqrt{\eps}+2\eps+\underbrace{2\sqrt{\eps_\mttx}+2\sqrt{\eps_\mtty}}_{\leq_{\textrm{Jensen}}4\sqrt{\frac{\eps_\mttx+\eps_{\mtty}}{2}}}\Big]\\
	&\leq_{\textrm{Jensen}}4\sqrt\eps+2\eps+4\sqrt{\Es{\mttx\mtty\sim \mu}\left[\frac{\eps_\mttx+\eps_\mtty}{2}\right]}\\
	&\leq_\eqref{eq:abcdefg0}8\sqrt\eps+2\eps\leq 10\sqrt\eps\ .
	\end{align*}

\end{proof}

We end this subsection by recalling a standard definition of \emph{robustness} for games. This notion is commonlu used in the soundness analysis of games, which generally uses the condition $\val(\game,\strategy)\geq 1-\eps$ to deduce many constraints on the structure of $\strategy$. 

\begin{definition}\label{defn:robustness}
	A game $\game$ is said to be $\delta$-\emph{robust} (or \emph{rigid} or \emph{stable}), where $\delta\colon [0,1]\to [0,1]$ is a non-decreasing function with $\delta(\eps)\xrightarrow{\eps\to 0}0$, if for every strategy $\strategy$ with $\val(\game,\strategy)\geq 1-\eps$, there is a perfect strategy $\strategy'$ where $d(\strategy,\strategy')\leq \delta(\eps)$.
	
	A game $\game$ is a \emph{self test} if all perfect strategies for it are the same up to isometries and corners (Definition \ref{defn:projections_isometries_partial_isometries}). Namely, it has essentially one perfect strategy.
\end{definition}

\begin{remark}
	An example of an $O(\eps)$-robust self test is the magic square game from Example \ref{example:magic-square}.
\end{remark}

\subsection{Data processing}
\label{sec:toolbox-dp}

\emph{Data processing} refers to the process of ``coarse-graining'' a POVM by applying a (generally non-injective) function to its output to define a new POVM. To formalize this, we introduce the following notation. 

\begin{definition}[Data processing POVM]\label{defn:Data_proccessed_PVM}
	Let $\{\cal{P}_a\}_{a\in A}$ be a POVM, and $f\colon A\to A'$ be a function. The $f$-evaluated POVM $\{\cal{P}_{[f(\cdot)=a']}\}_{a'\in A'}$ is defined to be 
	\[
	\cal{P}_{[f(\cdot)=a']}=\sum_{a\in A\colon f(a)=a'}\cal{P}_a.
	\]
	One can think of this POVM procedurally as first measuring $a\in A$ and then outputting $f(a)$ --- clarifying the term data processing. If $A=\FF_2^S$, $A'=\FF_2^{S'}$ for finite sets $S$ and $S'$, and $\cal{P}$ is projective,  then $\cal{P}_a$ and $\cal{P}_{[f(\cdot)=a']}$ have an observable and representation form. If $\cal{U}$ is the observable (or representation) form of $\cal{P}$, then we denote by $\cal{U}_{[f]}$ the observable form of $\cal{P}_{[f(\cdot)=a']}$.
	
	A common function that we data process along is restriction to a substring. For this case, we use the following notation. Let
	$S=S'\sqcup S''$, and let $f\colon\FF_2^S\to \FF_2^{S'}$ be the restriction to the $S'$ substring, namely $f(\gamma)=\gamma|_{S'}$. In this case, we commonly denote $\cal{P}_{[f(\cdot)=a]}$ by $\cal{P}^{S'}_a$.
	So, sampling $a\colon S'\to \FF_2$ according to $\cal{P}^{S'}$ is the same as sampling $\gamma\colon S\to \FF_2$ according to $\cal{P}$ and returning the restriction of $\gamma$ to $S'$, namely $a=\gamma|_{S'}$. This restriction operation makes sense also in the observable and representation form $\cal{U}$ of the PVM: Let $\cal{U}^{S'}\colon \{0,1\}^{S'}\to U(n)$ be the composition of the embedding $\iota\colon \{0,1\}^{S'}\to \{0,1\}^S$ --- defined by extending every function to be zero outside of $S'$ --- with $\cal{U}$, i.e.,
	\[
	\forall \alpha\colon S'\to \FF_2\ \colon \ \ \cal{U}^{S'}(\alpha)=\prod_{\sX\in S'}\cal{U}(\sX)^{\alpha(\sX)}.
	\]
	It is straightforward to check that, indeed, $\cal{U}^{S'}$ is the Fourier transform of $\cal{P}^{S'}$.
\end{definition}

\begin{remark}
	Note that the general data processing operation is very natural on PVMs in projective form, and usually unnatural in observable or representation forms (except for special cases, such as the restriction). This is a recurrent theme. Some operations and arguments are easier in the projective viewpoint and others in the observable viewpoint. It is good to remember that the object is the same, whether it is viewed in projective or observable (or representation) form, and thus one can apply operations and arguments in the more convenient form.
\end{remark}

\begin{observation}\label{obs:data_processing1}
	Let $\cal{P}$ be a POVM with outcomes in $A$, $\cal{Q}$ a POVM of the same dimension with outcomes in $B$, and $f\colon A\to C, g\colon B\to C$ two functions. Then, in the spirit of Remark \ref{rem:inconsistency_and_joint_sampling}, the probability a jointly sampled pair $(a,b)\sim(\cal{P},\cal{Q})$ {does not} satisfy $f(a)=g(b)$ is exactly the inconsistency of the data processed POVMs $\cal{P}_{[f(\cdot)=\cdot]}$ and $\cal{Q}_{[g(\cdot)=\cdot]}$. As small inconsistency implies small distance (clause $4.$ in Proposition \ref{prop:properties_of_distance_and_inconsistency}), this will be a useful tool for deducing that the POVMs a strategy associates to the endpoints of an edge are close (after data processing them).
	
	In the other direction, as PVMs are self-consistent,  a jointly sampled pair $(a,a')\sim(\cal{P},\cal{P}_{[f(\cdot)=\cdot]})$  where $\cal{P}$ is a PVM always satisfies $f(a)=a'$.
\end{observation}

\begin{claim}[Inconsistency can only decrease by data processing. Cf.\ Fact 4.25 in \cite{NW19}]\label{claim:data_processing3}
	Let $\cal{P}$ and $\cal{Q}$ be POVMs of the same dimension with outcomes in the set $A$, and let $f\colon A\to A'$ be a function. Then $\cal{P}\simeq_\eps\cal{Q}$ implies $\cal{P}_{[f(\cdot)=\cdot]}\simeq_\eps \cal{Q}_{[f(\cdot)=\cdot]}$.
\end{claim}
\begin{proof}
	This is immediate from the fact that applying a function on a pair of answers may only increase the probability of them agreeing. Let us provide the calculation in any case:
	\begin{align*}
	\sum_{a'\neq b'\in A'}\tau(\cal{P}_{[f(\cdot)=a']}\cal{Q}_{[f(\cdot)=b']})&=\sum_{a'\neq b'\in A'}\tau\Big(\Big(\sum_{a\in A\colon f(a)=a'}\cal{P}_a\Big)\Big(\sum_{b\in A\colon f(b)=b'}\cal{Q}_b\Big)\Big)\\
	&=\sum_{a,b\in A\colon f(a)\neq f(b)}\tau(\cal{P}_a\cal{Q}_b)\\
	&\leq \sum_{a\neq b\in A}\tau(\cal{P}_a\cal{Q}_b)\ .
	\end{align*}
\end{proof}

As in tailored games the comparisons along edges are linear, the following special case of Observation \ref{obs:data_processing1} and Claim \ref{claim:data_processing3} will be repeatedly used. We add the proof for clarity.

\begin{claim}[Consistency of linear checks]\label{claim:perfect_3Lin_commuting_observables}
	Let $S_\mttx,S_\mtty$ be finite sets, $\cal{U}^\mttx\colon \FF_2^{S_\mttx}\to U(n)$ and $\cal{U}^\mtty\colon \FF_2^{S_\mtty}\to U(n)$ two representations.  Fix $\alpha\colon S_\mttx\to \FF_2$ and $\beta\colon S_{\mtty}\to \FF_2$. Then, the probability  that $\sum_{\sX\in S_\mttx}\alpha(\sX)\gamma(\sX)\neq\sum_{\sY\in S_\mtty}\beta(\sY)\gamma(\sY)$  when $\gamma\colon S_\mttx\sqcup S_\mtty\to \FF_2$ is jointly sampled (Definitions \ref{defn:joint_measurement} and \ref{defn:projective_representation_and_observable_form_PVM}) according to $(\cal{U}^\mttx,\cal{U}^\mtty)$  is exactly $\nicefrac{1}{4}\cdot\|\cal{U}^\mttx(\alpha)-\cal{U}^\mtty(\beta)\|_{hs}^2$.
\end{claim}
\begin{proof}
	Denote, as usual, $\gamma=(a,b)$. First, note that $\sum_{\sX\in S_\mttx}\alpha(\sX)\gamma(\sX)\neq\sum_{\sY\in S_\mtty}\beta(\sY)\gamma(\sY)$ if and only if $\langle (\alpha,\beta),\gamma\rangle=1$.
	Then,
	\[
	\begin{split}
	\Pro{\gamma\sim (\cal{U}^\mttx,\cal{U}^\mtty)}[\langle (\alpha,\beta),\gamma\rangle=0]&=\sum_{\gamma=(a,b)\colon \langle \alpha,a\rangle=\langle \beta,b\rangle} \tau\left(\cal{P}^{\mttx}_{a}\cal{P}^\mtty_{b}\right)\ ,\\
	\Pro{\gamma\sim (\cal{U}^\mttx,\cal{U}^\mtty)}[\langle (\alpha,\beta),\gamma\rangle=1]&=\sum_{\gamma=(a,b)\colon \langle \alpha,a\rangle\neq \langle \beta,b\rangle} \tau\left(\cal{P}^{\mttx}_{a}\cal{P}^\mtty_{b}\right)\ ,\\
	\cal{U}^\mttx(\alpha)&=\sum_{a\colon \langle a,\alpha\rangle=0}\cal{P}^\mttx_a-\sum_{a\colon \langle a,\alpha\rangle=1}\cal{P}^\mttx_a\ ,\\
	\cal{U}^\mtty(\beta)&=\sum_{b\colon \langle b,\beta\rangle=0}\cal{P}^\mtty_b-\sum_{b\colon \langle b,\beta\rangle=1}\cal{P}^\mtty_b\ .
	\end{split}
	\]
	So,
	\[
	\begin{split}
	\tau\left(\cal{U}^\mttx(\alpha)\cal{U}^\mtty(\beta)\right)&=\tau\Big( \sum_{\gamma=(a,b)\colon \langle{a,\alpha}\rangle =\langle b,
		\beta\rangle }\cal{P}^\mttx_a\cal{P}^\mtty_b-\sum_{\gamma=(a,b)\colon \langle{a,\alpha}\rangle \neq \langle b,\beta\rangle}\cal{P}^\mttx_a\cal{P}^\mtty_b\Big)\\
	&=\Pro{\gamma\sim (\cal{U}^\mttx,\cal{U}^\mtty)}[\langle (\alpha,\beta),\gamma\rangle=0]-\Pro{\gamma\sim(\cal{U}^\mttx,\cal{U}^\mtty)}[\langle (\alpha,\beta),\gamma\rangle=1]\\
	&=1-2\Pro{\gamma\sim (\cal{U}^\mttx,\cal{U}^\mtty)}[\langle (\alpha,\beta),\gamma\rangle=1]\ ,
	\end{split}
	\]
	and
	\[
	\begin{split}
	\left\|\cal{U}^\mttx(\alpha)-\cal{U}^\mtty(\beta)\right\|_{hs}^2&=\left\|\Id- \cal{U}^\mttx(\alpha)\cal{U}^\mtty(\beta)\right\|_{hs}^2\\
	&=2-2\mathrm{Re}\left(\tau\left(\cal{U}^\mttx(\alpha)\cal{U}^\mtty(\beta)\right)\right)\\
	&=2-2\mathrm{Re}\left(1-2\Pro{\gamma\sim (\cal{U}^\mttx,\cal{U}^\mtty)}[\langle (\alpha,\beta),\gamma\rangle=1]\right)\\
	&=4\Pro{\gamma\sim (\cal{U}^\mttx,\cal{U}^\mtty)}[\langle (\alpha,\beta),\gamma\rangle=1]\ ,
	\end{split}
	\]
	as claimed.
\end{proof}

\subsection{Data processing of  permutation strategies}\label{sec:About_perm_strategies}
As opposed to $f$-evaluating (Definition \ref{defn:Data_proccessed_PVM}) POVMs and PVMs, which clearly remain POVMs and PVMs respectively, it is not true for  a general function $f\colon \FF_2^S\to \FF_2^{S'}$ that the $f$-evaluation  of a {signed permutation PVM} (Definition \ref{defn:perm_PVM}) remains a signed permutation PVM. A simple example for this phenomenon is the bit product function (which is the arithmetic version of the boolean $\rm{AND}$), namely $f(a,b)=a\cdot b$. One can check that if we have two formal variables $\sX,\sY$ that are sent by $\cal{U}$ to the commuting involutive permutation matrices \begin{equation}\label{eq:example_of_permutations_with_non_perm_AND}
\begin{bmatrix}
0 & 1 & 0 & 0 \\
1 & 0 & 0 & 0 \\
0 & 0 & 0 & 1 \\
0 & 0 & 1 & 0
\end{bmatrix}\quad,\quad \begin{bmatrix}
0 & 0 & 1 & 0 \\
0 & 0 & 0 & 1 \\
1 & 0 & 0 & 0 \\
0 & 1 & 0 & 0
\end{bmatrix}\ ,
\end{equation}
then the observable form of the $f$-evaluated PVM $\cal{U}_{[f]}$ is 
\[
\frac{1}{2}\begin{bmatrix}
1 & 1 & 1 & -1 \\
1 & 1 & -1 & 1 \\
1 & -1 & 1 & 1 \\
-1 & 1 & 1 & 1
\end{bmatrix}\ ,
\]
which is not a signed permutation matrix.
As the completeness in Theorem \ref{thm:tailored_MIP*=RE} requires working with $Z$-aligned permutation strategies that commute along edges ($\ZPC$ strategies), we need to characterize the functions $f$ for which $f$-evaluation preserves the $\ZPC$-property.

To that end, we need the following to claims:
\begin{claim}[Data processing diagonal PVMs]\label{calim:diagonal_data_processing}
	If $\cal{P}$ is a diagonal PVM (Definition \ref{defn:diagonal_PVM}) with outcomes in $A$, and $f\colon A\to A'$ is a function, then $\cal{P}_{[f(\cdot)=\cdot]}$ is a diagonal PVM.
\end{claim}

\begin{proof}
	As diagonal matrices are closed under addition, the claim follows. 
\end{proof}
Given a finite set $S$, there is a one to one correspondence between functions $\alpha\colon S\cup \{\sJ\}\to \FF_2$ and affine maps (i.e., linear maps plus a constant) $\beta\colon \FF_2^S\to \FF_2$ given by 
\begin{equation}\label{eq:correspondence_affine_and_functions}
\forall a\colon S\to \FF_2\ \colon \ \ \beta(a)=\alpha(\sJ)+\sum_{\sX\in S}\alpha(\sX)a(\sX)\ .
\end{equation}
\begin{claim}[Affine data processing]\label{claim:affine_data_processing}
	Let $\cal{U}\colon S\to U(n)$ be a PVM in observable form, let $\alpha\colon S\cup \{\sJ\}\to \FF_2$ be a function, and let $\beta$ be the corresponding affine map as in \eqref{eq:correspondence_affine_and_functions}.
	Let $\cal{U}_{[\beta]}$ be the $\beta$-evaluated PVM (Definition \ref{defn:Data_proccessed_PVM}), which consists of a single observable $\cal{O}$. Then, 
	\[
	\cal{O}=(-\Id)^{\alpha(\sJ)}\cdot \prod_{\sX\in S} \cal{U}(\sX)^{\alpha(\sX)}\ .
	\]
	In particular, $\cal{O}$ is in the group generated by $-\Id$ and $\Img(\cal{U})$.
\end{claim}

\begin{proof}
	This follows from Observation \ref{obs:data_processing1} and Claim \ref{claim:perfect_3Lin_commuting_observables}.
\end{proof}

\begin{corollary}\label{cor:linear_data_processed_PVM_is_ZPC_and_left_multiplication}
	If $\beta\colon \FF_2^S\to \FF_2$ is an affine function, and $\cal{U}\colon S\to \Sym_{\pm}(\Omega)$ is a signed permutation PVM in observable form (Definition \ref{defn:perm_PVM}), then the $\beta$-evaluated $\cal{U}_{[\beta]}$ is also a signed permutation PVM. Moreover, if $\gamma\colon \FF_2^S\to \FF_2^{S'}$ is a linear map, then 
	\begin{equation}
	\forall         \alpha\colon S'\to \FF_2\ \colon \ \ \cal{U}_{[\gamma]}(\alpha)=\cal{U}(\gamma^*(\alpha))\ ,
	\end{equation}
	where $\gamma^*\colon \FF_2^{S'}\to \FF_2^S$ is the dual map with respect to the bilinear form $\langle\cdot,\cdot\rangle$. In other words, if we fix the standard bases of $\FF_2^S,\ \FF_2^{S'}$ to be the indicators ${\bf 1}_\sX$, then $\gamma$ is a matrix and $\gamma^*(\alpha)$ is the left multiplication 
	\[
	\alpha\cdot \gamma=\Big(\sum_{\sX'\in S}\alpha_{\sX'}\cdot \gamma_{\sX' \sX}\Big)_{\sX\in S}\ .
	\]
\end{corollary}

\begin{observation}\label{obs:Z-aligned-perm_PVM-structure}
	A readably $Z$-aligned (Definition \ref{defn:readably_Z-aligned_PVM}) signed permutation PVM (Definition \ref{defn:perm_PVM}) in observable form 
	\[
	\cal{U}\colon S^\frR\cup S^\frL\to \Sym_\pm(\Omega)\subseteq U(W^-)\ ,
	\] when represented with respect to the standard basis $B^-$ from \eqref{eq:bases_for_symmetric_and_anti-symmetric_functions}, consists of block diagonal matrices, where in each block the readable observables are constant (and diagonal) --- this is immediate from the fact  that the image of $\cal{U}$ consists of commuting matrices for which the readable variables are diagonal $\pm 1$ matrices.
\end{observation}

\begin{corollary}\label{cor:encodings}
	Let $\cal{U}\colon S^\frR\cup S^\frL\to \Sym_\pm(\Omega)$ be a readably  $Z$-aligned (Definition \ref{defn:readably_Z-aligned_PVM}) signed permutation PVM (Definition \ref{defn:perm_PVM}). For every $a^\frR\in \FF_2^{S^\frR}$ let $\frS_{1,
		a^\frR},...,\frS_{k,a^\frR}$ be a sequence of affine maps  from $\FF_2^{S^\frR\cup S^\frL}\to \FF_2$ and let $f_1,...,f_t$ be a collection of functions from $\FF_2^{S^\frR}\to \FF_2$. Then, by adding a set $\{\sX_{f_i}\}_{\i=1}^t$ of readable variables and $\{\sY_{\frS_j}\}_{j=1}^k$ of unreadable variables, we can define a readably $Z$-aligned signed  permutation PVM $\cal{V}$ that extends\footnote{For the joint sampling to make sense, the formal variables of $\cal{U}$ and $\cal{V}$ should be disjoint. So, the notion of extension is a bit misleading, but it is easier to follow notationally. What we mean is that $\cal{V}$ has a {copy} of the variables at $\cal{U}$ and associates the same observables to them.} $\cal{U}$ to the new variables and satisfies the following: Given a sampled pair $(a,b)\sim (\cal{U},\cal{V})$, denoting $a^\frR=a|_{S^\frR}$,  we have
	\begin{align}
	\forall \sX\in S^\frR\cup S^\frL\ \colon \ \ a(\sX)&=b(\sX)\ ,\label{eq:ZZZZZ1}\\
	\forall 1\leq i\leq t\ \colon\ \ b(\sX_{f_i})&=f_i(a^\frR)\ ,\label{eq:ZZZZZ2}\\
	\forall 1\leq j\leq k\ \colon \ \ b(\sY_{\frS_j})&=\frS_{j,a^\frR}(a)\ .\label{eq:ZZZZZ3}
	\end{align}
	In words, we can replace a readably $Z$-aligned signed permutation PVM with outcomes in $\FF_2^S$ with a new readably $Z$-aligned signed permutation PVM with outcomes in $\FF_2^{S'}$, where $S\subseteq S'$, and the new PVM samples the same strings as the original PVM (as part of its output), and in addition has bits which are either functions on the values of the readable variables, or affine combinations of all values, where the specific combinations depend on the values of the readable variables. 
\end{corollary}

\begin{proof}
	Condition \eqref{eq:ZZZZZ1} is guaranteed by choosing $\cal{V}$ to be an extension of $\cal{U}$. For \eqref{eq:ZZZZZ2}, the restriction of $\cal{U}$ to the readable variables is diagonal (by the definition of a readably $Z$-aligned measurement), and thus it is immediate from Claim \ref{calim:diagonal_data_processing}. Finally, due to Observation \ref{obs:Z-aligned-perm_PVM-structure}, Claim \ref{claim:affine_data_processing} can be applied to each block individually with the appropriate affine map, as the value of the readable variables there is constant.
\end{proof}

\begin{remark}
	Corollary~\ref{cor:encodings} is very useful, as it allows one to take perfect strategies and encode each output of them in various ways. Specifically, one can encode the outputs using error correcting codes, which is an important step in the construction of PCPs. See Section \ref{sec:answer_reduction} for more on that. It essentially characterizes the types of functions $f\colon \FF_2^S\to \FF_2^{S'}$ for which the $f$-evaluated PVM remains a signed permutation PVM  and preserves the readably $Z$-aligned structure.
\end{remark}

\subsection{Transformations of games}\label{sec:composition_of_games}

Compression consists of applying various transformations to the input normal form verifier $\verifier$. Some of these transformations, when observed  as acting on the associated games  $\verifier_n$, are applying some form of \emph{game composition}. As it sounds, composing games is just a process that takes two (or more) games, and generates a new game out of them. 

\subsubsection{Product and sum of games}

We describe two straightforward examples: \emph{product} and \emph{sum} of games.  The product  is the parallel play in both games --- namely, each round of the product consists of a round from both games --- while the sum is the barycenter  of them --- namely, with probability $\nicefrac{1}{2}$ it plays a round of one game, and with probability $\nicefrac{1}{2}$ it plays a round in the other game.\footnote{The exact probabilities will not necessarily be $\nicefrac{1}{2}$, but it is a good example to hold in mind --- so, the sum is more of a convex combination than exactly the barycenter.} On the level of the underlying graphs, the product of games has the tensor product of the graphs underlying it, and the sum of games has a disjoint union of the graphs underlying it.

\begin{definition}[Product of games]\label{def:product-game}
	Given two games $\game_1$ and $\game_2$, their product $\game_1\otimes\game_2$ is defined as follows. If $G_1$ and $G_2$ are the underlying graphs of $\game_1$ and $\game_2$ respectively, then the underlying graph of the product is $G_1\otimes G_2$. An  edge in $G_1\otimes G_2$ is of the form $(\mttx_1,\mttx_2)(\mtty_1,\mtty_2)$, where $\mttx_1\mtty_1$ is an edge in $G_1$ and $\mttx_2\mtty_2$ is an edge in $G_2$. The probability of sampling $(\mttx_1,\mttx_2)(\mtty_1,\mtty_2)$ is $\mu_1(\mttx_1\mtty_1)\cdot \mu_2(\mttx_2\mtty_2)$. For the length function (and in the tailored category, functions), $\ell(\mttx_1,\mttx_2)=\ell_1(\mttx_1)+\ell_2(\mttx_2)$, and we think of $S_{(\mttx_1,\mttx_2)}$ as the disjoint union of (its own copies of) $S_{\mttx_1}$ and $S_{\mttx_2}$ (respectively for readable and unreadable variables in the tailored category). Hence, we can think of the answer to $(\mttx_1,\mttx_2)$ as being a pair of answers $(a_1,a_2)$.  Finally, $D_{(\mttx_1,\mttx_2)(\mtty_1,\mtty_2)}((a_1,a_2)(b_1,b_2))=D^1_{\mttx_1\mtty_1}(a_1b_1)\cdot D^2_{\mttx_2\mtty_2}(a_2b_2)$, where $D^1$ (respectively $D^2$) is the decision function of $\game_1$ (respectively $\game_2$). This again works out in the tailored category by letting $L_{(\mttx_1,\mttx_2)(\mtty_1,\mtty_2)}((a^\frR_1,a^\frR_2),(b^\frR_1,b^\frR_2))=L^1_{\mttx_1\mtty_1}(a^\frR_1,b^\frR_1)\sqcup L^2_{\mttx_2\mtty_2}(a^\frR_2,b^\frR_2)$, where  $L^1$ and $L^2$ are the respective controlled linear constraints functions of $\game_1$ and $\game_2$.\footnote{It is straightforward to check that indeed, by taking the disjoint union of linear constraints, the canonical decider will accept the answers only when it would have accepted them in each game separately.} Note that the above union of linear constraints makes sense only because the formal generating sets at $\mttx_1$ and $\mttx_2$ are ``embedded'' in $S_{(\mttx_1,\mttx_2)}$ (and similarly for $\mtty$). 
\end{definition}

Though it is fitting to start and analyze the completeness and soundness properties of the product game, we leave it to the parallel repetition section, Section \ref{sec:parallel_rep}, in which it is used. 

\begin{definition}[Sum of games]\label{defn:sum_of_games}
	Given two games $\game_1$ and $\game_2$, their sum $\game_1\oplus\game_2$ is defined as follows. The underlying graph is the disjoint union of the underlying graphs $G_1$ and $G_2$ of $\game_1$ and $\game_2$. The distribution on edges is $$\mu(e)=\begin{cases}
	\frac{\mu_1(e)}{2} & e\in G_1\ ,\\
	\frac{\mu_2(e)}{2} & e\in G_2\ ,
	\end{cases}$$
	where $\mu_1$ and $\mu_2$ are the respective distributions on edges in $\game_1$ and $\game_2$.\footnote{As we remarked before, the distribution may change from exactly $\nicefrac{1}{2}-\nicefrac{1}{2}$ to some other one.} 
	The length function is 
	$$\ell(\mttx)=\begin{cases}
	\ell_1(\mttx) & \mttx\in G_1\ ,\\
	\ell_2(\mttx) & \mttx\in G_2\ ,
	\end{cases}$$
	with $\ell_1$ and $\ell_2$ being the respective length functions (and similarly in the tailored category). We presume $S_\mttx$ remains the same in this case. Finally, every edge $\mttx\mtty$ is either in $G_1$ or in $G_2$. If it is in $G_1$, then $D_{\mttx\mtty}(\gamma)=D^1_{\mttx\mtty}(\gamma)$, and if it is in $G_2$, then $D_{\mttx\mtty}(\gamma)=D^2_{\mttx\mtty}(\gamma)$.  Furthermore, in the tailored category, we assume $L_{\mttx\mtty}(\gamma)=L^1_{\mttx\mtty}(\gamma)$ and $L_{\mttx\mtty}(\gamma)=L^2_{\mttx\mtty}(\gamma)$ with respect to whether $\mttx\mtty\in G_1$ or $\mttx\mtty\in G_2$.\footnote{This again can be checked to work the same on the level of canonical deciders.}
\end{definition}

Since the sum has a disconnected underlying graph, it is natural to  \emph{augment} it, so that the games are forced to be related in some way. There are many ways to do so. Usually, the augmentation involves the addition of  vertices and edges between the graphs that check various  consistencies between the answers. Both the Pauli basis game (Section \ref{sec:Pauli_basis_definition}) and the question reduction game (Section \ref{sec:augmentation_in_the_CLM_case}) are augmented sums of smaller games.

\begin{definition}[Augmentation of a game]\label{defn:augmentation_of_a_game}
	Given a game $\game$, we say that another game $\game'$ is \emph{an augmentation of} $\game$, or conversely that $\game$ is \emph{contained} or a \emph{sub-game} of $\game'$, if there is a subgraph of the underlying graph of $\game'$ such that the restriction of $\game'$ to this subgraph is an instance of $\game$ (up to the distribution on edges $\mu$). In more words, $\game'$ is defined by adding vertices and edges to the underlying graph of $\game$, such that the lengths and decision procedure on the ``original'' edges stays the same.
\end{definition}

\subsubsection{Product and Sum of PVMs}

Recall the notion of the Kronecker tensor product of matrices, which we denote by $\otimes$.
Similar to composition of games, we can also compose PVMs (and thus strategies), which results in new PVMs with some new properties. 
\begin{definition}[Product of PVMs]\label{claim:tensor-permutation-strategies}
	Given two PVMs (in observable form) $\cal{U}^1\colon S\to U(N)$ and $\cal{U}^2\colon S\to U(M)$ over {the same} variable set $S$,  we define their product to be the PVM $\cal{U}^1\otimes\cal{U}^2\colon S\to U(NM)$ satisfying \[
	(\cal{U}^1\otimes \cal{U}^2)(\sX)=\cal{U}^1(\sX)\otimes\cal{U}^2(\sX)\ .
	\]
\end{definition}

\begin{definition}[Sum of PVMs]\label{def:sum-pvm}
	Given two PVMs (in observable form) $\cal{U}^1\colon S^1\to U(N)$ and $\cal{U}^2\colon S^2\to U(M)$ over \textbf{different} variable sets $S$,  we define their sum to be the PVM $\cal{U}^1\oplus\cal{U}^2\colon S^1\sqcup S^2\to U(NM)$ defined by 
	\[
	\cal{U}^1\oplus\cal{U}^2(\sX)=\begin{cases}
	\cal{U}^1(\sX)\otimes \Id_M & \sX\in S^1\ ,\\
	\Id_N\otimes \cal{U}^2(\sX) & \sX\in S^2\ .
	\end{cases}
	\]
\end{definition}

\begin{remark}\label{rk:sum-product}
	It is straightforward to check that given two signed permutation matrices their Kronecker tensor product is a signed permutation as well, and that the tensor product of diagonal matrices is diagonal.  Hence, the above two operations on PVMs (sum and product) preserve readable $Z$-alignment as well as being a signed permutation PVM.
	
	As signed permutations act on a signed sets, when one performs the tensor product of two signed permutations, one acting on $\Omega^1_\pm$ and one on $\Omega^2_\pm$, the resulting signed permutation acts on $(\Omega^1\times \Omega^2)_\pm$. For explanatory  reasons, we   define an equivalence relation on $\Omega^1_\pm\times \Omega^2_\pm$ which bijects it on $(\Omega^1\times \Omega^2)_{\pm}$ by letting 
	\begin{equation}\label{eq:equivalence_relation_product_of_sets}
	\forall \star\in \Omega^1,\diamond\in \Omega^2\ \colon \ \ (+\star,+\diamond)=(-\star,-\diamond)=+(\star,\diamond)\quad \textrm{and}\quad (-\star,+\diamond)=(+\star,-\diamond)=-(\star,\diamond)\ .
	\end{equation}
	In this guise, the tensor product acts as expected: Given two signed permutation strategies $\sigma_1\colon S\to \Sym(\Omega^1_\pm)$ and $\sigma_2\colon S\to \Sym(\Omega^2_\pm)$, we have
	\begin{equation}\label{eq:def_tensor_of_perm_sterategies}
	\begin{split}
	\forall \sX\in S,\spadesuit\in \Omega^1_\pm,\diamondsuit\in \Omega^2_\pm\ \colon \ \ \sigma_1\otimes \sigma_2(\sX).(\spadesuit,\diamondsuit)=\sigma_1(\sX)\times \sigma_2(\sX).(\spadesuit,\diamondsuit)=(\sigma_1(\sX).\spadesuit,\sigma_2(\sX).\diamondsuit)\ .
	\end{split}
	\end{equation}
\end{remark}
\begin{remark}
	The product and sum operations should be familiar to graph theorists, as these are PVM analogs of the \emph{tensor product and cartesian product of graphs}. Indeed, if one applies these transformations to signed permutation PVMs, and look at the resulting Schreier graph induced by the new PVMs (in observable form), then it is respectively the tensor product and cartesian product of the original Schreier graphs.
\end{remark}

\begin{lemma}\label{lem:sum-zpc}
	Let $\game_1$ and $\game_2$ be two (tailored) games, and let $\game=\game_1\otimes \game_2$ be their product taken according to Definition~\ref{def:product-game}. For $i\in\{1,2\}$ let
	$\cal{U}^i : S \to U(N_i)$
	be a strategy for $\game_i$, in observable form. Let $\cal{U} = \cal{U}^1 \oplus \cal{U}^2$ be their sum, taken according to Definition~\ref{def:sum-pvm}. Then the following hold:
	\begin{enumerate}
		\item $\cal{U}$ is a valid strategy for $\game$. 
		\item If both $\cal{U}^1$ and $\cal{U}^2$ have value $1$, then so does $\cal{U}$. 
		\item If both $\cal{U}^1$ and $\cal{U}^2$ are $Z$-aligned, then so is $\cal{U}$. 
		\item If both $\cal{U}^1$ and $\cal{U}^2$ are commuting along edges, then so is $\cal{U}$. 
	\end{enumerate}
	As a consequence, if $\cal{U}^1$ and $\cal{U}^2$ are perfect $\ZPC$ strategies then so is $\cal{U}$. 
\end{lemma}

\begin{proof}
	The first item follows because, according to the definition, the set of generators $S$ for $\game_1\otimes \game_2$ is the disjoint union $S_1\sqcup S_2$. To show the second item, fix a question pair $(\mttx_1,\mttx_2),(\mtty_1,\mtty_2)$ in $\game$. Then the strategy $\cal{U}$ samples answers $\gamma$ according to a product distribution, i.e.
	\begin{align*}
	&\Pr[\gamma=((a_1,a_2),(b_1,b_2))\text{ is sampled by }\cal{U}\mid (\mttx_1,\mttx_2),(\mtty_1,\mtty_2)\text{ were sampled}] \\
	&= \Pr[\gamma_1=(a_1,b_1)\text{ is sampled by }\cal{U}^1\mid \mttx_1,\mtty_1\text{ were sampled}] \cdot\Pr[\gamma_2=(a_2,b_2)\text{ is sampled by }\cal{U}^2\mid \mttx_2,\mtty_2\text{ were sampled}] \ .
	\end{align*}
	Item 2 follows since the decision function of the product game simply checks the conjunction of the decision functions of each individual game. Regarding item 3, its validity was already observed in Remark~\ref{rk:sum-product} above. Finally, item 4 follows because given an edge $(\mttx_1,\mttx_2),(\mtty_1,\mtty_2)$, the associated permutations are either associated to the edge $\mttx_1\mtty_1$ from $\game_1$ and act on the first tensor factor in $U(N_1)\otimes U(N_2)$, or associated with the  edge $\mttx_2\mtty_2$ from $\game_2$ and act on the first tensor factor. Since both $\cal{U}^1$ and $\cal{U}^2$ are assumed to commute along edges, and since unitaries acting on different tensor factors commute, the conclusion follows. 
\end{proof}

\subsubsection{Double cover of a game}\label{sec:double-cover}

Another natural transformation on games is their \emph{double cover}. A double cover of a graph $G=(V,E)$ is the  graph $G_\pm=(V_\pm,E_\pm)$ defined as follows:  $V_\pm=\{\pm \}\times V$, and we denote, as usual,  $+v$ instead of $(+,v)$ and $-v$ instead of $(-,v)$; for any (oriented) edge $e=(v,w)\in E$, there are two appropriate (oriented) edges $+e=(+v, -w)$ and $-e= (-v,+w)$ in $E_\pm$. As its name suggest, the double cover is indeed a combinatorial covering space (cf.\ \cite{bilu2006lifts} under the name of \emph{lifts}) of $G$, and the covering map $\pi\colon G'\to G$ is the one which removes the signs. 
The following are easy to verify facts about the double cover of a graph. 
\begin{fact}
	\ 
	\begin{enumerate}
		\item The double cover of a graph is always bipartite.
		\item The double cover of a bipartite graph is a disjoint union of two copies of the original graph.
	\end{enumerate}
\end{fact}

\begin{definition}[Double cover of a game]\label{defn:double_cover}
	Let $\game$ be a tailored game. Its double cover $\frak{DoubleCover}(\game)=\game'$ is a game whose underlying graph is the double cover $G_\pm$ of the underlying graph $G$ of the game $\game$. The distribution over edges in $\game'$ is defined to be 
	\[
	\forall \pm e\in E'\ \colon \ \ \mu'(\pm e)=\nicefrac{\mu(e)}{2}\ ,
	\]
	where $\mu$ is the distribution of $\game$ over the edges in $G$. Namely, the sampling scheme of the double cover game is as follows: Sample $e\in E$ according to $\mu$, and choose a sign $\eps\in \{\pm\}$ uniformly; output $\eps \cdot e\in E_\pm$.
	The lengths of the vertex $\eps \cdot \mttx\in V_\pm$ are the same as the lengths of $\mttx$ in $\game$.  In addition, the elements of the formal generating set $S_{+\mttx}$ will be of the form $+\sX$ for $\sX\in S_{\mttx}$, and similarly $-\sX$ will be the form of elements in $S_{-\mttx}$.
	If $+e=(+\mttx,-\mtty)$ (respectively $-e=(-\mttx,+\mtty)$) is sampled, then $L_{+ e}$ (respectively $L_{-e}$) treats $S_{+ \mttx}$ (respectively $S_{-\mttx}$) as $S_\mttx$ and $S_{- \mtty}$ (respectively $S_{+y}$) as $S_\mtty$ and outputs the appropriate linear constraints (given the restriction $\gamma^\frR\colon S^\frR_{\pm \mttx}\sqcup S^\frR_{\mp \mtty}\to \FF_2$). If $\mttx=\mtty$, then in addition to the above constraints, it also outputs the consistency checks
	\[
	\forall \sX\in S_{\mttx}\ \colon \ \ \gamma(+\sX)=\gamma(-\sX)\ .
	\]
	
	On the combinatorial level, the double cover acts as follows: If $a^\frR,a^\frL$ are the answers associated to $+\mttx$ and $b^\frR,b^\frL$ are the answers associated to $-\mtty$, for $\mttx\neq \mtty$,  then the double cover will accept these answers if and only if the original game would accept these answers for $\mttx$ and $\mtty$ respectively. In the case $\mttx=\mtty$, the double cover needs (in addition to the checks induced by the original game $\game$) to check  \emph{consistency}, namely that $a^\frR=b^\frR$ and $a^\frL=b^\frL$.
\end{definition}

\begin{remark}
	The definition of the double cover is natural when trying to relate non-synchronous strategies to synchronous strategies of the same game (see Section \ref{sec:non-synch_setup}). In addition, it is used in the \emph{detyping transformation} (Definition \ref{defn:combi_detyping}). 
\end{remark}

\begin{claim}\label{claim:completeness_soundness_double_cover}
	Let $\game$ be a tailored game, such that in its underlying graph $G=(V,E)$, all loops $\mttx\mttx$ for $\mttx\in V$ appear as edges in $E$. Assume in addition that there is some constant $c>0$, such that for every $\mttx\in V$ we have
	\begin{equation}\label{eq:loops_have_constant_mass}
	\frac{\mu(\mttx\mttx)}{\mu(\mttx)}\geq c\ ,        
	\end{equation}
	where $\mu$ is the distribution over edges in $\game$, and $\mu(\mttx)$ is (as before) the marginal on vertices, namely $\mu(\mttx)=\sum_{\mtty\in V}\frac{\mu(\mttx\mtty)+\mu(\mtty\mttx)}{2}$. Then:
	\begin{itemize}
		\item (Completeness) if $\game$ has a perfect $\ZPC$ strategy, then so does $\frak{DoubleCover}(\game)$;
		\item (Soundness) if $\frak{DoubleCover}(\game)$ has a strategy $\strategy$ with value $1-\eps$, then $\game$ has a strategy with value of at least $1-O(\nicefrac{\sqrt \eps}{c})$. In particular, 
		\[
		\Ent(\frak{DoubleCover}(\game),1-\eps)\geq \Ent(\game,1-O(\nicefrac{\sqrt \eps}{c}))\ .
		\]
	\end{itemize}
\end{claim}
\begin{proof}
	For completeness, note that if $\sigma\colon S\to \Sym(\Omega_\pm)$ is a perfect $\ZPC$ strategy, then $\sigma'\colon S_{\pm}\to \Sym(\Omega_\pm)$ defined by $\sigma'(\pm \sX)=\sigma(\sX)$ is a perfect $\ZPC$ strategy for $\frak{DoubleCover}(\game)$.

	For soundness, let $\strategy=\{\cal{U}\}$ pass $\frak{DoubleCover}(\game)$ with probability $1-\eps$. By \eqref{eq:loops_have_constant_mass}, 
	\[
	\sum_{\mttx\in V}\mu(\mttx\mttx)\leq \sum_{\mttx\in V}\mu(\mttx)\leq \frac{1}{c}\cdot \sum_{\mttx\in V}\mu(\mttx\mttx)\ ,
	\]
	and hence
	\[
	\Pro{}[\strategy\ {\rm loses}\mid {\rm a\ loop\ was\ sampled}]\leq \frac{\Pro{}[\strategy\ {\rm loses}]}{\Pro{}[{\rm a\ loop\ was\ sampled}]}\leq \frac{\eps}{\sum_{\mttx\in V}\mu(\mttx\mttx)}\leq \frac{\eps}{c\cdot\sum_{\mttx\in V}\mu(\mttx)}= \nicefrac{\eps}{c}\ .
	\]
	Let $\eps_{\mttx}$ be the probability $\strategy$ loses when $(+\mttx,-\mttx)$ or $(-\mttx,+\mttx)$ is sampled. Then by the above derivations and using \eqref{eq:loops_have_constant_mass} again,
	\begin{equation}\label{eq:pob_S_losses_given_loop}
	c\cdot \Es{\mttx\sim \mu}[\eps_\mttx]= c\cdot\sum_{\mttx\in V}\mu(\mttx)\eps_\mttx\leq \sum_{\mttx\in V}\mu(\mttx\mttx)\eps_\mttx\leq \frac{\sum_{\mttx\in V}\mu(\mttx\mttx)\eps_\mttx}{\sum_{\mtty\in V}\mu(\mtty\mtty)}=\Pro{}[\strategy\ {\rm loses}\mid {\rm a\ loop\ was\ sampled}]\leq \nicefrac{\eps}{c}\ .
	\end{equation}
	On the other hand, whenever $(+\mttx,-\mttx)$  or $(-\mttx,+\mttx)$ is sampled, the answers must be consistent;  by the equivalence between inconsistency and distance for projective measurements \eqref{eq:consistency_vs_distance_PVM_case}, and the distance notion for PVMs in representation form (Claim \ref{claim:dist_in_projection_and_observable_are_the_same}), one deduces
	\begin{equation}\label{eq:eps_x_lower_bound}
	\eps_\mttx\geq \Pro{(a,b)\sim (\cal{U}^{+\mttx},\cal{U}^{-\mttx})}[a\neq b]=\nicefrac{1}{2}\cdot\Es{\alpha\colon S_\mttx\to\FF_2 }\|\cal{U}^{+\mttx}(\alpha)-\cal{U}^{-\mttx}(\alpha)\|_{hs}^2\ .        
	\end{equation}
	Let $\strategy'=\{\cal{U}'\}$ be the strategy that uses the observables of the positive side for both sides of the double cover, namely satisfy
	\begin{equation}\label{eq:symmetrized_strategy}
	\forall \sX\in S\ \colon \ \ \cal{U}'(\pm \sX)=\cal{U}(+\sX)\ .
	\end{equation}
	Combining \eqref{eq:pob_S_losses_given_loop} and \eqref{eq:eps_x_lower_bound},  we deduce that the distance between the strategy $\strategy$ and $\strategy'$ is at most $\frac{\eps}{c^2}$. 
	As close by strategies produce similar values (Claim \ref{claim:close_strat_implies_close_correlations}), the value of $\strategy'$ is at least $1-\eps-10\nicefrac{\sqrt{\eps}}{c}$. Moreover, it is straightforward to check that the strategy $\strategy'$ for $\frak{DoubleCover}(\game)$ has the same value as the strategy $\strategy''=\{\cal{U}''\}$ for $\game$ that is defined by
	$\cal{U}''(\sX)=\cal{U}'(\pm\sX)=\cal{U}(+\sX)$, which proves the claim.
\end{proof}

\begin{remark}\label{rem:completeness_soundness_of_double_cover_bipartite_case}
	As the double cover of a bipartite graph is just a disjoint union of two copies of the underlying graph, the double cover  is {the same} game as the original one (with just two copies of the underlying graph instead of one). So, in this case, the double cover is complete and sound without any extra assumptions on self loops.
\end{remark}

\subsection{Non-synchronous strategies, values and entanglement lower bounds}\label{sec:non-synch_setup}
When defining quantum strategies (Definition \ref{defn:quantum_strategy}), we marked that our definition is commonly called in the literature ``synchronous''; namely, our definition is some specialization of the more general notion of a quantum strategy, which is the topic of this subsection. This notion of ``synchronicity'' encapsulates three properties of the given strategy:  The strategy is ``projective'', i.e.,  associates a projective measurement (PVM, Definition \ref{defn:PVM}), and not the more general notion of a measurement (POVM), to every vertex in the game. The strategy is ``maximally entangled'', i.e., the  state of the bipartite system on which the measurements are defined is the maximally entangled one. The strategy is ``symmetric'', i.e., the measurements the strategy associates with each vertex are   the same on both sides of the bipartite system. Let us make this discussion formal.

\begin{definition}[Measuring with respect to a general state. Compare to Definition \ref{defn:PVM}]\label{defn:generalized_measurement}
	Let $\cal{P}$ be an $n$-dimensional POVM with outcomes in $A$, and $\psi\in \complex^n$ a unit vector. Recall also, from Remark \ref{rem:procedural_joint_sampling}, that $\langle u|v\rangle=u^*\cdot v=\sum_{i=1}^n \overline{u_i}\cdot  v_i$ is the standard inner product on $\complex^n$, where $\overline{(\cdot)}$ is the complex conjugate. Then, the probability distribution induced by $(\psi,\cal{P})$ is 
	\[
	\Pro{}[a\ \textrm{is sampled}]:=\psi^*\cal{P}_a \psi=\langle \psi|\cal{P}_a \psi\rangle\ .
	\]
	Sampling $a\in A$ as above is often called ``{measuring} according to $(\psi,\cal{P})$'', and is denoted by $a\sim (\psi,\cal{P})$. 
	
	In a similar manner to Definition \ref{defn:joint_measurement}, given two $n$-dimensional POVMs, $\cal{P}$ with outcomes in $A$ and $\cal{Q}$ with outcomes in $B$, the tensor product $\cal{P}\otimes \cal{Q}^T$ is a POVM with outcomes in $A\times B$, where $(\cdot)^T$ is the transposition of matrices. Given a unit vector $\psi\in \complex^n\otimes\complex^n$, we get the probability distribution 
	\begin{equation}\label{eq:general_joint_measurement_formula}
	\Pro{}[a,b\ \textrm{are sampled}]:=\psi^*(\cal{P}_a\otimes\cal{Q}^T_b) \psi=\langle \psi|\cal{P}_a\otimes\cal{Q}^T_b \psi\rangle\ ,
	\end{equation}
	and again, we call this jointly sampling  mechanism ``measuring according to $(\psi,\cal{P},\cal{Q})$'', and denote it by $(a,b)\sim( \psi,\cal{P},\cal{Q})$.
\end{definition}
\begin{claim}\label{claim:general_joint_sampling_max_entangled_is_the_same_as_joint_sampling}
	Let $\cal{P},\cal{Q}$ be $n$-dimensional POVMs as in Definition \ref{defn:generalized_measurement}. Assume $\psi$ is the maximally entangled state, namely, that $\psi=\frac{1}{\sqrt n}\sum_{i=1}^n e_i\otimes e_i$, where $\{e_i\}$ is the standard basis of $\complex^n$.\footnote{Note that the maximally entangled state is equal to $\frac{1}{\sqrt n}\sum_{i=1}^n u_i^*\otimes u_i$ for \textbf{any} orthonormal basis $\{u_i\}$ of $\complex^n$, and not only with respect to the standard basis --- this is a useful fact which is often used in the analysis of measurements.} Then, jointly  measuring according to $(\psi,\cal{P},\cal{Q})$ as defined in \eqref{eq:general_joint_measurement_formula} is {the same} as jointly measuring according to $(\cal{P},\cal{Q})$ as in \eqref{eq:joint_sampling_according_PVM}.
\end{claim}
\begin{proof}
	This is immediate, because for the maximally entangled state $\psi$, $ \psi^*  P\otimes Q^T \psi $ is equal to $\tau(PQ)$ for any two $n\times n$ matrices $P,Q$.
\end{proof}

\begin{definition}[General quantum strategies. Compare to Definition \ref{defn:quantum_strategy}]\label{defn:general_quantum_strategies}
	Let $\game$ be a (tailored) non local game with underlying graph $G=(V,E)$ and length function $\ell\colon V\to \mathbb{N}$.  A (generalized) $n$-dimensional quantum strategy $\strategy$ consists of a unit vector $\psi\in \complex^n\otimes \complex^n$, together with {two} mappings $\cal{P},\cal{Q}$, that given a vertex $\mttx\in V$, associate to it POVMs $\cal{P}^\mttx,\cal{Q}^\mttx$ acting on $\complex^n$ and with outcomes in $\FF_2^{\ell(\mttx)}$. As in Remark \ref{rem:induced_correlations}, such a strategy induces a correlation 
	\begin{equation}\label{eq:correlation_induced_by_general_strategy}
	p_{\strategy}(a,b|\mttx,\mtty)= \psi^* \cal{P}^\mttx_a\otimes (\cal{Q}^\mtty_b)^T \psi \ .    
	\end{equation}
	A generalized strategy $\strategy=(\psi,\cal{P},\cal{Q})$ is called:
	\emph{projective} if $\cal{P}^\mttx,\cal{Q}^\mttx$ are PVMs for every vertex $\mttx$;
	\emph{symmetric} if $\cal{P}^\mttx=(\cal{Q}^\mttx)^T$ for every vertex $\mttx$;
	\emph{maximally entangled} if $\psi$ is the maximally entangled state $\frac{1}{\sqrt n}\sum_{i=1}^n e_i\otimes e_i$. A projective, symmetric, maximally entangled strategy is called \emph{synchronous}.
\end{definition}

The way we defined a game beforehand (Definition \ref{defn:non-local_game}), there was a single generating set at each vertex, and thus when the edge sampled in the game was a loop $\mttx\mttx$, the answer $\gamma$ was a bit string parametrized by $S_\mttx$ and not $S_\mttx\sqcup S_\mttx$. Once one allows general strategies, it is not clear how to decide about an answer $\gamma$ in such a case, as $\psi^* \cal{P}^\mttx_a\otimes (\cal{Q}^\mttx_b)^T \psi$ may be positive for $a\neq b$ --- namely, there are two answers $a,b\colon S_\mttx\to \FF_2$, which one should be $\gamma$? 
Though our discussion on double covers (Definition \ref{defn:double_cover}) was motivated by other constructions along this paper, the resolution to the aforementioned issue is in it. 
Usually in the literature, the (synchronous) game $\game$ \emph{is} its double cover, namely at each vertex there are two distinct sets of formal variables $S^+_\mttx$ and $S^-_\mttx$ of size $\ell(\mttx)$, and given that the sampled edge was $\mttx\mtty$  the assignment $\gamma$ is from $S^+_\mttx\sqcup S^-_\mtty$ to  $\FF_2$ and not from $S_\mttx\sqcup S_\mtty$, and it is sampled to be $ab$ with probability \eqref{eq:correlation_induced_by_general_strategy}.
Let us define this more general notion of a game.
\begin{definition}[General game. Compare to Definition \ref{defn:non-local_game}]\label{defn:general_non-local_game}
	A general game  $\game$ consists of an underlying graph $G=(V,E)$, a length function $\ell\colon V\to \mathbb{N}$ (or two length $\ell^\frR,\ell^\frL$ functions in the case of a tailored game), \textbf{two distinct} formal generators sets $S^+_\mttx$ and $S^-_\mttx$ at each vertex $\mttx\in V$, a distribution $\mu$ over $E$, and for every $e=\mttx\mtty\in E$ a decision predicate $D_{\mttx\mtty}\colon \FF_2^{S_\mttx^+}\times \FF_2^{S_{\mtty}^-}\to \FF_2$. Such a game is called \emph{synchronous} if $D_{\mttx\mttx}(ab)=0$ whenever $a\neq b$.
	
\end{definition}
\begin{remark}\label{rem:genera_synch_as_double_cover}
	Indeed, a general synchronous game as in Definition  \ref{defn:general_non-local_game} is exactly the double cover (Definition \ref{defn:double_cover}) of a game $\game$ as in Definition \ref{defn:non-local_game}. So, when discussing general strategies, there is no reason to distinguish between the double cover and the game itself.
\end{remark}

\begin{definition}[Non-synchronous value of a general game, and non-synchronous entanglement. Compare to Definitions \ref{defn:value_of_a_game} and \ref{def:ent}]\label{defn:non-synch_value}
	The value of a general strategy $\strategy=(\psi,\cal{P},\cal{Q})$ in a general game $\game$ (Definition \ref{defn:general_non-local_game}) is the same as it was in Definition \ref{defn:value_of_a_game}; the only difference is due to the way the correlation is induced by the strategy, namely using \eqref{eq:correlation_induced_by_general_strategy} instead of \eqref{eq:correlation_induced_by_synch_quantum_strat}. Namely,
	\begin{equation}\label{eq:value_of_general_strat_versus_game}
	\val(\game,\strategy)=\sum_{\mathtt{xy}\in E}\sum_{\substack{{a\colon{S^+_\mathtt{x}}\to \FF_2}\\ b\colon {S^-_\mathtt{y}}\to \FF_2}} \mu(\mathtt{xy})D_{\mathtt{xy}}(ab)\cdot \psi^* \cal{P}^\mathtt{x}_a\otimes(\cal{Q}^\mathtt{y}_b)^T \psi\ .
	\end{equation}
	So, taking the supremum of the value of a general game over \textbf{general} strategies gives a new notion of a value which we call the \emph{non-synchronous} value of $\game$, and denote it by $\valns(\game)$. In addition, let $\Entns(\game,1-\eps)$ be the smallest $n$ such that there is a an $n$-dimensional \textbf{general} strategy with value of at least $1-\eps$. This quantity is the non-synchronous entanglement lower bound of $\game$ (with parameter $1-\eps$).
\end{definition}

\begin{remark}\label{rem:synch_strategy_as_a_general_one}
	Given an $n$-dimensional quantum strategy $\strategy=\{\cal{P}\}$ as in Definition \ref{defn:quantum_strategy}, one can define a general synchronous $n$-dimensional strategy $\strategy'=(\frac{1}{\sqrt n}\sum_{i=1}^n e_i\otimes e_i,\cal{P},\cal{P}^T)$. This mapping `embeds' our notion of a quantum strategy as a special case of general quantum strategies. By Claim \ref{claim:general_joint_sampling_max_entangled_is_the_same_as_joint_sampling}, this mapping  preserves the correlations induced by the appropriate strategies, and thus the  value of $\strategy'$ versus $\game$ is the same as that of $\strategy$.
\end{remark}

\begin{fact}[Translating non-synchronous bounds to synchronous bounds]\label{fact:non-synch_high_implies_synch_high}
	Let $\game$ be a general synchronous (tailored) game, such that in its underlying graph $G=(V,E)$, all loops $\mttx\mttx$ for $\mttx\in V$ appear as edges in $E$. Assume in addition that there is some constant $c>0$, such that for every $\mttx\in V$ we have
	\begin{equation}\label{eq:loops_have_constant_mass_general_setup}
	\frac{\mu(\mttx\mttx)}{\mu(\mttx)}\geq c\ ,        
	\end{equation}
	where $\mu$ is the distribution over edges in $\game$, and $\mu(\mttx)$ is (as before) the marginal on vertices, namely $\mu(\mttx)=\sum_{\mtty\in V}\frac{\mu(\mttx\mtty)+\mu(\mtty\mttx)}{2}$. Then, from every $n$-dimensional general strategy $\strategy$ for $\game$ with value $1-\eps$, one can extract an $n$-dimensional synchronous strategy (i.e., projective, symmetric and maximally entangled) for $\game$ with value of at least $1-\poly(\nicefrac{\eps}{c^2})$. This in particular says that $\Entns(\game,1-\eps)\geq \Ent(\game,1-\poly(\nicefrac{\eps}{c^2}))$.
\end{fact}

\begin{proof}[Proof idea]
	By  analysing the given strategy $\strategy=(\psi,\cal{P},\cal{Q})$ with value $1-\eps$ in a similar fashion to the strategy that had high probability of winning in  the double cover (cf.\ \eqref{eq:pob_S_losses_given_loop} and \eqref{eq:eps_x_lower_bound}), we can deduce that $\strategy$ is $(\nicefrac{\eps}{c^2})$-self inconsistent --- which is the generalized quantity of inconsistency between $\cal{P}^\mttx$ and $(\cal{Q}^\mttx)^T$ (on average over all $\mttx\in V$, see \cite[Equation (4)]{vidick2022almost}). Once the strategy has low self inconsistency, it is close to being projective as well as symmetric. Hence, using orthonormalization (Fact \ref{fact:orthogonalization}) and naive symmetrization (similar to the choice of $\cal{U}'$ in the soundness of the double cover \eqref{eq:symmetrized_strategy}), we can perturb it to being projective and symmetric without enlarging the dimension. The fact that close by strategies provide close by value (even in the general setup, see \cite[Lemma 2.10]{vidick2022almost}), means that the value degrades only by some polynomial in $\nicefrac{\eps}{c^2}$, as required. Once this is done, we are only left to make it maximally entangled. It turns out that this cannot be done naively --- maybe it is genuinely far from a maximally entangled strategy. But, there is a convex combination of projective, symmetric maximally entangled strategies of dimension at most the dimension of $\strategy$  that is close (in terms of correlations produced) to it (see \cite[Corollary 3.3]{vidick2022almost}). In particular, as the value of the game is linear, one of these strategies provides a value that is at most  polynomial in $\nicefrac{\eps}{c^2}$ lower that that of $\strategy$, finishing the proof (see the paragraph immediately after \cite[Corollary 3.3]{vidick2022almost}).
\end{proof}

\subsection{The Pauli group}\label{sec:Pauli_gp}

The Pauli matrices $\PXm$, $\PZm$ (Definition \ref{defn:Pauli_matrices_acting_on_C^2}) are ubiquitous in quantum information theory; together with $\PYm=i\PXm\PZm$ and $\Id$ they form a linear basis of all observables that can be performed on a qubit, and $\PXm$, $\PZm$ are generally interpreted as the observables associated with two fundamental incompatible degrees of freedom such as the angular momentum, along two orthogonal directions, of an electron, or the position and momentum of a particle (in the infinite-dimensional case). 

It turns out that these matrices are characterized, among all $2$-dimensional complex observables and up to a global unitary rotation, by the anti-commutation relation $\PXm\PZm=-\PZm \PXm$. In this section we take a (classic) group-theoretic perspective and introduce the generalized Pauli group acting on $k$ qubits. This perspective will be used in the next section, where we introduce a nonlocal game that essentially forces any good strategy to make use of these matrices as observables --- namely, it is  a robust self-test (Definition \ref{defn:robustness}) with the single optimal strategy being induced by the Pauli matrices. 

The resulting non-local game, which we call the \emph{generalized Pauli basis game} and is introduced in the next section (following~\cite{NatarajanVidick,NW19,MIPRE,de_la_Salle_spectral_gap,CVY_efficient}), will later enable us to modify the naive introspection game  $\Introspect(\game)$ (Section \ref{sec:the_introspection_game}) so as to force the pair of questions sampled by the strategy to conform to the question distribution $\mu$ of the game $\game$. 
\\



Recall that $\mathbb{F}_{2}=\{0,1\}$ is the field with two elements, 
$\mathbb{F}_{2}^{k}$ is the $k$-dimensional vector space over $\mathbb{F}_{2}$, and $\langle \cdot,\cdot\rangle\colon \FF_2^k\times \FF_2^k \to \FF_2$ is the bilinear form
\begin{equation}\label{eq:inner_product_F^k}
\forall v,w\in\FF_2^k \ \colon\ \ \langle v,w\rangle=\sum_{i=1}^k v_iw_i.
\end{equation}
This  bilinear form induces an isomorphism between $\FF_2^k$ and its dual space, $(\FF_2^k)^*=\{f\colon \FF_2^k\to \FF_2\mid f\ \textrm{is\ linear}\},$  by defining 
$v\mapsto v^*=\langle v,\cdot\rangle$. 
Under this isomorphism, the standard basis $\{e_1,...,e_k\}$ is dual to itself, namely 
\[
\forall i,j\in[k]\ \colon\ \ 
e_i^*(e_j)=\begin{cases}
1 & i=j,\\
0 & i\neq j.
\end{cases}
\]
All of these choices allow us to think of $\FF_2^k$ as column vectors, $(\FF_2^k)^*$ as row vectors, the $*$ operation as transposition of matrices, and the bilinear form $\langle\cdot ,\cdot \rangle$ as matrix multiplication between row and column vectors. 
\begin{definition}\label{defn:Pauli_group_acting_on_k}
	The \emph{Pauli group acting on $k$ qubits} (also known as the Weyl--Heisenberg group, or the $k$-dimensional Heisenberg group over $\FF_2$) is the collection of triples $\WH_k=\{(v,w,a)\mid v,w\in \FF_2^k, a\in \FF_2\}$ with multiplication
	\[
	\forall v,v',w,w'\in \FF_2^k,\ a,a'\in \FF_2\ \colon\ \ (v,w,a)\cdot (v',w',a')=(v+v',w+w',a+a'+\langle w,v'\rangle).
	\]
\end{definition}
\begin{remark}
	Note that $\{(v,0,0)\}$  and  $\{(0,w,0)\}$ are subgroups of $\WH_k$ isomorphic to $\FF_2^k$. We usually call them the $X$-subgroup and $Z$-subgroup for reasons that will soon be clear.
\end{remark}
There is a faithful $\FF_2$-representation of $\WH_k$ as $(k+2)\times (k+2)$ matrices by mapping 
\[
(v,w,a)\mapsto \left(\begin{array}{ccc}
1 & w^* & a\\
\vec{0}_k &  \Id_k & v\\
0 &  (\vec{0}_k)^* & 1
\end{array}\right)\in GL_{k+2}(\FF_2),
\]
where $\vec{0}_k$ is the length $k$ all zero column vector, and $\Id_k$ is the $k\times k$ identity matrix. 
In this guise, the group is commonly called the $k$-dimensional Heisenberg group over $\FF_2$.

\subsubsection{Complex representations of the Pauli group}\label{sec:cpx_rep_Pauli_gp}
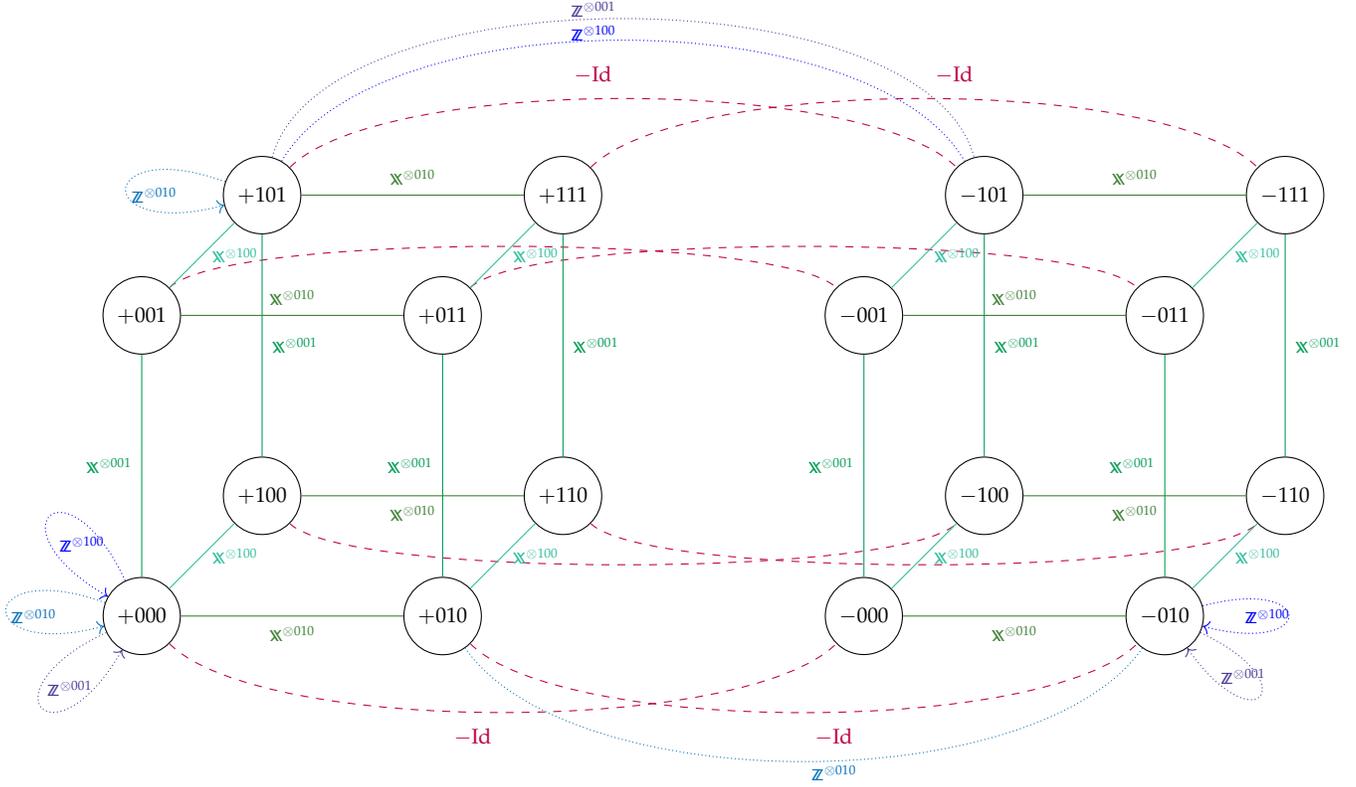
\begin{figure}[h]
	\begin{adjustwidth*}{0em}{0em}
		\begin{tikzpicture}[scale=0.8]
		
		\node[draw, color=black, shape=circle] (x000) at (0,0) {\scriptsize $+000$}; 
		\node[draw, color=black, shape=circle] (x001) at (0,5) {\scriptsize $+001$};
		\node[draw, color=black, shape=circle] (x010) at (5,0) {\scriptsize $+010$};
		\node[draw, color=black, shape=circle] (x100) at (2,2) {\scriptsize $+100$};
		\node[draw, color=black, shape=circle] (x111) at (7,7) {\scriptsize $+111$}; 
		\node[draw, color=black, shape=circle] (x011) at (5,5) {\scriptsize $+011$};
		\node[draw, color=black, shape=circle] (x110) at (7,2) {\scriptsize $+110$};
		\node[draw, color=black, shape=circle] (x101) at (2,7) {\scriptsize $+101$};
		
		\node[draw, color=black, shape=circle] (y000) at (12,0) {\scriptsize $-000$}; 
		\node[draw, color=black, shape=circle] (y001) at (12,5) {\scriptsize $-001$};
		\node[draw, color=black, shape=circle] (y010) at (17,0) {\scriptsize $-010$};
		\node[draw, color=black, shape=circle] (y100) at (14,2) {\scriptsize $-100$};
		\node[draw, color=black, shape=circle] (y111) at (19,7) {\scriptsize $-111$}; 
		\node[draw, color=black, shape=circle] (y011) at (17,5) {\scriptsize $-011$};
		\node[draw, color=black, shape=circle] (y110) at (19,2) {\scriptsize $-110$};
		\node[draw, color=black, shape=circle] (y101) at (14,7) {\scriptsize $-101$};
		
		\draw[purple,-, dashed] (x000) to  [out=315,in=225,looseness=0.5] (y000) ; \node[purple] at (5.5,-2) {\scriptsize $-\Id$};
		\draw[purple,-, dashed] (x010) to  [out=315,in=225,looseness=0.5] (y010) ; \node[purple] at (11.5,-2) {\scriptsize $-\Id$};
		\draw[purple,-, dashed] (x100) to  [out=315,in=225,looseness=0.3] (y100) ;
		\draw[purple,-, dashed] (x110) to  [out=315,in=225,looseness=0.3] (y110) ;
		\draw[purple,-, dashed] (x101) to  [out=45,in=135,looseness=0.5] (y101) ; \node[purple] at (7.5,9) {\scriptsize $-\Id$};
		\draw[purple,-, dashed] (x111) to  [out=45,in=135,looseness=0.5] (y111) ; \node[purple] at (13.5,9) {\scriptsize $-\Id$};
		\draw[purple,-, dashed] (x001) to  [out=45,in=135,looseness=0.3] (y001) ;
		\draw[purple,-, dashed] (x011) to  [out=45,in=135,looseness=0.3] (y011) ;
		
		\draw[ForestGreen, -, solid] (x000)--(x001) node[midway,left]{\tiny $\mathds{X}^{\otimes 001}$};
		\draw[ForestGreen, -, solid] (x100)--(x101) node[midway,right]{\tiny $\mathds{X}^{\otimes 001}$};
		\draw[ForestGreen, -, solid] (x010)--(x011) node[midway,left]{\tiny $\mathds{X}^{\otimes 001}$};
		\draw[ForestGreen, -, solid] (x110)--(x111) node[midway,right]{\tiny $\mathds{X}^{\otimes 001}$};
		
		\draw[ForestGreen, -, solid] (y000)--(y001) node[midway,left]{\tiny $\mathds{X}^{\otimes 001}$};
		\draw[ForestGreen, -, solid] (y100)--(y101) node[midway,right]{\tiny $\mathds{X}^{\otimes 001}$};
		\draw[ForestGreen, -, solid] (y010)--(y011) node[midway,left]{\tiny $\mathds{X}^{\otimes 001}$};
		\draw[ForestGreen, -, solid] (y110)--(y111) node[midway,right]{\tiny $\mathds{X}^{\otimes 001}$};
		
		\draw[OliveGreen, -, solid] (x000)--(x010) node[midway,below]{\tiny $\mathds{X}^{\otimes 010}$};
		\draw[OliveGreen, -, solid] (x100)--(x110) node[midway,below]{\tiny $\mathds{X}^{\otimes 010}$};
		\draw[OliveGreen, -, solid] (x001)--(x011) node[midway,above]{\tiny $\mathds{X}^{\otimes 010}$};
		\draw[OliveGreen, -, solid] (x101)--(x111) node[midway,above]{\tiny $\mathds{X}^{\otimes 010}$};
		
		\draw[OliveGreen, -, solid] (y000)--(y010) node[midway,below]{\tiny $\mathds{X}^{\otimes 010}$};
		\draw[OliveGreen, -, solid] (y100)--(y110) node[midway,below]{\tiny $\mathds{X}^{\otimes 010}$};
		\draw[OliveGreen, -, solid] (y001)--(y011) node[midway,above]{\tiny $\mathds{X}^{\otimes 010}$};
		\draw[OliveGreen, -, solid] (y101)--(y111) node[midway,above]{\tiny $\mathds{X}^{\otimes 010}$};
		
		\draw[SeaGreen, -, solid] (x000)--(x100) node[midway,right]{\tiny $\mathds{X}^{\otimes 100}$};
		\draw[SeaGreen, -, solid] (x010)--(x110) node[midway,right]{\tiny $\mathds{X}^{\otimes 100}$};
		\draw[SeaGreen, -, solid] (x001)--(x101) node[midway,right]{\tiny $\mathds{X}^{\otimes 100}$};
		\draw[SeaGreen, -, solid] (x011)--(x111) node[midway,right]{\tiny $\mathds{X}^{\otimes 100}$};
		
		\draw[SeaGreen, -, solid] (y000)--(y100) node[midway,right]{\tiny $\mathds{X}^{\otimes 100}$};
		\draw[SeaGreen, -, solid] (y010)--(y110) node[midway,right]{\tiny $\mathds{X}^{\otimes 100}$};
		\draw[SeaGreen, -, solid] (y001)--(y101) node[midway,right]{\tiny $\mathds{X}^{\otimes 100}$};
		\draw[SeaGreen, -, solid] (y011)--(y111) node[midway,right]{\tiny $\mathds{X}^{\otimes 100}$};
		
		\draw[blue,->, densely dotted] (x000) to  [out=115,in=150,looseness=15] (x000) ; 
		\node[blue] at (-1,1.2) {\tiny $\mathds{Z}^{\otimes 100}$};
		
		\draw[NavyBlue,->, densely dotted] (x000) to  [out=160,in=195,looseness=15] (x000) ; 
		\node[NavyBlue] at (-1.8,0) {\tiny $\mathds{Z}^{\otimes 010}$};
		\draw[BlueViolet,->, densely dotted] (x000) to  [out=205,in=240,looseness=15] (x000) ; 
		\node[BlueViolet] at (-1.2,-1.2) {\tiny $\mathds{Z}^{\otimes 001}$};
		
		\draw[blue,-, densely dotted] (x101) to  [out=60,in=120,looseness=0.7] (y101) ; 
		\node[blue] at (7.5,9.7) {\tiny $\mathds{Z}^{\otimes 100}$};
		
		\draw[NavyBlue,->, densely dotted] (x101) to  [out=160,in=195,looseness=15] (x101) ; 
		\node[NavyBlue] at (0.2,7) {\tiny $\mathds{Z}^{\otimes 010}$};
		\draw[BlueViolet,-, densely dotted] (x101) to  [out=75,in=105,looseness=0.7] (y101) ; 
		\node[BlueViolet] at (7.5,10.1) {\tiny $\mathds{Z}^{\otimes 001}$};

		\draw[blue,->, densely dotted] (y010) to  [out=15,in=-15,looseness=15] (y010) ; 
		\node[blue] at (18.7,0) {\tiny $\mathds{Z}^{\otimes 100}$};
		
		\draw[NavyBlue,-, densely dotted] (y010) to  [out=235,in=305,looseness=0.7] (x010) ; 
		\node[NavyBlue] at (11.5,-2.6) {\tiny $\mathds{Z}^{\otimes 010}$};
		\draw[BlueViolet,->, densely dotted] (y010) to  [out=-25,in=-55,looseness=15] (y010) ; 
		\node[BlueViolet] at (18.3,-1) {\tiny $\mathds{Z}^{\otimes 001}$};
		
		\end{tikzpicture}
	\end{adjustwidth*}
	\caption{This is an example of the actions of the Pauli matrices on $Y_\pm$, which consists of all signed bit strings of length $3$ in this case.  The {\color{purple} purple} dashed lines are the actions of {\color{purple}$-\Id$}. The shades of {\color{green}green} solid lines are the actions of the {\color{green}$\mathds{X}$}-generators for the standard basis elements. The shades of {\color{blue} blue} dotted lines are the actions of the {\color{blue}$\mathds{Z}$}-generators. We intentionally did not include all  actions of  $\mathds{Z}$-generators, for visual clarity.}
	\label{tikz:Schreier_graph_WHk_mod_X}
	
\end{figure}

The map $(v,w,a)\mapsto (v,w)$ is an epimorphism of $\WH_k$ onto $\FF_2^{2k}$. Hence, all complex irreducible representations of $\FF_2^{2k}$ are also irreducible representations of $\WH_k$. There are $2^{2k}$ such $1$-dimensional representations. It turns out $\WH_k$ has only one extra irreducible representation of dimension $2^k$, which we will describe shortly. 
Let $U(\cal{H})$ be the group of  unitary operators acting on a Hilbert space $\cal{H}$. 
\begin{definition}\label{defn:Pauli_matrices_acting_on_C^2}
	The  $X$ and $Z$ \emph{Pauli matrices} are  the following \textbf{signed permutation matrices} \[ \PXm=\left(\begin{array}{cc}
	0 & 1\\
	1 & 0
	\end{array}\right)\ ,\ \PZm=\left(\begin{array}{cc}
	1 & 0\\
	0 & -1
	\end{array}\right) \in U(\complex^2)\ .
	\] 
\end{definition}
By viewing $\complex^2$ as $\complex^{\FF_2}$, and letting ${\bf 1}_a$ be the indicator of $a\in \FF_2$, we can see that $\PXm{\bf 1}_a={\bf 1}_{a+1}$ and that $\PZm{\bf 1}_a=(-1)^a{\bf 1}_a$.
For every $v=(v_1,...,v_k)$ and $w=(w_1,...,w_k)$ in ${\FF}_{2}^{k}$,
let
\[ 
\PXm^{\otimes v}=\bigotimes_{i=1}^{k}\PXm^{v_{i}}\; ,\quad \PZm^{\otimes w}=\bigotimes_{i=1}^{k}\PZm^{w_{i}}\in U(\complex^{2^{k}})\ ,
\]
where $\PXm^{0}=\PZm^0={\rm Id}=\left(\begin{array}{cc}
1 & 0\\
0 & 1
\end{array}\right)$ and $\otimes$ is (again) the Kronecker tensor product of matrices.  These matrices act naturally on $(\complex^{\FF_2})^{\otimes k}\cong \complex^{\FF_2^k}$ as follows. Let ${\bf 1}_v \in \complex^{\FF_2^k}$ be the indicator function of $v\in \FF_2^k$. Then
\begin{equation}\label{eq:action_of_Pauli_matrices}
\forall v,w\in \FF_2^k\ \colon\ \ \PXm^{\otimes v}{\bf 1}_w={\bf 1}_{w+v}\;,\quad \PZm^{\otimes v}{\bf 1}_w=(-1)^{\langle v,w\rangle}{\bf 1}_w\ .    
\end{equation}
As the tensor product of signed permutation matrices is  a signed permutation matrix,  the matrices $\PXm^{\otimes v}$ and $\PZm^{\otimes v}$  (and their products) are signed permutation matrices; the signed set on which they naturally act is  the signed standard basis $$Y_{\pm} =\{\pm {\bf 1}_{v} \mid v\in \FF_2^k\}\ .$$ See Figure \ref{tikz:Schreier_graph_WHk_mod_X} for a visualization of the action of these matrices in case $k=3$. 

\begin{definition}\label{defn:F^Z_v}
	Let  $\mathscr{F}_v^{\PZm}$ be the (orthogonal) projection on the $1$-dimensional subspace in $\complex^{\FF_2^k}$ spanned by ${\bf 1}_v$,  and $\mathscr{F}_v^{\PXm}$ the (orthogonal) projection on the $1$-dimensional subspace in $\complex^{\FF_2^k}$ spanned by $\sum_{w\in \FF_2^k}(-1)^{\langle v,w\rangle}{\bf 1}_w$.
\end{definition} 
Then, $\{\mathscr{F}_v^{\PXm}\}$ (respectively $\{\mathscr{F}_v^{\PZm}\}$) is a PVM with outcomes in $\FF_2^k$, and its observable form is $i\mapsto \PXm^{\otimes e_i}$ (respectively $i\mapsto \PZm^{\otimes e_i}$) for $i\in [k]$. Moreover, if $z$ is sampled according to $\{\mathscr{F}_v^{\PZm}\}$ (or $\{\mathscr{F}_v^{\PXm}\}$), then it is a uniform bit string of length $k$.

\begin{definition}[The unique non-commuting unitary irreducible representation of the Pauli group]\label{defn:rho}
	The map $\rho \colon \WH_k\to\Sym_{\pm}(Y)\subseteq  U(2^k) $ defined by 
	\begin{equation}\label{eq:def_of_rho_for_P_k}
	\rho(v,w,a)=(-1)^a\PXm^{\otimes v}\PZm^{\otimes w}
	\end{equation}
	is a faithful irreducible signed permutation representation of $\WH_k$. In particular, 
	given $v,v',w,w'\in\FF_2^k$, we have
	\[
	(\PXm^{\otimes v})^2=\Id\ ,\quad (\PZm^{\otimes w})^2=\Id\ ,\quad \PXm^{\otimes v}\PXm^{\otimes v'}=\PXm^{\otimes v+v'}\ ,\quad \PZm^{\otimes w}\PZm^{\otimes w'}=\PZm^{\otimes w+w'}   
	\]
	and
	\[
	\PXm^{\otimes v}\PZm^{\otimes w}=(-1)^{\langle v,w\rangle} \PZm^{\otimes w}\PXm^{\otimes v}\ .
	\]
	Moreover,  $\{\PXm^{\otimes v}\mid v\in \FF_2^k\}$  and $\{\PZm^{\otimes w}\mid w\in\FF_2^k\}$ are isomorphic to the $X$ and $Z$ subgroups in $\WH_k$.
	For later use, we let $\rho^\PZm$ and $\rho^\PXm$ be the restrictions of $\rho$ to the $Z$ and $X$ subgroups, namely
	\begin{equation}\label{eq:defn_restriction_rho_to_X_and_Z_subgroups}
	\forall \alpha\in \FF_2^k\ \colon \ \ \rho^\PZm(\alpha)=\PZm^{\otimes\alpha}\quad\mathrm{and}\quad\rho^\PXm(\alpha)=\PXm^{\otimes \alpha}\ .
	\end{equation}
\end{definition} 
\begin{remark}[The $\mathscr{F}$-projections as inverse Fourier transform]\label{rem:inverse_Fourier_for_F_projections}
	As usual, the PVM $\{\mathscr{F}_v^{\PXm}\}$ (respectively $\{\mathscr{F}_v^{\PZm}\}$) is the Fourier transform (Definition \ref{defn:Fourier_transform_reps}) of the representation $\rho^\PXm$ of $\FF_2^k$ defined in \eqref{eq:defn_restriction_rho_to_X_and_Z_subgroups} by $v\mapsto \PXm^{\otimes v}$ (respectively $\rho^\PZm$ defined by  $v\mapsto \PZm^{\otimes v}$), namely
	\[
	\forall v\in \FF_2^k\ \colon \ \ \mathscr{F}^\PZm_v=\Es{w\in \FF_2^k}\left[(-1)^{\langle w,v\rangle}\PZm^{\otimes w}\right]\quad\textrm{and}\quad \PZm^{\otimes v}=\sum_{w\in \FF_2^k}(-1)^{\langle w,v\rangle}\mathscr{F}^\PZm_w\;,
	\]
	and similarly for $\PXm$ and $\mathscr{F}^\PXm_v$.
\end{remark}

\begin{remark}
	The $2^k$-dimensional representation $\rho$ is 
	the unique non-commuting irreducible representation of $\WH_k$ (up to isomorphism). This is because there are  $2^{2k}$ one-dimensional representations, and the squares of the dimensions of the irreducible representations of $\WH_k$ should sum up to its order, which is $2^{2k+1}$.
\end{remark}

\begin{corollary}
	\label{claim:Pauli_rep_implies_uniform}
	The representation $\rho^\PXm$ from \eqref{eq:defn_restriction_rho_to_X_and_Z_subgroups} is a signed permutation PVM (Definition \ref{defn:perm_PVM}). The representation $\rho^\PZm$ is a diagonal PVM (Definition \ref{defn:diagonal_PVM}).
	Finally, a vector $z\in \FF_2^k$ sampled according to any of these PVMs (as was defined in \eqref{eq:sampling_according_POVM}) is uniformly distributed.
\end{corollary}

\subsubsection{Error correcting codes and stability of the Pauli group} \label{subsec:error_correctoin_prelims}
Let $k\leq n$ be positive integers. Let $C$ be a $k$-dimensional linear error correcting code of length $n$, i.e.\ a linear subspace $C\subseteq \FF_2^n$ of dimension $k$. Let $E\in M_{n\times k}(\FF_2)$ be a matrix whose columns span $C$. We call such matrices \emph{encoding matrices}, since they induce an encoding of $\FF_2^k$ as vectors in $C$ via the mapping $$\forall v\in \FF_2^k\ \colon\ \ v\mapsto Ev\ .$$ 
The \emph{(Hamming) weight} of a vector $u=(u_1,...,u_n)\in \FF_2^n$ is the number of non-zero entries in it, namely
\[
\omega_H(u):=|\{1\leq i\leq n \mid u_i\neq 0\}|\  .
\]
We say that $C$ has \emph{distance} $d$ if 
$$\forall 0\neq c\in C\ \colon \ \ \omega_H(c)\geq d\ .$$ 
All in all, $C$ is called a (binary) \emph{linear $[n,k,d]$-code}.
Let $A\in M_{m\times n}(\FF_2)$ be a matrix whose (right) kernel is $C$, namely $$C=\{u\in \FF_2^n\mid Au=0\}\ .$$ 
Such matrices are called \emph{parity check matrices} of $C$.
Every ordered set $\mathscr{B}=\{w^1,...,w^n\}\subseteq \FF_2^k$ defines an encoding matrix $E$  by letting $w^i$ be the $i^{\rm th}$ row of $E$, namely $E_{ij}=w^i_j$. We refer to the image of $E$ in this case as the \emph{code induced by} $\mathscr{B}$.

For the purpose of this section we can use any binary linear code that has linear dimension and distance, and whose encoding matrix $E$ can be efficiently constructed. The existence of such codes is guaranteed by the following well-known fact.

\begin{fact}\label{fact:good_efficiently_calculable_codes}
	For any $R\in (0,1)$ there is a $\delta >0$ and a family of binary linear codes $(C_n)_{n \geq 1}$ of dimension $k=\lfloor R n \rfloor$, length $n$, and distance $d\geq \delta n$ such that furthermore an encoding matrix $E_n$ for $C_n$ can be computed in time polynomial in $n$. 
\end{fact}

\begin{proof}
	An example construction is given by the Justesen codes~\cite{justesen1972class}, which can be obtained from the concatenation of a Reed--Solomon code over $\F_q$ and a suitably chosen inner code. Better constructions are possible if one is interested in a specific range of $(R,\delta)$; for us it suffices that $\delta>0$ can be guaranteed for any $R<1$.
\end{proof}

\begin{fact}[Semi-stability of  $\WH_k$, cf.\ Corollary 2.6 in \cite{de_la_Salle_spectral_gap}]\label{fact:semi-stability_P_k}
	Let $\chi$ and $\zeta$ be two $N$-dimensional unitary representations of $\FF_2^k$. Let $\mathscr{B}=\{w^1,...,w^n\}\subseteq \FF_2^k$ be an ordered  set which induces an $[n,k,d]$-code.  Assume that $\chi$ and $\zeta$ satisfy the following ``almost (anti-)commutation relations''
	\begin{equation}\label{eq:anti-com_almost_satisfied}
	\Ex_{i,j\in [n]}\left[\Vert\chi(w^i)\zeta(w^j)-(-1)^{\langle w^i,w^j\rangle }\zeta(w^j)\chi(w^i)\Vert_{hs}^2 \right]\leq \eps\ ,
	\end{equation}
	where $\Vert \cdot \Vert_{hs}$ is the normalized Hilbert--Schmidt norm (Definition \ref{defn:normalized_p_norms}).
	Then, there exists an integer $m$ and a $C(\nicefrac{k}{d})^2\eps$-near bijection (Definition \ref{defn:near_bijections}) $\omega\colon \complex^N\to \complex^{\FF_2^k}\otimes\complex^{m}$ for which $\chi$ is  $C(\nicefrac{k}{d})^2\eps$-close (Definition \ref{defn:distance_inconsistency_POVMs}  and Claim \ref{claim:dist_in_projection_and_observable_are_the_same}) to $\omega^*(\rho^\PXm\otimes\Id_m)\omega$    and similarly $\zeta$ is ${C(\nicefrac{k}{d})^2\eps}$-close to  $\omega^*(\rho^\PZm\otimes \Id_m)\omega$, where  $C$ is a universal constant (independent of any other parameter). Namely,
	\[
	\begin{split}
	\Es{v\in \FF_2^k}[\Vert\chi(v)-\omega^*\cdot\PXm^{\otimes v}\otimes \Id_m\cdot\omega\Vert_{hs}^2]&\leq C\left(\nicefrac{k}{d}\right)^2\eps\;,\\
	\Es{v\in \FF_2^k
	}[\Vert\zeta(v)-\omega^*\cdot\PZm^{\otimes v}\otimes \Id_m\cdot\omega\Vert_{hs}^2]&\leq C\left(\nicefrac{k}{d}\right)^2\eps\;,\\
	\tau(\Id_{N}-\omega^*\omega)\ ,\ \tau(\Id_{2^k}\otimes \Id_m-\omega\omega^*)&\leq C\left(\nicefrac{k}{d}\right)^2\eps\;.
	\end{split}
	\]
\end{fact}

In words, any $\chi$ and $\zeta$ which almost satisfy the appropriate (anti-)commutation relations of $\WH_k$  are close to (a direct sum of $m$ copies of) the respective restrictions $\rho^\PXm,\rho^\PZm$ to the $X$ and $Z$ subgroups of the  unique non-commuting representation $\rho$ of $\WH_k$ from Definition \ref{defn:rho}.

\begin{remark}
	The proof of the Fact~\ref{fact:semi-stability_P_k} is due to de la Salle \cite{de_la_Salle_spectral_gap}.  It uses a combination of ideas. The first is a method of Natarajan--Vidick \cite{natarajan2018two} which translates anti-commutation to commutation. The second is  a spectral gap argument, standard in the analysis of groups with property $(T)$, that allows to translate almost invariance against a generating sets to almost invariance against the whole group --- this is sometimes called the $L^2$-Poincare inequality of spectral expanders (cf.\ Theorem 13.9 in \cite{Hoory_Linial_Wigderson}). Lastly, an ``on average'' version of a stability result of finite groups due to Gowers--Hatami \cite{GowersHatami} is used. Though this description may seem intimidating, all the ingredients are quite straightforward (see \cite{CVY_efficient} for more on this). 
\end{remark}

\subsection{The  generalized Pauli basis game}\label{sec:gen_pauli_basis}

\begin{figure}[!htbp]
	\centering
	\begin{gamespec}
		\setlength{\tabcolsep}{1em}
		
		Throughout this description, $n$ and $k$ are positive integers,  $i,j\in [n]$,  $\mathscr{B}=\{w^1,...,w^n\}\subseteq \FF_2^k$ and $a,b\in [3]$. The game $\PauliBasis_k(\mathscr{B})$ is an augmented sum (Definitions \ref{defn:sum_of_games} and \ref{defn:augmentation_of_a_game}) of $\frC^{i,j}$ --- each of which is either a commutation game (Section \ref{sec:com_game}) or a  null-commutation game --- and $\frM^{i,j}$ --- each of which is either an anti-commutation game   (Section \ref{sec:Acom_game}) or a null-anti-commutation game.
		
		\vspace{1em}
		\begin{tabularx}{\textwidth}{ l    X X X  }
			\toprule
			Sub-Structure & Question &   Variables (all unreadable)  \\
			\midrule
			Augmentation: & $\Pauli_\PXm$  &  $\{\sP\sX^\alpha\}_{\alpha=1}^k$ \\
			& $\Pauli_\PZm$ &   $\{\sP\sZ^\beta\}_{\beta=1}^k$  \\
			&$\mttX^i$ & $\sX^i$    \\
			&$\mttZ^j$    & $\sZ^j$\\
			(null-)Commutation game:  &$\mathtt{First}^{i,j}$ &
			$\mathsf{First}^{i,j}$
			\\
			& $  \mathtt{Second}^{i,j}$& $\mathsf{Second}^{i,j}$ \\
			&   $\mathtt{Both}^{i,j}$  &  $\{\mathsf{Both}_1^{i,j},\mathsf{Both}_2^{i,j}\}$\\
			(null-)Anti-commutation game:  &$\mathtt{var}^{i,j}_{ab}$ &
			$\mathsf{Var}^{i,j}_{ab}$,
			\\
			& $\mathtt{row}^{i,j}_{a}$ & $\{\mathsf{Row}^{i,j}_{a1},\mathsf{Row}^{i,j}_{a2},\mathsf{Row}^{i,j}_{a3}\}$\\
			& $\mathtt{col}^{i,j}_{b}$  &    $\{\mathsf{Col}^{i,j}_{1b},\mathsf{Col}^{i,j}_{2b},\mathsf{Col}^{i,j}_{3b}\}$\\
			\bottomrule
		\end{tabularx}
		
		\vspace{1em}
		
		\begin{enumerate}
			\setlength\itemsep{1pt}
			\item \textbf{Commutation}: The (null-)commutation game $\frC^{i,j}$ involves the questions $\mathtt{First}^{i,j}$, $\mathtt{Second}^{i,j}$ and $\mathtt{Both}^{i,j}$.  If $\langle w^i,w^j\rangle=0$, then it forces the observable of $\mathsf{First}^{i,j}$ to commute with the observable of $\mathsf{Second}^{i,j}$.
			\item \textbf{Anti-commutation} The (null)-anti-commutation game $\frM^{i,j}$ involves the questions $\mathtt{var}^{i,j}_{ab}$, $\mathtt{row}^{i,j}_{a}$  and $\mathtt{col}^{i,j}_{b}$.  If $\langle w^i,w^j\rangle=1$, then it forces the observable of $\mathsf{Var}^{i,j}_{11}$ to anti-commute with the observable of $\mathsf{Var}^{i,j}_{22}$.
			\item \textbf{Consistency of $\sX$}: The observable of $\sX^i$ is forced to be consistent with the observables  $\mathsf{First}^{i,j}$ and $\mathsf{Var}^{i,j}_{11}$ for all $j$.
			\item \textbf{Consistency of $\sZ$}: The observable of $\sZ^j$ is forced to be consistent with the observables $\mathsf{Second}^{i,j}$ and $\mathsf{Var}^{i,j}_{22}$ for all $i$.
			\item\textbf{Linear conditions on $\sX$}: The observable of  $\sX^i$ is forced to be consistent with the observable of  the product $\prod_{\alpha=1}^k(\sP\sX^\alpha)^{w^i_\alpha}$.
			\item\textbf{Linear conditions on $\sZ$}: The observable of  $\sZ^j$ is forced to be consistent with the observable of  the product $\prod_{\beta=1}^k(\sP\sZ^\beta)^{w^j_\beta}$.
		\end{enumerate}
	\end{gamespec}
	\caption{Questions and answers in the generalized Pauli basis game  $\PauliBasis_k(\mathscr{B})$. Since the game is tailored as an LCS, all answers are unreadable. We also list the conditions on a strategy's observables that the game enforces.}
	\label{fig:genpauli-summary}
\end{figure}

We can now describe the generalized Pauli basis game $\PauliBasis_k$. The version provided here is due to de la Salle \cite{de_la_Salle_spectral_gap}. A group theoretic perspective on the Pauli basis game (and the following generalization of it) appears in \cite{CVY_efficient}.

In $\PauliBasis_k$, there are two special questions, $\Pauli_ \PXm$  and $\Pauli_\PZm$. Their length will be $k$, and we expect that a perfect strategy restricted to $S_{\Pauli_\PXm}$ and $S_{\Pauli_\PZm}$ induces (up to isometry and direct sums)  the unique non-commuting irreducible representation $\rho$ of $\WH_k$ defined in \eqref{eq:def_of_rho_for_P_k} --- namely, it is a self test (Definition \ref{defn:robustness}). 
Note that every strategy $\strategy$, when restricted to $S_{\Pauli_\PXm}$ (or $S_{\Pauli_\PZm}$), is a representation of $\FF_2^k$. These restrictions will play the role of $\chi$ and $\zeta$ in the semi-stability result in Fact \ref{fact:semi-stability_P_k}. So, we need to find a way to force \eqref{eq:anti-com_almost_satisfied} to be satisfied with a small enough $\eps$, namely for the commutator of $\chi(w^i)$ and $\zeta(w^j)$ to be $\eps$-close, on average, to $(-1)^{\langle w^i,w^j\rangle}\Id$ --- that will ensure that the Pauli basis game is robust (Definition \ref{defn:robustness}). 

To that end, for every $i,j\in [n]$, there will be questions $\mttX^i$ and $\mttZ^j$ of length $1$, whose observables are (expected to be) corresponding to $\chi(w^i)$ and $\zeta(w^j)$ respectively. This is achieved by a consistency check of $\mttX^i$'s vs.\ $\Pauli_\PXm$ and $\mttZ^j$ vs.\ $\Pauli_\PZm$.
Then, we check that the observables at the vertices $\mttX^i$ and $\mttZ^j$ (anti-)commute, according to whether $\langle w^i,w^j\rangle=0$ or $1$. This is done using ``small'' games that force either commutation or anti-commutation between two observables.  
See Figure \ref{fig:PB_k_unerlying_graph} for a partial representation of the underlying graph of $\PauliBasis_k$.\footnote{Note that for every $\mttX^i,\mttZ^j$ there is both a commutation game and an anti-commutation game attached to them. As we see later, the irrelevant one will be ignored. This is a quirk of the way compression works: The running time of the question reduced verifier needs to be exponentially faster, but calculating $\langle w^i,w^j\rangle$ may take a long time. Thus, we delegate this check to the decision process --- i.e., linear constraints processor combined with the canonical decider --- (which may still run in the original running time), and let it a posteriori ignore irrelevant (anti-)commutation checks that are not part of the presentation of $\WH_k$.}

To implement this last step, we need a game that forces commutation, and a game that forces anti-commutation.

\subsubsection{Commutation game}\label{sec:com_game}
The commutation game $\frak{C}$ has three questions (vertices) in its underlying graph: $\mathtt{First},\mathtt{Second}$ and $\mathtt{Both}$. The vertex  $\mathtt{First}$ is of length $1$ and has the associated formal generator $\mathsf{First}$, the vertex  $\mathtt{Second}$ is of  length $1$ and has the associated formal generator $\mathsf{Second}$, and the vertex $\mathtt{Both}$ is of length $2$ and has associated formal generators $\mathsf{Both}_1,\mathsf{Both}_2$. The edges in the underlying graph  are  $\mathtt{First}-\mathtt{Both}$ and $\mathtt{Second}-\mathtt{Both}$. Then, $D_{\mathtt{First\ Both}}$ checks that $\gamma(\mathsf{First})=\gamma(\mathsf{Both}_1)$, and $D_{\mathtt{Second\ Both}}$ checks that $\gamma(\mathsf{Second})=\gamma(\mathsf{Both}_2)$. Note that this is a linear constraint system game, and thus can be tailored a la Example \ref{example:LCSs}, in particular without readable variables. The distribution over edges is uniform. As described formally in the next fact, perfect strategies for this game imply commutation of observables, and almost perfect strategies imply almost commutation of observables. 

\begin{figure}[httb!]
	\centering
	\begin{tikzpicture}
	\node[draw, color=black, shape=circle] (First) at (-3,0) {\scriptsize $\mathtt{First}$}; 
	\node[draw, color=black, shape=circle] (Both) at (0,0) {\scriptsize $\mathtt{Both}$}; 
	\node[draw, color=black, shape=circle] (Second) at (3,0) {\scriptsize $\mathtt{Second}$};

	\draw[black, -, solid] (First)--(Both);
	\draw[black, -, solid] (Both)--(Second);

	\end{tikzpicture}
	\caption{The underlying graph of the commutation game $\frC$.
	}
	\label{fig:commutation_Game}
\end{figure}

\begin{fact} [Completeness and soundness of the comutation game, cf.\ Lemma 3.5 in \cite{de_la_Salle_spectral_gap}]\label{fact:soundness_com}
	If $\strategy$ is a perfect strategy for the commutation game $\frak{C}$, and $\cal{U}$ is $\strategy$ in observable form, then $\cal{U}(\mathsf{First}),\cal{U}(\mathsf{Second})$ are commuting involutions, i.e.,
	\[
	\cal{U}(\mathsf{First})\cal{U}(\mathsf{Second})=\cal{U}(\mathsf{Second})\cal{U}(\mathsf{First})\ .
	\] 
	Moreover, if $\strategy$ has value $1-\eps$, then $\Vert \cal{U}(\mathsf{First})\cal{U}(\mathsf{Second})-\cal{U}(\mathsf{Second})\cal{U}(\mathsf{First})\Vert^2_{hs}\leq 64\eps.$ 
\end{fact}
\begin{remark}
	Fact \ref{fact:soundness_com} can be deduced almost immediately from Claim \ref{claim:perfect_3Lin_commuting_observables}.
\end{remark}
\begin{claim}[Extending commuting observables to perfect strategies]\label{claim:extending_to_perfect_com}
	Given two commuting involutions $O_1,O_2\in U(n)$, there is a perfect strategy $\strategy=\{\cal{U}\}$ for $\frak{C}$ that commutes along edges such that $\cal{U}(\mathsf{First})=O_1$ and $\cal{U}(\mathsf{Second})=O_2$.
\end{claim}
\begin{proof}
	We are  left to define the observables associated to the $\mathtt{Both}$ vertex. For the strategy to be perfect, they need to be consistent with the observables at the other vertices, so we are forced to let $\cal{U}(\mathsf{Both}_1)=O_1$ and $\cal{U}(\mathsf{Both}_2)=O_2$. This is a well defined strategy, as indeed the observables at $\mathtt{Both}$  are commuting (by assumption) which induces a PVM in observable form at this vertex.
\end{proof}
Let the \emph{null-commutation game} $\frak{C}_{null}$ be the game whose underlying graph, length functions and sets of formal variables are the same as in the commutation game, but it always accepts. This is also a linear constraint system game.

\subsubsection{Anti-commutation game}\label{sec:Acom_game}
We have already seen the anti-commutation game: The magic square game from Example \ref{example:magic-square}. Again, we note that this game is an LCS, and thus can be tailored such that all variables are unreadable. As the next fact shows, it has the property that in a perfect strategy the observables of $\mathsf{Var}_{11}$ and $\mathsf{Var}_{22}$ anti-commute, and in an almost perfect strategy they almost anti-commute.

\begin{figure}[httb!]
	\centering
	\begin{tikzpicture}[scale=1.5]
	\node[draw, color=black, shape=circle] (row1) at (-3,3) {\scriptsize $\mathtt{row}_1$}; 
	\node[draw, color=black, shape=circle] (row2) at (-3,0) {\scriptsize $\mathtt{row}_2$}; 
	\node[draw, color=black, shape=circle] (row3) at (-3,-3) {\scriptsize $\mathtt{row}_3$}; 
	\node[draw, color=black, shape=circle] (Col1) at (3,3) {\scriptsize $\mathtt{col}_1$}; 
	\node[draw, color=black, shape=circle] (Col2) at (3,0) {\scriptsize $\mathtt{col}_2$}; 
	\node[draw, color=black, shape=circle] (Col3) at (3,-3) {\scriptsize $\mathtt{col}_3$};

	\node[draw, color=black, shape=circle] (V11) at (0,4) {\scriptsize $\mathtt{var}_{11}$}; 
	\node[draw, color=black, shape=circle] (V12) at (0,3) {\scriptsize $\mathtt{var}_{12}$}; 
	\node[draw, color=black, shape=circle] (V13) at (0,2) {\scriptsize $\mathtt{var}_{13}$}; 
	\node[draw, color=black, shape=circle] (V21) at (0,1) {\scriptsize $\mathtt{var}_{21}$}; 
	\node[draw, color=black, shape=circle] (V22) at (0,0) {\scriptsize $\mathtt{var}_{22}$}; 
	\node[draw, color=black, shape=circle] (V23) at (0,-1) {\scriptsize $\mathtt{var}_{23}$}; 
	\node[draw, color=black, shape=circle] (V31) at (0,-2) {\scriptsize $\mathtt{var}_{31}$}; 
	\node[draw, color=black, shape=circle] (V32) at (0,-3) {\scriptsize $\mathtt{var}_{32}$}; 
	\node[draw, color=black, shape=circle] (V33) at (0,-4) {\scriptsize $\mathtt{var}_{33}$};

	\draw[olive, -, solid] (row1)--(V11);
	\draw[olive, -, solid] (row1)--(V12);
	\draw[olive, -, solid] (row1)--(V13);
	\draw[olive, -, solid] (row2)--(V21);
	\draw[olive, -, solid] (row2)--(V22);
	\draw[olive, -, solid] (row2)--(V23);
	\draw[olive, -, solid] (row3)--(V31);
	\draw[olive, -, solid] (row3)--(V32);
	\draw[olive, -, solid] (row3)--(V33);
	
	\draw[cyan, -, solid] (Col1)--(V11);
	\draw[cyan, -, solid] (Col1)--(V21);
	\draw[cyan, -, solid] (Col1)--(V31);
	\draw[cyan, -, solid] (Col2)--(V12);
	\draw[cyan, -, solid] (Col2)--(V22);
	\draw[cyan, -, solid] (Col2)--(V32);
	\draw[cyan, -, solid] (Col3)--(V13);
	\draw[cyan, -, solid] (Col3)--(V23);
	\draw[cyan, -, solid] (Col3)--(V33);

	\end{tikzpicture}
	\caption{The underlying graph of the anti-commutation game $\frM$.
	}
	\label{fig:Acommutation_Game}
\end{figure}

\begin{fact}[Completeness and soundness of the magic square game, cf.\ Lemma 3.6 in \cite{de_la_Salle_spectral_gap}]\label{fact:soundness_Acom}
	Let $\frak{M}$ be the magic square game from Example \ref{example:magic-square}. Let $\strategy$ be a perfect strategy for $\frak{M}$, and let $\cal{U}$ be $\strategy$ in observable form. Then $\cal{U}(\mathsf{Var}_{11})$ and $\cal{U}(\mathsf{Var}_{22})$ are anti-commuting involutions, namely $$\cal{U}(\mathsf{Var}_{11})\cal{U}(\mathsf{Var}_{22})=-\cal{U}(\mathsf{Var}_{22})\cal{U}(\mathsf{Var}_{11})\ .$$
	Moreover, if $\strategy$ has value $1-\eps$, then 
	$$\Vert\cal{U}(\mathsf{Var}_{11})\cal{U}(\mathsf{Var}_{22})+\cal{U}(\mathsf{Var}_{22})\cal{U}(\mathsf{Var}_{11})\Vert_{hs}^2\leq 432\eps\ .$$
\end{fact}

\begin{remark}
	Fact \ref{fact:soundness_Acom} can also be deduced almost immediately from Claim \ref{claim:perfect_3Lin_commuting_observables}.
\end{remark}
\begin{claim}[Extending anti-commuting observables to perfect strategies]\label{claim:extending_to_perfect_anti_com}
	Let $O_{11},O_{12},O_{21},O_{22}\in U(n)$ be involutions satisfying the following four commutation conditions
	\[
	O_{11}O_{12}=O_{12}O_{11}\ ,\ O_{11}O_{21}=O_{21}O_{11}\ ,\ O_{22}O_{12}=O_{12}O_{22}\ ,\ O_{22}O_{21}=O_{21}O_{22}\ ,
	\]
	as well as the following two anti-commutation conditions
	\[
	O_{11}O_{22}=-O_{22}O_{11}\ ,\ O_{12}O_{21}=-O_{21}O_{12}\ .
	\]
	Then, there exists a perfect strategy $\strategy=\{\cal{U}\}$ for the magic square game $\frak{M}$ satisfying for every $a,b\in \{1,2\}$ that $\cal{U}(\mathsf{Var}_{ab})=O_{ab}$.
\end{claim}
\begin{proof}
	For every $a,b\in [3]$, let $\cal{U}(\mathsf{Var}_{ab})=\cal{U}(\mathsf{Row}_{ab})=\cal{U}(\mathsf{Col}_{ab})$ be the  $ab^{\rm th}$ entry in  Table \ref{tab:induced_perfect_strat_magic_square}.
	\begin{table}[httb!]
		\centering
		\begin{tabular}{|c|c|c|}
			\hline
			$O_{11}$ & $O_{12}$  & $O_{11}O_{12}$ \\
			\hline
			$O_{21}$ & $O_{22}$  & $O_{21}O_{22}$ \\
			\hline
			$-O_{11}O_{21}$ & $-O_{21}O_{22}$ & $-O_{11}O_{12}O_{21}O_{22}$ \\
			\hline
		\end{tabular}
		\caption{Perfect strategy for $\frak{M}$ induced by the quadruple $O_{11},O_{12},O_{21},O_{22}$.}
		\label{tab:induced_perfect_strat_magic_square}
	\end{table}
	We leave it to the reader to verify that this is a well defined, perfect strategy that commutes along edges. We encourage the reader to compare the above general strategy to the one we described in Example \ref{example:magic-square}.
\end{proof}

Let the \emph{null-anti-commutation game} $\frak{M}_{null}$ be the game whose underlying graph, length functions and sets of formal variables are the same as in the magic square game, but it always accepts. This is also a linear constraint system game.

\subsubsection{Pauli basis game --- See Figure \ref{fig:genpauli-summary} for a summary}\label{sec:Pauli_basis_definition}
For any set $\mathscr{B}=\{w^1,...,w^n\}\subseteq \FF_2^k$ we define an appropriate Pauli basis game $\PauliBasis_k=\PauliBasis_k(\mathscr{B})$.
For every $i$ and $j$ in $[n]$, we let $\frak{C}^{i,j}$ be either a copy of the commutation game $\frak{C}$ (Section \ref{sec:com_game}) or the null-commutation game $\frak{C}_{null}$: It will be a copy of $\frak{C}$ in case $\PXm^{w^i}$ should commute with $\PZm^{w^j}$, namely when $\langle w^i,w^j\rangle=0$, and $\frak{C}_{null}$ otherwise. Similarly,  we let $\frak{M}^{i,j}$ be either a copy of the anti-commutation game $\frak{M}$ (Section \ref{sec:Acom_game}) or the null-anti-commutation game $\frak{M}_{null}$: It will be a copy of $\frak{M}$ if $\PXm^{w^i}$ should anti-commute with $\PZm^{w^j}$, namely when  $\langle w^i,w^j\rangle=1$, and  $\frak{M}_{null}$ otherwise.
For clarity of notation, the vertices in $\frak{C}^{i,j}$ will be $\mathtt{First}^{i,j},\mathtt{Second}^{i,j}$ and $\mathtt{Both}^{i,j}$, while the vertices in $\frak{M}^{i,j}$ will be $\{\mathtt{var}^{i,j}_{ab}\}_{a,b=1}^3$, $\{\mathtt{row}^{i,j}_{a}\}_{a=1}^3$ and $\{\mathtt{col}^{i,j}_{b}\}_{b=1}^3$ and they are connected as in Figures \ref{fig:commutation_Game} and \ref{fig:Acommutation_Game} respectively. 

The Pauli basis game  $\PauliBasis_k$ is an augmented sum (Definitions \ref{defn:sum_of_games} and \ref{defn:augmentation_of_a_game}) of the $2n^2$ games $\frak{C}^{i,j}$ and $\frak{M}^{i,j}$. It is augmented with $2n+2$ extra vertices ---  $\{\mttX^i\}_{i=1}^n,\{\mttZ^j\}_{j=1}^n, \Pauli_{\PXm}$ and $\Pauli_{\PZm}$. The lengths of $\mttX^i$ and $\mttZ^j$ are $1$ with associated generators $\sX^i$ and $\sZ^j$, while $ \Pauli_{\PXm}$ and $\Pauli_{\PZm}$ have length $k$ with associated generators $\{\sP\sX^i\}_{i=1}^k$ and $\{\sP\sZ^j\}_{j=1}^k$.
The vertex $\mttX^i$  is connected to $\Pauli_\PXm$,   $\mathtt{First}^{i,j}$ for every $j\in[n]$, and  $\mathtt{var}^{i,j}_{11}$ for every $j\in[n]$.  The vertex $\mttZ^j$  is connected to $\Pauli_\PZm$,  $\mathtt{Second}^{i,j}$ for every $i\in[n]$ and  $\mathtt{var}^{i,j}_{22}$ for every $i\in[n]$ (see Figure \ref{fig:PB_k_unerlying_graph} for a partial view).
\begin{figure}[httb!]
	\centering
	\begin{tikzpicture}
	\node[draw, color=black, shape=circle] (PX) at (0,7) {\scriptsize $\Pauli_\PXm$}; 
	\node[draw, color=black, shape=circle] (PZ) at (0,-7) {\scriptsize $\Pauli_\PZm$}; 
	
	\node[draw, color=black, shape=circle] (Xw1) at (-5,4) {\scriptsize $\mttX^{1}$}; 
	\node[draw, color=black, shape=circle] (Xw2) at (-2,4) {\scriptsize $\mttX^{2}$}; 
	
	\node[ color=black] (Xdots) at (0,4) {\scriptsize $\dots$}; 
	\node[draw, color=black, shape=circle] (Xwn-1) at (2,4) {\scriptsize $\mttX^{{n-1}}$}; 
	
	\node[draw, color=black, shape=circle] (Xwn) at (5,4) {\scriptsize $\mttX^{n}$};

	\node[draw, color=black, shape=circle] (Zw1) at (-5,-4) {\scriptsize $\mttZ^{1}$}; 
	\node[draw, color=black, shape=circle] (Zw2) at (-2,-4) {\scriptsize $\mttZ^{2}$}; 
	
	\node[ color=black] (Zdots) at (0,-4) {\scriptsize $\dots$}; 
	\node[draw, color=black, shape=circle] (Zwn-1) at (2,-4) {\scriptsize $\mttZ^{{n-1}}$}; 
	
	\node[draw, color=black, shape=circle] (Zwn) at (5,-4) {\scriptsize $\mttZ^{n}$}; 
	

	\node [ draw] (C_11) at (-4,0) {\scriptsize $
		\frC^{1,1} $};
	
	\node [ draw] (C_22) at (-1.5,0) {\scriptsize $
		\frC^{2,{n-1}} $};

	\node [ draw] (AC_11) at (-7.5,0) {\scriptsize $
		{\frM}^{1,1} $};
	

	\node [ draw] (C_n11) at (4,0) {\scriptsize $
		\frC^{n,n} $};
	
	\node [ draw] (AC_22) at (1.5,0) {\scriptsize $
		\frM^{2,{n-1}} $};

	\node [ draw] (AC_n11) at (7.5,0) {\scriptsize $
		\frM^{n,n} $};

	\draw[black, -, solid] (PX)--(Xw1);
	\draw[black, -, solid] (PX)--(Xw2);
	\draw[black, -, solid] (PX)--(Xwn-1);
	\draw[black, -, solid] (PX)--(Xwn);
	\draw[black, -, solid] (PZ)--(Zw1);
	\draw[black, -, solid] (PZ)--(Zw2);
	\draw[black, -, solid] (PZ)--(Zwn-1);
	\draw[black, -, solid] (PZ)--(Zwn);
	
	\draw[Cyan, -, dashed] (C_11)--(Xw1);
	
	\draw[Cyan, -, dashed] (C_22)--(Xw2);
	
	\draw[TealBlue, -, dashed] (C_11)--(Zw1);
	\draw[TealBlue, -, dashed] (C_22)--(Zwn-1);

	\draw[Cyan, -, dashed] (C_n11)--(Xwn);

	\draw[TealBlue, -, dashed] (C_n11)--(Zwn);

	\draw[Magenta, -, dotted] (AC_11)--(Xw1);

	\draw[BrickRed, -, dotted] (AC_11)--(Zw1);
	
	\draw[Magenta, -, dotted] (AC_22)--(Xw2);

	\draw[BrickRed, -, dotted] (AC_22)--(Zwn-1);

	\draw[Magenta, -, dotted] (AC_n11)--(Xwn);
	
	\draw[BrickRed, -, dotted] (AC_n11)--(Zwn);

	\end{tikzpicture}
	\caption{This is a partial picture of the underlying graph of $\PauliBasis_k$. Note that the nodes $\frC^{i,j}$ and $\frM^{i,j}$ are not single vertices in the graph, but some (constant sized) subgraphs associated with the commutation \ref{sec:com_game} and anti-commutation \ref{sec:Acom_game} games (respectively). The $\mttX^i$ and $\mttZ^j$ vertices are attached to $\frC^{i,j}$ and $\frM^{i,j}$ in a specific way: $\mttX^i$ is connected to $\mathtt{First}^{i,j}$ in $\frC^{i,j}$ and to $\mathtt{var}^{i,j}_{11}$ in $\frM^{i,j}$, while $\mttZ^j$  is connected to $\mathtt{Second}^{i,j}$ in $\frC^{i,j}$ and to $\mathtt{var}^{i,j}_{22}$ in $\frM^{i,j}$.
		There is a commutation (or null commutation)  and anti-commutation (or null anti-commutation) game between every  $\mttX^i$ and $\mttZ^j$, but we have only drawn the local picture for the pairs $(\mttX^{1},\mttZ^{1})$, $(\mttX^{2},\mttZ^{{n-1}})$ and $(\mttX^{{n}},\mttZ^{{n}})$. 
	}
	\label{fig:PB_k_unerlying_graph}
\end{figure}

Now, if an edge within $\frak{C}^{i,j}$ or $\frak{M}^{i,j}$ is sampled, the decision procedure is already defined. When $\mttX^i$ (respectively $\mttZ^j$) is sampled against $\mathtt{First}^{i,j}$ (respectively $\mathtt{Second}^{i,j}$), we check consistency between their values, namely 
$\gamma(\sX^i)=\gamma(\mathsf{First}^{i,j})$ (respectively $\gamma(\sZ^j)=\gamma(\mathsf{Second}^{i,j})$). Similarly, when $\mttX^i$ (respectively $\mttZ^j$) is sampled against $\mathtt{var}^{i,j}_{11}$ (respectively $\mathtt{var}^{i,j}_{22}$), we check consistency between their values, namely 
$\gamma(\sX^i)=\gamma(\mathsf{Var}^{i,j}_{11})$ (respectively $\gamma(\sZ^j)=\gamma(\mathsf{Var}^{i,j}_{22})$). Lastly, if $\mttX^i$ (respectively $\mttZ^j$)  is sampled against $\Pauli_{\PXm}$ (respectively $\Pauli_{\PZm}$), then we check that 
$\gamma(\sX^i)=\sum_{j=1}^k w^i_j\gamma(\sP\sX^j)$ (respectively $\gamma(\sZ^j)=\sum_{i=1}^k w^j_i\gamma(\sP\sZ^i)$).

Note that this is an LCS game, as was defined in Example \ref{example:LCSs}. In particular, by tailoring it as described in the aforementioned example, all variables are linear, and the linear constraint processor is implicitly defined by the decision procedure above.

Finally, we need to describe the distribution used in $\PauliBasis_k$. For now,\footnote{The distribution we actually use needs to be induced by a conditionally linear sampling scheme (Definition \ref{defn:sampling_induced_by_CLMs}). See Example \ref{example:quered_has_typed_CL_sampling_scheme} for the actual distribution we use. We note that in the resulting distribution, the probability each edge is sampled is at least some constant times the distribution we provided here. So, all of our arguments, which anyway use the asymptotic $O(\cdot)$-notation, stay the same.} let us assume it is the following --- with probability $\nicefrac{1}{8}$ do one of the following: sample a uniform edge from a uniformly random $\frak{M}^{i,j}$; sample a uniform edge from a uniformly random $\frak{C}^{i,j}$; sample a uniform edge of the form $\mttX^i-\mathtt{First}^{i,j}$;  sample a uniform edge of the form $\mttZ^j-\mathtt{Second}^{i,j}$; sample a uniform edge of the form $\mttX^i-\mathtt{Var}_{11}^{i,j}$; sample a uniform edge of the form $\mttZ^j-\mathtt{Var}_{22}^{i,j}$; sample a uniform edge of the form $\mttX^i-\Pauli_{\PXm}$; sample a uniform edge of the form $\mttZ^j-\Pauli_{\PZm}$.

\begin{remark}\label{rem:goal_of_checks_PB_k}
	Let us briefly motivate the structure and checks of $\PauliBasis_k$. 
	As discussed in Definition \ref{defn:PVM}, every strategy $\strategy=\{\cal{U}\}$ induces two representations of $\FF_2^k$ --- $\chi=\cal{U}^{\Pauli_\PXm}$ associated with the image of $S_{\Pauli_{\PXm}}$ and $\zeta=\cal{U}^{\Pauli_\PZm}$ associated with the image of $S_{\Pauli_{\PZm}}$. 
	The check $\mttX^i-\Pauli_{\PXm}$ forces $\strategy$ to satisfy $\cal{U}(\sX^i)=\chi(w^i)$, and the check $\mttZ^j-\Pauli_{\PZm}$ forces $\strategy$ to satisfy $\cal{U}(\sZ^j)=\zeta(w^j)$. 
	Then, for $i,j$ such that $\langle w^i,w^j\rangle=0$, the consistency  checks $\mttX^i-\mathtt{First}^{i,j}$ and $\mttZ^j-\mathtt{Second}^{i,j}$ together with running $\frak{C}^{i,j}$ forces $\strategy$ to satisfy $\chi(w^i)\zeta(w^j)=\zeta(w^j)\chi(w^i)$. 
	Finally, for $i,j$ such that $\langle w^i,w^j\rangle=1$, the consistency  checks $\mttX^i-\mathtt{var}_{11}^{i,j}$ and $\mttZ^j-\mathtt{var}_{22}^{i,j}$ together with running $\frak{M}^{i,j}$ forces $\strategy$ to satisfy $\chi(w^i)\zeta(w^j)=-\zeta(w^j)\chi(w^i)$. 
	Hence, by taking all product of images of $\chi$ and $\zeta$, we get a representation of the Pauli group $\WH_k$, and since some of the images anti-commute, all irreducible components of this representation are copies of  the unique non-commuting representation $\rho$ defined in \eqref{eq:def_of_rho_for_P_k} ---  which was our goal.
\end{remark}

\begin{claim}[Completeness of the Pauli basis game]\label{claim:completeness_PB_k}
	Let $m$ and $k$ be positive integers with $k\geq 2$. Let $\rho\colon \WH_k\to U(\complex^{\FF_2^k})$ be the representation of the Pauli group acting on $k$ qubits defined in \eqref{eq:def_of_rho_for_P_k}, and let $\mathscr{U}$ be a unitary in $U(\complex^{\FF_2^k}\otimes \complex^m)$. Then, there is a perfect strategy that commutes along edges $\strategy=\{\cal{U}\}$ for the Pauli basis game $\PauliBasis_k$ such that the representations $\cal{U}^{\Pauli_\PXm}$ and $\cal{U}^{\Pauli_\PZm}$, which $\strategy$ associates to the vertices $\Pauli_\PXm$ and $\Pauli_\PZm$, are $\mathscr{U}^{-1}\rho^\PXm\otimes\Id_m \mathscr{U}$ and $\mathscr{U}^{-1}\rho^\PXm\otimes\Id_m \mathscr{U}$ respectively (see \eqref{eq:defn_restriction_rho_to_X_and_Z_subgroups} in Definition \ref{defn:rho}).
	In particular, if $\mathscr{U}$ is the identity, then this strategy is a  permutation strategy, and the images $\cal{U}^{\Pauli_\PZm}$ are diagonal in the standard basis.
\end{claim}
\begin{proof}[Proof sketch]
	We mainly follow the restrictions of the Pauli basis game, as described in Remark \ref{rem:goal_of_checks_PB_k}.
	We are forced, by the claim, to let 
	\[
	\forall v,w\in \FF_2^k\ \colon \ \ \cal{U}^{\Pauli_\PXm}(v)=\mathscr{U}^{-1}(\PXm^{\otimes v}\otimes\Id_m) \mathscr{U}\ ,\ \cal{U}^{\Pauli_\PZm}(w)=\mathscr{U}^{-1}(\PZm^{\otimes w}\otimes\Id_m) \mathscr{U}\ .
	\]
	By claim \ref{claim:perfect_3Lin_commuting_observables}, for the checks incident to the $\mttX^i$ and $\mttZ^j$ vertices to perfectly be satisfied, we need 
	\[
	\cal{U}(\sX^i)=\cal{U}(\mathsf{First}^{i,j})=\cal{U}(\mathsf{Var}_{11}^{i,j})=\mathscr{U}^{-1}(\PXm^{\otimes w^i}\otimes\Id_m)\mathscr{U}\ ,\ \cal{U}(\sZ^i)=\cal{U}(\mathsf{Second}^{i,j})=\cal{U}(\mathsf{Var}_{22}^{i,j})=\mathscr{U}^{-1}(\PZm^{\otimes w^j}\otimes \Id_m)\mathscr{U}\ ,
	\]
	where $w^i$ and $w^j$ are the $i^{\rm th}$ and $j^{\rm th}$ vectors from the fixed set $\mathscr{B}$. Now, we can use Claim \ref{claim:extending_to_perfect_com} to extend $\cal{U}$ to  $\mathsf{Both}_1^{i,j}$ and $\mathsf{Both}_2^{i,j}$ in case  
	$\cal{U}(\sX^i)$ and $\cal{U}(\sZ^j)$ commute, which is exactly the case where $\langle w^i,w^j\rangle=0$; otherwise, we let  $\cal{U}(\mathsf{Both}_1^{i,j})=\cal{U}(\mathsf{Both}_2^{i,j})=\Id$. This verifies that indeed $\cal{U}$ is perfect when restricted to 
	the (null-)commutation games $\frak{C}^{i,j}$.
	For the (null-)anti-commutation games $\frak{M}^{i,j}$ we have more flexibility, as a perfect strategy for them requires a quadruple of observables and we fixed only two, namely $\cal{U}(\mathsf{Var}^{i,j}_{11})$ and $\cal{U}(\mathsf{Var}^{i,j}_{22})$. In case $\langle w^i,w^j\rangle=1$, we fix a pair of vectors $v^{i,j}_1,v^{i,j}_2\in \FF_2^k$ satisfying $\langle v^{i,j}_1,w^j\rangle=0\ ,\  \langle v^{i,j}_1,v^{i,j}_2\rangle=1$ and $\langle w^i,v^{i,j}_2\rangle=0$ --- 
	we leave it to the reader to check that such vectors exist whenever $k\geq 2$. Then,  the quadruple 
	\[
	\cal{U}(\sX^i)\ ,\ \mathscr{U}^{-1} (\PXm^{v^{i,j}_1}\otimes \Id_m)\mathscr{U}\ ,\  \mathscr{U}^{-1} (\PZm^{v^{i,j}_2}\otimes \Id_m)\mathscr{U}\ ,\ \cal{U}(\sZ^j)\ ,
	\]
	satisfies the conditions of Claim \ref{claim:extending_to_perfect_anti_com}, and can thus be extended to a perfect strategy for $\frak{M}^{i,j}$. In case $\langle w^i,w^j\rangle=0$, $\frak{M}^{i,j}$ is a null-anti-commutaion game, and we can thus extend $\cal{U}$ to it such that it is $\Id$ for all variables  not yet defined.

	In case $\mathscr{U}$ is the identity, the image of $\cal{U}$ consists of products and tensor products of signed permutation matrices, and is thus  a permutation strategy. In particular, $\cal{U}^{\Pauli_\PZm}=\rho^\PZm\otimes \Id_m$ is a diagonal representation, as needed.
\end{proof}

\begin{claim}[Characterization of almost-perfect strategies of the Pauli basis game]\label{claim:almost_perfect_strategies_of_PB_k}
	Let $\mathscr{B}=\{w^1,...,w^n\}\subseteq \FF_2^k$ be an ordered set, and let $\PauliBasis_k=\PauliBasis_k(\mathscr{B})$ be the appropriate generalized Pauli basis game. Let $\strategy$ be an $N$-dimensional strategy  satisfying $\val(\PauliBasis_k,\strategy)\geq 1-\eps$, and $\cal{U}$ be $\strategy$ in observable (and representation) form.
	Then, for some universal constant $C>0$, we have that $\cal{U}^{\Pauli_\PXm}(w^i)\cal{U}^{\Pauli_\PZm}(w^j)$ is $C\eps$-close to $(-1)^{\langle w^i,w^j\rangle }\cal{U}^{\Pauli_\PZm}(w^j)\cal{U}^{\Pauli_\PXm}(w^i)$ on average over uniform $i,j\in [n]$; namely
	\[
	\Es{i,j\in [n]}\left[\left\Vert\cal{U}^{\Pauli_\PXm}(w^i)\cal{U}^{\Pauli_\PZm}(w^j)-(-1)^{\langle w^i,w^j\rangle }\cal{U}^{\Pauli_\PZm}(w^j)\cal{U}^{\Pauli_\PXm}(w^i)\right\Vert_{hs}^2 \right]\leq C\eps \ .
	\]
\end{claim}

\begin{proof}[Proof sketch]
	By the fact that $\strategy$ passes the game with probability of at least $1-\eps$, and the collection of edges of type $\Pauli_{\PXm}-\mttX^i$ and $\Pauli_{\PZm}-\mttZ^j$ have a constant probability of being sampled under the game distribution, $\strategy$ passes a uniformly random edges of this type with probability of at least $1-O(\eps)$. Hence, $\cal{U}^{\Pauli_\PXm}(w^i)$ is $O(\eps)$-inconsistent (Definition \ref{defn:distance_inconsistency_POVMs}) with $\cal{U}(\sX^i)$ on average over $i\in [n]$, and similarly that $\cal{U}^{\Pauli_\PZm}(w^j)$ is $O(\eps)$-inconsistent with $\cal{U}(\sZ^j)$ on average over $j\in [n]$. Hence, by Proposition \ref{prop:properties_of_distance_and_inconsistency} and Claim \ref{claim:dist_in_projection_and_observable_are_the_same}, we have\footnote{This can also be deduced from Claim \ref{claim:perfect_3Lin_commuting_observables}.}
	\[
	\Ex_{i\in[n]}\left[\Vert\cal{U}^{\Pauli_\PXm}(w^i)-\cal{U}(\sX^{i})\Vert_{hs}^2\right]\;,\quad \Ex_{j\in[n]}\left[\Vert\cal{U}^{\Pauli_\PZm}(w^j)-\cal{U}(\sZ^{j})\Vert_{hs}^2\right]\ \leq  O(\eps)\ .
	\]
	Since $\strategy$ passes the edges of  type $\mttX^i-\mathtt{First}^{i,j}$, $\mttX^i-\mathtt{var}_{11}^{i,j}$, $\mttZ^j-\mathtt{Second}^{i,j}$ and $\mttZ^j-\mathtt{var}_{22}^{i,j}$ in $\PauliBasis_k$ with probability $1-O(\eps)$ (on average over uniform pairs $i,j\in [n]$), we can deduce by Claim \ref{claim:perfect_3Lin_commuting_observables} that 
	\[
	\begin{split}
	\Ex_{i,j\in[n]}\left[\Vert\cal{U}(\mathsf{First}^{i,j})-\cal{U}(\sX^{i})\Vert_{hs}^2\right]\;,\quad \Ex_{i,j\in[n]}\left[\Vert\cal{U}(\mathsf{Second}^{i,j})-\cal{U}(\sZ^{j})\Vert_{hs}^2\right]\ &\leq  O(\eps)\ ,\\
	\Ex_{i,j\in[n]}\left[\Vert\cal{U}(\mathsf{Var}^{i,j}_{11})-\cal{U}(\sX^{i})\Vert_{hs}^2\right]\;,\quad \Ex_{i,j\in[n]}\left[\Vert\cal{U}(\mathsf{Var}^{i,j}_{22})-\cal{U}(\sZ^{j})\Vert_{hs}^2\right]\ &\leq  O(\eps)\ .
	\end{split}
	\]
	Since $\strategy$ passes the copies of the commutation games $\frak{C}^{i,j}$ and anti-commutation games $\frak{M}^{i,j}$ in $\PauliBasis_k$ with probability $1-O(\eps)$ (on average over uniform $i,j\in [n]$), we can deduce using Facts \ref{fact:soundness_com} and \ref{fact:soundness_Acom} that  
	\[
	\begin{split}
	\Pro{i,j\in [n]}[\langle w^i,w^j\rangle=0] \cdot \Es{i,j\in [n]\colon \langle w^i,w^j\rangle=0}\left[\Vert\cal{U}(\mathsf{First}^{i,j})\cal{U}(\mathsf{Second}^{i,j})-\cal{U}(\mathsf{Second}^{i,j})\cal{U}(\mathsf{First}^{i,j})\Vert_{hs}^2\right]&\leq O(\eps)\ ,\\
	\Pro{i,j\in [n]}[\langle w^i,w^j\rangle=1] \cdot \Es{i,j\in [n]\colon \langle w^i,w^j\rangle=1}\left[\Vert\cal{U}(\mathsf{Var}^{i,j}_{11})\cal{U}(\mathsf{Var}^{i,j}_{22})+\cal{U}(\mathsf{Var}^{i,j}_{22})\cal{U}(\mathsf{Var}^{i,j}_{11})\Vert_{hs}^2\right]&\leq O(\eps)\ .
	\end{split}
	\]
	By combining all of the above observations, the claim is deduced. 
\end{proof}

\begin{corollary}[The Pauli basis game is a (semi)-robust self test]
	For every $N$-dimensional strategy $\strategy=\{\cal{U}\}$ for the Pauli basis game $\PauliBasis_k$ with value $1-\eps$, there is a  perfect strategy $\strategy'=\{\cal{V}\}$ for the game such that: 
	\begin{enumerate}
		\item The representations $\cal{V}^{\Pauli_\PXm}$ and $\cal{V}^{\Pauli_\PZm}$ are respective direct sums of $\rho^\PXm$ and $\rho^\PZm$  from \eqref{eq:defn_restriction_rho_to_X_and_Z_subgroups}. Namely, there is a positive integer $m$ such that 
		\[
		\forall v,w\in \FF_2^k\ \colon \ \ \cal{V}^{\Pauli_\PXm}(v)=\PXm^{\otimes v}\otimes \Id_m\quad{\rm and}\quad \cal{V}^{\Pauli_\PZm}(w)=\PZm^{\otimes w}\otimes \Id_m\ .
		\]
		\item The representations $\cal{V}^{\Pauli_\PXm}$ and $\cal{V}^{\Pauli_\PZm}$ are $O(\eps\cdot\nicefrac{k^2}{d^2})$-flexibly-close to $\cal{U}^{\Pauli_\PXm}$ and $\cal{U}^{\Pauli_\PZm}$ respectively. Namely, there is a universal constant $C>0$, and a $(C\cdot\eps\cdot\nicefrac{k^2}{d^2})$-near bijection $\omega\colon  \complex^N\to  \complex^{\FF_2^k}\otimes \complex^m$ such that 
		\[
		\Es{v\in \FF_2^k}\left[\|\cal{U}^{\Pauli_\PXm}(v)-\omega^*\cal{V}^{\Pauli_\PXm}(v)\omega\|^2_{hs}\right]\ ,\ \Es{w\in \FF_2^k}\left[\|\cal{U}^{\Pauli_\PZm}(w)-\omega^*\cal{V}^{\Pauli_\PZm}(w)\omega\|^2_{hs}\right]\leq C\cdot\eps\cdot\nicefrac{k^2}{d^2}\ .
		\]
	\end{enumerate}
\end{corollary}

\begin{proof}
	By the characterization of almost perfect strategies for $\PauliBasis_k$ (Claim \ref{claim:almost_perfect_strategies_of_PB_k}), $\chi=\cal{U}^{\Pauli_\PXm}$ and $\zeta=\cal{U}^{\Pauli_\PZm}$ satisfy the conditions of the semi-stability result for $\WH_k$ (Fact \ref{fact:semi-stability_P_k}). Applying the semi-stability result provides a near bijection $\omega$ so that conjugating $\cal{U}^{\Pauli_\PXm}$  and $\cal{U}^{\Pauli_\PZm}$ by it brings them close to $\rho^\PXm\otimes \Id_m$ and $\rho^\PZm\otimes \Id_m$ respectively. Finally, Claim \ref{claim:completeness_PB_k} says that this representation $\rho\otimes\Id_m$ can be extended to a perfect strategy for $\PauliBasis_k$, which we denote by $\cal{V}$.
\end{proof}

\section{Question reduction via introspection}\label{sec:quered}

The goal of this section is to devise an algorithm $\mathsf{QuestionReduction}$ that takes as input a tailored normal form verifier and outputs a new tailored normal form verifier whose $n^{\rm th}$ game \emph{simulates} the $(2^{n})^{\rm th}$ game of the original verifier. Though this new verifier is not as time efficient as needed for compression (Theorem \ref{thm:compression}), its sampling procedure is. Recall  the asymptotic notation from Remark \ref{rem:asymptotic_notation}.

\begin{theorem}[Informal Question Reduction, see Theorem \ref{thm:h_level_question_reduciton} for the formal version]\label{thm:informal_question_reduciton}

	There exists a polynomial time  $2$-input Turing machine $\QueRed$ that takes as input a
	TNFV $\verifier=(\sampler,\length,\linproc,\decider)$ and a positive integer $\lambda$, and outputs  a  TNFV 
	\[
	\QueRed(\verifier,\lambda)=\verifier'= (\sampler',\length',\linproc',\decider)\ ,
	\]
	such that  $\sampler'$ runs in $\poly(n,\lambda)$-time, $\length'$ and $\linproc'$ run in $\exp(n,\lambda)$-time, and given that $\verifier$ is $\lambda$-bounded, the output 
	$\verifier'$  satisfies: For all $n \geq 2$,
	\begin{enumerate}
		\item \textbf{Completeness}: If $\verifier_{2^n}$ has a perfect $Z$-aligned permutation strategy, then so does $\verifier'_n$.
		\item \textbf{Soundness}: For every $\eps>0$, if $\verifier'_{n}$ has a value $1-\eps$ strategy, then $\verifier_{2^n}$  has a value $1-O(\eps^{\nicefrac{1}{16}})$ strategy.
		\item \textbf{Entanglement}: For every $\eps>0$, 
		\[
		\Ent(\verifier'_n,1-\eps)\geq (1-O(\eps))\cdot 2^{2^{\lambda n}}\cdot\Ent(\verifier_{2^n},1-O(\eps^{\nicefrac{1}{16}}))\ .
		\]
	\end{enumerate}
\end{theorem}

The combinatorial transformation underlying the question reduction algorithm stems from the straightforward idea of \emph{introspection} --- ``let the provers sample their own questions''.
This can be done naively, by letting $\Introspect(\game)$ be as in Definition \ref{defn:Introspection}.  But, for this transformation to be helpful for compression,  we are going to take an augmented sum (see Definitions \ref{defn:sum_of_games} and \ref{defn:augmentation_of_a_game}) of $ \Introspect(\game)$  with the \emph{generalized Pauli basis} game $\PauliBasis_k$ (see Section \ref{sec:gen_pauli_basis}). As we previously showed, the game $\PauliBasis_k$ is robust, and has essentially one perfect strategy, which induces the non-commutative representation $\rho$ \eqref{eq:def_of_rho_for_P_k} of the Pauli group acting on $k$ qubits (Definition \ref{defn:Pauli_group_acting_on_k}). By connecting the total $\PXm$ and total $\PZm$ measurements guaranteed by $\PauliBasis_k$ --- namely, the vertices $\Pauli_\PXm$ and $\Pauli_\PZm$ --- in a clever way to $\Introspect(\game)$ we can ensure that any almost perfect strategy for the introspection game is close to being honest (Definition \ref{defn:honest_strat_for_Intro}), and thus induces an almost perfect strategy of $\game$.

\begin{figure}[!htbp]
	\centering
	\begin{gamespec}
		\begin{enumerate}
			\setlength\itemsep{1pt}
			\item \textbf{Introspection $\Introspect(\cdot)$.} This transformation takes a game  $\game$ with (possibly) a very intricate underlying graph and replaces it by a game  $\Introspect(\game)$ with an underlying graph containing a single edge between two vertices. The provers in the transformed game are expected to both \emph{sample} their own questions and then \emph{answer} them accordingly. It is introduced in Section~\ref{sec:the_introspection_game} (see Figure~\ref{fig:introspect-summary} for a summary). Completeness and Soundness for ``honest'' strategies is sketched in the same section, specifically in Claim  \ref{claim:complete_sound_honest_intro}. 
			\item \textbf{Commutation game $\frC$.} This game is described in Section~\ref{sec:com_game}. It is a small, independent game on $3$ questions that aims to verify commutation between two observables. Its completeness and soundness are stated in Fact~\ref{fact:soundness_com}.
			\item \textbf{Anti-commutation game $\frM$.} This game is described in Section~\ref{sec:Acom_game}. It is a small, independent game on $15$ questions that aims to verify anti-commutation between two observables. Its completeness and soundness are stated in Fact~\ref{fact:soundness_Acom}.
			\item \textbf{Pauli basis game $\PauliBasis_k(\mathscr{B})$.} This game depends on an integer $k$ and a set $\mathscr{B}=\{w^1,...,w^n\}\subseteq \FF_2^k$. Its goal is, essentially, to verify that a subset of a strategy's observables induce a non-commutative representation of the Pauli group $\WH_k$ (introduced in Section~\ref{sec:Pauli_gp}). A combinatorial description of $\PauliBasis_k(\mathscr{B})$ is given in Section~\ref{sec:Pauli_basis_definition}, see also Figure~\ref{fig:genpauli-summary} for a summary. Completeness of this game is shown in Claim~\ref{claim:completeness_PB_k} and soundness in Claim~\ref{claim:almost_perfect_strategies_of_PB_k}. An algorithmic implementation of $\PauliBasis_k(\mathscr{B})$ as a tailored normal form verifier is implicitly given in Section~\ref{thm:h_level_question_reduciton}, where a tailored normal form verifier for the larger game $\frak{QueRed}(\game)$ is given. 
			\item \textbf{``Baby'' question reduction $\frak{Baby}(\cdot)$.} This transformation takes as input a game $\game$ whose question distribution is a pushforward of the uniform measure over $\F_2^k$ through a \emph{linear} function, and returns a game $\frak{Baby}(\game)$ that has exponentially fewer questions and yet perfect $\ZPC$ completeness and soundness are preserved. The game is introduced in Section~\ref{sec:baby_question_reduction} (see Figure~\ref{fig:babqr-summary} for a summary). Completeness and soundness are shown in Theorem~\ref{thm:bab-qr}. (This game is introduced for illustrative purposes and results about it are not formally used.)
			\item \textbf{Question reduced $\frak{QueRed}(\cdot)$.} This transformation takes as input a game $\game$ whose question distribution is \emph{conditionally linear}, a generalization of the previous case described in Section~\ref{sec:CLMs}, and returns a game $\frak{QueRed}(\game)$ that has exponentially fewer questions and yet perfect $\ZPC$ completeness and soundness are preserved. The game is introduced in Section~\ref{sec:augmentation_in_the_CLM_case}, see also Figure~\ref{fig:quered-summary} for a summary. Completeness and soundness are shown in Theorem~\ref{thm:complet_sound_quered}. A tailored normal form verifier implementing this game (given as input a tailored normal verifier for $\game$) is described in Section~\ref{sec:quered-nf}, resulting in the proof of the main theorem of this section, Theorem~\ref{thm:h_level_question_reduciton}, in  Section~\ref{sec:prf-442}. 
		\end{enumerate}
	\end{gamespec}
	\caption{We list the main games, or transformations thereof, used and introduced in this section, and where to find the most important statements about them.}
	\label{fig:intro-section-summary}
\end{figure}

\subsection{The introspection game}\label{sec:the_introspection_game}

Throughout this section, let $\game$ be a tailored game with vertex set $\FF_2^r$, and assume the distribution $\mu$ over its edges is a pushforward of the uniform measure on $\FF_2^k$.\footnote{This is always the case for games defined via normal form verifiers, as $\sampler$ calculates a pushforward of this form. It is not clear that all the vertices have the same bit length description ($r$ in this case), but up to some encoding it can be assumed  as well.} Namely, there is a function 
$ \frS\colon \FF_2^k\to \FF_2^r\times \FF_2^r$ such that 
\begin{equation}\label{eq:pushforward_of_uniform_distribution}
\forall \mttx,\mtty\in \FF_2^r\ \colon \ \ \mu(\mttx\mtty)=\frac{|\frS^{-1}(\mttx,\mtty)|}{2^k}\ .
\end{equation}
Given $z\in \FF_2^k$, we  use $\frS^A(z)=\mttx$ to denote the first coordinate of the output of $\frS$, and similarly $\frS^B(z)=\mtty$ for the second coordinate.
Assume furthermore that the readable and unreadable answer length functions of $\game$ are constant and equal  $\Lambda\in \mathbb{N}$.\footnote{As will be seen in Section \ref{sec:Padding}, this assumption is not much of a constraint.} Finally, let us denote by $S_{\mttx}^\frR=\{\sX^{\frR,i}\mid 1\leq i\leq\Lambda\}$ and $S_{\mttx}^\frL=\{\sX^{\frL,i}\mid 1\leq i\leq\Lambda\}$ the formal generator sets at $\mttx\in \FF_2^r$, and similarly $S_\mtty^\cdot=\{\sY^{\cdot,i}\}$ for $\mtty\in \FF_2^r$.

\begin{figure}[!htbp]
	\centering
	\begin{gamespec}
		\setlength{\tabcolsep}{1em}
		Let $\game$ be a tailored game whose readable and unreadable answer length functions are constant and equal to $\Lambda$.
		
		\vspace{1em}
		\begin{tabularx}{\textwidth}{  X X X }
			\toprule
			Question &   Readable variables & Unreadable variables  \\
			\midrule
			$\Intro_\cdot$ & $\{\sQue^{\cdot,i}\}_{i=1}^r$ & $\{\sAns^{\cdot,\frL,j}\}_{j=1}^\Lambda$  \\
			& $\{\sAns^{\cdot,\frR,j}\}_{j=1}^\Lambda$ & \\
			\bottomrule
		\end{tabularx}
		
		\vspace{1em}
		
		For any $\gamma\colon S_{\Intro_A}\cup S_{\Intro_B}\to \FF_2$,  denote 
		\[
		\mttx=\gamma|_{\sQue^{A}} ,\ 
		\mtty=\gamma|_{\sQue^{B}}\ ,\ a^\frR=\gamma|_{\sAns^{A,\frR}}\ ,\ a^\frL=\gamma|_{\sAns^{A,\frL}}\ ,\ b^\frR=\gamma|_{\sAns^{B,\frR}}\ ,\  b^\frL=\gamma|_{\sAns^{B,\frL}}\;.
		\]
		Then $\Introspect(\game)$ accepts $\mttx,(a^\frR,a^\frL),\mtty,(b^\frR,b^\frL)$, if and only if $\game$ accepts $(a^\frR,a^\frL),(b^\frR,b^\frL)$ given that $\mttx\mtty$ were asked.
	\end{gamespec}
	\caption{Questions and answers in the game $\frak{Intro}(\game)$.   }
	\label{fig:introspect-summary}
\end{figure}

\begin{definition}[The introspection transformation of a tailored game]\label{defn:Introspection}
	Let $\game$ be a (tailored) game with the above fixed properties.  The \emph{introspection game} $\Introspect(\game)$ consists of only two vertices $\Intro_A$ and $\Intro_B$, with a single edge between them (see Figure~\ref{fig:introspect-summary} for a summary). As there is only one edge, it is always chosen by the question distribution of $\Introspect(\game)$. The readable length of both $\Intro_A$ and  $\Intro_B$  is $r+\Lambda$, and their unreadable length is $\Lambda$. Define 
	\[
	\sQue^A=\{\sQue^{A,i}\mid 1\leq i\leq r\}\quad , \quad \sAns^{A,\frR}=\{\sAns^{A,\frR,j}\mid 1\leq j \leq \Lambda\}\quad \textrm{and}\quad \sAns^{A,\frL}=\{\sAns^{A,\frL,j}\mid 1\leq j \leq \Lambda\},
	\]
	and similarly $\sQue^B,\sAns^{B,\frR}$ and $\sAns^{B,\frL}$. Then, let the formal readable variables  at $\Intro_A$ and $\Intro_B$ be
	\[
	S^\frR_{\Intro_A}=\sQue^{A}\sqcup\sAns^{A,\frR}\quad\textrm{and}\quad S^\frR_{\Intro_B}=\sQue^{B}\sqcup\sAns^{B,\frR}
	\]
	respectively, and let the formal unreadable variables at these vertices be
	$$S^\frL_{\Intro_A}=\sAns^{A,\frL}\quad\textrm{and}\quad S^\frL_{\Intro_B}=\sAns^{B,\frL}.$$
	The naming scheme is $\sQue$ for ``question'' and $\sAns$ for `` answer''. I.e., the assignment to the variable $\sQue^{A,i}$ (respectively $\sQue^{B,i}$) is expected to be the $i^{\rm th}$ bit of a question $\mttx\in \FF_2^r$ (respectively  $\mtty\in \FF_2^r$),  the assignment to $\sAns^{A,\frR,j}$ (respectively $\sAns^{B,\frR,j}$) is expected to be the $j^{\rm th}$ bit of the readable part of an answer in $\game$ to the question $\mttx$ (respectively  $\mtty$), and the assignment to $\sAns^{A,\frL,j}$ (respectively $\sAns^{B,\frL,j}$) is expected to be the $j^{\rm th}$ bit of the unreadable part of the answer to $\mttx$ (respectively  $\mtty$).
	The controlled linear constraint function $L_{\Intro_A\ \Intro_B}(\gamma)$ works as follows. 
	Let $\gamma\colon S_{\Intro_A}\cup S_{\Intro_B}\to \FF_2$,  and denote its restrictions as follows
	\[
	\begin{split}
	\mttx=\gamma|_{\sQue^{A}} ,\ 
	\mtty=\gamma|_{\sQue^{B}}\ &,\ a^\frR=\gamma|_{\sAns^{A,\frR}}\ ,\ a^\frL=\gamma|_{\sAns^{A,\frL}}\ ,\ b^\frR=\gamma|_{\sAns^{B,\frR}}\ ,\quad\textrm{and} \quad b^\frL=\gamma|_{\sAns^{B,\frL}}\ .
	\end{split}
	\]
	Note that $\mttx$ and $\mtty$ are $r$-long bit strings, and can thus be viewed as vertices in the underlying graph of the original game $\game$. Recall that we denoted $S_\mttx^\cdot=\{\sX^{\cdot,j}\}$ and $S_\mtty^\cdot=\{\sY^{\cdot,j}\}$ for  the formal generator sets associated with $\mttx$ and $\mtty$ in $\game$. 
	Then, if $\mathtt{xy}$ is not an edge in the underlying graph of $\game$, then $L_{\Intro_A\ \Intro_B}(\gamma)$ will output the singleton $\{\sJ\}$ --- which induces the linear constraint $1=0$, i.e., rejection. 
	Otherwise, for every $c\colon S_\mttx\cup S_\mtty\cup \{\sJ\}\to \FF_2$ in $L_{\mathtt{xy}}(a^\frR,b^\frR)$, we add the constraint coefficients function $c'\colon S_{\Intro_A}\cup S_{\Intro_B}\cup \{\sJ\}\to \FF_2$ to $L_{\Intro_A\ \Intro_B}(\gamma)$, where 
	\[
	\begin{split}
	\forall 1\leq i\leq r,1\leq  j\leq\Lambda\ \colon \ \ c'(\sQue^{A,i})&=c'(\sQue^{B,i})=0\ ,\\  
	c'(\sAns^{A,\frR,j})&=c(\sX^{\frR,j})\ ,\\
	c'(\sAns^{A,\frL,j})&=c(\sX^{\frL,j})\ ,\\
	c'(\sAns^{B,\frR,j})&=c(\sY^{\frR,j})\ ,\\
	c'(\sAns^{B,\frL,j})&=c(\sY^{\frL,j})\ ,\\
	c'(\sJ)&=c(\sJ)\ .
	\end{split}
	\]
	In words, $\Introspect(\game)$ treats $\sAns^{A,\cdot,j}$ as if they were the generators $\sX^{\cdot,j}$ of the sampled vertex $\mttx$, and similarly for $\sAns^{B,\cdot,j}$ and $\sY^{\cdot,j}$ for the other sampled vertex $\mtty$.
\end{definition}
\begin{remark}
	Though the above definition is a bit technical, it can be explained in plain words: The answer to $\Intro_A$ is of the form $(\mttx,a^\frR,a^\frL)$ and to $\Intro_B$ is of the form $(\mtty,b^\frR,b^\frL)$. Then, $\Introspect( \game)$ accepts this pair of answers if and only if $\game$ would accept $(a^\frR,a^\frL,b^\frR,b^\frL)$ given that $\mathtt{xy}$ was the sampled edge.
\end{remark}

We now define the notion of an \emph{honest} strategy for $\Introspect(\game)$. Colloquially, such a strategy $\strategy$ is derived from a strategy $\strategy'$ for $\game$ as follows: First, it samples a bit string $z\in \FF_2^k$ uniformly at random. Then, it calculates $\frS(z)=\mttx\mtty$. Then, only depending on $\mttx$ it performs the measurements induced by $\strategy'$ given that $\mttx$ was asked, which yields the answers $a^\frR,a^\frL\in \FF_2^\Lambda$; similarly, only depending on $\mtty$, using the measurements of $\strategy'$, it obtains $b^\frR,b^\frL\in \FF_2^\Lambda$. Finally, it replies $(\mttx,a^\frR,a^\frL)$ as the assignment to the $\Intro_A$ variables and $(\mtty,b^\frR,b^\frL)$ as the assignment to the $\Intro_B$ variables. It is straightforward that the value of this honest strategy $\strategy$ is the same as the value of the associated strategy $\strategy'$ for the original game. 
\begin{definition}[Honest strategies for the introspection game]\label{defn:honest_strat_for_Intro}
	Given an $N$-dimensional  strategy $\strategy=\{\cal{P}\}$ to the original game $\game$, one can construct the following \emph{honest} strategy $\strategy'=\{\cal{Q}\}$ to $\Introspect(\game)$ acting on $\complex^{\FF_2^k}\otimes\complex^N$: As the length functions of all the vertices in $\game$ are $\Lambda$ (both readable and unreadable), $\cal{P}^\mttx\colon \FF_2^{\Lambda}\times \FF_2^{\Lambda}\to M_{N}(\complex)$. For $\Intro_A$ and $\Intro_B$, their readable length is $r+\Lambda$ and unreadable length is $\Lambda$, so $\cal{Q}^{\Intro_\cdot}\colon \FF_2^r\times \FF_2^{\Lambda}\times \FF_2^{\Lambda}\to \textrm{End}(\complex^{\FF_2^k})\otimes M_{N}(\complex)$. Recall the notation $\mathscr{F}^\PZm_z\in \textrm{End}(\complex^{\FF_2^k})$ for the orthogonal projection on the subspace spanned by the indicator ${\bf 1}_z$ in $\complex^{\FF_2^k}$ (Definition \ref{defn:F^Z_v}). Then, for every $\mttx\in \FF_2^r, a^\frR,a^\frL\in \FF_2^\Lambda$, let 
	\begin{equation}\label{eq:defn_honest_strategy1}
	\cal{Q}^{\Intro_A}_{\mttx,a^\frR,a^\frL}=\sum_{z\in \FF_2^k\colon \frS^A(z)=\mttx}\mathscr{F}^{\PZm}_z\otimes \cal{P}^\mttx_{a^\frR,a^\frL}\ ,
	\end{equation}
	and similarly for every $\mtty\in \FF_2^r, b^\frR,b^\frL\in \FF_2^\Lambda$, let
	\begin{equation}\label{eq:defn_honest_strategy2}
	\cal{Q}^{\Intro_B}_{\mtty,b^\frR,b^\frL}=\sum_{z\in \FF_2^k\colon \frS^B(z)=\mtty}\mathscr{F}^{\PZm}_z\otimes \cal{P}^\mtty_{b^\frR,b^\frL}\ .
	\end{equation}
\end{definition}
\begin{remark}\label{rem:structure_of_F^PZm_tensor_something_matrices}
	Note that the matrix $\mathscr{F}^\PZm_z$ has a single $1$ on the diagonal at the position $(z,z)$ (the matrix's rows and columns are parameterized by $\FF_2^k$) and $0$ everywhere else --- this matrix is often denoted by $e_{zz}$ or ${\bf 1}_{zz}$. Hence, for every collection of same sized square matrices $\mathscr{A}^z$, the matrix $\sum_z \mathscr{F}^\PZm_z\otimes\mathscr{A}^z$ is a block diagonal matrix with the $\mathscr{A}^z$'s on the diagonal. In particular, if $\mathscr{A}^z$ are diagonal then also $\sum_z \mathscr{F}^\PZm_z\otimes\mathscr{A}^z$ is diagonal, and similarly if $\mathscr{A}^z$ are signed permutation matrices then also $\sum_z \mathscr{F}^\PZm_z\otimes\mathscr{A}^z$ is a signed permutation matrix.
\end{remark}
\begin{claim}[Completeness and soundness of honest strategies for the introspection game]\label{claim:complete_sound_honest_intro}
	Let $\strategy=\{\cal{P}\}$ be a strategy for $\game$ and $\strategy'=\{\cal{Q}\}$ the honest strategy for $\Introspect(\game)$ associated with $\strategy$ (as defined in \eqref{eq:defn_honest_strategy1} and \eqref{eq:defn_honest_strategy2}). Then,
	\begin{enumerate}
		\item $\val(\game,\strategy)=\val(\Introspect(\game),\strategy')$;
		\item $\strategy$ being $\ZPC$ implies $\strategy'$ is $\ZPC$.
	\end{enumerate}
\end{claim}
\begin{proof}
	For item 1,  jointly sampling $((\mttx,a^\frR,a^\frL)(\mtty,b^\frR,b^\frL))\sim (\cal{Q}^{\Intro_A},\cal{Q}^{\Intro_B})$ (Definition \ref{defn:joint_measurement})   gives the same distribution on six-tuples as first sampling $\mttx\mtty\sim\mu$ (as was defined in \eqref{eq:pushforward_of_uniform_distribution}) and then jointly sampling $((a^\frR,a^\frL),(b^\frR,b^\frL))\sim (\cal{P}^\mttx,\cal{P}^\mtty)$. This means that indeed the value of the honest strategy against the introspection game is the same as that of the original strategy against the original game.
	
	For item 2, we need to view both $\strategy$ and $\strategy'$ in their observable forms, which we denote by $\cal{U}$ and $\cal{V}$ respectively. As $\cal{P}^\mttx$ is a PVM for every $\mttx\in \FF_2^r$, the marginalization (i.e., restriction, cf.\ Definition \ref{defn:Data_proccessed_PVM}) of $\cal{Q}$ to the $\sQue^A$-variables  satisfies
	\[
	\cal{Q}^{\mathsf{Que}^A}_\mttx=\sum_{a^\frR,a^\frL\in \FF_2^\Lambda} \cal{Q}^{\Intro_A}_{\mttx,a^\frR,a^\frL}=_{\eqref{eq:defn_honest_strategy1}}\sum_{z\in \FF_2^k\colon \frS^A(z)=\mttx}\mathscr{F}^{\PZm}_z\otimes \Id = \mathscr{F}^\PZm_{[\frS^A(\cdot)=\mttx]}\ ,
	\]
	and similarly $\cal{Q}^{\sQue^B}_\mtty=\mathscr{F}^\PZm_{[\frS^B(\cdot)=\mtty]}$. As $\mathscr{F}^\PZm$ is a diagonal PVM, it remains diagonal under data processing, and thus the inverse Fourier transformed (Definition \ref{defn:Fourier_transform_reps}) representations $\cal{V}^{\sQue^A}$ and $\cal{V}^{\sQue^B}$  are also diagonal --- this shows that the observables associated with the readable variables $\sQue^A$ and $\sQue^B$ are indeed $Z$-aligned and consist of signed permutations.
	
	When marginalizing $\cal{Q}$ to the $\sAns^{A}$ variables we get
	\[
	\cal{Q}^{\sAns^A}_{a^\frR,a^\frL}=\sum_{z\in \FF_2^k}\mathscr{F}^\PZm_z\otimes \cal{P}^{\frS^A(z)}_{a^\frR,a^\frL} \ .
	\]
	As described in Remark \ref{rem:structure_of_F^PZm_tensor_something_matrices}, these are block diagonal matrices whose $zz$-block (for $z\in \FF_2^k$) contains the PVM $\cal{P}^{\frS^A(z)}$. As the inverse Fourier transform for block diagonal matrices works block by block, we deduce that the representation $\cal{V}^{\sAns^A}$ satisfies 
	\begin{equation}\label{eq:honest_strategy_observable_form}
	\forall \alpha^\frR,\alpha^\frL\in \FF_2^\Lambda\ \colon \ \ \cal{V}(\alpha^\frR,\alpha^\frL)=\sum_{z\in \FF_2^k} \mathscr{F}^\PZm_z\otimes \cal{U}^{\frS^A(z)}(\alpha^\frR,\alpha^\frL).    
	\end{equation}
	In particular, if $\cal{U}$ consists of only signed permutation matrices, then so does $\cal{V}$, and similarly if the marginalization to the readable variables is diagonal for $\cal{U}$, then it is diagonal for $\cal{V}$. We can thus deduce that $\strategy$ being a $Z$-aligned permutation strategy implies $\strategy'$ is such. 
	
	We are left to prove that $\strategy$ commuting along edges implies $\strategy'$ is also commuting along edges. By  \eqref{eq:defn_honest_strategy1} and \eqref{eq:defn_honest_strategy2} we have
	\begin{align}
	\cal{Q}^{\Intro^A}_{\mttx,a^\frR,a^\frL}\cal{Q}^{\Intro^B}_{\mtty,b^\frR,b^\frL}=\left(\sum_{z\in \FF_2^k\colon \frS^A(z)=\mttx}\mathscr{F}^{\PZm}_z\otimes \cal{P}^\mttx_{a^\frR,a^\frL}\right)\left(\sum_{z'\in \FF_2^k\colon \frS^B(z')=\mtty}\mathscr{F}^{\PZm}_{z'}\otimes \cal{P}^\mtty_{b^\frR,b^\frL}\right)\ .\label{eq:product_along_edge_honest_strategy}
	\end{align}
	As $\mathscr{F}^\PZm$ is projective, when one distributes the above product, the only summands that  are potentially non-zero are those indexed by $z\in \FF_2^k$ for which both $\frS^A(z)=\mttx$ and $\frS^B(z)=\mtty$; in particular, this product is zero if $\mttx\mtty$ is not an edge in the original game $\game$. This is true for the reversed product $\cal{Q}^{\Intro^B}_{\mtty,b^\frR,b^\frL}\cal{Q}^{\Intro^A}_{\mttx,a^\frR,a^\frL}$, and thus if $\mttx\mtty$ is not an edge, then $\cal{Q}^{\Intro^B}_{\mtty,b^\frR,b^\frL}$ and $\cal{Q}^{\Intro^A}_{\mttx,a^\frR,a^\frL}$ commute (as their product in both orders is equal to zero). 
	In case $\mttx\mtty$ is an edge in the original game, the product in \eqref{eq:product_along_edge_honest_strategy} is equal to 
	\begin{align*}
	\sum_{\substack{z\in \FF_2^k\\\frS(z)=\mttx\mtty}}\mathscr{F}^\PZm_z\otimes (\cal{P}^\mttx_{a^\frR,a^\frL}\cal{P}^\mtty_{b^\frR,b^\frL})&=\sum_{\substack{z\in \FF_2^k\\\frS(z)=\mttx\mtty}}\mathscr{F}^\PZm_z\otimes (\cal{P}^\mtty_{b^\frR,b^\frL}\cal{P}^\mttx_{a^\frR,a^\frL})=\cal{Q}^{\Intro^B}_{\mtty,b^\frR,b^\frL}\cal{Q}^{\Intro^A}_{\mttx,a^\frR,a^\frL}\ ,
	\end{align*}
	where the first equation is due to $\strategy=\{\cal{P}\}$ being commuting along edges, and the second equation is, again, from the projectivity of $\mathscr{F}^\PZm$. All in all, $\strategy'=\{\cal{Q}\}$ commutes along edges, as needed.
\end{proof}

Although, given that the original game has a perfect $\ZPC$-strategy, one can extract a perfect honest $\ZPC$-strategy for $\Introspect(\game)$ (as desecribed above), there are many perfect strategies for $\Introspect(\game)$ that are \textbf{not honest}. For example, if there is any edge $\mathtt{xy}$ and answer $(a^\frR,a^\frL,b^\frR,b^\frL)$  for it which is accepted by the decision predicate of $\game$, then a strategy for $\Introspect(\game)$ can always assign the values  $\mttx$ and $\mtty$ to the $\sQue$ variables, and assign the accepting answer $(a^\frR,a^\frL,b^\frR,b^\frL)$ to the $\sAns$ variables. A further dismotivating observation is the following: even if $\strategy$ indeed samples a string $z\in \FF_2^k$ as a random seed and uses it to calculate $\mttx$ and $\mtty$ appropriately --- namely samples an edge in $\game$ according to the correct distribution ---   there is no guarantee that the observables it associates with $\sAns^{A,\cdot,\cdot}$ depend only on $\mttx$ and disregard $\mtty$. This means that the strategy can choose for every edge $\mttx\mtty$ a fixed accepting answer $(a^\frR,a^\frL,b^\frR,b^\frL)$ and provide it given that $\frS(z)=\mttx\mtty$. In plain words, the fact that the strategy sampled its own edge is the same as for the provers to be able to share their questions before providing their answers (in the dramatized version, Remark \ref{rem:dramatization_non_local_game}) --- which usually collapses everything to a single prover interactive proof.

Therefore,  as the above two non-honest perfect strategies suggest, $\Introspect(\game)$ is not very useful on its own. We amend this by taking the sum of  $\Introspect(\game)$  and the Pauli basis game $\PauliBasis_k$ (defined in Section \ref{sec:gen_pauli_basis}) and augment it by connecting the total $\PZm$-measurement (vertex $\Pauli_\PZm$) to the $\sQue$ variables so that $\strategy$ is forced to sample an edge according to the suitable distribution induced by $\frS$. Then, we are going to use the $\PXm$-measurements (vertex $\Pauli_\PXm$) of $\PauliBasis_k$ to ensure that the observables $\strategy$ associates to  $\sAns^{A,\cdot,\cdot}$ depend only on $\mttx$ and that the observables $\strategy$ associates to  $\sAns^{B,\cdot,\cdot}$ depend only on $\mtty$ --- which forces any almost perfect strategy for this augmentation to be close to an honest strategy for the introspective game. This second amendment uses the fact that ``depending only on $\mttx$'' is the same as providing the same answer for any two seeds $z_1,z_2$ such that $\frS^A(z_1)=\frS^A(z_2)=\mttx$, and there is  an $\PXm$-Pauli matrix that moves from the seed $z_1$ to the seed $z_2$ --- namely, this independence boils down to certain commutation relations with $\PXm$-Pauli matrices.

\subsection{Motivational interlude --- Question Reduction in the linear sampler case}\label{sec:baby_question_reduction}

\textbf{Note,} this section provides a simpler version of the final (combinatorial) transformation of question reduction. The full version is described in Section \ref{sec:augmentation_in_the_CLM_case}.
\\

Let $\game$ be a tailored game with the properties fixed in the beginning of the section, namely, its vertex set is $\FF_2^r$, the distribution on edges is induced by the pushforward of the uniform distribution on $\FF_2^k$ through $\frS\colon \FF_2^k\to \FF_2^r\times \FF_2^r$, and its length functions are constant and equal to $\Lambda$. In addition, assume that $\frS$  is \textbf{linear}.
As $\frS^A,\frS^B\colon \FF_2^k\to \FF_2^r$ are linear, we think about them as $r\times k$ matrices (with respect to the standard basis), and we assume given two additional  $k\times k$ matrices $(\frS^A)^\perp,(\frS^B)^\perp$ (also thought of as linear operators on $\FF_2^k$)  whose rows span the respective kernels $\ker\frS^A,\ker\frS^B\leq \FF_2^k$.  
\\

\begin{figure}[!htbp]
	\centering
	\begin{gamespec}
		\setlength{\tabcolsep}{1em}
		Let $\game$ be a tailored game with vertex set $\F_2^r$ and whose distribution on edges is the pushforward of the uniform distribution on $\F_2^k$ through a \textbf{linear map} $\frS=(\frS^A,\frS^B)\colon \FF_2^k\to \FF_2^r\times \FF_2^r$; let $(\frS^A)^\perp,(\frS^B)^\perp$ be $k\times k$ matrices whose rows span $\ker(\frS^A)$ and $\ker(\frS^B)$. In addition, the game $\game$ is assumed to have  readable and unreadable answer lengths equal to some constant $\Lambda$.
		
		\vspace{1em}
		\begin{tabularx}{\textwidth}{ l    X l l }
			\toprule
			Sub-Structure & Question &   Readable answer & Unreadable answer   \\
			\midrule
			$\PauliBasis_k$ &  $\Pauli_\PZm$ &  &  $z\in \FF_2^k$\\
			& $\Pauli_\PXm$ & & $\chi\in \FF_2^k$\\
			& See Figure~\ref{fig:genpauli-summary} for rest& &\\
			$\Introspect(\game)$ & $\Intro_A$ & $(\mttx,a^\frR)\in \FF_2^r\times \FF_2^\Lambda$ &  $a^\frL\in \FF_2^\Lambda$  \\
			Sampling apparatus   &$\Sample_A$  & $(z_{sam},a^\frR_{sam})\in \FF_2^k\times \FF_2^\Lambda $&
			$a^\frL_{sam}\in \FF_2^{\Lambda}$\\
			Hiding apparatus& $\Read_A$   &  
			$(\mttx_{read},a^\frR_{read})\in \FF_2^r\times \FF_2^\Lambda$  & $(\nu_{read},a^\frL_{read})\in \FF_2^k\times\FF_2^\Lambda$\\
			& $\Hide A$  &
			& $\nu\in \FF_2^k$  \\
			\bottomrule
		\end{tabularx}
		
		\vspace{1em}
		
		The following tests are performed when the corresponding augmented edge is sampled --- the checks for $B$ are similar:
		
		\begin{enumerate}
			\setlength\itemsep{1pt}
			\item  $\Pauli_\PZm-\Sample_A$: Check that $z=z_{sam}$.
			\item $\Intro_A-\Sample_A$: Check that $\mttx=\frS^A(z_{sam})$, $a^\frR=a^\frR_{sam}$ and $a^\frL=a^\frL_{sam}$.
			\item $\Intro_A-\Read_A$: Check that $\mttx=\mttx_{read}$, $a^\frR=a^\frR_{read}$ and $a^\frL=a^\frL_{read}$.
			\item $\Hide A-\Read_A$: Check that $\nu_{read}=\nu$.
			\item $\Hide A-\Pauli_{\PXm}$: Check that $\nu=(\frS^A)^\perp(\chi)$.
		\end{enumerate}
	\end{gamespec}
	\caption{Questions and answers in the game $\frak{Baby}(\game)$. Since the game is an augmentation of the sum of $\PauliBasis_k$ and $\Introspect(\game)$ we only list new questions and answers, and additional tests, and refer to Figure~\ref{fig:genpauli-summary} and Figure~\ref{fig:introspect-summary} for questions and answers of the latter.  Note that all the augmented checks are linear and independent of the values associated to readable variables --- this will not be the case in the general question reduction transformation described in Section \ref{sec:augmentation_in_the_CLM_case}.
	}
	\label{fig:babqr-summary}
\end{figure}

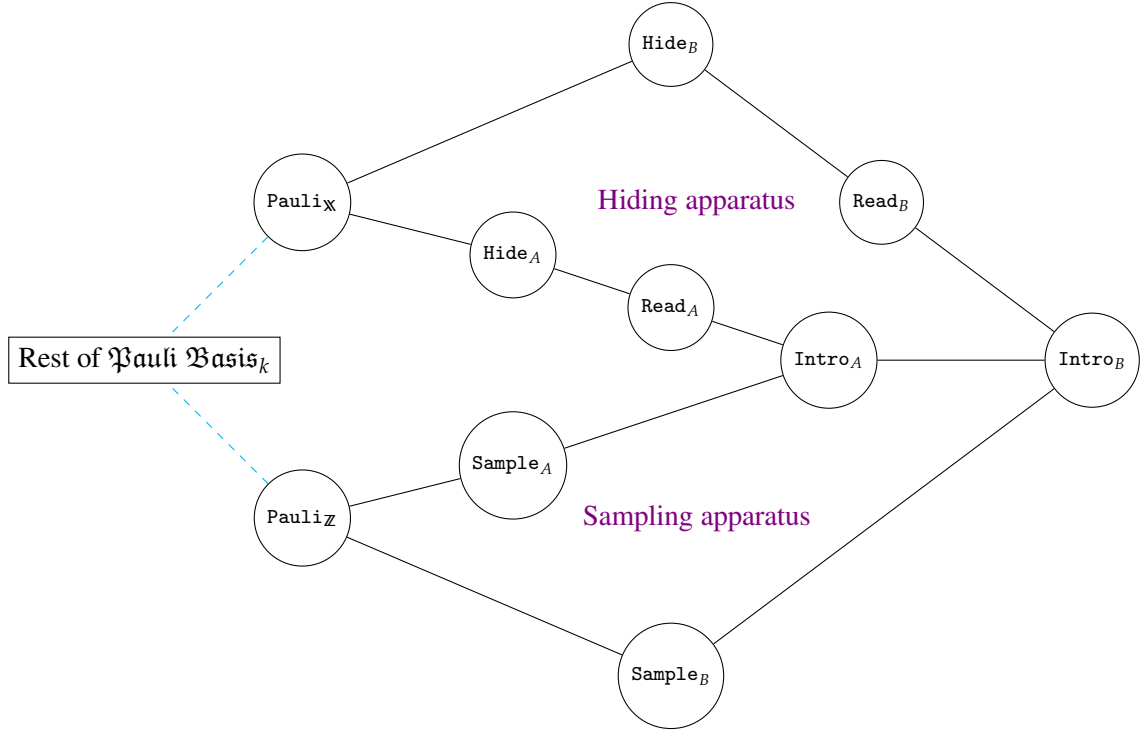
\begin{figure}[h]
	\centering
	\begin{tikzpicture}[scale=0.7]
	\node[draw, color=black, shape=circle] (PX) at (0,3) {\scriptsize $\Pauli_\PXm$}; 
	\node[draw, color=black, shape=circle] (PZ) at (0,-3) {\scriptsize $\Pauli_\PZm$}; 
	
	\node[draw, color=black] (PB) at (-3,0) {Rest of $\PauliBasis_k$}; 
	\node[draw, color=black, shape=circle] (IntA) at (10,0) {\scriptsize $\Intro_A$}; 
	
	\node[color=violet] (Sampling) at (7.5,-3) {Sampling apparatus}; 
	\node[color=violet] (Hiding) at (7.5,+3) {Hiding apparatus}; 
	
	\node[draw, color=black, shape=circle] (IntB) at (15,0) {\scriptsize $\Intro_B$}; 
	
	\node[draw, color=black, shape=circle] (SamA) at (4,-2) {\scriptsize $\Sample_A$};

	\node[draw, color=black, shape=circle] (SamB) at (7,-6) {\scriptsize $\Sample_B$};
	
	\node[draw, color=black, shape=circle] (ReadA) at (7,1) {\scriptsize $\Read_A$};

	\node[draw, color=black, shape=circle] (ReadB) at (11,3) {\scriptsize $\Read_B$};
	
	\node[draw, color=black, shape=circle] (HidA) at (4,2) {\scriptsize $\Hide A$};

	\node[draw, color=black, shape=circle] (HidB) at (7,6) {\scriptsize $\Hide B$};

	\draw[cyan, -, dashed] (PX)--(PB);
	\draw[cyan, -, dashed] (PZ)--(PB);
	\draw[black, -, solid] (PZ)--(SamA);
	\draw[black, -, solid] (PZ)--(SamB);
	\draw[black, -, solid] (IntA)--(IntB);
	\draw[black, -, solid] (SamA)--(IntA);
	\draw[black, -, solid] (SamB)--(IntB);
	\draw[black, -, solid] (PX)--(HidA);
	\draw[black, -, solid] (PX)--(HidB);
	\draw[black, -, solid] (HidA)--(ReadA);
	\draw[black, -, solid] (HidB)--(ReadB);
	\draw[black, -, solid] (IntA)--(ReadA);
	\draw[black, -, solid] (IntB)--(ReadB);

	\end{tikzpicture}
	\caption{The underlying graph of $\frak{Baby}(\game)$, where most of the embedded Pauli basis game is hidden.
	}
	\label{fig:Bad_augmentation}
\end{figure}

\textbf{The baby question reduction transformation
	$\frak{Baby}(\game)=\frak{Baby}(\game,k,\mathscr{B})$} (see  Figure~\ref{fig:babqr-summary} for a summary): First, as hinted in the notation, this transformation depends on three inputs --- a positive integer $k$, a subset $\mathscr{B}$ of $n$ vectors in $\FF_2^k$ which induces an $[n,k,d]$-code as was defined Section \ref{subsec:error_correctoin_prelims}, and a game $\game$ with the properties fixed in the previous paragraph.
With these inputs, the Pauli basis game (Section \ref{sec:Pauli_basis_definition}) $\PauliBasis_k=\PauliBasis_k(\mathscr{B})$  and the introspection game (Definition \ref{defn:Introspection}) $\Introspect(\game)$  can be defined. The baby question reduction game  $\frak{Baby}(\game)$ is an augmentation (Definition \ref{defn:augmentation_of_a_game}) of  the sum (Definition \ref{defn:sum_of_games}) of $\PauliBasis_k$ and $\Introspect(\game)$; the augmentation consists of two apparatuses:
\begin{enumerate}
	\item A \textbf{Sampling apparatus} which connects the   introspection game vertices to the total $Z$-measurement of the Pauli basis game (i.e., the vertex $\Pauli_\PZm$). The goal of this apparatus is twofold --- first, to verify that the ``questions'' part of the  players' answers when the copy of $\Introspect(\game)$ is played is distributed according to the question distribution of $\game$; second, to verify that the observables associated with the ``answers`` part of the players' answers in $\Introspect(\game)$ commute with the total $Z$-measurement. 
	\item  A \textbf{Hiding apparatus} which connects the   introspection game vertices to the total $X$-measurement of the Pauli basis game (i.e., the vertex $\Pauli_\PXm$). The goal of this apparatus is to verify that the ``answers'' part of the players' answers in $\Introspect(\game)$ commute with certain $X$-measurements (though not the total one).
\end{enumerate}
In the sampling apparatus, two additional vertices $\Sample_A$, $\Sample_B$ are added, and they are connected as follows
\[
\Intro_A-\Sample_A-\Pauli_\PZm-\Sample_B-\Intro_B\ .
\]
In the hiding apparatus, $4$ additional vertices are added 
\[
\Read_A\ ,\  \Read_B\ ,\  \Hide A\ ,\  \Hide B\ ,
\] 
and they are connected as follows
\[
\Intro_A-\Read_A-\Hide A-\Pauli_\PZm-\Hide B-\Read_B-\Intro_B\ .
\]
For a graphical view of the underlying graph of $\frak{Baby}(\game)$, see Figure \ref{fig:Bad_augmentation}.
\\

\textbf{Question distribution of the baby question reduced game}:
With probability $\nicefrac{1}{4}$ do one of the following:
\begin{itemize}
	\item  Sample an edge from  $\PauliBasis_k$ according to the appropriate distribution therein.
	\item Sample the single edge $\Intro_A-\Intro_B$ from $\Introspect(\game)$.
	\item Sample a uniformly random edge from the Sampling apparatus.
	\item Sample a uniformly random edge from the Hiding apparatus.
\end{itemize}

\textbf{Lengths and formal generating sets for the augmented vertices of baby question reduction}: 

\emph{Sampling apparatus} --- The readable length of $\Sample_A$ (and $\Sample_B$) is $k+\Lambda$, and its unreadable length is $\Lambda$. We associate with it the formal generators
\[\begin{split}
\mathsf{SamZ}^A=\{\mathsf{SamZ}^{A,i}&\mid 1\leq i\leq k\}\;,\\
\mathsf{SamAns}^{A,\frR}=\{\mathsf{SamAns}^{A,\frR,j}\mid  ,1\leq j\leq \Lambda\}\;,&\quad\mathsf{SamAns}^{A,\frL}=\{\mathsf{SamAns}^{A,\frL,j}\mid  ,1\leq j\leq \Lambda\}\;,\\
S^\frR_{\Sample_A}=\mathsf{SamZ}^{A}\sqcup \mathsf{SamAns}^{A,\frL}\;,&\quad 
S^\frL_{\Sample_A}=\mathsf{SamAns}^{A,\frL}\;.
\end{split}
\]
and similarly for $\Sample_B$. Namely, answers are formatted as $(z_{sam},a_{sam}^\frR,a_{sam}^\frL)$, where $z_{sam}\in \FF_2^k$, and $a_{sam}^\frR,a_{sam}^\frL\in \FF_2^\Lambda$.
\\

\emph{Hiding apparatus} ---
\begin{itemize}
	\item The readable length of $\Read_A$ (respectively $\Read_B$) is $r+\Lambda$, and its unreadable length is $k+\Lambda$. We associate with it the formal generators
	\[
	\begin{split}
	\mathsf{ReadQue}^{A}=\{\mathsf{ReadQue}^{A,i}\mid 1\leq i\leq r\}\;, & \quad \mathsf{ReadPerp}^{A}=\{\mathsf{ReadPerp}^{A,i}\mid 1\leq i\leq k\}\;,\\
	\mathsf{ReadAns}^{A,\frR}=\{\mathsf{ReadAns}^{A,\frR,j}\mid 1\leq j\leq \Lambda\}\;,&\quad \mathsf{ReadAns}^{A,\frL}=\{\mathsf{ReadAns}^{A,\frL,j}\mid 1\leq j\leq \Lambda\}\;,\\
	S^\frR_{\Read_A}=\mathsf{ReadQue}^{A}\sqcup\mathsf{ReadAns}^{A,\frR}\;,&\quad
	S^\frL_{\Read_A}=\mathsf{ReadPerp}^{A}\sqcup \mathsf{ReadAns}^{A,\frL}\;,
	\end{split}
	\]
	and similarly for $\Read_B$. Namely, answers are formatted as $(\mttx_{read},a_{read}^\frR,\nu_{read},a_{read}^\frL)$, where $\mttx_{read}\in \FF_2^r,a_{read}^\frR,a_{read}^\frL\in \FF_2^\Lambda$ and $\nu_{read}\in \FF_2^{k}$ (and for $B$, $(\mtty_{read},b_{read}^\frR,\mu_{read},b_{read}^\frL)$ in the appropriate spaces).
	\item The readable length of $\Hide A$ (respectively $\Hide B$) is $0$, and its unreadable length is $k$. We associate with it the formal generators
	\[
	\begin{split}
	S^\frL_{\Hide A}&=\mathsf{Perp}^{A}=\{\mathsf{Perp}^{A,i}\mid 1\leq i\leq k\}\;,
	\end{split}
	\]
	and similarly for $\Hide B$. Namely, the answer is formatted as $\nu\in \FF_2^k$ (respectively $\mu\in \FF_2^{k}$).
\end{itemize}

\textbf{The decision process for the augmented edges of the baby question reduced game}:\footnote{This is essentially the description of the controlled linear constraints function.}

\begin{itemize}
	\item \emph{Sampling apparatus} ---
	\begin{enumerate}[label=\textcolor{black}{(\arabic*)}, ref= (\arabic*)]
		\item In case $\Pauli_\PZm-\Sample_A$ (respectively $\Pauli_\PZm-\Sample_B$) is sampled, then check that 
		\[
		\forall 1\leq i\leq k\  \colon \ \ \gamma(\sP\sZ^i)=\gamma(\mathsf{SamZ}^{A,i})\;.
		\]
		In other words, if $z$ is the answer to $\Pauli_\PZm$, then it checks that $z=z_{sam}$.
		\item  In case $\Intro_A-\Sample_A$ (respectively $\Intro_B-\Sample_B$) is sampled, first check that 
		\[
		\forall  1\leq j\leq\Lambda \  \colon \ \ \gamma(\sAns^{A,\cdot,j})=\gamma(\mathsf{SamAns}^{A,\cdot,j})\;,
		\]
		and then check that 
		\[
		\forall 1\leq i\leq r\ \colon \ \ \gamma(\sQue^{A,i})=\sum_{j=1}^k \frS^A_{ij}\gamma(\mathsf{SamZ}^{A,j})\; .
		\]
		In other words, if $(z_{sam},a_{sam}^\frR,a_{sam}^\frL)$ is the answer to $\Sample_A$, and $(\mttx,a^\frR,a^\frL)$ is the answer to $\Intro_A$, then it checks that $\mttx=\frS^A(z_{sam}), a^\frR=a_{sam}^\frR$ and $a^\frL=a_{sam}^\frL$.
	\end{enumerate}
	
	\item \emph{Hiding apparatus} ---
	\begin{enumerate}[label=\textcolor{black}{(\arabic*)}, ref= (\arabic*)]
		\item  In case $\Intro_A-\Read_A$ (respectively $\Intro_B-\Read_B$) is sampled,  check that 
		\[
		\forall  1\leq i\leq r, 1\leq j\leq\Lambda  \  \colon \ \ \gamma(\sAns^{A,\cdot,j})=\gamma(\mathsf{ReadAns}^{A,\cdot,j})\;,\quad \gamma(\sQue^{A,i})=\gamma(\mathsf{ReadQue}^{A,i})\;.
		\]
		In other words, if $(\mttx_{read},a_{read}^\frR,\nu_{read},a_{read}^\frL)$ is the answer to $\Read_A$, and $(\mttx,a^\frR,a^\frL)$ is the answer to $\Intro_A$, check that $\mttx_{read}=\mttx,a^\frR_{read}=a^\frR$ and $a^\frL_{read}=a^\frL$.
		\item  In case $\Hide A-\Read_A$ (respectively $\Hide B-\Read_B$) is sampled,  check that 
		\[
		\forall  1\leq i\leq k \  \colon \ \ \gamma(\mathsf{Perp}^{A,i})=\gamma(\mathsf{ReadPerp}^{A,i})\;.
		\]
		In other words, if $(\mttx_{read},a_{read}^\frR,\nu_{read},a_{read}^\frL)$ is the answer to $\Read_A$, and $\nu$ is the answer to $\Hide A$, check that $\nu=\nu_{read}$.
		\item \label{clause:hide_vs_PauliX_in_Baby} In case $\Hide A-\Pauli_{\PXm}$ (respectively $\Hide B-\Pauli_{\PXm}$) is sampled,  check that 
		\[
		\forall  1\leq i\leq k \  \colon \ \ \gamma(\mathsf{Perp}^{A,i})=\sum_{j=1}^k(\frS^A)^\perp_{ij}\gamma(\mathsf{PX}^{j})\;,
		\]
		where $(\frS^A)^\perp$  was the matrix whose rows span $\ker\frS^A$. 
		In other words, if  $\nu$ is the answer to $\Hide A$ and $\chi$ is the answer to $\Pauli_{\PXm}$, check that $\nu=(\frS^A)^\perp (\chi)$.
	\end{enumerate}
\end{itemize}
\begin{remark}[Informal analysis of $\frak{Baby}(\game)$]
	Before proving rigorously that this game is complete and sound, let us  discuss the role of the various checks described above in forcing strategies to behave appropriately in this augmented version of $\PauliBasis_k\oplus\Introspect(\game)$. Similarly to the way the ultimate goal of all the checks in $\PauliBasis_k$ was for the observables at $\Pauli_{\PZm}$ and $\Pauli_{\PXm}$ to induce a (specific) representation of the Pauli group, the ultimate goal of all the above checks is to force the strategy to play \emph{honestly} (Definition \ref{defn:honest_strat_for_Intro}) when the copy of $\Introspect(\game)$ is played ---  namely, it samples a seed $z\in \FF_2^k$ uniformly at random, calculates $\frS(z)=\mttx\mtty$, calculates $(a^\frR,a^\frL)$ depending only on $\mttx$ and $(b^\frR,b^\frL)$ depending only on $\mtty$, and finally replies with $(\mttx,a^\frR,a^\frL,\mtty,b^\frR,b^\frL)$.
	
	This is achieved as follows: First, the copy of the Pauli basis game forces the answer $z$ in $\Pauli_\PZm$ to be uniformly distributed over $\FF_2^k$, using the PVM (in representation form) $\rho^\PZm\otimes \Id$  (and  $\rho^\PXm\otimes \Id$ for $\Pauli_\PXm$), where $\rho$ is the representation from Definition~\ref{defn:rho}. The checks $\Pauli_\PZm-\Sample_A$ and $\Intro_A-\Sample_A$ force $\mttx$ to be $\frS^A(z)$ and similarly $\Pauli_\PZm-\Sample_B$ and $\Intro_B-\Sample_B$ force $\mtty$ to be $\frS^B(z)$. Now, $\Intro_A-\Sample_A$ and $\Intro_A-\Read_A$ force $a^\cdot=a_{sam}^\cdot=a_{read}^\cdot$, and similarly $\Sample_B-\Intro_B-\Read_B$ force $b^\cdot=b_{sam}^\cdot=b_{read}^\cdot$. Furthermore, since they are mutually measured with the seed $z$ (in $\Sample_A$) and with $\nu_{read}$ (in $\Read_A$), they are forced to commute with the observables associated with them. 
	The checks $\Read_A-\Hide A$ and $\Hide A-\Pauli_\PXm$, force $\nu_{read}=\nu=(\frS^A)^\perp(\chi)$. Hence, the observables of $\sAns^{A,\cdot,\cdot}$ are forced to commute with $\PXm^{\otimes \alpha}\otimes \Id$  for every $\alpha\in \ker \frS^A$, and with  $\PZm^{\otimes v}\otimes \Id$ for every  $v\in \FF_2^k$. 
	The second commutation means that the mutual orthonormal eigenbasis for the observables of $\sAns^{A,\cdot,\cdot}$ is of the form $\{{\bf 1}_z\otimes u_{z,t}\mid z\in \FF_2^k, t\in \FF_2^{2\Lambda}\}$, and the first commutation means that they act the same on  ${\bf 1}_z\otimes u$ and $\PXm^{\alpha}\otimes \Id\cdot {\bf 1}_z\otimes u={\bf 1}_{z+\alpha}\otimes u$. This  means that the response $a^\frR,a^\frL$ is the same for every two $z$'s that differ by an element of $\ker\frS^A$, which implies that they depend only on their $\frS^A$-image, and this is what we wanted! 
	
	A reader may notice that to achieve the above goal, we could have dropped the $\Sample_\cdot$ and $\Hide \cdot$ vertices and applied a more direct check (simplifying the augmentation).  Though this is true, it will hinder the perfect completeness case, as we seek perfect $\ZPC$ strategies, in particular strategies that commute along edges, which is problematic without these ``buffer'' questions.
\end{remark}

\begin{theorem}\label{thm:bab-qr}
	Let $k\geq 2$ be an integer, $\mathscr{B}$ a tuple of $n$ vectors in $\FF_2^k$ that induce an $[n,k,d]$-code, and $\game$ a game with the properties fixed in the beginning of this subsection. Then, the baby question reduction game $\mathfrak{Baby}(k,\mathscr{B},\game)=\mathfrak{Baby}(\game)$ has the following properties:
	\begin{enumerate}[label=\textcolor{black}{(\arabic*)}, ref= (\arabic*)]
		\item \label{clause:completeness_of_baby}\emph{Perfect $\ZPC$ Completeness}: If $\game$ has a perfect $\ZPC$ strategy, then so does $\frak{Baby}(\game)$.
		\item \label{clause:soundness_of_baby}\emph{Soundness}: If $\frak{Baby}(\game)$ has a strategy with value $1-\eps$, then $\game$ has a strategy with value of at least $1-O(\sqrt{(1+\nicefrac{k^2}{d^2})\eps})$.\footnote{The $O$-notation is genuinely some universal constant that can be extracted from all the approximations we are doing.}
		\item \label{clause:entanglement_of_baby}\emph{Entanglement}: For every $\eps>0$, 
		\[
		\Ent(\frak{Baby}(\game),1-\eps)\geq 2^{k}\cdot\left(1-O\left({(1+\nicefrac{k^2}{d^2})\eps}\right)\right)\cdot\Ent\Big(\game,1-O\Big(\sqrt{(1+\nicefrac{k^2}{d^2})\eps}\Big)\Big).
		\]
	\end{enumerate}
\end{theorem}

\subsubsection*{Proof of perfect completeness \labelcref{clause:completeness_of_baby}}

Recall the notation $\cal{V}^{S'}$ for the restriction  to $\FF_2^{S'}$  of  $\cal{V}$  with outcomes in $\FF_2^{S'\sqcup S''}$ (Definition \ref{defn:Data_proccessed_PVM}). 

Assume $\game$ has a perfect $m$-dimensional $\ZPC$ strategy  $\strategy=\{\cal{U}\}$. Then, it induces an honest $\ZPC$ strategy $\strategy'=\{\cal{V}\}$ which is perfect for $\Introspect(\game)$ (Definition \ref{defn:honest_strat_for_Intro} and Claim \ref{claim:complete_sound_honest_intro}). 
Let us extend $\cal{V}$ to the other vertices so it becomes a perfect $\ZPC$ strategy for $\frak{Baby}(\game)$. First, let $\cal{V}^{\Pauli_\PXm}$ and $\cal{V}^{\Pauli_\PZm}$ be the appropriate restrictions of $\rho\otimes \Id_m$, where $\rho$ is the representation from Definition~\ref{defn:rho}, to the $X$ and $Z$ subgroups of $\WH_k$, namely
\[
\forall \alpha\in \FF_2^k\ \colon \ \ \cal{V}^{\Pauli_\PXm}(\alpha)=\rho^\PXm(\alpha)\otimes \Id_m=\PXm^{\otimes \alpha}\otimes \Id_m\quad,\quad \cal{V}^{\Pauli_\PZm}(\alpha)=\rho^\PZm(\alpha)\otimes\Id_m=\PZm^{\otimes \alpha}\otimes \Id_m\ .
\] 
By Claim \ref{claim:completeness_PB_k}, it can be extended to the rest of the $\PauliBasis_k$ vertices in a $\ZPC$ manner such that on the copy of $\PauliBasis_k$ in $\frak{Baby}(\game)$ it has value $1$.
As described in Corollary \ref{cor:linear_data_processed_PVM_is_ZPC_and_left_multiplication}, by calculating the inverse Fourier transform of the data processed $\mathscr{F}^\PZm_{[\frS^\cdot(\cdot)=\cdot]}$, and by denoting $\alpha\cdot \frS^\cdot$ for the multiplication from the left of (the row vector) $\alpha\in \FF_2^r$ with the matrix $\frS^\cdot\in M_{r\times k}(\FF_2)$,  we have
\[
\forall \alpha,\beta\in \FF_2^r\ \colon \ \ \cal{V}^{\sQue^A}(\alpha)=\PZm^{\otimes \alpha \cdot \frS^A}\otimes  \Id_{m}\ ,\ \cal{V}^{\sQue^B}(\beta)=\PZm^{\otimes \beta \cdot \frS^B}\otimes  \Id_{m}\ \in\  U(\complex^{\FF_2^k}\otimes \complex^m)\ .
\]
Note that in particular we have the following relationship through $\frS^\cdot$-evaluation (data processing, Definition \ref{defn:Data_proccessed_PVM}) in representation form
\[
\cal{V}^{\sQue^A}=\cal{V}^{\Pauli_\PZm}_{[\frS^A]}\quad,\quad \cal{V}^{\sQue^B}=\cal{V}^{\Pauli_\PZm}_{[\frS^B]}\ .
\]
As the rest of the checks in $\frak{Baby}(\game)$ are linear consistency checks, by Claim \ref{claim:perfect_3Lin_commuting_observables}, they are forcing us to choose the following PVMs (in representation form) to be the same ---
\begin{align*}
\cal{V}^{\sAns^A}= \cal{V}^{\mathsf{ReadAns}^A}= \cal{V}^{\mathsf{SamAns}^A}\;, &\quad \cal{V}^{\sAns^B}= \cal{V}^{\mathsf{ReadAns}^B}= \cal{V}^{\mathsf{SamAns}^B}\ ,\ \\
\cal{V}^{\Pauli_\PZm}=&\cal{V}^{\mathsf{SamZ}^A}=\cal{V}^{\mathsf{SamZ}^B}\ ,\\
\cal{V}^{\sQue^A}=\cal{V}^{\mathsf{Read}\sQue^{A}}=\cal{V}^{\Pauli_\PZm}_{[\frS^A]}\;, &
\quad
\cal{V}^{\sQue^B}=\cal{V}^{\mathsf{Read}\sQue^{B}}=\cal{V}^{\Pauli_\PZm}_{[\frS^B]}
,\\
\cal{V}^{\mathsf{Perp}^{A}}=\cal{V}^{\mathsf{ReadPerp}^{A}}=\cal{V}^{\Pauli_\PXm}_{[(\frS^A)^\perp]}\;, &\quad
\cal{V}^{\mathsf{Perp}^{B}}=\cal{V}^{\mathsf{ReadPerp}^{B}}=\cal{V}^{\Pauli_\PXm}_{[(\frS^B)^\perp]}\ .
\end{align*}
By choosing the above extension of $\cal{V}$, we are guaranteed that it passes all the augmented edges perfectly (both from the Sampling apparatus and the Hiding apparatus). We already described why $\cal{V}$ passes the copies of $\PauliBasis_k$ and $\Introspect(\game)$ perfectly, so, if it is well defined, then it is a perfect strategy for the game  $\frak{Baby}(\game)$.
Furthermore, it is straightforward to check that all the observables we chose are signed permutation matrices, and the readable ones are diagonal; hence, if this strategy is well defined, it is a perfect $Z$-aligned permutation strategy.

We are left to argue why $\cal{V}$ is well defined --- we chose restrictions of the PVMs at each vertex in a well defined manner, but it may be that these restricted PVMs do not amount to a single global one at the vertex, as they may be non-commuting. In addition, we need to check that $\cal{V}$ is commuting along edges. These are all quite straightforward checks (or, are corollaries of Claims \ref{claim:completeness_PB_k} and \ref{claim:complete_sound_honest_intro}), except for the $\Read_\cdot$-variables. --- as the images of $\cal{V}^{\mathsf{ReadPerp}^{\cdot}}$ are of the form $\PXm^\cdot\otimes \Id_m$, they may not commute with the images of $\cal{V}^{\mathsf{ReadQue}^{\cdot,}}$ which are of the form $\PZm^\cdot\otimes \Id_m$, and the images of  $\cal{V}^{\mathsf{ReadAns}^{\cdot}}$ which are of the form $\sum \mathscr{F}^\PZm_z \otimes \mathscr{A}^z$.  Let us demonstrate why they are commuting nonetheless --- we focus on $\Read_A$, but the proof for $\Read_B$ is almost identical. For every $\alpha\in \FF_2^r$ and $\beta\in \FF_2^k$, we have
\begin{equation}\label{eq:readque^A_and_readperp^A_observables}
\cal{V}^{{\mathsf{Read}\sQue}^A}(\alpha)=\PZm^{\otimes\alpha\cdot \frS^A}\otimes \Id_m\;,\quad \cal{V}^{\mathsf{ReadPerp}^A}(\beta)=\PXm^{\otimes \beta\cdot (\frS^A)^\perp}\otimes \Id_m\ ,
\end{equation}
and thus they commute if and only if $\langle \alpha\cdot \frS^A,\beta\cdot(\frS^A)^\perp\rangle=0$, where we think of both as row vectors.
By recalling the notation $*$ for transposition of matrices with coefficients in $\FF_2$ (Section \ref{sec:Pauli_gp}), and by the choice of $(\frS^A)^\perp$ having rows in $\ker\frS^A$, we have 
\[
\langle \alpha\cdot \frS^A,\beta\cdot(\frS^A)^\perp \rangle=\alpha \cdot \frS^A\cdot ((\frS^A)^\perp)^*\cdot \beta^*=0\ ,
\]
as $\frS^A\cdot ((\frS^A)^\perp)^*$ is the matrix whose columns are the $\frS^A$-evaluation of the rows of $(\frS^A)^\perp$. Now, for every pair $\alpha^\frR,\alpha^\frL\in \FF_2^\Lambda$,  by \eqref{eq:honest_strategy_observable_form} and the fact we chose $\cal{V}^{\sAns^A}=\cal{V}^{\mathsf{ReadAns}^A}$, we have
\[
\cal{V}^{\mathsf{ReadAns}^A}(\alpha^\frR,\alpha^\frL)=\sum_{z\in \FF_2^k}\mathscr{F}^\PZm_z\otimes \cal{U}^{\frS^A(z)}(\alpha^\frR,\alpha^\frL)\ .
\]
Hence, 
\begin{align*}
\cal{V}^{\mathsf{ReadPerp}^A}(\beta)\cdot\cal{V}^{\mathsf{ReadAns}^A}(\alpha^\frR,\alpha^\frL)\cdot\cal{V}^{\mathsf{ReadPerp}^A}(\beta)&=\sum_{z\in \FF_2^k}\left(\PXm^{\otimes \beta\cdot(\frS^A)^\perp}\cdot \mathscr{F}^\PZm_z\cdot \PXm^{\otimes \beta\cdot(\frS^A)^\perp}\right)\otimes \cal{U}^{\frS^A(z)}(\alpha^\frR,\alpha^\frL)\\
&=\sum_{z\in \FF_2^k} \mathscr{F}^\PZm_{z+\beta\cdot(\frS^A)^\perp}\otimes \cal{U}^{\frS^A(z)}(\alpha^\frR,\alpha^\frL)\\
&=\sum_{z\in \FF_2^k} \mathscr{F}^\PZm_{z+\beta\cdot(\frS^A)^\perp}\otimes \cal{U}^{\frS^A(z+\beta\cdot(\frS^A)^\perp)}(\alpha^\frR,\alpha^\frL)\\
&=\cal{V}^{\mathsf{ReadAns}^A}(\alpha^\frR,\alpha^\frL)\ ,
\end{align*}
where the first and last equations are by definition, the second equation is due to $\PXm^{\otimes\gamma}\mathscr{F}^\PZm_z\PXm^{\otimes \gamma}=\mathscr{F}^\PZm_{z+\gamma}$, and the third equation is since $\beta\cdot(\frS^A)^\perp$ is a linear combination of rows of $(\frS^A)^\perp$, which means it is in the kernel of $\frS^A$ and thus satisfies $\frS^A(z+\beta\cdot(\frS^A)^\perp)=\frS^A(z)$ for every $z\in \FF_2^k$.
\qed
\subsubsection*{Proof of soundness \labelcref{clause:soundness_of_baby} and entanglement lower bound \labelcref{clause:entanglement_of_baby}}

The proof idea is as follows: Given an almost perfect strategy of $\frak{Baby}(\game)$, we are going to perturb it so that  it passes perfectly all non-$\Introspect(\game)$ edges. We are then going to show that the resulting strategy is honest (Definition \ref{defn:honest_strat_for_Intro}) when restricted to the copy of $\Introspect(\game)$, and thus has the same value as some strategy of $\game$ (in a much smaller dimension). As we did not perturb the strategy too much, its value did not change too much, and we can deduce the soundness and entanglement lower bound claims.

Let us move to the formal proof.
Assume $\strategy=\{\cal{U}\}$ is an $N$-dimensional strategy in observable form for $\mathfrak{Baby}(\game)$ with value $1-\eps$. For notational simplicity, let $\eps'=(1+\nicefrac{k^2}{d^2})\eps$.
All $O$-notations in this proof are universal constants, and in particular are independent of $k,\mathscr{B},\game$ or the strategy $\strategy$ --- genuinely universal. We repeatedly use Claim \ref{claim:L1_closeness_of_unirep_implies_Linfty}, replacing previous bounds on expectations by the same bounds on the maxima (up to a constant factor that is absorbed into the $O$ and $\approx$ notations).

As the copy of $\PauliBasis_k$ is played with probability $\nicefrac{1}{4}$ when running $\mathfrak{Baby}(\game)$, the restriction of $\strategy$ to the vertices of $\PauliBasis_k$ passes it with value of at least $1-4\eps$. 
Hence, by Claim \ref{claim:almost_perfect_strategies_of_PB_k} and Fact \ref{fact:semi-stability_P_k}, there is a $\nicefrac{k^2\eps}{d^2}$-near bijection (Definition \ref{defn:near_bijections}) $\omega\colon \complex^N\to \complex^{\FF_2^k}\otimes \complex^m$ for which the corner POVM $\omega \cal{U}^{\Pauli_\PXm}\omega^*$ is $\nicefrac{k^2\eps}{d^2}$-close to $\rho^\PXm\otimes \Id_m$, and the corner POVM $\omega \cal{U}^{\Pauli_\PZm}\omega^*$ is $\nicefrac{k^2\eps}{d^2}$-close to $\rho^\PZm\otimes \Id_m$; namely, using Claim \ref{claim:L1_closeness_of_unirep_implies_Linfty}, we have 
\begin{equation}\label{eq:Pauli_close_to_Pauli_matrices_in_Baby}\begin{split}
\forall \alpha\in \FF_2^k\ \colon \ \ \left\|\omega\cal{U}^{\Pauli_\PXm}(\alpha)\omega^*-\PXm^{\otimes \alpha}\otimes \Id_m\right\|_{hs}^2\;,&\quad\left\|\omega\cal{U}^{\Pauli_\PZm}(\alpha)\omega^*-\PZm^{\otimes \alpha}\otimes \Id_m\right\|_{hs}^2\leq O(\nicefrac{k^2\eps}{d^2})\;,\\
&\textrm{and}\\
1-\tau(\omega^*\omega)\;,&\quad 1-\tau(\omega\omega^*)\leq O(\nicefrac{k^2\eps}{d^2})\;.
\end{split}\end{equation}
Let $\eps'=(1+\nicefrac{k^2}{d^2})\eps$. So, the above quantities are all $O(\eps')$. Moreover, by orthonormalization (Fact \ref{fact:orthogonalization}), the corner PVMs $\omega\cal{U}^{\sAns^A}\omega^*$ and $\omega\cal{U}^{\sAns^B}\omega^*$ are $O(\eps')$-close to genuine  representations that we denote by $\theta^A$ and $\theta^B$; namely, using Claim \ref{claim:L1_closeness_of_unirep_implies_Linfty}, we have
\begin{align}
\forall \alpha^\frR,\alpha^\frL\in \FF_2^\Lambda\  \colon \ \ \left\| \omega \left(\cal{U}^{\sAns^A}(\alpha^\frR,\alpha^\frL)\right)\omega^* -\theta^A(\alpha^\frR,\alpha^\frL)\right\|_{hs}^2\leq O(\eps')\ ,\label{eq:thetaA_close_to_AnsA_baby}\\ 
\forall \beta^\frR,\beta^\frL\in \FF_2^\Lambda\  \colon \ \ \left\| \omega \left(\cal{U}^{\sAns^B}(\beta^\frR,\beta^\frL)\right)\omega^* -\theta^B(\beta^\frR,\beta^\frL)\right\|_{hs}^2\leq O(\eps')\ .\label{eq:thetaB_close_to_AnsA_baby}
\end{align}
The following few calculations aim to show that $\theta^\cdot$ almost commutes with $\PZm^{\otimes z}\otimes \Id_m$ for every $z\in \FF_2^k$ and with $\PXm^{\otimes \alpha}\otimes \Id_m$ for every $\alpha\in \ker\frS^\cdot$.
As $\strategy$ has value $1-\eps$, and the augmented edges are sampled with probability of at least $\nicefrac{1}{24}$, we can deduce from  the definition of inconsistency (Definition \ref{defn:distance_inconsistency_POVMs}) and the equivalence of inconsistency and distance for projective measurements (item 1 of Proposition \ref{prop:properties_of_distance_and_inconsistency}) various results:
\begin{enumerate}
	\item By the comparison along $\Intro_A-\Sample_A$, we can deduce that $\cal{U}^{\sQue^A}\simeq _{O(\eps)} \cal{U}^{\mathsf{SamZ}^A}_{[\frS^A]}$ and $\cal{U}^{\sAns^A}\simeq_{O(\eps)}\cal{U}^{\mathsf{SamAns}^A}$; namely, using   Claim \ref{claim:L1_closeness_of_unirep_implies_Linfty}, Claim \ref{claim:affine_data_processing} and the notation $\alpha\cdot \frS^A$ for the product from the left of the row vector $\alpha\in \FF_2^r$ with the $r\times k$ matrix $\frS^A$,
	\begin{align}
	\forall \alpha\in \FF_2^r\ &\colon \ \ \|\cal{U}^{\sQue^A}(\alpha)-\cal{U}^{\mathsf{SamZ}^A}(\alpha\cdot \frS^A)\|_{hs}^2\leq O(\eps)\ ,\label{eq:Que_close_to_SamZ_Baby}\\
	\forall \alpha^\frR,\alpha^\frL\in \FF_2^\Lambda\ &\colon \ \ \|\cal{U}^{\sAns^A}(\alpha^\frR,\alpha^\frL)-\cal{U}^{\mathsf{SamAns}^A}(\alpha^\frR,\alpha^\frL)\|_{hs}^2\leq O(\eps)\ .\label{eq:Ans_close_to_SamAns_in_Baby}
	\end{align}
	\item By the comparison along $\Pauli_\PZm-\Sample_A$, we can deduce that $\cal{U}^{\Pauli_\PZm}\simeq_{O(\eps)} \cal{U}^{\mathsf{SamZ}^A}$, and using Claim \ref{claim:L1_closeness_of_unirep_implies_Linfty} this implies
	\begin{align}
	\forall \alpha\in \FF_2^k\ \colon \ \ \|\cal{U}^{\Pauli_\PZm}(\alpha)-\cal{U}^{\mathsf{SamZ}^A}(\alpha)\|_{hs}^2\leq O(\eps)\ .\label{eq:Pauli_close_to_SamZ_in_Baby}
	\end{align}
	\item By the comparison $\Intro_A-\Read_A$, we can deduce that $\cal{U}^{\sQue^A}\simeq_{O(\eps)}\cal{U}^{\mathsf{ReadQue}^A}$ and $\cal{U}^{\sAns^A}\simeq_{O(\eps)}\cal{U}^{\mathsf{ReadAns}^A}$, and using Claim \ref{claim:L1_closeness_of_unirep_implies_Linfty},
	\begin{align}
	\forall \alpha\in \FF_2^r\ &\colon \ \ \|\cal{U}^{\sQue^A}(\alpha)-\cal{U}^{\mathsf{ReadQue}^A}(\alpha)\|_{hs}^2\leq O(\eps)\ ,\\
	\forall \alpha^\frR,\alpha^\frL\in \FF_2^\Lambda\ &\colon \ \ \|\cal{U}^{\sAns^A}(\alpha^\frR,\alpha^\frL)-\cal{U}^{\mathsf{ReadAns}^A}(\alpha^\frR,\alpha^\frL)\|_{hs}^2\leq O(\eps)\ .\label{eq:Ans_close_to_ReadAnd_Baby}
	\end{align}
	\item  By the comparison $\Hide A-\Read_A$, we can deduce that $\cal{U}^{\Hide A}\simeq_{O(\eps)}\cal{U}^{\mathsf{ReadPerp}^A}$, and using Claim \ref{claim:L1_closeness_of_unirep_implies_Linfty},
	\begin{align}
	\forall \alpha\in \FF_2^k\ &\colon \ \ \|\cal{U}^{\Hide A}(\alpha)-\cal{U}^{\mathsf{ReadPerp}^A}(\alpha)\|_{hs}^2\leq O(\eps)\ .\label{eq:Perp_close_to_ReadPerp_Baby}
	\end{align}
	\item  By the comparison $\Hide A-\Pauli_\PXm$, we can deduce that $\cal{U}^{\Hide A}\simeq_{O(\eps)} \cal{U}^{\Pauli_\PXm}_{[(\frS^A)^\perp]}$, and using Claim \ref{claim:L1_closeness_of_unirep_implies_Linfty} and Claim \ref{claim:affine_data_processing}, 
	\begin{align}
	\forall \alpha\in \FF_2^k\ &\colon \ \ \left\|\cal{U}^{\Hide A}(\alpha)-\cal{U}^{\mathsf{Pauli}_{\PXm}}\left(\alpha\cdot (\frS^A)^\perp\right)\right\|_{hs}^2\leq O(\eps)\ .\label{eq:perp_close_to_PauliX_Baby}
	\end{align}
\end{enumerate}
Thus, using the notation $\square\approx_\eps\heartsuit$   whenever $\|\square-\heartsuit\|_{hs}^2\leq \eps$ (similar to the distance notation, Definition \ref{defn:distance_inconsistency_POVMs}), we have that 
\begin{equation}\label{eq:Ans_and_PauliZ_almost_commute_baby}
\begin{split}
\forall \alpha^\frR,\alpha^\frL\in \FF_2^\Lambda, z\in \FF_2^k\ \colon \ \ \cal{U}^{\sAns^A}(\alpha^\frR,\alpha^\frL)\cal{U}^{\Pauli_\PZm}(z)&\approx_{O(\eps)} \cal{U}^{\mathsf{SamAns}^A}(\alpha^\frR,\alpha^\frL)\cal{U}^{\mathsf{SamZ}}(z)\\
&=\cal{U}^{\mathsf{SamZ}}(z)\cal{U}^{\mathsf{SamAns}^A}(\alpha^\frR,\alpha^\frL)\\
&\approx_{O(\eps)} \cal{U}^{\Pauli_\PZm}(z)\cal{U}^{\sAns^A}(\alpha^\frR,\alpha^\frL)\ ,
\end{split}
\end{equation}
where both approximations use \eqref{eq:Ans_close_to_SamAns_in_Baby} and \eqref{eq:Pauli_close_to_SamZ_in_Baby}. Therefore, 
\begin{equation}
\begin{split}
\forall \alpha^\frR,\alpha^\frL\in \FF_2^\Lambda, z\in \FF_2^k\ \colon \ \ \theta^A(\alpha^\frR,\alpha^\frL)\cdot\PZm^{\otimes z}\otimes \Id_m&\approx_{O(\eps')} \omega \cal{U}^{\sAns^A}(\alpha^\frR,\alpha^\frL)\omega^*\omega\cal{U}^{\Pauli_\PZm}(z)\omega^*\\
&\approx_{O(\eps)}  \omega \cal{U}^{\sAns^A}(\alpha^\frR,\alpha^\frL)\cal{U}^{\Pauli_\PZm}(z)\omega^*\\
&\approx_{O(\eps)}  \omega \cal{U}^{\Pauli_\PZm}(z)\cal{U}^{\sAns^A}(\alpha^\frR,\alpha^\frL)\omega^*\\
&\approx_{O(\eps)}  \omega \cal{U}^{\Pauli_\PZm}(z)\omega^*\omega\cal{U}^{\sAns^A}(\alpha^\frR,\alpha^\frL)\omega^*\\
&\approx_{O(\eps')} \PZm^{\otimes z}\otimes \Id_m\cdot\theta^A(\alpha^\frR,\alpha^\frL)\ ,
\end{split}    
\end{equation}
where the first and last approximations are by \eqref{eq:thetaA_close_to_AnsA_baby} and \eqref{eq:Pauli_close_to_Pauli_matrices_in_Baby},  and the middle one is by \eqref{eq:Ans_and_PauliZ_almost_commute_baby}. For the second and fourth approximations, note that $1-\tau(\omega^*\omega),1-\tau(\omega\omega^*)\leq O(\eps)$ from \eqref{eq:Pauli_close_to_Pauli_matrices_in_Baby} and that $\|\square\heartsuit\|_{hs}\leq \|\square\|_{op}\|\heartsuit\|_{hs}$ for square matrices, therefore using Claim \ref{claim:hs_norm_of_corner_is_close_to_original}, we have
\begin{equation}\label{eq:argument_for_eliminating_omegaomegastar}
\begin{split}
\|\omega \cal{U}_{\sAns^A}(\alpha^\frR,\alpha^\frL)(\Id-\omega^*\omega)\cal{U}_{\Pauli_\PZm}(z)\omega^*\|_{hs}^2
&\leq \|\cal{U}_{\sAns^A}(\alpha^\frR,\alpha^\frL)(\Id-\omega^*\omega)\cal{U}_{\Pauli_\PZm}(z)\|_{hs}^2+4\eps\\
&\leq \underbrace{\|\cal{U}_{\sAns^A}(\alpha^\frR,\alpha^\frL)\|_{op}^2}_{=1}\underbrace{\|\Id-\omega^*\omega\|_{hs}^2}_{=1-\tau(\omega^*\omega)}\underbrace{\|\cal{U}_{\Pauli_\PZm}(z)\|_{op}^2}_{=1}+4\eps\\
&\leq 5\eps\ .
\end{split}   
\end{equation}
As the rows of $(\frS^A)^\perp$ span the kernel of $\frS^A$,  for every $\alpha\in \ker\frS^A$ there is a $\beta\in \FF_2^k$ such that $\alpha=\beta\cdot(\frS^A)^\perp$. Hence,
\begin{equation}\label{eq:Ans_and_PauliX_for_kerSA_almost_commute_baby}
\begin{split}
\forall \alpha^\frR,\alpha^\frL\in \FF_2^\Lambda\ \colon \ \ \cal{U}^{\sAns^A}(\alpha^\frR,\alpha^\frL)\cal{U}^{\Pauli_\PXm}(\alpha)&\approx_{O(\eps )}\cal{U}^{\mathsf{ReadAns}^A}(\alpha^\frR,\alpha^\frL)\cal{U}^{\Hide A}(\beta)\\
&\approx_{O(\eps)} \cal{U}^{\mathsf{ReadAns}^A}(\alpha^\frR,\alpha^\frL)\cal{U}^{\mathsf{ReadPerp}^A}(\beta)\\
&=\cal{U}^{\mathsf{ReadPerp}^A}(\beta)\cal{U}^{\mathsf{ReadAns}^A}(\alpha^\frR,\alpha^\frL)\\
&\approx_{O(\eps)} \cal{U}^{\Hide A}(\beta)\cal{U}^{\mathsf{ReadAns}^A}(\alpha^\frR,\alpha^\frL)\\
&\approx_{O(\eps)} \cal{U}^{\Pauli_\PXm}(\alpha)\cal{U}^{\sAns^A}(\alpha^\frR,\alpha^\frL)\ ,
\end{split}    
\end{equation}
where the first and last approximations  use \eqref{eq:Ans_close_to_ReadAnd_Baby} and \eqref{eq:perp_close_to_PauliX_Baby}, and the middle ones use \eqref{eq:Perp_close_to_ReadPerp_Baby}. Therefore, 
\begin{equation}
\begin{split}
\forall \alpha^\frR,\alpha^\frL\in \FF_2^\Lambda, \alpha\in \ker\frS^A\ \colon \ \ \theta^A(\alpha^\frR,\alpha^\frL)\cdot\PXm^{\otimes \alpha}\otimes \Id_m&\approx_{O(\eps')} \omega \cal{U}^{\sAns^A}(\alpha^\frR,\alpha^\frL)\omega^*\omega\cal{U}^{\Pauli_\PXm}(\alpha)\omega^*\\
&\approx_{O(\eps)}  \omega \cal{U}^{\sAns^A}(\alpha^\frR,\alpha^\frL)\cal{U}^{\Pauli_\PXm}(\alpha)\omega^*\\
&\approx_{O(\eps)}  \omega \cal{U}^{\Pauli_\PXm}(\alpha)\cal{U}^{\sAns^A}(\alpha^\frR,\alpha^\frL)\omega^*\\
&\approx_{O(\eps)}  \omega \cal{U}^{\Pauli_\PXm}(\alpha)\omega^*\omega\cal{U}^{\sAns^A}(\alpha^\frR,\alpha^\frL)\omega^*\\
&\approx_{O(\eps')} \PXm^{\otimes \alpha}\otimes \Id_m\cdot\theta^A(\alpha^\frR,\alpha^\frL)\ ,
\end{split}    
\end{equation}
where the first and last approximations are due to \eqref{eq:thetaA_close_to_AnsA_baby} and \eqref{eq:Pauli_close_to_Pauli_matrices_in_Baby}, the second and fourth are using   \eqref{eq:argument_for_eliminating_omegaomegastar}, and the middle approximation is by \eqref{eq:Ans_and_PauliX_for_kerSA_almost_commute_baby}.

All in all, we deduced that the images of $\theta^A$ (and similarly for $\theta^B$) almost commute with all the $\PZm$-matrices and certain $\PXm$-matrices. This is (essentially) the end of the proof, in a similar manner to that of the Pauli basis game, in which getting to an approximate relations situation allows one to apply a group stability result to finish the argument. Here we also need to analyze why commuting with these specific matrices completes the proof, but this is quite straightforward.

\begin{claim}\label{claim:strict_stability_with_one_rep_fixed}
	Let $\rho\colon G\to U(N)$ be a (unitary) representation of a finite group $G$, and $\phi\colon A\to U(N)$ be a representation of a  finite abelian group. Assume 
	\[
	\forall g\in G,a\in A\ \colon \|\rho(g)\phi(a)-\phi(a)\rho(g)\|_{hs}^2\leq \eps.
	\]
	Then, there is another representation $\xi\colon A\to U(N)$, such that $\xi(a)\phi(g)=\phi(g)\xi(a)$ for every $a\in A$ and $g\in G$, and 
	\[
	\forall a\in A\ \colon \ \ \|\xi(a)-\phi(a)\|_{hs}^2\leq O(\eps).
	\]
\end{claim}

\begin{proof}
	The proof is a combination of an averaging trick common in the study of property $(T)$ groups (cf.\ \cite{Ioana,de_la_Salle_spectral_gap}), and a strict version of the Gowers--Hatami theorem due to Akhtiamov--Dogon \cite{akhtiamov2022uniform}. It can also be deduced directly from orthonormalization (Fact \ref{fact:orthogonalization}), but we show a different argument. 
	
	First, let $\tilde \phi(a)=\Es{g\in G}[\rho(g)\phi(a)\rho(g)^{-1}]\in M_{N}(\complex)$. 
	By our assumption, $\|\tilde \phi(a)-\phi(a)\|_{hs}\leq \Es{g\in G}\|\rho(g)\phi(a)\rho(g)^{-1}-\phi(a)\|\leq \sqrt{\eps}. $ 
	Moreover, $\tilde{\phi}$ commutes with $\rho$. 
	Now, denote by $\mathscr{M}$ the commutant of $\rho(G)$, namely the collection of matrices that commute with all the $\rho$-images of $G$. Then, $\Img\tilde \phi$ is in $\mathscr{M}$, and we can apply the rest of our arguments in this von-Neumann algebra. Note that 
	\[
	\begin{split}
	\|\tilde \phi(a)\tilde \phi^*(a)-\Id\|_{hs}&\leq\Es{g,h\in G}[\|\rho(g)\phi(a)\rho(g)^{-1}\rho(h)\phi(a)^{-1}\rho(h)^{-1}-\underbrace{\Id}_{\rho(g)\rho(g)^{-1}\rho(h)\phi(a)\phi(a)^{-1}\rho(h)^{-1}}\|_{hs}]\\
	&= \Es{g,h\in G}[\|\phi(a)\rho(g^{-1}h)-\rho(g^{-1}h)\phi(a)\|_{hs}]\\
	&\leq \sqrt{\eps}.
	\end{split}
	\]
	Then, by Lemma 2.2 in \cite{akhtiamov2022uniform}, there is a map $\zeta\colon A\to U(\mathscr{M})$, namely to unitaries in the von-Neumann algebra $\mathscr{M}$ such that $\|\zeta(a)-\tilde \phi(a)\|_{hs}\leq \|\tilde \phi(a)\tilde \phi^*(a)-\Id\|_{hs}$ (this is quite straightforward from the SVD decomposition).  Note in addition that 
	\[\begin{split}
	\forall a,b\in A\ \colon \ \ \|\zeta(a)\zeta(b)-\zeta(ab)\|_{hs}&\leq \|\zeta(a)-\tilde \phi(a)\|_{hs}+\|\zeta(b)-\tilde \phi(b)\|_{hs}\\
	&+\|\zeta(ab)-\tilde \phi(ab)\|_{hs}+ \|\tilde\phi(a)\tilde\phi(b)-\tilde\phi(ab)\|_{hs}\\
	\forall a,b\in A\ \colon \ \ \|\tilde\phi(a)\tilde\phi(b)-\tilde\phi(ab)\|_{hs}&\leq \|\phi(a)-\tilde \phi(a)\|_{hs}+\|\phi(b)-\tilde \phi(b)\|_{hs}\\
	&+\|\phi(ab)-\tilde \phi(ab)\|_{hs}+ \|\phi(a)\phi(b)-\phi(ab)\|_{hs}\\
	\end{split}
	\]
	Now, $\phi(a)\phi(b)=\phi(ab)$ since $\phi$ is a representation, and all other summands are bounded by $\sqrt\eps$. Hence, 
	\[
	\forall a,b\in A\ \colon \ \ \|\zeta(a)\zeta(b)-\zeta(ab)\|_{hs}\leq 6\sqrt\eps.
	\]
	By \cite[Corollary 1.7 and Claim 3.3]{akhtiamov2022uniform}, there is a unitary representation $\xi\colon A\to U(\mathscr{M})$\footnote{Note that $\mathscr{M}$ is the same! This is the main difference between Akhtiamov--Dogon to the standard Gowers--Hatami. This is possible thanks to the additional assumption that $A$ is abelian.} such that 
	\[
	\forall a\in A\ \colon \ \ \|\xi(a)-\zeta(a)\|_{hs}\leq O(\sqrt \eps).
	\]
	Applying several triangle inequalities shows that 
	\[
	\forall a\in A\ \colon \ \ \|\xi(a)-\phi(a)\|_{hs}\leq O(\sqrt \eps),
	\]
	which in turn finishes the proof.
\end{proof}
By applying Claim \ref{claim:strict_stability_with_one_rep_fixed} where $G$ is the group generated by $\PZm^{\otimes z}\otimes\Id_m$ and $\PXm^{\otimes \alpha}\otimes\Id_m$ for all $z\in \FF_2^k,\alpha\in \ker\frS^A$, $\rho$ is the identity map, and with $A=\FF_2^\Lambda\times\FF_2^\Lambda$ and $\phi=\theta^A$, we deduce that there is a representation $\xi^A\colon \FF_2^\Lambda\times \FF_2^\Lambda\to U(\complex^{\FF_2^k}\otimes \complex^m)$ such that $\xi^A$ commutes with all $\PZm^{\otimes z}\otimes\Id_m$ and $\PXm^{\otimes \alpha}\otimes\Id_m$  and also 
\begin{equation}\label{eq:perturbed_rep_that_commutes_with_Paulis}
\forall a^\frR,a^\frL\in\FF_2^\Lambda\ \colon \ \ \|\theta^A(a^\frR,a^\frL)-\xi^A(a^\frR,a^\frL)\|_{hs}^2\leq O(\eps')\;.
\end{equation}
Moreover, everything can be done similarly for $B$ resulting with a $\xi^B$ that commutes with all $\PZm^z\otimes \Id_m$ and $\PXm^{\beta}\otimes \Id_m$ for every $\beta\in \ker\frS^B$
.

Now, we can define an $2^k\times m$-dimensional strategy $\strategy'=\{\cal{V}\}$ almost as we did in the completeness proof:
\begin{itemize}
	\item We let $\cal{V}^{\Pauli_\PXm}(\alpha)= \PXm^{\otimes\alpha}\otimes \Id_m,\cal{V}^{\Pauli_\PZm}(\alpha)=\PZm^{\otimes \alpha}\otimes \Id_m$, 
	and extend it to a perfect strategy of the Pauli basis game using Claim   \ref{claim:completeness_PB_k}.
	\item For the other variables, let
	\begin{align*}
	\cal{V}^{\sAns^A}= \cal{V}^{\mathsf{ReadAns}^A}= \cal{V}^{\mathsf{SamAns}^A}=\xi^A\ &,\ \cal{V}^{\sAns^B}= \cal{V}^{\mathsf{ReadAns}^B}= \cal{V}^{\mathsf{SamAns}^B}=\xi^B\ ,\ \\
	\cal{V}^{\Pauli_\PZm}=&\cal{V}^{\mathsf{SamZ}^A}=\cal{V}^{\mathsf{SamZ}^B}\ ,\\
	\cal{V}^{\sQue^A}=\cal{V}^{\mathsf{Read}\sQue^{A}}=\cal{V}^{\Pauli_\PZm}_{[\frS^A]}\ &
	,
	\cal{V}^{\sQue^B}=\cal{V}^{\mathsf{Read}\sQue^{B}}=\cal{V}^{\Pauli_\PZm}_{[\frS^B]}
	,\\
	\cal{V}^{\mathsf{Perp}^{A}}=\cal{V}^{\mathsf{ReadPerp}^{A}}=\cal{V}^{\Pauli_\PXm}_{[(\frS^A)^\perp]}\ &,
	\cal{V}^{\mathsf{Perp}^{B}}=\cal{V}^{\mathsf{ReadPerp}^{B}}=\cal{V}^{\Pauli_\PXm}_{[(\frS^B)^\perp]}\ .
	\end{align*}
	
\end{itemize}
Now, $\strategy'$  is indeed a strategy --- namely, all images are order $2$ unitaries that commute for every fixed vertex --- and it passes by design all checks in $\frak{Baby}(\game)$ with probability $1$, except for maybe $\Intro_A-\Intro_B$. For that edge, we note that by \eqref{eq:perturbed_rep_that_commutes_with_Paulis}, \eqref{eq:Pauli_close_to_SamZ_in_Baby}, \eqref{eq:Que_close_to_SamZ_Baby}, \eqref{eq:thetaA_close_to_AnsA_baby} and \eqref{eq:thetaB_close_to_AnsA_baby} the strategy $\strategy'=\{\cal{V}\}$ is $O(\eps')$-close on this edge to the original strategy $\{\cal{U}\}$,  and  Claim \ref{claim:close_strat_implies_close_correlations} states that this means they produce $\sqrt {\eps'}$-close correlations, and thus $\cal{V}$ passes this edge with probability close to that of  $\cal{U}$. As this edge is sampled with probability $\nicefrac{1}{4}$ in $\frak{Baby}(\game)$, $\cal{U}$ passes it with probability $1-O(\eps)$, which means $\cal{V}$ passes it with probability $1-O(\sqrt{\eps'})$.

We are left to show that $\cal{V}$ is an honest strategy, and thus (by definition) a strategy with the same value can be extracted for $\game$. This is immediate by analyzing the commutant of $\{\PZm^{\otimes z}\otimes \Id_m,\PXm^{\otimes \alpha}\otimes \Id_m\}_{z\in \FF_2^k,\alpha\in \ker\frS^A}$. A matrix that commutes with all $\PZm^{\otimes z}\otimes \Id_m$ is of the form $\sum_{z\in \FF_2^k} \mathscr{F}^\PZm_z\otimes \mathscr{A}^z$ for $\mathscr{A}^z\in M_m(\complex)$ and $\mathscr{F}^\PZm_z$ the projections on the indicators ${\bf 1}_z$ in $\complex^{\FF_2^k}$ (Definition  \ref{defn:F^Z_v}). 
For such matrices, commuting with $\PXm^{\otimes \alpha}\otimes \Id_m$ is the same as requiring $\mathscr{A}^z=\mathscr{A}^{z+\alpha}$ for every $z\in \FF_2^k$. 
Hence, the commutant consists of all matrices of the form $\sum_z(\sum_{\alpha\in \ker\frS^A}\mathscr{F}^\PZm_{z+\alpha})\otimes\mathscr{A}^z$ where the sum over $z$'s takes a representative from every coset of $\ker\frS^A$ in $\FF_2^k$. But, this is the same as writing every matrix as $\sum_{\mttx\in \FF_2^r}(\sum_{z\colon \frS^A(z)=\mttx}\mathscr{F}^\PZm_{z})\otimes\mathscr{A}^\mttx$. As this is true, in particular, for any projection in the commutant,  we can write the PVM associated with the images of $\cal{V}$ at $\Intro_A$ as $\cal{Q}^{\Intro_A}_{\mttx,a^\frR,a^\frL}=\sum_{z\in \FF_2^k:\frS^A(z)=\mttx}\mathscr{F}^\PZm_{z}\otimes\cal{P}^{\mttx}_{a^\frR,a^\frL}$, and similarly for $\Intro_B$.  The resulting $\cal{P}\colon \FF_2^r\times \FF_2^\Lambda\times \FF_2^\Lambda\to M_m(\complex)$ is a PVM strategy for $\game$ that passes it with the same probability as $\cal{V}$ passes $\Intro_A-\Intro_B$ (as in Definition \ref{defn:honest_strat_for_Intro} on honest strategies), which is $1-O(\sqrt{\eps'})$. This finishes the proof of soundness. 

Note that by \eqref{eq:Pauli_close_to_Pauli_matrices_in_Baby}, the normalized dimension difference $1-\frac{N}{2^k\cdot m}\leq 1-\tau (\omega\omega^*)\leq  O(\eps').$
Furthermore, as we extracted from the honest $\cal{V}$ an $m$-dimensional strategy 
$\cal{P}$ for $\game$ with value $1-O(\sqrt{\eps'})$, we deduce that $m\geq \Ent(\game,1-O(\sqrt{\eps'}))$, which proves the entanglement lower bound \labelcref{clause:entanglement_of_baby}.

\subsection{Conditionally linear maps}\label{sec:CLMs}

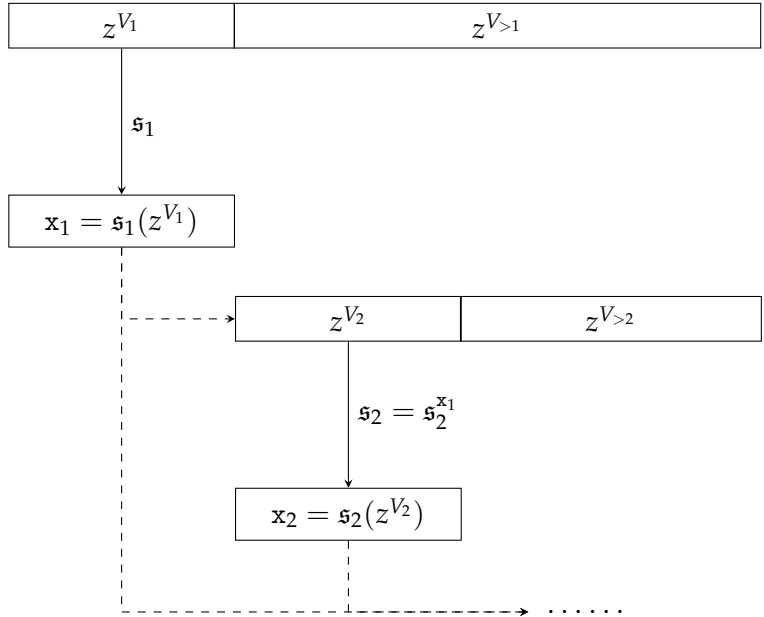
\begin{figure}[!htbp]
	\centering
	\begin{tikzpicture}[scale=1.3]

	\node[draw, minimum width=3cm, anchor=east] at (0,0) (xv1) {$z^{V_{1}}$};
	\node[draw, minimum width=7cm, anchor=west] at (-.4pt,0) {$z^{V_{>1}}$};
	\node[draw, minimum width=3cm, anchor=east] at (0,-2) (xl1) {$\mttx_1=\frS_1(z^{V_1})$};
	\draw[-stealth] (xv1.south) -- ++(0,-1) node[above right] {$\frS_{1}$} -- (xl1.north);
	
	\node[draw, minimum width=3cm, anchor=west] at (0,-3) (xv2) {$z^{V_{2}}$};
	\node[draw, minimum width=4cm, anchor=west] at (2.3cm,-3) {$z^{V_{>2}}$};
	
	\node[draw, minimum width=3cm, anchor=west] at (0,-5) (xl2) {$\mttx_2=\frS_2(z^{V_2})$};
	
	\draw[-stealth] (xv2.south) -- ++(0,-1) node[above right]
	{$\frS_2=\frS_{2}^{\mttx_1}$} -- (xl2.north);
	\draw[dashed, -stealth, to path={|- (\tikztotarget)}] (xl1) edge (xv2.west);
	
	\draw[dashed, -stealth, to path={|- (\tikztotarget)}] (xl2) edge (3cm,-6);
	\draw[dashed, -stealth, to path={|- (\tikztotarget)}] (xl1) edge (3cm,-6);
	\node[right] at (3cm,-6) {$\;\cdots\cdots$};
	
	\end{tikzpicture}
	\caption{An illustration of an $h$-level CLM  $\frS$ (adapted from \cite[Figure 1]{MIPRE}). Let $z\in \FF_2^k$ be the input.
		First a register subspace $V_{1}$ and a linear map $\frS_{1}$ are chosen and
		applied on the restriction of $z$ to $V_1$ to obtain $\mttx_1 = \frS_{1}(z^{V_{1}}) \in V_{1}$.
		Then depending the value of $\mttx_1$, a register subspace $V_{2}=V_{2}^{\mttx_1}$ and a
		linear map $\frS_2=\frS_{2}^{\mttx_1}$ are chosen and applied on the restriction of $z$ to $V_2$ to obtain
		$\mttx_2 = \frS_{2}(z^{V_{2}}) \in V_{2}$ and so on.
		Finally, $\frS(z)$ is defined to be
		$\sum_{j=1}^{h} \mttx_j$.}\label{fig:cl-functions}
\end{figure}

The collection of linear maps $\frS\colon \FF_2^k\to \FF_2^r\times \FF_2^r$ induces a family of samplers which is too restrictive for us to  prove compression with. But, a certain generalization of linear maps, called \emph{conditionally linear maps} \cite[Definition 4.1]{MIPRE}, are rich enough to deduce compression. This section is devoted to this generalized setup.

Intuitively, conditionally linear maps that act on $\FF_2^k$ apply a sequence of linear maps on subspaces of it, where each map in the sequence depends on the value which the previous linear maps produced. These maps are in a sweet spot, being rich enough so that all the samplers that we will need can be described as pushforwards of the uniform measure along them, while being amenable to a  construction similar to $\frak{Baby}(\game)$ from Section \ref{sec:baby_question_reduction}.

Conditionally linear maps are not complicated objects, yet the notation associated with them can take some time getting used to. We recommend that the reader attempt to follow the visual explanation given in Figure \ref{fig:cl-functions} first, to form their own intuition; which can then be matched to the formal definitions that follow. 

\begin{definition}\label{defn:register_subspace}
	A \emph{register subspace} of $\FF_2^k$ is one which is spanned by some subset of the standard basis $\{e_1,...,e_k\}$.  As there is a bijection between register subspaces and subsets of $[k]$, we often associate with such a subspace  the appropriate subset of indices $I\subseteq [k]$.  Given a register subspace $V=\mathrm{Span}\{e_{i_1},...,e_{i_m}\}$ (in which case $I=\{i_1,...,i_m\}$) and a vector $z\in \FF_2^k$ we denote by $z^V$ the restriction of $z$ to the coordinates of $V$, namely $z^V=\sum_{j=1}^m \langle z,e_{i_j}\rangle e_{i_j}$. As $V$ is canonically isomorphic to $\FF_2^I$, we often treat vectors in $V$ as parameterized by $I$ instead of $[k]$.
	
	Two register subspaces are said to be \emph{disjoint} if their intersection is trivial. Two register subspaces are said to be \emph{complementary} if they are disjoint and sum up to the whole space $\FF_2^k$.
\end{definition}

\begin{definition}[Conditionally linear map --- recursive definition]\label{def:CLM_rec}
	Let $k\geq 1$ and $h\geq 0$ be integers. The collection of $h$-level conditionally linear maps (CLMs) on $\F_2^k$ is defined inductively on $h$ as follows.  
	\begin{itemize}
		\item A $0$-level CLM over $\FF_2^k$ is the zero map. 
		\item Assume $(h-1)$-level CLMs were already defined. An $h$-level CLM $\frS$ over $\FF_2^k$ consists of  the following data:
		\begin{itemize}
			\item a register subspace (Definition  \ref{defn:register_subspace}) $V_1\subseteq\FF_2^k$, whose complement is denoted by $V_{>1}\subseteq\FF_2^k$;
			\item a linear map $\frS_1\colon V_1\to V_1$; 
			\item for every $u\in V_1$, an $(h-1)$-level CLM $\frS_{> 1}^{u}$ on $V_{>1}$.
		\end{itemize}
	\end{itemize}
	The data of an $h$-level CLM defines a function $\frS\colon \FF_2^k\to \FF_2^k$ as follows:
	\begin{itemize}
		\item For the $0$-level case the function is the zero function. 
		\item Assuming we defined evaluation along $(h-1)$-level CLMs, the evaluation along an $h$-level CLM is:
		\begin{itemize}
			\item Given $z\in \FF_2^k$, recall that $z^{V_1}\in V_1$ is its restriction to $V_1$. One can thus use the linear map $\frS_1\colon V_1\to V_1$ to evaluate $\mttx_1=\frS_1(z^{V_1})$.
			\item As there is an $(h-1)$-level CLM associated to $\mttx_1$, $\frS^{\mttx_1}_{>1}\colon V_{>1}\to V_{>1}$, and as we assumed evaluation along $(h-1)$-level CLMs was already defined, we let $\mttx_{>1}=\frS_{>1}^{\mttx_1}(z^{V_{>1}})\in V_{>1}$, where again $z^{V_{>1}}$ is the restriction of the input $z$ to the complementary subspace $V_{>1}$.
			\item Finally, the $\frS$-evaluation of $z$ is defined to be $\frS(z)=\mttx_1+\mttx_{>1}\in V_1\oplus V_{>1}= \FF_2^k$.
		\end{itemize}
	\end{itemize}
\end{definition}

\begin{remark}
	Note that $1$-level CLMs are exactly linear functions from $\FF_2^k$ to itself.  An example of a $2$-level CLM is $(x_1,x_2,x_3)\mapsto (0,x_1x_3+x_2x_3+x_1,x_3)$ --- this example can be used as a sanity check, and a proof of it being a $2$-level CLM appears in \cite[Example 4.3]{MIPRE}. 
\end{remark}
The following is a more intricate definition of $h$-level CLMs, which avoids the recursive nature of Definition \ref{def:CLM_rec}.  A proof that the two definitions are equivalent is given in \cite[Lemma 4.6]{MIPRE} (with somewhat different notation).

\begin{definition}[Conditionally linear map --- direct definition]\label{def:CLM}
	Let $h\geq 0$ and $k\geq 1$ be integer. To describe an \emph{$h$-level conditionally linear map} (CLM) $\frS\colon \FF_2^k\to \FF_2^k$ we need the following structure. 
	First, there is a collection of register subspaces defined inductively: \begin{itemize}
		\item $V_1\subseteq \FF_2^k$ is a fixed register subspace, and we denote by $V_{>1}$ its complement. 
		\item   For every $u_1\in V_1$, there is a register subspace $V_{2}^{u_1}\subseteq V_{>1}$, and we denote the sum $V_1\oplus V^{u_1}_2$ by $V_{\leq 2}^{u_1}$, and its complement by $V_{>2}^{u_1}$.
		\item Then, for every $u_2\in V_{2}^{u_1}$, there is a register subspace $V_{3}^{u_1,u_2}\subseteq V^{u_1}_{>2}$, which gives rise to the subspaces $V_{\leq 3}^{u_1,u_2}=V^{u_1}_{\leq 2}\oplus V^{u_1,u_2}_3$ and its complement $V_{>3}^{u_1,u_2}$. 
		\item  This keeps on, so that in the $j^{\rm th}$ step, for every $u_1\in V_1,u_2\in V_{2}^{u_1},\dots,u_{j-1}\in V_{j-1}^{u_1,...,u_{j-2}}$ there is a register subspace $V_{j}^{u_1,...,u_{j-1}}$ disjoint of $V_1\oplus V_{2}^{u_1}\oplus \dots\oplus V_{j-1}^{u_1,...,u_{j-2}}$, giving rise to the appropriate $V_{\leq j}=V_{\leq j}^{u_1,...,u_{j-1}}$ and complement $V_{>j}^{u_1,...,u_{j-1}}$.
		\item    No matter the process, we are guaranteed to reach $\FF_2^k$ after $h$ steps, namely $V_1\oplus V_{2}^{u_1}\oplus \dots\oplus V_{h}^{u_1,...,u_{h-1}}=\FF_2^k$ for every $u_1\in V_1,u_2\in V_{2}^{u_1},\dots,u_{h-1}\in V_{h-1}^{u_1,...,u_{h-1}}$.
	\end{itemize}
	Now, in addition to the above collections of register subspaces, there is a collection of linear maps on them:
	\begin{itemize}
		\item On $V_1$ there is a fixed $\frS_1\colon V_1\to V_1$.
		\item   For every $u_1\in V_1$ there is  a {linear map}  $\frS_{2}^{u_1}\colon V_{2}^{u_1}\to V_{2}^{u_1}$.
		\item  This continues in a similar manner to before, where in the $j^{\rm th}$ step, for every $u_1\in V_1,u_2\in V_{2}^{u_1},\dots,u_{j-1}\in V_{j-1}^{u_1,...,u_{j-2}}$ there is a {linear map} $\frS_j=\frS_{j}^{u_1,...,u_{j-1}}\colon V_{j}^{u_1,...,u_{j-1}}\to V_{j}^{u_1,...,u_{j-1}}$. 
	\end{itemize} 
	Finally, the function $\frS$ is calculated as follows:
	\begin{itemize}
		\item Let $z\in \FF_2^k$ be the input to $\frS$.
		\item Calculate $\mttx_1=\frS_1(z^{V_1})$, and let $V_2=V_2^{\mttx_1}$.
		\item  Calculate $\mttx_2=\frS_{2}^{\mttx_1}(z^{V_{2}})$, and let $V_3=V_3^{\mttx_1,\mttx_2}$.
		\item In the $j^{\rm th}$ step, calculate $\mttx_j=\frS_{j}^{\mttx_1,...,\mttx_{j-1}}(z^{V_{j}})$, and let $V_{j+1}=V_{j+1}^{\mttx_1,...,\mttx_j}$.
		\item After $h$ steps are completed, resulting in $\mttx_1,...,\mttx_h$, output $\frS(z)=\mttx=\mttx_1+\mttx_2+...+\mttx_h\in V_1\oplus\dots\oplus V_h=\FF_2^k$. 
	\end{itemize}
	We denote by $\frS_j(z)$ the value of $\mttx_j$ in the above computation of $\frS(z)$, and let 
	\begin{equation}\label{eq:def_frS_leqj}
	\frS_{\leq j}(z)=\mttx_{\leq j}=\mttx_1+...+\mttx_j
	\end{equation}
	be the cumulative output up to the $j^{\rm th}$ computation.
\end{definition}

\begin{definition}[Seeded conditionally linear maps]\label{defn:seeded_CLM}
	Let $\frS$ be an $h$-level CLM with all the data from Definition \ref{def:CLM}. Let $u\in \FF_2^k$ be a vector which plays the role of a \emph{seed}.  Then, $u$ induces a decomposition of $\FF_2^k$ into $h$  disjoint register subspaces, which we call the \emph{$u$-seeded register subspaces}, as follows:
	\begin{itemize}
		\item Regardless of $u$, we let $W_{1}^u=V_1$. As $W_1^u$ is a register subspace, it has an associated subset of indices from $[k]$, which we denote by $I^u_1\subseteq [k]$. Furthermore,  denote the restriction of $u$ to this register subspace by  $u_1=u^{W_1^u}$.
		\item Then, we let $W_{2}^{u}=V_{2}^{u_1}$ with $I^u_2\subseteq [k]$ the associated subset of indices, and denote  the restriction of $u$ to it by $u_2=u^{W^{u}_2}$. 
		\item More generally, given that we have already defined  $W^u_1,...,W^u_{j-1}$ and  thus the respective restrictions  $u_1,...,u_{j-1}$ of $u$, we let 
		\begin{equation}\label{eq:reg_subspaces_depending_on_prefixes}
		W^u_j=V_j^{u_1,...,u_{j-1}}\quad \textrm{and}\quad u_j=u^{W^u_j}\ ,
		\end{equation}
		with $I^u_j\subseteq[k]$ being again the associated subset of indices.
	\end{itemize}
	
	We use $W_{\leq j}^u,W_{< j}^u,W_{\geq j}^u$  and $W_{> j}^u$ in a similar way to before, and $I^u_{\leq j},I^u_{< j},I^u_{\geq j},I^u_{> j}$ for the indices supporting each of these register  subspaces. We call 
	\begin{equation}\label{eq:def_of_leqj_prefix}
	u_{\leq j}=u_1+...+u_j\in W^u_1\oplus...\oplus W^u_j
	\end{equation} 
	the $j^{\rm th}$ \emph{prefix of} $u$, and note that $W^u_{\leq j+1}$, $W^u_{>j+1}$ depend only on this $j^{\rm th}$ prefix and not all of $u$. Note also that $\frS_{\leq j}(z)$ as in \eqref{eq:def_frS_leqj} is \textbf{equal} to the $j^{\rm th}$ prefix of $\frS(z)$, namely to $(\frS(z))_{\leq j}$, since $\frS$ uses its partial computation as the seed to the rest of it. We also have the notions of \emph{$u$-seeded $j^{\rm th}$-position, prefix and suffix} of any vector, which are defined by 
	\begin{equation}\label{eq:u-seeded_prefix_suffix}
	\forall z\in \FF_2^k\ \colon\  \  z^u_j=z^{W^u_j}\;,\quad z^u_{\leq j}=z^{W^u_{\leq j}}\;,\quad z^u_{\geq j}=z^{W^u_{\geq j}}\ .
	\end{equation}
	
	The \emph{$u$-seeded CLM}, denoted  by $\frS^u$, is the following {linear} map: Letting $u_i=u^{W^u_i}$ be as in \eqref{eq:reg_subspaces_depending_on_prefixes}, we define the $j^{\rm th}$ $u$-seeded linear map $\frS_{j}^u\colon W^u_j\to W^u_j$ by 
	\begin{equation}\label{eq:seeded_CLM}
	\forall z\in W^u_j\  \colon \ \ \frS^u_j(z)=\frS_{j}^{u_1,...,u_{j-1}}(z)\ .    
	\end{equation}
	Namely, $\frS_j^u$ uses the $(j-1)^{\rm th}$ prefix of $u$ as a seed, which defines both the appropriate $j^{\rm th}$ subspace $W^u_j$ (and thus the restriction of $z$ to this subspace) as well as the $j^{\rm th}$ linear map  $\frS^u_j=\frS_j^{u_{< j}}$ acting on this subspace. Then, we let $\frS^u_{\leq j}\colon W^u_{\leq j}\to W^u_{\leq j}$  be 
	\begin{equation}\label{eq:seeded_CLM_complete}
	\frS_{\leq j}^u=\frS^u_1\oplus \frS^u_2 \oplus...\oplus\frS^u_j\quad\textrm{and} \quad \frS^u=\frS_{\leq h}^u\colon \FF_2^k\to \FF_2^k\ .
	\end{equation} 
	
	As $\frS^u_j\colon W^u_j\to W^u_j$ are linear maps, and $I^u_j$ is the set of indices associated with $W^u_j$, we think of them as matrices represented in the standard basis supported on $I^u_j$, i.e., for every $i,t\in I^u_j$ we have $(\frS^u_j)_{it}=\langle e_i,\frS^u_j (e_t)\rangle$.
	
	We often use the following natural extension of the $j^{\rm th}$ $u$-seeded linear map to all of $\FF_2^k$: As $W^u_j$ has a complement register subspace $W^u_{\neq j}$, which is the sum of all $W^u_i$ such that $i\neq j$, we can let the \emph{extended $j^{\rm th}$ $u$-seeded linear map} $\frSS^u_j\colon \FF_2^k\to \FF_2^k$ by letting 
	\begin{equation}\label{eq:extended_u_seeded_CLM}
	\forall z\in \FF_2^k\ \colon \ \ \frSS^u_j(z)=\frS^u_j(z^u_j)\ ,
	\end{equation}
	namely, it acts the same way as $\frS^u_j$ on $W^u_j$, and sends everything else to zero. We use a similar  notation as before,  $\frSS^u_{\leq j}=\bigoplus_{i=1}^j \frSS^u_i$, and note that $\frSS^u_{\leq h}=\frS^u_{\leq h}=\frS^u$.
\end{definition}


\begin{corollary}[Seeded versus unseeded CLMs]\label{cor:seeded_and_unseeded_CLMs}
	Let $h$ and $k$ be positive integers, and let $\frS$ be an $h$-level CLM on $\FF_2^k$. Let $j\in [h]$, $u\in \FF_2^k$ a seed and $z\in \FF_2^k$ a vector. Then, $\frS_{\leq j}(z)=u_{\leq j}$ as defined in \eqref{eq:def_frS_leqj} if and only if $\frS^u_{\leq j}(z^u_{\leq j})=u_{\leq j}$ as defined in \eqref{eq:seeded_CLM_complete}. In particular,  $\frS(z)=u$ if and only if   $\frS^u(z)=u$.
\end{corollary}

\begin{definition}[Sampling scheme induced by $h$-level CLMs]\label{defn:sampling_induced_by_CLMs}
	A tailored game $\game$ is said to have a sampling scheme induced by $h$-level CLMs, if there exist a pair of $h$-level CLMs $\frS=(\frS^A,\frS^B)\colon \FF_2^k\to \FF_2^k \times \FF_2^k$ (Definition \ref{def:CLM}), such that the vertex set of the underlying graph of $\game$ is $\FF_2^k$, and the distribution over edges is the pushforward of the uniform distribution over $\FF_2^k$ through $\frS$. Namely, for $\mttx,\mtty\in \FF_2^k$, 
	\[
	\mu(\mttx\mtty)=\frac{|\{z\in \FF_2^k\mid \frS^A(z)=\mttx,\frS^B(z)=\mtty\}|}{2^k}\ .
	\]
\end{definition}

\subsubsection*{Perpendicular maps}
As seen in the baby question reduction transformation from Section \ref{sec:baby_question_reduction}, we need a notion of a ``perpendicular map''. Specifically for the full question reduction transformation, we will need a perpendicular map for every seeded CLM (Definition \ref{defn:seeded_CLM}).
\begin{definition}\label{defn:perpendicular_maps}
	Let $f\colon \FF_2^k\to \FF_2^k$ be a linear map. A \emph{perpendicular map} to $f$, is a linear map $f^\perp\colon \FF_2^k\to \FF_2^k$ whose rows span $\ker(f)$, namely $\Img((f^\perp)^*)=\ker(f)$, where $*$ is the dual map with respect to the bilinear form $\langle \cdot,\cdot\rangle$ --- see the beginning of Section \ref{sec:Pauli_gp}.
\end{definition}

\begin{remark}
	As defined, the perpendicular map is not unique. There is an efficient algorithmic way of extracting a perpendicular map given the matrix representation of a linear map using Gaussian elimination --- see, e.g., \cite[Definition 3.11]{MIPRE}.
\end{remark}

\begin{claim}\label{claim:inverse_fourier_of_data_processed_is_structured}
	Let $f\colon \FF_2^k\to \FF_2^k$ be a linear map, and let $f^\perp\colon \FF_2^k\to \FF_2^k$ be a perpendicular map to $f$ (Definition \ref{defn:perpendicular_maps}), namely a matrix whose rows span the kernel of $f$. Then, the $f$-evaluated (Definition \ref{defn:Data_proccessed_PVM})   PVM $\mathscr{F}^\PZm$ commutes with the $f^\perp$-evaluated PVM $\mathscr{F}^\PXm$. Namely,
	\[
	\forall \nu,\mttx\in \FF_2^k\ \colon \ \  \mathscr{F}^\PZm_{[f(\cdot)=\mttx]}\cdot \mathscr{F}^\PXm_{[f^\perp(\cdot)=\nu]}= \mathscr{F}^\PXm_{[f^\perp(\cdot)=\nu]}\cdot\mathscr{F}^\PZm_{[f(\cdot)=\mttx]}\ .
	\]
	In words, one can measure the $f$-evaluation according to the $\PZm$-basis simultaneously with the $f^\perp$-evaluation according to the $\PXm$-basis.
\end{claim}
\begin{proof}
	These two PVMs commute in projective form if and only if they commute in representation form. Let $\rho^\PZm$ and $\rho^\PXm$ be the representation forms of $\mathscr{F}^\PZm$ and $\mathscr{F}^\PXm$ respectively (as they were defined in \eqref{eq:defn_restriction_rho_to_X_and_Z_subgroups}), namely $\rho^\PZm(\alpha)=\PZm^{\otimes \alpha}$ and $\rho^\PXm(\beta)=\PXm^{\otimes\beta}$. Then, by Corollary \ref{cor:linear_data_processed_PVM_is_ZPC_and_left_multiplication}, 
	\[
	\rho^\PZm_{[f]}(\alpha)=\PZm^{\otimes\alpha \cdot f}\quad,\quad \rho^\PXm_{[f^\perp]}(\beta)=\PXm^{\otimes\beta\cdot f^\perp}\ .
	\]
	Now, for every $\alpha,\beta\in \FF_2^k$, thought of as row vectors,  we have
	\[
	\PZm^{\otimes \alpha \cdot f}\PXm^{\otimes \beta \cdot f^\perp}=(-1)^{\langle  \alpha \cdot f, \beta \cdot f^\perp\rangle }\PXm^{\otimes \beta \cdot f^\perp}\PZm^{\otimes \alpha \cdot f}\ .
	\]
	But, 
	\[
	\langle  \alpha \cdot f, \beta \cdot f^\perp\rangle=\alpha \underbrace{f (f^\perp)^*}_{=0} \beta^*=0\ ,
	\]
	where $*$ is  transposition in the above calculation. Hence the PVMs commute as claimed.
\end{proof}

\begin{remark}\label{rem:extended_perpendicular_maps_along_a_CLM}
	In the next section, we assume to be given in addition to a CLM $\frS$ acting on $\FF_2^k$, a collection of perpendicular maps $(\frS_j^u)^\perp\colon W^u_j\to W^u_j$ for each $j^{\rm th}$ $u$-seeded CLM. In the same spirit as before, we use the notation $(\frS^u_{\leq j})^\perp$ for $\bigoplus_{i=1}^j (\frS^u_i)^\perp$, and it is indeed perpendicular to the map $\frS^u_{\leq j}$.
	
	By extending $(\frS^u_j)^\perp$ to $(\frSS^u_j)^\perp\colon \FF_2^k\to \FF_2^k$ to be {the identity} on $W^u_{\neq j}$ we indeed obtain a function which is perpendicular to the extended $j^{\rm th}$ $u$-seeded CLM defined in \eqref{eq:extended_u_seeded_CLM} --- so the notation is fitting.  This extension satisfies that 
	\begin{equation}\label{eq:defn_frSS_perp_leqj}
	(\frSS_{\leq j}^u)^\perp= (\frSS^u_{j})^\perp\circ...\circ(\frSS^u_{2})^\perp\circ(\frSS^u_{1})^\perp\ .
	\end{equation}
\end{remark}

\subsection{Question Reduction in the conditionally linear sampler case}\label{sec:augmentation_in_the_CLM_case}

\textbf{Note}, this is the proper augmentation that is used in compression, as opposed to the simplified case described and analyzed in Section \ref{sec:baby_question_reduction}. The sections are structured in a {very} similar manner, in the hope that by first reading Section \ref{sec:baby_question_reduction}, the following description and analysis of the proper augmentation become clear.
\\

Let $\game$ be a tailored game with the following properties:
\begin{enumerate}[label=\textcolor{black}{(\arabic*)}, ref= (\arabic*)]
	\item \label{clause1:prop_of_input_game_to_quered}Its sampling scheme is induced by $h$-level CLMs (Definition \ref{defn:sampling_induced_by_CLMs}), for some positive integer $h$. Namely, its vertex set is $\FF_2^k$, and the distribution on edges is induced by the pushforward of the uniform distribution on $\FF_2^k$ through a pair of $h$-level CLMs (Definition \ref{def:CLM})  $\frS=(\frS^A,\frS^B)\colon \FF_2^k\to \FF_2^k\times \FF_2^k$.
	\item  \label{clause2:prop_of_input_game_to_quered}Its length functions are constant and equal to some positive integer $\Lambda$.
	\item  \label{clause3:prop_of_input_game_to_quered} In a similar manner  to Section \ref{sec:baby_question_reduction}, we need a basis for the kernel of the  $j^{\rm th}$ $u$-seeded CLM $\frS^{A,u}_{j}\colon W^{A,u}_{j}\to W^{A,u}_{j}$ (Definition \ref{defn:seeded_CLM}), which is a linear map, for every $j$ and $u$. So, we assume to be given perpendicular maps (Definition \ref{defn:perpendicular_maps})  $(\frS^{A,u}_{j})^\perp\colon W^{A,u}_j\to W^{A,u}_j$, namely  their rows, parametrized by the indices in $I^{A,u}_{j}\subset [k]$, span $\ker\frS^{A,u}_j$. We mainly use the extensions $(\frSS^{A,u}_j)^\perp$ of these maps to all of $\FF_2^k$ as in Remark \ref{rem:extended_perpendicular_maps_along_a_CLM}, and specifically those defined in \eqref{eq:defn_frSS_perp_leqj}.
\end{enumerate}

\begin{figure}[!htbp]
	\centering
	\begin{gamespec}
		\setlength{\tabcolsep}{1em}
		Let $\game$ be a tailored game with vertex set $\F_2^k$ and whose distribution on edges is induced by the pushforward of the uniform distribution on $\F_2^k$ through a pair of $h$-level CLMs $\frS=(\frS^A,\frS^B)\colon \FF_2^k\to \FF_2^k\times \FF_2^k$. As usual, it is assumed that $\game$ has constant readable and unreadable answer lengths both equal to $\Lambda$. In addition, a collection of perpendicular maps $(\frS_j^{A,\mttx})^\perp$ to the seeded CLMs $\frS_j^{A,\mttx}$ are assumed to be provided, and we use the notation $(\frSS_j^{A,\mttx})^\perp$ for their extensions as in Remark \ref{rem:extended_perpendicular_maps_along_a_CLM}.
		
		\vspace{1em}
		\begin{tabularx}{\textwidth}{ l    X l l }
			\toprule
			Sub-Structure & Question &   Readable answers & Unreadable answers  \\
			\midrule
			$\PauliBasis_k$ &  $\Pauli_\PZm$ &  &  $z\in \FF_2^k$\\
			& $\Pauli_\PXm$ & & $\chi\in \FF_2^k$\\
			& See Figure~\ref{fig:genpauli-summary} for rest& &\\
			$\Introspect(\game)$ & $\Intro_A$ & $(\mttx,a^\frR)\in \FF_2^k\times \FF_2^\Lambda$ &  $a^\frL\in \FF_2^\Lambda$  \\
			Sampling apparatus   &$\Sample_A$  & $(z_{sam},a^\frR_{sam})\in \FF_2^k\times \FF_2^\Lambda $&
			$a^\frL_{sam}\in \FF_2^{\Lambda}$\\
			Hiding apparatus& $\Read_A$   &  
			$(\mttx_{read},a^\frR_{read})\in \FF_2^k\times \FF_2^\Lambda$  & $(\nu_{read},a^\frL_{read})\in \FF_2^k\times\FF_2^\Lambda$\\
			& $\Hide A^j$  & $\mttx_{hide\ j}\in \FF_2^k$
			& $\nu_{hide\ j}\in \FF_2^k$  \\
			\bottomrule
		\end{tabularx}
		
		\vspace{1em}
		
		The following tests are performed when the corresponding augmented edge is sampled:
		
		\begin{enumerate}
			\setlength\itemsep{1pt}
			\item  $\Pauli_\PZm-\Sample_\cdot$: Check that $z=z_{sam}$.
			\item $\Intro_\cdot-\Sample_\cdot$: Check that $\mttx=\frS^A(z_{sam})$, $a^\frR=a^\frR_{sam}$ and $a^\frL=a^\frL_{sam}$.
			\item $\Intro_\cdot-\Read_\cdot$: Check that $\mttx=\mttx_{read}$, $a^\frR=a^\frR_{read}$ and $a^\frL=a^\frL_{read}$.
			\item  $\Hide \cdot^h-\Read_\cdot$: Check that $\nu_{hide\ h}=\nu_{read}$ and that $\mttx_{hide\ h}=(\mttx_{read})_{<h}$, where $(\cdot)_{<h}$ is the $(h-1)^{\rm th}$ prefix \eqref{eq:def_of_leqj_prefix}.
			\item $\Hide \cdot^1-\Pauli_{\PXm}$: Check that $\mttx_{hide\ 1}=0$ and that $\nu_{hide\ 1}=(\frSS_1^{A,\mttx_{hide\ 1}})^\perp(\chi)$.
			\item $\Hide\cdot^j-\Hide\cdot^{j-1}$: Fixing $\mttx=\mttx_{hide\ j}$, we  check two things. First, that $\mttx_{hide\ j-1}=\mttx_{<j-1}$, where  $(\cdot)_{<j-1}$ is the $(j-2)^{\rm th}$ prefix \eqref{eq:def_of_leqj_prefix}. Second, that $(\frSS^{A,\mttx}_{j})^\perp(\nu_{hide\ j-1})=\nu_{hide\  j}$.
		\end{enumerate}
	\end{gamespec}
	\caption{Questions and answers in the game $\frak{QueRed}_h(\game,k,\mathscr{B})$. Since the game is an augmentation of the sum of $\PauliBasis_k(\mathscr{B})$ and $\Introspect(\game)$ we only list new questions and answers, and additional tests, and refer to Figure~\ref{fig:genpauli-summary} and Figure~\ref{fig:introspect-summary} for questions and answers of the latter.} 
	\label{fig:quered-summary}
\end{figure}

\textbf{The question reduction transformation $\frak{QueRed}(\game)=\frak{QueRed}_h(\game,k,\mathscr{B})$} (See Figure \ref{fig:quered-summary} for an overview): The inputs are expected to be a positive integer $k$, a tuple $\mathscr{B}$ of $n$  vectors in $\FF_2^k$ that induce and $[n,k,d]$-code, and a tailored game $\game$ satisfying \labelcref{clause1:prop_of_input_game_to_quered}, \labelcref{clause2:prop_of_input_game_to_quered} and \labelcref{clause3:prop_of_input_game_to_quered} from the beginning of the section.
The game $\frak{QueRed}(\game)$ is then an augmented (Definition \ref{defn:augmentation_of_a_game}) sum (Definition \ref{defn:sum_of_games}) of  the Pauli basis game  $\PauliBasis_k=\PauliBasis_k(\mathscr{B})$ (Section \ref{sec:Pauli_basis_definition}) and the introspection game $\Introspect(\game)$ (Definition \ref{defn:Introspection}). The augmentation consists of two apparatuses:
\begin{enumerate}
	\item A \textbf{Sampling apparatus} which connects the   introspection game vertices to the total $Z$-measurement of the Pauli basis game (i.e., the vertex $\Pauli_\PZm$). The goal of this apparatus is two-fold --- first, to verify that the ``questions'' part of the  players' answers when the copy of $\Introspect(\game)$ is played is distributed according to the question distribution of $\game$ --- namely the pushforward of the uniform distribution along the CLMs $(\frS^A,\frS^B)$; second, to verify that the observables associated with the ``answers`` part of the players' answers in $\Introspect(\game)$ commute with the total $Z$-measurement. 
	\item  A \textbf{Hiding apparatus} which connects the   introspection game vertices to the total $X$-measurement of the Pauli basis game (i.e., the vertex $\Pauli_\PXm$). The goal of this apparatus is to verify that the ``answers`` part of the players' answers in $\Introspect(\game)$ commute with a certain data processing of the $X$-measurements, specifically through the map $(\frS^{\cdot,\cdot})^\perp$.
\end{enumerate}

\begin{figure}[httb!]
	\centering
	\begin{tikzpicture}[scale=0.9]
	\node[draw, color=black, shape=circle] (PX) at (0,8) {\scriptsize $\Pauli_\PXm$}; 
	\node[draw, color=black, shape=circle] (PZ) at (0,-6) {\scriptsize $\Pauli_\PZm$}; 
	
	\node[draw, color=black,] (PB) at (-3,0) {\small Rest of $\PauliBasis_k$}; 
	\node[draw, color=black, shape=circle] (IntA) at (10,0) {\scriptsize $\Intro_A$}; 
	\node[color=violet] (Sampling) at (7,-3) {Sampling apparatus}; 
	\node[color=violet] (Hiding) at (7,5) {Hiding apparatus};

	\node[draw, color=black, shape=circle] (IntB) at (15,0) {\scriptsize $\Intro_B$}; 
	
	\node[draw, color=black, shape=circle] (SamA) at (4,-2) {\scriptsize $\Sample_A$};

	\node[draw, color=black, shape=circle] (SamB) at (7,-6) {\scriptsize $\Sample_B$};
	
	\node[draw, color=black, shape=circle] (ReadA) at (8,1) {\scriptsize $\Read_A$};

	\node[draw, color=black, shape=circle] (ReadB) at (14,2) {\scriptsize $\Read_B$};
	
	\node[draw, color=black, shape=circle] (HidA1) at (2,6) {\scriptsize $\Hide A^1$}; 
	
	\node[draw, color=black, shape=circle] (HidAh) at (6,2) {\scriptsize $\Hide A^h$}; 
	\node[draw, color=black, shape=circle] (HidAdots) at (4,4) {\scriptsize $\ddots$};

	\node[draw, color=black, shape=circle] (HidB1) at (6,10) {\scriptsize $\Hide B^1$};
	
	\node[draw, color=black, shape=circle] (HidBh) at (12,4) {\scriptsize $\Hide B^h$};
	
	\node[draw, color=black, shape=circle] (HidBdots) at (9,7) {\scriptsize $\ddots$};

	\draw[cyan, -, dashed] (PX)--(PB);
	\draw[cyan, -, dashed] (PZ)--(PB);
	\draw[black, -, solid] (PZ)--(SamA);
	\draw[black, -, solid] (PZ)--(SamB);
	\draw[black, -, solid] (IntA)--(IntB);
	\draw[black, -, solid] (SamA)--(IntA);
	\draw[black, -, solid] (SamB)--(IntB);
	\draw[black, -, solid] (PX)--(HidA1);
	\draw[black, -, solid] (PX)--(HidB1);
	\draw[black, -, solid] (HidAh)--(ReadA);
	\draw[black, -, dashed] (HidAh)--(HidAdots);
	\draw[black, -, dashed] (HidA1)--(HidAdots);
	\draw[black, -, solid] (HidBh)--(ReadB);
	\draw[black, -, dashed] (HidBh)--(HidBdots);
	\draw[black, -, dashed] (HidB1)--(HidBdots);
	\draw[black, -, solid] (IntA)--(ReadA);
	\draw[black, -, solid] (IntB)--(ReadB);

	\end{tikzpicture}
	\caption{The underlying graph of $\frak{QueRed}(\game)$, where most of the embedded Pauli basis game is hidden. Also, there are $h-2$  extra vertices between $\Hide \cdot^1$ and $\Hide \cdot ^h$.
	}
	\label{fig:Good_augmentation}
\end{figure}
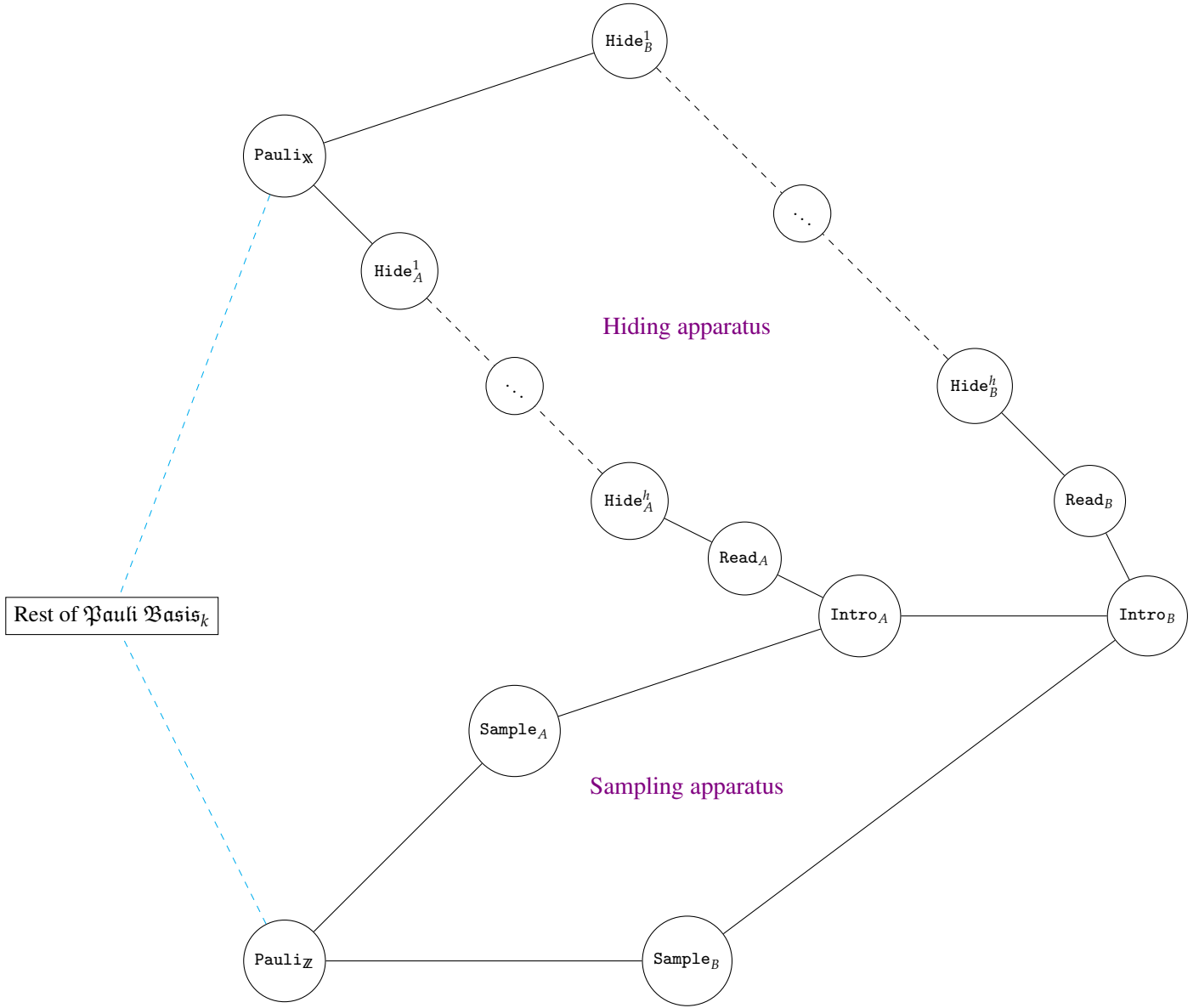

For the sampling apparatus, two vertices $\Sample_A,\Sample_B$ are added and are connected as follows
\[
\Intro_A-\Sample_A-\Pauli_\PZm-\Sample_B-\Intro_B\ .
\]
For the hiding apparatus, $2h+2$ vertices are added   --- $\Read_A,\Read_B,$ and for every $1\leq j\leq h$ the vertices $\Hide A^j,\Hide B^j$ --- and are connected as follows
\begin{equation}\label{eq:augmented_edges_quered_version}
\begin{split}
\Intro_A-\Read_A-\Hide A^h -...-\Hide A^1 -\Pauli_\PXm - \Hide B^1 -...-\Hide B^h -\Read_B-\Intro_B\ .
\end{split}
\end{equation}
See Figure \ref{fig:Good_augmentation} for a graphical view of the underlying graph of $\frak{QueRed}(\game)$.

\textbf{Question distribution of the question reduced game}:\footnote{The  distribution that we eventually use is slightly different, as the sampler of this game needs to be induced by CLMs so that we can iterate compression. This is handled in Section \ref{sec:typed_CL_samplings}, where we provide  a distribution such that for every edge the probability is the same as this one up to some global constant factor independent of $k,\mathscr{B}$ or $\game$, though it may depend on $h$.}
With probability $\nicefrac{1}{4}$ do one of the following ---
\begin{itemize}
	\item  Sample an edge from  $\PauliBasis_k$ according to the appropriate distribution therein.
	\item Sample the single edge $\Intro_A-\Intro_B$ from $\Introspect(\game)$.
	\item Sample a uniformly random edge from the Sampling apparatus.
	\item Sample a uniformly random edge from the Hiding apparatus.
\end{itemize}

\textbf{Lengths and formal generating sets for the augmented vertices of question reduction} (which are almost the same as in Section \ref{sec:baby_question_reduction}):

\emph{Sampling apparatus} ---  The readable length of $\Sample_A$ (and $\Sample_B$) is $k+\Lambda$, and its unreadable length is $\Lambda$. We associate with it the formal generators
\[\begin{split}
S^\frR_{\Sample_A}&=\mathsf{SamZ}^{A}\sqcup \mathsf{SamAns}^{A,\frR}=\{\mathsf{SamZ}^{A,i},\mathsf{SamAns}^{A,\frR,j}\mid 1\leq i\leq k ,1\leq j\leq \Lambda\},\\
S^\frL_{\Sample_A}&=\mathsf{SamAns}^{A,\frL}=\{\mathsf{SamAns}^{A,\frL,j}\mid 1\leq j\leq \Lambda\}.
\end{split}
\]
and similarly for $\Sample_B$. Namely, answers are formatted as $(z_{sam},a_{sam}^\frR,a_{sam}^\frL)$, where $z_{sam}\in \FF_2^k$, and $a_{sam}^\frR,a_{sam}^\frL\in \FF_2^\Lambda$.

\emph{Hiding apparatus} ---
\begin{itemize}
	\item The readable length of $\Read_A$ (respectively $\Read_B$) is $k+\Lambda$, and its unreadable length is $k+\Lambda$ as well. We associate with it the formal generators
	\[
	\begin{split}
	S^\frR_{\Read_A}&=\mathsf{ReadQue}^{A}\sqcup\mathsf{ReadAns}^{A,\frR}=\{\mathsf{ReadQue}^{A,i},\mathsf{ReadAns}^{A,\frR,j}\mid 1\leq i\leq k ,1\leq j\leq \Lambda\},\\
	S^\frL_{\Read_A}&=\mathsf{ReadPerp}^{A}\sqcup\mathsf{ReadAns}^{A,\frL}=\{\mathsf{ReadPerp}^{A,i},\mathsf{ReadAns}^{A,\frL,j}\mid 1\leq i\leq k,1\leq j\leq \Lambda\},
	\end{split}
	\]
	and similarly for $\Read_B$. Namely, answers are formatted as $(\mttx_{read},a_{read}^\frR,\nu_{read},a_{read}^\frL)$, where $\mttx_{read}\in \FF_2^k,a_{read}^\frR,a_{read}^\frL\in \FF_2^\Lambda$ and $\nu_{read}\in \FF_2^{k}$ (and for $B$, $(\mtty_{read},b_{read}^\frR,\mu_{read},b_{read}^\frL)$ in the appropriate spaces).
	\item The readable length of $\Hide A^j$ (respectively $\Hide B^j$) is $k$, and its unreadable length is $k$. We associate with it the formal generators
	\[
	\begin{split}
	S^\frR_{\Hide A^j}&=\mathsf{Hide}^j\sQue^{A}=\{ \mathsf{Hide}^j\sQue^{A,i}\mid 1\leq i\leq k\},\\
	S^\frL_{\Hide A^j}&=\mathsf{Hide}^j\mathsf{Perp}^{A}=\{\mathsf{Hide}^j\mathsf{Perp}^{A,i}\mid 1\leq i\leq k\},
	\end{split}
	\]
	and similarly for $\Hide B^j$. Namely, the answer is formatted as $(\mttx_{hide\ j},\nu_{hide\ j})\in \FF_2^{2k}$ (respectively \\
	$(\mtty_{hide\ j},\mu_{hide\ j})\in \FF_2^{2k}$).
\end{itemize}

\textbf{Decision procedure of the augmented edges in the question reduced game}: This is essentially the description of the controlled linear constraints function $L_{\mttx\mtty}$ of $\frak{QueRed}(\game)$, but we use the phrase ``check that'' repeatedly, by which we mean ``add this sequence of linear constraints to the image of $L_{\mttx\mtty}$ and the canonical verifier will check them''.
\\

\emph{Sampling apparatus} ---
\begin{enumerate}[label=\textcolor{black}{(\arabic*)}, ref= (\arabic*)]
	\item  In case $\Pauli_\PZm-\Sample_A$ (respectively $\Pauli_\PZm-\Sample_B$) is sampled,  check that 
	\begin{equation}\label{eq:PZ_vs_Sample_quered_defn}
	\forall 1\leq i\leq k\  \colon \ \ \gamma(\sP\sZ^i)=\gamma(\mathsf{SamZ}^{A,i})\ .  
	\end{equation}
	In other words, if $z$ is the answer to $\Pauli_\PZm$, then check that $z=z_{sam}$.
	\item  In case $\Intro_A-\Sample_A$ (respectively $\Intro_B-\Sample_B$) is sampled, first check that 
	\begin{equation}\label{eq:linear_checks_over_que_vs_sam_quered}
	\forall  1\leq j\leq\Lambda \  \colon \ \ \gamma(\sAns^{A,\cdot,j})=\gamma(\mathsf{SamAns}^{A,\cdot,j})
	\end{equation}
	and then check that 
	\begin{equation}\label{eq:que_vs_samZ_check_quered}
	\frS^A(\gamma(\mathsf{SamZ}^{A,1}),...,\gamma(\mathsf{SamZ}^{A,k}))=(\gamma(\sQue^{A,1}),...,\gamma(\sQue^{A,k}))\ .
	\end{equation}
	In other words, if $(z_{sam},a_{sam}^\frR,a_{sam}^\frL)$ is the answer to $\Sample_A$, and $(\mttx,a^\frR,a^\frL)$ is the answer to $\Intro_A$, then check that $\mttx=\frS^A(z_{sam}), a^\frR=a_{sam}^\frR$ and $a^\frL=a_{sam}^\frL$.
	
	Note that the check \eqref{eq:que_vs_samZ_check_quered} is {not} linear (because $\frS^A$ is only \emph{conditionally} linear, not linear). But, as both $\mathsf{SamZ}^{\cdot,\cdot}$ and $\sQue^{\cdot,\cdot}$ are readable variables, it can be tailored  as follows:  $L_{\Intro_A\ \Sample_A}(\mttx,a^\frR,z_{sam},a^\frR_{sam})$ contains all constraints induced by \eqref{eq:linear_checks_over_que_vs_sam_quered}, adds no additional linear constraints if \eqref{eq:que_vs_samZ_check_quered} is satisfied, and adds $\{\sJ\}$ as  a constraint if \eqref{eq:que_vs_samZ_check_quered} is not satisfied (which translates to definite rejection).  
\end{enumerate}

\emph{Hiding apparatus} ---
\begin{enumerate}[label=\textcolor{black}{(\arabic*)}, ref= (\arabic*)]
	\item  In case $\Intro_A-\Read_A$ (respectively $\Intro_B-\Read_B$) is sampled,  check that 
	\begin{equation}\label{eq:intro_vs_read_quered_defn}
	\forall  1\leq i\leq k, 1\leq j\leq\Lambda  \  \colon \ \ \gamma(\sAns^{A,\cdot,j})=\gamma(\mathsf{ReadAns}^{A,\cdot,j})\;,\quad \gamma(\sQue^{A,i})=\gamma(\mathsf{ReadQue}^{A,i})\ .
	\end{equation}
	In other words, if $(\mttx_{read},a_{read}^\frR, \nu_{read}, a_{read}^\frL)$ is the answer to $\Read_A$, and $(\mttx,a^\frR,a^\frL)$ is the answer to $\Intro_A$, check that $\mttx_{read}=\mttx$, $a^\frR_{read}=a^\frR$ and $a^\frL_{read}=a^\frL$.
	\item  In case $\Hide A^h-\Read_A$ (respectively $\Hide B^h-\Read_B$) is sampled:   First check that 
	\begin{equation}\label{eq:read_vs_hide^h_quered_defn1}
	\forall  1\leq i\leq k \  \colon \ \ \gamma(\mathsf{Hide}^h\mathsf{Perp}^{A,i})=\gamma(\mathsf{ReadPerp}^{A,i})\ .
	\end{equation}
	Namely, if we denote by $(\mttx_{read},a_{read}^\frR,\nu_{read},a_{read}^\frL)$ the answer to $\Read_A$, and by $(\mttx_{hide\ h},\nu_{hide\ h})$ the answer to $\Hide A^h$, \eqref{eq:read_vs_hide^h_quered_defn1} checks that $\nu_{hide\ h}=\nu_{read}$. In addition, check that $\mttx_{hide\ h}$ is the $(h-1)$-prefix \eqref{eq:def_of_leqj_prefix} of $\mttx_{read}$, namely that $(\mttx_{read})_{\leq h-1}=\mttx_{hide\ h}$, or equivalently
	\begin{equation}\label{eq:read_vs_hide^h_quered_defn2}
	\begin{split}
	\forall i\in I^{\mttx_{read}}_{<h}\ \colon \ \ \gamma(\mathsf{Hide}^h\sQue^{A,i})&=\gamma(\mathsf{Read}^h\sQue^{A,i})\ ,\\
	\forall i\in I^{\mttx_{read}}_{h}\ \colon\ \ \gamma(\mathsf{Hide}^h\sQue^{A,i})&=0\ .
	\end{split}
	\end{equation}
	Note that although these are linear checks, they  depend on the value of $\mttx_{read}$. But, as $\mathsf{ReadQue}^{\cdot,\cdot}$ are readable, this is allowed in the tailored category. 
	
	\item \label{clause:hide_vs_PauliX_in_quered} In case $\Hide A^1-\Pauli_{\PXm}$ (respectively $\Hide B^1-\Pauli_{\PXm}$) is sampled, let $\mttx:=\mttx_{hide\ 1}=(\gamma(\mathsf{Hide}^1\sQue^{A,i}))_{i=1}^k$, and check that 
	\begin{equation}\label{eq:hide^1_vs_PauliX_quered_defn1}
	(\frSS^{A,\mttx}_1)^\perp(\gamma(\sP\sX^i))_{i=1}^k=(\gamma(\mathsf{Hide}^1\mathsf{Perp}^i))_{i=1}^k\ ,
	\end{equation}
	where $(\frSS^{A,\mttx}_1)^\perp$ is the extended perpendicular $1^{\rm st}$ $\mttx$-seeded CLM defined in \cref{clause3:prop_of_input_game_to_quered}. In addition, check that 
	\begin{equation}\label{clause:hide_vs_PauliX_in_quered2}
	\forall i\in [k]\ \colon\ \ \gamma(\mathsf{Hide}^1\sQue^{A,i})=0\ .  
	\end{equation}
	In other words, if  $(\mttx_{hide\ 1},\nu_{hide\ 1})$ is the answer to $\Hide A^1$ and $\chi$ is the answer to $\Pauli_{\PXm}$, check that $\mttx_{hide\ 1}=\vec 0$, and that 
	\begin{equation}\label{clause:hide_vs_PauliX_in_quered3}
	(\frSS^{A,\mttx_{hide\ 1}}_1)^\perp(\chi)=\nu_{hide\ 1}\ .
	\end{equation}
	This is the same as for $\chi^{V_{>1}}=\nu_{hide\ 1}^{V_{>1}}$ and $(\frS^A_1)^\perp(\chi^{V_1})=\nu_{hide\ 1}^{V_1}$. Note that, as $(\frSS_1^{A,\mttx})^\perp\colon \FF_2^k\to \FF_2^k$ is a linear map, this check can be tailored appropriately. 
	\item \label{clause:hide_vs_hide_in_quered} In case $\Hide A^j-\Hide A^{j-1}$ (respectively $\Hide B^j-\Hide B^{j-1}$) is sampled for $2\leq j\leq h$:  Let $(\mttx_{hide\ j},\nu_{hide\ j})$ be the answer to $\Hide A^j$,  and $(\mttx_{hide\ j-1},\nu_{hide\ j-1})$ the answer to $\Hide A^{j-1}$. Fix $\mttx:=\mttx_{hide\ j}$ as the seed, and note that $\mathsf{Hide}^j\sQue^{A,\cdot}$ are readable variables so we may perform checks that depend non-linearly on them. First check that the $(j-2)^{\rm th}$ prefix  of $\mttx$ (see \eqref{eq:def_of_leqj_prefix}) is equal to  $\mttx_{hide\ j-1}$, namely that $\mttx_{\leq j-2}=\mttx_{hide\ j-1}$, or equivalently
	\begin{equation}\label{eq:hide_vs_hids_quered_defn1}
	\begin{split}
	\forall i\in I^{\mttx}_{<j-1}\ &\colon \ \
	\gamma(\mathsf{Hide}^{j-1}\sQue^{A,i})=\gamma(\mathsf{Hide}^j\sQue^{A,i})\ ,\\ 
	\forall i\in I^{\mttx}_{\geq j-1}\ &\colon \ \ \gamma(\mathsf{Hide}^{j-1}\sQue^{A,i})=0\ ,
	\end{split}
	\end{equation}
	where $I^\mttx_\cdot$ is the set of indices associated with the seeded register subspace $W^\mttx_\cdot$ which was defined in \eqref{eq:reg_subspaces_depending_on_prefixes}.
	In addition, check that $(\frSS^{A,\mttx}_{j})^\perp(\nu_{hide\ j-1})=\nu_{hide\ j}$, namely that 
	\begin{equation}\label{eq:hide_vs_hids_quered_defn2}
	\forall i\in I^\mttx_{\neq j}\ \colon \ \ \gamma(\mathsf{Hide}^j\mathsf{Perp}^{A,i})=\gamma(\mathsf{Hide}^{j-1}\mathsf{Perp}^{A,i})\ ,
	\end{equation}
	and 
	\begin{equation}\label{eq:hide_vs_hids_quered_defn3}
	\forall i\in I^\mttx_j\ \colon\ \ \gamma(\mathsf{Hide}^j\mathsf{Perp}^{A,i})=\sum_{t\in I_j} (\frS^{A,\mttx}_j)^\perp_{it}\gamma(\mathsf{Hide}^{j-1}\mathsf{Perp}^{A,t})\ .
	\end{equation}
	Again, as $(\frSS^{A,\mttx}_j)^\perp$ is linear for every seed $\mttx$,  and $\mttx$ is decided by readable variables, these checks can be tailored appropriately.
\end{enumerate}
\begin{remark}[Restriction notation]\label{rem:recalling_restriction_notation}
	Recall the notation $\cal{V}^{S'}$ (in observable form) and $\cal{Q}^{S'}$ (in projective form) for the restriction  to $\FF_2^{S'}$  of the respective PVMs $\cal{V}$ and $\cal{Q}$  with outcomes in $\FF_2^{S'\sqcup S''}$ (Definition \ref{defn:Data_proccessed_PVM}). 
	For example, in the case of $\frak{QueRed}(\game)$,  recall that 
	\[
	{\sAns^A}={\sAns^{A,\frR}\sqcup \sAns^{A,\frL}}=\{\sAns^{A,\frR,j},\sAns^{A,\frL,j}\}_{j=1}^\Lambda 
	\]
	is the set of $\sAns^A$-variables, which is a subset of $S_{\Intro_A}$. Then, the representation $\cal{V}^{\sAns^A}\colon \FF_2^{{\sAns^A}}\to U(N)$ is induced by the observable form $\cal{V}$ of some strategy $\strategy$ by letting
	\[
	\forall \alpha^\frR,\alpha^\frL \in \FF_2^\Lambda\ \colon \ \ \cal{V}^{\sAns^A}(\alpha^\frR,\alpha^\frL)=\prod_i\cal{V}(\sAns^{A,\frR,i})^{\alpha^\frR_i}\prod_j\cal{V}(\sAns^{A,\frL,j})^{\alpha^\frL_j}\ ,
	\]
	and the PVM $\cal{Q}^{\sAns^A}$ is induced by the projective form $\cal{Q}$ by letting 
	\[
	\cal{Q}^{\sAns^A}_{a^\frR,a^\frL}=\sum_{\mttx\in \FF_2^r}\cal{Q}^{\Intro_A}_{\mttx,a^\frR,a^\frL} \ .
	\]
	We use a similar notation for restrictions to various  subsets of generators at different vertices, such as 
	\[
	\mathsf{ReadQue}^\cdot,\mathsf{ReadPerp}^\cdot,\mathsf{SamZ}^\cdot\;,
	\]
	and so on.
\end{remark}
\begin{remark}[Analysis of the question reducred game $\frak{QueRed}(\game)$]\label{rem:analysis_que_red}
	Let us briefly analyze the properties of a strategy $\strategy$, with $\cal{U}$ being its observable form and $\cal{P}$ its projective form, that passes all edges but $\Intro_A-\Intro_B$ in $\frak{QueRed}(\game)$ {perfectly}:
	\begin{enumerate}[label=\textcolor{black}{(\arabic*)}, ref= (\arabic*)]
		\item Since it passes the copy of $\PauliBasis_k$ perfectly, we can deduce by Claim \ref{claim:almost_perfect_strategies_of_PB_k} and Fact \ref{fact:semi-stability_P_k} (in the case $\eps=0$) that the observables $\strategy$ associates with the generators at the vertices $\Pauli_\PXm$ and $\Pauli_\PZm$ induce the unique (up to direct sums) representation of the Pauli group $\WH_k$ defined in \eqref{eq:def_of_rho_for_P_k}.  Namely, there is some natural number $m\in \mathbb{N}$ such that $\strategy$ acts on $\complex^{\FF_2^k}\otimes \complex^m$, and
		\begin{equation}\label{eq:perfect_Pauli_variables_are_Pauli_measurements}
		\begin{split}
		\forall 1\leq i\leq k\ \colon \ \ \cal{U}(\sP\sX^i)=\PXm^{\otimes e_i}\otimes \Id_m\;,\quad \cal{U}(\sP\sZ^i)=\PZm^{\otimes e_i}\otimes \Id_m\ ,
		\end{split}
		\end{equation}
		or equivalently in representation form
		\[
		\forall \alpha\in \FF_2^k\ \colon \ \ \cal{U}^{\Pauli_\PXm}(\alpha)=\rho^{\PXm}(\alpha)\otimes \Id_m=\PXm^{\otimes \alpha}\otimes \Id_m\;,\quad \cal{U}^{\Pauli_\PZm}(\alpha)=\rho^{\PZm}(\alpha)\otimes \Id_m=\PZm^{\otimes \alpha}\otimes \Id_m\ .
		\]
		\item As $\strategy$ passes the check along the edges $\Pauli_\PZm-\Sample_\cdot$ perfectly, the assignments to the $\sP\sZ$ variables and $\mathsf{SamZ}$ variables are  consistent. Namely,
		\[
		\forall 1\leq i\leq k\ \colon\ \ \cal{U}(\mathsf{SamZ}^{\cdot,i})=\cal{U}(\sP\sZ^i)=\PZm^{\otimes e_i}\otimes \Id_m\ ,
		\]
		and equivalently in projective form 
		\begin{equation}\label{eq:P^SamZ_z_in_perfect_strategy}
		\forall z\in \FF_2^k \ \colon \ \ \cal{P}^{\mathsf{SamZ}^\cdot}_{z}=\mathscr{F}^\PZm_{z}\otimes \Id_m\ .
		\end{equation}
		\item As $\strategy$ passes the checks $\Sample_A -\Intro_A-\Read_A$ perfectly, we can deduce that the assignments to the $\sAns$, $\mathsf{ReadAns}$ and $\mathsf{SamAns}$ variables are consistent:
		\begin{equation}\label{eq:Ans_values_in_perfect_strat_are_consistent}
		\cal{U}^{\mathsf{SamAns}^A}=\cal{U}^{\sAns^A}=\cal{U}^{\mathsf{Read}\sAns^A},     
		\end{equation}
		namely that for every $1\leq j\leq \Lambda$,  $\cal{U}(\mathsf{SamAns}^{A,\cdot,j})=\cal{U}({\sAns^{A,\cdot,j}})=\cal{U}(\mathsf{ReadAns}^{A,\cdot,j})$ (and similarly for $B$). In addition, as the $\sQue^A$-variables are checked to be the $\frS^A$ image of the $\mathsf{SamZ}^A$-variables, and $\mathsf{ReadQue}^A$ are checked to be consistent with $\sQue^A$, we can deduce that 
		\[
		\forall \mttx\in \FF_2^k\ \colon \ \ \cal{P}^{\mathsf{Read}\sQue^A}_{\mttx}=\cal{P}^{\sQue^A}_{\mttx}=\cal{P}^{\mathsf{SamZ}^A}_{[\frS^A(\cdot)=\mttx]}=\sum_{z\colon \frS^A(z)=\mttx} \mathscr{F}^\PZm_z\otimes \Id_m\ ,
		\]
		where $\cal{P}^{\mathsf{SamZ}^A}_{[\frS^A(\cdot)=\mttx]}$ is the $\frS^A$-evaluated PVM (Definition \ref{defn:Data_proccessed_PVM}) associated to $\cal{P}^{\mathsf{SamZ}^A}_z$, and the last equality is due to \eqref{eq:P^SamZ_z_in_perfect_strategy}.
		\item As $\strategy$ passes the checks $\Read_A-\Hide A^h-...-\Hide A^2-\Hide A^1$ perfectly, we can deduce that for every $1\leq r\leq h-1$ and $\mttx\in \FF_2^k$,
		\begin{equation}\label{eq:Que_var_in_perfect_strat_are_consistent_quered}
		\cal{P}^{\mathsf{Hide}^{r+1}\sQue^A}_{\mttx}=\cal{P}^{\mathsf{ReadQue}^A}_{[(\cdot)_{\leq r}=\mttx]}=\sum_{z\colon \frS^A_{\leq r}(z)=\mttx}\mathscr{F}^\PZm_z\otimes\Id_m\;,
		\end{equation}
		where $(\cdot)_{\leq r}$ is the $r$-prefix function \eqref{eq:def_of_leqj_prefix} --- which is part of the data of the CLM $\frS^A$ --- and $\frS^A_{\leq r}(z)$  is as   in \eqref{eq:def_frS_leqj}. 
		\item As $\strategy$ passes the checks $\Pauli_\PXm-\Hide A^1-...-\Hide A ^{h}-\Read$ perfectly, and using \eqref{eq:perfect_Pauli_variables_are_Pauli_measurements} and \eqref{eq:Que_var_in_perfect_strat_are_consistent_quered}, we get for every $1\leq r\leq h$ and $ \mttx,\nu\in \FF_2^k$ that
		\[
		\cal{P}^{\Hide A^{r}}_{\mttx,\nu}=\mathscr{F}^{\PZm}_{[\frS^A_{<r}(\cdot)=\mttx]}\cdot \mathscr{F}^{\PXm}_{[(\frSS^{A,\mttx}_{\leq r})^\perp(\cdot)=\nu]}=\sum_{\substack{z\in \FF_2^k \\ \frS^A_{<r}(z)=\mttx}}\sum_{\substack{w\in \FF_2^k\\ (\frSS_{\leq r}^{A,\mttx})^\perp(w)=\nu}}\mathscr{F}^\PZm_z\mathscr{F}^\PXm_w\otimes\Id_m\ ,
		\]
		where $\frSS^{A,\mttx}_{\leq j}$ are again the extensions of the perpendicular maps that were defined in \cref{clause3:prop_of_input_game_to_quered};  also
		\[
		\forall \mttx,\nu \in \FF_2^k\ \colon\ \ \cal{P}^{\mathsf{ReadQue}^A\sqcup\mathsf{ReadPerp}^A}_{\mttx,\nu}=\mathscr{F}^\PZm_{[\frS(\cdot)=\mttx]}\mathscr{F}^\PXm_{[(\frS^{A,\mttx})^\perp(\cdot)=\nu]}\otimes \Id_m\ .
		\]
		\item Finally, according to \eqref{eq:Ans_values_in_perfect_strat_are_consistent}, the $\sAns^A$-observables commute with the $\mathsf{ReadPerp}^A$-observables and the $\mathsf{SamZ}^A$-observables. Commuting with the $\mathsf{SamZ}^A$-observables translates to the $\sAns^A$-observables  being of the form $\sum \mathscr{F}^{\PZm}_z\otimes \mathscr{A}^z$. Commuting with the $\mathsf{ReadPerp}^A$-observables implies that $\mathscr{A}^z$ is equal to $\mathscr{A}^{z'}$ whenever $\frS(z)=\frS(z')$. Namely (by repeating all these arguments for $B$ and $\frS^B$ as well), the strategy  $\strategy$ is {honest} (Definition \ref{defn:honest_strat_for_Intro}) when restricted to the copy of $\Introspect(\game)$ in $\mathfrak{QueRed}(\game)$. In particular, its value on $\Introspect(\game)$ is the same as some quantum strategy for $\game$ itself.
	\end{enumerate}

\end{remark}

\begin{remark}
	A reader may notice that to achieve the above goals, we could have dropped the $\Sample_\cdot$ and $\Hide \cdot^\cdot $ vertices altogether, and applied a more direct check (simplifying the augmentation).  Though this is true, it will hinder the perfect completeness case, as we seek perfect $\ZPC$ strategies, in particular strategies that commute along edges, which is problematic without these buffer questions.
\end{remark}

\begin{theorem}[Completeness and Soundness of Question Reduction]\label{thm:complet_sound_quered}
	Let $k$ be a positive integer, $\mathscr{B}$ a tuple of $n$ vectors in $\FF_2^k$ that induce an $[n,k,d]$-code (Section \ref{subsec:error_correctoin_prelims}), and  $\game$ a tailored game satisfying \labelcref{clause1:prop_of_input_game_to_quered},\labelcref{clause2:prop_of_input_game_to_quered} and \labelcref{clause3:prop_of_input_game_to_quered} from the beginning of this section. Then, the question reduced  game $\mathfrak{QueRed}(\game)=\mathfrak{QueRed}_h(\game,k,\mathscr{B})$ has the following properties:
	\begin{enumerate}[label=\textcolor{black}{(\arabic*)}, ref= (\arabic*)]
		\item \label{clause:completeness_of_quered}\emph{Completeness}: If $\game$ has a perfect $\ZPC$ strategy, then so does $\frak{QueRed}(\game)$.
		\item \label{clause:soundness_of_quered}\emph{Soundness}: If $\frak{QueRed}(\game)$ has a strategy with value $1-\eps$, then $\game$ has a strategy with value  at least 
		$$1-O(h^2\cdot 2^h\cdot(1+\nicefrac{k^2}{d^2})\cdot \eps^{\nicefrac{1}{8}})\;.$$ 
		\item \label{clause:entanglement_of_quered}\emph{Entanglement}: For every $\eps>0$, 
		\[
		\Ent(\frak{QueRed}(\game),1-\eps)\geq 2^{k}\cdot\left(1-O((1+\nicefrac{k^2}{d^2})\eps)\right)\cdot\Ent\big(\game,1-O\big(h^2\cdot 2^h\cdot(1+\nicefrac{k^2}{d^2})\cdot \eps^{\nicefrac{1}{8}}\big)\big)\;.
		\]
	\end{enumerate}
\end{theorem}

\begin{remark}
	The underlying combinatorial game (Definition \ref{defn:combinatorial_game_underlying_tailored}) of $\frak{QueRed}(\game)$ is the same (when $\mathscr{B}$ for $\PauliBasis_k$ is chosen appropriately) as the question reduction applied in \cite{MIPRE}. This means that the soundness proof therein already covers the soundness of Theorem \ref{thm:complet_sound_quered}. Although this is true, and the reader familiar with \cite{MIPRE} may even prefer their soundness proof, we include our own proof here. They are essentially the same, up to our proof leaning more on the observable perspective, which may be easier for readers who approach this result from the group stability community.
	
	Although we construct a perfect $\ZPC$ strategy  $\varphi$ in our completeness proof, which is not something the authors of \cite{MIPRE} were concerned about, this strategy is (essentially) the same as their perfect complete strategy (Section 8.3.2 therein). So, if one seeks more details regarding the perfect value of our strategy, then they can seek there as well.
	
\end{remark}

\subsubsection*{Proof of perfect completeness \labelcref{clause:completeness_of_quered}}

Assume that $\game$ has a perfect $m$-dimensional $\ZPC$ strategy  $\strategy$, and let $\cal{U}$ be its observable (and representation) form and $\cal{P}$ be its projective form. Then, we can induce from it a perfect honest $\ZPC$ strategy $\strategy'$ for $\Introspect(\game)$, with $\cal{V}$  being its observable form and $\cal{Q}$ its projective form, 
acting on $\complex^{\FF_2^k}\otimes\complex^m$ (see \eqref{eq:defn_honest_strategy1} and \eqref{eq:defn_honest_strategy2} in Definition \ref{defn:honest_strat_for_Intro}, and Claim \ref{claim:complete_sound_honest_intro}). Let us extend $\strategy'$ to the other vertices so it becomes a perfect $\ZPC$ strategy for $\frak{QueRed}(\game)$. First, let $\cal{V}^{\Pauli_\PXm}$ and $\cal{V}^{\Pauli_\PZm}$ be $\rho^{\PXm}\otimes\Id_m$ and $\rho^{\PXm}\otimes\Id_m$, namely
\[
\forall \alpha\in \FF_2^k\ \colon \ \ \cal{V}^{\Pauli_\PXm}(\alpha)=\PXm^{\otimes \alpha}\otimes \Id_m\quad,\quad \cal{V}^{\Pauli_\PZm}(\alpha)=\PZm^{\otimes \alpha}\otimes \Id_m\ .
\] 
By Claim \ref{claim:completeness_PB_k}, it can be extended to the rest of the $\PauliBasis_k$ vertices in a $\ZPC$-manner such that on the copy of $\PauliBasis_k$ in $\frak{QueRed}(\game)$ it has value $1$.
The rest of the PVMs are forced on us via the consistency checks along the augmented edges (see the  analysis in Remark \ref{rem:analysis_que_red}); recall the restriction notations from Remark \ref{rem:recalling_restriction_notation} and the data processing notation from Definition \ref{defn:Data_proccessed_PVM}. For the sampling apparatus, we have 
\begin{align}
\cal{Q}^{\sAns^A}= \cal{Q}^{\mathsf{SamAns}^A}\ &,\ \cal{Q}^{\sAns^B}= \cal{Q}^{\mathsf{SamAns}^B}\ ,\ \\
\cal{Q}^{\Pauli_\PZm}=&\cal{Q}^{\mathsf{SamZ}^A}=\cal{Q}^{\mathsf{SamZ}^B}\ ,\\
\cal{Q}^{\sQue^A}=\cal{Q}^{\mathsf{SamZ}^A}_{[\frS^A(\cdot)=\cdot]}\ &
,
\cal{Q}^{\sQue^B}=\cal{Q}^{\mathsf{SzmZ}^B}_{[\frS^B(\cdot)=\cdot]}
\ .
\end{align}
For the PVMs in the hiding apparatus, we elaborate more as the notation may be confusing. For the $\Read_\cdot$ vertices, recalling the linear map $(\frS^{A,\mttx})^\perp=(\frS^{A,\mttx}_{\leq h})^\perp$ from  \cref{clause3:prop_of_input_game_to_quered} and Remark \ref{rem:extended_perpendicular_maps_along_a_CLM}, we have 
\begin{equation}\label{eq:read_PVMs_completeness_quered}
\forall \mttx,\nu\in \FF_2^k\ ,\  a^\frR,a^\frL\in \FF_2^\Lambda\ \colon \ \ \cal{Q}^{\Read_A}_{\mttx,a^\frR,\nu,a^\frL}=\sum_{z\in \FF_2^k\colon \frS^A(z)=\mttx}\mathscr{F}^\PZm_z\otimes \cal{P}^\mttx_{a^\frR,a^\frL} \cdot \sum_{\chi\in \FF_2^k\colon (\frS^{A,\mttx})^\perp(\chi)=\nu}\mathscr{F}^\PXm_\chi\otimes \Id_m\ ,
\end{equation}
where $\cal{P}$ is the projective form of the original perfect strategy for $\game$ (and similarly for $B$). Following the definition of an honest strategy, \eqref{eq:defn_honest_strategy1} and \eqref{eq:defn_honest_strategy2}, 
it is straightforward to check that the restrictions to the $\mathsf{ReadQue}^\cdot$ and $\mathsf{ReadAns}^\cdot$ variables satisfy
\[
\begin{split}
\cal{Q}^{\mathsf{ReadQue}^A}=\cal{Q}^{\mathsf{Que}^A}\ &,\ \cal{Q}^{\mathsf{ReadQue}^B}=\cal{Q}^{\mathsf{Que}^B}\ ,\ \\
\cal{Q}^{\mathsf{ReadAns}^A}=\cal{Q}^{\mathsf{Ans}^A}\ &,\ \cal{Q}^{\mathsf{ReadAns}^B}=\cal{Q}^{\mathsf{Ans}^B}\ .
\end{split}
\]
Though we claim that \eqref{eq:read_PVMs_completeness_quered} is a PVM, this is not obvious from its definition --- one needs to be convinced that the right sum commutes with the left sum for it to be an orthogonal projection. Let us prove that. By Corollary \ref{cor:seeded_and_unseeded_CLMs}, $\sum_{z\colon \frS^{A}(z)=\mttx}=\sum_{z\colon \frS^{A,\mttx}(z)=\mttx}$, and thus by Claim \ref{claim:inverse_fourier_of_data_processed_is_structured} the product of the two sums does commute. Moreover, as the readable variables are data processed versions of $\cal{V}^{\Pauli_\PZm}=\rho^\PZm\otimes \Id_m$, which is diagonal, we deduce that this PVM is readably $Z$-aligned. For the $\mathsf{ReadAns}$ and $\mathsf{ReadQue}$ observables, as they are consistent with $\sAns$ and $\sQue$ at $\Intro$, they agree with an honest strategy induced by a $\ZPC$ one and hence consist of signed permutations. For the $\mathsf{ReadPerp}$ observables, taking the inverse Fourier transform of \eqref{eq:read_PVMs_completeness_quered}, we have
\[
\cal{V}^{\mathsf{ReadPerp}^A}(\alpha)=\sum_{\mttx}\mathscr{F}^\PZm_{[\frS^A(\cdot)=\mttx]}\PXm^{\otimes \alpha \cdot (\frS^{A,\mttx})^\perp}\mathscr{F}^\PZm_{[\frS^A(\cdot)=\mttx]}\otimes \Id_m\ ,
\]
where the sum is over all $\mttx$ in the image of $\frS^A$ --- namely, this is a block diagonal matrix whose blocks are corners of the permutation matrices $\PXm^{\otimes \cdot}$. Hence, if this matrix is invertible, it is a permutation matrix.  As $\mathscr{F}^\PZm_{[\frS^A(\cdot)=\mttx]}$ commutes with $\PXm^{\otimes \alpha \cdot (\frS^{A,\mttx})^\perp}$ (by Corollary \ref{cor:seeded_and_unseeded_CLMs} and Claim \ref{claim:inverse_fourier_of_data_processed_is_structured}), it is its own inverse, which proves this is indeed a signed permutation PVM.

We are left to define the PVMs at the $\Hide \cdot^j$ vertices for $1\leq j\leq h$. This is done in a similar way to the $\Read_\cdot$ vertices, and is forced on us by the consistency checks along the augmented edges. In projective form, 
\begin{equation}\label{eq:PVM_at_Hide_vertex}
\begin{split}
\forall \mttx,\nu\in \FF_2^k\ \colon \ \ \cal{Q}^{\Hide A^j}_{\mttx,\nu}&=\mathscr{F}^\PZm_{[\frS^A_{<j}(\cdot)=\mttx]}\cdot \mathscr{F}^\PXm_{[(\frSS^{A,\mttx}_{\leq j})^\perp(\cdot)=\nu]}\otimes \Id_m\\
&=\sum_{\substack{z\in \FF_2^k\\ \frS^A_{<j }(z)=\mttx}}\sum_{\substack{\alpha\in \FF_2^k \\  (\frSS^{A,\mttx}_{\leq j})^\perp(\alpha)=\nu}}\mathscr{F}^\PZm_z\mathscr{F}^\PXm_\alpha\otimes \Id_m\ .
\end{split}
\end{equation}
By Corollary \ref{cor:seeded_and_unseeded_CLMs}, $\frS^A_{<j}(z)=\mttx$  if and only if $\frSS^{A,\mttx}_{<j}(x)=\mttx$. Hence, by Claim \ref{claim:inverse_fourier_of_data_processed_is_structured}, this is indeed a PVM.  Actually, Claim \ref{claim:inverse_fourier_of_data_processed_is_structured} shows that  this strategy commutes along all $\Hide \cdot^j-\Hide \cdot^{j+1}$ edges, as $\mathscr{F}^\PZm_{[\frS^{A}_{\bf \leq j}(\cdot)=\mttx]}=\mathscr{F}^\PZm_{[\frSS^{A,\mttx}_{\bf \leq j}(\cdot)=\mttx]}$ commutes with $\mathscr{F}^\PXm_{[(\frSS^{A,\mttx}_{\leq j})^\perp(\cdot)=\nu]}$ according to it. Furthermore, in a similar manner to the $\Read$-vertex PVM,  we can deduce that this is a readably $Z$-aligned signed permutation PVM.
All in all,   $\strategy'=\{\cal{Q}\}$ is indeed a $\ZPC$-strategy (all non-hide to non-hide edges are clearly commuting). The fact that this strategy is perfect can be checked by the reader, with the help of the analysis of perfect strategies in Remark \ref{rem:analysis_que_red}.
This finishes the perfect completeness proof.

\subsubsection*{Proof of soundness \labelcref{clause:soundness_of_quered} and entanglement lower bound \labelcref{clause:entanglement_of_quered}}

The idea in the soundness proof is to perturb an almost perfect strategy for $\frak{QueRed}(\game)$ to become a strategy that passes all edges of $\frak{QueRed}(\game)$ perfectly, except for the single edge of $\Introspect(\game)$.
The way we perturb $\strategy$ to be perfect on all edges (except for  $\Intro
_A-\Intro_B$), roughly follows the  analysis of such strategies in Remark \ref{rem:analysis_que_red}. 
\\

\begin{claim}\label{claim:1_soundness_quered}
	Let $\strategy=\{\cal{U}\}$ be an $N$-dimensional strategy for $\mathfrak{QueRed}(\game)$ with value $1-\eps$. Let $\eps'=(1+\nicefrac{k^2}{d^2})\eps$. Then, there is a strategy $\strategy'=\{\cal{W}\}$, acting on $\complex^{\FF_2^k}\otimes\complex^m$, such  that: 
	\begin{enumerate}
		\item (Almost Perfect) $\strategy'$ has value of at least $1-O(\sqrt{\eps'})$; 
		\item (Perfect on Pauli basis) $\strategy'$ passes the edges from the copy of $\PauliBasis_k$ perfectly; 
		\item (Uses a specific representation of the Pauli group) $\strategy'$ satisfies 
		\[
		\cal{W}^{\Pauli_\PXm}=\rho^\PXm\otimes \Id_m\quad \textrm{and}\quad \cal{W}^{\Pauli_\PZm}=\rho^\PZm\otimes \Id_m\ ,
		\]
		where $\rho$ is the representation specified in Definition~\ref{defn:rho}.
	\end{enumerate}
\end{claim}

\begin{proof}
	As the copy of $\PauliBasis_k$ is played with probability $\nicefrac{1}{4}$ when running $\mathfrak{QueRed}(\game)$, the restriction of $\strategy$ to the vertices of $\PauliBasis_k$ passes it with probability of at least $1-4\eps$. Hence, by Claim \ref{claim:almost_perfect_strategies_of_PB_k}, Fact \ref{fact:semi-stability_P_k} and Claim \ref{claim:L1_closeness_of_unirep_implies_Linfty},   there is a partial isometry $\omega\colon \complex^N\to \complex^{\FF_2^k}\otimes \complex^m$ such that 
	\begin{equation}\label{eq:Pauli_close_to_Pauli_matrices_in_quered}\begin{split}
	\forall \alpha\in \FF_2^k\ \colon \ \ \left\|\omega\cal{U}^{\Pauli_\PXm}(\alpha)\omega^*-\PXm^{\otimes \alpha}\otimes \Id_m\right\|_{hs}^2&\; ,\quad\left\|\omega\cal{U}^{\Pauli_\PZm}(\alpha)\omega^*-\PZm^{\otimes \alpha}\otimes \Id_m\right\|_{hs}^2\leq O(\nicefrac{k^2\eps}{d^2})\\
	&\textrm{and}\\
	1-\tau(\omega^*\omega)&\; ,\quad \ 1-\tau(\omega\omega^*)\leq O(\nicefrac{k^2\eps}{d^2})\;
	\end{split}\end{equation}
	As $\eps'=(1+\nicefrac{k^2}{d^2})\eps$,  the above quantities are all $O(\eps')$. Letting $\cal{W}^{\Pauli_\PZm}(\alpha)=\PZm^{\otimes \alpha}\otimes \Id_m$ and $\cal{W}^{\Pauli_\PXm}(\alpha)=\PXm^{\otimes \alpha}\otimes \Id_m$,  we can use Claim \ref{claim:completeness_PB_k} to extend  $\cal{W}$  to a perfect strategy  for $\PauliBasis_k$. As $1-\tau(\omega^*\omega),1-\tau(\omega\omega^*)\leq O(\eps')$, we can use orthogonalization (Fact \ref{fact:orthogonalization}) and extend $\cal{W}$ such that for every non-$\PauliBasis_k$ vertex $\mttx$,
	\begin{align}
	\forall \alpha\colon S_\mttx\to \FF_2\ \colon \ \ \|\omega\cal{U}^\mttx(\alpha)\omega^*-\cal{W}^\mttx(\alpha)\|_{hs}^2\leq O(\eps').\label{eq:W_is_close_on_non_PB_vertices}
	\end{align}
	Now $\cal{W}$ induces a representation on all vertices of $\frak{QueRed}(\game)$, and is thus a strategy for it. We already assured that $\strategy'=\{\cal{W}\}$ passes the copy of $\PauliBasis_k$ perfectly, and we chose it such that $\cal{W}^{\Pauli_{\PXm}}=\rho^\PXm\otimes \Id$ and  $\cal{W}^{\Pauli_{\PZm}}=\rho^\PZm\otimes \Id$. Hence, conditions $2.$ and $3.$ are satisfied.  Finally, equations \eqref{eq:Pauli_close_to_Pauli_matrices_in_quered} and \eqref{eq:W_is_close_on_non_PB_vertices} imply that across any non-$\PauliBasis_k$ edge, the observables of $\cal{W}$ are $O(\eps')$-close to those of $\cal{U}$, hence they pass all these edges with probability at most $O(\sqrt{\eps'})$ worse than $\cal{U}$ (Claim \ref{claim:close_strat_implies_close_correlations}). 
	This implies $\val(\strategy',\frak{QueRed}(\game))\geq 1-O(\sqrt{\eps'})$, finishing the proof.
\end{proof}

\begin{claim}\label{claim:2_soundness_quered}
	Let $\strategy$ be a strategy for $\frak{QueRed}(\game)$ with value $1-\eps$, where $\cal{U}$ is its observable form and $\cal{P}$ its projective from, that acts on $\complex^{\FF_2^k}\otimes \complex^{m}$. Moreover, assume it passes the copy of $\PauliBasis_k$ with probability $1$, and satisfies $\cal{U}^{\Pauli_\PXm}=\rho^\PXm\otimes \Id_m$ and  $\cal{U}^{\Pauli_\PZm}=\rho^\PZm\otimes \Id_m$, or equivalently in projective form $\cal{P}^{\Pauli_\PZm}=\mathscr{F}^\PZm\otimes\Id_m$ and $\cal{P}^{\Pauli_\PXm}=\mathscr{F}^\PXm\otimes\Id_m$. Then, there is a strategy $\strategy'$, with $\cal{Q}$ being its projective form and $\cal{W}$ its observable form, such that:
	\begin{enumerate}
		\item (Almost Perfect) $\strategy'$ has value $1-O(\sqrt{h^3\cdot \eps})$;
		\item (Agrees on Pauli basis) $\strategy'$ agrees with $\strategy$ on the copy of $\PauliBasis_k$;
		\item (Readable variables are consistent with $\PZm$-measurements) $\strategy'$ satisfies
		\begin{align*}
		\forall z,\mttx,\mtty\in \FF_2^k\ ,\ r\in [h]\ \colon\quad \ \cal{Q}^{\mathsf{SamZ}^A}_z&=\cal{Q}^{\mathsf{SamZ}^B}_z=\mathscr{F}^\PZm_z\otimes \Id_m\ ,\\            
		\cal{Q}^{\mathsf{Read}\sQue^A}_{\mttx}&=\cal{Q}^{\sQue^A}_{\mttx}=\mathscr{F}^\PZm_{[\frS^A(\cdot)=\mttx]}\otimes \Id_m=\sum_{z\colon \frS^A(z)=\mttx} \mathscr{F}^\PZm_{z}\otimes \Id_m\ ,\\
		\cal{Q}^{\mathsf{Read}\sQue^B}_{\mtty}&=\cal{Q}^{\sQue^B}_{\mtty}=\mathscr{F}^\PZm_{[\frS^B(\cdot)=\mtty]}\otimes \Id_m=\sum_{z\colon \frS^B(z)=\mtty} \mathscr{F}^\PZm_{z}\otimes \Id_m\ ,\\
		\cal{Q}^{\mathsf{Hide}^r\sQue^A}_{\mttx}&=\mathscr{F}^\PZm_{[\frS^A_{<r(\cdot)=\mttx}]}=\sum_{z\colon \frS^A_{<r}(z)=\mttx}\mathscr{F}^\PZm_z\otimes \Id_m\ ,\\
		\cal{Q}^{\mathsf{Hide}^r\sQue^B}_{\mtty}&=\mathscr{F}^\PZm_{[\frS^B_{<r(\cdot)=\mtty}]}=\sum_{z\colon \frS^B_{<r}(z)=\mtty}\mathscr{F}^\PZm_z\otimes \Id_m\ .
		\end{align*}
		Namely, all consistency checks on the readable variables in $\frak{QueRed}(\game)$ are satisfied.
	\end{enumerate}
\end{claim}

\begin{proof}
	As $\strategy$ has value $1-\eps$, and each edge in the sequence $\Pauli_\PZm-\Sample_\cdot-\Intro_\cdot -\Read_\cdot -\Hide \cdot ^h-...-\Hide \cdot ^1$ is sampled with probability $\Omega(\frac{1}{h})$, we can deduce that $\strategy$ passes each such edge with probability of at least $1-O(h\eps)$. Specifically, from the consistency check along $\Pauli_\PZm-\Sample_\cdot$ together with the fact that small inconsistency implies closeness of PVMs (Proposition \ref{prop:properties_of_distance_and_inconsistency}), we can deduce that 
	\begin{equation}\label{eq:SamZ_close_to_F^Z_z_in_quered}
	\cal{P}^{\mathsf{SamZ}^\cdot}\approx_{O(h\eps)} \cal{P}^{\Pauli_\PZm} = \mathscr{F}^\PZm\otimes \Id_m\ .
	\end{equation}
	By the comparison along $\Sample_\cdot-\Intro_\cdot$ together with Observation \ref{obs:data_processing1}, we can deduce that 
	\begin{equation}
	\cal{P}^{\mathsf{SamZ}^\cdot}_{[\frS^\cdot(\cdot)=\cdot]}\approx_{O(h\eps)} \cal{P}^{\sQue^\cdot}\ .
	\end{equation}
	In combination with \eqref{eq:SamZ_close_to_F^Z_z_in_quered} and the semi-triangle inequality for PVMs (Item $2.$ in Proposition \ref{prop:properties_of_distance_and_inconsistency}), we get
	\begin{equation}\label{eq:Que_variables_are_data_processed_F^Z_z_measurements_quered}
	\cal{P}^{\sQue^\cdot}\approx_{O(h\eps)} \mathscr{F}^\PZm_{[\frS^\cdot(\cdot)=\cdot]}\otimes \Id_m\ .
	\end{equation}
	Just as a sanity check for the reader, the  closeness claim on PVMs in \eqref{eq:Que_variables_are_data_processed_F^Z_z_measurements_quered} means, for the $A$ vertices, that
	\[
	\sum_{\mttx\in \FF_2^k} \Big\|\cal{P}^{\sQue^A}_\mttx-\sum_{\substack{z\in \FF_2^k\\ \frS^A(z)=\mttx}}\mathscr{F}^\PZm_z\otimes \Id_m\Big\|_{hs}^2 \leq O(h\eps)\ .
	\]
	By the comparison along $\Intro_\cdot-\Read_\cdot$, the $\mathsf{Que}^\cdot$ and $\mathsf{ReadQue}^\cdot$ observables are highly consistent with each other, and as high consistency implies closeness, we have
	\begin{equation}
	\cal{P}^{\mathsf{ReadQue}^\cdot}\approx_{O(h\eps)} \cal{P}^{\sQue^\cdot}\ .
	\end{equation}
	In combination with \eqref{eq:Que_variables_are_data_processed_F^Z_z_measurements_quered} and the semi-triangle inequality, we get
	\begin{equation}\label{eq:ReadQue_are_data_processed_F^Z_z_measurements_quered}
	\cal{P}^{\mathsf{Read}\sQue^\cdot}\approx_{O(h\eps)} \mathscr{F}^\PZm_{[\frS^\cdot(\cdot)=\cdot]}\otimes \Id_m\ .
	\end{equation}
	By the consistency checks along $\Read_\cdot -\Hide\cdot^h-...-\Hide\cdot^2-\Hide\cdot^1$ combined with \eqref{eq:ReadQue_are_data_processed_F^Z_z_measurements_quered}, and using $h$-many times the semi-triangle inequality, we get that 
	\begin{equation}\label{eq:QueHide_variables_are_close_to_F^Z_z_evaluated_approp_quered}
	\forall 1\leq r\leq h\ \colon \ \ \cal{P}^{\mathsf{Hide}^r\sQue^\cdot}\approx_{O(h^3\cdot\eps)} \mathscr{F}^\PZm_{[\frS^\cdot_{<r}(\cdot)=\cdot]}\otimes \Id_m
	\  .
	\end{equation}
	
	We can now describe the perturbed strategy $\strategy'$, with projective form $\cal{Q}$ and observable form $\cal{W}$. First, it agrees with $\strategy$ on $\PauliBasis_k$ vertices (and thus satisfies condition $2.$). 
	Then, let 
	\begin{align}
	\forall z\in \FF_2^k\ &\colon\ \ \cal{Q}^{\mathsf{SamZ}^\cdot}_z=\mathscr{F}^\PZm_z\otimes \Id_m\ ,   \label{eq:QSamZ_value_quered} \\
	\forall \mttx\in \FF_2^k\ &\colon \ \ \cal{Q}^{\sQue^\cdot}_\mttx=\cal{Q}^{\mathsf{Read}\sQue^\cdot}_\mttx=\mathscr{F}^\PZm_{[\frS^\cdot(\cdot)=\mttx]}\otimes \Id_m=\sum_{z\colon \frS^\cdot(z)=\mttx}\mathscr{F}^\PZm_z\otimes \Id_m\ , \label{eq:QQue_value_quered}\\
	\forall 1\leq r\leq h,\ \forall\mttx\in \FF_2^k\ &\colon \ \ \cal{Q}^{\mathsf{Hide}^r\sQue^\cdot}_\mttx=\mathscr{F}^\PZm_{[\frS^\cdot_{<r}(\cdot)=\mttx]}\otimes \Id_m=\sum_{z\colon \frS^\cdot_{<r}(z)=\mttx}\mathscr{F}^\PZm_z\otimes \Id_m\ . \label{eq:QHiderQue_value_quered}
	\end{align}
	Note that if we extend this choice of $\strategy'$ to a quantum strategy, then condition $3.$ is satisfied, which means the only condition left to be verified (after $\strategy'$ is fully defined) is that it has value at least $1-O(\sqrt{h^3\cdot \eps})$.  Let us complete the definition of $\strategy'$: For the rest of the $\cal{W}$-observables in the vertices where changes were made, we are going to use Claim \ref{claim:strict_stability_with_one_rep_fixed} to change the $\cal{U}$-observables so they commute with our choices in \eqref{eq:QSamZ_value_quered}, \eqref{eq:QQue_value_quered} and \eqref{eq:QHiderQue_value_quered}. Let us demonstrate this analysis for two  vertices, $\Sample_\cdot$ and $\Read_\cdot$, as for the rest it is essentially the same type of argument. For $\Sample_\cdot$, as closeness in $L^1$ for representations implies closeness in $L^\infty$ (Claim \ref{claim:L1_closeness_of_unirep_implies_Linfty}), we have
	\[
	\begin{split}
	\forall \alpha^\frR,\alpha^\frL\in \FF_2^\Lambda,\ \beta\in \FF_2^k\ \colon\quad \PZm^{\otimes \beta}\otimes\Id_m \cdot \cal{U}^{\mathsf{SamAns}^\cdot}(\alpha^\frR,\alpha^\frL)&\approx_{O(h\eps)} \cal{U}^{\mathsf{SamZ}^\cdot}(\beta)\cdot\cal{U}^{\mathsf{SamAns}^\cdot}(\alpha^\frR,\alpha^\frL)\\
	&=\cal{U}^{\mathsf{SamAns}^\cdot}(\alpha^\frR,\alpha^\frL)\cdot\cal{U}^{\mathsf{SamZ}^\cdot}(\beta)\\
	&\approx_{O(h\eps)} \cal{U}^{\mathsf{SamAns}^\cdot}(\alpha^\frR,\alpha^\frL)\cdot \PZm^{\otimes \beta}\otimes\Id_m\;,
	\end{split}
	\]
	where the approximations are due to \eqref{eq:SamZ_close_to_F^Z_z_in_quered} and Claim \ref{claim:L1_closeness_of_unirep_implies_Linfty}, while the middle equality is due to the fact $\cal{U}$ is a quantum strategy (and thus the observables at the same vertex commute). 
	Hence, by Claim \ref{claim:strict_stability_with_one_rep_fixed} (or by Orthonormaliztion \ref{fact:orthogonalization}), there are observables 
	\[
	\forall \alpha^\frR,\alpha^\frL\in \FF_2^\Lambda\ \colon \ \ \cal{W}^{\mathsf{SamAns}^\cdot}(\alpha^\frR,\alpha^\frL)\ ,
	\]
	which induce a representation of $\FF_2^{2\Lambda}$ that commutes with $\PZm^\beta\otimes \Id_m$ for every $\beta\in \FF_2^k$, and 
	\[
	\cal{W}^{\mathsf{SamAns}^\cdot}(\alpha^\frR,\alpha^\frL)\approx_{O(h\eps)}\cal{U}^{\mathsf{SamAns}^\cdot}(\alpha^\frR,\alpha^\frL).
	\]
	For $\Read_\cdot$, we have that for every $\alpha^\frR,\alpha^\frL\in \FF_2^\Lambda$ and $\beta,\gamma\in \FF_2^k$,
	\[
	\begin{split}
	\rho^\PZm_{[\frS^\cdot]}(\gamma)\otimes \Id_m \cdot \cal{U}^{\mathsf{ReadAns}^\cdot\sqcup\mathsf{ReadPerp}^\cdot}(\alpha^\frR,\alpha^\frL,\beta)&\approx_{O(h\eps)} \cal{U}^{\mathsf{ReadQue}^\cdot}(\gamma)\cdot \cal{U}^{\mathsf{ReadAns}^\cdot\sqcup\mathsf{ReadPerp}^\cdot}(\alpha^\frR,\alpha^\frL,\beta)\\
	&=\cal{U}^{\mathsf{ReadAns}^\cdot\sqcup\mathsf{ReadPerp}^\cdot}(\alpha^\frR,\alpha^\frL,\beta)\cdot \cal{U}^{\mathsf{ReadQue}^\cdot}(\gamma)\\
	&\approx_{O(h\eps)} \cal{U}^{\mathsf{ReadAns}^\cdot\sqcup\mathsf{ReadPerp}^\cdot}(\alpha^\frR,\alpha^\frL,\beta)\cdot \rho^\PZm_{[\frS^\cdot]}(\gamma)\otimes \Id_m \ .
	\end{split}
	\]
	Thus,  we can apply Claim \ref{claim:strict_stability_with_one_rep_fixed} again  to obtain a representation $\cal{W}$ on $\FF_2^{2\Lambda+k}$ that commutes with the PVM $\mathscr{F}^\PZm_{[\frS(\cdot)=\cdot]}\otimes \Id_m$ and satisfies 
	\[
	\cal{U}^{\mathsf{ReadAns}^\cdot\sqcup\mathsf{ReadPerp}^\cdot}(\alpha^\frR,\alpha^\frL,\beta)\approx_{O(h\eps)} \cal{W}^{\mathsf{ReadAns}^\cdot\sqcup\mathsf{ReadPerp}^\cdot}(\alpha^\frR,\alpha^\frL,\beta)
	\]
	for every $\alpha^\frR,\alpha^\frL$ and $\beta$. The change in the $\Hide {}$ vertices depends on the closeness parameter achieved in \eqref{eq:QueHide_variables_are_close_to_F^Z_z_evaluated_approp_quered}, which is $O(h^3\cdot  \eps)$. All in all, the constructed strategy $\strategy'$ is $O(h^3\cdot \eps)$-close to the original one $\strategy$, and thus by Claim \ref{claim:close_strat_implies_close_correlations} it has value of at least $1-O(\sqrt{h^3\cdot \eps})$ against $\frak{QueRed}(\game)$, proving clause $1.$ and completing  the proof.
\end{proof}

\begin{claim}\label{claim:3_soundness_quered}
	Let $\strategy=\{\cal{P}\}$ be a strategy for $\frak{QueRed}(\game)$, acting on $\complex^{\FF_2^k}\otimes \complex^m$,  satisfying:
	\begin{enumerate}
		\item (Almost Perfect) $\val(\strategy;\frak{QueRed}(\game))\geq 1-\eps$;
		\item (Perfect on Pauli basis and all  readable variables) $\strategy$ passes all $\PauliBasis_k$ edges perfectly, and in addition satisfies
		\begin{equation}\label{eq:assumptions_on_calP_in_last_claim_soundness_quered}
		\begin{split}
		\forall z,\chi\in \FF_2^k\ &\colon \ \ \cal{P}^{\Pauli_\PZm}_{z}=\cal{P}^{\mathsf{SamZ}^{\cdot}}_{z}=\mathscr{F}_z^\PZm\otimes \Id_m\quad,\quad \cal{P}^{\Pauli_\PXm}_\chi=\mathscr{F}^\PXm_\chi\otimes \Id_m\ ,\\
		\forall \mttx\in \FF_2^k\ &\colon \ \ \cal{P}^{\mathsf{Read}\sQue^\cdot}_{\mttx}=\cal{P}^{\sQue^\cdot}_{\mttx}=\mathscr{F}^\PZm_{[\frS^\cdot(\cdot)=\mttx]}\otimes \Id_m=\sum_{v\colon \frS^\cdot(v)=\mttx} \mathscr{F}^\PZm_{v}\otimes \Id_m\ ,    \\
		\forall 1\leq r\leq h,\ \forall\mttx\in \FF_2^k\  &\colon \ \ \cal{P}^{\mathsf{Hide}^r\sQue^\cdot}_\mttx=\mathscr{F}^\PZm_{[\frS^\cdot_{<r}(\cdot)=\mttx]}=\sum_{z\colon \frS_{<r}^\cdot(z)=\mttx}\mathscr{F}^\PZm_z\otimes \Id_m\ .
		\end{split}              
		\end{equation}
	\end{enumerate}
	Then, there is another strategy $\strategy'=\{\cal{Q}\}$ for $\frak{QueRed}(\game)$ acting on the same space $\complex^{\FF_2^k}\otimes\complex^m$, satisfying:
	\begin{enumerate}
		\item (Almost Perfect) $\val(\strategy',\frak{QueRed}(\game))\geq 1-O(h\cdot 2^h\cdot\sqrt{\eps})$;
		\item (Agrees with $\strategy$ on $\PauliBasis_k$ vertices as well as $\mathsf{SamZ}$ and $\sQue$-variables) For all $\mttx\in \PauliBasis_k$,  $\cal{P}^\mttx=\cal{Q}^\mttx$, and in addition the PVM $\mathcal{Q}^\cdot_\cdot$ satisfy the same conditions as $\mathcal{P}^\cdot_\cdot$ in~\eqref{eq:assumptions_on_calP_in_last_claim_soundness_quered}.
		\item (Passes all non-$\Intro_A-\Intro_B$ edges perfectly) 
		\begin{equation}\label{eq:Q_new_properties}
		\begin{split}
		\forall a^\frR,a^\frL\in \FF_2^\Lambda \ \colon \ \ \cal{Q}^{\sAns^\cdot}_{a^\frR,a^\frL}&=\cal{Q}^{\mathsf{Sam}\sAns^\cdot}_{a^\frR,a^\frL}=\cal{Q}^{\mathsf{Read}\sAns^\cdot}_{a^\frR,a^\frL},\\
		\forall \mttx,\nu\in \FF_2^k\ \colon \ \ \cal{Q}^{\mathsf{ReadQue}^\cdot\sqcup \mathsf{ReadPerp}^\cdot}_{\mttx,\nu}&=\mathscr{F}^\PZm_{[\frS^\cdot(\cdot)=\mttx]}\cdot \mathscr{F}^\PXm_{[(\frS^{\cdot,\mttx})^\perp(\cdot)=\mttx]}\otimes \Id_m\\
		&=\sum_{\substack{z,\chi\in\FF_2^k\colon \\ \frS^\cdot(z)=\mttx,\ (\frS^{\cdot,\mttx})^\perp(\chi)=\nu}}\mathscr{F}^\PZm_z\mathscr{F}^\PXm_{\chi}\otimes \Id_m,\\
		\forall r,\ \forall \mttx,\nu\in \FF_2^k\ \colon \cal{Q}^{\Hide \cdot^r}_{\mttx,\nu}&=\mathscr{F}^\PZm_{[\frS_{<r}^\cdot(\cdot)=\mttx]}\cdot \mathscr{F}^\PXm_{[(\frSS_{\leq r}^{\cdot,\mttx})^\perp(\cdot)=\nu]}\otimes \Id_m\\
		&=\sum_{\substack{z,\chi\in\FF_2^k\colon \\ \frS_{<r}^\cdot(z)=\mttx,\ (\frSS_{\leq r}^{\cdot,\mttx})^\perp(\chi)=\nu}}\mathscr{F}^\PZm_z\mathscr{F}^\PXm_{\chi}\otimes \Id_m\ ,
		\end{split}    
		\end{equation}
		where $(\frSS^{\cdot,\mttx}_{\leq j})^\perp$ are the extended perpendicular maps to the seeded CLMs as defined in \cref{clause3:prop_of_input_game_to_quered} and Remark \ref{rem:extended_perpendicular_maps_along_a_CLM}.
	\end{enumerate}
\end{claim}

\begin{proof}
	Recall that every augmented edge in $\frak{QueRed}(\game)$ is sampled with probability $\Omega(\frac{1}{h})$. For ease of following the proof, let us denote by $C>0$ the universal constant induced by $\Omega(\frac{1}{h})$, namely the probability of sampling any augmented edge is at least $\frac{1}{Ch}$  and hence $\strategy$ passes every such edge with probability of at least $1-Ch\eps$. The reader can check that for the probability distribution on edges we fixed for $\frak{QueRed}(\game)$, each augmented edge is sampled with probability of at least $\frac{1}{2h+4}$, so $C=6$ is enough. The reason we use an abstract constant instead of $6$, is that we later change the distribution over edges in $\frak{QueRed}(\game)$ (see Example \ref{example:quered_has_typed_CL_sampling_scheme} and specifically Figure \ref{fig:typed_graph_quered}) so that every augmented edge is sampled with probability of at least $\frac{1}{4h+63}$, in which case $C=67$ is enough. In any case, let us treat $C$ as an unknown constant.
	
	First, we have
	\[
	\begin{split}
	\cal{P}^{\Hide\cdot ^1}_{\mttx,\nu}&=\cal{P}^{\mathsf{Hide}^1\mathsf{Que}^\cdot }_{\mttx}\cal{P}^{\mathsf{Hide}^1\mathsf{Perp}^\cdot }_{\nu}\\
	&=\mathscr{F}^\PZm_{[(\cdot)_{<1}=\mttx]}\otimes \Id_m\cdot\cal{P}^{\mathsf{Hide}^1\mathsf{Perp}^\cdot }_{\nu}\\
	&\simeq_{Ch\eps}\mathscr{F}^\PZm_{[(\cdot)_{<1}=\mttx]}\otimes \Id_m\cdot\cal{P}^{\Pauli_\PXm}_{[(\frSS_{\leq 1}^{\cdot,\mttx})^\perp(\cdot)=\nu]}\\
	&=\mathscr{F}^\PZm_{[(\cdot)_{<1}=\mttx]}\mathscr{F}^\PXm_{[(\frSS^{\cdot,\mttx}_{\leq 1})^\perp(\cdot)=\nu]}\otimes \Id_m\ ,
	\end{split}
	\]
	where the first equation is by definition for a projective measurement, the second and last equations use the assumptions from \eqref{eq:assumptions_on_calP_in_last_claim_soundness_quered} on the PVMs at $\Pauli_\PXm$ and $\Hide\cdot^1$, and the inconsistency in the middle is due to $\strategy$ passing $\Pauli_\PXm-\Hide \cdot^1$ with probability $1-Ch\eps$. Hence, by the translation of consistency to closeness (Proposition \ref{prop:properties_of_distance_and_inconsistency}), we deduce that
	\begin{equation}\label{eq:base_case_base_case_base_case_base_case}
	\cal{P}^{\Hide\cdot ^1}_{\mttx,\nu} \approx_{2Ch\eps} \mathscr{F}^\PZm_{[(\cdot)_{<1}=\mttx]}\mathscr{F}^\PXm_{[(\frSS^{\cdot,\mttx}_{\leq 1})^\perp(\cdot)=\nu]}\otimes \Id_m\ .           
	\end{equation}
	We now establish the following inductive  step. Assume that for some $1\leq j<r$ it holds that
	\begin{equation}\label{eq:abcdefghijk}
	\cal{P}^{\Hide\cdot^{j-1}}_{\mttx,\nu} \approx_\delta  \mathscr{F}^\PZm_{[\frS_{<j-1}(\cdot)=\mttx]}\mathscr{F}^\PXm_{[(\frSS^{\cdot,\mttx}_{\leq j-1})^\perp(\cdot)=\nu]}\otimes \Id_m\ ,
	\end{equation}
	for some $\delta\geq 2Ch\eps$. Then it follows that
	\begin{equation}\label{eq:abcdefghijk22222}
	\cal{P}^{\Hide\cdot^{j}}_{\mttx,\nu} \approx_{4\delta}  \mathscr{F}^\PZm_{[\frS_{<j}(\cdot)=\mttx]}\mathscr{F}^\PXm_{[(\frSS^{\cdot,\mttx}_{\leq j})^\perp(\cdot)=\nu]}\otimes \Id_m\ .
	\end{equation}
	To show the implication~$\eqref{eq:abcdefghijk}\implies \eqref{eq:abcdefghijk22222}$, we first use the fact that $\strategy$ passes $\Hide \cdot^{j-1}-\Hide \cdot ^j$ with probability of at least $1-Ch\eps$ to write 
	\begin{equation}\label{eq:Hidej-Hidej-1_comparison}
	\sum_{\substack{\mttx,\mttx',\nu,\nu'\\(\mttx)_{<j-1}=\mttx'\\\nu=(\frSS_j^{\cdot,\mttx})^\perp(\nu')}}\tau\Big(\cal{P}^{\Hide\cdot^j}_{\mttx,\nu}\cal{P}^{\Hide\cdot^{j-1}}_{\mttx',\nu'}\Big)\geq 1-Ch\eps\ .
	\end{equation}
	Using the projectivity of $\cal{P}^{\Hide\cdot ^r}$ (for every $r\in [h]$) and our assumptions in \eqref{eq:assumptions_on_calP_in_last_claim_soundness_quered}, 
	\[
	\begin{split}    \cal{P}^{\Hide\cdot^r}_{\mttx,\nu}=\cal{P}^{\mathsf{Hide}^r\mathsf{Que}^\cdot}_{\mttx}\cal{P}^{\mathsf{Hide}^r\mathsf{Perp}^\cdot}_{\nu}=\mathscr{F}^\PZm_{[\frS^\cdot_{<r}(\cdot)=\mttx]}\cal{P}^{\mathsf{Hide}^r\mathsf{Perp}^\cdot}_{\nu} \ .
	\end{split}
	\]
	For every $\mttx\in \FF_2^k$, let 
	\[
	A^\mttx_{\nu}=\cal{P}^{\mathsf{Hide}^{j-1}\mathsf{Perp}^\cdot}_{[(\frSS^{\cdot,\mttx}_{ j})^\perp(\cdot)=\nu]}=\sum_{\nu'\colon (\frSS_j^{\cdot,\mttx})^\perp(\nu')=\nu}\cal{P}^{\mathsf{Hide}^{j-1}\mathsf{Perp}^\cdot}_\nu\ .
	\]
	This is a data processed version of $\cal{P}^{\mathsf{Hide}^{j-1}\mathsf{Perp}^\cdot}$, but the exact function through which we are evaluating depends on the seed $\mttx$. Then, \eqref{eq:Hidej-Hidej-1_comparison} can be rewritten as
	\[
	\sum_{\mttx,\nu}\tau\left(\mathscr{F}^\PZm_{[\frS_{< j}(\cdot)=\mttx]}\cal{P}^{\mathsf{Hide}^{j}\mathsf{Perp}^\cdot}_{\nu}A^\mttx_\nu\right)\geq 1-Ch\eps\ ,
	\]
	which means that $\cal{P}^{\Hide\cdot^{j}}_{\mttx,\nu}$ is $Ch\eps$-inconsistent  with $\mathscr{F}^\PZm_{[\frS_{< j}(\cdot)=\mttx]}A^\mttx_\nu\mathscr{F}^\PZm_{[\frS_{< j}(\cdot)=\mttx]}$ --- note that this is a POVM but not necessarily a projective one.   In general, if one has three PVMs, $B$ and $C$ with outcomes in $X$, and $D$ with outcomes in $Y$, then $B\approx_\delta C$ implies $DBD\approx_\delta DCD$. Hence, using \eqref{eq:abcdefghijk}, we can deduce that 
	\[
	\mathscr{F}^\PZm_{[\frS_{< j}(\cdot)=\mttx]}A^\mttx_\nu\mathscr{F}^\PZm_{[\frS_{< j}(\cdot)=\mttx]}\approx_\delta \mathscr{F}^\PZm_{[\frS_{< j}(\cdot)=\mttx]}\mathscr{F}^\PXm_{[(\frSS^{\cdot,\mttx}_j)^\perp\circ (\frSS^{\cdot,\mttx}_{\leq j-1})^\perp(\cdot)=\nu]}\mathscr{F}^\PZm_{[\frS_{< j}(\cdot)=\mttx]}=\mathscr{F}^\PZm_{[\frS_{< j}(\cdot)=\mttx]}\mathscr{F}^\PXm_{[(\frSS^{\cdot,\mttx}_{\leq j})^\perp(\cdot)=\nu]}\ ,
	\]
	where the equation uses both that, as defined in Remark \ref{rem:extended_perpendicular_maps_along_a_CLM}, $(\frSS^{\cdot,\mttx}_j)^\perp\circ (\frSS^{\cdot,\mttx}_{\leq j-1})^\perp=(\frSS^{\cdot,\mttx}_{\leq j})^\perp$, as well as the fact that $\mathscr{F}^\PZm_{[\frS_{< j}(\cdot)=\mttx]}$ and $\mathscr{F}^\PXm_{[(\frSS^{\cdot,\mttx}_{\leq j})^\perp(\cdot)=\nu]}$  commute.
	Combined with the above, we have
	\[
	\cal{P}^{\Hide\cdot^j}_{\mttx,\nu}\approx_{2Ch\eps} \mathscr{F}^\PZm_{[\frS_{< j}(\cdot)=\mttx]}A^\mttx_\nu\mathscr{F}^\PZm_{[\frS_{< j}(\cdot)=\mttx]}\approx_\delta \mathscr{F}^\PZm_{[\frS_{< j}(\cdot)=\mttx]}\mathscr{F}^\PXm_{[(\frSS^{\cdot,\mttx}_{\leq j})^\perp(\cdot)=\nu]}\ ,
	\]
	which implies \eqref{eq:abcdefghijk22222} using the semi-triangle inequality for closeness of POVMs and the fact $\delta\geq 2Ch\eps$.
	
	This establishes the desired implication $\eqref{eq:abcdefghijk}\implies \eqref{eq:abcdefghijk22222}$. Together with the base case \eqref{eq:base_case_base_case_base_case_base_case}, we deduce that 
	\begin{equation}\label{eq:distance_on_hide_P_from_Q_soundness_quered}
	\forall j\in [h]\ \colon \ \ \cal{P}^{\Hide\cdot^{j}}_{\mttx,\nu} \approx_{4^jCh\eps}  \mathscr{F}^\PZm_{[\frS_{<j}(\cdot)=\mttx]}\mathscr{F}^\PXm_{[(\frSS^{\cdot,\mttx}_{\leq j})^\perp(\cdot)=\nu]}\otimes \Id_m\ .
	\end{equation}
	By the consistency check along $\Hide\cdot ^h-\Read_\cdot$, we deduce from~\eqref{eq:distance_on_hide_P_from_Q_soundness_quered} that 
	\begin{equation}\label{eq:closeness_of_Read_vertex_to_what_we_need_quered}
	\cal{P}^{\mathsf{ReadQue}^\cdot\sqcup\mathsf{ReadPerp}^\cdot}_{\mttx, \nu}\approx_{4^{h+1}Ch\eps }\mathscr{F}^\PZm_{[\frS^\cdot(\cdot)=\mttx]}\mathscr{F}^\PXm_{[(\frS^{\cdot,\mttx})^\perp(\cdot)=\nu]}\otimes \Id_m\ .
	\end{equation}
	Let us choose the following signed permutation representations of $\FF_2^k$,
	\[
	B^\cdot(\alpha)=\sum_{\mttx\in \FF_2^k}\mathscr{F}^\PZm_{[\frS^\cdot (\cdot)=\mttx]}\PXm^{\otimes \alpha \cdot (\frS^{\cdot,\mttx})^\perp}\ ,
	\]
	where, as usual, $\alpha \cdot (\frS^{\cdot,\mttx})^\perp$ is the left multiplication of the row vector $\alpha$ with the matrix $(\frS^{\cdot,\mttx})^\perp$ --- this is indeed a matrix as $(\frS^{\cdot,\mttx})^\perp$ is a linear map for every fixed $\mttx$. By applying the inverse Fourier transform to \eqref{eq:closeness_of_Read_vertex_to_what_we_need_quered}, one can deduce that 
	\begin{equation}\label{eq:ReadPerp_is_close_to_B}
	\cal{U}^{\mathsf{ReadPerp}^\cdot}\approx_{4^{h+1}Ch\eps} B^\cdot\ .
	\end{equation}
	By the consistency checks along $\Sample_\cdot-\Intro_\cdot-\Read_\cdot$, we deduce that 
	\begin{equation}\label{eq:answers_at_sample_intro_read_are_close}
	\cal{U}^{\mathsf{Sam}\sAns^\cdot}\approx_{Ch\eps} \cal{U}^{\sAns^\cdot}\approx_{Ch\eps} \cal{U}^{\mathsf{Read}\sAns^\cdot}\ .
	\end{equation}
	Hence, using  the ``small $L^1$-distance between representations implies small $L^\infty$ distance'' proved in Claim \ref{claim:L1_closeness_of_unirep_implies_Linfty} twice, combined with \eqref{eq:assumptions_on_calP_in_last_claim_soundness_quered} and \eqref{eq:answers_at_sample_intro_read_are_close}, we have
	\begin{align*}
	\forall \alpha^\frR,\alpha^\frL\in \FF_2^\Lambda \ ,\  \beta\in \FF_2^k\ \colon \ \ \PZm^{\otimes \beta}\otimes \Id_m \cdot\cal{U}^{\mathsf{Ans}^\cdot}(\alpha^\frR,\alpha^\frL)&\approx_{6Ch\eps} \cal{U}^{\mathsf{SamZ}^\cdot}(\beta)\cdot \cal{U}^{\mathsf{SamAns}^\cdot}(\alpha^\frR,\alpha^\frL)\\
	&= \cal{U}^{\mathsf{SamAns}^\cdot}(\alpha^\frR,\alpha^\frL)\cdot \cal{U}^{\mathsf{SamZ}^\cdot}(\beta)\\
	&\approx_{6Ch\eps} \cal{U}^{\mathsf{Ans}^\cdot}(\alpha^\frR,\alpha^\frL)\cdot  \PZm^{\otimes \beta}\otimes \Id_m\ .
	\end{align*}
	Similarly, using \eqref{eq:ReadPerp_is_close_to_B}, \eqref{eq:answers_at_sample_intro_read_are_close} and Claim \ref{claim:L1_closeness_of_unirep_implies_Linfty}, we have
	\begin{align*}
	\forall \alpha^\frR,\alpha^\frL\in \FF_2^\Lambda \ ,\  \beta\in \FF_2^k\ \colon \ \ B^\cdot(\beta)\cdot\cal{U}^{\mathsf{Ans}^\cdot}(\alpha^\frR,\alpha^\frL)&\approx_{6\cdot 4^{h+1}Ch\eps}\cal{U}^{\mathsf{ReadPerp}^\cdot}(\beta)\cdot \cal{U}^{\mathsf{Ans}^\cdot}(\alpha^\frR,\alpha^\frL)\\
	&\approx_{6Ch\eps}\cal{U}^{\mathsf{ReadPerp}^\cdot}(\beta)\cdot \cal{U}^{\mathsf{ReadAns}^\cdot}(\alpha^\frR,\alpha^\frL)\\
	&= \cal{U}^{\mathsf{ReadAns}^\cdot}(\alpha^\frR,\alpha^\frL)\cdot \cal{U}^{\mathsf{ReadPerp}^\cdot}(\beta)\\
	&\approx_{6Ch\eps} \cal{U}^{\mathsf{Ans}^\cdot}(\alpha^\frR,\alpha^\frL)\cdot \cal{U}^{\mathsf{ReadPerp}^\cdot}(\beta)\\
	&\approx_{6\cdot 4^{h+1}Ch\eps} \cal{U}^{\mathsf{Ans}^\cdot}(\alpha^\frR,\alpha^\frL)\cdot B^\cdot(\beta)\ .
	\end{align*}
	Note that, as $B^{\cdot}$ and $\rho^\PZm\otimes \Id_m$ are both signed permutation representations, the group generated by their images is finite. We can thus apply Claim \ref{claim:strict_stability_with_one_rep_fixed} where $G$ is the group generated by the images of $B^\cdot$ and $\rho^\PZm\otimes \Id_m$ (which  also fixes the representation of $G$ in the claim), and $A$ is $\FF_2^{2\Lambda}$ with representation $\psi=\cal{U}^{\sAns^\cdot}$; this gives us  two representations $\theta^A,\theta^B$ of $\FF_2^{2\Lambda}$ such that $\theta^A\approx_{O(4^h h\eps)} \cal{U}^{\sAns^A}$, $\theta^B\approx_{O(4^h h\eps)} \cal{U}^{\sAns^B}$ and $\theta^\cdot$ perfectly commutes with  $B^\cdot$ and $\rho^\PZm\otimes \Id_m$.
	
	Combining all of the above, if we let $\strategy'=\{\cal{V}\}=\{\cal{Q}\}$ satisfy \eqref{eq:assumptions_on_calP_in_last_claim_soundness_quered}, \eqref{eq:Q_new_properties} and 
	\[
	\cal{V}^{\sAns^A}=\theta^A\quad,\quad \cal{V}^{\sAns^B}=\theta^B\ ,
	\]
	then $\strategy'$ satisfies clause $2.$ and $3.$ from the requirements of this claim, as well as being $O(4^h\cdot h\cdot\eps)$-close to $\strategy$. By applying Claim \ref{claim:close_strat_implies_close_correlations}, we can  deduce clause $1.$ as well.
\end{proof}

To conclude the soundness proof, we need to combine the three preceding claims:
\begin{enumerate}
	\item Given a strategy $\strategy$ acting on $\complex^N$ that has value $1-\eps$, we can apply Claim \ref{claim:1_soundness_quered} on it to get a strategy $\strategy'$ that acts on $\complex^{\FF_2^k}\otimes\complex^m$, passes the $\PauliBasis_k$-vertices perfectly and has value of $1-O(\sqrt{(1+\nicefrac{k^2}{d^2})\eps})$. Moreover, the strategies are $O((1+\nicefrac{k^2}{d^2})\eps)$-close on $\Pauli_\PXm$ and $\Pauli_\PZm$ observables, which means in particular that $1-\frac{N}{2^k\cdot m}\leq O((1+\nicefrac{k^2}{d^2})\eps)$ and thus $N\geq (1-O((1+\nicefrac{k^2}{d^2})\eps))2^k\cdot m$.
	\item The strategy $\strategy'$ satisfies the assumptions of Claim \ref{claim:2_soundness_quered}, and thus there is a strategy $\strategy''$ for $\frak{QueRed}(\game)$ that behaves well on all $\sQue$-variables and has value $1-O\left(h^{3/2}\cdot((1+\nicefrac{k^2}{d^2})\eps)^{\nicefrac{1}{4}}\right)$.
	\item The strategy $\strategy''$ satisfies the assumptions of Claim \ref{claim:3_soundness_quered}, and thus there is a strategy $\strategy'''$ which passes all edges of $\frak{QueRed}(\game)$ perfectly (except for maybe $\Intro_A-\Intro_B$), and has value of at least $1-O(h^2\cdot 2^h\cdot(1+\nicefrac{k^2}{d^2})\cdot \eps^{\nicefrac{1}{8}})$, which proves the soundness in \cref{clause:soundness_of_quered}.
	In addition, the resulting strategy is {honest} (Definition \ref{defn:honest_strat_for_Intro}). Hence, by Claim \ref{claim:complete_sound_honest_intro}, such a strategy induces a strategy for $\game$ with the same value which acts on $\complex^m$. Hence, $m\geq \Ent(\game,1-O(h^2\cdot 2^h\cdot(1+\nicefrac{k^2}{d^2})\cdot \eps^{\nicefrac{1}{8}}))$, and  we can conclude that 
	\[
	\Ent(\frak{QueRed}(\game),1-\eps)\geq 2^{k}\cdot\left(1-O((1+\nicefrac{k^2}{d^2})\eps)\right)\cdot\Ent(\game,1-O(h^2\cdot 2^h\cdot(1+\nicefrac{k^2}{d^2})\cdot \eps^{\nicefrac{1}{8}}))\ ,
	\]
	which proves the entanglement lower bound \cref{clause:entanglement_of_quered}.
\end{enumerate}

\subsection{Applying question reduction to a tailored normal form verifier}
\label{sec:quered-nf}
Up until now, we discussed a certain combinatorial transformation that takes as input an integer $k$, a set $\mathscr{B}=\{w_1,...,w_N\}\subseteq\FF_2^k$, and a game $\game$ (with certain assumptions on its sampling mechanism and answer length functions), and outputs a new game $\frak{QueRed}(\game)$ (defined in Section \ref{sec:augmentation_in_the_CLM_case}), which is a specific augmented sum of $\PauliBasis_k(\mathscr{B})$ from Section \ref{sec:Pauli_basis_definition} and $\Introspect(\game)$ from Section \ref{sec:the_introspection_game}. 

For the proof of Compression (Theorem \ref{thm:compression}), one needs a way of applying this combinatorial transformation on the level of tailored normal form verifiers (TNFVs), as was described in Theorem \ref{thm:informal_question_reduciton}. Namely, we seek a transformation on a pair consisting of an integer $\lambda$ and a TNFV $\cal{V}=(\sampler,\length,\linproc,\decider)$ that outputs a new TNFV  $\QueRed(\verifier,\lambda)=\verifier'=(\sampler^\lambda_{\qr},\length^\lambda_{\qr},\linproc',\decider)$, such that on the combinatorial level the $n^{\rm th}$ game of $\verifier'$ is the question reduced ${2^{n}}$-${\rm th}$ game of $\verifier$. So, 
$$\verifier'_n=\frak{QueRed}(\verifier_{2^n},k(n,\lambda), \mathscr{B}(n,\lambda))$$
for some integer-valued function $k(n,\lambda)$ and  function $\mathscr{B}(n,\lambda)$ valued in tuples of vectors in $\FF_2^{k(n,\lambda)}$.

Recall that for $\frak{QueRed}_h(\game,k,\mathscr{B})$ to have the desired properties of Theorem \ref{thm:complet_sound_quered}, several non-trivial assumptions about the inputs need to be satisfied:
\begin{enumerate}[label=\textcolor{black}{(\arabic*)}, ref= (\arabic*)]
	\item \label{clause:1_problems_quered_on_NFV} The game $\game$ is tailored, with an underlying $h$-level CL sampling scheme (Definition \ref{defn:sampling_induced_by_CLMs}); namely, its underlying graph's vertex set is $\FF_2^r$ and its distribution over edges $\mu$ is the pushforward of the uniform distribution over $\FF_2^r$ through a fixed pair of $h$-level conditionally linear maps (Definition \ref{def:CLM}) $\frS=(\frS^A,\frS^B)\colon \FF_2^r\to \FF_2^r\times \FF_2^r$. 
	
	Note that in the definition of $\frak{QueRed}(\game)$ we always insisted that this parameter $r$, controlling the number of vertices in $\game$, to be equal to $k$, which is the length of answers at the $\Pauli_\PXm$ and $\Pauli_\PZm$ vertices in the generalized Pauli basis game $\PauliBasis_k(\mathscr{B})$. But, and it is straightforward to check, we only needed $k$ to be larger or equal to $r$ --- that is because we can pre-compose $(\frS^A,\frS^B)$ with the restriction to the first $r$ coordinates ${\rm rest}\colon \FF_2^k\to \FF_2^r$ and proceed accordingly. All in all, we need the sampling procedure of $\game$ to be induced by two CLMs $\frS^A,\frS^B\colon \FF_2^r\to \FF_2^r$ and for $r\leq k$.
	
	\label{clause:4_problems_quered_on_NFV} As the input game $\game$ is assumed to have a sampling procedure induced by an $h$-level CLM, we need  the TNFV $\verifier$ that we manipulate to be such that for every $n\in \mathbb{N}$, the $n^{\rm th}$ game $\verifier_{n}$ has an underlying $h$-level sampling scheme. In addition, every transformation that we apply on $\verifier$, e.g.\  $\mathsf{QuestionReduction}$, should retain this property. Hence, we are going to move to a subcategory of TNFVs that have this property. 
	\item \label{clause:2_problems_quered_on_NFV} In addition to the sampling procedure, we needed $\game$ to have constant length functions $\ell^\frR,\ell^\frL\colon \FF_2^r\to \mathbb{N}$. Namely, there is some integer $\Lambda$ such that for all $\mttx\in \FF_2^r$ the functions satisfy $\ell^\frR(\mttx)=\ell^\frL(\mttx)=\Lambda$. This turns out to be an easy condition to satisfy, and even if the normal form verifier does not satisfy it, a \emph{padding} transformation can be applied on it so that it does (see Section \ref{sec:Padding}).
	\item \label{clause:3_problems_quered_on_NFV} We aim to ``question reduce'', which translates to making the underlying graph of $\frak{QueRed}(\game)$ exponentially smaller than that of $\game$, i.e.,  the number of questions in it needs to be $\poly(r)$.\footnote{Actually, it is enough to be quasi-polynomial in $r$, which  is the parameter setup used in \cite{MIPRE}.} As $\Introspect(\game)$ contributes $2$ vertices, and the augmentation adds on $2h+4$ more\footnote{For the proof of Compression, $h$ can be upper bounded by $\LevelConstant$ (see Remark \ref{rem:LevelConstant}). So this can be thought of as a constant number of vertices.}, most of the vertices in the underlying graph of $\frak{QueRed}(\game)$ come from $\PauliBasis_k$. In $\PauliBasis_k$, there are $2+2N+18N^2$ vertices, where $N=|\mathscr{B}|$. 
	In addition, for the soundness and entanglement lower bounds proved before (\cref{clause:soundness_of_quered} and \cref{clause:entanglement_of_quered} of Theorem \ref{thm:complet_sound_quered}) to have any meaning, we need the parameter $\nicefrac{k}{d}$ to be bounded, where $d$ is the distance of the code induced by the set $\mathscr{B}$. 
	A tradeoff arises: For the distance of the code defined by $\mathscr{B}$ to be large enough, the set itself needs to be large enough (in particular larger than $k$). But, on the other hand, the larger $\mathscr{B}$ is the larger the underlying graph of $\frak{QueRed}(\game)$ is.  Finally, there should be an efficient way of calculating, given $i\in [N]$, the vector $w_i\in \mathscr{B}$. This point is solved completely by the existence of good error correcting codes with an efficient algorithm to calculate their encoding matrix (Fact \ref{fact:good_efficiently_calculable_codes}).
	\item \label{clause:5_problems_quered_on_NFV} As part of the application of $\frak{QueRed}$, we used quite a lot of structure regarding the CLMs $\frS^A$ and $\frS^B$. We assumed to have access to the spaces $W^\mttx_r$, the $j^{\rm th}$ step $\frS^{\cdot,\mttx}_{j}$ of the CLMs calculation for every $1\leq j\leq h$ and  seed $\mttx$, and also to the \emph{perpendicular} functions $(\frS^{\cdot,\mttx}_j)^\perp$. To handle this intricacy,  we are going to change the definition of a sampler (Definition \ref{def:sampler}) so that it can describe further its ``inner working''.
\end{enumerate}

\subsubsection{The category of  $h$-level tailored normal form verifiers}

\begin{definition}[$h$-level conditionally linear sampler]\label{defn:h-level_CL_Sampler}
	Let $h$ be a positive integer. An $h$-level conditionally linear sampler (CL sampler) is a $6$-input {deterministic} Turing machine $\sampler$ that satisfies the following restrictions. 
	First, the input to $\sampler$ is expected to be 
	\[
	(n,{\rm Action},{\rm Player},j,\mttx,z)\ ,
	\]
	where $n$  and $j$ are positive integers in binary, ${\rm Action}$ is taken from the set $$\{{\rm Dimension,\ Register,\ Marginal,\ Evaluate,\ Perpendicular }\}\ ,$$ ${\rm Player}$ is taken from the set $\{A,B\}$, and $\mttx,z$ are bit strings (interpreted as vectors in some finite vector space). 
	Second, for every positive integer $n$, there must exist an integer $r=r(n)$ and  two $h$-level conditionally linear maps (CLMs, Definition \ref{def:CLM}) $\frS^A(n),\frS^B(n)=\frS^A,\frS^B\colon \FF_2^r\to \FF_2^r$, such that $\sampler$ \emph{encodes} this pair of functions: 
	\begin{enumerate}
		\item If $\sampler$ gets as input $(n,{\rm Dimension},\cdot,\cdot,\cdot,\cdot)$, then it outputs (the binary encoding of) $r(n)$.
		\item If $\sampler$ gets as input $(n,{\rm Register},{\rm Player},j,\mttx,\cdot)$, then it outputs the $j^{\rm th}$ register subspace with respect to the seed $\mttx$, namely $W^{\mttx}_j$ 
		\eqref{eq:reg_subspaces_depending_on_prefixes} with respect to $\frS^{\rm Player}$. As $W^\mttx_j$ is a register subspace, the way it is encoded is by providing the indicator of the set $I^\mttx_j\subseteq [r]$, which says what are the standard basis vectors that span $W^\mttx_j$. Note that the indicator of $I^\mttx_j$ is just  a bit string of length $r$.
		\item If $\sampler$ gets as input $(n,{\rm Marginal},{\rm Player},j,\cdot,z)$, then it outputs the $j^{\rm th}$ prefix of $\frS^{\rm Player}$'s evaluation of $z$, namely  $\frS^{\rm Player}_{\leq j}(z)$.
		\item If $\sampler$ gets as input $(n,{\rm Evaluate},{\rm Player},j,\mttx,z)$, then it outputs the $j^{\rm th}$-register output of $\frS^{\rm Player}$  evaluated on $z$ given the seed $\mttx$, namely $\frS^{\rm Player,\mttx}_{j}(z^{W^\mttx_j})$. Recall that $\frS^{{\rm Player},\mttx}_j\colon W^\mttx_j\to W^\mttx_j$ is the linear function which controls the $j^{\rm th}$ step in the calculation of $\frS^{\rm Player}$, given that the calculation up to this point produced $\mttx_{<j}$.
		\item If $\sampler$ gets as input $(n,{\rm  Perpendicular},{\rm Player},j,\mttx,z)$, then it outputs the $j^{\rm th}$-register output of $(\frS^{\rm Player,\mttx})^{\perp}$  evaluated on $z$ given the seed $\mttx$, namely $(\frS^{\rm Player,\mttx}_{j})^{\perp}(z^{W^\mttx_j})$. Recall that the maps $(\frS^{{\rm Player},\mttx}_j)^{\perp}\colon W^\mttx_j\to W^\mttx_j$ are some fixed linear function whose rows are spanning the subspace perpendicular to the rows of $\frS^{{\rm Player},\mttx}_j$. \footnote{Note that these maps were not part of the definition of a CLM, but were assumed to be part of the data needed for question reduction in the beginning of Section \ref{sec:augmentation_in_the_CLM_case}. As Remark \ref{rem:on_our_h-level_samplers_compared_to_MIPRE} notes, there is a canonical way of extracting such maps from the rest of the CLMs data.}
	\end{enumerate}
\end{definition}

\begin{remark}\label{rem:on_our_h-level_samplers_compared_to_MIPRE}
	A few things to note about the differences between the above definition and \cite[Definition 4.14]{MIPRE}: They use the Action name ``Linear'' instead of ``Evaluate''. Furthermore, they do not include the Perpendicular action --- This is because there is a canonical (and efficient) way of calculating linear maps $(\frS^{{\rm Player},\mttx}_j)^{\perp}$ from the rest of the possible outputs of the sampler, as described in \cite[Section 8.2]{MIPRE} clause $6.$ in page $94$, which is the detailed description of the decider of their introspective verifier.
\end{remark}
\begin{remark}[Dimension of CL sampler bounded by running time]\label{rem:dim_of_CL_sampler_bounded_by_run_time}
	Since, given  inputs such as $(n,{\rm Marginal,\ Player},j,\cdot,z)$, the CL sampler $\sampler$ needs to output an $r(n)$-long bit string, where $r(n)=\sampler(n,{\rm Dimension},\cdot,\cdot,\cdot,\cdot)$, it is immediate that $r(n)\leq \TIME(\sampler;{n},\cdot,\cdot,\cdot,\cdot,\cdot)$.
\end{remark}
\begin{remark}
	Our original sampler from Definition \ref{def:sampler} is a randomized TM, while the $h$-level sampler is a deterministic one. To extract the output of the original sampler out of an $h$-level sampler, we can --- though it is not part of the definition of a CL sampler --- attach a ${\rm Sample}$ action to the list. Given $(n,{\rm Sample},\cdot,\cdot,\cdot,\cdot)$, $\sampler$ runs as follows:
	\begin{enumerate}
		\item First, it calls $\sampler(n,{\rm Dimension},\cdot,\cdot,\cdot,\cdot)$ to obtain $r$.
		\item Then, it samples $r$ random bits to obtain $z\in \FF_2^r$ (which makes $\sampler$ back to a randomized TM, but only with respect to this ${\rm Sample}$ action).
		\item It then calls $\sampler(n,{\rm Marginal},A,h,\cdot,z)$ to obtain $\mttx=\frS^A_{\leq h}(z)=\frS^A(z)$, and $\sampler(n,{\rm Marginal},B,h,\cdot,z)$ to obtain $\mtty=\frS^B_{\leq h}(z)=\frS^B(z)$.
		\item Finally, it outputs (the encoding of) $\mttx\sqcup \mtty$.
	\end{enumerate}
	Note that, indeed, this is what we expect a TM to do to be able to sample from the distribution induced by the pair $(\frS^A,\frS^B)$.
\end{remark}

\begin{definition}[Tailored $h$-level  normal form verifier]\label{defn:h-level_NFV}
	An \emph{$h$-level tailored normal form verifier} ($h$-level TNFV) is a quadruple of Turing machines $\verifier = (\sampler,\length, \linproc, \decider)$,
	where $\sampler$ is an $h$-level conditionally linear sampler as in Definition \ref{defn:h-level_CL_Sampler}, $\length$ is an answer length calculator as in Definition \ref{defn:answer_length_calc}, $\linproc$ is a linear constraint processor as in Definition \ref{def:linear_constraints_processor}, and $\decider$ is the canonical decider as in Definition \ref{def:canonical-decider}.
	
	Such a TNFV is \emph{$\lambda$-bounded}, for a positive integer $\lambda$, if
	\begin{itemize}
		\item The running times (Definition \ref{def:running_time}) of $\sampler,\length$ and $\linproc$ are all bounded by $n^\lambda$ , namely
		\[
		\forall {n}\in \{0,1\}^*\ \colon \  \ \TIME(\sampler;{n},\cdot,\cdot,\cdot,\cdot,\cdot)\ ,\ \TIME(\length;{n},\cdot,\cdot)\ ,\ \TIME(\linproc;{n},\cdot,\cdot,\cdot,\cdot)\ \leq\ n^\lambda.
		\]
		\item The description length (Definition \ref{def:Description_length}) of $\verifier$ is bounded by $\lambda$, namely $|\verifier|\leq \lambda$.
	\end{itemize}
	Similar to Definition \ref{def:normal-game}, when $\verifier$ is a $\lambda$-bounded  $h$-level TNFV, then there is an associated $n^{\rm th}$ game to it for every $n\geq 2$. Similar to Remark \ref{rem:n^th_game_less_restrictive_setup}, the $n^{\rm th}$ game of an $h$-level TNFV may be well defined even if it is not $\lambda$-bounded. Actually, all we need is for 
	\begin{itemize}
		\item $\length(n,\mttx,\kappa)$ to halt whenever $\mttx$ is of length $r(n)=\sampler(n,{\rm Dimension},\cdot,\cdot,\cdot,\cdot)$ and $\kappa\in \{\frR,\frL\}$;
		\item $\linproc(n,\mttx,\mtty,a^\frR,b^\frR)$ needs to halt whenever $\mttx,\mtty$ are of length $r(n)$, and $a^\frR$ and $b^\frR$ are of length $|\dec(\length(n,\mttx,\frR))|$ and\\ $|\dec(\length(n,\mtty,\frR))|$ respectively. 
	\end{itemize}
	Note that by assuming $\sampler$ is an $h$-level CL sampler (Definition \ref{defn:h-level_CL_Sampler}), we are guaranteed that it behaves well, in particular it always halts (on relevant inputs), and there are associated CLMs underlying it.  So, no additional assumptions on $\sampler$ are needed.

\end{definition}

We now have all the definitions required to formulate  the  version of compression (Theorem \ref{thm:compression}) which is proved in this paper. Recall  the asymptotic notation from Remark \ref{rem:asymptotic_notation}.

\begin{theorem}[Compression of  $h$-level tailored normal form verifiers] \label{thm:h_level_compression}

	For every positive integer $h$, there exist  two positive integers 
	\[
	c=c(h)\quad{\rm and}\quad C=C(h)
	\]
	that depend only on $h$,  and a $2$-input Turing machine $\Compress_h$, that takes as input a $h$-level
	TNFV $\verifier=(\sampler,\length,\linproc,\decider)$ and a positive integer $\lambda$ (in binary), and outputs  a \textbf{$\LevelConstant$-level} TNFV $\Compress_h(\verifier,\lambda)=\verifier'= (\sampler^\lambda, \length^\lambda, \linproc',\decider)$, such that:
	\begin{itemize}
		\item \underline{Sampler properties}:  The $5$-level CL sampler $\sampler^\lambda$ depends only on $\lambda$ and $h$, (but not the specific $\verifier$), and  $\Compress_h$ can calculate its description in time $\polylog_h (\lambda)$;\footnote{Recall our asymptotic notation from Remark \ref{rem:asymptotic_notation} to parse $\polylog_h$.} in particular, $|S^\lambda|\leq c\log^c\lambda$. In addition, $\sampler^\lambda$ runs in   $\poly_h(n,\lambda)$-time, namely
		\[
		\forall n\in \mathbb{N}\ \colon \ \ \TIME(\sampler^\lambda;n)\leq c\cdot (n^c+\lambda^c)\ .
		\]

		\item \underline{Answer length calculator properties}: $\length^\lambda$  depends only on $\lambda$ and $h$,  and $\Compress_h$ can calculate its description in time $\polylog_h (\lambda)$; in particular $|\length^\lambda|\leq c\log^c\lambda$. In addition, $\length^\lambda$ runs in   $\poly_h(n,\lambda)$-time, namely 
		\[
		\forall n\in \mathbb{N}\ \colon \ \ \TIME(\length^\lambda;n,\cdot,\cdot)\leq c\cdot (n^c+\lambda^c)\ .
		\]
		Finally,  given that $\mttx\in \FF_2^{r(n)}$, where $r(n)=\sampler^\lambda(n,{\rm Dimension},\cdot,\cdot,\cdot,\cdot)$,  and that $\kappa\in \{\frR,\frL\}$,  the output of $\length^\lambda(n,\mttx,\kappa)$  never decodes (Definition \ref{defn:the_alphabet}) to an $\frak{error}$ sign.
		
		\item \underline{Linear constraints process properties}: $\linproc'$ depends on both $\lambda$ and $\verifier$, and $\Compress_h$ can calculate its description in time $\poly_h(\log \lambda, |\verifier|)$; in particular, $|\linproc'|\leq c\cdot (\log^c\lambda +|\verifier|^c)$. In addition, $\linproc'$ runs in $\poly_h(n,\lambda)$-time, namely 
		\[
		\forall {n}\in \mathbb{N}\ \colon \ \ \TIME(\linproc';{n},\cdot,\cdot,\cdot,\cdot) \ \leq\ 
		c\cdot (n^c+\lambda^c)\ .
		\]
		\item \underline{Decider properties}: The canonical decider $\decider$ (Definition \ref{def:canonical-decider}) is fixed  and runs in  time  which is linear in its input length.
		
		\item  \underline{Value properties}: If $\verifier$ is $\lambda$-bounded, then 
		$\verifier'$, the output of $\Compress_h$,  satisfies: For all $n \geq C$,
		\begin{enumerate}
			\item \label{enu:h_level_compr-completeness} \textbf{Completeness}: If $\verifier_{2^n}$
			has a perfect $Z$-aligned permutation strategy that commutes along edges ($\ZPC$ strategy), then so does $\verifier'_n$.
			\item \label{enu:h_level_compr-soundness} \textbf{Soundness}:
			$\Ent(\verifier_{n}',\frac{1}{2}) \geq \max \left \{
			\Ent(\verifier_{2^n}, \frac{1}{2}), 2^{2^{\lambda n}-1}
			\right \}$.
		\end{enumerate}
	\end{itemize}
\end{theorem}

\begin{remark}\label{rem:LevelConstant}
	The above version of compression, Theorem \ref{thm:h_level_compression}, is the one proven in this paper. By choosing $h=\LevelConstant$, and using $\Compress_\LevelConstant$ from the above theorem instead of $\Compress$ from Theorem \ref{thm:compression},   $\TMIP^*=\RE$ (Theorem \ref{thm:tailored_MIP*=RE}) can still be deduced exactly as in Section \ref{sec:proving_Tailored-MIP*=RE}.
\end{remark}

In a similar way, we now describe the  version of Theorem \ref{thm:informal_question_reduciton} which is proved in this section. Recall again the asymptotic notation from Remark \ref{rem:asymptotic_notation}.

\begin{theorem}[$h$-level Question Reduction]\label{thm:h_level_question_reduciton}
	Let $h$ be a positive integer. There exists a positive integer 
	\begin{equation}\label{eq:defn_c_qr}
	c=c_\qr(h)
	\end{equation} 
	that depends only on $h$, and a  $2$-input Turing machine $\QueRed_h$ that takes as input an $h$-level TNFV $\verifier=(\sampler,\length,\linproc,\decider)$ and a positive integer $\lambda$ (in binary), and outputs a new  \textbf{$3$-level}  TNFV
	\[
	\QueRed_h(\verifier,\lambda)= \verifier'=(\sampler^\lambda_{\qr},\length^\lambda_{\qr},\linproc',\decider)
	\]
	such that:
	\begin{itemize}
		\item \underline{Sampler properties}:  $\sampler^\lambda_{\qr}$ depends only on $\lambda$ and $h$ (and not the specific $\verifier$), and  $\QueRed_h$ can calculate its description in time $\polylog_h (\lambda)$; in particular, $|S_\qr^\lambda|\leq c\log^c\lambda$. In addition, $\sampler^\lambda$ runs in   $\poly_h(n,\lambda)$-time, namely
		\[
		\forall n\in \mathbb{N}\ \colon \ \ \TIME(\sampler_\qr^\lambda;n)\leq c\cdot (n^c+\lambda^c)\ .
		\]

		\item \underline{Answer length calculator properties}: $\length_\qr^\lambda$  depends only on $\lambda$ and $h$,  and $\QueRed_h$ can calculate its description in time $\polylog_h (\lambda)$; in particular $|\length_\qr^\lambda|\leq c\log^c\lambda$. In addition, $\length^\lambda$ runs in   $\exp_h(n,\lambda)$-time, namely 
		\[
		\forall n\in \mathbb{N}\ \colon \ \ \TIME(\length^\lambda;n,\cdot,\cdot)\leq 2^{c\cdot (n^c+\lambda^c)}\ .
		\]
		Finally,  given that $\mttx\in \FF_2^{r(n)}$, where $r(n)=\sampler^\lambda_\qr(n,{\rm Dimension},\cdot,\cdot,\cdot,\cdot)$,  and that $\kappa\in \{\frR,\frL\}$,  the output of $\length^\lambda_\qr(n,\mttx,\kappa)$  never decodes (Definition \ref{defn:the_alphabet}) to an $\frak{error}$ sign.
		
		\item \underline{Linear constraints process properties}: $\linproc'$ depends on  $\lambda,h$ and $\verifier$, and $\QueRed_h$ can calculate its description in time $\poly_h(\log \lambda, |\verifier|)$; in particular, $|\linproc'|\leq c\cdot (\log^c\lambda +|\verifier|^c)$. In addition, $\linproc'$ runs in $\exp_h(n,\lambda)$-time, namely 
		\[
		\forall {n}\in \mathbb{N}\  \colon \ \ \TIME(\linproc';{n},\cdot,\cdot,\cdot,\cdot) \ \leq\ 
		2^{c\cdot (n^c+\lambda^c)}\ .
		\]
		Note that the running time upper bound itself is independent of the specific $\verifier$. 
		
		\item  \underline{Value properties}: If $\verifier$ is $\lambda$-bounded, then 
		$\verifier'$, the output of $\QueRed_h$,  satisfies: For all $n \geq 2$,
		\begin{enumerate}
			\item \textbf{Completeness}: If $\verifier_{2^n}$ has a perfect $Z$-aligned permutation strategy, then so does $\verifier'_n$.
			\item \textbf{Soundness}: For every $\eps>0$, if $\verifier'_{n}$ has a value $1-\eps$ strategy, then $\verifier_{2^n}$  has a value $1-c\cdot\eps^{\nicefrac{1}{16}}$ strategy.
			\item \textbf{Entanglement}: For every $\eps>0$, 
			\[
			\Ent(\verifier'_n,1-\eps)\geq (1-c\cdot\eps)\cdot 2^{2^{\lambda n}}\cdot\Ent(\verifier_{2^n},1-c\cdot\eps^{\nicefrac{1}{16}})\ .
			\]
		\end{enumerate}
	\end{itemize}
	
\end{theorem}

\begin{remark}
	As claimed in the theorem, the output of $\mathsf{QuestionReduction}_h$ is always a  \textbf{$3$-level} normal form verifier, regardless of what $h$ was. The original level $h$ plays a role in the underlying graph of $\verifier'$, as well as in the exact soundness guarantees, governed by $c_\qr(h)$ \eqref{eq:defn_c_qr}.
	
\end{remark}

\subsubsection{Typed conditionally linear sampling schemes}\label{sec:typed_CL_samplings}
Though we focus on the category of games induced by (two) $h$-level CLMs --- and respectively on  $h$-level TNFVs --- it will often be easier to describe the underlying sampling scheme of our games in a slightly different manner. 
\begin{definition}\label{defn:typed_h-level_sampling_scheme}
	A game $\game$ is said to have a \emph{$h$-level typed conditionally linear sampling scheme}  if its  underlying graph $(V,E)$ and measure $\mu$ on edges are defined as follows: There is a positive integer $r$ and a set $\type$ --- which we call the \emph{type set}, and its elements are called \emph{types} --- such that $V=\type \times \FF_2^r$. In addition, there is a subset $\cal{E}$ of $\type\times \type$ which induces a graph structure $(\type,\cal{E})$, and an $h$-level CLM $\frS^t\colon \FF_2^r\to \FF_2^r$ associated to every type $t\in \type$. Then, a pair in $V=\type\times \FF_2^r$ is sampled by taking a uniformly random seed $z\in \FF_2^r$ and a uniformly random edge $tt'\in \cal{E}$ and outputting the pair $(t,\frS^t(z)),(t',\frS^{t'}(z))$ --- namely, the pair of types are chosen uniformly from $\cal{E}$, and the seed defines the right coordinates by evaluating the CLM of the chosen types on it.
\end{definition}

\begin{example}\label{example:quered_has_typed_CL_sampling_scheme}
	Let $\game$ be a game with $h$-level CLMs acting on $\FF_2^k$ controlling its sampling scheme, and let  $\mathscr{B}\subseteq\FF_2^k$ be a set of size $2^m$ for some $m\geq \log k$.
	Then, the sampling scheme of $\frak{QueRed}(\game)=\frak{QueRed}_h(\game,k,\mathscr{B})$ which was described in Section \ref{sec:augmentation_in_the_CLM_case}, is (up to some constant factor that depends only on $h$) a  $1$-level typed conditionally linear sampling scheme. To see that, the type set $\type$ consists of 
	\[
	\begin{split}
	\forall 1\leq j\leq h\ \colon \ \ &\Hide A^j\ ,\ \Hide B^j\ ,\\
	&\Intro_A\ ,\ \Intro_B\ ,\ \Read_A\ ,\ \Read_B\ ,\ \Sample_A\ ,\ \Sample_B\ ,\\
	&\Pauli_\PZm\ ,\ \Pauli_\PXm\ ,\ \mttX\ ,\ \mttZ\ ,\ \mathtt{First}\ ,\ \mathtt{Second}\ ,\ \mathtt{Both}\ ,\\
	\forall 1\leq a\leq 3\ ,\  1\leq b\leq 3\ \colon \ \ &\mathtt{var}_{ab}\ ,\ \mathtt{row}_a\ ,\ \mathtt{col}_b\ .
	\end{split}
	\]
	All in all, $2h+28$ types. They are connected according to Figure \ref{fig:typed_graph_quered} --- though not drawn in the figure, we assume \textbf{all loops} appear in the type graph, namely for any type $t$ of the above $2h+28$ types, in addition to the edges in the figure, the edge $tt$ is also present. 
	
	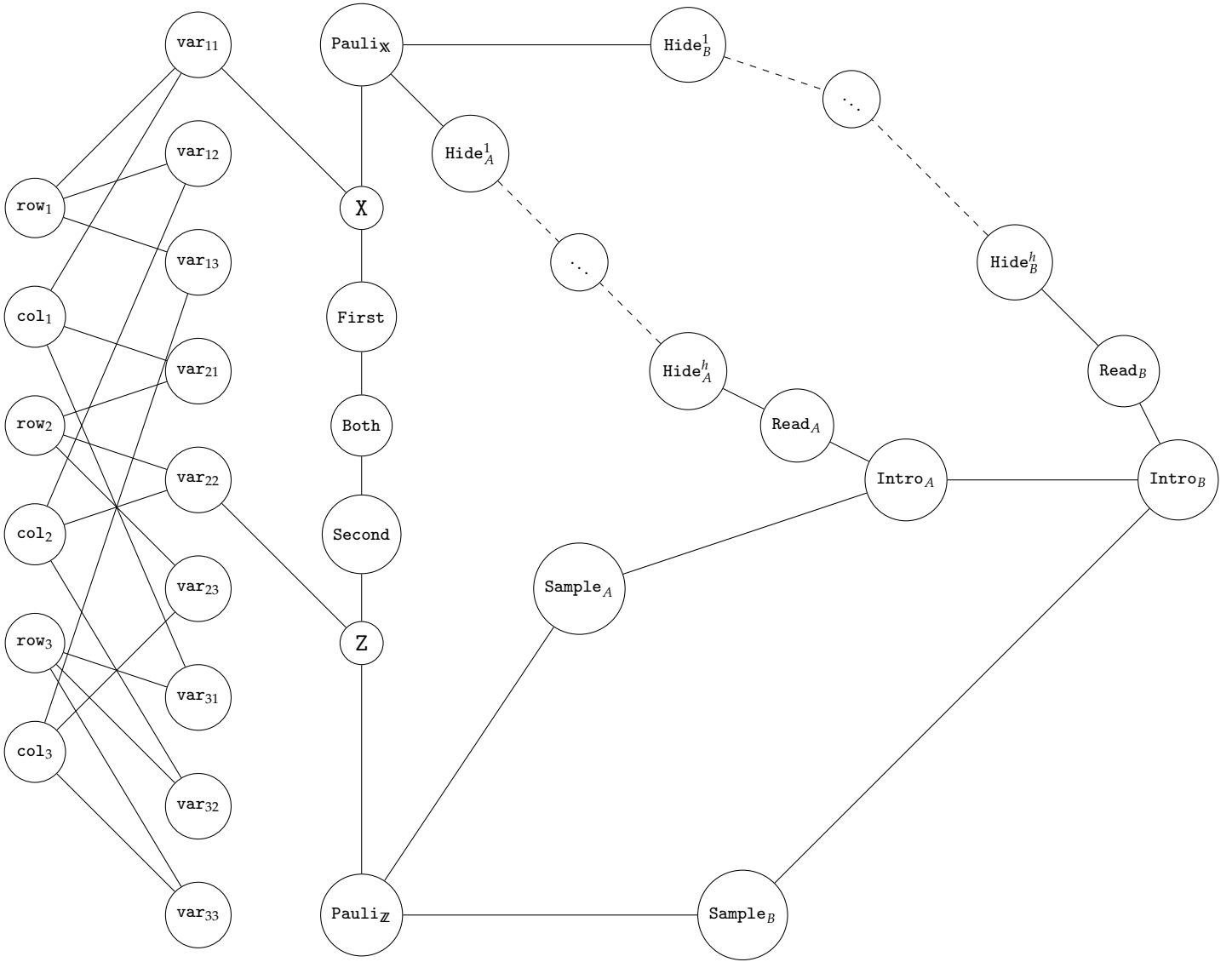
\begin{figure}[httb!]
		\centering
		\begin{tikzpicture}[scale=0.85]
		\node[draw, color=black, shape=circle] (PX) at (0,8) {\scriptsize $\Pauli_\PXm$}; 
		\node[draw, color=black, shape=circle] (PZ) at (0,-8) {\scriptsize $\Pauli_\PZm$}; 
		\node[draw, color=black, shape=circle] (X) at (0,5) { $\mathtt{X}$}; 
		\node[draw, color=black, shape=circle] (First) at (0,3) {\scriptsize $\mathtt{First}$};
		\node[draw, color=black, shape=circle] (Both) at (0,1) {\scriptsize $\mathtt{Both}$};
		\node[draw, color=black, shape=circle] (Second) at (0,-1) {\scriptsize $\mathtt{Second}$};
		\node[draw, color=black, shape=circle] (Z) at (0,-3) { $\mathtt{Z}$}; 
		\node[draw, color=black, shape=circle] (Var11) at (-3,8) {\scriptsize $\mathtt{var}_{11}$}; 
		\node[draw, color=black, shape=circle] (Var12) at (-3,6) {\scriptsize $\mathtt{var}_{12}$}; 
		\node[draw, color=black, shape=circle] (Var13) at (-3,4) {\scriptsize $\mathtt{var}_{13}$}; 
		\node[draw, color=black, shape=circle] (Var21) at (-3,2) {\scriptsize $\mathtt{var}_{21}$}; 
		\node[draw, color=black, shape=circle] (Var22) at (-3,0) {\scriptsize $\mathtt{var}_{22}$}; 
		\node[draw, color=black, shape=circle] (Var23) at (-3,-2) {\scriptsize $\mathtt{var}_{23}$}; 
		\node[draw, color=black, shape=circle] (Var31) at (-3,-4) {\scriptsize $\mathtt{var}_{31}$}; 
		\node[draw, color=black, shape=circle] (Var32) at (-3,-6) {\scriptsize $\mathtt{var}_{32}$}; 
		\node[draw, color=black, shape=circle] (Var33) at (-3,-8) {\scriptsize $\mathtt{var}_{33}$}; 
		
		\node[draw, color=black, shape=circle] (Row1) at (-6,5) {\scriptsize $\mathtt{row}_{1}$}; 
		\node[draw, color=black, shape=circle] (Col1) at (-6,3) {\scriptsize $\mathtt{col}_{1}$}; 
		\node[draw, color=black, shape=circle] (Row2) at (-6,1) {\scriptsize $\mathtt{row}_{2}$}; 
		\node[draw, color=black, shape=circle] (Col2) at (-6,-1) {\scriptsize $\mathtt{col}_{2}$}; 
		\node[draw, color=black, shape=circle] (Row3) at (-6,-3) {\scriptsize $\mathtt{row}_{3}$}; 
		\node[draw, color=black, shape=circle] (Col3) at (-6,-5) {\scriptsize $\mathtt{col}_{3}$};

		\node[draw, color=black, shape=circle] (IntA) at (10,0) {\scriptsize $\Intro_A$};

		\node[draw, color=black, shape=circle] (IntB) at (15,0) {\scriptsize $\Intro_B$}; 
		
		\node[draw, color=black, shape=circle] (SamA) at (4,-2) {\scriptsize $\Sample_A$};

		\node[draw, color=black, shape=circle] (SamB) at (7,-8) {\scriptsize $\Sample_B$};
		
		\node[draw, color=black, shape=circle] (ReadA) at (8,1) {\scriptsize $\Read_A$};

		\node[draw, color=black, shape=circle] (ReadB) at (14,2) {\scriptsize $\Read_B$};
		
		\node[draw, color=black, shape=circle] (HidA1) at (2,6) {\scriptsize $\Hide A^1$}; 
		
		\node[draw, color=black, shape=circle] (HidAh) at (6,2) {\scriptsize $\Hide A^h$}; 
		\node[draw, color=black, shape=circle] (HidAdots) at (4,4) {\scriptsize $\ddots$};

		\node[draw, color=black, shape=circle] (HidB1) at (6,8) {\scriptsize $\Hide B^1$};
		
		\node[draw, color=black, shape=circle] (HidBh) at (12,4) {\scriptsize $\Hide B^h$};
		
		\node[draw, color=black, shape=circle] (HidBdots) at (9,7) {\scriptsize $\ddots$};

		\draw[black, -, solid] (PX)--(X);
		\draw[black, -, solid] (First)--(X);
		\draw[black, -, solid] (Var11)--(X);
		\draw[black, -, solid] (Var22)--(Z);
		\draw[black, -, solid] (First)--(Both);
		\draw[black, -, solid] (Second)--(Both);
		\draw[black, -, solid] (Second)--(Z);
		\draw[black, -, solid] (PZ)--(Z);
		\draw[black, -, solid] (PZ)--(SamA);
		\draw[black, -, solid] (PZ)--(SamB);
		\draw[black, -, solid] (IntA)--(IntB);
		\draw[black, -, solid] (SamA)--(IntA);
		\draw[black, -, solid] (SamB)--(IntB);
		\draw[black, -, solid] (PX)--(HidA1);
		\draw[black, -, solid] (PX)--(HidB1);
		\draw[black, -, solid] (HidAh)--(ReadA);
		\draw[black, -, dashed] (HidAh)--(HidAdots);
		\draw[black, -, dashed] (HidA1)--(HidAdots);
		\draw[black, -, solid] (HidBh)--(ReadB);
		\draw[black, -, dashed] (HidBh)--(HidBdots);
		\draw[black, -, dashed] (HidB1)--(HidBdots);
		\draw[black, -, solid] (IntA)--(ReadA);
		\draw[black, -, solid] (IntB)--(ReadB);
		
		\draw[black, -, solid] (Var11)--(Row1);
		\draw[black, -, solid] (Var11)--(Col1);
		\draw[black, -, solid] (Var12)--(Row1);
		\draw[black, -, solid] (Var12)--(Col2);
		\draw[black, -, solid] (Var13)--(Row1);
		\draw[black, -, solid] (Var13)--(Col3);
		
		\draw[black, -, solid] (Var21)--(Row2);
		\draw[black, -, solid] (Var21)--(Col1);
		\draw[black, -, solid] (Var22)--(Row2);
		\draw[black, -, solid] (Var22)--(Col2);
		\draw[black, -, solid] (Var23)--(Row2);
		\draw[black, -, solid] (Var23)--(Col3);

		\draw[black, -, solid] (Var31)--(Row3);
		\draw[black, -, solid] (Var31)--(Col1);
		\draw[black, -, solid] (Var32)--(Row3);
		\draw[black, -, solid] (Var32)--(Col2);
		\draw[black, -, solid] (Var33)--(Row3);
		\draw[black, -, solid] (Var33)--(Col3);

		\end{tikzpicture}
		\caption{The type graph of $\frak{QueRed}(\game)$. Though not drawn, all self loops are also edges in this type graph. So, in total, there are $2h+28$ vertices and $4h+63$ edges in this type graph.
		}
		\label{fig:typed_graph_quered}
	\end{figure}
	Now, the $1$-level CLMs (i.e., linear maps) associated to the types act on the space $\FF_2^{2m}$, which has (not surprisingly) the same cardinality as $\mathscr{B}\times \mathscr{B}$. For all non Pauli basis types, i.e., 
	\[
	\Hide A^j\ ,\ \Hide B^j\ ,\ 
	\   \Intro_A\ ,\ \Intro_B\ ,\ \Read_A\ ,\ \Read_B\ ,\ \Sample_A\ ,\ \Sample_B\ ,
	\]
	and for $\Pauli_\PXm$ and $\Pauli_\PZm$, the associated CLM is the $0$-function. Namely, for every such type $t$,
	\[
	\forall u,v\in \FF_2^m\ \colon \ \ \frS^t(u,v)=(\vec 0,\vec 0)\ .
	\]
	For all other types $t$ except for $\mttX,\mttZ$, i.e., types from the list
	\[
	\mathtt{row}_a\ ,\ \mathtt{col}_b\ ,\ \mathtt{var}_{ab}\ ,\ \mathtt{First}\ ,\ \mathtt{Second}\ ,\ \mathtt{Both}\ ,
	\]
	the associated CLM is the identity map. Namely, 
	\[
	\forall u,v\in \FF_2^m\ \colon \ \ \frS^t(u,v)=(u,v).
	\]
	Finally, for $\mttX$ we have 
	\[
	\forall u,v\in \FF_2^m\ \colon \ \ \frS^\mttX(u,v)=(u,\vec 0).
	\]
	and for $\mttZ$ we have 
	\[
	\forall u,v\in \FF_2^m\ \colon \ \ \frS^\mttZ(u,v)=(\vec 0, v).
	\]
	In this perspective, the vertices of $\frak{QueRed}(\game)$ are from $\type\times \FF_2^m\times \FF_2^m$. But the vertices with  a positive probability to be sampled belong to the following strict subset:
	\[
	\begin{split}
	\forall 1\leq j\leq h\ \colon \ \ &(\Hide A^j,\vec 0,\vec 0)\ ,\ (\Hide B^j,\vec 0,\vec 0)\ ,\\
	&(\Intro_A,\vec 0,\vec 0)\ ,\ (\Intro_B,\vec 0,\vec 0)\ ,\ (\Read_A,\vec 0,\vec 0)\ ,\ (\Read_B,\vec 0,\vec 0)\ ,\\
	&(\Sample_A,\vec 0,\vec 0)\ ,\ (\Sample_B,\vec 0,\vec 0)\ ,\ (\Pauli_\PZm,\vec 0,\vec 0)\ ,\ (\Pauli_\PXm,\vec 0,\vec 0)\ \\
	\forall u,v\in \FF_2^m\ \colon \ \ &(\mttX,u,\vec 0)\ ,\ (\mttZ,\vec 0,v)\ ,\\
	\forall 1\leq a,b\leq 3\ ,\ u,v\in \FF_2^m\ \colon \ \ &(\mathtt{var}_{ab},u,v)\ ,\ (\mathtt{row}_a,u,v)\ ,\ (\mathtt{col}_b,u,v)\ ,\\
	\forall u,v\in \FF_2^m\ \colon\ \ &(\mathtt{First},u,v)\ ,\ (\mathtt{Second},u,v)\ ,\ (\mathtt{Both},u,v)\ .
	\end{split}
	\]
	It is straightforward to compare these vertices to the ones we originally had when defining $\frak{QueRed}(\game)$:
	\begin{enumerate}
		\item The vertices 
		\[
		\begin{split}
		\forall 1\leq j\leq h\ \colon \ \ &(\Hide A^j,\vec 0,\vec 0)\ ,\ (\Hide B^j,\vec 0,\vec 0)\ ,\ (\Intro_A,\vec 0,\vec 0)\ ,\ (\Intro_B,\vec 0,\vec 0)\ ,\ (\Read_A,\vec 0,\vec 0)\ \\
		&(\Read_B,\vec 0,\vec 0)\ ,\ (\Sample_A,\vec 0,\vec 0)\ ,\ (\Sample_B,\vec 0,\vec 0)\ ,\ (\Pauli_\PZm,\vec 0,\vec 0)\ ,\ (\Pauli_\PXm,\vec 0,\vec 0)\ ,
		\end{split}
		\]
		are the same vertices as 
		\[
		\begin{split}
		\forall 1\leq j\leq h\ \colon \ \ &\Hide A^j\ ,\ \Hide B^j\ ,\ \Intro_A\ ,\ \Intro_B\ ,\ \Read_A\ ,\\
		&\Read_B\ ,\ \Sample_A\ ,\ \Sample_B\ ,\ \Pauli_\PZm\ ,\ \Pauli_\PXm\ ,
		\end{split}
		\]
		in the original description.
		\item The vertices 
		\[
		\forall u,v\in \FF_2^m\ \colon \ \ (\mttX,u,\vec 0)\ ,\ (\mttZ,\vec 0,v)\ ,
		\]
		are the same as the vertices 
		\[
		\forall u,v\in \FF_2^m\ \colon \ \ \mttX^u ,\ \mttZ^v\ ,
		\]
		in the original description of $\PauliBasis_k$ in Section \ref{sec:Pauli_basis_definition}. This makes sense as the set $\mathscr{B}$ is of size $N=2^m$ which can be parametrized by $\FF_2^m$.\footnote{When we defined $\PauliBasis_k$ we used $i,j$ to parametrize the elements of $\mathscr{B}$ instead of $u,v$.}
		\item The vertices 
		\[
		\forall 1\leq a,b\leq 3\ ,\ u,v\in \FF_2^m\ \colon \ \ (\mathtt{var}_{ab},u,v)\ ,\ (\mathtt{row}_a,u,v)\ ,\ (\mathtt{col}_b,u,v)\ ,
		\]
		are the same as the vertices 
		\[
		\forall 1\leq a,b\leq 3\ ,\ u,v\in \FF_2^m\ \colon \ \ \mathtt{var}_{ab}^{u,v}\ ,\ \mathtt{row}_a^{u,v}\ ,\ \mathtt{col}_b^{u,v}\ ,
		\]
		from the $(u,v)$-th copy of the anti-commutation game (or its nullified version) $\frak{M}^{u,v}$  used in the definition of $\PauliBasis_k$.
		\item The vertices 
		\[
		\forall u,v\in \FF_2^m\ \colon\ \ (\mathtt{First},u,v)\ ,\ (\mathtt{Second},u,v)\ ,\ (\mathtt{Both},u,v)\ ,
		\]
		are the same as the vertices 
		\[
		\forall u,v\in \FF_2^m\ \colon\ \ \mathtt{First}^{u,v}\ ,\ \mathtt{Second}^{u,v}\ ,\ \mathtt{Both}^{u,v}\ ,
		\]
		from the $(u,v)$-th copy of the commutation game (or its nullified version) $\frak{C}^{u,v}$  used in the definition of $\PauliBasis_k$.
	\end{enumerate}
	The distribution on edges of $\frak{QueRed}(\game)$ induced by this typed $1$-level CL sampling scheme is not \emph{exactly} the one we used before, but the probabilities are the same up to some constant factor (which depends on $h$). For example, the probability of sampling $\Intro_A-\Intro_B$ in this setup is $\frac{1}{|\cal{E}|}=\frac{1}{63+4h}$, which is lower than the $\nicefrac{1}{4}$ we had before, while  the probability of sampling $\mathtt{var}^{u,v}_{11}-\mathtt{row}_1^{u,v}$ is  $\frac{1}{(63+4h)\cdot 2^{2m}}$, which may be lower or higher than the  $\frac{1}{72\cdot 2^{2m}}$ it was before (depending on $h$).
	Though these distributions differ, the new distribution samples an edge with probability of at  least $\frac{1}{63+4h}$ times the old probability. Thus, all of our soundness arguments (in which the distribution played a role) are the same up to a constant depending on $h$.
\end{example}

\subsubsection{Detyping}\label{sec:DeTyping}
As Compression (Theorem \ref{thm:h_level_compression}) begins and ends with an $h$-level normal form verifier, and {not} with typed ones, we ought to have a method of \emph{detyping} a sampling scheme in a way that preserves most of the properties of the original game. To that end we make the following definition.

\begin{definition}[Combinatorial Detyping]\label{defn:combi_detyping}
	Let $\game$ be a game with a  $h$-level typed CL sampling scheme (Definition \ref{defn:typed_h-level_sampling_scheme}), where the type graph is $(\cal{T},\cal{E})$, and where the CLMs $\frS^t$ act on $\FF_2^r$. The \emph{detyped version} of $\game$, which we denote by $\frak{DeType}(\game)$, is a game with an $(h+2)$-level {non-typed} CL sampling scheme  (Definition \ref{defn:sampling_induced_by_CLMs}), with CLMs $\frS^A,\frS^B$ acting on $\FF_2^{4|\cal{T}|+r}$, and which are defined as follows: 
	First, the type $t\in \type$ is encoded as a string $$\enc(t)=({\bf 1}_t,\sum_{t'\sim t}{\bf 1}_{t'})\in \FF_2^\type\times \FF_2^\type\ ,$$
	namely the first vector is the indicator of $t$, and the second vector is the indicator of neighbors of $t$ in the graph $(\type,\cal{E})$. Given 
	$$(u_{type},u_{neighbors},v_{type},v_{neighbors},z)\in (\FF_2^{\type})^4\times \FF_2^r\ ,$$
	$\frS^A$ operates as follows:
	\begin{enumerate}
		\item First, $\frS^A$  applies the identity on $u_{type},u_{neighbors}$. 
		\item Then, there are two options --- either there exists a $t\in \type$ such that $(u_{type},u_{neighbors})=\enc(t)$, or not. If not, then $\frS^A$ zeros out the rest of the registers, namely outputs $(u_{type},u_{neighbors},\vec 0,\vec 0,\vec 0)$. If there is such a $t\in \type$,
		then  it zeroes out all $v_{type}$ and all coordinates of $v_{neighbors}$ except the $t^{\rm th}$ one. 
		\item Lastly, if the partial computation of the first two steps resulted in  $(\enc(t),\vec 0,{\bf 1}_t)\in (\FF_2^{\type})^4$, then it applies $\frS^t$ on the seed $z\in \FF_2^r$, and otherwise zeroes $z$ out. 
	\end{enumerate}
	The CLM $\frS^B$ acts similarly with the roles of $u$ and $v$ swapped. Namely:
	\begin{enumerate}
		\item First, $\frS^B$  applies the identity on $v_{type},v_{neighbors}$. 
		\item Then, there are two options --- either there exists a $t'\in \type$ such that $(v_{type},v_{neighbors})=\enc(t')$, or not. If not, then $\frS^B$ zeros out the rest of the registers, namely outputs $(\vec 0,\vec 0,v_{type},v_{neighbors},\vec 0)$. If there is such a $t'\in \type$,
		then  it zeroes out all $u_{type}$ and all coordinates of $u_{neighbors}$ except the $t^{\rm th}$ one. 
		\item Lastly, if the partial computation of the first two steps resulted in  $(\vec 0,{\bf 1}_{t'},\enc(t'))\in (\FF_2^{\type})^4$, then it applies $\frS^{t'}$ on the seed $z\in \FF_2^r$, and otherwise zeroes $z$ out. 
	\end{enumerate}
	Given $t\in \cal{T}$, the vector $(\enc(t),\vec 0,{\bf 1}_t,\mttx)$ is called \emph{the $A$-copy of $(t,\mttx)$}, and similarly $(\vec 0,{\bf 1}_{t'},\enc(t'),\mtty)$ is called \emph{the $B$-copy of $(t',\mtty)$}.
	A vector of the form $(u_{type},u_{neighbors},\vec 0,\vec 0,\vec 0)$ is called an $A$-player anchor vertex, and a vector of the form  $(\vec 0,\vec 0,v_{type},v_{neighbors},\vec 0)$ is called a $B$-player anchor vertex.\footnote{As will soon be described, whenever an anchor vertex is sampled, the game always accepts. 
		In the case of detyping, these vertices are added just for the CL sampling structure to be attained. Later in this paper, the idea of anchoring a game is used for parallel repetition --- see Section \ref{sec:parallel_rep}.} 
	Note that the above sampling procedure produces a pair with at least one anchor vertex, unless  there is an edge $tt'\in \cal{E}$  in the type graph such that $(u_{type},u_{neighbors},v_{type},v_{neighbors})=(\enc(t),\enc(t'))$, in which case 
	\[
	\frS^A(\enc(t),\enc(t'),z)=(\enc(t),\vec 0,{\bf 1}_{t},\frS^t(z))\quad, \quad \frS^B(\enc(t),\enc(t'),z)=(\vec 0,{\bf 1}_{t'},\enc(t'),\frS^{t'}(z))\ ,
	\]
	namely this is an edge between the $A$-copy of $(t,\frS^t(z))$ and the $B$-copy of $(t',\frS^{t'}(z))$. All in all, the only vertices that have a positive probability of being sampled are those $\cdot$-player ``anchor'' and ``copy'' vertices. 
	
	For the length functions, if $\mttw$ is a vertex of $\frak{DeType}(\game)$ which is the $A$ or $B$ copy of a vertex $(t,\mttx)$ in $\game$, then $\ell_{\frak{DeType}(\game)}^\cdot(\mttw)=\ell_{\game}^\cdot(t,\mttx)$. Namely, the length of the copies of any vertex  in the detyped game is the same as in the original game. For the rest of the vertices, the length functions are zero.
	
	For the decision procedure: If one of the vertices of the sampled edge is an anchor vertex, then the game accepts no matter what the answers are. Otherwise, the seed was of the form $(\enc(t),\enc(t'),z)$, and thus the sampled edge is between the $A$-copy of $(t,\frS^z(t))$ and the $B$-copy of $(t',\frS^{t'}(z))$; in this case,  $\frak{DeType}(\game)$ checks the answers at these vertices as if they were from $\game$, and decides accordingly. In addition, if $t=t'$, then the detyped game will also check consistency, namely that the answers at the $A$-copy of  $(t,\frS^z(t))$  and the $B$-copy of the same vertex  $(t,\frS^z(t))$ are the same.
\end{definition}

\begin{remark}[The double cover embeds in the detyped game]\label{rem:double_cover_embeds_in_detyping}
	Note that the game  $\game$ does not embed in  $\frak{DeType}(\game)$ --- every vertex $(t,\frS^t(z))\in \type\times \FF_2^r$ is mapped to two vertices, 
	$$(\enc(t),\vec 0,{\bf 1}_t,\frS^t(z))\quad \textrm{and} 
	\quad    (\vec 0,{\bf 1}_t,\enc(t),\frS^t(z))\ .$$ 
	But, by restricting the detyped game $\frak{DeType}(\game)$ only to such vertices (i.e.,  $A$-copies and $B$-copies of vertices from $\game$), we get a  copy of the \textbf{double cover} $\frak{DoubleCover}(\game)$ (Definition \ref{defn:double_cover}) of the game $\game$. Moreover,  this copy is played with probability of at least $2^{-4|\type|}$ (and otherwise, one of the sampled vertices is an anchor one, which means it automatically accepts). 
\end{remark}

The above remark leads us to the following immediate corollary.
\begin{corollary}[Completeness and soundness of the detyped game. Cf.\ Lemma 6.18 in \cite{MIPRE}]\label{cor:comp_sound_combi_detyping}
	Let $\game$ be a tailored game with a $h$-level conditionally linear sampling scheme, type graph $(\type,\cal{E})$, and  CLMs which act on a space of dimension $r$. Then, letting  $\frak{DeType}(\game)$ be the detyping of $\game$ as in Definition \ref{defn:combi_detyping}, we have the following:
	\begin{itemize}
		\item $\frak{DeType}(\game)$ is a tailored game with an $(h+2)$-level conditionally linear sampling scheme.
		\item (Completeness) If $\game$ has a perfect $\ZPC$ strategy, then so does its detyping $\frak{DeType}(\game)$.
		\item (Soundness) If the detyped game  $\frak{DeType}(\game)$ has a value $1-\eps$ strategy, then the double cover $\frak{DoubleCover}(\game)$ (Definition \ref{defn:double_cover}) of the original game $\game$ has a value $1-O(2^{4|\type|}\eps)$ strategy. 
		\item (Entanglement) In addition, 
		\[
		\Ent(\game',1-\eps)\geq\Ent(\frak{DoubleCover}(\game),1-O(2^{4|\type|}{\eps}))\ .
		\]
	\end{itemize}
\end{corollary}
\begin{proof}[Proof sketch]
	The claim item is clear and follows by inspection. In particular, the description of $\frS^A$ and $\frS^B$ in Definition~\ref{defn:combi_detyping} make them clearly $(h+2)$-level.
	The soundness and entanglement lower bound are immediate from Remark \ref{rem:double_cover_embeds_in_detyping}. For completeness, recall that the double cover has a perfect $\ZPC$ strategy given that $\game$ has one (Claim \ref{claim:completeness_soundness_double_cover}). Every such  strategy extends,  in a $\ZPC$ manner, to the rest of the vertices of $\frak{DeType}(\game)$ --- as the lengths at all the other vertices is $0$. As the decision in $\frak{DeType}(\game)$ always accepts when one of the endpoints of the sampled edges  is an anchor, the resulting $\ZPC$ strategy for  $\frak{DeType}(\game)$ is indeed perfect.
\end{proof}

\begin{corollary}\label{cor:completeness_soundness_detyping_when_all_loops_occur}
	If in the type graph $(\type,\cal{E})$ underlying the sampling scheme of $\game$,  all self loops $tt$ are edges in $\cal{E}$, then the detyped game  $\frak{DeType}(\game)$ satisfies the following strengthened soundness and entanglement lower bound conditions: If $\frak{DeType}(\game)$  has a value $1-\eps$ strategy, then the {original game} $\game$ (and not its double cover) has a value $1-O(|\type|\cdot 2^{2|\type|}\cdot \sqrt{\eps})$ strategy of the same dimension, which implies 
	\[
	\Ent(\game',1-\eps)\geq\Ent(\game,1-O(|\type|\cdot 2^{2|\type|}\cdot {\sqrt\eps}))\ .
	\]
\end{corollary}
\begin{proof}
	Recall that if $\game$ has a typed $h$-level sampling scheme (Definition \ref{defn:typed_h-level_sampling_scheme}), then it samples edges as follows: It chooses a uniform edge of types $tt'\in \cal{E}$, and a uniform $z\in \FF_2^r$, and returns the edge $(t,\frS^t(z))(t',\frS^{t'}(z))$. This means that the marginal distribution $\mu(t,\frS^t(z))$ is $$\Pro{z'\in  \FF_2^r}[\frS^t(z')=\frS^t(z)]\cdot \frac{|\{t'\neq t\mid tt'\in \cal{E}\}|+|\{t'\neq t\mid t't\in \cal{E}\}|+1}{|\cal{E}|}\;, $$
	while the probability $\mu((t,\frS^t(z))(t,\frS^t(z)))$ of choosing this loop is
	$$\Pro{z'\in  \FF_2^r}[\frS^t(z')=\frS^t(z)]\cdot \frac{1}{|\cal{E}|}\ . $$
	Hence, for every $\mttx\in \type\times \FF_2^r$ (with positive probability of being sampled) we have
	\[
	\frac{\mu(\mttx\mttx)}{\mu(\mttx)}\geq \frac{1}{2|\type|+1}\ .
	\]
	Combining Corollary \ref{cor:comp_sound_combi_detyping} and Claim \ref{claim:completeness_soundness_double_cover} finishes the proof.
\end{proof}

As for all of our combinatorial transformations, we need to implement them on the level of {normal form verifiers} to prove compression. Thus, we need to define the \emph{typed} version of an  $h$-level CL sampler, which will underlie some normal form verifier instead of the \emph{usual} $h$-level sampler (Definition \ref{defn:h-level_CL_Sampler}). {The definitions are very similar}. The main difference is that the following encodes a sequence of CLMs parametrized by the vertices of a {fixed} finite type graph $(\type,\cal{E})$, instead of just two sequences of CLMs. Actually, the following can be seen as a generalization of the non-typed $h$-level sampler, where in that case the type graph consists of two vertices $A$ and $B$ with a single oriented edge $AB$ between them.
\begin{definition}[Typed $h$-level conditionally linear sampler]\label{defn:typed_h-level_CL_Sampler}
	Let $h$ be a positive integer. Colloquially, a \emph{$h$-level  typed conditionally linear sampler} (typed CL sampler) $\sampler$ with underlying type graph $(\type,\cal{E})$ is, essentially, an $h$-level conditionally linear sampler (Definition \ref{defn:h-level_CL_Sampler}), but instead of the ${\rm Player}$ action having only $2$ possible inputs, it has $|\type|$ inputs.
	
	Formally, $\sampler$ is a $6$-input deterministic Turing machine that satisfies some additional properties. First, its input is expected to be 
	\[
	(n,{\rm Action},{\rm Type},j,\mttx,z)\ ,
	\]
	where $n$  and $j$ are positive integers in binary, ${\rm Action}$ is taken from the set $$\{{\rm Graph},\ 
	{\rm Dimension,\ Register,\ Marginal,\ Evaluate,\ Perpendicular }\}\ ,$$ ${\rm Type}$ is taken from the set $\type$, and $\mttx,z$ are bit strings (interpreted as vectors in some finite vector space).  
	Second, for every positive integer $n$, there exist an integer $r=r(n)$, and for every type $t\in \type$ there is  an $h$-level conditionally linear map (CLM, Definition \ref{def:CLM}) $\frS^t(n)=\frS^t\colon \FF_2^r\to \FF_2^r$, such that $\sampler$ \emph{encodes} the appropriate typed $h$-level sampling scheme (Definition \ref{defn:typed_h-level_sampling_scheme}):
	\begin{enumerate}
		\item If $\sampler$ gets as input $(\cdot,{\rm Graph},\cdot,\cdot,\cdot,\cdot)$, then it outputs the graph $(\type,\cal{E})$ in the following way: It provides a list of all the types in $\type$ according to some order, and then the adjacency matrix associated to $\cal{E}$ with respect to the order induced on $\type$.
		\item If $\sampler$ gets as input $(\on,{\rm Dimension},\cdot,\cdot,\cdot,\cdot)$, then it outputs (the binary encoding of) $r(n)$.
		\item If $\sampler$ gets as input $(\on,{\rm Register},{\rm Type},j,\mttx,\cdot)$, then it outputs the $j^{\rm th}$ register subspace with respect to the seed $\mttx$, namely $W^{\mttx}_j$ \eqref{eq:reg_subspaces_depending_on_prefixes} with respect to  $\frS^{\rm Type}$.
		\item If $\sampler$ gets as input $(\on,{\rm Marginal},{\rm Type},j,\cdot,z)$, then it outputs the $j^{\rm th}$ prefix of $\frS^{\rm Type}$'s evaluation of $z$, namely  $\frS^{\rm Type}_{\leq j}(z)$.
		\item If $\sampler$ gets as input $(\on,{\rm Evaluate},{\rm Type},j,\mttx,z)$, then it outputs the $j^{\rm th}$-register output of $\frS^{\rm Type}$  evaluated on $z$ given the seed $\mttx$, namely $\frS^{\rm Type,\mttx}_{j}(z^{W^{\mttx}_j})$. Recall that $\frS^{{\rm Type},\mttx}_j\colon W^\mttx_j\to W^\mttx_j$ is the linear function which controls the $j^{\rm th}$ step in the calculation of $\frS^{\rm Type}$, given that the calculation up to this point produced $\mttx_{<j}$.
		\item If $\sampler$ gets as input $(\on,{\rm  Perpendicular},{\rm Type},j,\mttx,z)$, then it outputs the $j^{\rm th}$-register output of $(\frS^{\rm Type})^{\perp}$  evaluated on $z$ given the seed $\mttx$, namely $(\frS^{\rm Type,\mttx}_{j})^{\perp}(z^{W^\mttx_j})$. Recall that the maps $(\frS^{{\rm Type},\mttx}_j)^{\perp}\colon W^\mttx_j\to W^\mttx_j$ are some fixed linear function whose rows are spanning the subspace perpendicular to the rows of $\frS^{{\rm Type},\mttx}_j$. 
	\end{enumerate}
	Conditions 3. to 6. above are essentially identical to conditions 2. to 5. in Definition Definition \ref{defn:h-level_CL_Sampler}, except that the ``{\rm Player}'' input is replaced by the ``{\rm Type}'' input, which may have a bigger range (i.e.\ the set $\mathcal{T}$).
\end{definition}

\begin{definition}[Typed $h$-level tailored normal form verifier]\label{defn:typed_h-level_NFV}
	A \emph{{typed} $h$-level TNFV} is a quadruple of Turing machines $\verifier = (\sampler,\length, \linproc, \decider)$,
	where $\sampler$ is a {typed} $h$-level CL sampler as in Definition \ref{defn:typed_h-level_CL_Sampler}, $\length$ is an answer length calculator as in Definition \ref{defn:answer_length_calc}, $\linproc$ is a linear constraint processor as in Definition \ref{def:linear_constraints_processor}, and $\decider$ is the canonical decider as in Definition \ref{def:canonical-decider}.
	
	Such a  typed normal form verifier is said \emph{$\lambda$-bounded}, for a positive integer $\lambda$, if
	\begin{itemize}
		\item The running times (Definition \ref{def:running_time}) of $\sampler,\length$ and $\linproc$ are all bounded by $n^\lambda$ , namely
		\[
		\forall \overline{n}\in \{0,1\}^*\ \colon \  \ \TIME(\sampler;\overline{n},\cdot,\cdot,\cdot,\cdot,\cdot)\ ,\ \TIME(\length;\overline{n},\cdot,\cdot)\ ,\ \TIME(\linproc;\overline{n},\cdot,\cdot,\cdot,\cdot)\ \leq\ n^\lambda\ .
		\]
		\item The description length of $\verifier$ is bounded by $\lambda$, namely $|\verifier|\leq \lambda$ (Definition \ref{def:Description_length}).
	\end{itemize}
	Similar to Definition \ref{def:normal-game}, Remark \ref{rem:n^th_game_less_restrictive_setup} and Definition \ref{defn:h-level_NFV}, when $\verifier$ is a $\lambda$-bounded tailored typed $h$-level normal form verifier, then there is an associated $n^{\rm th}$ game to it for every $n\geq 2$.  Furthermore, the $n^{\rm th}$ game of such a normal form verifier is \emph{well defined}, even if it is not $\lambda$-bounded, if the normal form verifier satisfies the following conditions:
	\begin{itemize}
		\item $\length(\on,(t,\mttx),\kappa)$  halts whenever $t\in \type$, where $(\type,\cal{E})$ is the appropriate decoding of $\sampler(\cdot,{\rm Graph},\cdot,\cdot,\cdot,\cdot)$,  $\mttx$ is of length $r(n)=\sampler(\on,{\rm Dimension},\cdot,\cdot,\cdot,\cdot)$, and  $\kappa\in \{\frR,\frL\}$.
		\item $\linproc(\on,(t,\mttx),(t',\mtty),a^\frR,b^\frR)$   halts whenever   $t,t'\in \type$, $\mttx,\mtty$ are of length $r(n)$, and  $a^\frR$ and $b^\frR$ are of length $|\dec(\length(\on,\mttx,\frR))|$ and $|\dec(\length(\on,\mtty,\frR))|$ respectively. 
	\end{itemize}
	Note that by assuming $\sampler$ is a typed $h$-level CL sampler (Definition \ref{defn:typed_h-level_CL_Sampler}), we are guaranteed that it behaves well, in particular it always halts (on relevant inputs), and there is a type graph and associated CLMs underlying it.  So, no additional assumptions on $\sampler$ are needed.
\end{definition}

\begin{claim}[Algorithmic detyping of normal form verifiers]\label{claim:DeTyping_NFV} 
	There exists a  Turing machine $\mathsf{DeType}_h$ that takes as input a  \textbf{typed} $h$-level TNFV $\verifier=(\sampler,\length,\linproc,\decider)$ and outputs a (\textbf{non-typed})  $(h+2)$-level TNFV $$\mathsf{DeType}_h(\verifier)=\verifier'=(\sampler',\length',\linproc',\decider)$$ such that:
	\begin{itemize}
		\item (Combinatorial DeTyping) For every $n\in \mathbb{N}$,  if $\verifier_n$ is well defined (Definition \ref{defn:typed_h-level_NFV}), then $\verifier'_n$ is well defined (Definition \ref{defn:h-level_NFV}) and satisfies $\verifier'_n=\frak{DeType}(\verifier_n)$ with respect to the underlying type graph $(\type,\cal{E})$ decoded from $\sampler(\cdot,{\rm Graph},\cdot,\cdot,\cdot,\cdot)$.
		
		\item (Sampler properties) The $(h+2)$-level sampler  $\sampler'(\on,\cdot,\cdot,\cdot,\cdot,\cdot)$ runs in time which is polynomial (with constants that may depend on $h$) in:
		\begin{itemize}[noitemsep,topsep=0pt,parsep=0pt,partopsep=0pt] 
			\item the number of types $|\type|$, where $(\type,\cal{E})$ is the type graph decoded from $\sampler(\cdot,{\rm Graph},\cdot,\cdot,\cdot,\cdot)$; 
			\item the running time $\TIME(\sampler;\on,\cdot,\cdot,\cdot,\cdot,\cdot).$
		\end{itemize}
		Moreover, $\mathsf{DeType}_h$ calculates the description  of $\sampler'$ in  polynomial time from   the description  of $\sampler$, and in particular $|\sampler'|=\poly_h(|\sampler|)$.
		\item (Answer length properties) The output answer length calculator $\length'(\on,\mttx,\kappa)$ runs in time which is polynomial  (with constants that may depend on $h$) in:
		\begin{itemize}[noitemsep,topsep=0pt,parsep=0pt,partopsep=0pt]
			\item the number of types $|\type|$; 
			\item
			the running time $\TIME(\sampler;\on,\cdot,\cdot,\cdot,\cdot,\cdot);$
			\item
			the running time $\TIME(\length; \on,\cdot,\cdot)$.
		\end{itemize}
		Moreover, $\mathsf{DeType}_h$ calculates the description  of $\length'$ in  polynomial time  in that of $\sampler$ and $\length$; in particular, $|\length'|=\poly_h(|\sampler|,|\length|)$.
		\item (Linear constraints processor properties) The ouput linear constraints processor TM $\linproc'(\on,\cdot,\cdot,\cdot,\cdot)$ runs in time polynomial (with constants that may depend on $h$)  in:
		\begin{itemize}[noitemsep,topsep=0pt,parsep=0pt,partopsep=0pt]
			\item       the number of types $|\type|$; 
			\item        the running time $\TIME(\sampler;\on,\cdot,\cdot,\cdot,\cdot,\cdot);$
			\item    the running time $\TIME(\length; \on,\cdot,\cdot)$;
			\item  the running time $\TIME(\linproc; \on,\cdot,\cdot,\cdot,\cdot)$.
		\end{itemize}
		Moreover, $\mathsf{DeType}_h$ calculates the description  of $\linproc'$ is polynomial time from  the descriptions of $\sampler,$ $\length$ and $\linproc$; in particular $|\linproc'|=\poly_h(|\sampler|,|\length|,|\linproc|)$.
	\end{itemize}
\end{claim}

\begin{proof}
	Throughout this proof, we use the notation $\enc(t)=({\bf 1}_t,\sum_{t'\sim t}{\bf 1}_{t'})\in \FF_2^{\type}\times\FF_2^{\type}$ from Definition \ref{defn:combi_detyping}.
	Let us describe the sampler  $\sampler'$, answer length calculator $\length'$ and linear constraints processor  $\linproc'$ of  $\verifier'=\mathsf{DeType}_h(\sampler,\length,\linproc,\decider)$.
	\\
	
	\textbf{The Sampler:} 
	We start with a detailed description of the operations of the detyped sampler $\sampler'$. The operation of the sampler follows Definition~\ref{defn:combi_detyping}. The details of the implementation are straightforward, but we include them for completeness. 
	\begin{enumerate}    
		\item $\sampler'(\on,{\rm Dimension},\cdot,\cdot,\cdot,\cdot)$ runs as follows: First, it calls $\sampler(\cdot,{\rm Graph},\cdot,\cdot,\cdot,\cdot)$ to extract the set of types $\type$, and thus the size of this set. Then, it calls $\sampler(\on,{\rm dimension},\cdot,\cdot,\cdot,\cdot)$ to extract $r(n)$. Finally, it outputs $r'(n)=4|\type|+r(n)$.
		\item $\sampler'(\on,{\rm Register},{\rm Player},j,\mttx,\cdot)$ runs as follows:
		First, it runs as $\sampler'(\on,{\rm Dimension},\cdot,\cdot,\cdot,\cdot)$ --- which was defined in the previous clause --- to extract the type set $\type$, the edge set $\cal{E}$, and the dimension $r'(n)=4|\type|+r(n)$. Then, it checks that $\mttx$ is a bit string of length $4|\type|+r(n)$ and that $1\leq j\leq h+2$, and outputs an $\frak{error}$ sign otherwise. 
		
		Now, if ${\rm Player}=A$ and $j=1$, then it outputs a bit string of length $4|\type|+r(n)$ whose first $2|\type|$ entries are $1$ and the rest are $0$, namely $$(\sum_{t\in \type} {\bf 1}_t,\sum_{t\in \type} {\bf 1}_t,\vec 0,\vec 0,\vec 0)\in (\FF_2^{\type})^4\times \FF_2^{r(n)}\ .$$ Recall that such a bit string is interpreted as the registers which define $W^\mttx_1$ with respect to the CLM $\frS^A$, and in this case it means that regardless of $\mttx$, this space consists of the first and second copies of $\FF_2^\cal{T}$. 
		
		If ${\rm Player}=A$ and $j=2$,  then it outputs a bit string of length $4|\type|+r(n)$ whose first $2|\type|$ entries are $0$, the following $2|\type|$ bits are $1$, and the rest are $0$, namely $$(\vec 0,\vec 0,\sum_{t\in \type} {\bf 1}_t,\sum_{t\in \type} {\bf 1}_t,\vec 0)\in (\FF_2^{\type})^4\times \FF_2^{r(n)}\ .$$ 
		This means that regardless of $\mttx$, the register subspace $W^\mttx_2$ with respect to the CLM $\frS^A$ is always spanned by the third and fourth copies of $\FF_2^\type$.
		
		If ${\rm Player}=A$ and $j\geq 3$,  then it reads the first $4|\type|$ bits of $\mttx$. If there is no type $t\in \type$  such that these bits are equal to $(\enc(t),\vec 0,{\bf 1}_t)$, then it splits into two cases --- if $j=3$, then it outputs a bit string of length $4|\type|+r(n)$ whose first $4|\type|$ entries are $0$, and the rest are $1$, namely $$(\vec 0,\vec 0,\vec 0,\vec 0,\vec 1)\in (\FF_2^{\type})^4\times \FF_2^{r(n)}\ .$$  This is interpreted as $W^3_\mttx$ being a copy of $\FF_2^{r(n)}$. And, if $j>3$, then it outputs a bit string of length $4|\type|+r(n)$ consisting of only zeros, namely $\vec 0\in (\FF_2^{\type})^4\times \FF_2^{r(n)}$. This is interpreted as $W^j_\mttx=\{\vec 0\}$ being the trivial space for any $3<j\leq h+2$. 
		
		Otherwise, there is a type $t\in \type$ such that $\mttx|_{\FF_2^{4\type}}=(\enc(t),\vec 0,{\bf 1}_t)$. In this case, $\sampler'$ calls $$\sampler(n,{\rm Register},t,j-2,\mttx|_{\FF_2^{r(n)}},\cdot)\ ,$$ whose output is denoted by $\vec i\in \FF_2^{r(n)}$; then, it outputs  a bit string of length $4|\type|+r(n)$ whose first $4|\type|$ entries are $0$, and the rest are $\vec i$, namely 
		\[
		(\vec 0,\vec 0,\vec 0,\vec 0,\vec i)\in (\FF_2^{\type})^4\times \FF_2^{r(n)}\ .
		\]
		This is interpreted as $W^j_{\mttx}$ with respect to $\frS^A$ being $\{(\vec 0,\vec 0,\vec 0,\vec 0)\}\times W^{j-2}_{\mttx|_{\FF_2^{r(n)}}}$, where $W^{j-2}_{\mttx|_{\FF_2^{r(n)}}}$ is the $(j-2)^{\rm th}$-register subspace given the seed $\mttx|_{\FF_2^{r(n)}}$  with respect to the CLM $\frS^t$.
		\\
		
		The case where ${\rm Player}=B$ is similar, and we omit its description.
		\item $\sampler'(\on,{\rm Marginal},{\rm Player},j,\cdot,z)$ runs as follows: First, it runs $\sampler'(\on,{\rm Dimension},\cdot,\cdot,\cdot,\cdot)$ --- which was defined in the first clause --- to extract the type set $\type$, the edge set $\cal{E}$, and the dimension $r'(n)=4|\type|+r(n)$.  Then, it checks that $z$ is a bit string of length $4|\type|+r(n)$ and that $1\leq j\leq h+2$, and returns an $\frak{error}$ sign otherwise. 
		\\
		
		Now, if ${\rm Player}=A$ and $j=1$, then $\sampler'$ outputs the first $2|\type|$ bits of $z$.  If ${\rm Player}=A$ and $j=2$, then $\sampler'$ reads the first $2|\type|$ bits of $z$. If $(z_1,...,z_{2|\type|})=\enc(t)$ for some $t\in \type$, then it zeroes out all coordinates in the third copy of $\FF_2^\type$, and all coordinates {but the $t^{\rm th}$ one} in the fourth copy of $\FF_2^\type$; otherwise, it zeroes out the third and fourth copy of $\FF_2^\type$ completely. If ${\rm Player}=A$ and $j\geq 3$, then it does the first two steps as described above, resulting in a vector $\mttx$ in $(\FF_2^{\type})^4$; if there is no $t\in \type$ such that $\mttx=(\enc(t),\vec 0,{\bf 1}_t)$, then it zeros out all coordinates of $\FF_2^{r(n)}$, resulting in $(\mttx,\vec 0)$. Otherwise, there is some $t\in \type$ such that $\mttx=(\enc(t),\vec 0,{\bf 1}_t)$, in which case it calls $\sampler(n,{\rm Marginal},t,j-2,\cdot,z|_{\FF_2^{r(n)}})$, whose output we denote by $\mttx'$, and it outputs $(\mttx,\mttx')\in (\FF_2^\type)^4\times \FF_2^{r(n)}$.
		\\
		
		Again, the case ${\rm Player}=B$ is similar, and we omit it.
		\item $\sampler'(\on,{\rm Evaluate},{\rm Player},j,\mttx,z)$ runs as follows: If $j=1$, then it runs $\sampler'(\on,{\rm Marginal},{\rm Player},j,\cdot,z)$ as defined in the previous clause. If ${\rm Player}=A$ and $j=2$, then it checks whether the restriction of $\mttx$ to the first and second copies of $\FF_2^\type$ agrees with $\enc(t)$ for some  $t\in \type$; if it does, then it outputs the vector  $(\vec 0,\mttx')\in \FF_2^\type\times \FF_2^\type$, where $\mttx'$ is the vector whose all coordinates are zero except for the $t^{\rm th}$ coordinate, which is the $t^{\rm th}$ coordinate in the fourth copy of $\FF_2^\type$ in $z$; otherwise, it outputs $\vec 0 \in \FF_2^\type\times \FF_2^\type$. If ${\rm Player}=A$ and $j\geq 3$, it first checks whether the restriction of $\mttx|_{\FF_2^{4|\type|}}=(\enc(t),\vec0, {\bf 1}_t)$ for some $t\in \type$; if so, it outputs the same output as $\sampler(\on,{\rm Evaluate},t,j-2,\mttx|_{\FF_2^{r(n)}},z|_{\FF_2^{r(n)}})$; otherwise, if $j=3$, then it outputs $\vec 0\in \FF_2^{r(n)}$, and if $j>3$ it outputs the empty string. For ${\rm Player}=B$, it acts similarly with the first and second copies of $\FF_2^\type$ swapping roles with the third and fourth copies.

		\item There is a \emph{canonical} way of extracting the perpendicular action out of the others. See Clause 6 on page 94 of \cite{MIPRE}, which explains how step 3c in Figure 10 is implemented. 
	\end{enumerate}
	
	Before describing the rest of the TMs, we note that indeed the sampler satisfies the conditions of the proof: 
	
	For the first condition, which says that this transformed sampler is indeed the detyping transformation on the combinatorial level, we leave for the reader to compare this algorithm to the description in Definition \ref{defn:combi_detyping}. 
	
	For efficient runtime, note that all the operations are either calling the original sampler on the same index $n$ namely $\sampler(\on,\cdot,\ldots,\cdot)$, or a previous subroutine which  was already defined (again on the same index $n$), or is some polynomial time operation on $\FF_2^{r(n)+4|\type|}$ which translates to $\poly(|\type|,r(n))$ number of operations. As both $|\type|$ and $r(n)$ are bounded by $\TIME(\sampler;\on,\cdot,\cdot,\cdot,\cdot,\cdot)$ (Remark \ref{rem:dim_of_CL_sampler_bounded_by_run_time}), the runtime bound is deduced.
	
	To deduce that $\sampler'$ can be described in length that is polynomial in that of $\sampler$, note that the (constant length) natural language description we provided above can be translated to a constant length code in some programming language (or more precisely, a formal description of a TM according to the encoding fixed in Section \ref{sec:encodings}). 
	Therefore, the description length of $\sampler'$ is  some constant, up to the appending of the description of $\sampler$ (for it to run the appropriate subroutines). 
	As the exact effect of  appending $\sampler$ to $\sampler'$ on the level of the description of $\sampler'$ depends on the specific choice of encodings of TMs, we  use  \cref{clause3:conditions_of_TM_encodings} of Fact \ref{fact:properties_of_TMs}, and deduce that  the description of $\sampler'$ can be calculated from $\sampler$ in  polynomial time and thus $|\sampler'|=\poly_h(|\sampler|)$.\footnote{Actually, the description is fixed up to appending $\sampler$ and $h$, which means the dependence is $\poly(\log h,|\sampler|)$. This is better than $\poly_h(|\sampler|)$, but we do not need this better bound.}
	\\
	
	\textbf{The Answer length calculator:}
	Recall that the input to $\length'$ is expected to be $(\on,\mttx,\kappa)$, where $\mttx\in \FF_2^{4|\type|+r(n)}$ and $\kappa\in \{\frR,\frL\}$. On the other hand, $\length$ is expecting an input of the form $(\on,(t,\mtty),\kappa)$, where $t\in \type$ and $\mtty\in \FF_2^{r(n)}$.
	
	So, $\length'(\on,\mttx,\kappa)$ runs as follows: It calls $\sampler$ to extract $\type$ and $r(n)$, and then if the conditions on the inputs are not satisfied, it  outputs an $\frak{error}$ sign. Otherwise, it checks whether there is a $t\in\type$ such that $\mttx|_{\FF_2^{4|\type|}}=(\enc(t),\vec 0,{\bf 1}_t)$ or  $\mttx|_{\FF_2^{4|\type|}}=(\vec 0,{\bf 1}_t,\enc(t))$. If not, it outputs the empty string (which is the unary representation of $0$). If there is such a $t$, it provides the output of $\length(\on,(t,\mttx|_{\FF_2^{r(n)}}),\kappa)$.
	
	Again, it is straightforward to check that this induces a length function which is compatible with the description of $\frak{DeType}(\game)$ in Definition \ref{defn:combi_detyping}. For description length, again the above description is constant up to fixing the appropriate inputs to $\mathsf{DeType}_h$. Lastly, $\length'$ calls $\sampler$ and $\length$, and does some polynomial time manipulations on vectors in $\FF_2^{4|\type|+r(n)}$. Hence, it runs in time polynomial in the above.
	\\
	
	\textbf{The Linear constraints processor:}
	$\linproc'(\on,\mttx,\mtty,a^\frR,b^\frR)$ runs as follows. First, it calls $\sampler$ to extract $\type$ and $r(n)$.  Then, it checks that $\mttx,\mtty\in \FF_2^{4|\type|+r(n)}$, and if not outputs the single constraint $\{\sJ\}$. Otherwise, it calls $\length'(\on,\mttx,\frR)$ and $\length'(\on,\mtty,\frR)$ and compares the lengths of $a^\frR$ and $b^\frR$ to their decoded outputs respectively. If they do not match,  it outputs the single constraint $\{\sJ\}$. Finally, if the input passed all the above checks, then $\linproc'$ checks the following: If there are no $t,t'\in  \type$ such that $\mttx=(\enc(t),\vec 0,{\bf 1}_t)$ and $\mtty=(\vec 0,{\bf 1}_{t'},\enc(t'))$, then it outputs the empty string (interpreted as no constraints, which is immediate acceptance).  Otherwise, it returns the output of $\linproc(\on,(t,\mttx|_{\FF_2^{r(n)}}),(t',\mtty|_{\FF_2^{r(n)}}),a^\frR,b^\frR)$.
	
	As for the sampler and answer length calculator, checking that this $\linproc'$ satisfies what we need is straightforward, and we leave it to the reader. This finishes the proof.
\end{proof}

\subsubsection{Padding}\label{sec:Padding}

The idea of padding is enlarging the length of answers artificially in a way that \emph{essentially} does not change the game. There are many ways of doing that, and  the following is a simple version which is \emph{restriction free} --- namely, we enlarge the answer length and do not require the appended bits to satisfy any requirements. 
\begin{definition}[Restriction free combinatorial padding]
	Let $\game$ be a tailored game, and $\Lambda$ a positive integer. The tailored game $$\widetilde\game=\frak{Padding}(\game,\Lambda)$$ has the same underlying graph as $\game$, and the same edge sampling distribution $\mu$. For every 
	vertex $\mttx\in V$, both the readable length and unreadable length of $\mttx$ are defined to be $\Lambda$, namely $\widetilde{\ell^\frR}(\mttx)= \widetilde{\ell^\frL}(\mttx)=\Lambda$. For the controlled linear constraints function $\widetilde {L_{\mttx\mtty}}$:
	\begin{itemize}
		\item If the readable or unreadable length of either $\mttx$ or $\mtty$ are larger than $\Lambda$, namely $$\Lambda<\max\{\ell^\frR(\mttx),\ell^\frR(\mtty),\ell^\frL(\mttx),\ell^\frL(\mtty)\}\ ,$$ then $\widetilde {L_{\mttx\mtty}}$ outputs no constraints regardless of what $\gamma^\frR$ is (which translates to automatic acceptance).
		\item Otherwise,  for every $\mttz\in V$ we let $\widetilde {S_\mttz^\frR}=S_\mttz^\frR\sqcup T^\frR_\mttz$, where $S^\frR_\mttz$ is the original formal set of generators at $\mttz$ and $T^\frR_\mttz$ is a set of $\Lambda-\ell^\frR(\mttz)$ many additional variables (and similarly for $\widetilde{S^\frL_\mttz}$). Namely,
		\[
		\widetilde {S^\cdot_\cdot}={S^\cdot_\cdot}\sqcup {T^\cdot_\cdot}
		\]
		where the subscripts can be $\mttx,\mtty$ or $\mttx\mtty$ (which indicates union of $\mttx$ and $\mtty$ variables) and the superscripts can be $\frR,\frL$ or none (which indicates the union of readable and unreadable variables).
		Recall that for $\gamma^\frR\colon S^\frR_{\mttx\mtty}\to \FF_2$, the output of $L_{\mttx\mtty}(\gamma^\frR)$ is a collection of indicators on the set $S_{\mttx\mtty}\sqcup\{\sJ\}$, representing linear constraints that should be checked on $\gamma\colon S_{\mttx\mtty}\to \FF_2$. 
		So, for  $\widetilde{\gamma^\frR}\colon \widetilde{S^\frR_{\mttx\mtty}}\to \FF_2$, letting $\gamma^\frR=\widetilde{\gamma^\frR}|_{S_{\mttx\mtty}^\frR}$, we can define $\widetilde{L_{\mttx\mtty}}(\widetilde{\gamma^\frR})$ to be the extension by zeros of the outputs of $L_{\mttx\mtty}(\gamma^\frR)$ to the  $T_{\mttx\mtty}$ variables. Namely, for every $c\colon S_{\mttx\mtty}\sqcup \{\sJ\}\to\FF_2$ in the output of $L_{\mttx\mtty}(\gamma^\frR)$, we  let $\tilde c\colon \widetilde{S_{\mttx\mtty}}\sqcup \{\sJ\}\to \FF_2$ be defined by 
		\[
		\tilde c(\sX)=\begin{cases}
		c(\sX) & \sX\in S_{\mttx\mtty}\sqcup \{\sJ\}\ ,\\
		0 & \sX\in T_{\mttx\mtty}\ .
		\end{cases}
		\]
	\end{itemize}
	
	On the combinatorial level, the padded game $\widetilde \game$ samples an edge from the original graph (according to the same distribution), and as long as $\Lambda$ is large enough, it disregards the added variables $T$ and plays the original game $\game$  only according to the assignments to $S$.
\end{definition}
The following is a straightforward fact to check.
\begin{fact}\label{fact:padding_properties}
	Assume $\Lambda\geq \max(\ell^\frR,\ell^\frL)$. Then,
	\begin{itemize}
		\item (Completeness) If $\game$ has a perfect $\ZPC$ strategy, then so does $\frak{Padding}(\game,\Lambda)$.
		\item (Soundness and entanglement) If $\frak{Padding}(\game,\Lambda)$ has a value $1-\eps$ strategy, then so does $\game$, and furthermore
		\[
		\Ent(\frak{Padding}(\game,\Lambda),1-\eps)=\Ent(\game,1-\eps)\ .
		\]
	\end{itemize}
\end{fact}

\begin{claim}\label{claim:Padding_NFV}
	There is a polynomial-time TM $\mathsf{Padding}$  that takes as input a tailored $h$-level normal form verifier $\verifier=(\sampler,\length,\linproc,\decider)$ and a $1$-input TM $\Lambda$, and outputs a new tailored $h$-level normal form verifier $\mathsf{Padding}(\verifier,\Lambda)=\verifier'=(\sampler,\length^\Lambda,\linproc',\decider)$ satisfying:
	\begin{itemize}
		\item (Combinatorial Padding) For every $n\in \mathbb{N}$, if $\verifier_n$ is well defined (Definition \ref{defn:h-level_NFV}), then $\verifier'_n$ is well defined, and $\verifier'_n=\frak{Padding}(\verifier_n,|\Lambda(n)|)$, where $|\cdot|$ is the length of words function.\footnote{This is the same as thinking of the output of $\Lambda(n)$ as representing a natural number in {unary}.}
		\item (Sampler properties) The output sampler is the same as the original one, and thus its running time and description lengths stay the same.
		\item (Answer length properties) The output answer length TM $\length^\Lambda$ depends only on $\Lambda$. Furthermore, $\length^\Lambda(n,\cdot,\cdot)$ runs in time which is linear in $\TIME(\Lambda;n)$. Finally, the description length of $\length^\Lambda$ is linear in that of $\Lambda$.

		\item (Linear constraints processor properties) The output linear constraints processor $\linproc'$ runs in time which is polynomial in:
		\begin{itemize}[noitemsep,topsep=0pt,parsep=0pt,partopsep=0pt]
			\item the running time $\TIME(\Lambda;\on)$;
			\item  the running time $\TIME(\sampler;\on,\cdot,\cdot,\cdot,\cdot,\cdot);$
			\item  the running time $\TIME(\length; \on,\cdot,\cdot)$;
			\item   the running time $\TIME(\linproc; \on,\cdot,\cdot,\cdot,\cdot)$.
		\end{itemize}    
		Moreover, the description length of $\linproc'$ is linear in that of $\Lambda,\sampler,\length$ and $\linproc$.
	\end{itemize}
\end{claim}
\begin{proof}\ 
	
	\textbf{The Sampler:} We keep $\sampler$ as the sampler. So, running time and description length stay the same.
	\\
	
	\textbf{The answer length calculator:} For every $n\in \mathbb{N}$, $\length^\Lambda(\on,\mttx,\kappa)=\enc(\Lambda(n))$ (Definition \ref{defn:the_alphabet}), regardless of $\mttx$ or $\kappa$. 
	Hence, $|\dec(\length^\Lambda(\on,\mttx,\kappa))|=|\Lambda(n)|$, as is needed for $\verifier'_n$ to be equal to $\frak{Padding}(\verifier_n,|\Lambda(n)|)$.  It is immediate that the running time and description length are linear in $\Lambda$'s. 
	\\
	
	\textbf{The linear constraints processor:}  $\linproc'(\on,\mttx,\mtty,a^\frR,b^\frR)$ runs as follows.
	First it calls $\sampler(\on,{\rm Dimension},\cdot,\cdot,\cdot,\cdot)$ to retrieve $r(n)$ and checks that $\mttx,\mtty\in \FF_2^{r(n)}$; if not, it outputs  $\{\sJ\}$\footnote{By outputting $\{\sJ\}$, we mean the encoding (as in Definition \ref{def:linear_constraints_processor}) of the bit string of length $4|\Lambda(n)|+1$ where all of its bits are zero except the last one which is $1$.} (which is instant rejection); if they do satisfy this condition, then it checks whether $|a^\frR|=|b^\frR|=|\Lambda(n)|$; if not, it outputs $\{\sJ\}$; otherwise, the inputs are well structured and $\linproc'$ can proceed.
	
	Now, $\linproc'$ calls $$\ell^\frR_a=|\dec(\length(\on,\mttx,\frR))|\ ,\ \ell^\frR_b=|\dec(\length(\on,\mtty,\frR))|\ ,\ \ell^\frL_a=|\dec(\length(\on,\mttx,\frL))| \quad {\rm and} \quad  \ell^\frL_b=|\dec(\length(\on,\mtty,\frL))|\ .$$ 
	If $|\Lambda(n)|$ is strictly smaller than either of the lengths of these outputs, then $\linproc'$ outputs the empty string (i.e., no constraints). Otherwise, let $a_0^\frR$ be the restriction of $a^\frR$ to its first $|\ell^\frR_a|$ bits and $b_0^\frR$ be  the restriction of $b^\frR$ to its first $|\ell^\frR_b|$ bits. 
	Then, $\linproc'$ calls $\linproc(\on,\mttx,\mtty,a^\frR_0,b^\frR_0)$ and gets as output a bit string. If this bit string is not (the encoding) of bit strings $(c^1,...,c^k)$, where every $c^i$ is of length $|\ell^\frR_a|+|\ell^\frR_b|+|\ell^\frL_a|+|\ell^\frL_b|+1$, then it outputs $\{\sJ\}$. Otherwise, it does the following operation on each $c^i$: First, it splits it to $5$ bit strings $c^i_{a,\frR},c^i_{b,\frR},c^i_{a,\frL},c^i_{b,\frL},c^i_\sJ$ of lengths $\ell^\frR_a,\ell^\frR_b,\ell^\frL_a,\ell^\frL_b$ and $1$ respectively. Then, it appends each of $c^i_{a,\frR},c^i_{b,\frR},c^i_{a,\frL},c^i_{b,\frL}$ with zeros until they are of length $|\Lambda (n)|$ --- we denote the resulting strings by $\tilde c^i_{\cdot,\cdot}$. Finally, the bit string $\tilde c^i$  of length $4|\Lambda(n)|+1$ is defined to be the concatenation of $\tilde c^i_{a,\frR},\tilde c^i_{b,\frR},\tilde c^i_{a,\frL},\tilde c^i_{b,\frL}$ and $c^i_\sJ$.
	After this operation was done for each string $c^i$, resulting with new strings $\tilde c^i$, $\linproc'$ outputs (the encoding of) $(\tilde c^1,...,\tilde c^k)$.
	
	The description  is again just the above finite  one, with the inputs $\verifier$ and $\Lambda$ fixed. Hence, by \cref{clause3:conditions_of_TM_encodings} in Fact \ref{fact:properties_of_TMs}, the description can be calculated in polynomial time from them and is thus of polynomial length. For running time, $\linproc'$ either calls $\length,\sampler$ or $\linproc$, or is applying polynomial time operations on bit strings of length at most $O(|\Lambda(n)|)$ --- where $|\Lambda(n)|$ is a quantity smaller than the running time of $\Lambda(n)$ --- or bit strings of length at most $r(n)$.  Recall that $r(n)$ is bounded by $\TIME(\sampler;\on,\cdot,\cdot,\cdot,\cdot,\cdot)$ by Remark \ref{rem:dim_of_CL_sampler_bounded_by_run_time}, which explains the time bounds of $\linproc'$.
\end{proof}

\subsection{Proving the main theorem of Question Reduction: Theorem \ref{thm:h_level_question_reduciton}}\label{sec:prf-442}
Let $\verifier=(\sampler,\length,\linproc,\decider)$ be a tailored $h$-level normal form verifier, and $\lambda$ a positive integer. The goal is to describe the question reduced verifier $\mathsf{QuestionReduction}_h(\verifier,\lambda)=\verifier'=(\sampler^\lambda_{\qr},\length^\lambda_{\qr},\linproc',\decider)$ which proves the theorem. The idea is, under the assumption that $\verifier$ is $\lambda$-bounded, to first choose for every $n$: an appropriate integer $k$ which will be larger than the dimension of the CLMs used in $\verifier_{2^n}$, namely larger than $\sampler(2^n,{\rm Dimension},\cdot,\cdot,\cdot,\cdot)$; an appropriate $\Lambda$ which will be larger than the lengths used in $\verifier_{2^n}$, namely larger than every possible $\length(2^n,\cdot,\cdot)$;  an appropriate $\mathscr{B}\subseteq \FF_2^k$ that would have size a power of $2$,  and induce a good error correcting code with some predetermined parameters. After these choices are made, the goal of $\verifier'$ is for its $n^{\rm th}$ game to be, combinatorially, 
$$\verifier'_n=\frak{DeType}(\frak{QueRed}(\frak{Padding}(\verifier_{2^n},\Lambda),k,\mathscr{B})\;.$$

We already described how to detype and to pad on the level of verifiers in Claims \ref{claim:DeTyping_NFV} and \ref{claim:Padding_NFV}. So, we are left  to describe a Turing machine that assumes the input is already padded, and outputs a \textbf{typed} normal form verifier that implements combinatorial question reduction. In the next claim, we let $\mathscr{B}$ be a (fixed)  TM that takes as input a bit string $\mttx$ of length $m$ and outputs a list of $2\cdot 2^{m}$ vectors in $\FF_2^{2^m}$ that induces an encoding matrix whose associated code has normalized distance $\delta$ for some universal constant $\delta>0$, and such that $\TIME(\mathscr{B};\mttx)=2^{O(|\mttx|)}$ (the existence of such a TM $\mathscr{B}$ and such a universal constant $\delta$ is guaranteed by Fact \ref{fact:good_efficiently_calculable_codes}).

\begin{claim}\label{claim:typed_question_reduction_on_NFV}
	There is a polynomial time TM $\mathsf{TypedQuestionReduction}_h$ that takes two inputs --- an $h$-level normal form verifier $\verifier=(\sampler,\length,\linproc,\decider)$; a $1$-input TM $\cal{K}$ --- and outputs a {typed} $1$-level normal form verifier $$\mathsf{TypedQuestionReduction}_h(\verifier,\cal{K})=\widetilde \verifier=( \sampler^\cal{K}, \widetilde\length,\widetilde \linproc,\decider)$$  such that 
	\begin{itemize}
		\item (Combinatorial Question Reduction) For every $n\in \mathbb{N}$, if $\verifier_{2^n}$ is well defined (Definition \ref{defn:h-level_NFV}),  and there is a function $\Delta\colon \mathbb{N}\to \mathbb{N}$ such that $|{\dec}(\length(2^n,\mttx,\kappa))|=\Delta(n)$ regardless of $\mttx$ and $\kappa$, and $2^{|\cal{K}(n)|}$ is larger than 
		\[\sampler(2^n,{\rm Dimension},\cdot,\cdot,\cdot,\cdot)\;,\] 
		then  $\widetilde\verifier_n$ is well defined (Definition \ref{defn:typed_h-level_NFV}), and 
		$$\widetilde \verifier_n=\frak{QueRed}(\verifier_{2^n},2^{|\cal{K}(n)|},\mathscr{B}(\cal{K}(n)))\ .$$ 
		\item (Sampler properties) The sampler $ \sampler^{\cal{K}}$ depends only on $\cal{K}$ (and $h$), but {not} on $\verifier$. Furthermore, it runs in time which is polynomial in that of $\cal{K}$.\footnote{Here, the constants depend on $h$,  namely this is $\poly_h(\TIME(\cal{K};n))$.} Finally,  its description  can be calculated from that of $\cal{K}$ in polynomial time, which means in particular $|\sampler^\cal{K}|=\poly_h(|\cal{K}|)$.  
		\item (Answer length calculator properties) The TM $ \widetilde\length$ depends only on $\cal{K}$ and on $\length$, and not on  $\sampler$ or $\linproc$. In addition, it runs in time  $$\TIME(\widetilde\length;n,\cdot,\cdot)=\poly_h(2^{|\cal{K}(n)|},\TIME(\cal{K};n),\TIME(\length;2^n,\cdot,\cdot))\ .$$
		Finally, its description can be calculated in polynomial time from the relevant inputs,  and in particular $|\widetilde{\length}|\leq \poly_h(|\cal{K}|,|\length|)$.
		\item (Linear constraints processor properties) $\widetilde\linproc$ runs in time which is polynomial in:
		\begin{itemize}[noitemsep,topsep=0pt,parsep=0pt,partopsep=0pt]
			\item  the integer $2^{|\cal{K}(n)|}$;
			\item the running time $\TIME(\cal{K};n)$;
			\item  the running time $\TIME(\sampler;2^n,\cdot,\cdot,\cdot,\cdot,\cdot)$;
			\item the running time $\TIME(\length;2^n,\cdot,\cdot)$;
			\item  the running time $\TIME(\linproc;2^n,\cdot,\cdot,\cdot,\cdot,\cdot)$;
		\end{itemize}
		Furthermore, its description can be calculated from  $\cal{K},\Lambda,\sampler$ and $\linproc$ in polynomial time, and in particular $|\widetilde{\linproc}|\leq \poly_h(|\cal{K}|,|\Lambda,|\sampler|,|\linproc|)$. 
	\end{itemize}
\end{claim}

\begin{proof}
	\ 
	\\
	
	\textbf{The typed CL sampler:}  Recall Example \ref{example:quered_has_typed_CL_sampling_scheme}, and  specifically that we aim to define a  typed $1$-level sampler. This means that, in particular, in the expected input $(n,{\rm Action},{\rm Type},j,\mttx,z)$, we can assume that $j=1$ always.
	\begin{enumerate}
		\item $ \sampler^{\cal{K}}(\cdot,{\rm Graph},\cdot,\cdot,\cdot,\cdot)$  outputs the list 
		\[
		\begin{split}
		\forall 1\leq j\leq h\ \colon \ \ &\Hide A^j\ ,\ \Hide B^j\ ,\\
		&\Intro_A\ ,\ \Intro_B\ ,\ \Read_A\ ,\ \Read_B\ ,\ \Sample_A\ ,\ \Sample_B\ ,\\
		&\Pauli_\PZm\ ,\ \Pauli_\PXm\ ,\ \mttX\ ,\ \mttZ\ ,\ \mathtt{First}\ ,\ \mathtt{Second}\ ,\ \mathtt{Both}\ ,\\
		\forall 1\leq i\leq 3\ ,\  1\leq j\leq 3\ \colon \ \ &\mathtt{var}_{ij}\ ,\ \mathtt{row}_i\ ,\ \mathtt{col}_j\;,
		\end{split}
		\]
		followed by the adjacency matrix of the graph depicted in Figure \ref{fig:typed_graph_quered}.
		\item $ \sampler^{\cal{K}}(n,{\rm Dimension},\cdot,\cdot,\cdot,\cdot)$ calls $\cal{K}(n)$ and outputs $2|\cal{K}(n)|+2=2\log|\mathscr{B}(\cal{K}(n))|$.
		\item $ \sampler^{\cal{K}}(n,{\rm Register},{\rm Type},1,\mttx,\cdot)$  outputs a bit string consisiting of only $1$'s of length $2|\cal{K}(n)|+2$, which indicates that the whole space is the first (and only) register subspace of $\frS^{\rm Type}$ ---  note that this requires it to first call $\cal{K}(n)$ as a subroutine.
		\item $ \sampler^{\cal{K}}(n,{\rm Marginal},{\rm Type},1,\cdot,z)$ runs as follows:
		\begin{itemize}
			\item If ${\rm Type}$ is one of 
			\[
			\Hide A^j\ ,\ \Hide B^j\ ,\ 
			\   \Intro_A\ ,\ \Intro_B\ ,\ \Read_A\ ,\ \Read_B\ ,\ \Sample_A\ ,\ \Sample_B\ ,\ \Pauli_\PZm\ ,\ \Pauli_\PXm\ ,
			\]
			then it zeroes out $z$ and outputs $\vec 0$ (which is a concatenation of $2|\cal{K}(n)|+2$ zeros in this case).
			\item If ${\rm Type}$ is one of 
			\[
			\mathtt{row}_i\ ,\ \mathtt{col}_j\ ,\ \mathtt{var}_{ij}\ ,\ \mathtt{First}\ ,\ \mathtt{Second}\ ,\ \mathtt{Both}\ ,
			\]
			then it acts as the identity on $z$, namely outputs $z$.
			\item If ${\rm Type}$ is either $\mttX$ or $\mttZ$, let us split $z$ into $z_1,z_2$, where $z_1$ is the first $|\cal{K}(n)|+1$ bits of $z$ and $z_2$ are its last $|\cal{K}(n)|+1$ bits. Then, given ${\rm Type}=\mttX$,  it outputs $(z_1,\vec 0)$, and given ${\rm Type}=\mttZ$, it outputs $(\vec 0,z_2)$ (here $\vec 0$ is a concatenation of $|\cal{K}(n)|+1$ zeros).
		\end{itemize}
		
		\item As, again, we can assume $j=1$, $ \sampler^{\cal{K}}(n,{\rm Evaluate},{\rm Type},1,\mttx,z)$ runs  exactly the same as $ \sampler^\cal{K}(n,{\rm Marginal},{\rm Type},1,\cdot,z)$ .
		\item  There is a \emph{canonical} way of extracting the perpendicular action out of the others. See Clause 6 on page 94 of \cite{MIPRE}, which explains how step 3c in Figure 10 is implemented. Albeit, in this case it is straightforward what the perpendicular maps are. $ \sampler^\cal{K}(n,{\rm Perpendicular},{\rm Type},1,\mttx,z)$ runs as follows: 
		\begin{itemize}
			\item If ${\rm Type}$ is one of 
			\[
			\Hide A^j\ ,\ \Hide B^j\ ,\ 
			\   \Intro_A\ ,\ \Intro_B\ ,\ \Read_A\ ,\ \Read_B\ ,\ \Sample_A\ ,\ \Sample_B\ ,\ \Pauli_\PZm\ ,\ \Pauli_\PXm\ ,
			\]
			then $(\frS^{\rm Type})^\perp$ should act as the identity, and $ \sampler^\cal{K}$ outputs $z$.
			\item If ${\rm Type}$ is one of 
			\[
			\mathtt{row}_i\ ,\ \mathtt{col}_j\ ,\ \mathtt{var}_{ij}\ ,\ \mathtt{First}\ ,\ \mathtt{Second}\ ,\ \mathtt{Both}\ ,
			\]
			then $(\frS^{\rm Type})^\perp$ should act as the zero map, and $ \sampler^\cal{K}$ outputs $\vec 0$.
			\item Finally, $(\frS^{\mttX})^\perp(z)=(\vec 0,z_2)$ and $(\frS^{\mttZ})^\perp(z)=( z_1,\vec 0)$,  which means $ \sampler^\cal{K}$ outputs $(\vec 0,z_2)$ in case ${\rm Type}=\mttX$ and $(z_1,\vec 0)$ in case ${\rm Type}=\mttZ$.
		\end{itemize}
		
	\end{enumerate}
	
	We verify the required properties of the typed sampler $\sampler^\cal{K}$. Note that the above description is constant, and the only thing that actually needs to be appended is the description length of $\cal{K}$. By \cref{clause3:conditions_of_TM_encodings} of Fact \ref{fact:properties_of_TMs}, this shows that the description of $\sampler^\cal{K}$ can be calculated from that of $\cal{K}$ in polynomial time, which in particular implies the description length bound. For runtime, note that all the operations done by $\sampler^\cal{K}$ are either writing down the type graph (which takes $\poly(h)$-time), calls to $\cal{K}(n)$, or manipulations of vectors in $\FF_2^{2|\cal{K}(n)|+2}$ --- that take time at most $\poly(|\cal{K}(n)|)$, which is polynomial in the running time of $\cal{K}$. All in all, the running time is polynomial in that of $h$ and $\cal{K}$.
	\\
	
	\textbf{The Answer length calculator:}
	Recall that the readable and unreadable lengths of a vertex in $\frak{QueRed}$ (Section \ref{sec:augmentation_in_the_CLM_case}) depend only on its {type}.
	Hence, $\widetilde\length(n,(t,\mttx),\kappa)$ runs as follows: First, it calls $\cal{K}(n)$ and $\sampler^\cal{K}(\cdot,{\rm Graph},\cdot,\cdot,\cdot,\cdot)$  to retrieve the type set $\type$  underlying  the typed $1$-level CL sampling scheme. If $t\notin \type$ or $\mttx\notin \FF_2^{2|\cal{K}(n)|+2}$ or $\kappa\notin \{\frR,\frL\}$, then $\widetilde\length$ outputs an $\frak{error}$ sign. Otherwise, it lets $\Delta(n)=|\dec(\length(2^n,0,\frR))|$,\footnote{Here, $\widetilde\length$ will work as expected only if $\length$ is indeed padded and disregards its second and third inputs altogether.} and follows the table:
	\begin{center}
		\begin{tabular}{ |c|c|c|} 
			\hline
			Type $t$ & Decoded output if $\kappa=\frR$  & Decoded output if $\kappa=\frL$ \\ 
			\hline\hline
			$\Hide \cdot ^j$ & $2^{|\cal{K}(n)|}$ ones  & $2\cdot 2^{|\cal{K}(n)|}$ ones  \\
			\hline
			$\Intro_\cdot$ & $2^{|\cal{K}(n)|}+\Delta(n)$ ones &  $\Delta(n)$  ones\\
			\hline 
			$\Read_\cdot$ & $2^{|\cal{K}(n)|}+\Delta(n)$ ones & $2^{|\cal{K}(n)|}+\Delta(n) $ ones\\
			\hline 
			$\Sample_\cdot$ & $2^{|\cal{K}(n)|}+\Delta(n)$ ones & $\Delta(n) $ ones\\
			\hline 
			$\Pauli_\cdot$ & empty string & $2^{|\cal{K}(n)|}$ ones\\
			\hline 
			$\mttX$  & empty string & single one\\
			\hline 
			$\mttZ$  & empty string & single one\\
			\hline 
			$\mathtt{First}$  & empty string & single one\\
			\hline 
			$\mathtt{Second}$  & empty string & single one\\
			\hline 
			$\mathtt{Both}$  & empty string & two ones\\
			\hline 
			$\mathtt{var}_{ij}$  & empty string & single one\\
			\hline 
			$\mathtt{row}_{i}$  & empty string & three ones\\
			\hline 
			$\mathtt{col}_{j}$  & empty string & three ones\\
			\hline 
		\end{tabular}
	\end{center}
	We verify the required properties of $\widetilde\length$. The above description is constant, up to appending the descriptions of $\cal{K}$ and $\length$. 
	For running time, note that:
	\begin{itemize}
		\item it calls $\sampler^\cal{K}(\cdot,{\rm Graph},\cdot,\cdot,\cdot,\cdot)$ which takes $\poly(h)$ time;
		\item it calls $\cal{K}(n)$ which takes $\TIME(\cal{K};n)$ time;
		\item it verifies certain properties on bit strings of length $O(|\cal{K}(n)|)$, which takes $\poly(|\cal{K}(n)|)$ time;
		\item it calls $\length(2^n,\cdot,\cdot)$ which takes $\TIME(\length;2^n,\cdot,\cdot)$ time, and its output is of length $\Delta(n)$ which by definition is smaller or equal to $\TIME(\length;2^n,\cdot,\cdot)$;
		\item it outputs bit strings of length $O(2^{|\cal{K}(n)|}+\Delta(n))$, which takes $\poly(2^{|\cal{K}(n)|},\TIME(\length;2^n,\cdot,\cdot))$ time.
	\end{itemize}
	All in all, it runs in time which is $\poly(2^{|\cal{K}(n)|},\TIME(\cal{K};n),\TIME(\length;2^n,\cdot,\cdot))$, as claimed.
	\\
	
	\textbf{The Linear constraints processor:} 
	$\widetilde \linproc(n,(t,\mtty),(t',\mtty'),a^\frR,b^\frR)$ runs as follows.
	First it calls $\cal{K}(n)$, and    $\sampler^{\cal{K}}(\cdot,{\rm Graph},\cdot,\cdot,\cdot,\cdot)$  --- to get the type graph $(\type,\cal{E})$. If $tt'\notin \cal{E}$ or $\mtty,\mtty'\notin \FF_2^{2|\cal{K}(n)|+2}$, then it outputs $\frak{error}$ (note that in this case, the canonical decider will reject as this sign is not a proper encoding of a sequence of bit strings).
	Then, it checks that $|a^\frR|=| \dec(\widetilde\length(n,(t,\mtty),\frR))|$ and that  $|b^\frR|=| \dec(\widetilde\length(n,(t',\mtty'),\frR))|$, and outputs $\frak{error}$ otherwise. Given that the input was well structured, it runs $\mathscr{B}(\cal{K}(n))$, which outputs a sequence $\mathscr{B}$ of  $2^{|\cal{K}(n)|+1}$-many vectors in $\FF_2^{2^{|\cal{K}(n)|}}$ --- this can be thought of as a matrix over $\FF_2$ with $2^{|\cal{K}(n)|}$ columns and $2^{|\cal{K}(n)|+1}$ rows --- and  thus the elements of $\mathscr{B}$ (which are the rows of the aforementioned matrix) can be parameterized by vectors in $\FF_2^{|\cal{K}(n)|+1}$. Finally, it recovers the value $\Delta(n)=|\dec(\length(2^n,0,\frR))|$. Then, $\widetilde \linproc$ acts as follows:\footnote{ The format is the following: Each enumerated clause is some sub graph of the typed graph of $\frak{QueRed}$, which should help navigate the checks more easily. The actual operation of $\widetilde\linproc$ is to check what is the relevant question format, and acting according to the appropriate bullet. We first go over edges from $\PauliBasis_{2^{|\cal{K}(n)|}}(\mathscr{B})$ which was described in Section \ref{sec:Pauli_basis_definition}, then the single edge from $\Introspect(\verifier_{2^n})$ which was described in Section \ref{sec:the_introspection_game}, and finally the augmented edges of $\frak{QueRed}(\verifier_{2^{|\cal{K}(n)|}})$  described in Section \ref{sec:augmentation_in_the_CLM_case}.}

	\begin{enumerate}
		\item  \textbf{Pauli Basis Test --- Consistency checks of $\sX$-variables}:
		\begin{itemize}
			\item \emph{Question format}: $(t,\mtty)=(\Pauli_\PXm,\vec 0,\vec 0)\ ,\  (t',\mtty')=(\mttX,u,\vec 0)$.
			
			\emph{Operation}: The bit string $u\in \FF_2^{|\cal{K}(n)|+1}$ is the index of some vector $w^u\in \mathscr{B}\subseteq \FF_2^{2^{|\cal{K}(n)|}}$. Then, $\widetilde\linproc$ outputs (the encoding) of the single bit string  $(w^u,1,0)\in \FF_2^{2^{|\cal{K}(n)|}+2}$.
			
			\emph{Interpretation}: In this case $S^\frL_{\Pauli_{\PXm}}=\{\sP\sX^i\}_{i=1}^{2^{|\cal{K}(n)|}}$ and $S^\frL_{\mttX^u}=\{\sX^u\}$, and the above bit string encodes the linear constraint 
			\[
			\Big( \sum_{i=1}^{2^{|\cal{K}(n)|}}w^u_i\gamma(\sP\sX^i)\Big)+\gamma(\sX^u)=0\;.
			\]
			
			\item  \emph{Question format}: $(t,\mtty)=(\mathtt{var}_{11},u,v)\ ,\  (t',\mtty')=(\mttX,u,\vec 0)$.
			
			\emph{Operation}:  $\widetilde\linproc$ outputs (the encoding) of the single bit string  $(1,1,0)\in \FF_2^{3}$.
			
			\emph{Interpretation}: In this case $S^\frL_{\mathtt{var}_{11}^{u,v}}=\{\mathsf{Var}_{11}^{u,v}\}$ and $S^\frL_{\mttX^u}=\{\sX^u\}$, and the above bit string encodes the linear constraint 
			\[
			\gamma(\mathsf{Var}_{11}^{u,v})+\gamma(\sX^u)=0\;.
			\]

			\item  \emph{Question format}: $(t,\mtty)=(\mathtt{First},u,v)\ ,\  (t',\mtty')=(\mttX,u,\vec 0)$.
			
			\emph{Operation}:  $\widetilde\linproc$ outputs (the encoding) of the single bit string  $(1,1,0)\in \FF_2^{3}$.
			
			\emph{Interpretation}: In this case $S^\frL_{\mathtt{First}^{u,v}}=\{\mathsf{First}^{u,v}\}$ and $S^\frL_{\mttX^u}=\{\sX^u\}$, and the above bit string encodes the linear constraint 
			\[
			\gamma(\mathsf{First}^{u,v})+\gamma(\sX^u)=0.
			\]
		\end{itemize}
		
		\item  \textbf{Pauli Basis Test --- Consistency checks of $\sZ$-variables:}\\
		This is similar to the previous case, with the obvious modifications. For completeness, we give the details:
		\begin{itemize}
			\item \emph{Question format}: $(t,\mtty)=(\Pauli_\PZm,\vec 0,\vec 0)\ ,\  (t',\mtty')=(\mttZ,\vec 0,v)$.
			
			\emph{Operation}: The bit string $v\in \FF_2^{|\cal{K}(n)|+1}$ is the parameter of some vector $w^v\in \mathscr{B}\subseteq \FF_2^{2^{|\cal{K}(n)|}}$. Then, $\widetilde\linproc$ outputs (the encoding) of the single bit string  $(w^v,1,0)\in \FF_2^{2^{|\cal{K}(n)|}+2}$.
			
			\emph{Interpretation}: In this case $S^\frL_{\Pauli_{\PZm}}=\{\sP\sZ^i\}_{i=1}^{2^{|\cal{K}(n)|}}$ and $S^\frL_{\mttZ^v}=\{\sZ^v\}$, and the above bit string encodes the linear constraint 
			\[
			\Big( \sum_{i=1}^{2^{|\cal{K}(n)|}}w^v_i\gamma(\sP\sZ^i)\Big)+\gamma(\sZ^v)=0\;.
			\]
			
			\item  \emph{Question format}: $(t,\mtty)=(\mathtt{var}_{22},u,v)\ ,\  (t',\mtty')=(\mttZ,\vec 0,v)$.
			
			\emph{Operation}:  $\widetilde\linproc$ outputs (the encoding) of the single bit string  $(1,1,0)\in \FF_2^{3}$.
			
			\emph{Interpretation}: In this case $S^\frL_{\mathtt{var}_{22}^{u,v}}=\{\mathsf{Var}_{22}^{u,v}\}$ and $S^\frL_{\mttZ^v}=\{\sZ^v\}$, and the above bit string encodes the linear constraint 
			\[
			\gamma(\mathsf{Var}_{22}^{u,v})+\gamma(\sZ^v)=0\;.
			\]

			\item  \emph{Question format}: $(t,\mtty)=(\mathtt{Second},u,v)\ ,\  (t',\mtty')=(\mttZ,\vec 0,v)$.
			
			\emph{Operation}: $\widetilde\linproc$ outputs (the encoding) of the single bit string  $(1,1,0)\in \FF_2^{3}$.
			
			\emph{Interpretation}: In this case $S^\frL_{\mathtt{Second}^{u,v}}=\{\mathsf{Second}^{u,v}\}$ and $S^\frL_{\mttZ^v}=\{\sZ^v\}$, and the above bit string encodes the linear constraint 
			\[
			\gamma(\mathsf{Second}^{u,v})+\gamma(\sZ^v)=0\;.
			\]
			
		\end{itemize}

		\item \textbf{Pauli Basis Test --- (null-)Commutation game} (Section \ref{sec:com_game}):
		\begin{itemize}
			\item  \emph{Question format}: $(t,\mtty)=(\mathtt{First},u,v)\ ,\  (t',\mtty')=(\mathtt{Both},u,v)$.
			
			\emph{Operation}: $\widetilde\linproc$ reads $w^u,w^v\in \mathscr{B}\subseteq \FF_2^{2^{|\cal{K}(n)|}}$. Then, it calculates $\langle w^u,w^v\rangle$. If the result is $1$, it outputs  the empty string (which translates to immediate acceptance). Otherwise, it outputs $(1,1,0,0)\in \FF_2^{4}$.
			
			\emph{Interpretation}: In this case $S^\frL_{\mathtt{First}^{u,v}}=\{\mathsf{First}^{u,v}\}$ and $S^\frL_{\mathtt{Both}^{u,v}}=\{\mathsf{Both}^{u,v}_1,\mathsf{Both}^{u,v}_2\}$. If $\langle w^u,w^v\rangle=1,$ then this is a copy of the null-commutation game $\frC_{null}$, which always accepts. Otherwise, $\langle w^u,w^v\rangle=0$ and the game is the commutation game $\frC$, in which case the single bit string encodes the linear constraint 
			\[
			\gamma(\mathsf{First}^{u,v})+\gamma(\mathsf{Both}_1^{u,v})=0\;.
			\]
			
			\item  \emph{Question format}: $(t,\mtty)=(\mathtt{Second},u,v)\ ,\  (t',\mtty')=(\mathtt{Both},u,v)$.
			
			\emph{Operation}: $\widetilde\linproc$ reads $w^u,w^v\in \mathscr{B}\subseteq \FF_2^{2^{|\cal{K}(n)|}}$. Then, it calculates $\langle w^u,w^v\rangle$. If the result is $1$, it outputs  the empty string (which translates to immediate acceptance). Otherwise, it outputs $(1,0,1,0)\in \FF_2^{4}$.
			
			\emph{Interpretation}: In this case $S^\frL_{\mathtt{Second}^{u,v}}=\{\mathsf{Second}^{u,v}\}$ and $S^\frL_{\mathtt{Both}^{u,v}}=\{\mathsf{Both}^{u,v}_1,\mathsf{Both}^{u,v}_2\}$. If $\langle w^u,w^v\rangle=1,$ then this is a copy of the null-commutation game $\frC_{null}$, which always accepts. Otherwise, $\langle w^u,w^v\rangle=0$ and the game is the commutation game $\frC$, in which case the single bit string encodes the linear constraint 
			\[
			\gamma(\mathsf{Second}^{u,v})+\gamma(\mathsf{Both}_2^{u,v})=0\;.
			\]
		\end{itemize}
		
		\item \textbf{Pauli Basis Test --- (null-)Anti-Commutation game} (Section \ref{sec:Acom_game}): For every $1\leq i,j\leq 3$,
		\begin{itemize}
			\item  \emph{Question format}: $(t,\mtty)=(\mathtt{var}_{ab},u,v)\ ,\  (t',\mtty')=(\mathtt{row}_a,u,v)$.
			
			\emph{Operation}: $\widetilde\linproc$ reads $w^u,w^v\in \mathscr{B}\subseteq \FF_2^{2^{|\cal{K}(n)|}}$. Then, it calculates $\langle w^u,w^v\rangle$. If the result is $0$, it outputs  the empty string (which translates to immediate acceptance). Otherwise, given that $e_b\in \FF_2^3$ is the $b^{\rm th}$ vector of the standard basis (i.e., the indicator of $b$), $\widetilde\linproc$ outputs (the encoding) of the two bit strings $(1,e_b,0),(0,1,1,1,0)\in \FF_2^5$.\footnote{Just as a sanity check, e.g. if $b=1$, we seek the encoding of $11000\sqcup01110$, which is $0101000000100001010100$ according to Definition \ref{defn:the_alphabet}.}

			\emph{Interpretation}: In this case $S^\frL_{\mathtt{var}_{ab}^{u,v}}=\{\mathsf{Var}_{ab}^{u,v}\}$ and $S^\frL_{\mathtt{row}_a^{u,v}}=\{\mathsf{Row}_{a1}^{u,v},\mathsf{Row}^{u,v}_{a2},\mathsf{Row}^{u,v}_{a3}\}$. If $\langle w^u,w^v\rangle=0,$ then this is a copy of the null-anti-commutation game $\frM_{null}$, which always accepts. Otherwise, $\langle w^u,w^v\rangle=1$ and the game is the anti-commutation game $\frM$ --- i.e., the magic square game (Example \ref{example:magic-square}) --- in which case the two  bit strings encode the linear constraints
			\[
			\gamma(\mathsf{Var}_{ab}^{u,v})+\gamma(\mathsf{Row}_{ab}^{u,v})=0\quad {\rm and}\quad \gamma(\mathsf{Row}_{a1}^{u,v})+\gamma(\mathsf{Row}_{a2}^{u,v})+\gamma(\mathsf{Row}_{a3}^{u,v})=0\;.
			\]
			
			\item  \emph{Question format}: $(t,\mtty)=(\mathtt{var}_{ab},u,v)\ ,\  (t',\mtty')=(\mathtt{col}_j,u,v)$.
			
			\emph{Operation}: (This is very similar to the previous case, with rows and columns swapped, and with the sum along columns needing to be $1$ instead of $0$.) $\widetilde\linproc$ reads $w^u,w^v\in \mathscr{B}\subseteq \FF_2^{2^{|\cal{K}(n)|}}$. Then, it calculates $\langle w^u,w^v\rangle$. If the result is $0$, it outputs  the empty string (which translates to immediate acceptance). Otherwise, given that $e_a\in \FF_2^3$ is the $a^{\rm th}$ vector of the standard basis, $\widetilde\linproc$ outputs (the encoding) of the two bit strings $(1,e_a,0),(0,1,1,1,1)\in \FF_2^5$.

			\emph{Interpretation}: In this case $S^\frL_{\mathtt{var}_{ab}^{u,v}}=\{\mathsf{Var}_{ab}^{u,v}\}$ and $S^\frL_{\mathtt{col}_b^{u,v}}=\{\mathsf{Col}_{1b}^{u,v},\mathsf{Col}^{u,v}_{2b},\mathsf{Col}^{u,v}_{3b}\}$. If $\langle w^u,w^v\rangle=0,$ then this is a copy of the null-anti-commutation game $\frM_{null}$, which always accepts. Otherwise, $\langle w^u,w^v\rangle=1$ and the game is the anti-commutation game $\frM$ --- i.e., the magic square game (Example \ref{example:magic-square}) --- in which case the two  bit strings encode the linear constraints
			\[
			\gamma(\mathsf{Var}_{ab}^{u,v})+\gamma(\mathsf{Col}_{ab}^{u,v})=0\quad {\rm and}\quad \gamma(\mathsf{Col}_{1b}^{u,v})+\gamma(\mathsf{Col}_{2b}^{u,v})+\gamma(\mathsf{Col}_{3b}^{u,v})=1\;.
			\]

		\end{itemize}
		
		\item \textbf{Introspection Game} (Section \ref{sec:the_introspection_game}):

		\emph{Question format}: $(t,\mtty)=(\Intro_A,\vec 0,\vec 0)\ ,\  (t',\mtty')=(\Intro_B,\vec 0,\vec 0)$.
		
		\emph{Operation}: Recall that $\widetilde\linproc$ gets as input $a^\frR,b^\frR$, which in this case are in $\FF_2^{2^{|\cal{K}(n)|}+\Delta(n)}$, namely bit strings of length $2^{|\cal{K}(n)|}+\Delta(n)$. Denote by $que_A\in \FF_2^{2^{|\cal{K}(n)|}}$ (resp. $que_B$) the restriction of $a^\frR$ (resp. $b^\frR$) to its first $2^{|\cal{K}(n)|}$ bits, and $ans_A^\frR\in \FF_2^{\Delta(n)}$ (resp. $ans_B^\frR$) the restriction to its last $\Delta(n)$ bits. Now, $\widetilde\linproc$ calls the {original} linear constraints processor $\linproc(n,que_A,que_B,ans^\frR_A,ans^\frR_B)$. If the decoding of $\linproc$'s output is not  a sequence $(c^1,...,c^m)$ of bit strings of length $4\Delta(n)+1$, then $\widetilde\linproc$ outputs $\frak{error}$. Otherwise, each $c^i$ is of the form  $c^i=(c^i_{A,\frR},c^i_{A,\frL},c^i_{B,\frR},c^i_{B,\frL},c^i_{\sJ})$, where the first four (sub-)bit strings are of length $\Delta(n)$, and $c^i_{\sJ}$ is of length $1$. Then, $\widetilde\linproc$ outputs (the encoding) of the sequence of bit strings $(\tilde{c}^1,...,\tilde c^m)$, where each $\tilde c^i=(\vec 0,c^i_{A,\frR},c^i_{A,\frL},\vec 0,c^i_{B,\frR},c^i_{B,\frL},c^i_{\sJ})$ and $\vec 0\in \FF_2^{2^{|\cal{K}(n)|}}$.
		
		\emph{Interpretation}: In this case $S^\frR_{\Intro_\cdot}=\{\sQue^{\cdot,i}\}_{i=1}^{2^{|\cal{K}(n)|}}\bigcup \{\sAns^{\cdot,\frR,j}\}_{j=1}^{\Delta(n)}$ and $S^\frL_{\Intro_\cdot}=\{\sAns^{\cdot,\frL,j}\}_{j=1}^{\Delta(n)}$. Then, $que_
		A=\gamma(\sQue^{A,i})_{i=1}^{2^{|\cal{K}(n)|}},que_
		B=\gamma(\sQue^{B,i})_{i=1}^{2^{|\cal{K}(n)|}}$ are treated as a pair of questions in the original game, and $ans^\frR_\cdot=\gamma(\sAns^{\cdot,\frR,j})_{j=1}^{\Delta(n)}$ as the respective readable parts of answers. Thus $L_{que_A,que_B}(ans^\frR_A,ans^\frR_B)$ induces linear constraints on the $4\Delta(n)$ variables at the {vertices} $que_A$ and $que_B$ in the original game, which are checked instead on $
		\{\sAns^{\cdot,\cdot,j}\}_{j=1}^{\Delta(n)}$ by $$\widetilde L_{\Intro_A,\Intro_B}(que_A,ans^\frR_A,que_B,ans^\frR_B)\;.$$
		
		\item \textbf{Augmentation --- Sampling apparatus}: For $\circ\in \{A,B\}$,\footnote{To be able to distinguish between a blank spot for a player (that needs to be consistent with $A$ or $B$) and inputs to Turing machines that are disregarded, both of which were denoted by $\cdot$ in the text, we use $\circ$ for the player notation. }
		\begin{itemize}
			\item  \emph{Question format}: $(t,\mtty)=(\Pauli_\PZm,\vec 0,\vec 0)\ ,\  (t',\mtty')=(\Sample_\circ,\vec 0,\vec 0)$.
			
			\emph{Operation}: $\widetilde \linproc$ outputs (the encoding of) the following $2^{|\cal{K}(n)|}$ strings
			\[
			\forall 1\leq i\leq 2^{|\cal{K}(n)|}\ \colon \ \ (e_i,e_i,\vec 0,\vec 0,0)\in \FF_2^{2^{|\cal{K}(n)|}}\times \FF_2^{2^{|\cal{K}(n)|}}\times \FF_2^{\Delta(n)}\times \FF_2^{\Delta(n)}\times \FF_2\;,
			\]
			where $e_i$ is the $i^{\rm th}$ standard basis vectors (i.e., indicator of $i$) of $\FF_2^{2^{|\cal{K}(n)|}}$.
			
			\emph{Interpretation} (Compare to \eqref{eq:PZ_vs_Sample_quered_defn}): In this case $S^\frL_{\Pauli_\PZm}=\{\sP\sZ^i\}_{i=1}^{2^{|\cal{K}(n)|}}$ and $S^\frR_{\Sample_\circ}$ contains $\{\mathsf{SamZ}^{\circ,i}\}_{i=1}^{2^{|\cal{K}(n)|}}$. Then, the sequence of bit strings induce the checks
			\[
			\forall 1\leq i\leq 2^{|\cal{K}(n)|}\ \colon \ \ \gamma(\sP\sZ^i)=\gamma(\mathsf{SamZ}^{\circ,i})\;.
			\]
			
			\item  \emph{Question format}: $(t,\mtty)=(\Intro_\circ,\vec 0,\vec 0)\ ,\  (t',\mtty')=(\Sample_\circ,\vec 0,\vec 0)$.
			
			\emph{Operation}:  Recall that $\widetilde\linproc$ gets as input $a^\frR,b^\frR$, and in this case $$a^\frR=(que_\circ,ans_{\circ }^\frR)\in \FF_2^{2^{|\cal{K}(n)|}}\times \FF_2^{\Delta(n)} \quad {\rm and}\quad b^\frR=(seed,ans_{sam \circ}^\frR)\in \FF_2^{2^{|\cal{K}(n)|}}\times \FF_2^{\Delta(n)}.$$ Then,  
			$\widetilde \linproc$ calls $\sampler(2^n,{\rm Dimension},\cdot,\cdot,\cdot,\cdot)$ and denotes its length by $r$. If $r>2^{|\cal{K}(n)|}$, then it outputs the empty string  (which translates to acceptance). Otherwise, it takes $que'_\circ$ (resp. $seed'$) to be the restriction of $que_\circ$ (resp. $seed$) to its first $r$ bits, and calls $\sampler(2^n,{\rm Marginal},\circ,h,\cdot,seed')$ and compares it to $que'_\circ$. If they disagree, $\widetilde\linproc$ returns (the encoding of) the single string $(\vec 0,1)\in \FF_2^{2\cdot 2^{|\cal{K}(n)|}+4\Delta(n)} \times \FF_2$ (which is the equation associated with the singleton $\{\sJ\}$, implying rejection). Otherwise, it outputs (the encoding of) the sequence of bit strings $(c^{1,\frR},c^{1,\frL},...,c^{2^{|\cal{K}(n)|},\frR},c^{2^{|\cal{K}(n)|},\frL})$, where 
			\[
			\begin{split}
			\forall 1\leq i\leq \Delta(n)\ \colon \ \ c^{i,\frR}&=(\vec 0,e_i,\vec 0,\vec 0,e_i,\vec 0,0)\in \FF_2^{2^{|\cal{K}(n)|}}\times \FF_2^{\Delta(n)}\times \FF_2^{\Delta(n)}\times\FF_2^{2^{|\cal{K}(n)|}}\times \FF_2^{\Delta(n)}\times \FF_2^{\Delta(n)}\times \FF_2\ ,\\
			c^{i,\frL}&=(\vec 0,\vec 0,e_i,\vec 0,\vec 0,e_i,0)\in \FF_2^{2^{|\cal{K}(n)|}}\times \FF_2^{\Delta(n)}\times \FF_2^{\Delta(n)}\times\FF_2^{2^{|\cal{K}(n)|}}\times \FF_2^{\Delta(n)}\times \FF_2^{\Delta(n)}\times \FF_2\ .    \end{split}
			\]
			
			\emph{Interpretation} (Compare to \eqref{eq:linear_checks_over_que_vs_sam_quered} and \eqref{eq:que_vs_samZ_check_quered}): In this case 
			\[
			\begin{split}
			S_{\Intro_\circ}^\frR&=\sQue^\circ\bigcup\sAns^{\circ,\frR}=\{\sQue^{\circ,i}\}_{i=1}^{2^{|\cal{K}(n)|}}\bigcup \{\sAns^{\circ,\frR,j}\}_{j=1}^{\Delta(n)}\ ,\\
			S_{\Intro_\circ}^\frL&=\sAns^{\circ,\frL}= \{\sAns^{\circ,\frL,j}\}_{j=1}^{\Delta(n)}\ ,\\
			S_{\Sample_\circ}^\frR&= \mathsf{SamZ}^\circ\bigcup\mathsf{SamAns}^{\circ,\frR}=\{\mathsf{SamZ}^{\circ,i}\}_{i=1}^{2^{|\cal{K}(n)|}}\bigcup \{\mathsf{SamAns}^{\circ,\frR,j}\}_{j=1}^{\Delta(n)}\ ,\\
			S_{\Sample_\circ}^\frL&= \mathsf{SamAns}^{\circ,\frL}=\ \{\mathsf{SamAns}^{\circ,\frL,j}\}_{j=1}^{\Delta(n)}\ .
			\end{split}
			\]
			First, if $2^{|\cal{K}(n)|}$ is smaller than the dimension of the CLM $\frS^\circ$ --- which  is the CLM induced by the sampler $\sampler$ when fixing the ${\rm Player}$ input to $\circ$ --- then no constraints are checked. Otherwise, we denote  $\gamma|_{\sQue^\circ}=\mttx$ and $\gamma|_{\mathsf{SamZ}^\circ}=z$, and  $\widetilde \linproc$  verifies that $\frS^\circ(z)=\mttx$, and outputs the certain rejection linear constraint $\gamma(\sJ)=0$ if not (recall that $\gamma(\sJ)=1$ by Definition \ref{defn:tailored_games}). If the above condition was held, it then outputs the linear constraints
			\[
			\forall 1\leq j\leq \Delta(n)\ \colon \ \ \gamma(\sAns^{\circ,\frR,j})=\gamma(\mathsf{SamAns}^{\circ,\frR,j})\quad,\quad \gamma(\sAns^{\circ,\frL,j})=\gamma(\mathsf{SamAns}^{\circ,\frL,j})\ .
			\]

		\end{itemize}
		
		\item \textbf{Augmentation --- Hiding apparatus}: For $\circ\in \{A,B\}$,
		\begin{itemize}
			\item  \emph{Question format}: $(t,\mtty)=(\Intro_\circ,\vec 0,\vec 0)\ ,\  (t',\mtty')=(\Read_\circ,\vec 0,\vec 0)$.
			
			\emph{Operation}:  $\widetilde\linproc$ outputs (the encoding of) the sequence of bit strings consisting of 
			\[
			\forall 1\leq i\leq 2^{|\cal{K}(n)|}\ \colon \ \  c^{i,\sQue}=(e_i,\vec 0,\vec 0,e_i,\vec 0,\vec 0,\vec 0,0),
			\]
			and 
			\[
			\forall 1\leq j\leq \Delta(n)\ \colon \ \ c^{j,\frR}=(\vec 0,e_j,\vec 0,\vec 0,e_j,\vec 0,\vec 0,0)\quad,\quad   c^{j,\frL}=(\vec 0,\vec 0,e_j,\vec 0,\vec 0,\vec 0,e_j,0)
			\]
			all of which are in 
			\[
			\begin{split}
			\FF_2^{2^{|\cal{K}(n)|}}\times \FF_2^{\Delta(n)}\times \FF_2^{\Delta(n)}\times\FF_2^{2^{|\cal{K}(n)|}}\times \FF_2^{\Delta(n)}\times\FF_2^{2^{|\cal{K}(n)|}}\times  \FF_2^{\Delta(n)}\times \FF_2\;,
			\end{split}
			\]
			with $e_i$ being the $i^{\rm th}$ standard basis in the respective space. 
			
			\emph{Interpretation} (Compare to \eqref{eq:intro_vs_read_quered_defn}): In this case 
			\[
			\begin{split}
			S_{\Intro_\circ}^\frR&=\sQue^\circ\bigcup\sAns^{\circ,\frR}=\{\sQue^{\circ,i}\}_{i=1}^{2^{|\cal{K}(n)|}}\bigcup \{\sAns^{\circ,\frR,j}\}_{j=1}^{\Delta(n)},\\
			S_{\Intro_\circ}^\frL&=\sAns^{\circ,\frL}= \{\sAns^{\circ,\frL,j}\}_{j=1}^{\Delta(n)},\\
			S_{\Read\circ}^\frR&= \mathsf{ReadQue}^\circ\bigcup\mathsf{ReadAns}^{\circ,\frR}=\{\mathsf{ReadQue}^{\circ,i}\}_{i=1}^{2^{|\cal{K}(n)|}}\bigcup \{\mathsf{ReadAns}^{\circ,\frR,j}\}_{j=1}^{\Delta(n)},\\
			S_{\Read_\circ}^\frL&= \mathsf{ReadPerp}^{\circ}\bigcup \mathsf{ReadAns}^{\circ,\frL}=\{\mathsf{ReadPerp}^{\circ,i}\}_{i=1}^{2^{|\cal{K}(n)|}}\bigcup\{\mathsf{ReadAns}^{\circ,\frL,j}\}_{j=1}^{\Delta(n)}.
			\end{split}
			\]
			The linear constraints induced by the above are 
			\[
			\begin{split}
			\forall 1\leq i\leq 2^{|\cal{K}(n)|}\ &\colon \ \ \gamma(\sQue^{\circ,i})=\gamma(\mathsf{ReadQue}^{\circ,i})\quad,\\
			\forall 1\leq j\leq \Delta(n)\ &\colon \ \ \gamma(\sAns^{\circ,\frR,j})=\gamma(\mathsf{ReadAns}^{\circ,\frR,j})\quad,\quad \gamma(\sAns^{\circ,\frL,j})=\gamma(\mathsf{ReadAns}^{\circ,\frL,j}).
			\end{split}
			\]
			
			\item  \emph{Question format}: $(t,\mtty)=(\Hide \circ^h,\vec 0,\vec 0)\ ,\  (t',\mtty')=(\Read_\circ,\vec 0,\vec 0)$.
			
			\emph{Operation}:   Recall that $\widetilde\linproc$ gets as input $a^\frR,b^\frR$, and in this case $$a^\frR=(que_{hide\ h\ \circ})\in \FF_2^{2^{|\cal{K}(n)|}} \quad {\rm and}\quad b^\frR=(que_{read \circ},ans_{read \circ}^\frR)\in \FF_2^{2^{|\cal{K}(n)|}}\times \FF_2^{\Delta(n)}\;.$$ Then,  
			$\widetilde \linproc$ calls $\sampler(2^n,{\rm Dimension},\cdot,\cdot,\cdot,\cdot)$ and denotes its length by $r$. If $r>2^{|\cal{K}(n)|}$, then it outputs the empty string (which translates to acceptance). Otherwise, it lets $que'_{read\circ}$ be the restriction of $que_{read\circ}$ to its first $r$ bits, and calls $\sampler(2^n,{\rm Register},\circ,j,que'_{read\circ},\cdot)$ for every $1\leq j\leq h-1$. The output of each of these runs should be a vector in $\FF_2^r$, and $\widetilde\linproc $ adds these outputs and get a bit string $I_{< h}\in \FF_2^r$. Finally, $\widetilde\linproc$ outputs (the encoding of) the following sequence of bit strings: For every $1\leq i\leq r$, if $I_{< h}(i)=1$, then
			\[
			c^{i,que}=(e_i,\vec 0,e_i,\vec 0,\vec 0,\vec 0,0)\;,
			\]
			and if $I_{<h}(i)=0$, then 
			\[
			c^{i,que}=(e_i,\vec 0,\vec 0,\vec 0,\vec 0,\vec 0,0)\;.
			\]
			For every $1\leq j\leq r$ it adds
			\[
			c^{j,perp}=(\vec 0,e_j,\vec 0,\vec 0,e_j,\vec 0,0)\;.
			\]
			All the constraints are in 
			\[
			\begin{split}
			\FF_2^{2^{|\cal{K}(n)|}}\times \FF_2^{2^{|\cal{K}(n)|}}\times\FF_2^{2^{|\cal{K}(n)|}}\times \FF_2^{\Delta(n)}\times\FF_2^{2^{|\cal{K}(n)|}}\times  \FF_2^{\Delta(n)}\times \FF_2\;,
			\end{split}
			\]
			with $e_i$ being the $i^{\rm th}$ standard basis in the respective space. 
			
			\emph{Interpretation} (Compare to \eqref{eq:read_vs_hide^h_quered_defn1} and \eqref{eq:read_vs_hide^h_quered_defn2}): In this case 
			\[
			\begin{split}
			S_{\Hide \circ^h}^\frR&=\mathsf{Hide}^h\sQue^\circ=\{\mathsf{Hide}^h\sQue^{\circ,i}\}_{i=1}^{2^{|\cal{K}(n)|}},\\
			S_{\Hide \circ^h}^\frL&=\mathsf{Hide}^h\mathsf{Perp}^{\circ}= \{\mathsf{Hide}^h\mathsf{Perp}^{\circ,j}\}_{j=1}^{2^{|\cal{K}(n)|}},\\
			S_{\Read\circ}^\frR&= \mathsf{ReadQue}^\circ\bigcup\mathsf{ReadAns}^{\circ,\frR}=\{\mathsf{ReadQue}^{\circ,i}\}_{i=1}^{2^{|\cal{K}(n)|}}\bigcup \{\mathsf{ReadAns}^{\circ,\frR,j}\}_{j=1}^{\Delta(n)},\\
			S_{\Read_\circ}^\frL&= \mathsf{ReadPerp}^{\circ}\bigcup \mathsf{ReadAns}^{\circ,\frL}=\{\mathsf{ReadPerp}^{\circ,i}\}_{i=1}^{2^{|\cal{K}(n)|}}\bigcup\{\mathsf{ReadAns}^{\circ,\frL,j}\}_{j=1}^{\Delta(n)}.
			\end{split}
			\]
			The  vector $I_{< h}$ calculated by $\widetilde\linproc$ is indeed the indicator of coordinates active in the register subspace $W^{que'_{read\circ}}_{< h}$ (see \eqref{eq:reg_subspaces_depending_on_prefixes}) associated with the CLM $\frS^\circ$ induced by $\sampler$ when fixing the ${\rm Player}$ input to be $\circ$. So, the linear constraints induced by the above are 
			\[
			\begin{split}
			\forall 1\leq i\leq r\ \ {\rm s.t.}\ \ I_{< h}(i)=1\ &\colon\ \ \gamma(\mathsf{Hide}^h\sQue^{\circ,i})=\gamma(\mathsf{ReadQue}^{\circ,i})\ ,\\
			I_{<h}(i)=0\ &\colon \ \  \gamma(\mathsf{Hide}^h\sQue^{\circ,i})=0\ ,\\
			\forall 1\leq j\leq 2^{|\cal{K}(n)|} \ &\colon \ \ \gamma(\mathsf{Hide}^h\mathsf{Perp}^{\circ,j})=\gamma(\mathsf{ReadPerp}^{\circ,j})\ . 
			\end{split}
			\]
			
			\item  \emph{Question format}: $(t,\mtty)=(\Hide \circ^j,\vec 0,\vec 0)\ ,\  (t',\mtty')=(\Hide \circ^{j-1},\vec 0,\vec 0)$.
			
			\emph{Operation}:   Recall that $\widetilde\linproc$ gets as input $a^\frR,b^\frR$, and in this case $$a^\frR=(que_{hide\ j\ \circ})\in \FF_2^{2^{|\cal{K}(n)|}} \quad {\rm and}\quad b^\frR=(que_{hide\ j-1\ \circ})\in \FF_2^{2^{|\cal{K}(n)|}}.$$ Then,  
			$\widetilde \linproc$ calls $\sampler(2^n,{\rm Dimension},\cdot,\cdot,\cdot,\cdot)$ and denotes its length by $r$. If $r>2^{|\cal{K}(n)|}$, then it outputs the empty string (which translates to acceptance). Otherwise, it lets $que'_{hide\ j\ \circ}$ be the restriction of $que_{hide\ j\ \circ}$ to its first $r$ bits, and calls $\sampler(2^n,{\rm Register},\circ,t,que'_{hide\ j\ \circ},\cdot)$ for every $1\leq t\leq j$. The output of each of these runs should be a vector in $\FF_2^r$, and $\widetilde\linproc $:
			\begin{itemize}
				\item [$\heartsuit$] sums the first $j-2$ of these outputs to  get a bit string $I_{<j-1}\in \FF_2^r$;
				\item [$\heartsuit$] denotes the output when $t=j$ as $I_j\in \FF_2^r$.
			\end{itemize}
			Now, $\widetilde\linproc$ calls $\sampler(2^n,{\rm Perpendicular},\circ,j,que'_{hide\ j\ \circ},e_i)$ for every $i$ in the support ${\rm Supp}(I_j)$ of $I_j$ (i.e., such that $I_j(i)=1$) --- where $e_i$ is the $i^{\rm th}$ standard basis vector (i.e., indicator of $i$) --- and denotes their output as $Col_i\in \FF_2^{r}$. By the definition of a CL sampler (Definition \ref{defn:h-level_CL_Sampler}),  the support of each $Col_i$ is contained in ${\rm Supp}(I_j)$. Let $Col_i=\vec 0\in \FF_2^r$ for every $i\notin {\rm Supp}(I_j)$. Then, collecting all of these $Col_i$'s as columns in a matrix gives an $r\times r$ matrix $\Psi$ which is supported on ${\rm Supp}(I_j)\times {\rm Supp}(I_j)$.  Let $Row_i\in \FF_2^r$ be the $i^{\rm th}$ row of $\Psi$.
			
			Finally, $\widetilde\linproc$ outputs (the encoding of) the following sequence of bit strings:
			\[
			\begin{split}
			\forall i\in {\rm Supp}(I_{<j-1})\ &\colon\ \ c^{i,que}=(e_i,\vec 0,e_i,\vec 0, 0)\;,\\
			\forall i\notin{\rm Supp}(I_{<j-1})\ &\colon\ \ c^{i,que}=(\vec 0,\vec 0,e_i,\vec 0, 0)\;,\\
			\forall i\notin {\rm Supp}(I_{j})\  &\colon \ \    c^{i,perp}=(\vec 0,e_i, \vec 0,e_i,0)\;,\\
			\forall i\in {\rm Supp}(I_j)\ &\colon\ \  c^{i,perp}=(\vec 0,e_i, \vec 0,Row_i,0)\;.
			\end{split}
			\]
			All of the above are in 
			\[
			\begin{split}
			\FF_2^{2^{|\cal{K}(n)|}}\times \FF_2^{2^{|\cal{K}(n)|}}\times  \FF_2^{2^{|\cal{K}(n)|}}\times \FF_2^{2^{|\cal{K}(n)|}}\times \FF_2\;.
			\end{split}
			\]

			\emph{Interpretation} (Compare to \eqref{eq:hide_vs_hids_quered_defn1}, \eqref{eq:hide_vs_hids_quered_defn2} and \eqref{eq:hide_vs_hids_quered_defn3}): In this case 
			\[
			\begin{split}
			S_{\Hide \circ^\cdot}^\frR&=\mathsf{Hide}^\cdot\sQue^\circ=\{\mathsf{Hide}^\cdot\sQue^{\circ,i}\}_{i=1}^{2^{|\cal{K}(n)|}},\\
			S_{\Hide \circ^\cdot}^\frL&=\mathsf{Hide}^\cdot\mathsf{Perp}^{\circ}= \{\mathsf{Hide}^\cdot\mathsf{Perp}^{\circ,i}\}_{i=1}^{2^{|\cal{K}(n)|}}.
			\end{split}
			\]
			The  vectors $I_{<j-1},I_j$ calculated by $\widetilde\linproc$ are indeed the indicators of coordinates active in the respective register subspaces $W^{que'_{hide\ j\ \circ}}_{<j-1}$ and $W^{que'_{hide\ j\ \circ}}_{j}$  associated with the CLM $\frS^\circ$ induced by $\sampler$ when fixing the ${\rm Player}$ input to be $\circ$. So, the linear constraints induced by the above are 
			\[
			\begin{split}
			\forall i\in {\rm Supp}(I_{<j-1})\ &\colon \ \ \gamma(\mathsf{Hide}^j\sQue^{\circ,i})=\gamma(\mathsf{Hide}^{j-1}\sQue^{\circ,i})\ ,\\
			\forall i\notin {\rm Supp}(I_{<j-1})\ &\colon \ \ 0=\gamma(\mathsf{Hide}^{j-1}\sQue^{\circ,i})\ ,\\
			\forall i\notin {\rm Supp}(I_{j})\ &\colon \ \ \gamma(\mathsf{Hide}^j\mathsf{Perp}^{\circ,i})=\gamma(\mathsf{Hide}^{j-1}\mathsf{Perp}^{\circ,i})\ ,\\
			\forall i\in {\rm Supp}(I_{j})\ &\colon \ \ \gamma(\mathsf{Hide}^j\mathsf{Perp}^{\circ,i})=\sum_{t=1}^r Row_i(t)\gamma(\mathsf{Hide}^{j-1}\mathsf{Perp}^{\circ,t})\ .\\ 
			\end{split}
			\]
			This makes sense, as $Row_i(t)$ is the $it$ entry of $\Psi$, whose ${\rm Supp}(I_j)\times {\rm Supp}(I_j)$ block is exactly $(\frS^{\circ,que'_{hide\ j\ \circ}}_j)^\perp$.
			
			\item  \emph{Question format}: $(t,\mtty)=(\Hide \circ^1,\vec 0,\vec 0)\ ,\  (t',\mtty')=(\Pauli_\PXm,\vec 0,\vec 0)$.
			
			\emph{Operation}:     
			$\widetilde \linproc$ calls $\sampler(2^n,{\rm Dimension},\cdot,\cdot,\cdot,\cdot)$ and denotes its length by $r$. If $r>2^{|\cal{K}(n)|}$, then it outputs the empty string (which translates to acceptance). Otherwise, it calls $\sampler(2^n,{\rm Register},\circ,1,\vec 0,\cdot)$ and denotes its  output  by $I_1\in \FF_2^r$ and its complement by $I_{>1}\in \FF_2^r$. 
			Now, $\widetilde\linproc$ calls $\sampler(2^n,{\rm Perpendicular},\circ,1,\vec 0,e_i)$ for every $i$ in  ${\rm Supp}(I_1)$, and denotes their output as $Col_i\in \FF_2^{r}$. By the definition of a CL sampler (Definition \ref{defn:h-level_CL_Sampler}),  the support of each $Col_i$ is contained in ${\rm Supp}(I_1)$. Let $Col_i=\vec 0\in \FF_2^r$ for every $i\notin {\rm Supp}(I_1)$. Then, collecting all of these $Col_i$'s as columns in a matrix gives an $r\times r$ matrix $\Psi$ which is supported on ${\rm Supp}(I_1)\times {\rm Supp}(I_1)$.  Let $Row_i\in \FF_2^r$ be the $i^{\rm th}$ row of $\Psi$.
			
			Finally, $\widetilde\linproc$ outputs (the encoding of) the following sequence of bit strings:
			\[
			\begin{split}
			\forall i \in [2^{\cal{K}(n)}]\ &\colon \ \ c^{i,que}=(e_i,\vec 0,\vec 0,0)\;,\\
			\forall i\in {\rm Supp}(I_{>1})\ &\colon \ \  c^{i,perp}=(\vec 0,e_i,e_i, 0)\;,\\
			\forall i\in {\rm Supp}(I_1)\ &\colon \ \ c^{i,perp}=(\vec 0,e_i,Row_i, 0)\;,
			\end{split}
			\]
			all of which are in 
			\[
			\begin{split}
			\FF_2^{2^{|\cal{K}(n)|}}\times \FF_2^{2^{|\cal{K}(n)|}}\times\FF_2^{2^{|\cal{K}(n)|}}\times \FF_2\ .
			\end{split}
			\]
			
			\emph{Interpretation} (Compare to \eqref{eq:hide^1_vs_PauliX_quered_defn1},~\eqref{clause:hide_vs_PauliX_in_quered2} and \eqref{clause:hide_vs_PauliX_in_quered3}): In this case 
			\[
			\begin{split}
			S_{\Hide \circ^1}^\frR&=\mathsf{Hide}^1\sQue^\circ=\{\mathsf{Hide}^1\sQue^{\circ,i}\}_{i=1}^{2^{|\cal{K}(n)|}},\\
			S_{\Hide \circ^1}^\frL&=\mathsf{Hide}^1\mathsf{Perp}^{\circ}= \bigcup \{\mathsf{Hide}^1\mathsf{Perp}^{\circ,j}\}_{j=1}^{2^{|\cal{K}(n)|}},\\
			S_{\Pauli_\PXm}^\frL&= \sP\sX=\{\sP\sX^i\}_{i=1}^{2^{|\cal{K}(n)|}}.
			\end{split}
			\]
			The  vectors $I_{1}$ and $I_{>1}$ calculated by $\widetilde\linproc$ are indeed the indicators of coordinates active in the register subspaces $W^{\vec 0}_{1}=V_1$ and $W^{\vec 0}_{>1}=V_{>1}$ (see \eqref{eq:reg_subspaces_depending_on_prefixes}) associated with the CLM $\frS^\circ$ induced by $\sampler$ when fixing the ${\rm Player}$ input to be $\circ$. So, the linear constraints induced by the above are 
			\[
			\begin{split}
			\forall i\in [2^{\cal{K}(n)}]\ &\colon \ \ \gamma(\mathsf{Hide}^1\mathsf{Que}^{\circ, i})=0\ ,\\
			\forall i\in {\rm Supp}(I_{>1})\  &\colon\ \ \gamma(\mathsf{Hide}^1\mathsf{Perp}^{\circ,i})=\gamma(\sP\sX^i)\ ,\\
			\forall i\in {\rm Supp}(I_1)\ &\colon \ \ \gamma(\mathsf{Hide}^1\mathsf{Perp}^{\circ,i})=\sum_{t=1}^r Row_i(t)\gamma(\sP\sX^t)\ .
			\end{split}
			\]
		\end{itemize}   
	\end{enumerate}
	For properties of $\widetilde \linproc$, note that all of its operations are either calls to $\cal{K}(n),\sampler(2^n,\cdot,\cdot,\cdot,\cdot,\cdot),\length(2^n,\cdot,\cdot)$ and $\linproc(2^n,\cdot,\cdot,\cdot,\cdot)$, or manipulations of vectors in a space $\FF_2^m$ where $m=O(2^{|\cal{K}(n)|}+\Delta(n))$. This proves the running time argument. Regarding description length, the above description is fixed up to appending the descriptions of $\mathscr{B}$ (which is constant length), $\cal{K},\sampler,\length$ and $\linproc$.
\end{proof}

\subsubsection{Proof of Theorem \ref{thm:h_level_question_reduciton}}\label{sec:proof_of_QR}

The input to $\mathsf{QuestionReduction}_h$ is an $h$-level verifier $\verifier$ and an integer $\lambda$ in binary. First, the TM $\mathsf{QuestionReduction}_h$ uses $\lambda$ to define two other  TMs, $\cal{K}^\lambda$ and $\Lambda^\lambda$, which act as follows:
\begin{enumerate}
	\item $\cal{K}^\lambda$ gets as input $n$ in binary (i.e., $\overline{n}$), and outputs a string of $n\cdot \lambda$ ones (i.e., $1^{*n\cdot\lambda}$).
	\item $\Lambda^\lambda$  gets as input $n$ in binary, and outputs a string of $n^\lambda$ ones (i.e., $1^{*{n^\lambda}}$).
\end{enumerate}
Note that the description lengths of both of these TMs is fixed up to appending $\lambda$, which requires $\log\lambda$-bits. Hence, by \cref{clause3:conditions_of_TM_encodings} of Fact \ref{fact:properties_of_TMs}, their description length is $\polylog(\lambda)$. Furthermore, $\TIME(\cal{K};\overline{n})=O(\lambda \cdot n)$ while $\TIME(\Lambda;\overline{n})=O(n^\lambda)$. Now:
\begin{enumerate}
	\item Let $\verifier^{(1)}=(\sampler^{(1)},\length^{(1)},\linproc^{(1)},\decider)$  be the output of $\mathsf{Padding}(\verifier,\Lambda^\lambda)$ (Claim \ref{claim:Padding_NFV}).
	\item Let $\verifier^{(2)}=(\sampler^{(2)},\length^{(2)},\linproc^{(2)},\decider)$ be the output of $\mathsf{TypedQuestionReduction}_h(\verifier^{(1)},\cal{K}^\lambda)$ (Claim \ref{claim:typed_question_reduction_on_NFV}).
	\item Let  $\verifier^{(3)}=(\sampler^{(3)},\length^{(3)},\linproc^{(3)},\decider)$ be the output of $\mathsf{DeType}_1(\verifier^{(2)})$ (Claim \ref{claim:DeTyping_NFV}).\footnote{The subscript $1$ in the $\mathsf{DeType}_1$ is not a mistake, as the output of $\mathsf{TypedQuestionReduction}_h$ is a typed tailored $1$-level normal form verifier regardless of $h$.}
\end{enumerate}

\textbf{The Sampler}:
By Claim \ref{claim:typed_question_reduction_on_NFV}, $\sampler^{(2)}$ depends only on $\cal{K}^\lambda$ (which itself depends only on $\lambda$) and $h$, and is a typed $1$-level CL sampler. 
Specifically, its description length is polynomial in that of $\cal{K}^\lambda$ (with the constants depending on $h$), i.e. it is $\poly_h(\log\lambda)$, and $\TIME(\sampler^{(2)};n,\cdot,\cdot,\cdot,\cdot,\cdot)=\poly_h(\TIME(\cal{K}^\lambda;n))=\poly_h(\lambda, n).$ 
By Claim \ref{claim:DeTyping_NFV}, $\sampler^{(3)}$ is a $3$-level CL sampler. The description length of $\sampler^{(3)}$ is polynomial in that of $\sampler^{(2)}$ (and depends only on it), which means it is $\poly_h(\log\lambda)$ as well. 
Furthermore, the running time of $\sampler^{(3)}$ is polynomial in $|\type|=2h+28$, and $\TIME(\sampler^{(2)};n,\cdot,\cdot,\cdot,\cdot,\cdot)=\poly_h(\lambda,n)$. 
All in all,  $\TIME(\sampler^{(3)};n,\cdot,\cdot,\cdot,\cdot,\cdot)=\poly_h(\lambda,n)$. 
Let $\sampler^\lambda_{\qr}=\sampler^{(3)}$, and note that it satisfies all the required conditions.
\\

\textbf{The Answer length function}:
By Claim \ref{claim:Padding_NFV}, $\length^{(1)}$ depends only on $\Lambda^\lambda$, which itself only depends on $\lambda$. 
Specifically, its description length is polynomial in that of $\Lambda^\lambda$, namely $\poly(\log\lambda)$, and its runtime satisfies $\TIME(\length^{(1)};n,\cdot,\cdot)=O(\TIME(\Lambda^\lambda;n))=O(n^\lambda)$. 
By Claim \ref{claim:typed_question_reduction_on_NFV}, $\length^{(2)}$ depends only on $\cal{K}^\lambda$, $h$ and $\length^{(1)}$. Specifically, its description length is polynomial in theirs (up to constants that depend on $h$), which is $\poly_h(\log\lambda)$. For running time, we have $$\TIME(\length^{(2)};n,\cdot,\cdot)=\poly_h(2^{\overbrace{|\cal{K}^\lambda(n)|}^{=\lambda n}},\overbrace{\TIME(\cal{K}^\lambda;n)}^{=O(\lambda 
	n)},\overbrace{\TIME(\length^{(1)};2^n,\cdot,\cdot)}^{=O(2^{\lambda n})})=\poly_h(2^{\lambda n}).$$
By Claim \ref{claim:DeTyping_NFV}, $\length^{(3)}$ runs in time which is polynomial in   $|\type|=2h+28$, $\TIME(\sampler^{(2)};n,\cdot,\cdot,\cdot,\cdot,\cdot)=\poly_h( n ,\lambda)$ and $\TIME(\length^{(2)};n,\cdot,\cdot)=\poly_h(2^{\lambda n })$. Namely, $\TIME(\length^{(3)};n,\cdot,\cdot)=\exp_h(\lambda, n )$. Letting $\length^\lambda_{\qr}=\length^{(3)}$ satisfies the required conditions. We leave it to the reader to verify that for well structured inputs, the output of $\length^\lambda_\qr$ never decodes to $\frak{error}$.
\\

\textbf{The Linear constraints processor}:
By Claim \ref{claim:Padding_NFV}, 
\[
\TIME(\linproc^{(1)};n,\cdot,\cdot,\cdot,\cdot)=\poly(\overbrace{\TIME(\Lambda^\lambda;n)}^{=O(n^{\lambda})},\TIME(\sampler;n,\cdot,\cdot,\cdot,\cdot,\cdot),\TIME(\length; n,\cdot,\cdot),\TIME(\linproc; n,\cdot,\cdot,\cdot,\cdot))\;.
\]
By Claim \ref{claim:typed_question_reduction_on_NFV},
\[
\TIME(\linproc^{(2)};n,\cdot,\cdot,\cdot,\cdot)=\poly(2^{\overbrace{|\cal{K}^\lambda(n)|}^{\lambda n}},\overbrace{\TIME(\cal{K}^\lambda;n)}^{=\lambda n },\TIME(\overbrace{\sampler^{(1)}}^{=\sampler};2^n,\cdot,\cdot,\cdot,\cdot,\cdot),\overbrace{\TIME(\length^{(1)}; 2^n,\cdot,\cdot)}^{=O(2^{\lambda n })},\TIME(\linproc^{(1)}; 2^n,\cdot,\cdot,\cdot,\cdot))\;.
\]
By Claim \ref{claim:DeTyping_NFV},
\[
\TIME(\linproc^{(3)};n,\cdot,\cdot,\cdot,\cdot)=\poly(\overbrace{|\type|}^{=2h+28},\overbrace{\TIME(\sampler^{(2)};n,\cdot,\cdot,\cdot,\cdot,\cdot)}^{=\poly(\lambda, n )},\overbrace{\TIME(\length^{(2)};n,\cdot,\cdot)}^{=\poly(2^{\lambda n })},\TIME(\linproc^{(2)};n,\cdot,\cdot,\cdot,\cdot))\;.
\]
If $\verifier$ is $\lambda$-bounded (Definition \ref{defn:h-level_NFV}), then 
\[
\TIME(\sampler;n,\cdot,\cdot,\cdot,\cdot,\cdot),\TIME(\length; n,\cdot,\cdot),\TIME(\linproc; n,\cdot,\cdot,\cdot,\cdot)\leq n^\lambda\;,
\]
which means 
\[
\begin{split}
\TIME(\linproc^{(1)};n,\cdot,\cdot,\cdot,\cdot)&=\poly(n^\lambda)\;,\\
\TIME(\linproc^{(2)};n,\cdot,\cdot,\cdot,\cdot)&=\poly(2^{\lambda n })\;,\\
\TIME(\linproc^{(3)};n,\cdot,\cdot,\cdot,\cdot)&=\poly(2^{\lambda n })\;.
\end{split}
\]
Namely, given that $\verifier$ is $\lambda$-bounded, there is a constant $c=c(h)>0$ such that $\TIME(\linproc^{(3)};n,\cdot,\cdot,\cdot,\cdot)\leq c2^{c\lambda n} $. 

Let $\linproc'$ be the following $5$-input TM: Given that $(n,\mttx,\mtty,a^\frR,b^\frR)$ was its input, it runs $\linproc^{(3)}(n,\mttx,\mtty,a^\frR,b^\frR)$ for $c2^{c\lambda n}$ time steps. If it halted, $\linproc'$ outputs the same output as $\linproc^{(3)}$ did. Otherwise, it outputs $\frak{error}$. Note that when $\verifier$ is $\lambda$-bounded, $\linproc'$ and $\linproc^{(3)}$ always operate in the same way (they produce the same outputs). Furthermore, the running time of $\linproc^{(3)}$ is $\exp_h(\lambda,n)$, which was required.
\\

\textbf{Completeness,  Soundness and Entanglement lower bound}:
Here, we can assume that $\verifier$ is $\lambda$-bounded. Thus, by Claim \ref{claim:Padding_NFV}, $\verifier^{(1)}_n=\frak{Padding}(\verifier_n,n^\lambda)$. Then, by Claim \ref{claim:typed_question_reduction_on_NFV}, $\verifier^{(2)}_n=\frak{QueRed}(\verifier^{(1)}_{2^n},2^{\lambda n},\mathscr{B}(\lambda n))$. Finally, by Claim \ref{claim:DeTyping_NFV}, $\verifier^{(3)}_n=\frak{DeType}(\verifier^{(2)}_n)$. Now, as $\verifier$ is $\lambda$-bounded, $\verifier'=(\sampler_{\qr}^\lambda,\length_{\qr}^\lambda,\linproc',\decider)$ defines the same games as $\verifier^{(3)}$, which means that
\[
\verifier'_n=\frak{DeType}(\frak{QueRed}(\frak{Padding}(\verifier_{2^n},2^{\lambda n}),2^{\lambda n},\mathscr{B}(\lambda n)))\ .
\]
By Fact \ref{fact:padding_properties}, if $\verifier_{2^n}$ has a perfect $\ZPC$ strategy, then so does $\frak{Padding}(\verifier_{2^n},2^{\lambda n})$. 
Since $\verifier$ is $\lambda$-bounded, the length of questions in the $2^n$-th game $\verifier_{2^n}$ (as well as in the padded version) is at most $(2^n)^\lambda=2^{\lambda n}$, which is the length of vectors returned by $\mathscr{B}(\lambda n)$. Therefore, we may apply Theorem \ref{thm:complet_sound_quered} to deduce that
$$\frak{QueRed}(\frak{Padding}(\verifier_{2^n},2^{\lambda n}),2^{\lambda n},\mathscr{B}(\lambda n))$$ 
has a perfect $\ZPC$ strategy as well. Finally, by Corollary \ref{cor:comp_sound_combi_detyping}, the detyping of the above has a perfect $\ZPC$ strategy, but this is exactly $\verifier'_n$, proving the completeness requirements.
\\

In the other direction, assume $\val^*(\verifier'_n)\geq 1-\eps$. Then, as the typed graph of $\frak{QueRed}$, described in Example \ref{example:quered_has_typed_CL_sampling_scheme} and  in Figure \ref{fig:typed_graph_quered}, contains all self loops, using Corollary \ref{cor:completeness_soundness_detyping_when_all_loops_occur} we can deduce that the value of 
\[
\frak{QueRed}(\frak{Padding}(\verifier_{2^n},2^{\lambda n}),2^{\lambda n},\mathscr{B}(\lambda n))
\]
is at least $1-O((2h+28)\cdot 2^{4h+56}\cdot \sqrt{\eps})=1-O_h(\sqrt{\eps})$. Furthermore, for entanglement lower bound we have
\[
\Ent(\verifier'_n,1-\eps)\geq \Ent(\frak{QueRed}(\frak{Padding}(\verifier_{2^n},2^{\lambda n}),2^{\lambda n},\mathscr{B}(\lambda n)),1-O_h(\sqrt{\eps}))\;.
\]

By Theorem \ref{thm:complet_sound_quered}, if $\frak{QueRed}(\frak{Padding}(\verifier_{2^n},2^{\lambda n}),2^{\lambda n},\mathscr{B}(\lambda n))$ has value $1-O_h(\sqrt{\eps})$, then $\frak{Padding}(\verifier_{2^n},2^{\lambda n})$ has value $1-O_h(h^2\cdot 2^h\cdot(1+\nicefrac{2^{2\lambda n}}{d^2})\cdot \eps^{\nicefrac{1}{16}})=1-O_h(\eps^{\nicefrac{1}{16}})$, and $$1+\nicefrac{2^{2\lambda n}}{d^2}\leq 1+\nicefrac{|\mathscr{B}|^2}{d^2}\leq 1+\delta^{-2}=O(1)\; ,$$
as $d$ was the (un-normalized) distance of the error correcting code of dimension $2^{\lambda n}$ induced by $\mathscr{B}=\mathscr{B}(\lambda n)$, and we chose $\mathscr{B}$ such that its distance is at least $\delta|\mathscr{B}|$  for a universal constant $\delta>0$ (all of this was guaranteed by Fact \ref{fact:good_efficiently_calculable_codes}). Furthermore, for entanglement lower bounds we have 
\[
\Ent(\frak{QueRed}(\frak{Padding}(\verifier_{2^n},2^{\lambda n}),2^{\lambda n},\mathscr{B}(\lambda n)),1-O_h(\sqrt{\eps}))\geq \Ent(\frak{Padding}(\verifier_{2^n},2^{\lambda n}),1-O_h(\eps^{\nicefrac{1}{16}}))\;.
\]

Since $\verifier$ is $\lambda$-bounded, $2^{\lambda n}$ is an upper bound on $|\length(2^n,\cdot,\cdot)|$ which is $\max\{\ell^\frR,\ell^\frL\}$ in $\verifier_{2^n}$. Hence, By Fact \ref{fact:padding_properties},  $\verifier_{2^n}$ has value $1-O_h(\eps^{\nicefrac{1}{16}})$, and 
\[
\Ent(\frak{Padding}(\verifier_{2^n},2^{\lambda n}),1-O_h(\eps^{\nicefrac{1}{16}}))\geq \Ent(\verifier_{2^n},1-O_h(\eps^{\nicefrac{1}{16}}))\;.
\]
Combining all of the above, gives the required soundness and entanglement lower bounds. 

By choosing $c_\qr(h)$ \eqref{eq:defn_c_qr} to be large enough to bound all the constants along this proof, we conclude the theorem.
\qed

\newcommand{\tot}{\mathsf{tot}}
\newcommand{\mI}{\mathcal{I}}
\newcommand{\mJ}{\mathcal{J}}
\newcommand{\tspcp}{\hat{\sampler}^\pcp}

\section{Answer reduction using probabilistically checkable proofs}\label{sec:answer_reduction}

The goal of this section is to devise an algorithm $\AnsRed_{h,h'}$ that takes as input a tailored $h$-level normal form verifier, whose sampler is efficient but answer length calculator and linear constraints processors run in exponential time (with the constants in the bounds depending on $h'$),\footnote{Not surprisingly, these assumptions are satisfied by the output of the $\mathsf{QuestionReduction}_{h'}$ algorithm (Theorem \ref{thm:h_level_question_reduciton})  in the previous section.} and transforms it in a complete and sound way to a normal form verifier with all components running efficiently. Recall  the asymptotic notation from Remark \ref{rem:asymptotic_notation}.

\begin{theorem}[Answer Reduction, proved in Section \ref{sec:proof_main_thm_AR}]\label{thm:main_ans_red}
	Let $h$ and $h'$ be positive integers. There is a positive integer constant
	\begin{equation}\label{eq:defn_c_ar}
	c=c_\ar(h,h')
	\end{equation} 
	depending only on $h$ and $h'$, 
	and a $2$-input TM $\AnsRed_{h,h'}$, that takes as input a tailored $h$-level normal form verifier $\verifier=(\sampler,\length,\linproc,\decider)$, and a positive integer $\lambda$,    and outputs a tailored, \textbf{typed}  $\max(3,h)$-level  normal form verifier 
	\[
	\verifier_\ar=\AnsRed_{h,h'}(\verifier, \lambda)=(\sampler_\ar,\length_\ar,\linproc_\ar,\decider)
	\]
	with $9$ types and type graph depicted in Figure \ref{fig:ans_red_type_graph},  and the following properties:

	\begin{itemize}
		\item \underline{Sampler properties}:  $\sampler_{\ar}$ depends only on $\lambda,h,h'$ and the original sampler $\sampler$ (but not on $\length$ or $\linproc$), and  $\AnsRed_{h,h'}$ can calculate its description in time $\poly_{h,h'} (\log\lambda, |\sampler|)$ from them; in particular, $|S_\ar|\leq c(\log^c\lambda+|\sampler|^c)$. In addition, $\sampler_\ar$ runs in time  $\poly_{h,h'}(n,\lambda,\TIME(\sampler;n,\cdot,\cdot,\cdot,\cdot,\cdot))$, namely
		\[
		\forall n\in \mathbb{N}\ \colon \ \ \TIME(\sampler_\ar;n,\cdot,\cdot,\cdot,\cdot,\cdot)\leq c\cdot(n^c+\lambda^c+\TIME(\sampler;n,\cdot,\cdot,\cdot,\cdot,\cdot)^c)\ ,
		\]
		where $c$ is from \eqref{eq:defn_c_ar}.
		
		\item \underline{Answer length calculator properties}: $\length_\ar$  depends only on  $h,h'$ and $\lambda$,  and $\AnsRed_{h,h'}$ can calculate its description in time $\polylog_{h,h'} (\lambda)$; in particular $|\length_\ar|\leq c\log^c\lambda$. In addition, $\length_\ar$ runs in   $\poly_{h,h'}(n,\lambda)$-time, namely 
		\[
		\forall n\in \mathbb{N}\ \colon \ \ \TIME(\length_\ar;n,\cdot,\cdot)\leq {c\cdot (n^c+\lambda^c)}\ .
		\]
		Finally,  given that $\mttx\in \FF_2^{r(n)}$, where $r(n)=\sampler_\ar(n,{\rm Dimension},\cdot,\cdot,\cdot,\cdot)$,  and that $\kappa\in \{\frR,\frL\}$,  the output of $\length_\ar(n,\mttx,\kappa)$  never decodes (Definition \ref{defn:the_alphabet}) to an $\frak{error}$ sign.
		
		\item \underline{Linear constraints process properties}: $\linproc_\ar$ depends on all inputs, namely $h,h',\verifier$ and  $\lambda$. Also,  $\AnsRed_{h,h'}$ can calculate its description in time $\poly_{h,h'}(\log \lambda, |\verifier|)$; in particular, $|\linproc_\ar|\leq c\cdot (\log^c\lambda +|\verifier|^c)$. In addition, $\linproc_\ar$ runs in $\poly_{h,h'}(n,\lambda)$-time, namely 
		\[
		\forall {n}\in \mathbb{N}\  \colon \ \ \TIME(\linproc_\ar;{n},\cdot,\cdot,\cdot,\cdot) \ \leq\ 
		c\cdot(n^c+\lambda^c+\TIME(\sampler;n,\cdot,\cdot,\cdot,\cdot,\cdot)^c)\ .
		\]
		Note that although $\linproc_\ar$ depends on all of $\verifier$, its running time is bounded only in terms of the above parameters. i.e., it may not even read all the description of $\verifier$ in its operation (if it is too long).
		\item  \underline{Value properties}: Let $c_\qr=c_\qr(h')>0$ be the constant in  \eqref{eq:defn_c_qr}, guaranteed by Theorem \ref{thm:h_level_question_reduciton}. Given that 
		\[
		|\verifier|\leq 5\ c_\qr\cdot\lambda^{c_\qr}\ ,
		\]
		and
		\[
		\forall {n}\in \mathbb{N}\ \colon \ \ \TIME(\sampler\ ;\ {n},\cdot,\cdot,\cdot,\cdot,\cdot)\leq c_\qr(n^{c_\qr}+\lambda^{c_\qr})\quad;\quad  \TIME(\length\ ;\ {n},\cdot,\cdot)\ ,\ \TIME(\linproc\ ;\ {n},\cdot,\cdot,\cdot,\cdot) \leq 2^{c_\qr(n^{c_\qr}+\lambda^{c_\qr})}\ ,
		\]
		we have that
		$\verifier_\ar$, the output of $\AnsRed_{h,h'}$,  satisfies for all $n \geq 2$:
		\begin{enumerate}
			\item \textbf{Completeness}: If $\verifier_n$ has a perfect $Z$-aligned permutation strategy that commutes along edges ($\ZPC$ strategy), then so does $(\verifier_\ar)_n$.
			\item \textbf{Soundness}: If $(\verifier_\ar)_n$ has quantum  value $1-\eps$, then the value of $\frak{DoubleCover}(\verifier_n)$ is at least 
			\[
			1-c\cdot\left((n\lambda)^c \eps^{\nicefrac{1}{c}}+(n\lambda)^{-\nicefrac{1}{c}}\right)\;,
			\]
			where $c$ is again $c_\ar(h,h')$ from \eqref{eq:defn_c_ar}.
			\item \textbf{Entanglement bound}: For the same constant $c$, we have
			\[
			\Ent((\verifier_\ar)_n,1-\eps)\geq \Ent\Big(\frak{DoubleCover}(\verifier_n),1-c\cdot\left((n\lambda)^c \eps^{\nicefrac{1}{c}}+(n\lambda)^{-\nicefrac{1}{c}}\right)\Big)\;.
			\]
		\end{enumerate}
	\end{itemize}
\end{theorem}

The main idea underlying Answer Reduction is to use techniques from the field of probabilistically checkable proofs (PCPs). The way this is implemented requires several steps. First, some preprocessing on the  verifier needs to be done, and specifically it needs to be \emph{padded and purified}.  Then, the task of deciding whether the tuple $(a^\frR,a^\frL,b^\frR,b^\frL)$ passes the checks at a specific edge $\mttx\mtty$ or not is replaced by a succinct SAT instance and a succinct LIN instance, which have a PCP format that is both $\ZPC$-complete and sound. 

Let us elaborate more on this last step. After question reduction, the decision procedure in the $n^{\rm th}$ game defined by $\verifier$  takes exponential time in $n$, but the sampling mechanism and thus the length of questions are already polynomial in $n$. So, using standard techniques, in particular the Cook--Levin Theorem \ref{thm:scaled_up_cook_leving} and Reed--Muller encoding (Definition \ref{defn:restriction_and_induction_polynomials}), this decision problem can be replaced by a list of low degree polynomial equations that need to be satisfied. 
Once this is done, there are standard ways of verifying that two provers answer according to a list of low degree polynomials (Section \ref{sec:ld-decider}), called ``the low-degree test''. Thus, the aforementioned polynomial equations can be checked for a random point. By the Schwartz--Zippel Lemma \ref{lem:Schwartz-Zippel}, passing such a check with high probability implies the polynomial equations are indeed satisfied, which in turn means that the polynomials encode a tuple of answers that should be accepted by $\verifier_n$.

The section is structured as follows:
\begin{enumerate}
	\item  Section \ref{sec:prelude_decision_problems_PCPs} is a Prelude, which contains both basic definitions needed for Answer Reduction --- circuits, low-degree encoding and PCPs --- as well as sketching  the classical $\MIP=\NEXP$ result.
	\item  Section \ref{sec:prerequisites_ans_red} describes a few transformations that are used as part of Answer Reduction. These include purification, oracularization  triangulation and decoupling. Combinatorially, oracularization means applying a barycentric subdivision to the underlying graph of the game --- there is also a more compelling dramatized perspective of this transformation, on which we elaborate in Remark \ref{rem:dramatization_oracle_game}. Triangulation is the standard triangulation of systems of linear equations (see \eqref{eq:initial_triangulation}, \eqref{eq:ind_traingulation} and \eqref{eq:terminal_triangulation}). Purification is a simple transformation that removes all readable variables from the controlled linear constraints (Definition \ref{defn:combi_purification}). Decoupling is a standard way of ``block dividing'' a triangulated system of linear equations or a $3$CNF formula. 
	\item Section \ref{sec:succinct-linproc} trnaslates the condition ``$a^\frR,a^\frL,b^\frR,b^\frL$ are accepted by $\verifier_n$ given $\mttx\mtty$ were asked'' to a list of $13$ polynomial equations that can be checked probabilistically. This procedure uses various techniques from the field of probabilistically checkable proofs (PCPs), and specifically a decoupled version of the scaled up Cook--Levin transformation (Proposition \ref{prop:explicit-padded-succinct-deciders}).
	\item Section \ref{sec:ld-decider} recalls the quantum low degree test and its soundness properties \cite{quantum_soundness_tensor_codes}. This test verifies that two provers  that use a quantum strategy answer according to a list of low degree polynomials. This is important, as the $13$ polynomial equations recovered in Section \ref{sec:succinct-linproc} can be checked to be satisfied probabilistically only when the constituting polynomials that are checked have low degree.  
	\item Section \ref{sec:combi_and_algo_answer_red} provides a complete description of the Answer Reduction transformation, both the combinatorial one and the algorithmic one, yet under some assumptions on the input TNFV. In it, the transformation is showed to be $\ZPC$-complete and sound. 
	\item Finally, Section \ref{sec:proof_main_thm_AR} collects all of the above to prove Theorem \ref{thm:main_ans_red}.
\end{enumerate}

\subsection{Prelude --- Decision Problems, Complexity Classes, Low-Degree Polynomials, Circuits and PCPs}\label{sec:prelude_decision_problems_PCPs}
The goal of this prelude is to provide the required definitions and {sketch} the proof of the influential result  ${\MIP}={\NEXP}$ due to Babai--Fortnow--Lund \cite{BFL91}. Familiarity with these ideas is crucial for better understanding the answer reduction transformation, and thus we encourage any reader not familiar with this result, and the techniques used in it, to read this section.  

\subsubsection{Decision Problems and Complexity Classes}\label{sec:Descision_problems_complexity_classes}

A \emph{language}  $L$ is a subset of bit strings, namely $L\subseteq \{0,1\}^*$. Every language defines a \emph{decision problem}: Given a bit string $\mttx$, decide whether $\mttx\in L$ or not. A slight generalization of a language is a \emph{promise language}, which consists of two disjoint subsets $L_{yes},L_{no}\subseteq \{0,1\}^*$, and the decision problem in this case is: Given $\mttx\in L_{yes}\sqcup L_{no}$, decide whether $\mttx\in L_{yes}$ or $\mttx\in L_{no}$. Because promise languages include regular languages, from now on whenever we say language we mean a promise one.

The Halting Problem (HP) is the decision problem induced by the following subset of $\{0,1\}^*$: A bit string $\mttx$ is in ${\rm HALT}$ if and only if there is a $1$-input Turing machine (TM) $\cal{M}$ such that $\mttx$ is the encoding of $\cal{M}$, namely $\mttx=\desc{M}$ (see Definition \ref{def:Description_length}), and $\cal{M}$ halts on the empty input. Famously, Turing \cite{turing1937computable} proved that this problem is undecidable, namely that there is no TM  that takes $\mttx\in \{0,1\}^*$ as input, always halts, and outputs $1$ if $\mttx\in {\rm HALT}$ and $0$ if $\mttx\notin {\rm HALT}$.

We say that there is a \emph{reduction} from language $L^1$ to language $L^2$, if there is an always halting $1$-input Turing machine $\cal{M}$, such that $\mttx\in L^1_{yes}$ implies $\cal{M}(\mttx)\in L^2_{yes}$, and $\mttx\in L^1_{no}$ implies $\cal{M}(\mttx)\in L^2_{no}$.\footnote{Recall that a $k$-input TM always defines a partial function from $(\{0,1\}^*)^k$ to $\{0,1\}^*$, and if the TM always halts this is a proper function. Thus, we abuse notation and write $\cal{M}(\mttx)$ for the output of $\cal{M}$ given the input $\mttx$.} Such a reduction is $f$-time, for a function $f\colon \mathbb{N}\to \mathbb{N}$, if for every instance $\mttx$ the running time of the reduction $\cal{M}$ satisfies $\TIME(\cal{M};\mttx)\leq f(|\mttx|)$.
A \emph{class} of problems is a set $\mathscr{C}$ of languages, namely $\mathscr{C}\subseteq \{0,1\}^{\{0,1\}^*}$. A language $L$ is said to be \emph{complete} for $\mathscr{C}$ if $L\in \mathscr{C}$ and there is a reduction from every $L'\in \mathscr{C}$ to $L$. A language is $f$-time complete if the reduction is always $f$-time. Moreover, we say that $L$ is polynomial-time complete for $\mathscr{C}$ if for every $L'\in \mathscr{C}$, there is a polynomial $f$ and a $f$-time reduction from $L'$ to $L$. 

\begin{definition}[$\RE$]\label{defn:RE}
	We say that a (promise) language $L$ is in the class $\RE$ if there is a $2$-input TM $\verifier$ such that:
	\begin{itemize}
		\item (Halting condition) $\verifier$ always halts and outputs a single bit;
		\item (Completeness) if $\mttx\in L_{yes}$, then there is a $\pi\in \{0,1\}^*$ such that $\cal{V}(\mttx,\pi)=1$;
		\item (Soundness) if $\mttx\in L_{no}$, then for every $\pi\in \{0,1\}^*$ we have $\cal{V}(\mttx,\pi)=0$.
	\end{itemize}
\end{definition}
\begin{remark}
	It is straightforward to check that ${\rm HALT}$ is in $\RE$, and that ${\rm HALT}$ is  complete for $\RE$.    
\end{remark}
\begin{definition}[$\NP$ and $\NEXP$]\label{defn:NP_and_NEXP}
	Let $f\colon \mathbb{N}\to \mathbb{N}$ be a function. The language $L$ is in the class ${\NTIME}(f(n))$ (non-deterministic $f$-time) if there is a $2$-input TM $\cal{V}$ such that: 
	\begin{itemize}
		\item (Time bound) $\TIME(\cal{V};\mttx,\pi)\leq f(|\mttx|)$, namely $\cal{V}$ runs in $f$-time in its first input;
		\item (Completeness) if $\mttx\in L_{yes}$, then there is a $\pi\in \{0,1\}^*$ such that $\cal{V}(\mttx,\pi)=1$;
		\item (Soundness) if $\mttx\in L_{no}$, then for every $\pi\in \{0,1\}^*$ we have $\cal{V}(\mttx,\pi)=0$.
	\end{itemize}
	The class ${\NP}$ (non-deterministic polynomial time) is $\bigcup_{C\in \mathbb{N}}{\NTIME}(Cn^C)$ and the class ${\NEXP}$ (non-deterministic exponential time) is $\bigcup_{C\in \mathbb{N}}{\NTIME}(2^{Cn^C})$.
\end{definition}
\begin{remark}[Dramatization of ${\NP}$]\label{rem:dramatization_NP}
	There is a resource restricted (in this case, polynomial time) entity, called the \emph{verifier} and which is denoted by $\cal{V}$, that wants to decide whether a bit string $\mttx$ is in the language of interest $L$. It asks an all knowing \emph{prover} $\cal{P}$ to provide a written proof that indeed $\mttx\in L$. The prover $\cal{P}$ generates (in a single time step for $\cal{V}$) such a proof $\pi\in \{0,1\}^*$, and sends it to $\verifier$. The verifier then reads the proof (unless it is too long, in which case it reads only part of it), and  decides (under its time restrictions) whether to accept (i.e., declare ``$\mttx$ is in $L$'') or to reject (i.e., declare ``$\mttx$ is not in $L$''). The language $L$ is in ${\NP}$ if  such a verifier will be convinced by \textbf{some} proof $\pi$ given that $\mttx\in L$, and will never be convinced by \textbf{any} proof $\pi$ given that $\mttx\notin L$.
\end{remark}

\begin{definition}[$\MIP$, $\MIP^*$, $\TMIP$ and $\TMIP^*$]\label{defn:MIP_MIP*_etc}
	Let $f\colon \mathbb{N}\to \mathbb{N}$ be a function. A language $L$ is in the class $\mathsf{MIPTIME}(f(n),2,1)$ (multi-prover interactive proofs with an $f$-time verifier, $2$-provers and $1$-round) if there is a (tailored) normal form verifier $\verifier$ (Definition \ref{def:normal-ver}), such that:
	\begin{itemize}
		\item (Time bound) the TMs $\sampler,\length$ and $\linproc$ run in $f$-time in their first input, namely 
		\[
		\forall {n}\in \mathbb{N}\ \colon \ \ \TIME(\sampler;\overline{n}),\TIME(\length;\overline{n},\cdot,\cdot),\TIME(\linproc;\overline{n},\cdot,\cdot,\cdot,\cdot)\leq f(|\overline{n}|)\approx f(\log n).
		\]
		\item (Completeness) if $\overline{n}\in L_{yes}$, then $\verifier_n$ has a value $1$ \textbf{classical} strategy (Example \ref{example:classical_perm_strategies});
		\item (Soundness) if $\overline{n}\in L_{no}$, then every classical strategy for $\verifier_n$ has value of at most $\nicefrac{1}{2}$.
	\end{itemize}
	The class ${\MIPSTIME}(f(n),2,1)$ is defined almost the same, but with the classical strategies in the completeness and soundness conditions being replaced by \textbf{quantum} strategies (Definition \ref{defn:quantum_strategy}). Furthermore, the class ${\TMIPSTIME}(f(n),2,1)$ is defined  the same as ${\MIPSTIME}(f(n),2,1)$, but with the extra condition in the completeness case that the perfect strategy needs to be a $\ZPC$ one. Finally, 
	\[
	{\MIPP}=\bigcup_{C\in \mathbb{N}}{\MIPTIME}(Cn^C,2,1)\quad,\quad {\MIPEXP}=\bigcup_{C\in \mathbb{N}}{\MIPTIME}(2^{Cn^C},2,1)\ ,
	\]
	and similarly one defines ${\MIPSP}, {\MIPSEXP}$, ${\TMIPSP}$ and ${\TMIPSEXP}$. When we write ${\MIP^*}$ or ${\TMIP^*}$ we  mean the polynomial time versions.    
\end{definition}

\begin{remark}[Dramatization of ${\MIP^*}$]\label{rem:dramatization_MIP*}
	There is a time bounded entity, the \emph{verifier}  denoted by $\cal{V}$, that wants to decide whether a bit string $\mttx$ is in the language of interest $L$. It devises a game $\game_\mttx$ (Definition \ref{defn:non-local_game}), and describes it to \textbf{two}  provers, $A$ and $B$. 
	It then plays one round of this game against the two  provers --- as was described in Remark \ref{rem:dramatization_non_local_game}, where the verifier is the referee and the provers are the players. If the provers win the round, then  $\verifier$ accepts (i.e., declares ``$\mttx$ is in $L$''), and if they lose the round, then  $\verifier$ rejects (i.e., declares ``$\mttx$ is not in $L$'').
	
	The language $L$ is in ${\MIP^*}$ (resp. ${\TMIP^*}$), if  given that $\mttx\in L$ the provers can win $\game_\mttx$ with probability $1$ using a quantum strategy  (resp. a $\ZPC$-strategy), and  given that $\mttx\notin L$, the players  lose with probability  at least $\nicefrac{1}{2}$, regardless of the quantum strategy they chose.
\end{remark}

\begin{remark}
	Now it should be clear why Theorem \ref{thm:tailored_MIP*=RE} is called ${\TMIP^*}={\RE}$. By choosing the appropriate encoding of games $\game_{\cal{M}}$ using normal form verifiers (Definition \ref{def:normal-ver}), the theorem states that there is a normal form verifier $\verifier$, such that 
	\[
	\TIME(\sampler;\overline{n}),\TIME(\length;\overline{n},\cdot,\cdot),\TIME(\linproc;\overline{n},\cdot,\cdot,\cdot)=\poly(|\overline{n}|)=\polylog(n)\;,
	\]
	and if  $\overline{n}=\desc{M}$ for a TM $\cal{M}$, then: $\cal{M}$ halting means $\verifier_n$ has a perfect $\ZPC$ strategy; $\cal{M}$ not halting means $\val^*(\verifier_n)\leq \nicefrac{1}{2}.$ This specific normal form verifier $\verifier$ can be extracted from the proof in Section \ref{sec:proving_Tailored-MIP*=RE}: Given $\desc{M}$, calculate $\lambda=\lambda(\cal{M})$ (Lemma \ref{lem:lambda}), and then play the game $\verifier^{\cal{M},\lambda}_C$ for the universal constant $C$ guaranteed by Theorems \ref{thm:compression} and  \ref{thm:h_level_compression}. This exactly shows that ${\rm HALT}$ is in ${\TMIP^*}$, and as ${\rm HALT}$ is complete for ${\RE}$, this proves $\RE\subseteq\TMIP^*$. The reverse inclusion was described in the beginning of Section \ref{sec:proving_Tailored-MIP*=RE}.
\end{remark}

\begin{remark}
	Let $L$ be a language. Note that if one finds a $\lambda$-bounded tailored normal form verifier $\verifier$ such that $\verifier_{n}$ has a value $1$ strategy if $\overline{n}\in L$, and otherwise $\verifier_{n}$ has value bounded from above by $\nicefrac{1}{2}$, then $L$ is only in ${\MIPSEXP}$ and not ${\MIP^*}$. This is because $\verifier$ runs in time $n^\lambda=2^{\lambda \log n}$, which is exponential in the input length $|\ol{n}|\approx \log n$.
\end{remark}

\subsubsection{The Cook--Levin theorem}
Recall that this prelude focuses on proving  $\MIP=\NEXP$ (Definitions \ref{defn:NP_and_NEXP} and \ref{defn:MIP_MIP*_etc}).  The containment $\MIP\subseteq \NEXP$ is quite straightforward, and we leave it to the reader.
For the other direction, we need the scaled up version of the  celebrated Cook--Levin theorem \cite{Cook71,levin1973universal}. In this section we describe the content, and sketch the proof, of the \textbf{standard} Cook--Levin theorem. Later, in Section \ref{sec:circuits_and_S3SAT}, we describe the scaled up version of this theorem and the adjustments needed to prove it (see Theorem \ref{thm:scaled_up_cook_leving}).
To that end, we first observe that there is a natural complete decision problem for $\NP$ (and respectively $\NEXP$):
\begin{definition}[Time Restricted Halting]\label{defn:time_restricted_halting}
	The decision problem $\UnaryTimeHalt$ (respectively, $\BinaryTimeHalt$) is the following:    Given a pair consisting of (the encoding of) a single\footnote{The $k$-input instance version is similar, and we actually use later the $3$-input version to implement answer reduction (see Remark \ref{rem:blarembla}).} input TM $\cal{M}$ and a positive integer $T$ in unary (respectively binary, see Definition \ref{def:unary_encoding_int}), decide whether there exists an input $\pi\in \{0,1\}^*$ such that $\cal{M}(\pi)$ halts and outputs $1$ in less than $T$ time steps.
\end{definition}

\begin{claim}[Time restricted halting is complete for non-deterministic time]\label{claim:timehalt_is_complete}
	The decision problem $\UnaryTimeHalt$ is polynomial time complete for $\NP$, and similarly $\BinaryTimeHalt$ is polynomial time complete for $\NEXP$.
\end{claim}

\begin{proof}
	We leave it for the reader to check that, indeed $\UnaryTimeHalt$ is in $\NP$ and $\BinaryTimeHalt$ is in $\NEXP$.  We also omit the reduction from every $\NEXP$ language to $\BinaryTimeHalt$, as it is virtually identical to the reduction from every $\NP$ language to $\UnaryTimeHalt$ (up to the encoding of the integer), which we now present. 
	
	By Definition \ref{defn:NP_and_NEXP}, a language is in $\NP$ if there is a constant $C$ and a $2$-input TM $\verifier$ that runs in time $\TIME(\verifier;\mttx,\pi)\leq C |\mttx|^{C}$ such that $\mttx\in L_{yes}$ implies the existence of a $\pi\in \{0,1\}^*$ for which $\verifier(\mttx,\pi)=1$, and $\mttx\in L_{no}$ implies that for every $\pi\in \{0,1\}^*$ we have $\verifier(\mttx,\pi)=0$. So, for every $\mttx\in L_{yes}\cup L_{no}$ we can define the single input TM $\cal{M}_\mttx=\verifier(\mttx,\cdot)$ and an integer (in unary) $T_\mttx=1^{*C|\mttx|^C}=\underbrace{1...1}_{C|\mttx|^C-\textrm{times}}$. 
	Hence, $\mttx\in L_{yes}$ exactly implies that $(\cal{M}_\mttx,T_\mttx)$ is in $\UnaryTimeHalt$, and $\mttx\in L_{no}$ implies  $(\cal{M}_\mttx,T_\mttx)$ is not in $\UnaryTimeHalt$. As, given $\verifier$,  calculating $(\cal{M}_\mttx,T_\mttx)$  takes $\poly(|\mttx|)$-time, the proof is finished.
\end{proof}

Though it is nice to have some complete language to the complexity class of interest, time restricted halting is not a very useful one. So, it is natural to  seek some other language in $\NP$ (respectively $\NEXP$) to which $\UnaryTimeHalt$ (respectively $\BinaryTimeHalt$) can be reduced to (in polynomial time).

\begin{definition}[CNF] \label{defn:CNF}

	A \emph{literal} is either a formal variable $\sX$ or its negation $\lnot \sX$. For $\eps\in \FF_2$, we  use the notation $$\sX^\eps=\begin{cases}
	\sX & \eps=0\ ,\\
	\lnot \sX  & \eps=1\ .
	\end{cases}$$
	A \emph{disjunction} is an OR of smaller formulas, namely $\bigvee_{i=1}^m \varphi_i$, and a \emph{conjunction} is an AND of smaller formulas, namely $\bigwedge_{i=1}^m \varphi_i$. A ${\rm CNF}$ formula is a conjunction of disjunctions of literals, namely if $\{\sX_1,...,\sX_n\}$ is the set of formal variables, then there is an integer $m$, integers $k_1,...,k_m$, and functions $i_j\colon [k_j]\to [n]$ and $\eps_j\colon [k_j]\to \FF_2$ for every $1\leq j\leq m$, such that the formula is of the form 
	\[
	\varphi(\sX_1,...,\sX_n)=\bigwedge_{j=1}^m\bigvee_{t=1}^{k_j} \sX_{i_j(t)}^{\eps_j(t)}\ .
	\]
	It is called a $k$-${\rm CNF}$, if all the $k_j$'s in the above formula are equal to the same integer $k$. A formula $\varphi(\sX_1,...,\sX_n)$ is \emph{satisfiable} if there is an \emph{assignment} $\psi\colon \{\sX_1,...,\sX_n\}\to \{{\rm True},{\rm  False}\}$ such that $\varphi(\psi(\sX_1),...,\psi(\sX_n))={\rm True}$.
	
	A ${\rm CNF}$ formula can be encoded in many ways. E.g., one can provide the list of integers $n,m,k_1,...,k_m$ and then the evaluation table of the functions $i_j$ and $\eps_j$.  A $k$-${\rm CNF}$ formula has even nicer encodings, e.g., by providing a matrix of size $m\times k$ with entries being pairs of a bit $\eps$ and an integer between $1$ and $n$. In any case, after fixing such an encoding scheme for $3$-${\rm CNF}$ formulas, the language $3{\rm SAT}$ is the following: A bit string $\mttx$ is in $3{\rm SAT}$ if and only if it encodes a satisfiable $3$-${\rm CNF}$ formula.  We describe Succinct-$3{\rm SAT}$ later in this section.
	
\end{definition}

\begin{theorem}[Cook--Levin \cite{Cook71,levin1973universal}. See also \cite{Karp1972}]\label{thm:Cook-Levin}
	The language $3{\rm SAT}$ is  polynomial time ${\NP}$-complete.
\end{theorem}
\begin{proof}[Proof sketch]
	By Claim \ref{claim:timehalt_is_complete}, it is enough to show that one can reduce $\UnaryTimeHalt$ to $3{\rm SAT}$ in polynomial time. Namely, given a single input TM $\cal{M}$ and an integer in unary $T$, translate them to a $3$-${\rm CNF}$ formula such that this formula is satisfiable if and  only if there is an input that will make $\cal{M}$ output $1$ in time at most $T$. 
	
	Recall how a TM operates: It has $k$ infinite tapes ($3$ in the case of a single input TM --- one input tape, one memory tape, and one output tape), whose cells are parametrized by $\mathbb{Z}$, and each cell contains either a bit or is empty (which we think of as containing the special symbol $\sqcup$). Each tape has a head, positioned initially at cell number $0$. It has a finite list $Q$ of internal states, two of them are the initial state $q_{initial}$ and the halting state $q_{halt}$; the TM  is always initialized to be in $q_{initial}$, and if it arrives at $q_{halt}$ it stops its operation. Finally, there is an instruction table, that tells the machine given the reads from its heads and its current non-halt state, which new values to write in the current position of the heads, which way should each head move (either not move, one step up or one step down), and what should be its new internal state; namely, the instruction table is a mapping from $\{0,1,\sqcup\}^k \times \left(Q\setminus\{q_{halt}\}\right)$ to $\{0,1,\sqcup\}^k \times \{-1,0,1\}^k\times Q$.
	
	Since we care about the operation of the TM $\cal{M}$ for only $T$ steps, the only cells of the tapes it may visit are in the interval $-T$ to $+T$. So, we can have the following finitely many variables that ``remember'' everything about the operation of $\cal{M}$:
	\begin{itemize}
		\item   For each index $i$ between $-T$ and $T$, each index $j$ between $1$ and $k$, and  each time $t$ between $0$ and $T$, there should be a variable that contains the content of the $i^{\rm th}$ cell in the $j^{\rm th}$ tape at time step $t$ of the operation of $\cal{M}$. As the content of a cell may be either $0,1$ or $\sqcup$, and in the end these variables should be part of a boolean formula, we need  {two variables} to encode this information; the combination of values of the two boolean variables will be interpreted as $\{0,1,\sqcup\}$ according to the encoding map from Definition \ref{defn:the_alphabet}.
		
		\item In addition, for the same range for $i,j,t$, there should be a variable whose boolean value answers the question ``is the $j^{\rm th}$ head in position $i$ at time $t$?''.
		\item Finally, for each $t$ in the above range, there should be a variable that indicates the state of $\cal{M}$ at time $t$. Again, as there are $|Q|$ states, we need to choose some encoding of them as length $\lceil\log|Q|\rceil$ bit strings, and then there are $\lceil\log|Q|\rceil$ many boolean variables whose combinations of values  encode the state of $\cal{M}$ at each time step. 
	\end{itemize}
	Then, we need to describe the clauses of the $3$-${\rm CNF}$ formula that uses the above variables. For example, there will be clauses that check that, initially, at time $t=0$, the machine was in state $q_{initial}$, and that all the heads were in position $0$ (and not in any other position, as there is a single head at each tape), and that all but the input tape cells are empty and so on. Then, there will be clauses that check that the variables of time $t$ were well deduced from those of the previous time $t-1$ according to the instructions table. For example, the head moved at most $1$ step from its previous position, and it moved correctly; the new value at each position is indeed what the instructions table says it should be; the new state of the machine is what it should be, and so on. Finally, the variables associated to the output tape at time $T$ need to encode the output $1$, and similarly the variables associated to the state of the machine should be the encoding of $q_{halt}$.
	
	Note that the only ``free variables'' in the formula are those associated with the content of the input tape(s) at time $0$; the content of all the other variables (in a satisfying assignment) is either fixed or can be deduced from the content of the free variables. Hence, a satisfying assignment is essentially an appropriate choice for the free variables, such that indeed the machine halted in less than $T$ steps and its  output  is $1$, which is exactly what we sought after. 
\end{proof}

\subsubsection{Low-degree polynomials and robust tests for them} \label{sec:ldt}
Polynomials play a major role in the construction of probabilistically checkable proofs, both as the form of encoding of the proof and as a tool to verify the proof's validity. Although we have not motivated the ``why'' yet, let us provide some definitions and facts regarding polynomials of low degree.

\begin{definition}[Polynomials and their degrees]\label{def:poly-1}\label{def:polynomial_degree}
	A polynomial  with $n$ variables $\vec\sX=(\sX_1,...,\sX_n)$ and coefficients in the field $\FF$ is a formal sum $$f=\sum_{\vec e\in (\mathbb{Z}_{\geq 0})^n}c_{\vec e} \cdot\sX_1^{e_1}\cdot...\cdot \sX_n^{e_n}\ ,$$
	where each $c_{\vec e}\in \FF$, and all but finitely many $c_{\vec e}$'s are $0$. The \emph{total degree} of $f$ is 
	\[
	\max_{\vec e\colon c_{\vec e}\neq 0}(e_1+...+e_n)\;,
	\]
	while its \emph{individual degree} is \[\max_{\vec e\colon c_{\vec e}\neq 0}\max_{1\leq i\leq n}(e_i)\;.\]
	For a given $i\in[m]$, the $\sX_i$-degree of $f$ is 
	\[
	\max_{\vec e\colon c_{\vec e}\neq 0} e_i\;;
	\]
	$f$ is said to be \emph{indifferent} to the $i^{\rm th}$ input if the $\sX_i$-degree of it is $0$.
	We often emphasize the variable set of the polynomial by denoting $f(\vec \sX)$ or $f(\sX_1,...,\sX_n)$ instead of just $f$.
\end{definition}

\begin{remark}
	Note that if $f\in \FF[\sX_1,...,\sX_m]$ has total degree at most $d$, then it has individual degree at most $d$. On the other hand, if $f$ has individual degree at most $d$, then its total degree is at most $md$.
\end{remark}

It is natural to associate a function with each polynomial via assignments. Namely, if $f\colon \FF[\sX_1,...,\sX_m]$ is a polynomial, it induces a function $\Phi_{\FF}(f)\colon \FF^m\to \FF$ that  takes as input $(x_1,...,x_m)\in \FF^m$ and outputs 
\begin{equation}\label{def:Phi_FF}
\Phi_\FF(f)(x_1,...,x_m)=\sum_{\vec e}c_{\vec e}x_1^{e_1}\cdot...\cdot x_m^{e_m}\in \FF\ ,
\end{equation} 
i.e., uses the assignment $\sX_i\mapsto x_i$. As this transformation is so entrenched in mathematics and computer science, we usually think of $f$ itself as a function and use the same notation for it and for $\Phi_\FF(f)$. In this paper we mostly do the same, but for the following discussion, we distinguish between the two. 

Both $\FF[\sX_1,...,\sX_m]$ and $\FF^{\FF^m}$ are vector spaces over $\FF$, and $\Phi_\FF$ is a linear map between them. In case $\FF$ is an infinite field, $\Phi_\FF$ is an injection (and not a surjection), and in case $\FF$ is a finite field, it is a surjection (and not an injection). Let us elaborate more on the finite field case.  Let $q=p^t$ be a prime power, and $\FF=\FF_q$ be the field with $q$ elements, on which we shall focus. 
Let $(\FF_q)_{\leq d}[\sX_1,...,\sX_m]$ be the collection of individual degree at most $d$ polynomials with $m$ variables $\sX_1,...,\sX_m$ and coefficients in $\FF_q$ (Definition \ref{def:polynomial_degree}). A standard basis for these polynomials is the set of monomials $\mathscr{M}_d=\{\sX_1^{\alpha_1}\cdot...\cdot\sX_m^{\alpha_m}\mid 0\leq \alpha_i\leq d\}$. On the other hand, the functions $\FF_q^{\FF_q^m}$ also have a natural basis of indicators $\mathscr{I}$, namely for every $\vec x\in \FF_q^m$ the indicator ${\bf 1}_{\vec x}\colon \FF_q^m\to \FF_q$ defined by 
\[
{\bf 1}_{\vec x}(\vec y)=\begin{cases}
1 & \vec y=\vec x\ ,\\
0 & \vec y\neq \vec x\ .
\end{cases}
\]
Every such indicator can be written as (the $\Phi_{\FF_q}$-image of) an individual degree at most $q-1$ polynomial using the (multi-variate) Lagrange polynomial 
\[
{\bf 1}_{\vec x}(\sX_1,...,\sX_m)=(-1)^m\cdot \prod_{i=1}^m\prod_{x_i\neq a\in \FF_q}\left(\sX_i-a\right)\ .
\]
When restricted to individual degree at most $q-1$ polynomials, the function $\Phi_{\FF_q}$ is a bijection --- it is a basis change on the polynomials, moving from the basis $\mathscr{M}$ to the basis $\mathscr{I}$. This basis change is an instance of a Fourier transform (and there is a very efficient algorithm that calculates it, called the \emph{fast Fourier transform} \cite{cooley1965algorithm,gauss1886theoria}).

\begin{definition}[Subcubes in $\FF_q^m$] \label{defn:subcubes}
	Let $q=p^t$ be a prime power, and let $A\subseteq\FF_q$ be a subset. We call the set $A^m\subseteq \FF_q^m$ a \emph{subcube}. If $A=\FF_p\subseteq \FF_q$, then we often call $\FF_p^m$ \emph{the subcube}, without referring to a specific $A$.
\end{definition}
\begin{definition}\label{defn:restriction_and_induction_polynomials}
	Let $m,t$ be positive integers, $p$ a prime number, and $q=p^t$. Given a function $f\colon \FF_q^m\to \FF_q$ we denote by $\Res(f)\colon \FF_p^m\to \FF_q$ the restriction of $f$ to the subcube $\FF_p^m\subseteq \FF_q^m$, namely $f|_{\FF_p^m}$. Given $g\colon \FF_p^m\to \FF_q$, let the induction of $g$, $\Ind(g)\colon \FF_q^m\to \FF_q$, be the individual degree at most $p-1$ interpolation  of $g$. Namely, $\Ind(g)$ is the unique $m$-variate individual degree at most  $p-1$ polynomial with coefficients in $\FF_q$ that agrees with $g$.\footnote{Both restriction and induction depend on $m,p$ and $q$. Yet, these parameters should be understood from context and are not included in the notation.}
\end{definition}

\begin{remark}\label{rem:induction_is_linear}
	First, note that $\Ind$ is a $\FF_q$-linear map.
	In addition, when $g\colon \FF_p^m\to \FF_p$, namely it outputs only elements in the base field $\FF_p$ and not its extension $\FF_q$, $\Ind$ is the composition of the following
	\[\FF_p^{\FF_p^m}\xrightarrow{\Phi^{-1}_{\FF_p}}(\FF_p)_{\leq p-1}[\sX_1,...,\sX_m]\subseteq (\FF_q)_{\leq p-1}[\sX_1,...,\sX_m]\xrightarrow{\Phi_{\FF_q}}\FF_q^{\FF_q^m}\ .\]
	Furthermore, it can be derived from this perspective that  the coefficients of $\Ind(g)$ (as a polynomial, namely $\Phi_{\FF_q}^{-1}(\Ind(g))$) are in $\FF_p$.
	
	Moreover, note that $\Res\circ\Ind=\Id$, and that $\Ind\circ\Res$ is the identity on $m$-variable individual degree $p-1$ polynomials over $\FF_q$. Given $f\colon \FF_p^m\to \FF_p$, $\Ind(f)$ is usually called the (individual degree $p-1$) \emph{Reed--Muller encoding} of $f$. This is an error correcting code (as was defined in Section \ref{subsec:error_correctoin_prelims}) over $\FF_q$ that  has good distance (yet not that good of a rate). 
\end{remark}
Though it is standard (and proving it is not a hard exercise), we include the Schwartz--Zippel Lemma, which proves that the individual degree at most $d$ Reed--Muller codes have good distance, as long as the individual degree $d$ and number of variables $m$ are sufficiently small compared to $q$:
\begin{lemma}[Schwartz--Zippel \cite{schwartz1980fast,zippel1979probabilistic}]\label{lem:Schwartz-Zippel}
	Let $f$ be a non-zero \textbf{total} degree at most $d$ polynomial over $m$ variables with coefficients in $\FF_q$. Then, the probability a uniformly random $u\in \FF_q$ is a zero of $f$ is bounded from above by $\frac{d}{q}$.
\end{lemma}

The goal of low degree tests is to verify that a given function is (the $\Phi_{\FF_q}$-image of) a low degree polynomial (in our context, an individual low degree polynomial, but the total degree case is also very useful and usually has better soundness parameters).
The basic idea is the following: If a function $f\colon \FF_q^n\to \FF_q$ is an individual degree  at most $d$ polynomial, then restricting it to an axis parallel line of $\FF_q^n$ (namely, fixing all the coordinates of the function except one of them) will result in a degree (at most) $d$ univariate polynomial. It turns out that this property works in the other direction as well, and in a somewhat robust manner. Namely, given a function $f\colon \FF_q^n\to \FF_q$, if its restriction to all of its axis parallel lines is of degree at most $d$, then $f$ is a polynomial of individual degree at most $d$. Let us define all of this formally.

\begin{definition}[Lines]\label{def:lines_in_finite_VS}
	A line $\mathscr{L}$ in $\FF_q^m$ is a $1$-dimensional affine subspace. Namely, there are $u\in \FF_q^m$ and $\vec 0\neq v\in \FF_q^m$ that induce a parametrization affine map $\alpha\mapsto u+\alpha v$ from $\FF_q$ to $\FF_q^m$, whose image is $\mathscr{L}$; this means $\mathscr{L}=\{u+\alpha v\mid \alpha\in \FF_q\}$.  Note that there are $q(q-1)$ many pairs $u,v$ that give rise to the same line $\mathscr{L}$, and thus $q(q-1)$ many parametrizations of the same line.  We denote by $\mathscr{L}(u,v)$ the \emph{parametrized} line $\{u+\alpha v\mid \alpha\in \FF_q\}$, and say that it is in direction $v$. An \emph{$i$-axis parallel line} is one whose direction is $e_i$ (or a scalar multiple thereof), and an axis parallel line is an $i$-axis parallel line for some $i\in [m]$.\footnote{We later discuss a (somewhat) canonical way of representing each line in $\FF_q^m$. See Definition \ref{def:line-representative}.}
\end{definition}

We record the following well-known fact, sometimes referred to as a \emph{local characterization} of low individual degree $m$-variate polynomials~\cite{rubinfeld1996robust}.

\begin{fact}[Characterizations of low degree polynomials]\label{fact:characterization_of_low_degreeness_by_restrictions_to_linew}
	The restriction of $f\in \FF_q[\sX_1,...,\sX_m]$ to the parametrized line $\mathscr{L}(u,v)$ is a polynomial $f|_{\mathscr{L}(u,v)}\in \FF_q[\alpha]$ which is derived from $f$ by the assignment $\sX_i\mapsto u_i+\alpha v_i$.
	\begin{enumerate}
		\item $f$ is of total degree at most $d$ if and only if $f|_{\mathscr{L}(u,v)}$ has degree at most $d$ for every line $\mathscr{L}(u,v)$.
		\item $f$ is of individual degree at most $d$ if and only if $f|_{\mathscr{L}(u,e_i)}$ has degree at most $d$ for every \textbf{axis parallel} line $\mathscr{L}(u,e_i)$.
		\item Fix $i\in [m]$. Then, $f$ has $\sX_i$-degree at most $d$ if and only if $f|_{\mathscr{L}(u,e_i)}$ for every $i$-axis parallel line $\mathscr{L}(u,e_i)$.
	\end{enumerate}
\end{fact}

\begin{definition}[The low individual degree test]\label{defn:classical_low_deg_test_prelude}
	Let $d,t$ and $m$ be  positive integers, and $q=2^t$.        Let $f$ and $\mathsf{AL}f$ (acronym for ``axis parallel lines of $f$'') be two functions. The input to $f$ is a single $u\in \FF_q^m$, and it outputs an element of $\FF_q$. The input to $\mathsf{AL}f$ is a pair, consisting of $\hat{u}\in \FF_q^{m-1}$ and $i\in [m]$, and it outputs a tuple $(c_0,c_1,...,c_d)$ of $d+1$ elements from $\FF_q$.
	Given $u\in \FF_q^m$, let $\hat{u}^i$ be the $(m-1)$-tuple that results from removing the $i^{\rm th}$ entry of $u$. 
	The individual degree $d$ test on $f$ and $\mathsf{AL}f$ runs as follows: Sample a pair $(u,i)$ uniformly at random, where $u$ is a point in $\FF_q^m$ and $i$ is an axis direction, namely $i\in [m]$. Evaluate $f(u)$ and $\mathsf{AL}f(\hat{u}^i,i)=(c_0,...,c_d)$. Accept if
	\begin{equation}
	f(u)= \sum_{j=0}^d c_j(u_i)^j,
	\end{equation}
	and reject otherwise.
\end{definition}
Note that by Fact \ref{fact:characterization_of_low_degreeness_by_restrictions_to_linew}, individual degree $d$ polyonomials and their restrictions to axis parallel lines pass the above test with certainty. The following is a quantitative reverse statement, which is a ``robust'' version of Fact \ref{fact:characterization_of_low_degreeness_by_restrictions_to_linew}:
\begin{theorem}[Classical soundness of the low individual degree test. Babai--Fortnow--Lund \cite{BFL91}\footnote{See also \cite{polishchuk1994nearly} and the introduction of \cite{ji2020quantum}.}]\label{thm:classical_soundness_individual_low_degree}
	Let $m,d,t,q,f$ and $\mathsf{AL}f$ be as in Definition  \ref{defn:classical_low_deg_test_prelude}, and let $\eps\geq 0$.
	Assume  the probability that $f,\mathsf{AL}f$ pass the individual degree $d$ test is at least $1-\eps$. Namely
	\[
	\Pro{u\in \FF_q^m,i\in [m]}\Big[ f(u)\neq \sum_{j=0}^d c_j(u_i)^j  \Big]\leq \eps\;,
	\]
	where $c_0,...,c_d\in \FF_q$ are the outputs of $\mathsf{AL}f(\hat{u}^i,i)$. Then, there exists an individual degree at most $d$ polynomial $F\colon \FF_q^n\to \FF_q$ and a universal constant $C>0$ such that 
	\[
	\Pro{u\in \FF_q^m}\left[  f(u)\neq F(u) \right]\leq Cm^C\Big(\eps^{\nicefrac{1}{C}}+\Big(\frac{d}{q}\Big)^{\nicefrac{1}{C}}\Big)\;.
	\]
\end{theorem}

Since we care about the time complexity of operations used in our protocols, and Turing machines manipulate bit strings, whenever we deal with a finite field we need to be able to do the arithmetic operations on it efficiently. The following fact guarantees this is possible (in the relevant case for us).
\begin{fact}[See Section 3.3 in \cite{MIPRE}]\label{fact:basis_F_q_over_F_2}
	For every odd positive integer $t$, there is a $\poly(t)$-time algorithm that chooses a  basis of $\FF_q$ over $\FF_2$, where $q=2^t$, such that  the arithmetic operations (products, inverses, sums)  and taking traces  all take $\poly(t)$-time when the elements of $\FF_q$ are represented according to this basis (namely, as elements of $\FF_2^t$).
\end{fact}

One last ``still to be motivated'' definition  is needed at this point:
\begin{definition}[Zero on a subcube and Assignments]\label{def:zero_on_subcube_and_assignments}
	Let $\FF$ be a finite field, and $A\subseteq \FF$ a subset. A function $f\colon \FF^m\to \FF$ is said to be \emph{zero on the subcube} $A^m$, if for every $u\in A^m$ the function evaluates to zero, namely $f(u)=0$.
	
	In case $\FF=\FF_q$ where $q=2^t$, and $A=\FF_2$, we often use the term $f$ is zero on the subcube, without extra information.
	Under the same assumptions, a function $f\colon \FF_q^m\to \FF_q$ is said to be an \emph{assignment} if $f(\FF_2^m) \subset \FF_2$. 
	
\end{definition}
\begin{claim}[Assignment condition]\label{claim:assignment_condition}
	Let $q=2^t$. A function $f\colon \FF_q^m\to \FF_q$ is an assignment, if and only if $f\cdot(f+1)$ is zero on the subcube.
\end{claim}
\begin{proof}
	This is immediate from the fact that the only zeros in $\FF_q$ of the polynomial $\sX(\sX+1)$ are the elements of $\FF_2$.
\end{proof}
\begin{claim}[Combinatorial Nullstellensatz]\label{claim:combi_null}\label{rem:helper_poly_are_of_low_degree}
	Let $q=2^t$.    The polynomial $f\colon \FF_q^m\to \FF_q$ is zero on the subcube if and only if there are polynomials  $c_i\colon \FF_q^m\to \FF_q$, often called \emph{the helper polynomials}, such that 
	\[
	f(\sX_1,...,\sX_m)=\sum_{i=1}^m c_i(\sX_1,...,\sX_m)\cdot(\sX_i+1)\sX_i.
	\]
	If $\vec \sX$ is a tuple of variables indexed by a set $I$, and $i\in I$, then we denote  
	\begin{equation}\label{eq:defn_of_zero_i}
	\zero_i(\vec \sX)=(\sX_i+1)\sX_i\;.
	\end{equation}
	In addition, both the coefficients and the evaluation table of the helper polynomials $c_i$ are linear in the coefficients (or evaluation table) of $f$, and the individual degree of the helper polynomials is  smaller or equal to that of $f$. 
\end{claim}
\begin{proof}

	This is a  simple division of polynomials argument \cite[Proposition 10.21]{MIPRE}, which is a special case of \emph{Combinatorial Nullstellensatz} \cite[Theorem 1.1]{alon1999combinatorial}.
	
	Recall that $q=2^t$.  Fixing a univariate polynomial $f\in \FF_q[\sX_i]$, every multivariate polynomial $g\in \FF_q[\sX_1,...\sX_m]$ can be written in a unique way as  $g=mf+r$, where the $\sX_i$-individual degree of $r$ is strictly smaller than the degree of $f$. The maps ${\rm Div}_f(g)$ --- which outputs the quotient polynomial $m$ --- and ${\rm Mod}_f(g)$ --- which outputs the remainder polynomial $r$ --- are both $\FF_q$-linear, and the individual degree of ${\rm Div}_f(g)$ is at most that of $g$.  
	In particular, for every polynomial  $f\colon \FF_q^m\to \FF_q$, if we let 
	\[
	\forall i\in [m]\ \colon \ \ c_i(\sX_1,...,\sX_m)={\rm Div}_{\sX_i(\sX_i+1)}\circ {\rm Mod}_{\sX_{i-1}(\sX_{i-1}+1)}\circ ...\circ{\rm Mod}_{\sX_1(\sX_1+1)}(f)
	\]
	and 
	\[
	c_0(\sX_1,...,\sX_m)= {\rm Mod}_{\sX_{m}(\sX_{m}+1)}\circ ...\circ {\rm Mod}_{\sX_1(\sX_1+1)}(f)\ ,
	\]
	then $(i)$ the coefficients of each $c_i$ are linear combinations of the coefficients of $f$, $(ii)$ for $i>0$ the individual degree of each $c_i$ is at most that of $f$, $(iii)$ the total degree of $c_0(\sX_1,...,\sX_m)$ is at most $1$, and $(iv)$ we have
	\begin{equation}\label{eq:combi_null_is_linear}
	\begin{split}
	f(\sX_1,...,\sX_m)&=c_0(\sX_1,...,\sX_m)+\sum_{i=1}^m \sX_i(\sX_i+1)\cdot c_i(\sX_1,...,\sX_m)\\
	&=c_0(\sX_1,...,\sX_m)+\sum_{i=1}^m \zero_i(\sX_1,...,\sX_m)\cdot c_i(\sX_1,...,\sX_m)\ . 
	\end{split}
	\end{equation}
	Note that for every $\vec x\in \FF_2^m$, $\zero_i(\vec x)=x_i(x_i+1)=0$, as $x_i\in \FF_2$. Hence, 
	\[
	\forall \vec x\in \FF_2^m\ \colon \ \ f(\vec x)=c_0(\vec x)\ .
	\]
	So, if $f$ is zero on the subcube, we deduce that $c_0(\vec x)=0$ for every $\vec x\in \FF_2^n$. As its total degree is at most $1$, this implies $c_0$ is the zero polynomial, and this in this case 
	\begin{equation}
	\begin{split}
	f(\sX_1,...,\sX_m)=\sum_{i=1}^m \sX_i(\sX_i+1)\cdot c_i(\sX_1,...,\sX_m)\ . 
	\end{split}
	\end{equation}
	as required.
\end{proof}

\subsubsection{Circuits and Succinct-$\mathbf{3}$SAT}\label{sec:circuits_and_S3SAT}

The Turing machine is a \emph{uniform} computational model, because a given Turing machine can in principle accept inputs of any given length. In contrast, circuits are a \emph{non-uniform} model: a given circuit has a fixed number of input wires, which determine a unique input length that the circuit accepts. For this reason, in complexity one usually considers families of circuits $(\circuit_n)_{n\geq 1}$ indexed by a growing input length $n$; the model is called non-uniform because without any further restrictions, each circuit in the family can be quite different from any other (e.g.\ we do not necessarily require that $n\mapsto \circuit_n$ is an efficiently computable mapping).

\begin{definition}\label{defn:Circuits}
	A (binary, Boolean) \emph{circuit} $\cal{C}$ is a finite, vertex labeled, directed and acyclic graph,\footnote{A directed graph is acyclic if there is no directed path that starts and ends at the same vertex.} whose label set is 
	\begin{equation}\label{def:set_of_logic_gates}
	\{\lnot,\oplus,\land, {\rm Input},{\rm Output},{\rm Copy},{\rm True}, {\rm False}\}.
	\end{equation}
	In addition, the label of a vertex determines its in-degree (i.e., number of edges oriented into it) and out-degree (i.e., number of edges oriented out of it) according to the table below. See Figure \ref{fig:circuit_diagram} for visualization.
	\begin{center}
		\begin{tabular}{|l|l|l|}
			\hline
			Label & in-degree & out-degree    \\
			\hline
			{\rm Input} & 0 & 1   \\
			\hline
			{\rm True} & 0 & 1 \\
			\hline
			{\rm False} &0 & 1\\
			\hline
			{\rm Output} & 1& 0\\
			\hline
			{\rm Copy} &1 & 2\\
			\hline
			$\lnot$ & 1 & 1 \\
			\hline
			$\land$ & 2 & 1 \\
			\hline
			$\oplus$ & 2 & 1 \\
			\hline
		\end{tabular}
	\end{center}
	The vertices of a circuit are often called \emph{gates} and its edges are often called \emph{wires}. Such a circuit is called \emph{linear} if it has no vertices labeled by $\land$, i.e., no AND gates are used in it.
	
	For later discussions, note that a circuit $\circuit$ can be encoded as bit string $\ol{n}\sqcup c_1\sqcup...\sqcup c_n$, where $\ol{n}$ is interpreted as the binary encoding of an integer $n$, which represents the number of gates in $\circuit$, and each $c_i$ is (the encoding of) a tuple $({\rm Type}, {\rm inwire}\ 1, {\rm inwire}\ 2, {\rm outwire}\ 1, {\rm outwire}\ 2)$, where ${\rm Type}$ is one of the possible logic gates from \eqref{def:set_of_logic_gates}, the two in-wires are integers between $1$ and $n$ that indicate the origins of the two wires that are fed into the gate (if there are less then $2$ wires feeding into the gate, then the extra ones are ignored), and the two out-wires are integers between $1$ and $n$ that indicate the endpoints of the two wires stemming out of the gate (again, if there are less than $2$ wires that stem  out of the gate, the extra ones are ignored). This provides an encoding of $\circuit$ of size $O(n\log n)$, where $n$ is the number of gates (vertices) in the circuit.\\
	
	Let $I$ (resp. $O$) be the set of vertices in $\circuit$ labeled by ${\rm Input}$ (resp. ${\rm Output}$). Then $\circuit$ encodes a function  $P_\circuit\colon \FF_2^I\to \FF_2^O$: Given $\iota\colon I\to \FF_2$, write in each vertex $x\in I$ the value $\iota(x)$. Moreover, write in each True vertex $1$ and in each False vertex $0$. Then, repeat the following --- for every vertex that contains a value, write down this value on all its outgoing edges; if there is a vertex all of whose in-going edges have values written on them, act as follows:
	\begin{itemize}
		\item if the vertex is labeled by ${\rm Output}$, then it has one in-going edge; write in it the value appearing on this single edge;
		\item if the vertex is labeled by  ${\rm Copy}$, then it has one in-going edge; write in it the value appearing on this single edge;
		\item if the vertex is labeled by  $\lnot$, then it has one in-going edge; write in it the value appearing on this single edge plus $1$ (in $\FF_2$, namely, flip the bit);
		\item if the vertex is labeled by $\land$ then it has two in-going edges; write in it the product of the values on these two edges;
		\item if the vertex is labeled by $\oplus$ then it has two in-going edges; write in it the sum (in  $\FF_2$) of the values on these two edges;
	\end{itemize}
	To summarize, if a vertex has a single in-going edge $e$ or two in-going edges $e_1,e_2$ then we assign it a value according to the following table:
	\begin{center}
		\begin{tabular}{|l|l|}
			\hline
			Label & value at the vertex    \\
			\hline
			{\rm Output} & {\rm value}$(e)$\\
			\hline
			{\rm Copy} &value$(e)$\\
			\hline
			$\lnot$ &value$(e)+1$ \\
			\hline
			$\land$ & $\textrm{value}(e_1)\cdot \textrm{value}(e_2)$ \\
			\hline
			$\oplus$ & $\textrm{value}(e_1)+\textrm{value}(e_2)$ \\
			\hline
		\end{tabular}
	\end{center}
	
	If we write in each input vertex $x\in I$ a formal variable $\sX_x$, then each output vertex $y\in O$ will contain some polynomial (with $\FF_2$-coefficients) $P_{\circuit,y}$ in these variables  --- or equivalently, some Boolean formula in them. Note that a linear circuit induces total degree one polynomials, namely $P_\circuit\colon\FF_2^I\to\FF_2^O$ is an affine map in this case.\\
	
	There is another way of associating a polynomial $T_{\circuit,y}$ with each output vertex $y\in O$ of the circuit $\circuit$, which is called the Tseitin polynomial (or Tseitin formula): Associate a formal variable $\sY_{e}$ to each edge $e$ in $\circuit$. For each vertex $z$ 
	with a directed path to $y$\footnote{This property is commonly phrased as ``$y$ is reachable from $z$''.},
	denote the in-going edges of $z$ by $e$ or $e_1,e_2$ and its out-going edges by $f$ or $f_1,f_2$. Then we define a polynomial $t_z$ according to the following table.
	\begin{center}
		\begin{tabular}{|l|l|}
			\hline
			Label of $z$&     \\
			\hline
			{\rm Input} & $t_z=1$   \\
			\hline
			{\rm True} &  $t_z=\sY_f$ \\
			\hline
			{\rm False} & $t_z=\sY_f+1$\\
			\hline
			{\rm Output} & $t_z=\sY_e$ \\
			\hline
			{\rm Copy} &$t_z=(\sY_e+\sY_{f_1}+1)(\sY_e+\sY_{f_2}+1)$ \\
			\hline
			$\lnot$ & $t_z=\sY_{e}+\sY_f$ \\
			\hline
			$\land$ & $t_z=\sY_{e_1}\sY_{e_2}+\sY_{f}+1$ \\
			\hline
			$\oplus$ & $t_z=\sY_{e_1}+\sY_{e_2}+\sY_{f}+1$ \\
			\hline
		\end{tabular}
	\end{center}
	Finally, let $T_{\circuit,y}=\prod_z t_z$, where the product runs over all $z$ with a directed path to $y$. 
	
	\begin{figure}[httb!]
		\centering
		\captionsetup{singlelinecheck=off}
		\begin{tikzpicture}[scale=1.2]
		\node[draw, color=black, shape=circle] (Input1) at (-4,0) {\scriptsize Input $1$}; 
		\node[draw, color=black, shape=circle] (Input2) at (0,0) { \scriptsize Input $2$}; 
		\node[draw, color=black, shape=circle] (Input3) at (4,0) {\scriptsize Input $3$}; 
		
		\node[draw, color=black, shape=circle] (Copy1) at (-4,-2) {\scriptsize Copy}; 
		
		\node[draw, color=black, shape=circle] (Copy3) at (4,-2) { \scriptsize Copy};
		
		\node[draw, color=black, shape=circle] (Not1) at (-5,-4) {$\lnot$}; 
		\node[draw, color=black, shape=circle] (Not2) at (5,-4) {$\lnot$}; 
		
		\node[draw, color=black, shape=circle] (And1) at (-2,-4) {$\land$}; 
		
		\node[draw, color=black, shape=circle] (XOR1) at (2,-4) {$\oplus$};

		\node[draw, color=black, shape=circle] (And2) at (2,-6) {$\land$}; 
		
		\node[draw, color=black, shape=circle] (XOR2) at (-2,-6) {$\oplus$}; 
		\node[draw, color=black, shape=circle] (XOR3) at (0,-8) {$\oplus$}; 
		\node[draw, color=black, shape=circle] (Output) at (0,-10) {\scriptsize Output}; 
		

		
		
		
		
		
		
		

		\draw[black,  ->, solid] (Input1)--(Copy1) node[midway, right] {$e_1$};
		\draw[black,  ->, solid] (Input3)--(Copy3) node[midway, right] {$e_4$};
		
		\draw[black,  ->, solid] (Copy1)--(Not1) node[midway, right] {$e_5$};
		\draw[black,  ->, solid] (Copy1)--(And1) node[midway, right] {$e_6$};
		\draw[black,  ->, solid] (Input2)--(And1) node[midway, right] {$e_2$};
		\draw[black,  ->, solid] (Input2)--(XOR1) node[midway, right] {$e_3$};
		\draw[black,  ->, solid] (Copy3)--(XOR1) node[midway, right] {$e_7$};
		\draw[black,  ->, solid] (Copy3)--(Not2) node[midway, right] {$e_8$};
		
		\draw[black,  ->, solid] (Not2)--(And2) node[midway, right] {$e_{12}$};
		\draw[black,  ->, solid] (XOR1)--(XOR2) node[pos=0.25, right] {$e_{11}$};
		
		\draw[black,  ->, solid] (And1)--(And2) node[pos=0.25, right] {$e_{10}$};
		\draw[black,  ->, solid] (Not1)--(XOR2) node[midway, right] {$e_9$};
		\draw[black,  ->, solid] (And2)--(XOR3) node[midway, right] {$e_{14}$};
		\draw[black,  ->, solid] (XOR2)--(XOR3) node[midway, right] {$e_{13}$};
		\draw[black,  ->, solid] (XOR3)--(Output) node[midway, right] {$e_{15}$};

		
		\end{tikzpicture}
		\caption[]{An example of a circuit $\circuit$ with three input gates, a single output gate, and $15$ wires. If we denote by $\sX_i$ the formal variable associated with ${\rm Input}_i$, then the polynomial the circuit induces at the single output vertex is 
			\begin{align*}
			P_{\circuit}(\sX_1,\sX_2,\sX_3)=1+\sX_1+\sX_2+\sX_3+\sX_1\sX_2+\sX_1\sX_2\sX_3\ .
			\end{align*}  On the other hand, if we let $\sY_i$ be the formal variable associated with the edge $e_i$, then the Tseitin polynomial of this circuit is
			\begin{align*}
			T_{\circuit}(\sY_1,...,\sY_{15})=&\left(\sY_1+\sY_5+1\right)\left(\sY_1+\sY_6+1\right)\left(\sY_4+\sY_7+1\right)\left(\sY_4+\sY_8+1\right)\\
			&\left(\sY_5+\sY_9\right)\left(\sY_2\sY_6+\sY_{10}+1\right)\left(\sY_3+\sY_7+\sY_{11}+1\right)\left(\sY_8+\sY_{12}\right)\\
			&\left(\sY_9+\sY_{11}+\sY_{13}+1\right)\left(\sY_{10}\sY_{12}+\sY_{14}+1\right)\left(\sY_{13}+\sY_{14}+\sY_{15}+1\right)\sY_{15}\ .
			\end{align*}
			Although the Tseitin polynomial $T_\circuit$ is ``more complicated'', it is guaranteed to have individual degree at most $3$, while the circuit polynomial $P_\circuit$ may have arbitrarily large individual degree.}
		\label{fig:circuit_diagram}
	\end{figure}
	
\end{definition}

\begin{remark}\label{rem:individual_degree_Tseitin_is_3}
	Note that the individual degree of the Tseitin polynomial is at most $3$, namely there is no monomial and a variable $\sY_e$ whose exponent in this monomial is larger than $3$. This is because when $e=xy$, the variable $\sY_e$ appears only in $t_x$ and $t_y$, and its exponent in $t_x$ is at most $1$ while its exponent in $t_y$ is at most $2$.
	
	As long as we consider functions from $\FF_2^n$ to $\FF_2$, this is not a useful property, as any such function can be written as a polynomial of individual degree at most $1$. But, we are going to view these polynomials over some finite field extension of $\FF_2$, namely as functions from $\FF_q^n$ to $\FF_q$ for some $q=2^t$, in which case this property of having low individual degree will become very handy.
\end{remark}
\begin{claim}\label{claim:prop_of_Tseitin}
	Let $\circuit$ be a circuit (Definition \ref{defn:Circuits}), $y$ an output vertex in $\circuit$, $P_{\circuit,y}$ the polynomial associated with $y$ (over formal variables $\{\sX_x\}_{x\in I})$, and $T_{\circuit,y}$ the Tseitin polynomial associated with $y$ (over formal variables $\{\sY_e\}_e$, where $e$ is running over all edges in $\circuit$). Then:
	\begin{itemize}
		\item (Completeness) There is an assignment to the $\sY_e$ variables as polynomials in the $\sX_x$ variables such that $T_{\circuit,y}(\{\sY_e\}_e)=P_{\circuit,y}(\{\sX_x\}_x)$.
		\item (Soundness) If $P_{\circuit,y}$ induces the constant $0$ function, then so does $T_{\circuit,y}$.
	\end{itemize}
\end{claim}
\begin{proof}
	For the completeness requirement, assign to the $\{\sY_e\}$ variables the following values inductively: For every $e$ whose initial vertex $x$ is labeled by ${\rm Input}$, let $\sY_e=\sX_x$. For $e$ whose initial vertex is ${\rm False}$, let $\sY_e=0$, and for $e$ whose initial vertex is ${\rm True}$ let $\sY_e=1$ . Then, if $e$ is an edge whose initial vertex is $x$, and all of the edges whose terminal vertex is $x$ were already assigned values, then assign to it ---  the input edge value if $x$ is labeled ${\rm Copy}$; the input edge value plus $1$ if $x$ is labeled ${\lnot}$; the product of the input edges' values  if $x$ is labeled ${\land}$; the sum of the input edges' values  if $x$ is labeled ${\oplus}$. It is straightforward to check that indeed, under this assignment, $t_z(\sY_e)=1$ for every non ${\rm Output}$ vertex $z$, and $t_y(\sY_e)=P_{\circuit,y}(\sX_x)$. Thus $$T_{\circuit,y}(\sY_e)=\prod t_z(\sY_e)=t_y(\sY_e)=P_{\circuit,y}(\sX_x).$$ 
	
	For the soundness requirement, note that for $T_{\circuit,y}(\sY_e)$ to be $1$, all $t_z(\sY_e)$ need to be $1$. But, if all the $t_z$'s for which $z$ is not labeled by ${\rm Output}$ evaluate to  $1$, then the $\sY_e$'s were assigned the values as in the complete case, which means $t_{y}(\sY_e)=P_{\circuit,y}(\sX_x)$, where $\sX_x=\sY_{xz}$ for every $x\in I$ and $xz$ its outgoing edge. But, by assumption, the output of $P_{\circuit,y}$ is always $0$,  which implies that $T_{\circuit,y}$ must evaluate to $0$ as well.
\end{proof}

\begin{definition}[Succinct encodings of $3{\rm CNF}$ formulas]\label{defn:Succinct_3SAT}
	A circuit $\circuit$ with $3n+3$ input gates and a single output is said to succinctly encode a $3{\rm CNF}$ formula $\varphi_\circuit$ on $2^n$ variables $\{\sX_u\}_{u\in \FF_2^n}$ if the following holds: $\varphi_\circuit$ is the conjunction of all formulas $\sX_{u_1}^{\eps_1}\lor \sX_{u_2}^{\eps_2}\lor \sX_{u_3}^{\eps_3}$ for which $P_{\circuit}(u_1,u_2,u_3,\eps_1,\eps_2,\eps_3)=1$.
	
	The language Succinct-$3{\rm SAT}$ is the one containing all (encodings, as described in Definition \ref{defn:Circuits}, of) circuits $\circuit$ such that $\varphi_{\circuit}$ is satisfiable.
\end{definition}

\begin{theorem}[Scaled up Cook--Levin. Compare to Theorem \ref{thm:Cook-Levin}. See also Section 10.2 in \cite{MIPRE}]\label{thm:scaled_up_cook_leving}
	The language Succinct-$3{\rm SAT}$ is $\NEXP$-complete.
\end{theorem}

\begin{proof}[Proof ideas]
	Recall again that $\BinaryTimeHalt$ is $\NEXP$-complete due to Claim \ref{claim:timehalt_is_complete}, so it is enough to reduce it to Succinct-$3{\rm SAT}$. Namely, given a TM $\cal{M}$ and integer $T$ \textbf{in binary}, translate it in polynomial time to a circuit $\circuit$, such that if  $\cal{M}$ has an input $\pi\in \{0,1\}^*$ for which it halts and outputs $1$ in $T$ steps, then the $3$CNF formula $\varphi_\circuit$ is satisfiable, and otherwise $\varphi_\circuit$ is unsatisfiable. 
	
	The idea is the same as the proof of Theorem \ref{thm:Cook-Levin}. Namely, having formal variables that collect the tape contents, the head position, and the machine's internal state at each time step. Then, to add restrictions on the variables at time $0$ so that they represent $\cal{M}$ in time $0$ (with some input), and restrictions that check that the next time step variables were calculated correctly from the previous time according to the instructions table of $\cal{M}$. The important thing is that these restrictions and variables are so well structured, that one can encode them succinctly using a circuit of size $\poly(|\cal{M}|,\log T)$.
\end{proof}

\begin{observation}\label{obs:sufficient_condition_S3SAT}
	Let $\circuit$ be a circuit which succinctly encodes a formula $\varphi_\circuit$ on $2^n$ variables. An assignment to $2^n$ variables can be thought of as a function $\psi\colon \FF_2^n\to \FF_2$ by letting $\psi(u)=1$ if $\sX_u$ was assigned ${\rm True}$ and $\psi(u)=0$ if it was assigned ${\rm False}$. Now, $\sX_{u_1}^{\eps_1}\lor \sX_{u_2}^{\eps_2}\lor \sX_{u_3}^{\eps_3}={\rm True}$ if and only if $(\psi(u_1)+\eps_1+1)(\psi(u_2)+\eps_2+1)(\psi(u_1)+\eps_3+1)=0$. Therefore, $\varphi_{\circuit}$ is satisfiable if and only if there is a function $\psi\colon \FF_2^n\to \FF_2$ such that for all 
	\[
	(u_1,u_2,u_3,\eps_1,\eps_2,\eps_3)\in \FF_2^n\times \FF_2^n\times \FF_2^n\times \FF_2\times \FF_2\times \FF_2\;,
	\]
	we have
	\[
	P_{\circuit}(u_1,u_2,u_3,\eps_1,\eps_2,\eps_3)(\psi(u_1)+\eps_1+1)(\psi(u_2)+\eps_2+1)(\psi(u_3)+\eps_3+1)= 0\;.
	\]
	Furthermore, 
	by Claim \ref{claim:prop_of_Tseitin}, if $s$ is the number of non-input wires (edges that do not stem from an ${\rm Input}$-labeled vertex) in $\circuit$, then $\psi$ satisfies the above condition if and only if for every 
	\[
	u=(u_1,u_2,u_3,\eps_1,\eps_2,\eps_3,z)\in \FF_2^n\times \FF_2^n\times \FF_2^n\times \FF_2\times \FF_2\times \FF_2\times \FF_2^{s}\;,
	\]
	we have 
	\begin{equation}\label{eq:Tseitin_condition_in_remark_on_satisfiability_Prelude}
	T_{\circuit}(u)(\psi(u_1)+\eps_1+1)(\psi(u_2)+\eps_2+1)(\psi(u_3)+\eps_3+1)=0\;.    
	\end{equation}
	Actually, by embedding this setup into a larger field of characteristic $2$ via induction (Definition \ref{defn:restriction_and_induction_polynomials}), we are able to get the following.
\end{observation}
\begin{proposition}\label{prop:prelude_on_satisfiability}
	Let $q=2^t$ for a positive integer $t$. Let $\circuit$ be a circuit that succinctly encodes a $3{\rm CNF}$ formula $\varphi_\circuit$ (Definition \ref{defn:Succinct_3SAT}), in particular, it has $3n+3$ input gates, a single output gate, and we denote by $s$ the number of non-input wires in it.  Then:
	\begin{itemize}
		\item (Completeness) If $\varphi_\circuit$ is satisfiable, then there is a sequence  of $4n+4+s$ individual degree at most $6$ polynomials \[
		\begin{split}
		g_\psi&\colon \FF_q^n\to \FF_q\ ,\\
		\forall 1\leq i\leq 3n+3+s\ \colon \ \ \alpha_i&\colon \FF_q^{3n+3+s}\to \FF_q\ ,\\ 
		\forall 1\leq i\leq n\ \colon \ \ \beta_i&\colon \FF_q^n\to \FF_q\ ,
		\end{split} 
		\]
		such that for every $u=(u_1,u_2,u_3,\eps_1,\eps_2,\eps_3,z)\in \FF_q^{3n+3+s}$ we have
		\begin{equation}\label{eq:condcondcond1}
		T_\circuit(u)\prod_{j=1}^3(g_\psi(u_j)+\eps_j+1)=\sum_{i=1}^{3n+3+s}\alpha_i(u)\zero_i(u)\;,    
		\end{equation}
		and for every $u_0\in \FF_q^n$ we have 
		\begin{equation}\label{eq:condcondcond2}
		g_\psi(u_0)(g_\psi(u_0)+1)=\sum_{i=1}^n\beta_i(u_0)\zero_i(u_0),    
		\end{equation}
		where $\zero_i(\vec \sX)=\sX_i(\sX_i+1)$ (as was defined in \eqref{eq:defn_of_zero_i}).
		\item (Soundness) If there is  a sequence of $4n+4+s$ individual degree at most $6$ polynomials \[
		\begin{split}
		g_\psi&\colon \FF_q^n\to \FF_q\ ,\\ 
		\forall 1\leq i\leq 3n+3+s\ \colon \ \ \alpha_i&\colon \FF_q^{3n+3+s}\to \FF_q\ ,\\ 
		\forall 1\leq i\leq n\ \colon \ \ \beta_i&\colon \FF_q^n\to \FF_q\ ,
		\end{split} 
		\]
		such that 
		\begin{equation}\label{eq:blablabla1}
		\Pro{\substack{u\in \FF_q^{3n+3+s}\\u=(u_1,u_2,u_3,\eps_1,\eps_2,\eps_3,z)}}\Big[T_\circuit(u)\prod_{j=1}^3(g_\psi(u_j)+\eps_j+1)=\sum_{i=1}^{3n+3+s}\alpha_i(u)\zero_i(u)\Big]> \frac{21(3n+3+s)}{q}\;,
		\end{equation}
		and 
		\begin{equation}\label{eq:blablabla2}
		\Pro{u_0\in \FF_q^n}\Big[g_\psi(u_0)(g_\psi(u_0)+1)=\sum_{i=1}^{n}\beta_i(u_0)\zero_i(u_0)\Big]> \frac{21n}{q}\;,
		\end{equation}
		then $\varphi_\circuit$ is satisfiable.
	\end{itemize}
\end{proposition}
\begin{proof}
	If $\varphi_\circuit$ is satisfiable, then by Observation \ref{obs:sufficient_condition_S3SAT}, there is a function $\psi\colon \FF_2^n\to \FF_2$ that satisfies \eqref{eq:Tseitin_condition_in_remark_on_satisfiability_Prelude}. Let $g_\psi=\Ind(\psi)\colon \FF_q^n\to \FF_q$, which has individual degree at most $1$, and let $\Psi\colon \FF_q^{3n+3+s}\to \FF_q$ be 
	\[
	\Psi(u)=T_\circuit(u)\prod_{j=1}^3(g_\psi(u_j)+\eps_j+1),
	\]
	which has individual degree at most $6$ --- as $T_\circuit$ has individual degree at most $3$ (Remark \ref{rem:individual_degree_Tseitin_is_3}).
	Since $\psi$ satisfies \eqref{eq:Tseitin_condition_in_remark_on_satisfiability_Prelude}, $\Psi$ is zero on the subcube $\FF_2^{3n+3+s}$ (Definition \ref{def:zero_on_subcube_and_assignments}), and by the Combinatorial Nullstellensatz (Claim \ref{claim:combi_null}), there are $3n+3+s$ individual degree at most $6$ helper polynomials $\alpha_i\colon \FF_q^{3n+3+s}\to \FF_q$ such that
	\[
	\forall u=(u_1,u_2,u_3,\eps_1,\eps_2,\eps_3,z)\in \FF_q^{3n+3+s}\ \colon \ \ T_\circuit(u)\prod_{j=1}^3(g_\psi(u_j)+\eps_j+1)=\sum_{i=1}^{3n+3+s}\alpha_i(u)\zero_i(u)\;.
	\]
	As $\Img(\psi)\subseteq\FF_2$, and $g_\psi=\Ind(\psi)$, $g_\psi$ is an assignment (Definition \ref{def:zero_on_subcube_and_assignments}). Therefore, by Claim \ref{claim:assignment_condition}, the polynomial $g_\psi(g_\psi+1)$ is zero on the subcube $\FF_2^n$ (and has individual degree at most $2$). Thus, again by the Combinatorial Nullstellensatz, there are $n$ individual degree at most $2$ polynomials $\beta_i\colon \FF_q^n\to \FF_q$ such that
	\[
	\forall u_0\in \FF_q^n\ \colon \ \ g_\psi(u_0)(g_\psi(u_0)+1)=\sum_{i=1}^n\beta_i(u_0)\zero_i(u_0)\;,
	\]
	which finishes the proof of completeness.\\
	
	On the other hand, assume there are $g_\psi,\alpha_i$ and $\beta_i$ of individual degree at most $6$ that satisfy \eqref{eq:blablabla1} and \eqref{eq:blablabla2}. By the assumptions on the individual degree of $g_\psi,\alpha_i,\beta_i$ and the fact that $T_\circuit$ has individual degree at most $3$, the polynomials
	\[
	\heartsuit(u)=T_\circuit(u)\prod_{j=1}^3(g_\psi(u_j)+\eps_j+1)-\sum_{i=1}^{3n+3+s}\alpha_i(u)\zero_i(u)
	\]
	and
	\[
	\clubsuit(u_0)=g_\psi(u_0)(g_\psi(u_0)+1)-\sum_{i=1}^n\beta_i(u_0)\zero_i(u_0)
	\]
	have individual degree at most $21$. Hence, $\heartsuit$ has total degree at most $21(3n+3+s)$ and $\clubsuit$ has total degree at most $21n$. Equation  \eqref{eq:blablabla1} says that a uniformly random vector in $\FF_q^{3n+3+s}$ is a zero of  $\heartsuit$ with probability greater than $\frac{21(3n+3+s)}{q}$, which combined with the Schwartz--Zippel Lemma \ref{lem:Schwartz-Zippel} implies that $\heartsuit$ is the zero function. The same argument, using \eqref{eq:blablabla2}, implies $\clubsuit$ is the zero function. The fact $\clubsuit$ is the zero function implies that $g_\psi$ is an assignment, and we recover a potential boolean assignment $\psi=\Res(g_\psi)\colon \FF_2^n\to \FF_2$ to $\varphi_\circuit$. The fact $\heartsuit$ is the zero function implies that the $\psi$ recovered from $g_\psi$ satisfies \eqref{eq:Tseitin_condition_in_remark_on_satisfiability_Prelude}, and by Observation \ref{obs:sufficient_condition_S3SAT}, $\varphi_\circuit$ is  satisfied by $\psi$. In particular $\varphi_\circuit$ is satisfiable in this case.
\end{proof}

\subsubsection{Probabilistically checkable proofs}\label{sec:prelude_PCP}
Recall that a (promise) language $L$ is in ${\NP}$ (resp. ${\NEXP}$), if for every $\mttx\in L_{yes}$ there is a polynomial sized (resp. exponential sized) proof $\pi$, such that the polynomial time (resp. exponential time) verifier $\verifier$ will be convinced by $\pi$ that $\mttx\in L_{yes}$, and for every $\mttx\in L_{no}$ no proof $\pi$ would convince $\verifier$ that $\mttx\in L_{yes}$.
The goal of probabilistically checkable proofs (PCPs)  is to enable the verifier to read only a \textbf{small yet random} part of the proof, and still be able to distinguish with high probability between the cases where $\mttx\in L_{yes}$ and where $\mttx\in L_{no}$.

A \emph{black-box function} (also known as an \emph{oracle}) $\pi\colon \{0,1\}^*\to \{0,1\}^*$ is a function that a TM can interact with as follows: The TM can, as part of its operation, send to $\pi$ an input $\mttx$, and $\pi$ outputs (in a single time step) the value $\pi(\mttx)$ --- such an interaction is called a \emph{query} to the black-box function. 
\begin{definition}[$\PCP$]
	Let $f,r,qu\colon \mathbb{N}\to \mathbb{N}$ be functions. A language $L$ is in ${\PCP}(f(n),r(n),qu(n))$ if there is a $1$-input probabilistic TM $\verifier$  such that
	\begin{itemize}
		\item (Time bound) $\TIME(\verifier,\ol{n})\leq f(|\ol{n}|)$, namely $\verifier$ is $f$-time;
		\item (Randomness bound) $\verifier(\ol{n})$ uses at most $r(|\ol{n}|)$ many random bits --- these bits, together with $\ol{n}$, will determine what positions of the black-box function $\verifier$ will query;
		\item (Completeness) if $\ol{n}\in L_{yes}$, then there exists a  black-box function $\pi$  that $\verifier$ queries  at most $qu(|\ol{n}|)$-many times, and always decides to accept.
		\item (Soundness) if $\ol{n}\in L_{no}$, then for every black-box function $\pi$ that $\verifier$ queries at most $qu(|\ol{n}|)$-many times, the probability that $\verifier$ accepts is bounded from above by $\nicefrac{1}{2}$.
	\end{itemize}
	We denote by ${\PCP}$ the union of the classes ${\PCP}(f(n),r(n),qu(n))$ overall all polynomials $f,r$ and $qu$.
	
\end{definition}

\begin{remark}[Dramatization of ${\PCP}$]
	Again, a polynomial time verifier $\verifier$ wants to decide whether $\mttx\in L_{yes}$. It sends $\mttx$ to a prover $\cal{P}$. The prover then generates a black-box function $\pi$ that $\verifier$ can interact with. $\verifier$ reads some random bits, and according to them queries $\pi$ at several locations. According to the outputs of $\pi$, $\verifier$ needs to decide whether to accept or reject.
	
	The black-box function should be thought of as a proof to the claim $\mttx\in L_{yes}$. If $r$ is a polynomial, the length of this proof is at most exponential in $|\mttx|$. But, if $qu$ is polynomial, $\verifier$ reads only a logarithmic part of the proof. This indeed means that $\verifier$ does not have enough time to be convinced with certainty that the proof is correct. The point is that the verifier can ask the prover to format the proof in such a way that even this logarithmically sized view of the proof will enable it to reject it with constant probability $\nicefrac{1}{2}$ in case the claim is wrong (namely, when $\mttx\in L_{no}$).
\end{remark}
\begin{remark}
	For a nice historical survey of this field, and specifically of PCPs, see \cite{ODonellPCPhistory}.
\end{remark}

We are ready to describe the  ${\PCP}$ protocol for the language Succinct-$3{\rm SAT}$, which is the main step towards showing that ${\MIP=\NEXP}$.

\begin{observation}[Succinct-$3{\rm SAT}$ is in $\PCP$] \label{obs:PCPs}
	Let $\circuit$ be the instance received as input, namely it is (the encoding, as described in Definition \ref{defn:Circuits}, of) a circuit that succinctly encodes a $3{\rm CNF}$ formula $\varphi_\circuit$ (Definition \ref{defn:Succinct_3SAT}). In particular, it has $3n+3$ input gates, a single output gate, and let $s$ be the number of non-input wires of $\circuit$. 
	
	Then, the verifier $\verifier$ in the $\PCP$ protocol, which gets as input both $\circuit$ and a black box function $\pi$, acts as follows. 
	\begin{enumerate}
		\item First, $\verifier$ uses (the encoding of) $\circuit$ to recover the integers $n$ and $s$ --- this can be done in time linear in the encoding length of $\circuit$. It then chooses an odd positive integer $t$, according to a rule that we describe later, lets $q=2^t$, and fixes a basis of $\FF_q$ over $\FF_2$ (a la Fact \ref{fact:basis_F_q_over_F_2}). Thus, the notions of an $\FF_q$-input and an $\FF_q$-output are well defined, as every element of $\FF_q$ has now a fixed encoding as an element of $\FF_2^t$. 
		\item Then, $\verifier$ expects $\pi$ to be the evaluation table of functions $g_\psi\colon \FF_q^n\to \FF_q$, $\alpha_i\colon \FF_q^{3n+3+s}\to \FF_q$ and $\beta_i\colon \FF_q^n\to \FF_q$, as well as functions $\mathsf{AL}g_\psi\colon \FF_q^{n-1}\times [n]\to \FF_q^7$, $\mathsf{AL}\alpha_i\colon \FF_q^{3n+2+s}\times [3n+3+s]\to \FF_q^7$  and $\mathsf{AL}\beta_i\colon \FF_q^{n-1}\times [n]\to \FF_q^7$.\footnote{In general, a single black box function can encode any sequence of black box functions. E.g., in our context, $\verifier$ can verify that $\pi$ is structured that way by sending to it tuples of the form $({\rm Name}, {\rm Input})$ where ${\rm Name}$ is one of $g_\psi,{\mathsf{AL}g}_\psi,\alpha_i,{\mathsf{AL}\alpha}_i,\beta_i,\mathsf{AL}\beta_i$, and ${\rm Input}$ an appropriate input to the function, and seeing that indeed the outputs are from $\FF_q$ or $\FF_q^7$ respectively.} 
		
		The range of the $\mathsf{AL}\square$ functions is chosen to be $7$-dimensional so that their output encodes a degree $6$ univariate polynomial over $\FF_q$. All in all, $\verifier$ expects $g_\psi, \alpha _i, \beta_i$ to be individual degree at most $6$ polynomials as needed for Proposition \ref{prop:prelude_on_satisfiability},  and $\mathsf{AL}g_\psi, \mathsf{AL}\alpha _i, \mathsf{AL}\beta_i$ be their respective restrictions to axis parallel lines (which must be univariate polyonimals of degree $6$ according to Fact \ref{fact:characterization_of_low_degreeness_by_restrictions_to_linew}).
		\item The verifier $\verifier$ runs $r$ many --- where the procedure for choosing $r$ will be described later --- independent rounds of the (classical) individual degree at most $6$ test (Definition \ref{defn:classical_low_deg_test_prelude}) on each of the $4n+4+s$ pairs 
		\[ 
		(g_\psi,\mathsf{AL}g_\psi),(\alpha_i,\mathsf{AL}\alpha_i),(\beta_i,\mathsf{AL}\beta_i)\ .
		\] 
		If \textbf{any} of these rounds has rejected, then $\verifier$ rejects.
		\item If all of the low-degree test rounds have accepted, then $\verifier$ samples two additional points $u_0\in \FF_q^n$ and $u\in \FF_q^{3n+3+s}$, each of which uniformly at random, and asks $\pi$ to send the values 
		\begin{equation}\label{eq:sampling_points_PCP_prelude}
		g_\psi(u_0),g_\psi(u_1),g_\psi(u_2),g_\psi(u_3),\alpha_i(u),\beta_i(u_0)\;.
		\end{equation}
		It then evaluates $T_\circuit(u)$ on its own using $\circuit$. Finally, it verifies that \eqref{eq:condcondcond1} and \eqref{eq:condcondcond2} are satisfied. If so, then it accepts, and otherwise it rejects.
	\end{enumerate}
	
	As Proposition \ref{prop:prelude_on_satisfiability} shows, if $\varphi_\circuit$ is satisfiable, then a tuple of individual degree at most $6$ polynomials $g_\psi,\alpha_i,\beta_i$ that always  pass \eqref{eq:condcondcond1} and \eqref{eq:condcondcond2} exists. If we define $\mathsf{AL}g_\psi,\mathsf{AL}\alpha_i,\mathsf{AL}\beta_i$ according to their restrictions to axis parallel lines, then as the original polynomials were of low degree, according to Fact \ref{fact:characterization_of_low_degreeness_by_restrictions_to_linew}, the pairs $(g_\psi,\mathsf{AL}g_\psi),(\alpha_i,\mathsf{AL}\alpha_i),(\beta_i,\mathsf{AL}\beta_i)$ pass the individual degree at  most $6$ test with certainty. Thus, if the prover chooses $\pi$ such that it consists of these specific functions, then it passes the above protocol with certainty. Namely, the protocol is complete.\\

	Assume $\pi$ consisting of functions $g_\psi,\mathsf{AL}g_\psi,\alpha_i,\mathsf{AL}\alpha_i,\beta_i,\mathsf{AL}\beta_i$ passes the above protocol with probability strictly larger than $\nicefrac{1}{2}$.
	We are going to show that, under an appropriate choice of $t$ and $r$, this implies $\varphi_\circuit$ is satisfiable, proving the soundness of the protocol. 
	To pass the entire protocol with probability of at least $\nicefrac{1}{2}$, functions in the proof $\pi$ need to pass the $r$ rounds of individual degree at most $6$ test with probability of at least $\nicefrac{1}{2}$, which implies each pair of $(g_\psi,\mathsf{AL}g_\psi),(\alpha_i,\mathsf{AL}\alpha_i),(\beta_i,\mathsf{AL}\beta_i)$ passes the single round of the low-degree test with probability of at least $(\nicefrac{1}{2})^{\nicefrac{1}{r}}$, and as $r$ is a positive integer we have the inequalities 
	$(\nicefrac{1}{2})^{\nicefrac{1}{r}}\geq (\nicefrac{1}{e})^{\nicefrac{1}{r}}=e^{-\nicefrac{1}{r}} \ge 1-\frac{1}{r}$, where the last one is Bernoulli's inequality.
	Using the classical soundness of the low degree test (Theorem \ref{thm:classical_soundness_individual_low_degree}) on each pair $(\square,\mathsf{AL}\square)$ as above, there are individual degree at most $6$ polynomials $\widetilde{g}_\psi,\widetilde{\alpha}_i,\widetilde{\beta}_i$ such that 
	\begin{equation}\label{eq:close_to_individual_degree_PCP_prelude}
	\max\left\{\Pro{}[g_\psi(u)\neq \widetilde{g}_\psi(u) ]\ ,\ \Pro{}[\alpha_i(u)\neq \widetilde{\alpha}_i(u) ]\ ,\ \Pro{}[\beta_i(u)\neq \widetilde{\beta}_i(u) ]\right\}\leq C(3n+3+s)^C\left(\frac{1}{r^{\nicefrac{1}{C}}}+\frac{6}{q^{\nicefrac{1}{C}}}\right)\;,
	\end{equation}
	where $C\geq 0$ is a universal constant independent of everything.
	
	Coming back to $\pi$, for it to pass the protocol with probability of at least $\nicefrac{1}{2}$, it needs to pass the last check with this probability, namely
	\[
	\Pro{}\Big[T_\circuit(u)\prod_{j=1}^3(g_\psi(u_j)+\eps_j+1)=\sum\alpha_i(u)\zero_i(u)\Big]\ ,\ \Pro{}\left[g_\psi(u_0)(g_\psi(u_0)+1)=\sum\beta_i(u_0)\zero_i(u_0)\right]> \frac{1}{2}\ .\]
	These are the exact expressions as in \eqref{eq:blablabla1} and \eqref{eq:blablabla2}, which guarantee that $\varphi_\circuit$ is satisfiable according to Proposition \ref{prop:prelude_on_satisfiability}. Alas, we do not know that these functions are individual degree at most $6$ polynomials, which is needed for the soundness condition of the proposition to hold. But using a union bound   and \eqref{eq:close_to_individual_degree_PCP_prelude}, we can deduce that 
	\[
	\begin{split}
	\Pro{}\Big[T_\circuit(u)\prod_{j=1}^3(\widetilde{g}_\psi(u_j)+\eps_j+1)=\sum\widetilde{\alpha}_i(u)\zero_i(u)\Big]&\geq   \Pro{}\Big[T_\circuit(u)\prod_{j=1}^3(g_\psi(u_j)+\eps_j+1)=\sum\alpha_i(u)\zero_i(u)\Big]\\
	&-\sum_{j=1}^3\Pro{}[\widetilde{g}_\psi(u_j)\neq g_\psi(u_j)]-\sum_{i=1}^{3n+3+s}\Pro{}[\widetilde{\alpha}_i(u)\neq \alpha_i(u)]\\
	&> \frac{1}{2}-(3n+6+s)\cdot C(3n+3+s)^C\left(\frac{1}{r^{\nicefrac{1}{C}}}+\frac{6}{q^{\nicefrac{1}{C}}}\right)\ ,
	\end{split}
	\]
	and similarly 
	\[
	\Pro{}\left[\widetilde{g}_\psi(u_0)(\widetilde{g}_\psi(u_0)+1)=\sum\widetilde{\beta}_i(u_0)\zero_i(u_0)\right]> \frac{1}{2}-(2+n)C(3n+3+s)^C\left(\frac{1}{r^{\nicefrac{1}{C}}}+\frac{6}{q^{\nicefrac{1}{C}}}\right)\ .
	\]
	Hence, if we choose $r$ and $t$ such that 
	\[\frac{1}{2}-(3n+6+s)\cdot C(3n+3+s)^C\left(\frac{1}{r^{\nicefrac{1}{C}}}+\frac{6}{q^{\nicefrac{1}{C}}}\right)\geq \frac{21(3n+3+s)}{q}\ ,\]
	and
	\[\frac{1}{2}-(2+n)C(3n+3+s)^C\left(\frac{1}{r^{\nicefrac{1}{C}}}+\frac{6}{q^{\nicefrac{1}{C}}} \right)\geq \frac{21n}{q} \]
	then $\varphi_\circuit$ is indeed satisfiable due to the soundness condition of Proposition \ref{prop:prelude_on_satisfiability}. By choosing, for example, $r= (12C)^C(3n+3+s)^{3C^2}$ and  the smallest odd $t$ for which $q=2^t \geq (72C)^C(3n+3+s)^{3C^2}$, the above is satisfied and soundness of the protocol is proved.\\
	
	We leave for the reader to check that indeed  the number of random bits used in this protocol, the number of queries to the proof $\pi$, and the running time of it are all polynomial in $n$ and $s$, which in turn means they are polynomial in the input length (since it bounds both of these numbers from above), as needed.
\end{observation}

\subsubsection{What is missing for ${\MIP}={\NEXP}$?}

The above protocol assumed that the prover first fixed a function $\pi$, and only then the verifier queried it. In ${\MIP}$ (see Remark \ref{rem:dramatization_MIP*}), the provers see the questions before they commit to a certain answer. The way to overcome this is for one prover to provide {all} the needed values in the above protocol, and for the second prover to play a cross-checking role. Namely, the second prover gets just one of the functions and evaluation points, and its answers are checked to be consistent with the first prover.  This already demonstrates that one prover should be able to provide ``the whole proof'', which in terms of the underlying games will require oracularization --- namely, for one player to get both questions, and for the other player to get one of the original questions. See Section \ref{sec:orac} for more on that.

The other key component which is not clear in the two provers scenario is the parallel repetition part, namely step 3 in Observation~\ref{obs:PCPs}. In that step, the verifier executes $r$ ``independent rounds'' of a certain test. In our sketch for the soundness analysis we used that such repeated checking increases the likelihood of finding an error exponentially. However, in the setting of a two-prover interactive proof, all questions to each prover are sent simultaneously. By sending the prover multiple ``independent'' questions at once, the verifier potentially allows them to answer each individual question in a way that depends on the entire tuple, thus putting in question the exponential error improvement assumed above. It turns out that this is an intricate issue, which cannot be waved away by arguing using ``without loss of generality...'' type of arguments. 
In the case of classical two-prover interactive proofs, this problem was resolved by the celebrated parallel repetition theorem of Raz \cite{raz1995parallel}. There is no general parallel repetition theorem for quantum strategies, but there is a ``good enough'' version \cite{bavarian2017hardness} that assumes some simple transformation on the given game, called \emph{anchoring}, was applied before the repetition --- see Section \ref{sec:parallel_rep} for more on that.

\subsection{Prerequisites to Answer Reduction: Purification, Oracularization, Triangulation and Decoupling}\label{sec:prerequisites_ans_red}

Before we can apply PCP techniques to reduce the answer lengths of the tailored normal form verifier in mind, we need to define four transformations: \emph{purification}, \emph{oracularization}, \emph{triangulation}  and  \emph{Decoupling}. The first will be applied to the verifier before answer reduction, and the rest  are incorporated in the answer reduction transformation itself.

A game is said to be purely unreadable if the controlled linear constraints function $L_{\mttx\mtty}$ outputs only constraints on the unreadable variables. One could have {defined} a tailored non-local game to satisfy this property without hindering the expressiveness of the model; although this would have induced some complications to the presentation and description of games, hence we kept the looser definition. To purify, all one needs to note is that the values for the readable variables are known before choosing the controlled system of equations, so one can assign to the readable variables their already known values, which makes them non-variables, i.e.\ part of the constants in the equations. 

The idea behind oracularization is simple --- instead of sampling a pair of questions $\mttx,\mtty$ and sending a single question to each player, the oracularized version samples the same pair of questions, but sends one of the players {both} $\mttx$ and $\mtty$, and the other either $\mttx$ or $\mtty$. On the level of the underlying graph of the game, this boils down to a barycentric sub-division of the graph.  Though this seems to be a naive transformation, and indeed in the classical setup it is, in the quantum setup the completeness of this transformation is dependent on the capability of one player to always measure the needed values along every edge, which is exactly the \emph{commuting along edges} condition from the definition of a $\ZPC$ strategy.

Regarding triangulation, again, the transformation on the level of linear systems of equations is straightforward: Every equation which involves $k$ variables is transformed into $k+1$ equations that involve only three variables, by inductively defining a new variable at each step alongside the constraint that the sum of two of the variables currently appearing in the equation should equal the new variable. This is a standard trick, used  even in the non-commutative context, e.g.\ when showing that every group has a presentation with relations of length at most $3$ (cf.\ \cite{zuk2003property}). Though this is a straightforward transformation, similar to PCPs, it requires one of the parties to know what the system of equations is.\footnote{There are ways to avoid this assumption under some reasonable bounds on the degree in the underlying graph of the game, but the transformation is somewhat more complicated in this case.} This makes it natural to apply triangulation  in tandem with  oracularization.

Finally, decoupling is a method of transforming in a complete and sound way a triangulated system of linear equations into a system of linear equations whose variables come from $5$ blocks, and each equation in the new system contains at most one variable from each block. The same kind of transformation can be applied for $3$CNF formulas, and is essentially baked into the version of the scaled up Cook--Levin transformation that we use in Proposition \ref{prop:explicit-padded-succinct-deciders}.

\subsubsection{Purification}

\begin{definition}\label{defn:purely_unreadable_system}
	A system  of linear equations    $\mathscr{A}\vec S=\vec b$  over  a field $\FF$, where 
	\[
	S=S^\frR_\mttx\cup S^\frL_\mttx\cup S^\frR_\mtty\cup S^\frL_\mtty,
	\]
	(as is the case in the controlled linear constraint systems of tailored games), is said to be \emph{purely unreadable}, or \emph{purely linear}, if the columns of $\mathscr{A}$ associated with $S^\frR_\mttx\cup S^\frR_\mtty$ are all zero. This is the same as saying that no constraint is applied on the readable variables. 
\end{definition}
The idea in purification is  to assign to the readable variables in the controlled linear constraints of a tailored non-local game their already assigned values, and thereby change them from  variables to constants, which means their coefficients can be assumed to be zero. 
\begin{definition}[Combinatorial purification]\label{defn:combi_purification}
	Let $\game$ be a tailored game. 
	Define the purification $\game'=\frak{Pure}(\game)$ of $\game$ as follows:
	It has the same underlying graph and distribution along edges as $\game$. In addition, it has the same length functions and the same formal variable sets. 
	Recall that the controlled linear constraint function is $L_{\mttx\mtty}\colon \FF_2^{S_\mttx^\frR\cup S_\mtty^\frR}\to \FF_2^{\FF_2^{S_{\mttx\mtty}\cup \{\sJ\}}}$. 
	If $\gamma^\frR\colon S_\mttx^\frR\cup S_\mtty^\frR\to \FF_2$ is the readable variables assignment, then for every constraint $c\colon S_{\mttx\mtty}\cup \{\sJ\}\to \FF_2$   in $L_{\mttx\mtty}(\gamma^\frR)$,  $L'_{\mttx\mtty}(\gamma^\frR)$ will contain the  constraint $c'\colon S_{\mttx\mtty}\cup \{\sJ\}\to \FF_2$ defined by:
	\begin{equation}\label{eq:purified_equation}
	\begin{split}
	\forall \sX\in S_\mttx^\frR\cup S_\mtty^\frR\ \colon \ \ c'(\sX)&=0,\\
	\forall \sX\in S_\mttx^\frL\cup S_\mtty^\frL\ \colon \ \ c'(\sX)&=c(\sX),\\
	c'(\sJ)&=c(\sJ)+\sum_{\sX\in S_\mttx^\frR\cup S_\mtty^\frR} c(\sX)\gamma^\frR(\sX).
	\end{split}
	\end{equation}
\end{definition}

\begin{fact}[Completeness and soundness of purification]\label{fact:completeness_and_soundness_purification}
	As the underlying graph and length functions of $\game$ and $\frak{Pure}(\game)$ are the same, there is a one to one correspondence between the quantum strategies for them. This correspondence is value preserving.
\end{fact}

\begin{claim}[Algorithmic purification]\label{claim:algorithmic_purification}
	There is a polynomial time TM $\Purify$ that takes as input a tailored  (typed or non-typed) $h$-level normal form verifier $\verifier$, and outputs a tailored (typed or non-typed) $h$-level normal form verifier $\verifier'$, such that:
	\begin{itemize}
		\item \emph{Combinatorial purification}: If $\verifier_n$ is well defined, then $\verifier'_n$ is well defined and satisfies $\verifier'_n=\frak{Pure}(\verifier_n)$.
		\item \emph{Running times and description lengths}: The sampler and answer length remain the same (and so their running times and description lengths are the same). Moreover,
		\[
		\TIME(\linproc';n,\cdot,\cdot,\cdot,\cdot)=O\big(\TIME(\linproc;n,\cdot,\cdot,\cdot,\cdot)+\TIME(\length;n,\cdot,\cdot)\big)\;,
		\]
		and the description length of $\linproc'$ is linear in that of $\linproc$.
	\end{itemize}
\end{claim}

\begin{proof}
	We only need to describe the linear constraints processor. Given $(n,\mttx,\mtty,a^\frR,b^\frR)$ (the typed version is similar), $\linproc'$ first runs $\linproc(n,\mttx,\mtty,a^\frR,b^\frR)$ to obtain $(c^1,...,c^k)$, and calculates $\ell^\cdot_\cdot$  using $|\dec(\length(n,\cdot,\cdot))|$. Then, for every $i$, it replaces $c^i$ with $c'^i$ defined as in \eqref{eq:purified_equation}, which is possible as the positions in $c^i$ associated to $S^\cdot_\cdot$ can be deduced from the values $\ell^\cdot_\cdot$ previously calculated.
\end{proof}

\subsubsection{Oracularization}
\label{sec:orac}

\begin{definition}[Combinatorial Oracularization]\label{defn:combi_oracle}
	Let $\game$ be a tailored non local game. The \emph{oracularization} of $\game$, $\game'=\frak{Oracle}(\game)$, is defined as follows. If $G=(V,E)$ was the underlying graph of $\game$, then the underlying graph of the oracularization is the barycentric subdivision of $G$, namely $G'=(V',E')$, where $V'=V\sqcup E$, and for $\mttx\in V$ and $e\in E$, $(\mttx,e)\in E'$  if and only if $\mttx$ was one of the endpoints of $e$. The vertices of $G'$ coming from $E$ are called \emph{oracle player} questions, while those that come from $V$ are called \emph{isolated player} questions. Regarding the length functions $\ell^\frR,\ell^\frL$, they remain the same on $V$, and extended to $E$ as follows ---  for $e=\mttx\mtty$, $\ell^\cdot(e)=\ell^\cdot(\mttx)+\ell^\cdot(\mtty)$. For the distribution over questions, if $(\mttx,e)\in E'$, then $\mu'(\mttx,e)=\nicefrac{\mu(e)}{2}$ --- or in words, a pair of questions is sampled as before, both of which are sent to an oracle player, and then one of the other two is sent uniformly to the isolated player. 
	For the formal sets of variables, we leave those on isolated vertices as they were before, and if $e=\mttx\mtty$ is an edge with $S_\mttx=\{\sX^{\kappa,j}\mid \kappa\in \{\frR,\frL\}, j\in [\ell^\kappa(\mttx)]\}$ and $S_\mtty=\{\sY^{\kappa,j}\mid \kappa\in \{\frR,\frL\}, j\in [\ell^\kappa(\mtty)]\}$, 
	then 
	$
	S_e=S_{\mttx\in e}\sqcup S_{\mtty\in e}=\{\sO\sX_e^{\cdot,\cdot},\sO\sY_e^{\cdot,\cdot}\}.
	$
	Finally, the controlled linear constraints function $L'$ acts as follows: Assume  $(\mttx,e=\mttx\mtty)$ was sampled, and that $a^\frR$ was the assignment to $S^\frR_\mttx$ while $a^\frR_e,b^\frR_e$ was the assignment to $S^\frR_e$. Then, $L'$ first outputs the following $\ell(\mttx)$ equations:
	\begin{equation}\label{eq:consistency_isolated_and_oracle_in_oracle}
	\sO\sX_e^{\kappa,j}=\sX^{\kappa,j}\ .
	\end{equation}
	In addition, it outputs the same linear equations as $L_{\mttx\mtty}(a^\frR_e,b^\frR_e)$, but on the $S_e$ variables instead of the $S_\mttx\cup S_\mtty$. Namely, if $\sum \alpha_{\kappa,j}\sX^{\kappa,j}+\sum \beta_{\kappa,j}\sY^{\kappa,j}=b$ was an equation output by $L$, then $L'$ will output the equation 
	\begin{equation}\label{eq:linear_relations_checkes_on_oracle_player}
	\sum_{\kappa,j} \alpha_{\kappa,j}\sO\sX_e^{\kappa,j}+\sum_{\kappa,j} \beta_{\kappa,j}\sO\sY_e^{\kappa,j}=b\ .
	\end{equation}
\end{definition}

\begin{remark}\label{rem:dramatization_oracle_game}
	The oracularized game has a simple description: Sample a pair of questions as before, send both of them to an oracle player --- and expect it to reply with the appropriate answer to {both} questions -- and send only one of them to the isolated player. The answer of the oracle player should be accepted by the original game, while the answer of the isolated player should be consistent with the appropriate part of the answer of the oracle player.
\end{remark}
\begin{remark}[The pure part of an oracularized pure game]\label{rem:pure_part_of_oracularized_pure_game}
	Note that by oracularizing a purely unreadable game, you get a game which is not pure, but the equations~\eqref{eq:linear_relations_checkes_on_oracle_player} {are} purely unreadable. This will be sufficient for answer reduction to work.
\end{remark}

\begin{claim}[Completeness and soundness of the oracularized game]\label{claim:completeness_soundness_oracle}
	Let $\game$ be a tailored non-local game. Then,
	\begin{itemize}
		\item (\emph{Completeness}): If $\game$ has a perfect $\ZPC$ strategy, then so does $\frak{Oracle}(\game)$.
		\item (\emph{Soundness}): If $\frak{Oracle}(\game)$ has a value $1-\eps$ strategy, then $\game$ has a value $1-12\eps$ strategy of the same dimension, and $\Ent(\frak{Oracle}(\game),1-\eps)\geq \Ent(\game,1-12\eps)$.
	\end{itemize}
\end{claim}

\begin{proof}
	For completeness, assume $\game$ has a perfect $\ZPC$ strategy $\strategy=\{\cal{U}\}$.  Then, we can extend $\strategy$ to the variables $S_e$ at oracle player vertices $e$ in the straightforward manner $\cal{U}(\sO\sX_e^{\kappa,j})=\cal{U}(\sX^{\kappa,j})$. As $\strategy$ is commuting along edges, the observables at the oracle player vertices are commuting, which means this extension is a well-defined quantum strategy for $\frak{Oracle}(\game)$ (Definition \ref{defn:quantum_strategy}) --- this is a crucial point, and is the only reason the ``commuting along edges'' condition is always included in the completeness argument. In addition, once the observables at oracle vertices are commuting and are consistent with the isolated players' observables, the extended $\strategy$ is commuting along edges of $\frak{Oracle}(\game)$, $Z$-aligned and induced by a permutation strategy. It is left to be convinced that the extended $\strategy$  has value $1$, but this is also immediate as \eqref{eq:consistency_isolated_and_oracle_in_oracle} is satisfied because the observables at $e=\mttx\mtty$ are consistent with those of $\mttx$ and $\mtty$, and \eqref{eq:linear_relations_checkes_on_oracle_player} are satisfied because the original $\strategy$ was perfect for $\game$.
	
	For soundness, assume $\strategy=\{\cal{U}\}$ is a value $1-\eps$ strategy for $\frak{Oracle}(\game)$. The idea is to show that the restriction of $\cal{U}$ to the isolated vertices induces a strategy for $\game$ with value of at least $1-12\eps$. To that end, for $e=\mttx\mtty$, let $\eps_{\mttx,e}$ be the probability $\cal{U}$ fails the checks of the edge $(\mttx,e)$ in $\frak{Oracle}(\game)$, and  $\eps_{\mtty,e}$ is defined similarly. Then 
	\[
	\eps=\Es{(\mttx,e)\sim \mu'}[\eps_{\mttx,e}]=\Es{e\sim \mu}\left[\frac{\eps_{\mttx,e}+\eps_{\mtty,e}}{2}\right].
	\]
	By Claims \ref{claim:perfect_3Lin_commuting_observables} and \ref{claim:L1_closeness_of_unirep_implies_Linfty}, the fact that $\cal{U}$ passes the linear checks \eqref{eq:consistency_isolated_and_oracle_in_oracle} with probability $1-\eps_{\mttx,e}$ implies that 
	\[
	\forall \alpha\in \FF_2^{S_\mttx}\ \colon \ \ \Big\|\prod_{\sX\in S_\mttx}\cal{U}(\sX)^{\alpha(\sX)}-\prod_{\sO\sX_e\in S_{\mttx\in e}}\cal{U}(\sO\sX_e)^{\alpha(\sX)}\Big\|_{hs}^2\leq 6\eps_{\mttx,e}\;,
	\]
	and  similarly products of $\cal{U}(\sY)$'s are close to  products of $\cal{U}(\sO\sY_e)$'s. In addition,  as $\cal{U}$ passes the linear check \eqref{eq:linear_relations_checkes_on_oracle_player} with probability of at least $1-\min(\eps_{\mttx,e},\eps_{\mtty,e})\geq 1-\frac{\eps_{\mttx,e}+\eps_{\mtty,e}}{2}$, 
	we can deduce by Claim \ref{claim:perfect_3Lin_commuting_observables} that 
	\[
	\left\|(-\Id)^b \prod \cal{U}(\sO\sX_e^{\kappa,j})^{\alpha_{\kappa,j}}- \prod \cal{U}(\sO\sY_e^{\kappa,j})^{\beta_{\kappa,j}}\right\|_{hs}^2\leq 
	2\eps_{\mttx,e}+2\eps_{\mtty,e}.
	\]
	Combining the above and using the triangle inequality, 
	\[
	\begin{split}
	\left\| (-\Id)^b\prod \cal{U}(\sX^{\kappa,j})^{\alpha_{\kappa,j}}- \prod \cal{U}(\sY^{\kappa,j})^{\beta_{\kappa,j}}\right\|_{hs}^2&\leq   3\underbrace{\left\| (-\Id)^b\prod \cal{U}(\sX^{\kappa,j})^{\alpha_{\kappa,j}}-(-\Id)^b\prod \cal{U}(\sX\sO_e^{\kappa,j})^{\alpha_{\kappa,j}}\right\|_{hs}^2}_{\leq 6\eps_{\mttx,e}}\\
	&+3\underbrace{\left\| (-\Id)^b\prod \cal{U}(\sO\sX_e^{\kappa,j})^{\alpha_{\kappa,j}}- \prod \cal{U}(\sO\sY_e^{\kappa,j})^{\beta_{\kappa,j}}\right\|_{hs}^2}_{\leq 2\eps_{\mttx,e}+2\eps_{\mtty,e}}\\
	&+3\underbrace{\left\|\prod \cal{U}(\sO\sY_e^{\kappa,j})^{\beta_{\kappa,j}}-\prod \cal{U}(\sY^{\kappa,j})^{\beta_{\kappa,j}}\right\|_{hs}^2}_{\leq 6\eps_{\mtty,e}}\\
	&\leq 24(\eps_{\mttx,e}+\eps_{\mtty,e}),
	\end{split}
	\]
	which translates to $\cal{U}$ passing $\game$ with probability at least $1-6(\eps_{\mttx,e}+\eps_{\mtty,e})$ when $e$ is sampled. Therefore, the value of the restriction of $\cal{U}$ to the isolated vertices passes $\game$ with probability of at least $1-12\eps$, as claimed, and the entanglement lower bound is immediate from that.
\end{proof}

\begin{remark}[The sampling procedure of the oracularized game]\label{rem:sampling_scheme_underlying_oracularization}
	We do not know how to induce the sampling procedure described in Definition \ref{defn:combi_oracle} using CLMs, even when $\game$ has a sampling procedure induced by CLMs. But,  If the sampling procedure of $\game$ was induced by $h$-level CLMs $\frS^A,\frS^B$, then there is an $h$-level typed sampling scheme with type graph $\mttA-\oracle-\mttB$ that induces the aforementioned distribution for $\frak{Oracle}(\frak{DoubleCover}(\game))$, the oracularized \textbf{double cover} of the game (Definition \ref{defn:double_cover}). This is done by using the same dimension as before, letting $\frS^\mttA=\frS^A,\frS^\mttB=\frS^B$, and $\frS^\oracle=\frS^A\times \frS^B$, by which we mean, given input $z$, $\frS^\oracle$ outputs  $(\frS^A(z),\frS^B(z))$. As is common throughout this paper, given that $\game$ has enough consistency checks (or, given that it was already bipartite),  this move to the double cover does not hinder the desired conclusions.
	
	Though the above sampling scheme works, we later let $\frS^\oracle=\Id$ instead. This means that between every two isolated vertices $(\mttA,\mttx)$ and $(\mttB,\mtty)$, instead of having a single vertex $(\oracle,\mttx\mtty)$, there is a vertex $(\oracle,z)$ for every $z$ for which $\frS^A(z)=\mttx$ and $\frS^B(z)=\mtty$. 
	Although this allows the strategies in the oracularized (double cover) game more leniency, which presumably may elevate the value of the game compared to the case of a genuine barycentric subdivision, this turns out not to be the case and the above soundness argument works (essentially) the same.
\end{remark}

\subsubsection{Triangulation}\label{sec:triangulation}

\begin{definition}[Triangulated system]\label{defn:triangulated_system}
	A system  of linear equations    $\mathscr{A}\vec S=\vec b$  over  a field $\FF$ is said to be \emph{triangulated} if  every row of $\mathscr{A}$ has at most $3$ non-zero entries. 
\end{definition}

\begin{definition}[Triangulating a system  of linear equations]\label{defn:triangulating_linear_system}
	Triangulating a system of linear equations (or a system of word equations over a group) is a standard procedure. The idea is to replace an equation 
	\[
	a_0\sX_0+a_1\sX_1+a_2\sX_2+...+a_n\sX_n=b\ ,
	\]
	on $n+1$ variables, 
	by $n+2$ triangulated equations 
	\begin{align}
	a_0\sX_0&=\sY_0\ ,\label{eq:initial_triangulation}\\
	\forall 1\leq i\leq n \ \colon \ \ \sY_{i-1}+a_i\sX_{i}&=\sY_i\ ,\label{eq:ind_traingulation}\\
	\sY_n&=b\ .\label{eq:terminal_triangulation}
	\end{align}
	on $2n+2$ variables. On the level of the matrix representation of the system, this is the same as replacing the single row
	\[
	\begin{pmatrix}
	a_0 & a_1 & ... & a_n &|& b
	\end{pmatrix}
	\]
	with the system 
	\[
	\left(\begin{array}{ccccccccccc|c}
	a_0 & 0 & 0& ... & 0 & -1 &0 &0 &... &0 &0&0\\
	0 &a_1 & 0 &... &0 & 1 & -1 & 0& ... &0 &0&0\\
	0 &0 & a_2 &... &0 & 0 &1 &  -1& ... &0 &0&0\\
	\vdots &\vdots &\vdots &\ddots & \vdots &\vdots &\vdots &\vdots &\ddots &\vdots &\vdots & \vdots \\
	0 &0 & 0 &... &a_n & 0 &0 & 0& ... &1 & -1&0\\
	0 &0 & 0 &... &0 & 0 &0 & 0& ... &0 &1&b\\
	\end{array}\right)\ .
	\]
	When there is more than one equation in the system (which is usually the case), one adds $n+1$ new variables to {each} equation; namely, if there were $R$ equations in the system and $n+1$ variables, the triangulated system has $R(n+2)$ equations over $(R+1)(n+1)$ variables.
	
	More generally, given a system of linear equations with matrix representation $\mathscr{A}\cdot \vec S=\vec b$ and a non-negative integer $\Delta$, we define ${\rm Triangle}(\mathscr{A},\vec b, \Delta):=(\mathscr{A}_\triangle\mid \vec b_\triangle)$ to be the matrix representation $\mathscr{A}_{\triangle} \cdot \overrightarrow{S\sqcup  S_{\triangle}} =\vec b _\triangle$ of the triangulated system with $|S_\triangle|=\Delta$ more variables. Namely, if $\mathscr{A}$ is of size $R\times (n+1)$, and $\Delta\geq R(n+1)$, then  the system $\mathscr{A}_{\triangle} \cdot \overrightarrow{S\sqcup  S_{\triangle}} =\vec b _\triangle$ is the triangulated system (with extra $\Delta-R(n+1)$ variables that do not appear in any equation). Otherwise, it does not contain the last $R(n+1)-\Delta$ equations in the triangulated system (as there were not enough variables to fully triangulate).
\end{definition}

\begin{remark}[Properties of the triangulated system]\label{rem:prop_triangulated_system}
	\ 
	\begin{enumerate}
		\item The triangulation procedure increases both the number of variables and the number of equations. It is also complete and sound in the following sense: There is a one to one correspondence between solutions to the original system of equations and the triangulated system.  Namely, given a system $\mathscr{A}\cdot \vec S=\vec b$ with $R$ many equations, if we let  $\Delta= R(n+1)$, then the solutions to  ${\rm Triangle}(\mathscr{A},\vec b,\Delta)$ correspond perfectly to those of the original system. More generally, whenever $\Delta\geq R(n+1)$ there is still a perfect correspondence by assigning the value $0$  to all variables that do not participate in any equation. Given  an assignment $f\colon S \to \FF$ to the original variables of the system, let us denote by $f_\triangle\colon S_\triangle\to \FF$ the aforementioned unique extension to the triangulation variables $S_\triangle$ --- note that the values of $f_\triangle$ are affine combinations of the values of $f$.
		
		\item Triangulation is efficient. Namely, given a system $\mathscr{A} \vec x=\vec b$ with $\mathscr{A}$ of size $R\times (n+1)$, and an integer $\Delta\geq 0$, the system ${\rm Triangle}(\mathscr{A},\vec b,\Delta)$ is of size $R(n+2)\times(n+1+\Delta)$ and takes  $O(R(n+2)(n+1+\Delta))$-time to calculate it.
	\end{enumerate}
\end{remark}

\subsubsection{Decoupling}
It is much easier to implement a PCP protocol as a  non-local game if every polynomial in the PCP protocol is measured at a single point instead of several points; this should be contrasted with  the example in the Prelude Section \ref{sec:prelude_decision_problems_PCPs}, where $g_\psi$ is measured at $4$ potentially different points $u_0,u_1,u_2,u_3$, as described in  \eqref{eq:sampling_points_PCP_prelude}. To that end, we define a more restrictive format of systems of linear equations and $k$-SAT instances, so that the PCPs we construct measure every polynomial at a single point.
\begin{definition}[Decoupled systems of equations and decoupled CNFs]\label{defn:decoupled_equations_and_CNF}
	A  system of linear equations (over a field $\FF$) is said to be \emph{$k$-decoupled} if there are $k$ (disjoint)  sets of formal generators  $S_1,...,S_k$ --- each of which is called  \emph{a block of generators} --- such that each equation in the system contains at most one variable from each block; namely, the equations are of the form
	\[
	a_1\sX_1+...+a_k\sX_k=b,
	\]
	where $a_1,...,a_k,b\in \FF$ and $\sX_i\in S_i$ for $1\leq i\leq k$. Such an equation is uniquely defined by the tuple $(a_1,\sX_1,...,a_k,\sX_k,b)\in  \FF\times S_1\times...\times  \FF\times S_k\times  \FF$, and thus a  $k$-decoupled system of equations can be encoded as an indicator of a subset of $S_1\times...\times S_k\times  \FF^{k+1}$ (this encoding is finite when $S_i$ and $\FF$ are finite).
	\\
	
	Let $S_1,...,S_k$ be, again,  disjoint  sets of formal generators.  A Boolean formula is said to be a $k$-decoupled CNF (over the blocks $S_1,...,S_k$) if each disjunctive clause in the conjunction contains exactly one variable from each block; namely, the formula is a  conjunction of clauses of the form
	\[
	\sX_1^{\eps_1}\lor...\lor \sX_k^{\eps_l},
	\]
	where $\eps_i\in \FF_2$ and $\sX_i\in S_i$ for every $1\leq i\leq k$. Assume there are natural numbers $\{n_i\}_{i=1}^k$ such that  $S_i=\{\sX_{i,u}\}_{u\in \FF_2^{n_i}}$  for every $i\in[k]$. Then, a circuit $\circuit$ (Definition \ref{defn:Circuits}) with $k+\sum_{i=1}^k{n_i}$ input gates and a single output encodes a $k$-decoupled CNF $\varphi_\circuit$ by including the clause 
	\[
	\sX_{1,u_1}^{\eps_1}\lor...\lor \sX_{k,u_k}^{\eps_k}
	\]
	in the conjunction    whenever $P_{\circuit}(u_1,...,u_k,\eps_1,...,\eps_k)=1$, where $u_i\in \FF_2^{n_i}$ and $\eps_i\in \FF_2$ for every $i\in [k]$.
\end{definition} 
The following fact is immediate from the above definition.
\begin{fact}[Translating satisfiability conditions of decoupled systems and CNFs into polynomial equations. Cf.\ Observation \ref{obs:sufficient_condition_S3SAT}]\label{fact:polynomial_condition_for_satisfiability}
	Let $O\colon S_1\times S_2\times...\times S_k \times \FF^{k+1}\to \{0_\FF,1_\FF\}\subseteq \FF$ be the encoding of a $k$-decoupled system of linear equations with blocks $S_1,...,S_k$ over the field $\FF$ (Definition \ref{defn:decoupled_equations_and_CNF}); namely, the equation 
	\[
	a_1\sX_1+...+a_k\sX_k=b\ ,
	\]
	where each $\sX_i$ is in $S_i$, appears in the system induced by $O$ if and only if $O(\sX_1,...,\sX_k,a_1,...,a_k,b)=1_{\FF}$. Then, the assignments $f_i\colon S_i\to \FF$  for $i\in [k]$ satisfy the decoupled system of equations induced by $O$ if and only if 
	\begin{equation}
	\forall (\sX_1,...,\sX_k, a_1,...,a_k,b)\in S_1\times...\times S_k\times \FF^{k+1} \ \colon \ \ O(\sX_1,...,\sX_k,a_1,...,a_k,b)(a_1f_1(\sX_1)+...+a_kf_k(\sX_k)-b)=0\ .
	\end{equation}
	
	Similarly, let $\circuit$ be a circuit with $k+\sum_{i=1}^k{n_i}$ many input gates, where $k$ and $n_1,...,n_k$ are positive integers, which encodes a $k$-decoupled ${\rm CNF}$ $\varphi_\circuit$  on blocks $S_i=\{\sX_{i,u}\}_{\FF_2^{n_i}}$ as in Definition \ref{defn:decoupled_equations_and_CNF}. Then, the assignments $w_i\colon S_i\to \FF_2$ for $i\in [k]$ satisfy $\varphi_C$ if and only if 
	\begin{equation}
	\forall (u_1,...,u_k,\eps_1,...,\eps_k)\in \FF_2^{n_1}\times...\times \FF_2^{n_k}\times \FF_2^k\ \colon\ \  P_\circuit(u_1,...,u_k,\eps_1,...,\eps_k) \cdot \prod_{i=1}^k(w_i(\sX_{i,u_i})+\eps_i+1)=0\ ,
	\end{equation}
	where $P_\circuit$ is the function induced by $\circuit$ (Definition \ref{defn:Circuits}). Equivalently, if $\circuit$ has $s$ many non-input wires, then the aforementioned  $w_i$'s are a satisfying assignment to $\varphi_\circuit$ if and only if 
	\begin{equation}
	\forall (u_1,...,u_k,\eps_1,...,\eps_k,z)\in \FF_2^{n_1}\times...\times \FF_2^{n_k}\times \FF_2^k\times \FF_2^s\ \colon\ \  T_\circuit(u_1,...,u_k,\eps_1,...,\eps_k,z) \cdot \prod_{i=1}^k(w_i(\sX_{i,u_i})+\eps_i+1)=0\ ,
	\end{equation}
	where $T_\circuit$ is the Tseitin polynomial associated with $\circuit$ (Definition \ref{defn:Circuits}).
\end{fact}

\begin{remark}\label{rem:naive_decoupling}
	A triangulated system (Definition \ref{defn:triangulated_system}) of $m$ linear equations $\mathscr{A}\vec S=\vec b$  over $\FF$  can be $3$-decoupled in a straightforward manner: Let $\vec S_1,\vec S_2,\vec S_3$ be three disjoint copies of $\vec S$ --- namely, if $\vec S=(\sX_j)_{j=1}^n$, then $\vec S_i=(\sX^i_{j})_{j=1}^n$. First, regardless of what $\mathscr{A}$ was, add the $2n$ decoupled linear equations 
	\[
	\forall 1\leq j\leq n\ \colon\ \ \sX^1_j=\sX^2_j=\sX^3_j\ .
	\]
	Then, for every $1\leq k\leq m$, if the $k^{\rm th}$ equation of $\mathscr{A}\vec S=\vec b$ is $a_1\sX_{j_1}+a_2\sX_{j_2}+a_3\sX_{j_3}=b_k$ (where $j_1<j_2<j_3$),\footnote{Here we are using the arbitrary ordering $\vec S$  on  $S$ which is used to write down the system $\mathscr{A}\vec S=\vec b$.} then add the decoupled equation 
	\[
	a_1\sX^1_{j_1}+a_2\sX^2_{j_2}+a_3\sX^3_{j_3}=b_k\ .
	\]
	All in all, the new decoupled system has $2n+m$ equations over $3n$ variables, and it is straightforward to relate the set of solutions of the two systems of equations.
\end{remark}

For our purposes, we need some extra conditions on the new system to be able to answer reduce.  We want to decouple only the non consistency linear constraints at the oracle player vertices, under the assumption that they are triangulated and pure. In addition, we still want to be able to check the consistency in an easy manner. To that end, we define a $5$-decoupling instead of a $3$-decoupling, where the first two blocks should be the ``original variables'' that will be compared to the isolated player's answers, and three ``new blocks'' that play a similar role to the above naive decoupling variables --- namely, they are an aggregate of all original variables together with the  variables added in the triangulation phase.

\begin{definition} [Combinatorial $5$-decoupling of a triangulated linear system of equations]\label{defn:combi_decoupling_system_of_linear_equations}
	Let $S$ be the disjoint union of $3$ sets of formal variables 
	\[
	S_{\mttA}\ ,\ S_{\mttB}\ ,\  S_{\triangle}\ ,
	\]
	of respective sizes $\ell_\mttA,\ell_\mttB,\ell_\triangle$, and let $\ell=\ell_\mttA+\ell_\mttB+\ell_\triangle$.
	Let $\mathscr{A}\vec S=\vec b$ be a triangulated system of $m$ equations over $S$ --- note that here we assumed some ordering on $S$ was fixed, which is used later. The combinatorial $5$-decoupling  $\frak{DeCouple}(S_\mttA,S_\mttB,S_\triangle,(\mathscr{A}\mid\vec b))$  of $\mathscr{A}\vec S= \vec b$ has the following $5$ blocks of variables 
	\begin{center}
		\begin{tabular}{|c|c|c|c|c|c|}
			\hline
			Block number  & $1$ & $2$ & $3$ & $4$ & $5$  \\
			\hline
			Set of variables  & $S_{\mttA}$ & $S_{\mttB}$ & $ S_1$ & $S_2$ & $S_3$\\
			\hline
			Size  & $\ell_\mttA$ & $\ell_\mttB$ & $\ell$ & $\ell$ & $\ell$ \\
			\hline
		\end{tabular}    
	\end{center}
	where each $S_i$ is a copy of $S$, namely it is composed of a disjoint union of sets $S_{\mttA,i},S_{\mttB,i},S_{\triangle,i}$ each of which is a respective copy of   $S_\mttA,S_\mttB,S_\triangle$. For later use,  if $\mathsf{A},\mathsf{B},\mathsf{C}\in S$, then the corresponding formal variables in $S_i$ are denoted $\mathsf{A}^i,\mathsf{B}^i,\mathsf{C}^i$.
	
	For the equations, we have the following. Regardless of what $\mathscr{A}$ or $\vec b$ are, it has $3\ell_A+3\ell_B+2\ell_\Delta = (\ell_\mttA+\ell_{\mttB}+2\ell)$-many decoupled  equations 
	\begin{equation}\label{eq:X1=3=4=5,Y2=3=4=5}
	\begin{split}
	\forall \sX\in S_\mttA\ &\colon \ \ \sX=\sX^1=\sX^2=\sX^3\ ,\\
	\forall \sY\in S_\mttB\ &\colon \ \ \sY=\sY^1=\sY^2=\sY^3\ ,\\
	\forall \sZ\in S_\triangle\ &\colon \ \ \sZ^1=\sZ^2=\sZ^3\ .
	\end{split}
	\end{equation}
	In addition, as $\mathscr{A}\vec S=\vec b$ is triangulated,  an equation in it is of the form $a_1\mathsf{A}+a_2\mathsf{B}+a_3\mathsf{C}=b$, where  $\mathsf{A},\mathsf{B},\mathsf{C}\in S$ with $\mathsf{A}<\mathsf{B}<\mathsf{C}$ according to the ordering  $\vec S$, and 
	$a_1,a_2,a_3,b\in \FF$. For every such equation in $\mathscr{A}\vec S=\vec b$, add the decoupled equation  
	\begin{equation}\label{eq:decoupled_equations}
	a_1\mathsf{A}^1+a_2\mathsf{B}^2+a_3\mathsf{C}^3=b
	\end{equation}
	to $\frak{DeCouple}(S_\mttA,S_\mttB,S_\triangle,(\mathscr{A}\mid\vec b))$, 
	where $\mathsf{A}^1\in S_1,\mathsf{B}^2\in S_2$ and $\mathsf{C}^3\in S_3$ are the respective copies of $\mathsf{A},\mathsf{B},\mathsf{C}$.
	All in all, we defined a $5$-decoupled system of $m+2\ell+\ell_\mttA+\ell_\mttB$  linear equations over $3\ell+\ell_\mttA+\ell_\mttB$ variables. 
	
	The solutions to the original system $(\mathscr{A}\mid \vec b)$ and $\frak{DeCouple}(S_\mttA,S_\mttB,S_\triangle,(\mathscr{A}\mid\vec b))$  are in perfect correspondence: This correspondence is achieved by associating with any assignment $f\colon S_\mttA\sqcup S_\mttB\sqcup S_\triangle\to \FF$ the unique $5$-tuple $(f|_{S_\mttA},f|_{S_\mttB},f,f,f)$. Indeed, from \eqref{eq:X1=3=4=5,Y2=3=4=5}, every satisfying assignment for the decoupled system is of this form, and it is straightforward to check that the image of a satisfying assignment to the original triangulated system is satisfying the decoupled system.

\end{definition}

\begin{remark}[Algorithmic decoupling]\label{rem:alg_5-decoupling}
	Let $\mathscr{A}\vec S=\vec b$ be the matrix representation  of a \textbf{triangulated} linear system of equations over $\FF_2$, where $S=S_\mttA \sqcup S_\mttB \sqcup S_\triangle$. Then, the encoding (as in Definition \ref{defn:decoupled_equations_and_CNF}) of its $5$-decoupling $$\frak{DeCouple}(S_\mttA,S_\mttB,S_\triangle,(\mathscr{A}\mid \vec b))$$ (Definition \ref{defn:combi_decoupling_system_of_linear_equations}) as an indicator map on 
	\[
	S_\mttA\times S_\mttB\times S\times S\times S\times \FF_2^6 
	\]
	can be calculated efficiently, namely in time polynomial in the input length. We often denote by $O$ the resulting encoding, and may abuse notation and write $O=\frak{DeCouple}(S_\mttA,S_\mttB,S_\triangle,(\mathscr{A}\mid \vec b))$.
\end{remark}

\begin{corollary}[Combining triangulation and decoupling]    \label{cor:triang_and_decoupling_extend}
	Let $S=S_\mttA\sqcup S_\mttB$, and $\mathscr{A}\cdot \vec S=\vec b$ a system of linear equations over $\FF$ with $R$ many equations. Let $\Delta\geq R(|S|+1)$ be a positive integer, and $S_\triangle$ a set of $\Delta$-many formal variables (disjoint from $S$).  Then, there is an \textbf{affine} mapping $\mathsf{Extend}$\footnote{The mapping $\mathsf{Extend}$ depends on the decomposition of $S$ to $S_\mttA,S_\mttB$, the system $(\mathscr{A}\mid \vec b)$, and the chosen parameter $\Delta$, but we omit this dependence from the notation.} from assignments $f\colon S\to \FF$ to $5$-tuples $(f_1,...,f_5)$ which are assignments of 
	\[
	\frak{DeCouple}(S_\mttA,S_\mttB,S_\triangle,{\rm Triangle}(\mathscr{A},\vec b,\Delta))\ ,
	\]
	such that:
	\begin{itemize}
		\item \emph{Completeness}: Satisfying assignments are mapped to satisfying assignments.
		\item \emph{Soundness}: Non-satisfying assignments are sent to non-satisfying assignments.
		In addition, by ignoring the variables in $S_\triangle$ that do not appear in any equation of ${\rm Triangle}(\mathscr{A},\vec b,\Delta)$, every satisfying assignment to the decoupled system is the $\mathsf{Extend}$-image of a satisfying assignment to the original system. 
	\end{itemize}
\end{corollary}
\begin{proof}
	Given $f\colon S_\mttA\sqcup S_\mttB\to \FF$, the map $\mathsf{Extend}$ first defines the map $f_\triangle\colon S_\triangle\to \FF_2$ from clause $1.$~of Remark \ref{rem:prop_triangulated_system}. Hence, it retrieves a map $f'\colon  S_\mttA\sqcup S_\mttB\sqcup S_\triangle\to \FF$ by letting 
	\[
	f'(\sX)=\begin{cases}
	f(\sX) & \sX\in S_\mttA\sqcup S_\mttB\ ,\\
	f_\triangle(\sX) & \sX\in S_\triangle\ .
	\end{cases}
	\]
	Then, $\mathsf{Extend}$ outputs $(f|_{S_\mttA},f|_{S_\mttB},f',f',f')$. By combining Remark \ref{rem:prop_triangulated_system} and the observation at the end of Definition \ref{defn:combi_decoupling_system_of_linear_equations}, we deduce the corollary.
\end{proof}

\subsection{Translating the verifier's checks into polynomial equations}
\label{sec:succinct-linproc}

As described in the Prelude Section \ref{sec:prelude_decision_problems_PCPs},  PCP techniques are fit to decide whether a formula succinctly described by a circuit (Definition \ref{sec:circuits_and_S3SAT}) is satisfiable. So, after applying some prerequisite transformations --- namely padding and purification --- the next step towards answer reduction is to translate some of the checks in the game to succinct SAT (and LIN) instances. The succinct SAT instances described here are slightly different from those in the Prelude \ref{sec:prelude_decision_problems_PCPs}, and are adapted from~\cite[Sections 10.2 and 10.3]{MIPRE}.
The plan is as follows:
\begin{enumerate}
	\item First, we replace the check from \eqref{eq:decision_predicate_induced_by_TNFV}, which verifies that a tuple $a^\frR,a^\frL,b^\frR,b^\frL$ is accepted by the game $\verifier_n$ given $\mttx\mtty$ were asked, by two checks that verify the same thing. The first check verifies that a bit string $O$  encodes the (triangulated and decoupled) purely unreadable part of the linear system $L_{\mttx\mtty}(a^\frR,b^\frR)$. The second check verifies that (an appropriate extension of) $a^\frL,b^\frL$ solve the system $O$. This is done in Section \ref{sec:triangulated_output_indicator}, and the equivalence to $\verifier_n$ accepting this quadruple is stated in Claim \ref{claim:properties_of_L*}.
	\item Then, the check that $O$ encodes the purely unreadable part of $L_{\mttx\mtty}(a^\frR,b^\frR)$ is shown to be equivalent, using a version of the Cook--Levin transformation, to the satisfiability of a $6$-decoupled CNF formula  succinctly encoded by some  circuit $\circuit$. This  is done in Section \ref{sec:decoupled_Cook-Levin}, and the main take away from this section is  Corollary \ref{cor:functional_viewpoint_succinct_description}.
	\item At this point, the condition ``$a^\frR,a^\frL,b^\frR,b^\frL$ are accepted by the game $\verifier_n$ given $\mttx\mtty$ were asked'' was replaced by the satisfiability of a certain succinctly encoded formula  $\varphi_\circuit$ and an appropriate succinctly encoded system of linear equations $(\mathscr{A}_O\mid \vec b_O)$. At this point, PCP techniques allow to replace these two satisfiability conditions by  $13$ polynomial equations (see \eqref{eq:PCP_condition_1}, \eqref{eq:PCP_condition_2}, \eqref{eq:Tseitin_satisfiable_induced} and \eqref{eq:induced_system_is_satisfied}), whose satsifiability can be checked \emph{probabilistically} by reading only a logarithmic portion of the polynomias' values. This is done in Section \ref{sec:pcpverifier}.
\end{enumerate}

Let us elaborate  on why naively applying  the scaled up Cook--Levin theorem on \eqref{eq:decision_predicate_induced_by_TNFV} does not work in our case (as opposed to \cite{MIPRE}). The problem  is that the resulting PCP does not {behave well} with regards to permutation assignments to unreadable variables. 
More on that: Given two commuting unitary involutions, namely matrices $A,B\in U(n)$ such that $A^2=B^2=[A,B]=\Id$,  there is a well defined notion of their $\land$ (AND operation) --- As they are mutually diagonalizable with respect to some orthonormal basis, and on the diagonal there are only $\pm 1$'s, we can define $A\land B$ to be the diagonal matrix (with respect to the same orthonormal basis) whose $ii$ entry is $-1$ if and only if the $ii$ entries of $A$ and $B$ are both $-1$.\footnote{Similar to before, we interpret $-1=(-1)^1$ as True, and $1=(-1)^0$ as False.} A problem arises when the two matrices are  permutation matrices --- 
in this case, though their $\land$ is well defined, it is not necessarily  a permutation matrix (for example, the matrices in \eqref{eq:example_of_permutations_with_non_perm_AND}), which is problematic when the evaluation table of the PCP should be generated by measuring a  $\ZPC$-strategy. 
Thus,  constructing the PCP requires us to be careful with the exact operations applied to unreadable variables, so that in the complete case the proof {can} be induced by a $\ZPC$ strategy.

\begin{table}[!htbp]
	\centering
	\captionsetup{singlelinecheck=off}
	\begin{tabular}{|r|l|l|}
		\hline
		Name & Role & See\\
		\hline 
		$\linproc^*$ & The triangulated output indicator & Definition \ref{defn:decider-read}\\
		$\Lambda$ & A TM controlling the expected answer length, i.e. & Definition~\ref{defn:decider-read}\\       &\qquad\qquad\qquad\qquad$|a^\frL|=|a^\frR|=|b^\frL|=|b^\frR|=2^{\Lambda(n)}$ &\\
		$\Delta$ &  A TM controlling the padding   required for triangulation & Definition~\ref{defn:decider-read}\\
		$\diamondsuit$ & Number of bits required to specify a variable post-triangulation, i.e. & Definition~\ref{defn:decider-read}\\ 
		&\qquad\qquad\qquad\qquad   $\diamondsuit(n)=\lceil \log (2^{\Lambda(n)+1}+\Delta(n))\rceil$ & \\
		$T$ & A TM that bounds the running time of $\linproc^*$ with the first $6$ inputs fixed to  & Definition~\ref{def:succinct-bounded-perm} \\
		& \qquad\qquad\qquad\qquad $\verifier, \Lambda, \Delta, n, \mttx, \mtty$ & \\
		$M$ & Size of blocks in the circuit  representing $\linproc^*$ & Definition~\ref{def:succinct-bounded-perm} \\
		$s$ & Number of non-input wires in the circuit  representing $\linproc^*$ & Definition~\ref{def:succinct-bounded-perm} \\
		${Q}$ &  A TM that bounds the dimension of the CLM underlying $\sampler$ & Corollary~\ref{cor:functional_viewpoint_succinct_description}\\
		$D$ & A bound on the description lengths of $\verifier, \Lambda, \Delta, T$ and $Q$ & Proposition \ref{prop:explicit-padded-succinct-deciders}\\
		$h$ & The tailored normal form verifiers are $h$-level  & Definition \ref{defn:h-level_NFV}\\
		$m$ & Dimension of the PCPs, defined as  & Definition  \ref{defn:blocks_of_vars_circuit}\\
		&\qquad\qquad$m=|S|=4\Lambda(n)+3\diamondsuit(n)+3M(n)+s(n)+12$&\\
		$\heartsuit$ & Number of polynomials in a PCP, defined as & Definition \ref{defn:PCP_of_V_n_satisfied}\\
		&\qquad\qquad $\heartsuit(n)=12\Lambda(n)+12\diamondsuit(n)+6M(n)+s(n)+35$  &\\
		$q,t$ & The size of the field $\F_q=\FF_{2^t}$ which is used in the PCP& Definition \ref{defn:PCP_of_V_n_satisfied} \\
		
		\hline 
	\end{tabular}
	\caption[]{Summary of some relevant parameters used in the rest of Section~\ref{sec:answer_reduction}.}
	\label{tab:parameters_ans_red}
\end{table}

\subsubsection{The triangulated output indicator $\linproc^*$}\label{sec:triangulated_output_indicator}

Recall that our goal is to translate the decision problem ``given a tailored normal form verifier $\verifier$, are $a^\frR,a^\frL,b^\frR,b^\frL$  accepted in the game $\verifier_n$ assuming questions $\mttx,\mtty$ were asked?'' to a collection of polynomial equations on which PCP techniques can be applied. 
To that end, we first define a TM $\linproc^*$ called \emph{the triangulated output indicator} (Definition \ref{defn:decider-read}) which, under some padding and purification assumptions (Definition \ref{defn:padded_purified_NFV}), checks that a bit string $O$ is the encoding (as in Definition \ref{defn:decoupled_equations_and_CNF}) of the $5$-decoupling (Definition \ref{defn:combi_decoupling_system_of_linear_equations}) of the  triangulated system (Definition \ref{defn:triangulated_system}) of controlled linear constraints in the tailored game $\verifier_n$.    The reason for the name ``output indicator'' is that, when $\linproc^*$ halts, it   outputs either $0$ or $1$, and if it outputted $1$, then the aforementioned input $O$  is the expected output of the operation of the TM.

\begin{definition}[A padded, purified TNFV]\label{defn:padded_purified_NFV}
	Let $\verifier=(\sampler,\length,\linproc,\decider)$ be a  $h$-level tailored normal form verifier, and $\Lambda$ a single input TM that always halts. We say that $\verifier$ is $2^\Lambda$-padded if $|\dec(\length(n,\mttx,\kappa))|=2^{\Lambda(n)}$ regardless of $\mttx$ and $\kappa$. We say that $\verifier$ is purified if the controlled linear constraints in the game $\verifier_n$ (whenever it is well defined) are purely unreadable (Definition \ref{defn:purely_unreadable_system}).
\end{definition}

\begin{definition}[Triangulated output indicator of a linear constraint processor]\label{defn:decider-read}
	The \emph{triangulated output indicator} $\linproc^*$ is a $9$-input TM that takes as input: an $h$-level tailored normal form verifier $\verifier=(\sampler,\length,\linproc,\decider)$; two $1$-input TMs $\Lambda,\Delta$, which  induce (partial) functions $\Lambda,\Delta\colon \mathbb{N}\to \mathbb{N}$; an integer $n$  (in binary); five bit strings $\mttx,\mtty,a^\frR,b^\frR$ and $O$. 
	
	Let us first explain what $\linproc^*$ expects the given inputs to satisfy: $\verifier$ was already said to be an $h$-level normal form verifier. The inputs $n,\mttx,\mtty,a^\frR,b^\frR$ are expected to be, as usual, an index of a game, a pair of questions in this game and a pair of readable answers to these questions, all with respect to $\verifier_n$.  The TM $\Lambda$ is supposed to be  the padding parameter in   $\verifier$; namely, $\linproc^*$ expects the answer length calculator $\length$ to always imply $2^{\Lambda(n)}$-long (readable and linear) answers. The TM $\Delta$ controls the number of padding variables used in the triangulation of an intermediate system induced by $\linproc(n,\mttx,\mtty,a^\frR,b^\frR)$ --- namely, it is the parameter $\Delta$ as in Definition \ref{defn:triangulating_linear_system} for some system, or alternatively the size of the formal generating set $S_\triangle$ as in Definition \ref{defn:combi_decoupling_system_of_linear_equations}. Finally, $O$ is expected to be the encoding as an indicator (Definition \ref{defn:decoupled_equations_and_CNF} and Remark \ref{rem:alg_5-decoupling}) of some $5$-decoupled (Definition \ref{defn:combi_decoupling_system_of_linear_equations}) triangulated linear system induced by $\linproc(n,\mttx,\mtty,a^\frR,b^\frR)$.

	We now describe the operation of the TM $\linproc^*$, namely its  high-level description (Remark \ref{rem:high-level}). To make it easier to follow, we add remarks in each clause of the operation, as well as the running time bounds of the specific step.
	\begin{enumerate}
		\item \underline{The readable answers are of the appropriate length}: $\linproc^*$ calculates $\Lambda(n)$ (by running $\Lambda$ on input $n$), and checks  that 
		\[
		|a^\frR|,|b^\frR|=2^{\Lambda(n)}\ ;
		\]
		if not, it outputs $0$. 
		
		This ensures that the answers are of the length expected by a verifier that is $2^\Lambda$-padded (Definition~\ref{defn:padded_purified_NFV}).  This step takes $\poly(\TIME(\Lambda;n),2^{\Lambda(n)})$ time.\footnote{The TM $\Lambda$ with input $n$  may not halt; in this case $\linproc^*$ too will not halt, but this is consistent with our notation as $\TIME(\Lambda;n)=\infty$.}
		\item \underline{Calculate the system of controlled linear constraints}: In this step, $\linproc^*$ defines a system of linear equations $(\mathscr{A}
		\mid\vec b)$ over $2^{\Lambda(n)+2}$ many variables as follows. First, it calculates $\linproc(n,\mttx,\mtty,a^\frR,b^\frR)$, 
		and then decodes  it (Definition \ref{defn:the_alphabet}). If the result was well structured, namely of the form  $c^1 \sqcup...\sqcup c^k$ where each $c^i$ is a bit string of length $2^{\Lambda(n)+2}+1$, then it lets $(\mathscr{A}\mid  \vec b)$ be the system whose rows are $c^i$ (the first $2^{\Lambda(n)+2}$ bits of each $c^i$ belong to $\mathscr{A}$ and the last bit is the $i^{\rm th}$ value in $\vec b$) --- note that in this case this system has $k$-many equations. Otherwise, it lets $(\mathscr{A}\mid \vec b)$ be the system with $2^{\Lambda(n)+2}$-many variables, and a single equation $0=1$ (i.e., $\mathscr{A}$ is the zero matrix with a single row and $2^{\Lambda(n)+2}$-many columns, and $\vec b$ is the scalar $1$).

		The above choice ensures that $(\mathscr{A}\mid \vec b)$ agrees with the controlled linear constraints $L_{\mttx\mtty}(a^\frR,b^\frR)$ in the game $\verifier_n$ --- note that when the output of $\linproc$ is 
		\textbf{not} well formatted, the canonical decider will surely reject, which is the same as assuming $L_{\mttx\mtty}(a^\frR,b^\frR)$ is the never accepting system $0=1$ --- this is how we defined $\verifier_n$ in Definition \ref{def:normal-game}.\footnote{There, we used the formulation "$L$ outputs $\{\sJ\}$", but this is exactly the subset representation of the unsolvable system $0=1$.} 
		Hence, there is a natural association between the variables of this system and  $S=S^\frR_\mttx\sqcup S^\frL_\mttx\sqcup S^\frR_\mtty\sqcup S^\frL_\mtty$, where
		$S^\frR_\mttx,S^\frL_\mttx,S^\frR_\mtty,S^\frL_\mtty$ are the formal variables at the vertices $\mttx$ and $\mtty$ of $\verifier_n$.  
		This step takes $\poly(\TIME(\linproc;n,\cdot,\cdot,\cdot,\cdot),2^{\Lambda(n)})$ time.
		
		\item \underline{The system of linear equations is purely unreadable}: As $\linproc^*$  recovered a linear system  $(\mathscr{A}\mid \vec b)$  with variables $S^\frR_\mttx\sqcup S^\frL_\mttx\sqcup S^\frR_\mtty\sqcup S^\frL_\mtty$, it can check whether this system is purely unreadable (Definition \ref{defn:purely_unreadable_system}) --- namely, that the columns associated to the variables from $S^\frR_\mttx$ and $S^\frR_\mtty$ are all zeros --- and otherwise output $0$. 
		
		This step takes $O(k\cdot 2^{\Lambda(n)})$ time, and as $k\leq \TIME(\linproc;n,\cdot,\cdot,\cdot,\cdot)$,  the running time of this step is bounded by 
		$$\poly(\TIME(\linproc;n,\cdot,\cdot,\cdot,\cdot),2^{\Lambda(n)})\ .$$

		\item \underline{Extracting the pure system}: $\linproc^*$ removes the columns of  $(\mathscr{A}\mid \vec b)$ associated with $S^\frR_\mttx$ and $S^\frR_\mtty$, which were checked in the previous step to be zero, resulting in a new system $(\mathscr{A}^\frL\mid \vec b)$ on variables $S^\frL_\mttx\sqcup S^\frL_\mtty$. 
		
		As the original system ought to be pure, we do not lose any information by removing the readable columns --- it still encodes the same linear conditions on the unreadable variables that need to be satisfied in $\verifier_n$.  This step again takes time $O(k\cdot 2^{\Lambda(n)})$, which is $\poly(\TIME(\linproc;n,\cdot,\cdot,\cdot,\cdot),2^{\Lambda(n)})$.
		

		\item \underline{There are enough triangulation variables}: $\linproc^*$ calculates $\Delta(n)$ (by calling $\Delta$ on input $n$), and checks that $\Delta(n)\geq k\cdot2^{\Lambda(n)+1}$; otherwise it outputs $0$. 
		
		Note that $k$ is the number of rows and $2^{\Lambda(n)+1}$ is the number of  columns  in $\mathscr{A}^\frL$.  So, this check verifies that there are enough ``extra'' variables in $S_\triangle$ to completely triangulate the system. This check takes $\poly(\TIME(\Delta;n),k,2^{\Lambda(n)})\leq \poly(\TIME(\Delta;n),\TIME(\linproc;n,\cdot,\cdot,\cdot,\cdot),2^{\Lambda(n)})$ time.
		
		\item \underline{Calculate the triangulated system}: $\linproc^*$ calculates the triangulated system (Definition \ref{defn:triangulating_linear_system}) 
		\[
		{\rm Triangle}(\mathscr{A}^\frL,\vec b,\Delta(n))=\big(\mathscr{A}_\triangle\mid\vec b_\triangle\big)\  . 
		\]
		We interpret this system as having variables in $S^\frL_\mttx\sqcup S^\frL_\mtty\sqcup S_\triangle$, where $S^\frL_\mttx$ and $S^\frL_\mtty$ were the original variables, and $S_\triangle$ is a new set of $\Delta(n)$-many variables.
		
		Note that, as Remark \ref{rem:prop_triangulated_system} states, this computation takes at most $\poly(k,2^{\Lambda(n)},\Delta(n))$ time. 
		

		\item \underline{The decoupling of the triangulated system}: As $\linproc^*$ recovered the triangulated system $(\mathscr{A}_\triangle\mid\vec b_\triangle)$ with variables in  $S^\frL_\mttx\sqcup S^\frL_\mtty\sqcup S_\triangle$, it can apply $5$-decoupling $\frak{DeCouple}(S^\frL_\mttx,S^\frL_\mtty,S_\triangle,(\mathscr{A}_\triangle\mid\vec b_\triangle))$ (Definition \ref{defn:combi_decoupling_system_of_linear_equations}) to it,  resulting in a $5$-decoupled system of equations over blocks $S^\frL_\mttx,S^\frL_y,S_1,S_2,S_3$, where $S_1,S_2,S_3\cong S^\frL_\mttx\sqcup S^\frL_\mtty\sqcup S_\triangle$ (that is, each $S_i$ is a copy of $S^\frL_\mttx\sqcup S^\frL_\mtty\sqcup S_\triangle$). As described in Definition \ref{defn:decoupled_equations_and_CNF}, such $5$-decoupled system of equations can be encoded as an indicator map on $S^\frL_\mttx\times S^\frL_\mtty\times S_1\times S_2\times S_3\times \FF_2^6$.
		By recalling that $|S_\mttx^\frL|=|S_\mtty^\frL|=2^{\Lambda(n)}$ and letting 
		\begin{equation}\label{eq:defn_of_diamondsuit(n)}
		\diamondsuit(n)=\lceil \log |S_i|\rceil=\lceil \log (2^{\Lambda(n)+1}+\Delta(n))\rceil\ ,
		\end{equation}
		the set $S^\frL_\mttx\times S^\frL_\mtty\times S_1\times S_2\times S_3\times \FF_2^6$ naturally embeds into $\FF_2^{2\Lambda(n)+3\diamondsuit(n)+6}$. Hence, the decoupled system is encoded as a map $U\colon \FF_2^{2\Lambda(n)+3\diamondsuit(n)+6}\to \FF_2$, which is a bit string of length $2^{2\Lambda(n)+3\diamondsuit(n)+6}$. 
		
		In the notation of Remark \ref{rem:alg_5-decoupling}, $U=\frak{DeCouple}(S^\frL_\mttx,S^\frL_\mtty,S_\triangle,(\mathscr{A}_\triangle\mid\vec b_\triangle))$, and calculating it takes $\poly(k,2^{\Lambda(n)},\Delta(n))$-time.

		\item \underline{The input $O$ matches the expected calculation}: Finally, $\linproc^*$ outputs $1$ \textbf{only if} $O=U$.
		
		This ensures that the input $O$ matches the encoding of the decoupled, triangulated, purely unreadable controlled linear constraints of $\verifier_n$ given $\mttx,\mtty,a^\frR,b^\frR$. This takes at most $|\cal{U}|=2^{2\Lambda(n)+3\diamondsuit(n)+6}$ steps, which is $\poly(2^{\Lambda(n)},\Delta(n))$.
	\end{enumerate}
\end{definition}

\begin{claim}[Properties of $\linproc^*$]\label{claim:properties_of_L*}\label{cor:condition_1_for_Vn_accepting}
	Let 
	\begin{itemize}[noitemsep,topsep=0pt,parsep=0pt,partopsep=0pt]
		\item[--] $\Lambda$ be a single input TM that always halts;
		\item[--] $\verifier=(\sampler,\length,\linproc,\decider)$  a purified $2^\Lambda$-padded  $h$-level TNFV (Definition \ref{defn:padded_purified_NFV}) such that $\verifier_n$ is well defined for every $n$ (Definition \ref{defn:h-level_NFV});
		\item[--] $\Delta$ an always halting single input TM that satisfies $\Delta(n)\geq \TIME(\linproc;n,\cdot,\cdot,\cdot,\cdot)\cdot 2^{\Lambda(n)+1}$ and induces $\diamondsuit(n)$ as in \eqref{eq:defn_of_diamondsuit(n)};
		\item[--]  $n$ a positive integer;
		\item[--]  $\mttx,\mtty$ two bit strings of length $r(n)=\sampler(n,{\rm Dimension},\cdot,\cdot,\cdot,\cdot)$.
	\end{itemize}
	Then:
	\begin{enumerate}
		\item For every $a^\frR,b^\frR\colon \FF_2^{\Lambda(n)}\to \FF_2$, there \textbf{exists} a \textbf{unique} $O\colon \FF_2^{2\Lambda(n)+3\diamondsuit(n)+6}\to \FF_2$ such that  \begin{equation}\label{eq:L*=1_condition}
		\linproc^*(\verifier,\Lambda,\Delta,n,\mttx,\mtty,a^\frR,b^\frR,O)=1    \ .
		\end{equation}
		In addition,  $O$ is the encoding of the $5$-decoupled system of linear equations 
		\begin{equation}\label{eq:the_decoupled_system_O_encodes}
		\frak{DeCouple}(S^\frL_\mttx,S^\frL_\mtty,S_\triangle,{\rm Triangle}(\mathscr{A}^\frL,\vec b,\Delta(n)))\ ,
		\end{equation}
		where $(\mathscr{A}^\frL\mid \vec b)$ is the purely unreadable part of the system of controlled linear constraints  $L_{\mttx\mtty}(a^\frR,b^\frR)$ from $\verifier_n$.
		\item The quadruple $a^\frR,a^\frL,b^\frR,b^\frL\colon \FF_2^{\Lambda(n)}\to \FF_2$  is accepted in the game $\verifier_n$ given $\mttx,\mtty$ were asked if and only if the $5$-tuple $\mathsf{Extend}(a^\frL,b^\frL)$ (Corollary \ref{cor:triang_and_decoupling_extend}) satisfies the $5$-decoupled system of linear equations defined by (the unique) $O$ which satisfies \eqref{eq:L*=1_condition}.
	\end{enumerate}
	In addition the running time of the triangulated output indicator $\linproc^*$ satisfies 
	\begin{equation}\label{eq:time_bound_L*}
	\TIME(\linproc^*;\verifier,\Lambda,\Delta,n,\cdot,\cdot,\cdot,\cdot,\cdot)=\poly(\TIME(\Lambda;n),\TIME(\Delta;n),\TIME(\linproc;n,\cdot,\cdot,\cdot,\cdot),2^{\Lambda(n)},\Delta(n))\;.
	\end{equation} 
\end{claim}

\begin{proof}
	Let us start by proving the first clause.     By construction, $\linproc^*$ accepting (i.e., outputs $1$) implies $O$ is the encoding of the $5$-decoupling of the triangulation of the purely unreadable part of $L_{\mttx\mtty}(a^\frR,b^\frR)$  --- note that this uses the assumption  $\verifier$ is $2^{\Lambda}$-padded, as otherwise the recovered $(\mathscr{A}\mid \vec b)$ in step $2.$~of the operation of $\linproc^*$ is not $L_{\mttx\mtty}(a^\frR,b^\frR)$.
	In the other direction, note that as $a^\frR,b^\frR$ are of length $2^{\Lambda(n)}$ (which makes step $1.$~in the operation of $\linproc^*$  not reject, i.e., not output $0$), $\verifier$ is purified (which makes step $3.$~not reject), and $\Delta(n)\geq \TIME(\linproc;n,\cdot,\cdot,\cdot,\cdot)\cdot 2^{\Lambda(n)+1}$ (which makes step $5.$~not reject), the triangulated output indicator $\linproc^*$ will run step $7.$~and recover $U\colon \FF_2^{2\Lambda(n)+3\diamondsuit(n)+6}\to \FF_2$. Hence, by choosing $O=U$, the triangulated output indicator $\linproc^*$ will output $1$ on the chosen input. 
	
	Let us prove the second clause. The tuple $a^\frR,a^\frL,b^\frR,b^\frL$ is accepted in the tailored game $\verifier_n$ given $\mttx,\mtty$ were asked if and only if they satisfy the system $L_{\mttx\mtty}(a^\frR,b^\frR)$. As this system is purely unreadable, the quadruple satisfies it if and only if the unreadable parts $a^\frL,b^\frL$ satisfy the purely unreadable part of the system. Therefore, by Corollary \ref{cor:triang_and_decoupling_extend}, this happens if and only if the $5$-tuple $\mathsf{Extend}(a^\frL,b^\frL)$ satisfy the system \eqref{eq:the_decoupled_system_O_encodes} encoded by $O$.

	For the running time, one can follow the running time bounds calculated along the description.
\end{proof}

\subsubsection{Decoupled Cook--Levin: Generating a succinct description of the triangulated output indicator of a linear constraint processor}\label{sec:decoupled_Cook-Levin}

This section is dedicated to applying a version of the scaled up Cook--Levin transformation (Theorem \ref{thm:scaled_up_cook_leving}) to the triangulated output indicator $\linproc^*$ described in Definition \ref{defn:decider-read}, which results in a circuit $\circuit$ which encodes a decoupled succinct-6SAT instance that  describes the operation of $\linproc^*$, on which PCP techniques can be applied.

Let us elaborate. Recall the definition of a Boolean circuit $\circuit$ in Definition~\ref{defn:Circuits}, and of the function $P_\circuit:\F_2^I\to\F_2^O$ encoded by $\circuit$. Recall also the notion of a circuit succinctly encoding a formula from Definition~\ref{defn:Succinct_3SAT}, and more importantly the decoupled version from Definition \ref{defn:decoupled_equations_and_CNF}. The next definition describes what it means for a circuit $\circuit$ to succinctly describe the operation of the triangulated output indicator $\linproc^*(\verifier,\Lambda,\Delta,n,\mttx,\mtty,\cdot,\cdot,\cdot)$, where all the labeled inputs are considered fixed parameters (hardwired to the operation of $\linproc^*$) and the dotted-inputs are considered variables of the formula (and hence inputs to the circuit). 

The last three inputs of $\linproc^*$ are $a^\frR,b^\frR$ and $O$, which have length $2^{\Lambda(n)}$, $2^{\Lambda(n)}$ and $2^{2\Lambda(n)+3\diamondsuit(n)+6}$ respectively. 
The circuit $\circuit$ will have a block of input gates associated with each of these inputs,\footnote{The notion of a block of variables was described in Definition \ref{defn:decoupled_equations_and_CNF}.} of size $\Lambda(n)$, $\Lambda(n)$ and $2\Lambda(n)+3\diamondsuit(n)+6$ respectively. Each block of input gates can be used to address a single $\FF_2$-value of the corresponding function.
In addition we include three blocks of $M(n)$ input gates each, where $M\colon \mathbb{N}\to \mathbb{N}$ is some function to be fixed later, and $6$ blocks of $1$ input gate. The three blocks of $M(n)$ gates receive variables that are supposed to specify intermediate, internal values used in the computation of $\linproc^*$ --- e.g., the values of the variables described in Theorem \ref{thm:Cook-Levin}.\footnote{These last three  blocks of variables  in  $\varphi_\circuit$  play an analogous role to the three copies of the original variables  inserted in   the combinatorial decoupling from Definition~\ref{defn:combi_decoupling_system_of_linear_equations}.}
The resulting circuit $\circuit$ is a succinct encoding of a $6$-decoupled CNF $\varphi_\circuit$ that encodes the claim that $\linproc^*$ accepts a triple $(a^\rvar,b^\rvar,O)$, in the sense that this is the case if and only if $(a^\rvar,b^\rvar,O)$ can be completed to a ``proof'' $(a^\rvar,b^\rvar,O,w,w',w'')$ that satisfies the formula $\varphi_\circuit$. 
What we gain from this is that running $\linproc^*$ can take exponential time in principle, while the  succinct representation circuit has polynomial size, can be calculated in polynomial time, and can be verified to be satisfiable in polynomial time using PCP techniques (Section \ref{sec:prelude_PCP}).

\begin{definition}[Succinct description of $\linproc^*$]\label{def:succinct-bounded-perm}
	Let $\verifier=(\sampler,\length,\linproc,\decider)$ be a   $h$-level TNFV,
	$\Lambda,\Delta$  $1$-input TMs that induce (partial) functions $\mathbb{N}\to \mathbb{N}$, $n$ a natural number, and $\mttx,\mtty$ bit strings. In addition, let  $T,M,s\colon \mathbb{N}\to \mathbb{N}$ be functions.
	
	A circuit~$\circuit$ is said to \emph{succinctly describe}  $\linproc^*(\verifier,\Lambda,\Delta,n,\mttx,\mtty,\cdot,\cdot,\cdot)$
	with parameters $({T}(n),{M(n)},s(n))$ if: 
	\begin{enumerate}
		\item It has $s(n)$ many non-input wires, and $4\Lambda(n)+3\diamondsuit(n)+3{M}(n)+12$ many input gates (and thus input wires) collected as blocks of sizes 
		\[
		\Lambda(n),\Lambda(n),2\Lambda(n)+3\diamondsuit(n)+6,{M}(n),{M}(n),{M}(n),1,1,1,1,1,1
		\;,
		\]
		where $\diamondsuit(n)=\lceil \log(2^{\Lambda(n)+1}+\Delta(n))\rceil$, as it was  in Definition \ref{defn:decider-read}.
		
		Thus, $\circuit$ defines   a $6$-decoupled CNF $\varphi_\circuit$ (Definition \ref{defn:decoupled_equations_and_CNF}) on  $6$ blocks of  variables parametrized by 
		\[
		\FF_2^{\Lambda(n)},\FF_2^{\Lambda(n)},\FF_2^{2\Lambda(n)+3\diamondsuit(n)+6},\FF_2^{{M}(n)},\FF_2^{{M}(n)},\FF_2^{{M}(n)}\;.
		\]
		\item Fix $a^\frR,b^\frR\colon  \FF_2^{\Lambda(n)}\to \FF_2$ and $O\colon \FF_2^{2\Lambda(n)+3\diamondsuit(n)+6}\to \FF_2$.  Then, there are $w,w',w''\colon  \FF_2^{{M}(n)}\to \FF_2$ such that 
		\begin{equation}\label{eq:circuit_condition}
		\varphi_\circuit(a^\frR,b^\frR,O,w,w',w'')=1
		\end{equation}
		if and only if 
		the triangulated output indicator  $\linproc^*$ (Definition \ref{defn:decider-read}) outputs $1$ on input $(\verifier, \Lambda,\Delta,n,\mttx,\mtty,a^\frR,b^\frR,O)$  in \textbf{time at most} ${T}(n)$.  
	\end{enumerate}
	
\end{definition}
\begin{remark}\label{rem:blarembla}
	Unsurprisingly, the circuit in Definition \ref{def:succinct-bounded-perm} is very similar to the one produced by the scaled up Cook--Levin (Theorem \ref{thm:scaled_up_cook_leving})  to resolve $\BinaryTimeHalt$ (Definition \ref{defn:time_restricted_halting}) for the instance $(\linproc^*(\verifier,\Lambda,\Delta,n,\mttx,\mtty,\cdot,\cdot,\cdot),T(n))$ (with some bounds on the various parameters of the resulting circuit). Indeed, the next proposition is just a careful application of the scaled up Cook--Levin theorem to this specific instance of $\BinaryTimeHalt$, but which results, as needed in  Definition \ref{def:succinct-bounded-perm}, with the encoding of a $6$-decoupled CNF instead of a $3$CNF.
\end{remark}

\begin{proposition}[Algorithmic generation of a succinct description for the triangulated output indicator $\linproc^*$]
	\label{prop:explicit-padded-succinct-deciders}
	There is a TM  $\paddedsuccinctdecider$ (``succinct triangulated output indicator'') that takes as input
	\[
	(\verifier,\Lambda,\Delta,D,T,{Q}, n,\mttx,\mtty)\;,
	\]
	where $\verifier=(\sampler,\length,\linproc,\decider)$ is (the encoding of) an $h$-level TNFV,  $\Lambda,\Delta,T$ and ${Q}$ (the encodings of) always halting $1$-input TMs, $D$ and $n$  positive integers and $\mttx,\mtty$  bit-strings, and outputs a tuple $(M,s,\circuit)$ consisting of two (binary) integers    --- which we call the \emph{block size} $M$ and the \emph{number of non-input wires} $s$ for reasons to be understood later ---  and (the encoding of) a circuit $\circuit$, such that:
	\begin{enumerate}[label=\textcolor{black}{(\arabic*)}, ref= (\arabic*)]
		\item \underline{Properties of the block size and number of non-input wires}: The integers $M$ and $s$ are independent of the inputs  $\verifier$, $\Lambda$, $\Delta$, $\mttx$, $\mtty$, and we have 
		\begin{equation}\label{eq:size_bound_on_M_and_s}
		M,s =  \poly(\log(T(n)),Q(n),D)\ .
		\end{equation}
		\item \underline{Runtime bound}: We have
		\begin{equation}\label{eq:runtime_bound_succinctTOI}
		\begin{split}
		&\TIME(\paddedsuccinctdecider;\verifier,\Lambda,\Delta,D,T,Q, n,\mttx,\mtty)=\\
		&\poly(\TIME(\Lambda;n),\TIME(\Delta;n),\TIME(T;n),\TIME(Q;n),n,\log(T(n)) ,Q(n),D)\;.
		\end{split}
		\end{equation}
		In particular, the runtime is independent of $\verifier,\mttx$ and $\mtty$ (which means it may not even read them completely). 
		
		\item \label{clause3:prop_of_output_circuit_succinctTOI} \underline{Properties of the resulting circuit}: 
		If 
		\begin{equation}\label{eq:conditions_on_various_parameters_in_succinctTOI}
		|\mttx|,|\mtty|\leq Q(n)\quad ,\quad 2\Lambda(n)+3\diamondsuit(n)+6\leq \log(T(n))\quad {\rm and}\quad  |\verifier|,|\Lambda|,|\Delta|,|T|,|Q|\leq D\ ,
		\end{equation}
		where $\diamondsuit(n)=\lceil \log(2^{\Lambda(n)+1}+\Delta(n))\rceil$ as before, 
		then  the output circuit $\circuit$  succinctly describes
		$\linproc^*(\verifier,\Lambda,\Delta,n,\mttx,\mtty,\cdot,\cdot,\cdot)$ with parameters $(T(n),M,s)$  (Definition \ref{def:succinct-bounded-perm}). In particular, $\circuit$ has $s$ many non-input wires, and $4\Lambda(n)+3\diamondsuit(n)+3{M}+12$ many input gates collected as blocks of sizes 
		\[
		\Lambda(n),\Lambda(n),2\Lambda(n)+3\diamondsuit(n)+6,{M},{M},{M},1,1,1,1,1,1
		\;.\]
	\end{enumerate}
	
\end{proposition}

\begin{remark}
	Throughout this section, we keep the inputs $\Lambda,\Delta,D,T$ and $Q$ fixed, in which case the $M$ and $s$ calculated by $\paddedsuccinctdecider$ depend only on $n$, and we use the notation $M(n)$ and $s(n)$ for them. 
\end{remark}

\begin{proof}[Proof sketch of Proposition \ref{prop:explicit-padded-succinct-deciders}]
	Let us start with a short discussion on how $\circuit$ is constructed, so that the conditions from \eqref{eq:conditions_on_various_parameters_in_succinctTOI},  as well as the fact $M$ and $s$ can be taken to be of size \eqref{eq:size_bound_on_M_and_s}, will  become  clearer to the reader.  
	As mentioned in the proof of Theorem \ref{thm:Cook-Levin}, the first step in the Cook--Levin transformation is to describe the variables on which $\varphi_\circuit$ will be defined. 
	These variables control the contents of the tapes at each time step up to $T(n)$, as well as the head position and the internal state of the TM. The number of these variables is polynomial in the number of time steps --- $T(n)$ in our case --- and the description length of the appropriate TM --- $|\linproc^*(\verifier,\Lambda,\Delta,n,\mttx,\mtty,\cdot,\cdot,\cdot)|$ in our case. 
	As shown in \eqref{eq:size_bound_L*_with_fixed_vars}, this number is bounded by $\poly(T(n),Q(n),D)$. 
	There are more variables that need to be added to the formula, but it turns out that the number of them is also polynomial in $T(n)$ and $|\linproc^*(\verifier,\Lambda,\Delta,n,\mttx,\mtty,\cdot,\cdot,\cdot)|$. 
	All in all, the formula needs $\poly(T(n),Q(n),D)$-many variables,  
	and hence $\polylog(T(n),Q(n),D)$-sized blocks of input gates suffice.  As $\polylog(T(n),Q(n),D)$ is smaller than $\poly(\log(T(n)),Q(n),D)$, we can indeed choose $M$ as above and have enough flexibility to encode the required number of variables. 
	Similarly, the number of non-input wires in $\circuit$ can be taken to be $\poly(\log(T(n)),Q(n),D)$ as well. 
	The fact that the assignments $a^\frR,b^\frR,O$ for the variables induced by the first three blocks of $\circuit$ come from inputs that make $\linproc^*(\verifier,\Lambda,\Delta,n,\mttx,\mtty,\cdot,\cdot,\cdot)$ halt and output $1$ in at most $T(n)$ steps, can be enforced in a straightforward manner: There are variables of the formula $\varphi_\circuit$ that control the values of the input tapes at time $0$, and we can just force equality of these variables with those controlling the values of  $a^\frR,b^\frR$ and $O$ in a succinct way. 
	The high-level description (Remark \ref{rem:high-level}) of the algorithm $\paddedsuccinctdecider$ is thus as follows:
	\begin{itemize}
		\item First, it runs $\Lambda,\Delta,T$ and $Q$ on input $n$ to recover the values $\Lambda(n),\Delta(n),T(n)$ and $Q(n)$. This takes time at most 
		\[
		\poly(\TIME(\Lambda;n),\TIME(\Delta;n),\TIME(T;n),\TIME(Q;n))\ .
		\] 
		\item  
		The algorithm then calculates $M$ and $s$. 
		We do not describe this calculation in detail, but these are just fixed polynomials in $\log(T(n)),Q(n)$ and $D$ (as needed for \eqref{eq:size_bound_on_M_and_s} to hold), which are large enough to  play the role of the variables block size  and number of non-input wires in the output of a decoupled version of the scaled up Cook--Levin Theorem \ref{thm:scaled_up_cook_leving}, as sketched above. As this is just the calculation of fixed polynomials, it takes at most polynomial time in the \textbf{bit length} of the appropriate numbers to calculate them, namely 
		\[
		\polylog(\log(T(n)),Q(n),D)\ .
		\]
		Together with the previous step, this already shows that calculating $M$ and $s$ takes no more than the time bound from \eqref{eq:runtime_bound_succinctTOI}.
		\item After that, $\paddedsuccinctdecider$ checks whether the conditions from \eqref{eq:conditions_on_various_parameters_in_succinctTOI} are satisfied, namely \[
		|\mttx|,|\mtty|\leq Q(n)\quad ,\quad 2\Lambda(n)+3\diamondsuit(n)+6\leq \log(T(n))\quad {\rm and}\quad  |\verifier|,|\Lambda|,|\Delta|,|T|,|Q|\leq D\ .
		\]
		This takes at most $\poly(\log(T(n)),Q(n),D)$ time. If these conditions are not satisfied, then  there are no guarantees on the output circuit $\circuit$, and $\paddedsuccinctdecider$ can just output some fixed constant sized circuit (e.g., the empty circuit). 
		\item Otherwise, $\circuit$ needs to succinctly describe $\linproc^*$ with the inputs $\verifier,\Lambda,\Delta,n,\mttx,\mtty$ fixed, and with parameters $(T(n),M,s)$. To that end, it first needs to calculate the description of the $3$-input TM $\linproc^*(\verifier,\Lambda,\Delta,n,\mttx,\mtty,\cdot,\cdot,\cdot)$ --- by Fact \ref{fact:properties_of_TMs} and using $|\linproc^*|=O(1)$, we have
		\begin{equation}\label{eq:size_bound_L*_with_fixed_vars}
		|\linproc^*(\verifier,\Lambda,\Delta,n,\mttx,\mtty,\cdot,\cdot,\cdot)|=\poly(|\verifier|,|\Lambda|,|\Delta|,|n|,|\mttx|,|\mtty|)=\poly(D,\log(n),Q(n))\ ,
		\end{equation}
		and calculating this description takes $\poly(D,\log(n),Q(n))$-time as well. 
		
		As mentioned in Remark \ref{rem:blarembla}, the question of whether $\linproc^*(\verifier,\Lambda,\Delta,n,\mttx,\mtty,\cdot,\cdot,\cdot)$ will output $1$ in at most $T(n)$ time steps is a $3$-input version of the decision problem $\BinaryTimeHalt$ (Definition \ref{defn:time_restricted_halting}). 
		So, it is natural to apply on it some variant of the Cook--Levin theorem, as this theorem describes a transformation from pairs of a TM and time bound (in binary) to a circuit $\circuit$, such that satisfiability of the formula $\varphi_\circuit$  is associated with the TM indeed outputting $1$ in the respective number of time steps. 
		As opposed to the scaled up Cook--Levin theorem that we described in Theorem \ref{thm:scaled_up_cook_leving}, where the output circuit $\circuit$ succinctly encodes a 3SAT  instance  $\varphi_\circuit$ which is not decoupled, here we expect the resulting $\circuit$ to encode a $6$-decoupled formula with some extra properties, that we describe now:
		
		First of all,  the first three blocks of inputs in the circuit $\circuit$ (of sizes $\Lambda(n),\Lambda(n)$ and $2\Lambda(n)+3\diamondsuit(n)+6$) induce three blocks of variables in the decoupled formula $\varphi_\circuit$
		(of sizes $2^{\Lambda(n)},2^{\Lambda(n)}$ and $2^{2\Lambda(n)+3\diamondsuit(n)+6}$), and we expect the assignments to these blocks of variables to  ``remember'' the appropriate inputs to the TM that make it halt and output $1$.
		Namely, if $a^\frR,b^\frR,O$ are the assignments to these blocks of generators of $\varphi_\circuit$, then completing  them to  a satisfying assignment for the formula $\varphi_\circuit$ needs to be possible only if $\linproc^*(\verifier,\Lambda,\Delta,n,\mttx,\mtty,\cdot,\cdot,\cdot)$ halts in $T(n)$ time steps, where the three dots are replaced by $a^\frR,b^\frR,O$  --- this is exactly the condition phrased in \cref{clause3:prop_of_output_circuit_succinctTOI} of Definition \ref{def:succinct-bounded-perm}. 
		In addition, we need the number of input gates in $\circuit$ and the number of non-input wires in it to be of specific sizes, $4\Lambda(n)+3\diamondsuit(n)+3M+12$ and $s$ respectively. 
		Both conditions can be achieved by, for example, adapting the proofs  from~\cite[Sections 10.2 and 10.3]{MIPRE}.

	\end{itemize}

\end{proof}

\begin{remark}
	We decided not to include a full proof of the version of the Cook--Levin Theorem  \ref{thm:scaled_up_cook_leving} required for Proposition \ref{prop:explicit-padded-succinct-deciders}. This is mainly because this transformation is fairly standard, and we tried to provide enough information for the reader to be able to reconstruct this for themselves. Moreover,  the authors of \cite{MIPRE} {did} include a version, and anyone who seeks to prove Proposition \ref{prop:explicit-padded-succinct-deciders} can adapt their version to imply the above.
\end{remark}

Recall that in Claim \ref{cor:condition_1_for_Vn_accepting} we described an equivalent condition for the game $\verifier_n$ accepting the answers $a^\frR,a^\frL,b^\frR,b^\frL$ given the questions $\mttx,\mtty$ were asked. One of the conditions \eqref{eq:L*=1_condition} was for $\linproc^*$ to halt and output $1$ on a certain tuple of inputs. The whole goal of Proposition \ref{prop:explicit-padded-succinct-deciders} was to replace a check such as  \eqref{eq:L*=1_condition} by the satisfiability of a formula succinctly represented by a circuit. Hence, the following is obtained:

\begin{corollary} [Condition for $\verifier_n$ accepting]\label{cor:functional_viewpoint_succinct_description}
	Let 
	\begin{itemize}[noitemsep,topsep=0pt,parsep=0pt,partopsep=0pt]
		\item[--] $\Lambda$ be a single input TM that always halts;
		\item[--] $\verifier=(\sampler,\length,\linproc,\decider)$  a purified $2^\Lambda$-padded  $h$-level TNFV (Definition \ref{defn:padded_purified_NFV}) such that $\verifier_n$ is well defined for every $n$ (Definition \ref{defn:h-level_NFV});
		\item[--] $\Delta$ an always halting $1$-input TM that satisfies $\Delta(n)\geq \TIME(\linproc;n,\cdot,\cdot,\cdot,\cdot)\cdot 2^{\Lambda(n)+1}$ and induces $\diamondsuit(n)$ as in \eqref{eq:defn_of_diamondsuit(n)};
		\item [--] $Q$ an always halting $1$-input TM  satisfying 
		$\TIME(\sampler;{n},\cdot,\cdot,\cdot,\cdot,\cdot)\leq {Q}(n)$;
		\item [--] $T$ an always halting $1$-input TM  satisfying 
		\begin{equation}\label{eq:value_of_T(n)}    T(n)\geq c\cdot(\TIME(\Lambda;n)^c+\TIME(\Delta;n)^c+2^{c\cdot\Lambda(n)}+\Delta(n)^c+\TIME(\linproc;n,\cdot,\cdot,\cdot,\cdot)^c)\ ,
		\end{equation}
		where $c\geq 6$ is the positive integer implied by the $\poly$ notation in \eqref{eq:time_bound_L*};
		\item [--] $D$ a positive integer (in binary) satisfying 
		$|\verifier|,|\Lambda|,|\Delta|,|T|,|Q|\leq D$;
		\item[--]  $n$ a positive integer;
		\item[--]  $\mttx,\mtty$ two bit strings of length $r(n)=\sampler(n,{\rm Dimension},\cdot,\cdot,\cdot,\cdot)$.
		\item[--] $
		(M(n),s(n),\circuit)=\paddedsuccinctdecider(\verifier,\Lambda,\Delta,D,T,Q,n,\mttx,\mtty),
		$
		where $\paddedsuccinctdecider$ was defined in Proposition \ref{prop:explicit-padded-succinct-deciders}.
	\end{itemize}
	Then:
	\begin{enumerate}[label=\textcolor{black}{(\arabic*)}, ref= (\arabic*)]
		\item For every $a^\frR,b^\frR\colon \FF_2^{\Lambda(n)}\to \FF_2$, there are functions 
		\[
		O\colon \FF_2^{2\Lambda(n)+3\diamondsuit(n)+6}\to \FF_2\quad {\rm and}\quad w,w',w''\colon \FF_2^{M(n)}\to \FF_2 \ ,
		\]
		such that  \begin{equation}\label{eq:phi_circuit_is_satisfied_by_6_tuple}
		\varphi_\circuit(a^\frR,b^\frR,O,w,w',w'')=1    \  .
		\end{equation}
		In addition, the $O$ above is the same as in Claim \ref{claim:properties_of_L*} which encodes the $5$-decoupled system in \eqref{eq:the_decoupled_system_O_encodes}. Let us denote by $\mathsf{Prove}_\circuit$ the mapping that takes $a^\frR,b^\frR$ as inputs and outputs a  $6$-tuple $(a^\frR,b^\frR,O,w,w',w'')$ which satisfies $\varphi_\circuit$ (i.e., \eqref{eq:phi_circuit_is_satisfied_by_6_tuple}).
		
		\item \label{clause2:condition_for_V_n_accepting} \emph{Completeness}: If the quadruple $a^\frR,a^\frL,b^\frR,b^\frL\colon \FF_2^{\Lambda(n)}\to \FF_2$  is accepted in the game $\verifier_n$ given $\mttx,\mtty$, then the $6$-tuple $\mathsf{Prove}_\circuit(a^\frR,b^\frR)$ satisfies $\varphi_\circuit$ and the $5$-tuple $\mathsf{Extend}(a^\frL,b^\frL)$ (Corollary \ref{cor:triang_and_decoupling_extend}) satisfies the $5$-decoupled system of linear equations defined by $O$, the third entry of the tuple $\mathsf{Prove}_\circuit(a^\frR,b^\frR)$. 
		
		\emph{Soundness}: On the other hand, if the $6$-tuple $(f^\frR_1,f^\frR_2,f_0,f^\frR_3,f^\frR_4,f^\frR_5)$ satisfies $\varphi_\circuit$, and $(f^\frL_1,f^\frL_2,f^\frL_3,f^\frL_4,f^\frL_5)$ satisfies the $5$-decoupled linear system  induced by $f_0$, then the quadruple $f^\frR_1,f^\frL_1,f^\frR_2,f^\frL_2$ is accepted in the game $\verifier_n$ given $\mttx,\mtty$ were asked.
	\end{enumerate}
\end{corollary}
\begin{proof}
	This is immediate from Claim \ref{cor:condition_1_for_Vn_accepting}, Definition \ref{def:succinct-bounded-perm} and Proposition \ref{prop:explicit-padded-succinct-deciders}.
\end{proof}

\subsubsection{Converting the succinct descriptions into PCPs}
\label{sec:pcpverifier}

In this section we translate the condition from \cref{clause2:condition_for_V_n_accepting} of Corollary \ref{cor:functional_viewpoint_succinct_description}, which is equivalent to the tuple $a^\frR,a^\frL,b^\frR,b^\frL$ passing the  $n^{\rm th}$ game induced by a verifier $\verifier$ (given that it was padded and purified and satisfies certain bounds on its running time and output lengths) when asked $\mttx,\mtty$,  to somewhat longer conditions that can be checked probabilistically. Namely, we construct an appropriate PCP.

Compare this situation to the one we had  in Observation \ref{obs:sufficient_condition_S3SAT}. There, we had a single assignment $\psi\colon\FF_2^n\to \FF_2$ and we wanted to verify that it satisfies    the $3$CNF formula $\varphi_\circuit$ induced by some circuit $\circuit$. To that end,  we expected to have a single low degree polynomial $g_\psi$ over $\FF_q$ (some field extension of $\FF_2$) that plays the role of $\psi$, and a swath of helper polynomials of two types --- $\alpha$'s that verified that indeed the formula is satisfied, and $\beta$'s that verify that $g_\psi$ is an assignment (Definition \ref{def:zero_on_subcube_and_assignments}). 
Here, we have two conditions to be checked instead of one, and we have $11$ assignments instead of one --- a $6$-tuple $(f^\frR_1,f^\frR_2,f_0,f^\frR_3,f^\frR_4,f^\frR_5)$ needs to satisfy $\varphi_\circuit$, and  a $5$-tuple $(f^\frL_1,f^\frL_2,f^\frL_3,f^\frL_4,f^\frL_5)$ needs to satisfy the linear system induced by $f_0$. Still, we can have an $11$-tuple of polynomials $g^\cdot_\cdot$  over some field extension $\FF_q$ that play the role of the $f^\cdot_\cdot$, and helper polynomials of, again, two types --- one type checks that indeed the formula and system of linear equations are satisfied by the  tuple, and the other type checks that the $g^\cdot_\cdot$'s are assignments. 

As the PCP consists of polynomials, let us recall and describe some notations. The notation $\FF_q[S]$ is the space of polynomials with variables from $S$ and coefficients in $\FF_q$. Furthermore, the \emph{evaluation table} of $f\in \FF_q[S]$ is the $\Phi_{\FF_q}$-image of it \eqref{def:Phi_FF}, namely the function that takes an  $\FF_q$-assignment to the variables in $S$ and evaluates the polynomial accordingly. In our PCP, the different polynomials are expected to have various, not necessarily disjoint, variable sets $S$. 

\begin{definition}[Blocks of variables associated to the circuit $\circuit$]\label{defn:blocks_of_vars_circuit}
	Let $\Lambda,\diamondsuit,s$ and $M$ be positive integers. In addition, let $\circuit$ be a circuit with $s$ many non-input wires, as well as $4\Lambda+ 3\diamondsuit+3M+12$ input gates collected in blocks of size\footnote{This choice should be compared to Definition \ref{def:succinct-bounded-perm}. Note that we put the third block of the original circuit as the first block here. This is done so that the notations will be easier to follow. }
	\begin{equation}\label{eq:input_blocks_of_the_circuit}
	2\Lambda+3\diamondsuit+6,\Lambda,\Lambda,{M},{M},{M},1,1,1,1,1,1
	\;.
	\end{equation}
	Hence, the Tseitin polynomial $T_\circuit$ (Definition \ref{defn:Circuits}) has a total of 
	\begin{equation}\label{eq:size_of_m_PCPs}
	m=4\Lambda+3\diamondsuit+3M+12+s\    
	\end{equation}
	many variables, which we can collect in blocks of sizes 
	\begin{equation}\label{eq:decomposition_of_S}
	2\Lambda+3\diamondsuit+6,\Lambda,\Lambda,{M},{M},{M},1,1,1,1,1,1,s\;.
	\end{equation}
	We further decompose the first block, that consists of $2\Lambda+3\diamondsuit+6$ variables, into $11$ blocks of sizes 
	\begin{equation}\label{eq:decomposition_of_S_0}
	\Lambda,\Lambda,\diamondsuit,\diamondsuit,\diamondsuit,1,1,1,1,1,1\ .
	\end{equation}
	In total, we have $23$ blocks of variables. We now give to each block of variables a name. The block of size $2\Lambda+3\diamondsuit+6$ that we decomposed as in \eqref{eq:decomposition_of_S_0} will be denoted as $S_0$, and it is the disjoint union of the $11$ blocks of variables\footnote{For now it should not be clear why we use the linear part notation for these blocks.}
	\[
	S^\frL_1,S^\frL_2,S^\frL_3,S^\frL_4,S^\frL_5,S^\frL_6,S^\frL_7,S^\frL_8,S^\frL_9,S^\frL_{10},S^\frL_{11}\ .
	\]
	The other $12$ blocks from \eqref{eq:decomposition_of_S} will be denoted by 
	\[
	S^\frR_1,S^\frR_2,S^\frR_3,S^\frR_4,S^\frR_5,S^\frR_6,S^\frR_7,S^\frR_8,S^\frR_9,S^\frR_{10},S^\frR_{11},S^\frR_{12}\ .
	\]
	The disjoint union of all these blocks will be denoted by $S$. 
	All in all we have 
	\begin{equation}\label{eq:collection_of_all_blocks_of_circuit}
	\begin{split}
	|S^\frR_1|=|S^\frR_2|&=|S^\frL_1|=|S^\frL_2|=\Lambda\ ,\\
	|S^\frL_3|=|S^\frL_4|&=|S^\frL_5|=\diamondsuit\ ,\\
	|S^\frR_3|=|S^\frR_4|&=|S^\frR_5|=M\ ,\\ 
	|S^\frR_{12}|&=s\ ,\\
	|S^\frL_6|=|S^\frL_7|=|S^\frL_8|=|S^\frL_{9}|=|S^\frL_{10}|=|S^\frL_{11}|&=|S^\frR_6|=|S^\frR_7|=|S^\frR_8|=|S^\frR_{9}|=|S^\frR_{10}|=|S^\frR_{11}|=1\ ,\\
	S_0=\bigcup_{i=1}^{11} S^\frL_i\quad&,\quad S=S_0\cup \left(\bigcup _{i=1}^{12}S^\frR_i\right)\ .   
	\end{split}
	\end{equation}
\end{definition}
\begin{remark}[A sampled point for the PCP]\label{rem:point_for_PCP}
	The blocks of variables from Definition \ref{defn:blocks_of_vars_circuit} are expected to be variables of polynomials. As polynomials in $\FF_q[S]$ can be thought of as functions $\FF_q^S\to \FF_q$ (through the map $\Phi_{\FF_q}$ \eqref{def:Phi_FF}),  we will often sample points in $\FF_q^S$ and evaluate the polynomials at these points. We  use the notation 
	\begin{equation}\label{eq:point_p}
	p=(\underbrace{p^\frL_1,...,p^\frL_{11}}_{p_0},p^\frR_1,...,p^\frR_{12})\in \FF_q^{S}\ .
	\end{equation}
	Namely, $p$ is a function from $S$ to $\FF_q$, and $p^\kappa_i$ is the restriction of $p$ to the block $S^\kappa_i$. 
\end{remark}

\begin{definition}[Degree-$d$ PCP over $\FF_q$]\label{defn:PCP_of_V_n_satisfied}
	Let $d,t,\Lambda,\diamondsuit,s$ and $M$ be positive integers, and let $q=2^t$. Let $\circuit$ be a circuit with $s$ many non-input wires and $4\Lambda+ 3\diamondsuit+3M+12$ many input gates collected in blocks  as in \eqref{eq:input_blocks_of_the_circuit}. Recall the variable blocks from Definition \ref{defn:blocks_of_vars_circuit}.
	A (individual) \emph{degree}-$d$ \emph{probabilistically checkable proof over $\FF_q$} with parameters $(\Lambda,\diamondsuit,M,s,\circuit)$, denoted by $\Pi$  and referred to  as  just ``PCP'' from now onwards,  is the evaluation table of $\heartsuit=12\Lambda+12\diamondsuit+6M+s+35$ many individual degree at most $d$ polynomials over $\FF_q$ formatted as follows:
	\begin{align}
	\forall \kappa\in \{\frR,\frL\},\ i\in [5]\ ,\sX\in S^\kappa_{i}\ &\colon \ \ g^\kappa_i\ , \ \beta^\kappa_{i,\sX}\in \FF_q[S^\kappa_i]\ ,\label{eq:1_Pi}\\
	\forall \sX\in S_0\ &\colon \ \ g_0\ , \ \beta_{0,\sX}\ ,\ \alpha^\frL_\sX \in \FF_q[{S_0}]\ ,\label{eq:4_Pi}\\
	\forall \sX\in S\ &\colon \ \ \alpha^\frR_\sX\in \FF_q[{S}]\ .\label{eq:5_Pi}
	\end{align}
	The eleven $g$ polynomials are called the \emph{assignments} in $\Pi$, while the $\alpha$ and $\beta$ polynomials are called the \emph{helpers} in $\Pi$. 
	The linear part $\Pi^\frL$ of $\Pi$ consists of the ($\heartsuit^\frL=4\Lambda+6\diamondsuit+11$ many) polynomials $g^\frL_i,\alpha^\frL_{\sX},\beta^\frL_{i,\sX}$, and the readable part $\Pi^\frR$ consists of the rest ($\heartsuit^\frR=8\Lambda+6\diamondsuit+6M+s+24$ many)  of the polynomials, namely $g_0,\beta_{0,\sX},g^\frR_i,\alpha^\frR_\sX,\beta^\frR_{i,\sX}$.
	
	As every $S_i^\kappa$ is contained in $S$, every $\FF_q[S^\kappa_i]$ is contained in $\FF_q[S]$,\footnote{ When embedding all the polynomials of $\Pi$ in $\FF_q[S]$, they become indifferent (Definition \ref{def:polynomial_degree}) to the added variable-indexes.} and there is a well defined notion of evaluating each of the polynomials consisting of $\Pi$  --- \eqref{eq:1_Pi}, \eqref{eq:4_Pi} and \eqref{eq:5_Pi} --- at  $p\colon S\to \FF_q$ (cf.\ Remark \ref{rem:point_for_PCP}). 
	\emph{Evaluating $\Pi$ at $p$} is   reading the $p$-evaluation of all the polynomials consisting of $\Pi$, 
	and we denote  this tuple of $\heartsuit$-many values in $\FF_q$ by $\Pi(p)$. Namely, evaluating the PCP $\Pi$  induces a function 
	\[
	\Pi\colon \FF_q^{S}\to\FF_q^{\heartsuit}\ .
	\]
\end{definition}

\begin{observation}\label{obs:satisfiability_vs_PCP}
	Let $d,t,q,\Lambda,\diamondsuit,s,M$ and $\circuit$ be as in Definitions \ref{defn:blocks_of_vars_circuit} and \ref{defn:PCP_of_V_n_satisfied}. Such a circuit $\circuit$ induces a $6$-decoupled CNF $\varphi_\circuit$ (Definition \ref{defn:decoupled_equations_and_CNF}), whose six blocks of variables are parametrized by $\FF_2^{S_0},\FF_2^{S^\frR_1},...,\FF_2^{S^\frR_5}$.  By the structure assumed on $S_0$, a function $f_0\colon \FF_2^{S_0}\to  \FF_2$ encodes (according to Definition~\ref{defn:decoupled_equations_and_CNF}) a $5$-decoupled system $(\mathscr{A}_{f_0}\mid \vec b_{f_0})$ of linear equations, with the five blocks of variables being parametrized by $\FF_2^{S^\frL_1},...,\FF_2^{S^\frL_5}$.
	By Fact \ref{fact:polynomial_condition_for_satisfiability}, given an $11$-tuple
	\[
	f_0\colon \FF_2^{S_0}\to \FF_2\quad,\quad \forall \kappa\in \{\frR,\frL\}\ ,\ i\in [5]\ \colon \ \ f^\kappa_i\colon \FF_2^{S^\kappa_i}\to \FF_2\ ,
	\]
	the tuple $(f_0,f^\frR_1,...,f^\frR_5)$ satisfies the formula $\varphi_\circuit$  and the tuple $(f^\frL_1,...,f^\frL_5)$ satisfies the linear system $(\mathscr{A}_{f_0}\mid \vec b_{f_0})$ if and only if, for every 
	\begin{equation}\label{eq:point_PCP_again}
	u=(\underbrace{u^\frL_1,...,u^\frL_{11}}_{u_0},u^\frR_1,...,u^\frR_{12})\in \FF_2^{S}
	\end{equation}
	(cf.\ Remark \ref{rem:point_for_PCP}) the following two equations are satisfied:
	\begin{equation}\label{eq:formula_is_satisfied}
	T_\circuit(u)(f_0(u_0)+u^\frR_6+1)\prod_{i=1}^5\left( f^\frR_i(u^\frR_i)+u^\frR_{i+6}+1\right)=0\ ,
	\end{equation}
	and 
	\begin{equation}\label{eq:system_is_satisfied}
	f_0(u_0)\Big(u^\frL_{11}+\sum_{i=1}^5 u^\frL_{i+5}f_i^\frL(u^\frL_i)\Big)=0\ .
	\end{equation}
	
	The goal of a degree $d$ PCP $\Pi$ over $\FF_q$ (Definition \ref{defn:PCP_of_V_n_satisfied}) is to prove that the above two equations are indeed satisfied for every point in the subcube $\FF_2^S$, and thus that $\varphi_\circuit$ and $(\mathscr{A}_{f_0}\mid \vec b_{f_0})$ are satisfied by restricting (Definition \ref{defn:restriction_and_induction_polynomials}) the $g$ polynomials in the PCP $\Pi$ to the subcube. Namely, we expect the $g$ polynomials to be assignments, so that their restrictions to the subcube can play the role of the $f$ polynomials. To that end, the $\beta$ polynomials in $\Pi$ are expected to satisfy for every $p\in \FF_q^S$  as in \eqref{eq:point_p} --- note that this time the point is over $\FF_q$ and not $\FF_2$ as $u$ was in \eqref{eq:point_PCP_again} --- that 
	\begin{equation}\label{eq:PCP_condition_1}
	\forall \kappa\in \{\frR,\frL\}\ ,\ i\in[5]\ \colon \ \ g^\kappa_i(p^\kappa_i)(g^\kappa_i(p^\kappa_i)+1)=\sum_{\sX\in S^\kappa_i} \zero_\sX(p^\kappa_i)\cdot \beta^{\kappa}_{i,\sX}(p^\kappa_i)
	\end{equation}
	and
	\begin{equation}\label{eq:PCP_condition_2}
	g_0(p_0)(g_0(p_0)+1)=\sum_{\sX\in S_0} \zero_\sX(p_0)\cdot \beta_{0,\sX}(p_0)\ ,
	\end{equation}
	where $\zero_\sX(p)=p(\sX)(p(\sX)+1)$, as was defined in \eqref{eq:defn_of_zero_i}. In addition, the $\alpha$ polynomials verify that  the restrictions of the $g$ polynomials satisfy \eqref{eq:formula_is_satisfied} and \eqref{eq:system_is_satisfied}. Namely, for every $p\in \FF_q^S$ as in \eqref{eq:point_p},
	\begin{equation}\label{eq:Tseitin_satisfiable_induced}
	T_\circuit(p)(g_0(p_0)+p^\frR_6+1)\prod_{i=1}^5\left( g^\frR_i(p^\frR_i)+p^\frR_{i+6}+1\right)=\sum_{\sX\in S} \zero_{\sX}(p)\cdot \alpha^\frR_{\sX}(p)\ 
	\end{equation}
	and 
	\begin{equation}\label{eq:induced_system_is_satisfied}
	g_0(p_0)(p^\frL_{11}+\sum_{i=1}^5 p^\frL_{i+5}g_i^\frL(p^\frL_i))=\sum_{\sX\in S_0} \zero_{\sX}(p_0)\cdot \alpha^\frL_{\sX}(p_0)\ .
	\end{equation}
	The next proposition exactly relates the existence of such PCPs to the satisfiability of $\varphi_\circuit$ and $(\mathscr{A}_{f_0}\mid \vec b_{f_0})$, similar to the role of Proposition \ref{prop:prelude_on_satisfiability} in the Prelude.
\end{observation}   
\begin{definition}[Inducing to a PCP]\label{defn:inducing_to_a_PCP}
	Let $t,q,\Lambda,\diamondsuit,s,M,\circuit$ and $S^\cdot_\cdot$ be as in Definitions \ref{defn:blocks_of_vars_circuit}, \ref{defn:PCP_of_V_n_satisfied} and Observation \ref{obs:satisfiability_vs_PCP}. The map     $\mathsf{Induce}_{\circuit}$ takes an  $11$-tuple
	\[
	f_0\colon \FF_2^{S_0}\to \FF_2\quad,\quad \forall \kappa\in \{\frR,\frL\}\ ,\ i\in [5]\ \colon \ \ f^\kappa_i\colon \FF_2^{S^\kappa_i}\to \FF_2\ ,
	\]
	and outputs the following degree $9$ PCP $\Pi$ over $\FF_q$ with parameters $(\Lambda,\diamondsuit,M,s,\circuit)$. First, it lets 
	\begin{equation}\label{eq:the_g_polynomials_in_induce_circuit}
	g_0=\Ind(f_0)\quad,\quad \forall \kappa\in \{\frR,\frL\},i\in [5]\ \colon \ \ g^\kappa_i=\Ind(f^\kappa_i)\ ,
	\end{equation}
	with $\Ind(\cdot)$ being the induction from Definition \ref{defn:restriction_and_induction_polynomials}. Then, it defines the polynomials 
	\begin{equation}\label{eq:def_Psi_polynomials}
	\begin{split}
	\Psi^\frR(p)&= T_\circuit(p)(g_0(p_0)+p^\frR_6+1)\prod_{i=1}^5\left( g^\frR_i(p^\frR_i)+p^\frR_{i+6}+1\right)\ ,\\
	\Psi^\frL(p_0)&=g_0(p_0)(p^\frL_{11}+\sum_{i=1}^5 p^\frL_{i+5}g_i^\frL(p^\frL_i))\ ,\\
	\forall  \kappa \in \{\frL,\frR\}\ ,\  i \in [5]\ \colon \ \ \Psi^{\frak{A}}_{\kappa,i}(p^\kappa_i)&= g^\kappa_i(p^\kappa_i)(g^\kappa_i(p^\kappa_i)+1)\ ,\\
	\Psi^{\frak{A}}_{0}(p_0)&= g_0(p_0)(g_0(p_0)+1)\ ,
	\end{split}
	\end{equation}
	where $T_\circuit$ is the Tseitin polynomial associated with  $\circuit$ (Definition \ref{defn:Circuits}) and $p\in \FF_q^S$ is as in \eqref{eq:point_p}. As the $11$ $g$ polynomials have individual degree at most $1$, and $T_\circuit$ has individual degree at most $3$ (Remark \ref{rem:individual_degree_Tseitin_is_3}), all of the above $\Psi$ polynomials are of individual degree at most $9$. By fixing an order on each block $S^\cdot_\cdot$ from Definition \ref{defn:blocks_of_vars_circuit}, and using the notation $\sX_1<\sX_2<\sX_3<...$  for the list of variables in the block, we can define 
	\begin{equation}
	\begin{split}
	\forall \sX_i\in S\ \colon\ \ \alpha^\frR_{\sX_i}&={\rm Div}_{\sX_i(\sX_i+1)}\circ {\rm Mod}_{\sX_{i-1}(\sX_{i-1}+1)}\circ ...\circ{\rm Mod}_{\sX_1(\sX_1+1)}(\Psi^\frR)\ ,\\
	\forall \sX_i\in S_0\ \colon\ \ \alpha^\frL_{\sX_i}&={\rm Div}_{\sX_i(\sX_i+1)}\circ {\rm Mod}_{\sX_{i-1}(\sX_{i-1}+1)}\circ ...\circ{\rm Mod}_{\sX_1(\sX_1+1)}(\Psi^\frR)\ ,\\
	\forall\kappa \in \{\frR,\frL\}\ ,\ j\in [5]\ , \sX_i\in S^\kappa_j\ \colon\ \ \beta^\kappa_{j,\sX_i}&={\rm Div}_{\sX_i(\sX_i+1)}\circ {\rm Mod}_{\sX_{i-1}(\sX_{i-1}+1)}\circ ...\circ{\rm Mod}_{\sX_1(\sX_1+1)}(\Psi^{\frak{A}}_{\kappa,j})\ ,\\
	\forall \sX_i\in S_0\ \colon \ \ \beta_{0,\sX_i}&={\rm Div}_{\sX_i(\sX_i+1)}\circ {\rm Mod}_{\sX_{i-1}(\sX_{i-1}+1)}\circ ...\circ{\rm Mod}_{\sX_1(\sX_1+1)}(\Psi^{\frak{A}}_{0})\ ,
	\end{split}
	\end{equation}
	where ${\rm Div}$ and ${\rm Mod}$ were defined in the proof of Claim \ref{claim:combi_null}. As ${\rm Div}$   and ${\rm Mod}$ can only decrease the individual degree of the polynomial on which they are applied, the resulting $\alpha$ and $\beta$ polynomial have individual degree at most $9$. All in all, $\mathsf{Induce}_\circuit$ recovered a degree $9$ PCP over $\FF_q$.

\end{definition}
\begin{proposition}[A probabilistically checkable proof for the combined succinct $6$-decoupled ${\rm SAT}$ and $5$-decoupled linear system]\label{prop:completeness_and_soundness_of_PCP_for_V_n}
	Let $t,q,\Lambda,\diamondsuit,s,M,\circuit$ and $S^\cdot_\cdot$ be as in Definitions \ref{defn:blocks_of_vars_circuit}, \ref{defn:PCP_of_V_n_satisfied}, \ref{defn:inducing_to_a_PCP} and Observation \ref{obs:satisfiability_vs_PCP}. 
	Then:
	\begin{itemize}
		\item \emph{Completeness}: Given an $11$-tuple
		\[
		f_0\colon \FF_2^{S_0}\to \FF_2\quad,\quad \forall \kappa\in \{\frR,\frL\}\ ,\ i\in [5]\ \colon \ \ f^\kappa_i\colon \FF_2^{S^\kappa_i}\to \FF_2\ ,
		\]
		the degree $9$ PCP $\Pi=\mathsf{Induce}_\circuit(f_0,f_1^\frR,...,f_5^\frR,f_1^\frL,...,f_5^\frL)$ over $\FF_q$ with parameters $(\Lambda,\diamondsuit,M,s,\circuit)$ from Definition \ref{defn:inducing_to_a_PCP} satisfies that
		\begin{itemize}[noitemsep,topsep=0pt,parsep=0pt,partopsep=0pt]
			\item $g_{\circ}^\circ=\Ind(f_{\circ}^\circ)$;
			\item the equations \eqref{eq:PCP_condition_1},  \eqref{eq:PCP_condition_2}  are  satisfied by the polynomials in $\Pi$ for every $p\in \FF_q^{S}$;
			\item the readable part $\Pi^\frR$ of $\Pi$ depends only on $f_0,f^\frR_1,...,f^\frR_5$;
			\item for fixed $f_0,f^\frR_1,...,f^\frR_5$, the map $\mathsf{Induce}_\circuit(f_0,f^\frR_1,...,f^\frR_5,\cdot,\cdot,\cdot,\cdot,\cdot)$ from $\bigoplus_{i=1}^5(\FF_2)^{\FF_2^{S^\frL_i}}$ to $(\FF_q^\heartsuit)^{\FF_q^{S}}$, thought of as a vector space over $\FF_2$, is $\FF_2$-linear;
			\item if  $\varphi_\circuit(f_0,f_1^\frR,...,f_5^\frR)=1$, then \eqref{eq:Tseitin_satisfiable_induced} is satisfied by $\Pi$ for every $p\in \FF_q^S$;
			\item if $(f_1^\frL,...,f_5^\frL)$ satisfy the $5$-decoupled system of linear equations $(\mathscr{A}_{f_0}\mid \vec b_{f_0})$ induced by $f_0$,  then \eqref{eq:induced_system_is_satisfied} is satisfied by $\Pi$ for every $p\in \FF_q^S$.
		\end{itemize} 
		
		\item \emph{Soundness}: Let $d\geq 3$. If  $\Pi$ is a degree-$d$ PCP over $\FF_q$ with parameters  $(\Lambda,\diamondsuit,M,s,\circuit)$ that passes each of the checks \eqref{eq:PCP_condition_1},  \eqref{eq:PCP_condition_2}, \eqref{eq:Tseitin_satisfiable_induced} and \eqref{eq:induced_system_is_satisfied}  with probability strictly larger than $\frac{7dm}{q}$ (with $m=|S|$ from \eqref{eq:size_of_m_PCPs}) over the choice of a uniformly random $p\colon S\to \FF_q$, then it passes them with probability $1$. This in turn means that by taking $f_\circ^\circ=\Res(g^\circ_\circ)$, the resulting $11$-tuple satisfies both $\varphi_\circuit$ and $(\mathscr{A}_{f_0}\mid \vec b_{f_0})$.
	\end{itemize}
\end{proposition}

\begin{proof}
	\textbf{Completeness}:
	
	Let us begin by proving the first $4$ claimed properties of $\Pi=\mathsf{Induce}_\circuit(f_0,f_1^\frR,...,f_5^\frR,f_1^\frL,...,f_5^\frL)$, as they are non-conditional. The fact $g^\circ_\circ=\Ind(f^\circ_\circ)$ is by construction, see \eqref{eq:the_g_polynomials_in_induce_circuit}.  As the $g$ polynomials are inductions of functions from $\FF_2^\circ\to \FF_2$, they are assignments, and thus the polynomials $\Psi^{\frak{A}}_{\circ,\circ}$ from \eqref{eq:def_Psi_polynomials} are zero on the subcube (Definition \ref{def:zero_on_subcube_and_assignments}). Hence, by the proof of the Combinatorial Nullstellensatz (Claim \ref{claim:combi_null}),   equations \eqref{eq:PCP_condition_1},  \eqref{eq:PCP_condition_2}  are  satisfied by the polynomials in $\Pi$ for every $p\in \FF_q^{S}$. Recall that the readable part $\Pi^\frR$ of $\Pi$ consists of the polynomials  $g_0,g^\frR_i,\alpha^\frR_\sX,\beta _{0,\sX}$ and $\beta^\frR_{i,\sX}$. The fact the readable $g$ polynomials depend only on $f_0,f^\frR_1,...,f^\frR_5$ is immediate. For the other polynomials, note that $\Psi^\frR,\Psi^{\frak{A}}_{0}$ and $\Psi^{\frak{A}}_{\kappa,i}$ from \eqref{eq:def_Psi_polynomials} depend only on $f_0,f^\frR_1,...,f^\frR_5$, and thus taking ${\rm Div}$ and ${\rm Mod}$ with respect to fixed polynomials still depends only on them. Finally, let us address the required properties of the linear part $\Pi^\frL$ of $\Pi$, which needs to depend $\FF_2$-linearly on $f^\frL_i$ given $f_0,f^\frR_i$ were fixed. First, $\Ind$ is a $\FF_q$-linear map (Remark \ref{rem:induction_is_linear}), which guarantees that the values of the $g^\frL_i$ polynomials depend linearly on those of $f^\frL_i$. Now, as $f_0$ is fixed, $g_0$ is fixed as well and the values of $\Psi^\frL$ are by construction linear combinations of the values of $g^\frL_i$. As ${\rm Div}$ and ${\rm Mod}$ are $\FF_q$-linear functions, the values of the $\alpha^\frL_\sX$ polynomials depend linearly on  $f^\frL_i$. The map $\sY\mapsto\sY^2+\sY$ is $\FF_2$-linear over $\FF_q$, which means that the values of $\Psi^{\frak{A}}_{\frL,i}$ depend $\FF_2$-linearly on the values of $g^\frL_i$. Using again the fact that ${\rm Div}$ and ${\rm Mod}$ are $\FF_q$-linear proves that the $\beta^{\frL}_{i,\sX}$ polynomials depend $\FF_2$-linearly on the values of $f^\frL_i$. All in all, for a fixed $f_0,f^\frR_i$, the readable part $\Pi^\frR$ is fixed and the linear part $\Pi^\frL$ is an $\FF_2$-linear combination of the values of $f^\frL_i$.
	
	For the last two claimed properties of $\Pi$, as described in Observation \ref{obs:satisfiability_vs_PCP}, the fact that $\varphi_\circuit$ or $(\mathscr{A}_{f_0}\mid \vec b_{f_0})$ are satisfied by the $11$-tuple of  $f^\circ_\circ$'s, implies that \eqref{eq:formula_is_satisfied} or \eqref{eq:system_is_satisfied} are (respectively) satisfied for every $u\in \FF_2^S$ (as in \eqref{eq:point_PCP_again}). Hence, $\Psi^\frR$ or $\Psi^\frL$
	are  zero on their respective subcubes, and by the Combinatorial Nullstellensatz (Claim \ref{claim:combi_null}), our choice of  helper polynomials in $\Pi$ make  \eqref{eq:Tseitin_satisfiable_induced} or \eqref{eq:induced_system_is_satisfied} perfectly satisfied, as claimed. 
	
	\textbf{Soundness}:
	
	Let us analyze the $13$ polynomials
	\[
	\begin{split}
	\clubsuit^\frR(p)&= T_\circuit(p)(g_0(p_0)+p^\frR_6+1)\prod_{i=1}^5\left( g^\frR_i(p^\frR_i)+p^\frR_{i+6}+1\right)-\sum_{\sX\in S} \zero_{\sX}(p)\cdot \alpha^\frR_{\sX}(p)\ ,\\
	\clubsuit^\frL(p_0)&=g_0(p_0)(p^\frL_{11}+\sum_{i=1}^5 p^\frL_{i+5}g_i^\frL(p^\frL_i))-\sum_{\sX\in S_0} \zero_{\sX}(p_0)\cdot \alpha^\frL_{\sX}(p_0)\ ,\\
	\clubsuit^{\frak{A}}_{\circ,\circ}(p^\circ_\circ)&= g^\circ_\circ(p^\circ_\circ)(g^\circ_\circ(p^\circ_\circ)+1)-\sum_{\sX\in S^\circ_\circ} \zero_\sX(p^\circ_\circ)\cdot \beta^{\circ}_{\circ,\sX}(p^\circ_\circ)\ .
	\end{split}
	\]
	As $\Pi$ is of (individual) degree $d$, $\zero_\sX$ is of individual degree $2\leq d$, and $T_\circuit$ is of individual degree $3\leq d$, all of the $\clubsuit$ polynomials are of individual degree at most $7d$, and are thus of total degree at most $7md$ (as the number of variables in each of them is at most $m=|S|$). By the assumptions on $\Pi$ passing each of the checks \eqref{eq:PCP_condition_1},  \eqref{eq:PCP_condition_2}, \eqref{eq:Tseitin_satisfiable_induced} and \eqref{eq:induced_system_is_satisfied}  with probability strictly larger than $\frac{7dm}{q}$, the proportion of roots of each $\clubsuit$ polynomial is strictly larger than $\frac{7dm}{q}$. By the  Schwartz--Zippel Lemma \ref{lem:Schwartz-Zippel}, this implies all the $\clubsuit$ polynomials are identically zero. From the fact that $\clubsuit^{\frak{A}}_{\circ,\circ}$ is identically zero, we deduce that $\Psi^{\frak{A}}_{\circ,\circ}$ is zero on the subcube, and thus $g^\circ_\circ$ is an assignment (Definition \ref{def:zero_on_subcube_and_assignments}). Hence, $f^\circ_\circ=\Res(g^\circ_\circ)$ outputs only values in $\FF_2$. As $\clubsuit^\frR$ and $\clubsuit^\frL$ are identically zero, we deduce that  \eqref{eq:formula_is_satisfied} and \eqref{eq:system_is_satisfied} are satisfied, and thus the tuple of $f$ polynomials indeed satisfy $\varphi_\circuit$ and $(\mathscr{A}_{f_0}\mid \vec b_{f_0})$ (by Observation \ref{obs:satisfiability_vs_PCP}). This finishes the proof.
\end{proof}

\begin{corollary}[The functional viewpoint for $\verifier_n$ accepting]\label{cor:functional_viewpoint_final}
	Let 
	\begin{itemize}[noitemsep,topsep=0pt,parsep=0pt,partopsep=0pt]
		\item[--] $\Lambda$ be a single input TM that always halts;
		\item[--] $\verifier=(\sampler,\length,\linproc,\decider)$  a purified $2^\Lambda$-padded  $h$-level TNFV (Definition \ref{defn:padded_purified_NFV}) such that $\verifier_n$ is well defined for every $n$ (Definition \ref{defn:h-level_NFV});
		\item[--] $\Delta$ an always halting $1$-input TM that satisfies $\Delta(n)\geq \TIME(\linproc;n,\cdot,\cdot,\cdot,\cdot)\cdot 2^{\Lambda(n)+1}$ and induces $\diamondsuit(n)$ as in \eqref{eq:defn_of_diamondsuit(n)};
		\item [--] $Q$ an always halting $1$-input TM  satisfying 
		$\TIME(\sampler;{n},\cdot,\cdot,\cdot,\cdot,\cdot)\leq {Q}(n)$;
		\item [--] $T$ an always halting $1$-input TM  satisfying 
		\begin{equation}  T(n)\geq c\cdot(\TIME(\Lambda;n)^c+\TIME(\Delta;n)^c+2^{c\cdot\Lambda(n)}+\Delta(n)^c+\TIME(\linproc;n,\cdot,\cdot,\cdot,\cdot)^c)\ ,
		\end{equation}
		where $c\geq 6$ is the positive integer implied by the $\poly$ notation in \eqref{eq:time_bound_L*};
		\item [--] $D$ a positive integer (in binary) satisfying 
		$|\verifier|,|\Lambda|,|\Delta|,|T|,|Q|\leq D$;
		\item[--]  $n$ and $t$  positive integers, and $q=2^t$;
		\item[--]  $\mttx,\mtty$ two bit strings of length $r(n)=\sampler(n,{\rm Dimension},\cdot,\cdot,\cdot,\cdot)$;
		\item[--] $
		(M(n),s(n),\circuit)=\paddedsuccinctdecider(\verifier,\Lambda,\Delta,D,T,Q,n,\mttx,\mtty),
		$
		where $\paddedsuccinctdecider$ was defined in Proposition \ref{prop:explicit-padded-succinct-deciders}.
	\end{itemize}
	Then:
	\begin{enumerate}[label=\textcolor{black}{(\arabic*)}, ref= (\arabic*)]
		\item \label{clause:completeness_functional_viewpoint_for_V_n_accepting}\emph{Completeness}: For every quadruple $a^\frR,a^\frL,b^\frR,b^\frL\colon \FF_2^{\Lambda(n)}\to \FF_2$, there is a degree-$9$ PCP $\Pi$ over $\FF_q$ (Definition \ref{defn:PCP_of_V_n_satisfied})
		which satisfies the   equations \eqref{eq:PCP_condition_1}, \eqref{eq:PCP_condition_2} and 
		\eqref{eq:Tseitin_satisfiable_induced}  for every $p\in \FF_q^S$,\footnote{Note, and this is crucial, that \eqref{eq:induced_system_is_satisfied} may not be satisfied in this case}  and such that 
		\[
		\forall \kappa \in \{\frR,\frL\}\ \colon \ \ \Ind(a^\kappa)=g^\kappa_1\quad,\quad \Ind(b^\kappa)=g^\kappa_2\ ,
		\]
		where $g^\cdot_\cdot$ are the appropriate polynomials in $\Pi$. Furthermore, the readable part $\Pi^\frR$ of the PCP depends only on $a^\frR,b^\frR$, while the linear part $\Pi^\frL$ of the PCP depends  on $a^\frL,b^\frL$ in an $\FF_2$-affine manner (which may depend on $a^\frR,b^\frR$); namely,  by choosing a basis of $\FF_q$ over $\FF_2$, the bit representation of $\Pi^\frL$ is an $\FF_2$-affine combination of the bit representation of $a^\frL,b^\frL$. In addition, if $a^\frR,a^\frL,b^\frR,b^\frL$ are accepted in the game $\verifier_n$ given $\mttx,\mtty$ were asked, then \eqref{eq:induced_system_is_satisfied} is also satisfied by $\Pi$ for every $p\in \FF_q^S$.
		
		For later use, we denote by $\mathsf{PCP}_{\mttx\mtty}$ the mapping that takes $a^\frR,a^\frL,b^\frR,b^\frL$ as inputs and outputs this promised degree $9$ PCP $\Pi$. 
		
		\item \label{clause:soundness_functional_viewpoint_for_V_n_accepting} 
		\emph{Soundness}: If there is a degree $9$ PCP $\Pi$  over $\FF_q$  that passes each of the checks \eqref{eq:PCP_condition_1},  \eqref{eq:PCP_condition_2}, \eqref{eq:Tseitin_satisfiable_induced} and \eqref{eq:induced_system_is_satisfied}  with probability strictly larger than $\frac{63m}{q}$ (with $m=|S|$ as in \eqref{eq:size_of_m_PCPs}), then the quadruple $\Res(g^\frR_1),\Res(g^\frL_1),\Res(g^\frR_2),\Res(g^\frL_2)$ passes the game $\verifier_n$ given $\mttx,\mtty$ were asked.
	\end{enumerate}
\end{corollary}
\begin{proof}
	This is  a combination of Corollary \ref{cor:functional_viewpoint_succinct_description}  and Proposition \ref{prop:completeness_and_soundness_of_PCP_for_V_n}.  Let us just spell out   explicitly how $\mathsf{PCP}_{\mttx\mtty}$ operates. First, it lets  $\mathsf{Prove}_\circuit(a^\frR,b^\frR)=(f^\frR_1,f^\frR_2,f_0,f^\frR_3,f^\frR_4,f^\frR_5)$.  It then calculates $\Delta(n)$ and $\linproc(n,\mttx,\mtty,a^\frR,b^\frR)$ to retrieve $L_{\mttx\mtty}(a^\frR,b^\frR)=(\mathscr{A}\mid\vec b)$ and thus its unreadable part $(\mathscr{A}^\frL\mid \vec b)$. Then, the function $\mathsf{Extend}$ from Corollary \ref{cor:triang_and_decoupling_extend} is well defined with respect to $(\mathscr{A}^\frL\mid\vec b)$ and $\Delta(n)$, and we let $\mathsf{Extend}(a^\frL,b^\frL)=(f^\frL_1,f^\frL_2,f^\frL_3,f^\frL_4,f^\frL_5)$. Finally $\mathsf{PCP}_{\mttx\mtty}$ outputs 
	\[
	\Pi=\mathsf{Induce}_\circuit(f_0,f^\frR_1,...,f^\frR_5,f^\frL_1,...,f^\frL_5)\ .\qedhere
	\]
\end{proof}

\subsection{The low individual degree test for $\MIP^*$ protocols}
\label{sec:ld-decider}

A crucial, and technically involved, part of the proof of $\MIP=\NEXP$ (described in the Prelude \ref{sec:prelude_decision_problems_PCPs}) is to verify that the functions involved in the proof are indeed low individual degree polynomials (Theorem \ref{thm:classical_soundness_individual_low_degree}).
The  low individual degree test for $\MIP^*$ protocols,  or just low degree test from now onwards, is, as its name suggests, a non-local game analogue of the test described in Theorem \ref{thm:classical_soundness_individual_low_degree}. Namely, it is  designed to verify that a player in a $\MIP^*$ protocol returns an answer that can be interpreted as the simultaneous evaluation of a tuple of individual degree-$d$ polynomials $(g_1,\ldots,g_k)$ at a point $z\in \F_q^m$. The degree bound $d$, the number of polynomials $k$, the field size $q=2^t$ and the number of variables $m$ are all parameters of this game.
The soundness proof for this test is involved, and is the main theorem of \cite{Ji2022Quantum}.

\paragraph*{Preliminaries}
Recall the notions of polynomial degrees (Definition \ref{def:polynomial_degree}), lines in $\FF_q^m$ (Definition \ref{def:lines_in_finite_VS}) and the characterization of low degree polynomials via restrictions to lines (Fact \ref{fact:characterization_of_low_degreeness_by_restrictions_to_linew}).
There is no canonical way of choosing ``orthogonal projections'' and their ``complements'' in a finite vector space. So, we need to agree how to choose them in a consistent manner. The following definition provides a way of choosing such maps, which will be canonical for us.
\begin{definition}[Canonical linear maps]\label{defn:canonical_linear_maps}
	Let $\mathscr{B}=\{v_1,...,v_k\}\subseteq\FF_q^m$ be a set of linearly independent vectors. Complete them to a basis of $\FF_q^m$ as follows --- at each step, add to the set the standard basis vector $e_i$ with the largest possible $i$ so that the new sequence is still linearly independent.\footnote{This agrees with the canonical complement of a set defined in \cite[Definition 3.6]{MIPRE}.} So, we now have a basis $v_1,...,v_k,e_{i_1},...,e_{i_{m-k}}$. The canonical projection $\frak{null}_{\mathscr{B}}$ with kernel basis $\mathscr{B}$ is the map that takes a vector $v$, writes it as a linear combination  $\sum \alpha_i v_i+\sum \beta_j e_{i_j}$ (in the unique way), and returns $\sum \beta_j e_{i_j}$.\footnote{This agrees with \cite[Definition 3.10]{MIPRE}.} When $\mathscr{B}$ is a single vector $v$, we denote $\frak{null}_v$ instead of $\frak{null}_{\{v\}}$.
\end{definition}

\begin{fact}
	Let $u\in \FF_q^m$ and $\vec 0\neq v\in \FF_q^m$. Then, there is a point $u_0\in \mathscr{L}(u,v)$ such that for every $u'\in \mathscr{L}(u,v)$ we have $\frak{null}_v(u')=u_0$. 
\end{fact}

\begin{definition}[Canonical representation of a line]
	\label{def:line-representative}
	Let $\mathscr{L}\subseteq\FF_q^m$ be a line in direction $v\neq \vec 0$. Then, the canonical representation of $\mathscr{L}$ is $\mathscr{L}(u_0,v_0)$, where $u_0=\frak{null}_v(\mathscr{L})$ and $v_0$ is a non-zero scalar multiple of $v$ which is biggest in lexicographic order.\footnote{The reason to choose the biggest and not smallest element in the lexicographic order is that the number $1$ in $\FF_q$ is always maximal in lexicographic order when considering the basis from Fact \ref{fact:basis_F_q_over_F_2} --- this results in the vector $v_0=(0,\dots,0,1,\dots)$, which is a somewhat natural choice.} 
\end{definition}

\paragraph*{The low individual  degree game}

We begin by defining the game $\frak{LowDegree}(d,q,m,1)$, which is designed to check that the provers hold a single global $f\colon \FF_q^m\to \FF_q$ which is induced by a  polynomial of individual degree at most $d$, and answer according to its restriction to various points and lines. The general case of $\frak{LowDegree}(d,q,m,k)$, which checks that $k$ functions $f_1,...,f_k\colon \FF_q^m\to \FF_q$ are of individual degree at most $d$, will be explained afterwards.

\begin{figure}[!htbp]
	\centering
	\captionsetup{singlelinecheck=off}
	\begin{gamespec}
		\setlength{\tabcolsep}{1em}
		\begin{tabularx}{\textwidth}{ X   X  X   }
			\toprule
			Question Content & Formal variables (unreadable)  & Interpretation of answer    \\
			\midrule
			$\Point^u,\ u \in \F_q^m$&  
			$S_{\Point^u}^\lvar=\{\sPoint^{u,j,i}\mid$  & $f_1(u),...,f_k(u)$, where each  \\
			&$\qquad j\in [k]\ ,\ i\in[t]\}$ &  $f_j(u)\in \FF_q$\\
			$\ALine^\mathscr{L}$,\ $\mathscr{L}$ is an axis-parallel 
			&  $S_{\ALine^{\mathscr{L}}}^\lvar=\{\sALine^{\mathscr{L},j,s,i}\mid$ 
			& $(\mathsf{AL}f_1(\mathscr{L}),...\mathsf{AL}f_k(\mathscr{L}))$, where \\
			line in $\F_q^m$ &  $\qquad j\in[k],0\leq s\leq d,i\in[t]\}$ &  $\mathsf{AL}f_j(\mathscr{L})=(a_{j,0},...,a_{j,d})\in \FF_q^{d+1}$ encodes a degree $d$ univariate polynomial $\sum a_{j,s} \sX^s$. \\
			$\DLine^\mathscr{L}$,\ $\mathscr{L}$ is a line in $\F_q^m$ & $S_{\DLine^{\mathscr{L}}}^\lvar=\{\sDLine^{\mathscr{L},j,s,i}\mid $
			&
			$(\mathsf{DL}f_1(\mathscr{L}),...\mathsf{DL}f_k(\mathscr{L}))$, where \\
			& $\quad j\in[k]\ ,\ 0\leq s\leq md\ ,\ i\in[t]\}$ & $\mathsf{DL}f_j(\mathscr{L})=(a_{j,0},...,a_{j,md})\in \FF_q^{md+1}$ encodes a degree $md$ univariate polynomial $\sum a_{j,s}\sX^s$.\\
			\bottomrule
		\end{tabularx}
		
		\vspace{1em}
		
		\begin{enumerate}
			\setlength\itemsep{1pt}
			\item (\textbf{Axis-parallel line-versus-point test}) If $\ALine^{\mathscr{L}}-\Point^u$ was sampled, then $u\in \mathscr{L}=\mathscr{L}(u_0,e_i)$ and thus $u=u_0+\alpha e_i$. 
			For $s\in\{0,\ldots,d\}$ let $a_{j,s}=\mathsf{AL}f_j(\mathscr{L})_{s}$. Check that for every $j\in [k]$, $f_j(u)=\sum_{s=0}^d a_{j,s} \alpha^s$.
			
			\item (\textbf{Diagonal line-versus-point test}) If $\DLine^{\mathscr{L}}-\Point^u$ was sampled, then $u\in \mathscr{L}=\mathscr{L}(u_0,v_0)$ and thus $u=u_0+\alpha v_0$. For $s\in\{0,\ldots,d\}$ let $a_{j,s}=\mathsf{DL}f_j(\mathscr{L})_s$. Check that  for every $j\in [k]$, $f_j(u)=\sum_{s=0}^{md} a_{j,s} \alpha^s$.
		\end{enumerate}
	\end{gamespec}
	\caption[]{Description of the low degree game $\frak{LowDegree}(d,q,m,k)$.}
	\label{fig:ld-decider}
\end{figure}

The idea of the game is similar to the classical case (Theorem \ref{thm:classical_soundness_individual_low_degree}), namely to use the truth tables of restrictions of $f$ to lines and points,  checking that they are  consistent with one another, and that the lines satisfy condition 2 in Fact \ref{fact:characterization_of_low_degreeness_by_restrictions_to_linew}. In the quantum case, we need to add some diagonal line checks, where the diagonal lines are sampled according to a somewhat peculiar distribution. These checks  force commutation between all of the observables in the game. This should be seen as a quirk of the proof in \cite{Ji2022Quantum}, and we do not have much insight to it except that it allows the inductive step therein to work.

The vertices in the underlying graph are of three types: $\Point$, $\ALine$ (axis parallel line), and $\DLine$ (diagonal line). The $\Point$ vertices are parametrized by $\FF_q^m$, namely  $\{\Point^u\mid u\in \FF_q^m\}$. The $\ALine$ vertices are parametrized by axis parallel lines, namely $\{\ALine^{\mathscr{L}}\mid \mathscr{L}\  {\rm is\ an\  axis\ parallel\ line}\subseteq \FF_q^m\}$. Finally, the  $\DLine$ vertices are parametrized by diagonal lines, namely $\{\DLine^{\mathscr{L}}\mid \mathscr{L}\  {\rm is\ any\ line}\subseteq \FF_q^m\}$.
We now specify the  generators associated to each vertex, which also determines the length functions of the game, and we provide some notations that will clarify both this game as well as the Answer Reduced game (Section \ref{sec:combi_and_algo_answer_red}). The game $\frak{LowDegree}(d,q,m,1)$ is an LCS (recall Example~\ref{example:LCSs}), which means it can be tailored by making all variables unreadable. Recall that $q=2^t$, and using the basis from Fact \ref{fact:basis_F_q_over_F_2}, an element of $\FF_q$ is encoded as a length-$t$ bit string.
\begin{itemize}
	\item For $\Point^u$, we set $S_{\Point^u}^\lvar=\{\sPoint^{u,i}:\ i\in[t]\}$. Here, $\sPoint^{u,i}$ represents the $i^{\rm th}$ bit of the $\F_q$-value assigned to the point $u$ by the supposed low individual degree polynomial $f$ which controls the answers of the players. Hence, if $\gamma$ is the assignment to the variables,  we denote $f(u)=\gamma(\sPoint^{u,i})_{i=1}^t\in \FF_q$.
	\item For $\ALine^{\mathscr{L}}$, we set   $S_{\ALine^{\mathscr{L}}}^\lvar=\{\sALine^{\mathscr{L},j,i}:\ 0\leq j\leq d,i\in[t]\}$. As the restriction of an individual degree-$d$ polynomial to an axis parallel line is a univariate polynomial of degree at most $d$ (Fact \ref{fact:characterization_of_low_degreeness_by_restrictions_to_linew}), it can be written as $a_0+a_1\alpha +a_2\alpha ^2+...+a_d\alpha ^{d}$ with the $a_i$ being $\FF_q$-coefficients. The value of the variable $\sALine^{\mathscr{L},j,i}$ is interpreted as the $i^{\rm th}$ bit of the encoding of $a_j$ in the supposed restriction of the global low degree $f$ to the line $\mathscr{L}$.\footnote{Here it is  important that we  fixed a canonical representation to each line, as the restriction of a polynomial to a line depends on its representation (see Fact \ref{fact:characterization_of_low_degreeness_by_restrictions_to_linew}).} Hence, if $\gamma$ is the assignment to the variables, we denote $\mathsf{AL}f(\mathscr{L})=(a_0,...,a_d)=(\gamma(\sALine^{\mathscr{L},j,i})_{i=1}^t)_{j=0}^d\in \FF_q^{d+1}$.
	\item For $\DLine^{\mathscr{L}}$,  we set  $S_{\DLine^{\mathscr{L}}}^\lvar=\{\sDLine^{\mathscr{L},j,i}:\ 0\leq j\leq md,i\in[t]\}$. As  an individual degree-$d$ polynomial is a total degree at most $md$ polynomial, and the restriction of such a polynomial to a  line is a univariate polynomial of degree at most $md$ (Fact \ref{fact:characterization_of_low_degreeness_by_restrictions_to_linew}), it can be written as $a_0+a_1\alpha+a_2\alpha^2+...+a_{md}\alpha^{md}$ with the $a_i$ being $\FF_q$-coefficients. The value of the variable $\sDLine^{\mathscr{L},j,i}$ is interpreted as the $i^{\rm th}$ bit of the encoding of $a_j$ in the supposed restriction of the global low degree $f$ to the line $\mathscr{L}$.  Hence, if $\gamma$ is the assignment to the variables, we denote $\mathsf{DL}f(\mathscr{L})=(a_0,...,a_{md})=(\gamma(\sALine^{\mathscr{L},j,i})_{i=1}^t)_{j=0}^{md}\in \FF_q^{md+1}$.
\end{itemize}

The underlying graph of $\frak{LowDegree}(d,q,m,1)$ is induced by the incidence relation between points and lines. Namely, $\Point^u$ is connected to $\ALine^{\mathscr{L}}$ (respectively $\DLine^{\mathscr{L}}$) if and only if $u\in \mathscr{L}$. For the sampling scheme of edges,  let us provide  a typed $3$-level CLM (Definition \ref{defn:typed_h-level_sampling_scheme}) that describes it exactly.
The type set consists of three types $\Point$, $\ALine$ (axis parallel line), and $\DLine$ (diagonal line), and the type graph contains all loops as well as the edges $\Point-\ALine$ and $\Point-\DLine$. The dimension of the space the CLMs act on  is $(2m+1)\log q=(2m+1)t$, and by using the basis guaranteed by Fact \ref{fact:basis_F_q_over_F_2}, we can interpret each element from this space unambiguously as a triple in $\FF_q^m\times \FF_q\times \FF_q^m$.
\begin{itemize}
	\item The CLM $\frS^{\Point}$ is $1$-level, and is defined by 
	\begin{equation}\label{eq:cl-ptf}
	\forall u,v\in \FF_q^m,\ s\in \FF_q\ \colon \ \ \frS^\Point(u,s,v)=(u,0,0)\;.    
	\end{equation}
	Namely, the vertex $(\Point,u,0,0)$ corresponds to the vertex $\Point^u$ introduced above.
	\item The CLM $\frS^\ALine$ is $2$-level, and is defined by:
	\begin{equation}
	\label{eq:cl-alnf}
	\forall u,v\in \FF_q^m,\ s\in \FF_q\ \colon \ \ \frS^\ALine(u,s,v) = (\frak{null}_{e_{\chi(s)}}(u), s, 0)\ ,
	\end{equation}
	where $\frak{null}_\cdot$ is the canonical map with kernel $\cdot$ (Definition \ref{defn:canonical_linear_maps}), $e_\cdot$ is the appropriate standard basis element of $\FF_q^m$, and $\chi(s)$ is one more than the residue of the devision of $s$ by $m$, namely
	\begin{equation}
	\label{eq:chi-func}
	\chi(s) = 1+(s\pmod m)\in [m]\;,
	\end{equation}
	where  $s\in \FF_q$ is associated with the integer with the same binary representation (again, according to the fixed basis of $\FF_q$ over $\FF_2$ chosen in Fact \ref{fact:basis_F_q_over_F_2}). The resulting pair maps naturally to a canonical representation (Definition \ref{def:line-representative}) of an axis parallel line $\mathscr{L}=\mathscr{L}(u_0,e_i)$, where $i=\chi(s)$. Namely, the vertex $(\ALine,u_0,s,0)$ corresponds to (a copy of) the vertex $\ALine^{\mathscr{L}}$ defined above.
	\item The CLM $\frS^\DLine$ is $3$-level, and is defined by:
	\begin{equation}
	\label{eq:cl-dlnf}
	\forall u,v\in \FF_q^m,\ s\in \FF_q\ \colon \ \ \frS^\DLine(u,s,v) = (\frak{null}_{\pi_{\chi(s)-1}(v)}(u),s,\pi_{\chi(s)-1}(v))\ ,
	\end{equation}
	where $\pi_i\colon \FF_q^m\to \FF_q^m$ is the linear map that zeroes out the first $i$ coordinates of the input. It easily seen to be $3$-level CLM, as the first register space is the copy of $\FF_q$ on which $\frS^\DLine$ acts with the identity. Then, the second register space is the last copy of $\FF_q^m$, on which the linear map $\pi_{\chi(s)-1}$ is applied (and indeed, it depends only on the image of the previous linear map). And finally, the third register space is the first copy of $\FF_q^m$, on which $\frak{null}_{\pi_{\chi(s)-q}(v)}$ is applied, which is dependent on the result of the previous linear map. By ignoring $s$, we get a  representation of a diagonal line $\mathscr{L}=\mathscr{L}(\frak{null}_{\pi_{\chi(s)-1}(v)}(u),\pi_{\chi(s)-1}(v))$ --- note that the incidence point is canonical, while the direction may not be. So, each such $(\DLine,\frak{null}_{\pi_{\chi(s)-1}(v)}(u),s,\pi_{\chi(s)-1}(v))$ is (a copy of) the vertex $\DLine^{\mathscr{L}}$ introduced above.\footnote{It can already be noticed that the probability of sampling diagonal lines is far from being uniform over them. This is  a technical thing needed for the induction in the soundness proof in \cite{Ji2022Quantum} to work out. }
\end{itemize}

Finally we specify the decision procedure. Recall the canonical representation of lines from Definition \ref{def:line-representative}.
\begin{itemize}
	\item If $\Point^u-\ALine^\mathscr{L}$ is sampled, then $u\in \mathscr{L}=\mathscr{L}(u_0,e_i)$, and in particular $u=u_0+\alpha e_i$ for some $\alpha\in \FF_q$. Let  $\gamma$ be the answer of the players, and denote as before $\mathsf{AL}f(\mathscr{L})_j=a_j=\gamma(\sALine^{\mathscr{L},j,i})_{i=1}^t$ and $f(u)=(\sPoint^{u,i})_{i=1}^t$, which are elements of $\FF_q$. Then the decision procedure accepts if and only if $\sum_{j=0}^d a_j \alpha^j=f(u)$. This can easily be written as $t$ linear equations over $\FF_2$.
	\item If $\Point^u-\DLine^\mathscr{L}$ is sampled, then $u\in \mathscr{L}=\mathscr{L}(u_0,v_0)$, and in particular $u=u_0+\alpha v_0$ for some $\alpha\in \FF_q$.  Let  $\gamma$ be the answer of the players. The decision here is almost identical to the previous one --- we denote as before  $\mathsf{DL}f(\mathscr{L})_j=a_j=\gamma(\sDLine^{\mathscr{L},j,i})_{i=1}^t$ and $f(u)=(\sPoint^{u,i})_{i=1}^t$ as elements of $\FF_q$, and  accept if and only if $\sum_{j=0}^{md} a_j \alpha^j=f(u)$. This can  again be written as $t$ linear equations over $\FF_2$.
\end{itemize}

For the general case of $k>1$, $\frak{LowDegree}(d,q,m,k)$ uses the same question distribution, but now the sets of generators are $k$ times  larger. For example, for the vertex $\Point^u$, we have $S_{\Point^u}^\lvar=\{\sPoint^{u,j,i}:\ i\in[t],j\in[k]\}$ --- in this case, $\sPoint^{u,j,i}$ is supposed to be the $i^{\rm th}$ bit of the evaluation of a global function $f_j$, that is supposed to be of low degree, evaluated at $u$. In this case, given an assignment $\gamma$, we denote by $(f_1(u),...,f_k(u))$ the answer $(\gamma(\sPoint^{u,j,i})_{i=1}^t)_{j=1}^k$ (and similarly we denote $(\mathsf{AL}f_1(\mathscr{L}),...,\mathsf{AL}f_k(\mathscr{L}))$ and $(\mathsf{DL}f_1(\mathscr{L}),...,\mathsf{DL}f_k(\mathscr{L}))$ for the other types). The check performed is the same check as for the case $k=1$, executed independently $k$ times, once for each group of generators associated with the same $j\in[k]$.\footnote{Note that question types are \emph{not} mixed according the different copies of the test, e.g.\ a point or line is sampled simultaneously for all copies, not a mixture of points and lines.}

The following is based on~\cite{Ji2022Quantum}. We state the theorem for the case where the base code is the {Reed--Solomon} code with degree $d$ --- i.e., all {univariate} polynomials of degree at most $d$ over $\FF_q$ --- as this is the only case we use. The only fact about this code that is used in the theorem statement is that it has distance $1-\nicefrac{d}{q}$, by the Schwartz--Zippel Lemma \ref{lem:Schwartz-Zippel}. 

\begin{theorem}[Soundness of the low-degree game. See the main theorem in \cite{Ji2022Quantum} and  Theorem 4.43 in \cite{NW19}]\label{thm:ldc-soundness}
	There exists a universal positive integer constant 
	\begin{equation}\label{eq:defn_c_ld}
	c=c_\ld\ ,    
	\end{equation}
	and a function 
	\begin{equation}\label{eq:defn_delta_ld}
	\delta_\ld(m,d,k,\eps,q^{-1}) =c\cdot (m^{c}+d^{c}+k^{c})\cdot(\eps^{\nicefrac{1}{c}}+q^{-\nicefrac{1}{c}}+2^{-\nicefrac{md}{c}})
	\end{equation}
	such that the following holds. 
	Let $\strategy=\{\cal{P}\}$ be a  strategy  that is accepted in $\frak{LowDegree}(d,q,m,k)$ with probability $1 - \eps$. Then there exists 
	a PVM $\{\cal{G}_{f_1,...,f_k}\}$, {acting on the same Hilbert space as $\cal{P}$}, with outcomes in $k$-tuples $f_1,...,f_k\colon \FF_q^m\to \FF_q$ of polynomials of individual degree at most $d$, 
	such that  
	\begin{equation}\label{eq:eval_u_is_close_to_point}
	\cal{G}_{[\eval_u(\cdot)]}\approx_{\delta}\cal{P}^{\Point^u}\ ,    
	\end{equation}
	where $\delta=\delta_\ld(m,d,k,\eps,q^{-1})$ and $\eval_u$ is the ``evaluate at $u$'' function, namely $\eval_u(f_1,...,f_k)=(f_1(u),...,f_k(u))$. 
	In addition, by letting $\eval_{\mathscr{L}}$ be the function that restricts an individual degree at most $d$ polynomial to the line $\mathscr{L}$ and represents it in coefficient representation, we have that 
	\begin{equation}\label{eq:eval_L_is_close_to_line}
	\cal{G}_{[\eval_{\mathscr{L}}(\cdot)]}\approx_\delta \cal{P}^{\ALine^\mathscr{L}}\quad {\rm and}\quad \cal{G}_{[\eval_{\mathscr{L}}(\cdot)]}\approx_\delta \cal{P}^{\DLine^\mathscr{L}} .
	\end{equation}
\end{theorem}

\begin{proof}
	We first apply~\cite[Theorem 4.1]{Ji2022Quantum} to the  degree-$d$ Reed--Solomon code over $\FF_q$. The relative distance of this code is at least $(1-d/q)$. This gives the statement of the theorem for $k=1$. The extension to general $k$ can be done via a standard reduction, following the same steps as the derivation of Theorem 4.43 from Theorem 4.40 in~\cite{NW19}.
\end{proof}
\begin{remark}\label{rem:interpretation_of_the_distance_upper_bound}
	Let us recall the meaning of the notations in \eqref{eq:eval_u_is_close_to_point} and \eqref{eq:eval_L_is_close_to_line}. Recall the data processing notation (Definition \ref{defn:Data_proccessed_PVM}). Then \eqref{eq:eval_u_is_close_to_point} is equivalent to 
	\[
	\sum_{a_1,...,a_k\in \FF_q}\Es{u\sim \FF_q^m}\Big[\Big\|\cal{P}^{\Point^u}_{a_1,...,a_k}- \sum_{\substack{f_1,...,f_k\\ f_i(u)=a_i}}\cal{G}_{f_1,...,f_k}\Big\|^2_{hs}\Big]\leq \delta\ ,
	\]
	while \eqref{eq:eval_L_is_close_to_line} is equivalent to 
	\[
	\sum_{c_{i,j}\in \FF_q}\Es{\substack{\mathscr{L}=\mathscr{L}(u,e_i)\\ i\in [m], u\in \FF_q^m}}\Big[\Big\|\cal{P}^{\ALine^\mathscr{L}}_{(c_{1,0},...,c_{1,d}),...,(c_{k,0},...,c_{k,d}))}- \sum_{\substack{f_1,...,f_k\\ f_i|_{\mathscr{L}}=(c_{i,0},...,c_{i,d})}}\cal{G}_{f_1,...,f_k}\Big\|^2_{hs}\Big]\leq \delta
	\]
	and
	\[
	\sum_{c_{i,j}\in \FF_q}\Es{\substack{\mathscr{L}=\mathscr{L}(u,v)\\  u,v\in \FF_q^m}}\Big[\Big\|\cal{P}^{\DLine^\mathscr{L}}_{(c_{1,0},...,c_{1,md}),...,(c_{k,0},...,c_{k,md}))}- \sum_{\substack{f_1,...,f_k\\ f_i|_{\mathscr{L}}=(c_{i,0},...,c_{i,md})}}\cal{G}_{f_1,...,f_k}\Big\|^2_{hs}\Big]\leq \delta\ .
	\]
\end{remark}
\begin{fact}[Algorithmic Low Degree test]\label{fact:algorithmic_low_degree}
	There is a ($4$-input version of a) $3$-level tailored normal form verifier $\verifier^\ld=(\sampler^{\ld},\length^\ld,\linproc^\ld,\decider)$ with the following properties:\footnote{Though we did not define this $4$-input version, we hope it is clear from context what do we mean by that. Instead of having a sequence of games that are generated uniformly using a single input $n$, we have a sequence of games that are generated uniformly using $4$ inputs $d,t,m,k$.  We spell out explicitly the dependencies of each TM in the normal form verifier on each input, and so do not need the more intricate notion of being $\lambda$-bounded and so on.}
	\begin{enumerate}
		\item \emph{Combinatorial Low Degree test}: For every $d,t,m,k\in \mathbb{N}$, $\verifier^\ld_{d,t,m,k}=\frak{LowDegree}(d,2^t,m,k)$.
		\item \emph{Running time and description length}:  The runtimes of $\sampler^\ld,\length^\ld,\linproc^\ld$ are all bounded by $\poly(d,t,m,k)$. In addition, their description length is constant (up to appending the inputs $d,t,m,k$, which contributes length $O(\log(d\cdot t\cdot m\cdot k))$).
	\end{enumerate}
\end{fact}
\begin{proof}[Proof Sketch]
	Regardless of the rest, $\sampler^\ld,\length^\ld$ and $\linproc^\ld$ run the algorithm of Fact \ref{fact:basis_F_q_over_F_2} with respect to $t$, and recover a fixed basis of $\FF_q$ over $\FF_2$,  so that bit strings of length $t$ can be interpreted and manipulated as elements of the field $\FF_q$ in time $\poly(t)$.
	
	The sampler $\sampler^\ld$ follows the CLMs defined in equations \eqref{eq:cl-ptf},~\eqref{eq:cl-alnf} and~\eqref{eq:cl-dlnf}. Note that all the calculations are in $\FF_q^{2m+1}\cong \FF_2^{t(2m+1)}$, which takes time $\poly(t,m)$.
	
	The answer length calculator $\length^\ld$ outputs  a string of $1$'s of length $kt$ in case the type of question is $\Point$; $kt(d+1)$ in case the type of question is $\ALine$; $kt(md+1)$ in case the type of question is $\DLine$. All in all, this takes at most $kt(md+1)=\poly(d,t,m,k)$-time.
	
	The linear constraints processor $\linproc^\ld$, in case the sampled edge is $\Point^u-\ALine^\mathscr{L}$, calculates the canonical representation of $\mathscr{L}=\mathscr{L}(u_0,v_0)$ --- this takes $\poly(t,m)$-time. Then, it interprets $u$ as $u_0+\alpha v_0$ --- which takes again $\poly(t,m)$-time. Only according to that, it can write $kt$ many equations which amount to verifying that each bit of $f_i(u)$ is the appropriate bit of $\sum \mathsf{AL}f_i(\mathscr{L})_j \cdot \alpha^j$ --- note that the constants are coming from the powers of $\alpha$ and the variables are the bits of  $\mathsf{AL}f_i(\mathscr{L})_j$. All in all, this requires $\poly(d,t,m,k)$-time. 
	The case of $\Point^u-\DLine^\mathscr{L}$ is similar and its runtime is also bounded by $\poly(d,t,m,k)$.
\end{proof}

\subsection{Combinatorial and Algorithmic Answer Reduction}\label{sec:combi_and_algo_answer_red}

As opposed to the question reduction (Section \ref{sec:quered}) and parallel repetition (Section \ref{sec:parallel_rep}) transformations, which have non-complexity theoretic combinatorial descriptions, even the combinatorial transformation of answer reduction is tied to complexity theoretic aspects --- as should already be clear from the previous subsections. Regardless, before describing the answer reduction transformation on the level of normal form verifiers, we describe it on a combinatorial level, with the hopes it clarifies its operation as well as its completeness and soundness properties.

Let us describe the idea briefly. Given a previously $2^{\Lambda}$-padded and purified verifier $\verifier$, with certain bounds on the running times of its sampler, answer length calculator and linear constraint processor,  and fixing an index $n\in \mathbb{N}$, we aim to reduce the length of answers in $\verifier_n$ exponentially, as well as reducing the time it takes to decide whether to accept or reject them. At first, we choose  a field size $q=2^t$ where $t$ is odd.   
A question in the answer reduced game $\frak{AnsRed}(\verifier_n)$ would be a pair of questions, where the first is from the oracularization  (Section \ref{sec:orac}) of the double cover (Definition  \ref{defn:double_cover}) of $\verifier_n$, namely $\frak{Oracle}(\frak{DoubleCover}(\verifier_n))$ (Remark \ref{rem:sampling_scheme_underlying_oracularization} clarifies this point),  and the other from the low degree test $\frak{LowDeg}(9,q,m,\cdot)$, where $m$ is the same as in the definition of a PCP (Definition \ref{defn:PCP_of_V_n_satisfied}). 
The isolated player is assumed to answer with the evaluation at a point or restriction to a line of the multilinear encoding (i.e., individual degree at most $1$ Reed--Muller encoding) of its pair of answers.
The oracle player is assumed to answer with  the evaluation at a point or restriction to a line of a PCP as in Definition \ref{defn:PCP_of_V_n_satisfied}.
Then, the part of the PCP that is supposed to be consistent with the isolated player is checked to be so, and the PCP itself is checked to satisfy \eqref{eq:PCP_condition_1},  \eqref{eq:PCP_condition_2}, \eqref{eq:Tseitin_satisfiable_induced} and \eqref{eq:induced_system_is_satisfied}. 
Moreover, if in the edge sampled in $\frak{AnsRed}(\verifier_n)$, which is a pair of pairs, both pairs agree on the left coordinate, namely the question sampled from  $\frak{Oracle}(\frak{DoubleCover}(\verifier_n))$ is the same in both pairs, then the game is just an instance of $\frak{LowDeg}(9,q,m,\cdot)$. Namely, in such a case, the restriction of the polynomials to lines are checked to be consistent with their evaluations at a point. 

A minor problem arises with this idea, as  the low individual degree test guarantees that the polynomials are low-degree yet all have the same number of variables $m$, while the PCP needs to be with polynomials that depend only on subsets of the $m$ variables, namely, they should be {indifferent} to certain inputs (Definition \ref{def:polynomial_degree}). To that end, we add a ``constant on certain axis parallel lines'' condition that ensures that, indeed, the polynomials that should  depend only on a subset of the variables are such.

\begin{definition}[Combinatorial Answer Reduction]\label{defn:combi_ans_red}
	We defined the game  $\game=\frak{AnsRed}(\verifier,\Lambda,\Delta,D,{T},{Q},n,t)$  given the following provided data: Let
	\begin{itemize}[noitemsep,topsep=0pt,parsep=0pt,partopsep=0pt]
		\item[--] $\Lambda$ be a single input TM that always halts;
		\item[--] $\verifier=(\sampler,\length,\linproc,\decider)$  a purified $2^\Lambda$-padded  $h$-level TNFV (Definition \ref{defn:padded_purified_NFV}) such that $\verifier_n$ is well defined for every $n$ (Definition \ref{defn:h-level_NFV});
		\item[--] $\Delta$ an always halting $1$-input TM that satisfies $\Delta(n)\geq \TIME(\linproc;n,\cdot,\cdot,\cdot,\cdot)\cdot 2^{\Lambda(n)+1}$ and induces $\diamondsuit(n)$ as in \eqref{eq:defn_of_diamondsuit(n)};
		\item [--] $Q$ an always halting $1$-input TM  satisfying 
		$\TIME(\sampler;{n},\cdot,\cdot,\cdot,\cdot,\cdot)\leq {Q}(n)$;
		\item [--] $T$ an always halting $1$-input TM  satisfying 
		\begin{equation}
		T(n)\geq c\cdot(\TIME(\Lambda;n)^c+\TIME(\Delta;n)^c+2^{c\cdot\Lambda(n)}+\Delta(n)^c+\TIME(\linproc;n,\cdot,\cdot,\cdot,\cdot)^c)\ ,
		\end{equation}
		where $c\geq 6$ is the positive integer implied by the $\poly$ notation in \eqref{eq:time_bound_L*};
		\item [--] $D$ a positive integer (in binary) satisfying 
		$|\verifier|,|\Lambda|,|\Delta|,|T|,|Q|\leq D$;
		\item[--]  $n$ and $t$  positive integers, and $q=2^t$;
		
		\item[--] $
		(M(n),s(n),\cdot)=\paddedsuccinctdecider(\cdot,\cdot,\cdot,D,T,Q,n,\cdot,\cdot),
		$
		where $\paddedsuccinctdecider$ was defined in Proposition \ref{prop:explicit-padded-succinct-deciders}.
	\end{itemize}
	\quad\\
	\textbf{Underlying graph and sampling scheme of $\game$}: 
	
	The answer reduced game has a  typed $\max(h,3)$-level CL sampling scheme (Definition \ref{defn:typed_h-level_sampling_scheme}). First, the type graph is the tensor product of the graph $\mttA-\oracle-\mttB$ (including self loops) and the graph $\ALine-\Point-\DLine$ (including self loops) --- see Figure \ref{fig:ans_red_type_graph}. 
	
	For the CLMs associated with each type:
	\begin{itemize}
		\item Denote by $\frS^\mttA$ the $h$-level CLM induced by $\sampler(n,\cdot,A,\cdot,\cdot,\cdot)$, and similarly $\frS^\mttB$ for the one induced by $\sampler(n,\cdot,B,\cdot,\cdot,\cdot)$.
		\item Furthermore, let $r$ be the dimension of $\frS^\mttA$ and $\frS^\mttB$, namely the output of $\sampler(n,{\rm Dimension},\cdot,\cdot,\cdot,\cdot)$, and let $\frS^\oracle\colon \FF_2^r\to \FF_2^r$ be the identity map (which is a linear map, and hence a $1$-level CLM). 
		\item Finally, recall the CLMs $\frS^\Point,\frS^\ALine$ and $\frS^\DLine$ defined in \eqref{eq:cl-ptf},~\eqref{eq:cl-alnf} and \eqref{eq:cl-dlnf} respectively, which act on $\FF_2^{t(2m+1)}$, where $m=|S|=4\Lambda(n)+3\diamondsuit(n)+3M(n)+s(n)+12$ is the number of variables in a PCP (Definition \ref{defn:PCP_of_V_n_satisfied}).
	\end{itemize} 
	Then, the CLMs of $\game$ act on the space $\FF_2^{r}\times \FF_2^{t(2m+1)}$, and for every $$(\mathtt{Player},\mttSpace)\in \{\mttA,\mttB,\oracle\}\times \{\Point,\ALine,\DLine\}\quad {\rm and}\quad (z,(u,s,v))\in \FF_2^{r}\times \FF_2^{t(2m+1)}\ ,$$ we have 
	\[
	\frS^{(\mathtt{Player},\mttSpace)}(z,(u,s,v))=(\frS^{\mathtt{Player}}(z),\frS^\mttSpace(u,s,v))\ .
	\]
	By endowing $\frak{Oracle}(\frak{DoubleCover}(\verifier_n))$ with the appropriate typed CL sampling scheme (see Remark \ref{rem:sampling_scheme_underlying_oracularization}), the above is just the direct sum (\cite[Lemma 4.8]{MIPRE}) of it and the CL sampling scheme of $\frak{LowDegree}(9,q,m,\cdot)$ --- namely, a pair of questions is sampled in each game independently, and the resulting edge is the pair of pairs. Note that in particular, this typed CL sampling scheme has level $\max(h,3)$, as claimed.
	
	As the vertices of $\frak{LowDegree}(9,q,m,\cdot)$ are either $\Point^p$ for a point $p\in \FF_q^m$, $\ALine^\mathscr{L}$ for an axis parallel line $\mathscr{L}$ in $\FF_q^m$, or $\DLine^\mathscr{L}$ for any line $\mathscr{L}$ in $\FF_q^m$, we denote the vertices in $\game$ as $((\mathtt{Player},\mttw),\mttSpace^\rho)$, where $\mathtt{Player}\in \{\mttA,\mttB,\oracle\}$, $\mttw\in \FF_2^r$, $\mttSpace\in \{\Point,\ALine,\DLine\}$ and $\rho$ is either a point or a line in $\FF_q^m$. When $\mathtt{Player}$ is either $\mttA$ or $\mttB$, we call it an isolated player question, and when  $\mathtt{Player}$ is $\oracle$, we call it an oracle player question (compare to the naming convention in the oracularized game $\frak{Oracle}(\cdot)$ in Section \ref{sec:orac}).
	\\
	
	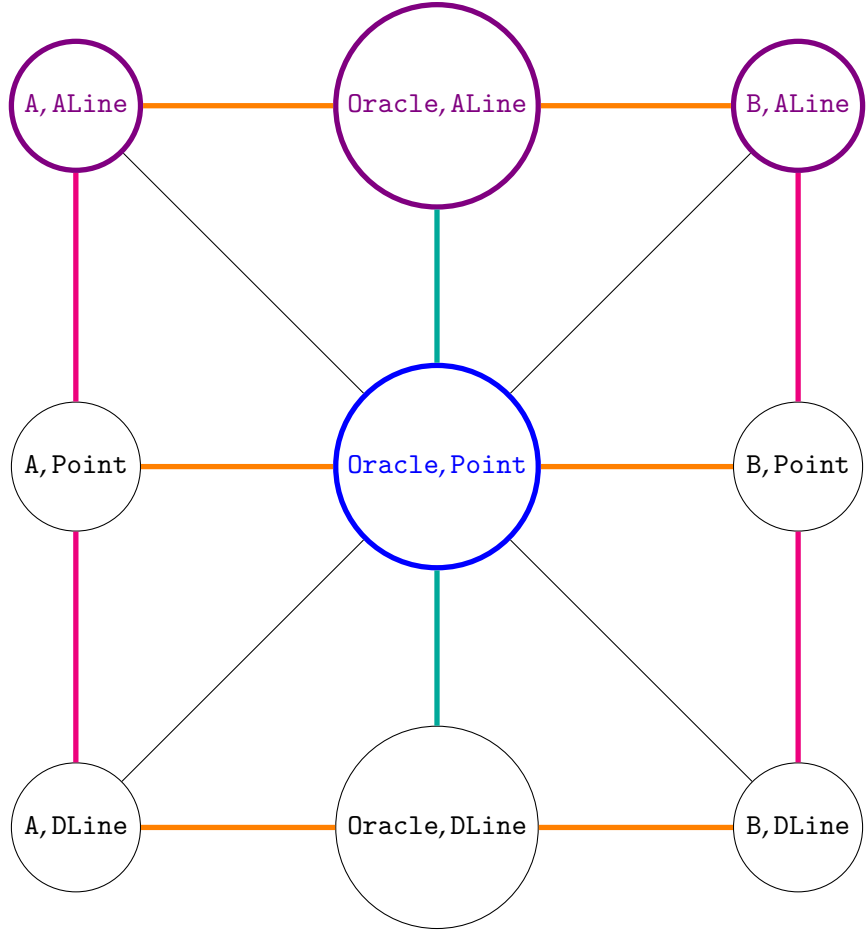
\begin{figure}[httb!]
		\centering
		\captionsetup{singlelinecheck=off}
		\begin{tikzpicture}[scale=1.2]
		\node[draw, line width=2, color=blue, shape=circle] (OP) at (0,0) { $\oracle,\Point$}; 
		
		\node[draw, color=black, shape=circle] (BP) at (4,0) { $\mttB,\Point$}; 
		
		\node[draw, color=black, shape=circle] (AP) at (-4,0) { $\mttA,\Point$}; 
		
		\node[draw, line width=2, color=violet, shape=circle] (OAL) at (0,4) { $\oracle,\ALine$}; 
		
		\node[draw, line width=2, color=violet, shape=circle] (BAL) at (4,4) { $\mttB,\ALine$}; 
		
		\node[draw, line width=2, color=violet, shape=circle] (AAL) at (-4,4) { $\mttA,\ALine$}; 
		
		\node[draw, color=black, shape=circle] (ODL) at (0,-4) { $\oracle,\DLine$}; 
		
		\node[draw, color=black, shape=circle] (BDL) at (4,-4) { $\mttB,\DLine$}; 
		
		\node[draw, color=black, shape=circle] (ADL) at (-4,-4) { $\mttA,\DLine$};

		\draw[orange,line width=2,  -, solid] (OP)--(AP);
		\draw[orange, line width=2, -, solid] (OP)--(BP);
		\draw[orange, line width=2, -, solid] (OAL)--(AAL);
		\draw[orange, line width=2, -, solid] (OAL)--(BAL);
		
		\draw[orange, line width=2, -, solid] (ODL)--(ADL);
		\draw[orange, line width=2, -, solid](ODL)--(BDL);
		\draw[JungleGreen, line width=2, -, solid] (OP)--(OAL);
		\draw[JungleGreen, line width=2, -, solid] (OP)--(ODL);
		\draw[RubineRed, line width=2, -, solid] (AP)--(AAL);
		\draw[RubineRed, line width=2, -, solid] (AP)--(ADL);
		\draw[RubineRed, line width=2, -, solid] (BP)--(BAL);
		\draw[RubineRed, line width=2, -, solid] (BP)--(BDL);
		\draw[black, -, solid] (OP)--(AAL);
		\draw[black, -, solid] (OP)--(BAL);
		\draw[black, -, solid] (OP)--(ADL);
		\draw[black, -, solid] (OP)--(BDL);
		\end{tikzpicture}
		\caption[]{The type graph of $\frak{AnsRed}(\verifier,\Lambda,\Delta,D,{T},{Q},n,t)$. Though not drawn, all self loops are also edges in this type graph. We added (five) colours that can be compared to the checks in the game. When the edge is either {\color{RubineRed} pink} or {\color{JungleGreen} green}, \cref{clause1:low_degree_check_in_ans_red}, which is a low-degree test,  is checked --- when  {\color{RubineRed} pink}, it is an instance of $\frak{LowDegree}(9,q,m,2)$, and when {\color{JungleGreen} green} it is an instance of $\frak{LowDegree}(9,q,m,\heartsuit(n))$ (with $\heartsuit$  being recalled in \eqref{eq:def_heartsuit}). When an edge is {\color{orange} orange},  \cref{clause2:consistency_checks_in_ans_red} is checked, which is a consistency check between the (evaluation of the) $g$ polynomials provided by the isolated player versus the respective  $g$ polynomials in the (evaluation of the) PCP proof $\Pi$ provided by the oracle player. When an edge incident to a {\color{violet} purple} vertex is sampled, \cref{clause3:Indifference_checks_in_ans_red} is checked, which is an indifference of the relevant polynomials in the direction the sampled axis parallel line is. Finally, when an edge incident to the (single) {\color{blue} blue} vertex is sampled, \cref{clause4:PCP_checks_in_ans_red} is checked, which is interpreting the answers at the $(\oracle,\Point)$-typed vertex as an evaluation of a PCP at a point (Definition \ref{defn:PCP_of_V_n_satisfied}), and verifying that these values satisfy \eqref{eq:PCP_condition_1},  \eqref{eq:PCP_condition_2}, \eqref{eq:Tseitin_satisfiable_induced} and \eqref{eq:induced_system_is_satisfied} with respect to the  circuit $\circuit$. 
		}
		\label{fig:ans_red_type_graph}
	\end{figure}

	\textbf{Answer lengths and structure of answers}:
	
	A full description appears in  Table \ref{tab:Answer_Lengths_combinatorial_ans_red}. Some guidance to parse the table: Whenever the second coordinate of the question is $\Point^p$, the answer consists of a sequence of values in $\FF_q$ (as length $t$ bit strings), that are supposed to be the evaluation of polynomials of degree at most $9$ in $m$ variables at the point $p\in \FF_q^m$ --- in case the first coordinate of the question is $(\mttA,\mttx)$ or $(\mttB,\mtty)$, evaluations of just two polynomials, one readable and one linear, and in case the first coordinate is $(\oracle,z)$,  the supposed evaluation of a PCP $\Pi$ (Definition \ref{defn:PCP_of_V_n_satisfied}) at a point $p\in \FF_q^m$, where $\Pi^\frR(p)$ is the readable part of the answer and $\Pi^\frL(p)$ the linear part. Similarly, when $\ALine^\mathscr{L}$ (respectively $\DLine^\mathscr{L}$) is the second coordinate of the question, the answer consists of a sequence of $10$-tuples  (respectively $( 9m+1)$-tuples) of values in $\FF_q$ that encode the restrictions of ``the same'' polynomials as before to the respective line $\mathscr{L}$.  As the answer when $\oracle$ is asked are restrictions of a PCP to a certain subspace (either a point or a line), this is a list of 
	\begin{equation}\label{eq:def_heartsuit}
	\heartsuit(n)=12\Lambda(n)+12\diamondsuit(n)+6M(n)+s(n)+35      
	\end{equation}
	many values (in $\FF_q$ when evaluating at a point, and tuples in $\FF_q$ when evaluating at a line). 
	\begin{table}[!htbp]
		\centering
		\captionsetup{singlelinecheck=off}
		\begin{tabular}{|c|c||c|c|c|}
			\hline
			& &$\Point^p$ & $\ALine^\mathscr{L}$ & $\DLine^\mathscr{L}$\\
			\hline
			\hline
			$(\mttA,\mttx)$& Readable & $g^\frR_{\mttA,\mttx}(p)$ & $\mathsf{AL}g^\frR_{\mttA,\mttx}(\mathscr{L})$ & $\mathsf{DL}g^\frR_{\mttA,\mttx}(\mathscr{L})$ \\
			& Linear & $g^\frL_{\mttA,\mttx}(p)$ & $\mathsf{AL}g^\frL_{\mttA,\mttx}(\mathscr{L})$ & $\mathsf{DL}g^\frL_{\mttA,\mttx}(\mathscr{L})$ \\
			\hline 
			$(\mttB,\mtty)$&Readable & $g^\frR_{\mttB,\mtty}(p)$ & $\mathsf{AL}g^\frR_{\mttB,\mtty}(\mathscr{L})$ & $\mathsf{DL}g^\frR_{\mttB,\mtty}(\mathscr{L})$ \\
			& Linear & $g^\frL_{\mttB,\mtty}(p)$ & $\mathsf{AL}g^\frL_{\mttB,\mtty}(\mathscr{L})$ & $\mathsf{DL}g^\frL_{\mttB,\mtty}(\mathscr{L})$ \\
			\hline 
			$(\oracle,z)$ & Readable & $\Pi_{\mttO,z}^\frR(p)$& $\mathsf{AL}\Pi_{\mttO,z}^\frR(\mathscr{L})$ & $\mathsf{DL}\Pi_{\mttO,z}^\frR(\mathscr{L})$ \\
			
			& Linear&$\Pi_{\mttO,z}^\frL(p) $& $\mathsf{AL}\Pi_{\mttO,z}^\frL(\mathscr{L})$ & $\mathsf{DL}\Pi_{\mttO,z}^\frL(\mathscr{L})$ \\
			\hline 
		\end{tabular}
		\caption[]{In the above table, $g^\circ_{\circ,\circ}(p)$ is an element of $\FF_q$ (namely a bit string of length $t$),  $\mathsf{AL}g^\circ_{\circ,\circ}(\mathscr{L})$ is a tuple of $10$ elements in $\FF_q$ (which encode a degree $9$ univariate polynomial), and $\mathsf{DL}g^\circ_{\circ,\circ}(\mathscr{L})$ is a tuple of $9m+1$ elements in $\FF_q$ (which encode a degree $9m$ univariate polynomial). 
			Recall the notion of a degree $9$ PCP $\Pi$ over $\FF_q$ (Definition \ref{defn:PCP_of_V_n_satisfied}), and specifically the number $\heartsuit^\frR(n)=8\Lambda(n)+6\diamondsuit(n)+6M(n)+s(n)+24$ of polynomials in the readable part $\Pi^\frR$ of $\Pi$ and the number $\heartsuit^\frL(n)=4\Lambda(n)+6\diamondsuit(n)+11$  of polynomials in the linear part $\Pi^\frL$ of $\Pi$.
			Then, the answer $\Pi^\frR_{\mttO,z}(p)$ consists of $\heartsuit^\frR(n)$-many values in $\FF_q$, which we denote according to the names of the polynomials in the readable part of a PCP; namely, $\Pi^\frR_{\mttO,z}(p)$ consists of the $\FF_q$-values
			\[
			\begin{split}
			\forall i\in [5]\ ,\ \sX\in S^\kappa_i\ &\colon  \ \ g^\frR_i(p)\ ,\ \beta^\frR_{i,\sX}(p)    \\
			\forall \sX\in S\ &\colon \ \ \alpha^\frR_\sX(p)\ ,\\ 
			\forall \sX\in S_0\  &\colon\ \  g_0(p)\ ,\ \beta_{0,\sX}(p)\ .
			\end{split}
			\]
			Similarly, the answer $\mathsf{AL}\Pi^\frR_{\mttO,z}(\mathscr{L})$ (respectively  $\mathsf{DL}\Pi^\frR_{\mttO,z}(\mathscr{L})$) consists of $\heartsuit^\frR(n)$-many  $10$-tuples (respectively $9m+1$-tuples)  of values in $\FF_q$, which we denote according to the names of the polynomials in the readable part of a PCP as well; namely, $\mathsf{AL}\Pi^\frR_{\mttO,z}(\mathscr{L})$  (respectively  $\mathsf{DL}\Pi^\frR_{\mttO,z}(\mathscr{L})$)  consists of the $10$-tuples  (respectively $9m+1$-tuples) of $\FF_q$-values
			\[
			\begin{split}
			\forall i\in [5]\ ,\ \sX\in S^\kappa_i\ &\colon  \ \ \mathsf{AL}g^\frR_i(\mathscr{L})\ ,\ \mathsf{AL}\beta^\frR_{i,\sX}(\mathscr{L})\ , \quad ({\rm resp.}\ \mathsf{DL}g^\frR_i(\mathscr{L})\ ,\ \mathsf{DL}\beta^\frR_{i,\sX}(\mathscr{L}))\ ,   \\
			\forall \sX\in S\ &\colon \ \ \mathsf{AL}\alpha^\frR_\sX(\mathscr{L})\ ,\quad ({\rm resp.}\ \mathsf{DL}\alpha^\frR_\sX(\mathscr{L}))\ ,\\ 
			\forall \sX\in S_0\  &\colon\ \  \mathsf{AL}g_0(\mathscr{L})\ ,\ \mathsf{AL}\beta_{0,\sX}(\mathscr{L})\ ,\quad ({\rm resp.}\ \mathsf{DL}g_0(\mathscr{L})\ ,\ \mathsf{DL}\beta_{0,\sX}(\mathscr{L}))\ .
			\end{split}
			\]
			For the linear part, the answer $\Pi^\frL_{\mttO,z}(p)$ consists of $\heartsuit^\frL(n)$-many values in $\FF_q$, which we denote according to the names of the polynomials in the linear part of a PCP; namely, $\Pi^\frL_{\mttO,z}(p)$ consists of the $\FF_q$-values
			\[
			\begin{split}
			\forall i\in [5]\ ,\ \sX\in S^\kappa_i\ &\colon  \ \ g^\frL_i(p)\ ,\ \beta^\frL_{i,\sX}(p)    \\
			\forall \sX\in S_0\ &\colon \ \ \alpha^\frL_\sX(p)\ .
			\end{split}
			\]
			Similarly, the answer $\mathsf{AL}\Pi^\frL_{\mttO,z}(\mathscr{L})$ (respectively  $\mathsf{DL}\Pi^\frL_{\mttO,z}(\mathscr{L})$) consists of $\heartsuit^\frL(n)$-many  $10$-tuples (respectively $9m+1$-tuples)  of values in $\FF_q$, which we denote according to the names of the polynomials in the linear part of a PCP as well; namely, $\mathsf{AL}\Pi^\frL_{\mttO,z}(\mathscr{L})$  (respectively  $\mathsf{DL}\Pi^\frL_{\mttO,z}(\mathscr{L})$)  consists of the $10$-tuples  (respectively $9m+1$-tuples) of $\FF_q$-values
			\[
			\begin{split}
			\forall i\in [5]\ ,\ \sX\in S^\kappa_i\ &\colon  \ \ \mathsf{AL}g^\frL_i(\mathscr{L})\ ,\ \mathsf{AL}\beta^\frL_{i,\sX}(\mathscr{L})\ , \quad ({\rm resp.}\ \mathsf{DL}g^\frL_i(\mathscr{L})\ ,\ \mathsf{DL}\beta^\frL_{i,\sX}(\mathscr{L}))\ ,   \\
			\forall \sX\in S\ &\colon \ \ \mathsf{AL}\alpha^\frL_\sX(\mathscr{L})\ ,\quad ({\rm resp.}\ \mathsf{DL}\alpha^\frL_\sX(\mathscr{L}))\ .
			\end{split}
			\]
		}
		\label{tab:Answer_Lengths_combinatorial_ans_red}
	\end{table}
	\\
	
	\textbf{Controlled linear constraints function}:
	
	We can now describe the game checks. They can be collected into four  (not mutually exclusive) checks, namely if a condition holds, then it is checked, and for certain pairs of questions more than one condition holds. Assume the sampled edge was $((\mathtt{Player},\mttw),\mttSpace_1^{\rho_1})-((\mathtt{Player}',\mttw'),\mttSpace_2^{\rho_2})$, where $\mathtt{Player},\mathtt{Player}'\in \{\mttA,\mttB,\oracle\}$, $\mttw,\mttw'\in \FF_2^r$, $\mttSpace_i\in \{\Point,\ALine,\DLine\}$ and $\rho_i$ is either a point $p\in \FF_q^m$ or a line $\mathscr{L}\subseteq \FF_q^m$.
	\begin{enumerate}[label=\textcolor{black}{(\arabic*)}, ref= (\arabic*)]
		\item \label{clause1:low_degree_check_in_ans_red} \underline{Low degree test}: In case the pair of questions agree on their first coordinate, namely $\mathtt{Player}=\mathtt{Player}'$ (and thus $\mttw=\mttw'$, as this is a CL sampling scheme), the second coordinates pair $\mttSpace_1^{\rho_1}-\mttSpace_2^{\rho_2}$ is an edge in the game $\frak{LowDeg}(9,q,m,\cdot)$. Hence, we run   the checks from $\frak{LowDeg}(9,q,m,k)$ on the respective answers of the players, where $k=2$ in case $\mathtt{Player}=\mttA$ or $\mttB$, and $k=\heartsuit(n)$ in case $\mathtt{Player}=\oracle$. In more words, assuming $\mttSpace_1=\Point$, $\rho_1=p$, $\mttSpace_2=\ALine$ (respectively $\DLine$) and $\rho_2=\mathscr{L}$, and letting $f$ be one of the presumed polynomials, the answer of the  first player is treated as $f(p)\in \FF_q$ (which also agrees with our notations in Table \ref{tab:Answer_Lengths_combinatorial_ans_red}) and the answer of the second player is treated as $\mathsf{AL}f(\mathscr{L})=(a_0,...,a_9)\in \FF_q^{10}$ (respectively $\mathsf{DL}f(\mathscr{L})=(a_0,...,a_{9m})\in \FF_q^{9m+1}$). As $p\in \mathscr{L}$, and letting $\mathscr{L}=\mathscr{L}(p_0,v_0)$ be its canonical representation (Definition \ref{def:line-representative}), there is a scalar $\alpha\in \FF_q$ such that $p_0+\alpha v_0=p$. Then, check that $f(p)=\sum a_i\alpha^i$. This can be implemented as $t$ many $\FF_2$-linear checks, and thus the tailored structure of the game is preserved. 
		
		\item \label{clause2:consistency_checks_in_ans_red}  \underline{Consistency check induced by the oracularized game}: In case the pair of questions agree on their second coordinate, namely  $\mttSpace_1=\mttSpace_2$ (and thus $\rho:=\rho_1=\rho_2$, as this is a CL sampling scheme), the first coordinates pair $(\mathtt{Player},\mttw)-(\mathtt{Player}',\mttw')$ is a valid pair of questions (see Remark \ref{rem:sampling_scheme_underlying_oracularization}) in the double cover of the oracularized game 
		\[
		\frak{Oracle}(\frak{DoubleCover}(\verifier_n))\ .
		\] 
		In such a case, we check consistency between the $g^\kappa_{\mathtt{Player},\mttw}$-polynomials provided by the isolated player and the appropriate part of the PCP $\Pi$ provided by the oracle player. Namely, if $(\mathtt{Player},\mttw)=(\mttA,\mttx)$ (respectively $(\mttB,\mtty)$) and $(\mathtt{Player}',\mttw')=(\oracle,z)$, then the first player responds with $\square g_{\mttA,\mttx}^\frR(\rho),\square g_{\mttA,\mttx}^\frL(\rho)$ (respectively $\square g_{\mttB,\mtty}^\frR(\rho),\square g_{\mttB,\mtty}^\frL(\rho)$), where $\square$ is either blank, $\mathsf{AL}$ or $\mathsf{DL}$,  and  part of the second player's response is $\square g_{1}^\frR(\rho),\square g_{1}^\frL(\rho)$ \quad (respectively  $\square g_{2}^\frR(\rho),\square g_{2}^\frL(\rho)$). Thus, it checks that 
		\[
		\square g_{\mttA,\mttx}^\frR(\rho)=\square g_{1}^\frR(\rho)\quad{\rm and}\quad \square g_{\mttA,\mttx}^\frL(\rho)=\square g_{1}^\frL(\rho)
		\]
		(respectively,
		\[
		\square g_{\mttB,\mtty}^\frR(\rho)=\square g_{2}^\frR(\rho)\quad{\rm and}\quad \square g_{\mttB,\mtty}^\frL(\rho)=\square g_{2}^\frL(\rho))\;.
		\]
		These can be realized as $2t$-many $\FF_2$-linear checks and thus this check preserves the tailored structure of the game.

		\item \label{clause3:Indifference_checks_in_ans_red} \underline{Indifference check}: Recall the notion of indifference of a polynomial to a certain index (Definition \ref{def:polynomial_degree}), and specifically which polynomials of the PCP $\Pi$ (Definition \ref{defn:PCP_of_V_n_satisfied}) are indifferent to which indexes (namely, which variables from $S$ do not appear in the polynomial).
		Assume $\mttSpace_1=\ALine$ (the case of $\mttSpace_2=\ALine$ is essentially the same, so we do not spell it out) $\rho_1=\mathscr{L}(p,e_\sX)$ where $\sX\in S$ --- here we use the bijection between $[m]$ and $S$ induced by \eqref{eq:point_p} --- and let $f$ be one of the polynomials provided by the player ($g^\circ_{\circ,\circ}$ in case of an isolated player, and one of $g^\circ_\circ,\alpha^\circ_\circ,\beta^\circ_{\circ,\circ}$ in case of an oracle player). If $f$ should be indifferent to $\sX$, as $\sX$ is not in the formal generating set defining $f$ (compare to Definition \ref{defn:blocks_of_vars_circuit}), then $f$ should be constant along the axis parallel line $\rho_1$. Hence, if $\mathsf{AL}f(\rho_1)=(a_0,...,a_9)$, then we check that for every $j\geq 1$, $a_j=0$. This is done for all $f$'s in the answer that are indifferent to the specific variable $\sX$. Also, equating to $0$  is clearly a linear check, and thus tailored (as a sanity check, note that these are $9t$-many $\FF_2$-linear equations).
		
		\item  \label{clause4:PCP_checks_in_ans_red} \underline{Proof check}: In case one of the questions is $((\oracle,z),\Point^p)$, where $p\in \FF_q^S$ is as in \eqref{eq:point_p},  calculate the following:
		\begin{itemize}
			\item $\mttx=\frS^\mttA(z)$ and $\mtty=\frS^\mttB(z)$;
			\item $\circuit$ which is the third  output of $\paddedsuccinctdecider(\verifier,\Lambda,\Delta,D,T,Q,n,\mttx,\mtty)$;
			\item $T_\circuit(p)\in \FF_q$, where $T_\circuit$ is the Tseitin polynomial (Definition \ref{defn:Circuits}) associated with $\circuit$.
		\end{itemize} 
		Treating the answer of the player who receives this question as the evaluation of a PCP $\Pi$ at the point $p$, we can check that the equations \eqref{eq:PCP_condition_1},  \eqref{eq:PCP_condition_2}, \eqref{eq:Tseitin_satisfiable_induced} and \eqref{eq:induced_system_is_satisfied} are satisfied at this specific point. 
		Namely, the following $13$ equations over $\FF_q$ are checked:
		\begin{align}
		\forall \kappa\in \{\frR,\frL\}\ ,\ i\in[5]\ \colon \ \ g^\kappa_i(p)(g^\kappa_i(p)+1)&=\sum_{\sX\in S^\kappa_i} \zero_\sX(p)\cdot \beta^{\kappa}_{i,\sX}(p)\ ,\label{eq:PCP_check_1_ansred}\\
		g_0(p)(g_0(p)+1)&=\sum_{\sX\in S_0} \zero_\sX(p)\cdot \beta_{0,\sX}(p)\ ,\label{eq:PCP_check_2_ansred}\\
		T_\circuit(p)(g_0(p)+p^\frR_6+1)\prod_{i=1}^5\left( g^\frR_i(p)+p^\frR_{i+6}+1\right)&=\sum_{\sX\in S} \zero_{\sX}(p)\cdot \alpha^\frR_{\sX}(p)\ ,\label{eq:PCP_check_3_ansred}\\
		g_0(p)(p^\frL_{11}+\sum_{i=1}^5 p^\frL_{i+5}g_i^\frL(p))&=\sum_{\sX\in S_0} \zero_{\sX}(p)\cdot \alpha^\frL_{\sX}(p)\ \label{eq:PCP_check_4_ansred}.
		\end{align}
		Let us briefly describe why  these equations are tailored, as this is not obvious. The right hand side in all of the equations is already a linear combination of the values provided by the oracle player.  For the left hand side of \eqref{eq:PCP_check_1_ansred} and \eqref{eq:PCP_check_2_ansred}, note that the map $\sX\mapsto \sX^2+\sX$ from $\FF_q$ to itself is $\FF_2$-linear, and thus these are also just linear combinations of the $t$-bits representing each $\FF_q$ value over $\FF_2$.  Hence, \eqref{eq:PCP_check_1_ansred} and \eqref{eq:PCP_check_2_ansred} are fixed linear checks that are independent of the value of the readable answers, and in particular are tailored. The left hand side of \eqref{eq:PCP_check_3_ansred} is some function of $p$ and the readable parts $g_0(p),g^\frR_1(p),...,g^\frR_5(p)$ of the answer, which makes it a constant once the readable part of the answer is given --- therefore this check is tailored as well. Finally, the left hand side of \eqref{eq:PCP_check_4_ansred} is a linear combination of the linear parts $g^\frL_1(p),...,g^\frL_5(p)$, with the exact coefficients depending on the value $g_0(p)$, which is readable --- this shows that, indeed, this check is properly tailored.
	\end{enumerate}
	
\end{definition}

\begin{proposition}[Completeness and Soundness of the Answer Reduced game]\label{prop:completeness_soundness_combi_ans_red}
	Recall all the data provided in Definition \ref{defn:combi_ans_red} and the assumptions on it, and recall the definition of the answer reduced game $\game=\frak{AnsRed}(\verifier,\Lambda,\Delta,D,{T},{Q},n,t)$.
	Then:
	\begin{enumerate}
		\item \emph{Completeness}: If the game $\verifier_n$ has a perfect $\ZPC$-strategy, then so does $\game$.
		\item \emph{Soundness}: If $\game$ has a strategy with value $1-\eps$, then $\frak{DoubleCover}(\verifier_n)$ has a strategy with value of at least \[1-\frac{10^6\cdot m}{1-\frac{72m}{q}}\cdot\delta^{\nicefrac{1}{8}}\ \] 
		of the same dimension, 
		where $\delta=\delta_\ld(m,9,\heartsuit(n),2\eps,q^{-1})$ and $\delta_\ld$ is as defined in~\eqref{eq:defn_delta_ld}.
		\item \emph{Entanglement lower bound}: Furthermore, 
		\[
		\Ent(\game,1-\eps)\geq \Ent\left(\frak{DoubleCover}(\verifier_n),1-\frac{10^6\cdot m}{1-\frac{72m}{q}}\cdot\delta^{\nicefrac{1}{8}}\right)\ .
		\]
	\end{enumerate}
\end{proposition}

\begin{proof}

	\textbf{Completeness}:
	\\
	
	Assume $\verifier_n$ has a perfect $\ZPC$-strategy. By the completeness of the double cover transformation (Claim \ref{claim:completeness_soundness_double_cover}) and the completeness of the oraculatrization transformation (Claim \ref{claim:completeness_soundness_oracle}), there is a perfect\footnote{Note that here we are using the commuting along edges condition, essentially, for the first and last time in the proof of compression.} $\ZPC$-strategy $\cal{U}$ for 
	\[
	\frak{Oracle}(\frak{DoubleCover}(\verifier_n)) \ .
	\]
	As discussed in Remark \ref{rem:sampling_scheme_underlying_oracularization}, the vertices in this game can be parametrized by $(\mathtt{Player},\mttw)$, where $\mathtt{Player}\in \{\mttA,\mttB,\oracle\}$ and $\mttw\in \FF_2^{r}$ with $r=\sampler(n,{\rm Dimension},\cdot,\cdot,\cdot,\cdot)$.
	Since $\verifier$ is  $2^\Lambda$-padded, the size of the generating sets at the isolated player vertices is $2^{\Lambda(n)}$  and at the oracle player vertices is $2\cdot 2^{\Lambda(n)}$; namely, the answers at an $\mttA$-player vertex consist of two functions $a^\frR,a^\frL\colon \FF_2^{\Lambda(n)}\to \FF_2$, the answers at a $\mttB$-player vertex consist of two functions $b^\frR,b^\frL\colon \FF_2^{\Lambda(n)}\to \FF_2$, and the  answers at an $\oracle$-player vertex consist of four functions $f^\frR_1,f^\frL_1,f^\frR_2,f^\frL_2\colon \FF_2^{\Lambda(n)}\to \FF_2$. 
	
	We will now construct the PVMs of the perfect $\ZPC$-strategy for $\game$ at the isolated player vertices. This is done by first inducing and then evaluating the outcomes of the PVM $\cal{U}$ associates with the vertices $(\mttA,\mttx)$ and $(\mttB,\mtty)$, namely data-processing; as both induction and evaluation are linear maps, this preserves the $Z$-alignment and permutation properties. Let us elaborate more.
	Recall the discussion from Remark \ref{rem:induction_is_linear}: Every  map $f\colon \FF_2^{\Lambda(n)}\to \FF_2$ can be thought of as a multilinear polynomial in $\FF_2[S_f]$, where $S_f$ is a formal set of $\Lambda(n)$-many variables. As $\FF_2[S_f]$ embeds naturally into $\FF_q[S_f]$, we associated with every function $f$ a polynomial in $\FF_q[S_f]$. As the variables of all of our polynomials are contained in the set $S$ described in Definition \ref{defn:blocks_of_vars_circuit}, $\FF_q[S_f]$ naturally embeds into $\FF_q[S]$. As every polynomial in $\FF_q[S]$ has an evaluation table (its $\Phi_{\FF_q}$-image \eqref{def:Phi_FF}), it induces a function $\FF_q^{m}\to\FF_q$, where as usual $m=|S|$. Let us denote by $\Ind(f)\colon \FF_q^{m}\to\FF_q$ this final function --- although this is a \textbf{variation} on the $\Ind(\cdot)$ map from Definition  \ref{defn:restriction_and_induction_polynomials} and Remark \ref{rem:induction_is_linear}, we use the same notation, and this map is linear as well. Note that the exact way $S_f$ is embeded in $S$ plays a role in the induction function. For example, $a^\kappa$ is associated with $S_1^\kappa$ from Definition \ref{defn:blocks_of_vars_circuit}, while $b^\kappa$ is associated with $S_2^\kappa$. Recall  also the evaluation at a point map $\eval_p(\cdot)$ and the evaluation at a line map $\eval_{\mathscr{L}}(\cdot)$ described in Theorem \ref{thm:ldc-soundness}, which are also $\FF_2$-linear. Define for every $\mttx,\mtty\in \FF_2^r$, $\mttSpace\in \{\Point,\ALine,\DLine\}$ and $\rho$ an appropriate point or line, the PVM
	\begin{equation}\label{eq:def_W_isolate}
	\cal{W}^{(\mttA,\mttx),\mttSpace^{\rho}}=\cal{U}^{(\mttA,\mttx)}_{[\eval_{\rho}\circ \Ind(\cdot)]}\quad,\quad \cal{W}^{(\mttB,\mtty),\mttSpace^{\rho}}=\cal{U}^{(\mttB,\mtty)}_{[\eval_{\rho}\circ \Ind(\cdot)]}\ ;
	\end{equation}
	namely,
	\[
	\begin{split}
	\forall \alpha^\frR,\alpha^\frL\in \FF_q\ \colon \ \ \cal{W}^{(\mttA,\mttx),\Point^{p}}_{\alpha^\frR,\alpha^\frL}&=\sum_{\substack{a^\frR,a^\frL\colon \FF_2^{\Lambda(n)}\to \FF_2\\ \Ind(a^\frR)(p)=\alpha^\frR\ ,\ \Ind(a^\frL)(p)=\alpha^\frL}} \cal{U}^{(\mttA,\mttx)}_{a^\frR,a^\frL}\\
	\forall \alpha^\frR,\alpha^\frL\in \FF_q^{10}\ \colon \ \ \cal{W}^{(\mttA,\mttx),\ALine^{\mathscr{L}}}_{\alpha^\frR,\alpha^\frL}&=\sum_{\substack{a^\frR,a^\frL\colon \FF_2^{\Lambda(n)}\to \FF_2\\ \Ind(a^\frR)(\mathscr{L})=\alpha^\frR\ ,\ \Ind(a^\frL)(\mathscr{L})=\alpha^\frL}} \cal{U}^{(\mttA,\mttx)}_{a^\frR,a^\frL}\\
	\forall \alpha^\frR,\alpha^\frL\in \FF_q^{9m+1}\ \colon \ \ \cal{W}^{(\mttA,\mttx),\DLine^{\mathscr{L}}}_{\alpha^\frR,\alpha^\frL}&=\sum_{\substack{a^\frR,a^\frL\colon \FF_2^{\Lambda(n)}\to \FF_2\\ \Ind(a^\frR)(\mathscr{L})=\alpha^\frR\ ,\ \Ind(a^\frL)(\mathscr{L})=\alpha^\frL}} \cal{U}^{(\mttA,\mttx)}_{a^\frR,a^\frL}
	\end{split}
	\]
	and similarly for the $(\mttB,\mtty)$ case. In words, these measurements are the following: First,  measure according to $\cal{U}^{(\mttA,\mttx)}$ and retrieve $a^\frR,a^\frL$. Then,  induce both function to get $\Ind(a^\frR),\Ind(a^\frL)$. Finally, evaluate these functions on the respective point $p$ or line $\mathscr{L}$ to get a value in $\FF_q$ or the coefficients of a low-degree polynomial. As both $\Ind$ and $\eval_{\rho}$ are linear maps, by Corollary \ref{cor:linear_data_processed_PVM_is_ZPC_and_left_multiplication} the resulting PVMs are readably $Z$-aligned and consist of signed permutations.

	We now aim to define the PVMs of the perfect $\ZPC$-strategy for $\game$ at the oracle player vertices. The process is similar to the isolated player vertices, but we need to retrieve more polynomials. Recall the mapping $\mathsf{PCP}_{\mttx\mtty}$ from Corollary \ref{cor:functional_viewpoint_final} that takes as input a quadruple $f_1^\frR,f_1^\frL,f_2^\frR,f_2^\frL\colon \FF_2^{\Lambda(n)}\to \FF_2$ (which  is  the format of measurement outcomes of the PVM $\cal{U}^{(\oracle,z)}$), and outputs a degree $9$ PCP $\Pi$ over $\FF_q$ with parameters $(\Lambda(n),\diamondsuit(n),M(n),s(n),\circuit)$ (Definition \ref{defn:PCP_of_V_n_satisfied}), such that: \begin{enumerate}
		\item  The values of readable part $\Pi^\frR$, i.e., the polynomials $g_0,\beta_{0,\sX},g^\frR_i,\alpha^\frR_\sX,\beta^\frR_\sX$, depends only on $f^\frR_1$ and $f^\frR_2$, and the rest of the polynomials in $\Pi$   depend in addition on the values of $f^\frL_1$ and $f^\frL_2$, but  in an affine manner; namely, the values of the additional polynomials are affine combinations of the values of $f^\frL_1$ and $f^\frL_2$, with the coefficients depending only on $f^\frR_1$ and $f^\frR_2$.
		\item  In addition,  if $f_1^\frR,f_1^\frL,f_2^\frR,f_2^\frL\colon \FF_2^{\Lambda(n)}\to \FF_2$ are accepted in $\verifier_n$ given  $\mttx=\frS^\mttA(z),\mtty=\frS^\mttB(z)$ were asked, then the PCP $\Pi$ passes \eqref{eq:PCP_condition_1},  \eqref{eq:PCP_condition_2}, \eqref{eq:Tseitin_satisfiable_induced} and \eqref{eq:induced_system_is_satisfied} for every $p\in \FF_q^S$.
	\end{enumerate}
	For ease of notation, given a seed $z$, instead of denoting $\mathsf{PCP}_{\frS^\mttA(z)\frS^\mttB(z)}$  we will use $\mathsf{PCP}_z$ to mean the same function. Then, recall the notion of evaluating a PCP $\Pi$ at a point (Definition \ref{defn:PCP_of_V_n_satisfied}), and denote this function by $\eval_p(\Pi)=\Pi(p)$. Similarly, one can evaluate a PCP $\Pi$ at a line $\mathscr{L}$, by providing the coefficient representation of the restriction of each polynomial in $\Pi$ to this line. We denote this function by $\eval_{\mathscr{L}}(\Pi)=\Pi(\mathscr{L})$. So, let 
	\begin{equation}\label{eq:def_W_oracle}
	\cal{W}^{(\oracle,z),\mttSpace^{\rho}}=\cal{U}^{(\oracle,z)}_{[\eval_{\rho}\circ \mathsf{PCP}_z(\cdot)]}\ ;
	\end{equation}
	namely,
	\[
	\begin{split}
	\forall \alpha\in \FF_q^{\heartsuit(n)}\ \colon \ \ \cal{W}^{(\oracle,z),\Point^{p}}_{\alpha}&=\sum_{\substack{f^\frR_1,f^\frL_1,f_2^\frR,f_2^\frL\colon \FF_2^{\Lambda(n)}\to \FF_2\\ \mathsf{PCP}_z(f^\frR_1,f^\frL_1,f_2^\frR,f_2^\frL)(p)=\alpha}} \cal{U}^{(\oracle,z)}_{f^\frR_1,f^\frL_1,f_2^\frR,f_2^\frL}\\
	\forall \alpha\in (\FF_q^{10})^{\heartsuit(n)}\ \colon \ \ \cal{W}^{(\oracle,z),\ALine^{\mathscr{L}}}_{\alpha}&=\sum_{\substack{f^\frR_1,f^\frL_1,f_2^\frR,f_2^\frL\colon \FF_2^{\Lambda(n)}\to \FF_2\\ \mathsf{PCP}_z(f^\frR_1,f^\frL_1,f_2^\frR,f_2^\frL)(\mathscr{L})=\alpha}} \cal{U}^{(\oracle,z)}_{f^\frR_1,f^\frL_1,f_2^\frR,f_2^\frL}\\
	\forall \alpha\in (\FF_q^{9m+1})^{\heartsuit(n)}\ \colon \ \ \cal{W}^{(\oracle,z),\DLine^{\mathscr{L}}}_{\alpha}&=\sum_{\substack{f^\frR_1,f^\frL_1,f_2^\frR,f_2^\frL\colon \FF_2^{\Lambda(n)}\to \FF_2\\ \mathsf{PCP}_z(f^\frR_1,f^\frL_1,f_2^\frR,f_2^\frL)(\mathscr{L})=\alpha}} \cal{U}^{(\oracle,z)}_{f^\frR_1,f^\frL_1,f_2^\frR,f_2^\frL}\ .
	\end{split}
	\]
	Though the notation is confusing, we can describe these measurements simply with words: First, measure according to $\cal{U}^{(\oracle,z)}$ to retrieve a quadruple $f^\frR_1,f^\frL_1,f_2^\frR,f_2^\frL\colon \FF_2^{\Lambda(n)}\to \FF_2$. Then, use $\mathsf{PCP}_z$ with input  $f^\frR_1,f^\frL_1,f_2^\frR,f_2^\frL$ to get a degree $9$ PCP $\Pi$. Finally, evaluate the resulting PCP $\Pi$ at $\rho$ to get $\heartsuit(n)$ many values --- in $\FF_q$ when $\rho$ is a point, in $\FF_q^{10}$ when $\rho$ is an axis parallel line, and in $\FF_q^{9m+1}$ when $\rho$ is a diagonal line. 
	
	Let us convince ourselves that the resulting strategy $\cal{W}$ is a perfect $\ZPC$ strategy for $\game$. As $\eval_\rho$ and $\Ind$ are linear functions, and $\mathsf{PCP}_z$ is $\FF_2$-affine for fixed readable values, by Corollary \ref{cor:encodings}, data processing along $\eval_\rho\circ \Ind$ and along $\eval_\rho\circ \mathsf{PCP}_z$ preserves $Z$-alignment and being a signed permutation PVM.  Since $\cal{U}^{(\mathtt{Player},\mttw)}$ are $Z$-aligned permutation PVMs, the same is true for $\cal{W}^{(\mathtt{Player},\mttw),\mttSpace^\rho}$. Also, data processing (Definition \ref{defn:Data_proccessed_PVM}) does not change the commuting along edges condition; namely, if $\cal{P},\cal{Q}$ are commmuting PVMs, then $\cal{P}_{[f(\cdot)=\cdot]}$ commutes with $\cal{Q}_{[g(\cdot)=\cdot]}$. Hence, as $\cal{U}$ is commuting along edges, $\cal{W}$ is commuting along edges as well. All in all, $\cal{W}$ is a $\ZPC$ strategy.
	Let us now elaborate on why $\cal{W}$ is perfect for $\game$:
	\begin{enumerate}
		\item As the isolated player  measurements are, by definition, the evaluation at lines and points of $\Ind(f)$ for measured functions of the form $f\colon\FF_2^m\to \FF_2$, they pass the low degree test $\frak{LowDeg}(d,q,m,2)$ perfectly for every $d\geq 1$ (and in particular for $9$). Similarly, the oracle player measurements are the evaluation at lines and points of $\mathsf{PCP}_z(\cdot,\cdot,\cdot,\cdot)$, which are of individual degree at most $9$, and will thus pass $\frak{LowDeg}(9,q,m,\heartsuit(n))$. This proves that $\cal{W}$ passes check \labelcref{clause1:low_degree_check_in_ans_red} perfectly (the {\color{RubineRed} pink} and {\color{JungleGreen} green} edges from Figure \ref{fig:ans_red_type_graph}).
		\item As $\cal{U}$ was perfect for $\frak{Oracle}(\frak{DoubleCover}(\verifier_n))$, given $z$ for which $\frS^\mttA(z)=\mttx$ and $\frS^\mttB(z)=\mtty$, the isolated player measurement $\cal{U}^{(\mttA,\mttx)}$ (respectively $\cal{U}^{(\mttB,\mtty)}$) is perfectly consistent with the first two outputs of the oracle player measurement $\cal{U}^{(\oracle,z)}$ (respectively the last two outputs of $\cal{U}^{(\oracle,z)}$); namely 
		\[
		\begin{split}
		\forall a^\frR,a^\frL\colon \FF_2^{\Lambda(n)}\to \FF_2\  \colon \ \ \cal{U}^{(\mttA,\mttx)}_{a^\frR,a^\frL}&=\sum_{b^\frR,b^\frL\colon \FF_2^{\Lambda(n)}\to \FF_2} \cal{U}^{(\oracle,z)}_{a^\frR,a^\frL,b^\frR,b^\frL}\ ,\\
		\forall b^\frR,b^\frL\colon \FF_2^{\Lambda(n)}\to \FF_2\ \colon \ \ \cal{U}^{(\mttB,\mtty)}_{b^\frR,b^\frL}&=\sum_{a^\frR,a^\frL\colon \FF_2^{\Lambda(n)}\to \FF_2} \cal{U}^{(\oracle,z)}_{a^\frR,a^\frL,b^\frR,b^\frL}\ .
		\end{split}
		\]
		Following the definition of $\mathsf{PCP}_z(a^\frR,a^\frL,b^\frR,b^\frL)=\Pi$ (Corollary \ref{cor:functional_viewpoint_final}), the polynomial $g^\kappa_1$ in $\Pi$ is $\Ind(a^\kappa)$ and $g^\kappa_2$ is $\Ind(b^\kappa)$, and thus evaluating these polynomials at a point or a line always agree. Recalling the definition of  $\cal{W}$ from  \eqref{eq:def_W_isolate} and \eqref{eq:def_W_oracle}, we deduce that $\cal{W}$ passes check \labelcref{clause2:consistency_checks_in_ans_red} perfectly (the {\color{orange} orange} edges from Figure \ref{fig:ans_red_type_graph}).
		\item Recalling the way we redefined $\Ind(\cdot)$ in this completeness proof, $\Ind(a^\kappa_i)\colon \FF_q^S\to \FF_q$ is  indifferent to every $\sX$ from $S$ which is not in $S^\kappa_1$ (and similarly $\Ind(b^\kappa_i)$ is indifferent to every $\sX$ not in $S^\kappa_2$). Also, by the definition of  a PCP $\Pi$ (Definition \ref{defn:PCP_of_V_n_satisfied}), the various polynomials are  indifferent to the variables not from their respective block. Hence, the $\cal{W}$ strategy passes check \labelcref{clause3:Indifference_checks_in_ans_red} perfectly (the {\color{violet} purple} vertices from Figure \ref{fig:ans_red_type_graph}).
		\item Now, as $\cal{U}$ was perfect for $\frak{Oracle}(\frak{DoubleCover}(\verifier_n))$, the measurement outcome of $\cal{U}^{(\oracle,z)}$ always passes $\verifier_n$ given $\frS^\mttA(z)\frS^\mttB(z)=\mttx\mtty$ were asked. Hence, by the completeness clause of Corollary \ref{cor:functional_viewpoint_final},  the measurement outcome of $\cal{U}^{(\oracle,z)}_{[\mathsf{PCP}_z(\cdot,\cdot,\cdot,\cdot)]}$ is a PCP $\Pi$ such that for every point $p\in \FF_q^S$, its evaluation at a point $\Pi(p)$  always passes \eqref{eq:PCP_check_1_ansred}, \eqref{eq:PCP_check_2_ansred}, \eqref{eq:PCP_check_3_ansred} and \eqref{eq:PCP_check_4_ansred}. Therefore, $\cal{W}$ passes check \labelcref{clause4:PCP_checks_in_ans_red} perfectly (the  {\color{blue} blue}  vertex from Figure \ref{fig:ans_red_type_graph}).
	\end{enumerate}

	\textbf{Soundness and entanglement lower bound}:
	
	The idea is similar to soundness of the question reduction transformation. Namely, we perturb the original almost perfect strategy bit by bit, so that in each step it passes more of the checks of $\game$ perfectly, while worsening the probability of passing the remaining checks. This is done in the following order: First, perturbing the strategy so as to  pass \cref{clause1:low_degree_check_in_ans_red} perfectly, which guarantees that there is a global PVM at any vertex of $\frak{Oracle}(\frak{DoubleCover}(\verifier_n))$ such that the strategy is close to restrictions of it to the appropriate lines and points. Then, using the Schwartz--Zippel Lemma~\ref{lem:Schwartz-Zippel}, these PVMs can be further purturbed to pass \cref{clause3:Indifference_checks_in_ans_red} perfectly. Then using the consistency checks of \cref{clause2:consistency_checks_in_ans_red} and the PCP check of \cref{clause4:PCP_checks_in_ans_red}, we deduce that the restrictions of the polynomials measured by the PVMs at isolated player vertices are almost perfect as a strategy for $\frak{DoubleCover}(\verifier_n)$.
	\\
	
	\emph{Preliminary analysis}:
	
	Recall that the vertices of $\game$ are a direct product of the vertices of $\frak{Oracle}(\frak{DoubleCover}(\verifier_n))$  and the vertices of $\frak{LowDegree}(9,t,m,\cdot)$, namely  a vertex in $\game$ is of the form 
	\[
	(\mathtt{Player},\mttw),\mttSpace^{\rho}\ ,
	\]
	where $(\mathtt{Player},\mttw)\in \{\mttA,\mttB,\oracle\}\times\FF_2^r$, $\mttSpace\in \{\Point,\ALine,\DLine\}$ and $\rho$ is either a point or a line in $\FF_q^m$. So, a sampled edge in $\game$ is of the form 
	\begin{equation}\label{eq:sampled_edge_in_ansred_game}
	e\ :=\ \ (\mathtt{Player},\mttw),\mttSpace_1^{\rho_1}-(\mathtt{Player}',\mttw'),\mttSpace_2^{\rho_2}\ . 
	\end{equation}
	For a  vertex $\mttv$ of $\frak{Oracle}(\frak{DoubleCover}(\verifier_n))$, we say that an edge $e$ of $\game$ as in \eqref{eq:sampled_edge_in_ansred_game} is first-$\mttv$-incident if  $(\mathtt{Player},\mttw)=\mttv$, second-$\mttv$-incident if $(\mathtt{Player}',\mttw')=\mttv$, and just $\mttv$-incident if it is either first-$\mttv$-incident or second-$\mttv$-incident (or both). Similarly, such an edge is said to be 
	$\mttv$-internal if it is both  first-$\mttv$-incident and second-$\mttv$-incident; namely, $e$ is an edge of the form 
	\begin{equation}\label{eq:edge_sampled_with_constant_left_coordinate}
	e\ :=\ \ \mttv,\mttSpace_1^{\rho_1}-\mttv,\mttSpace_2^{\rho_2}\ .
	\end{equation}
	Let $\mu(e)$ be the probability the edge $e$ is sampled in $\game$, and note that for each edge $e$ there is a unique $\mttv$ for which $e$ is first-$\mttv$-incident and a unique $\mttv'$ such that $e$ is second-$\mttv'$-incident. Now, $\mu$ induces a marginal distribution on the vertices of $\frak{Oracle}(\frak{DoubleCover}(\verifier_n))$ by the following sampling mechanism: sample $e\sim \mu$ as in \eqref{eq:sampled_edge_in_ansred_game}; output $\mttv=(\mathtt{Player},\mttw)$ or $\mttv=(\mathtt{Player}',\mttw')$ uniformly at random\footnote{Note that this agrees the marginal distribution from edges to vertices in the game $\frak{Oracle}(\frak{DoubleCover}(\verifier_n))$.} --- we denote the probability $\mttv$ is the output by $\mu(\mttv)$.
	Furthermore, let $\mu^\mttv$ be the probability $e$ was the sampled edge in the aforementioned sampling mechanism given $\mttv$ was the output, namely: if $e$ is not $\mttv$-incident, then $\mu^\mttv(e)=0$; if $e$ is $\mttv$-incident but not $\mttv$-internal, then $\mu^\mttv(e)=\frac{\mu(e)}{2\mu(\mttv)}$; if $e$ is $\mttv$-internal, then $\mu^\mttv(e)=\frac{\mu(e)}{\mu(\mttv)}$.

	Assume $\strategy=\{\cal{P}\}$ is a quantum strategy for the answer reduced game $\game$ with value $1-\eps$.  For an edge $e$ in $\game$, let $\eps^e$ be the probability $\cal{P}$ loses the round of the game given $e$ was sampled. Then,  
	\[
	\eps=\Es{e\sim \mu}[\eps^e]\ .
	\]
	For a vertex $\mttv$ of $\frak{Oracle}(\frak{DoubleCover}(\verifier_n))$,   
	let $\eps^\mttv$ be the probability that $\cal{P}$ loses the round of $\game$, given an edge was  sampled according to  $\mu^\mttv$; namely
	\[
	\eps^\mttv=\Es{e\sim \mu^\mttv}[\eps^e]\ .
	\]
	By our choices, sampling and edge $e$ by first sampling $\mttv\sim \mu$ and then $e\sim \mu^\mttv$ is the same as just sampling $e\sim \mu$, and thus $\eps=\Es{\mttv\sim \mu}[\eps^\mttv]$.
	\\
	
	\emph{First perturbation}:
	
	Recall that $\strategy=\{\cal{P}\}$ is a strategy for $\game$ with value $1-\eps$.  For the next discussion, let us fix a $\mttv=(\mathtt{Player},\mttw)$. By inspecting the type graph of $\game$ (Figure \ref{fig:ans_red_type_graph}), and recalling that the type graph contains all self loops, one can deduce that the probability $e\sim \mu^\mttv$ is $\mttv$-internal is at least $\nicefrac{1}{2}$. 
	Hence, the probability $\cal{P}$ loses versus $e$, given that the sampled $e$ is $\mttv$-internal \eqref{eq:edge_sampled_with_constant_left_coordinate}, 
	is at most $2\eps^\mttv$. 
	In case a $\mttv$-internal edge was sampled (which corresponds to a {\color{RubineRed} pink} or {\color{JungleGreen} green} edge in Figure~\ref{fig:ans_red_type_graph}), 
	\cref{clause1:low_degree_check_in_ans_red} in $\game$ is checked. 
	Namely, $\game$ runs  $\frak{LowDegree}(9,q,m,k)$, where $k=2$ in case $\mathtt{Player}=\mttA$ or $\mttB$ and  $k=\heartsuit(n)$ in case $\mathtt{Player}=\oracle$. Hence, by Theorem \ref{thm:ldc-soundness}, there is a PVM $\cal{G}^{\mttv}$ (of the same dimension as $\cal{P}$), with outcomes in $k$-tuples of individual degree at most $9$ polynomials from $\FF_q^m$ to $\FF_q$, that satisfies 
	\begin{align}
	\cal{P}^{\mttv,\mttSpace^\rho}\approx_{\delta^\mttv} \cal{G}^\mttv_{[\eval_\rho(\cdot)]}\ ,\label{eq:P_is_consistent_with_eval_of_global_polynomials_ansred}
	\end{align}
	where $\delta^\mttv=\delta_\ld(m,9,k,2\eps^\mttv,q^{-1})$ and $\delta_\ld$ is the function from \eqref{eq:defn_delta_ld}. Let $\strategy'=\{\cal{Q}\}$ be the strategy for $\game$ that satisfies $\cal{Q}^{\mttv,\mttSpace^\rho}=\cal{G}^\mttv_{[\eval_\rho(\cdot)]}$; namely, the way it answers the question $\mttv,\mttSpace^\rho$ is by first measuring a $k$-tuple $f_1,...,f_k$ according to $\cal{G}^\mttv$, and then evaluating it at the space $\rho$. By construction, $\cal{Q}$ always passes check \labelcref{clause1:low_degree_check_in_ans_red} of $\game$, and from \eqref{eq:P_is_consistent_with_eval_of_global_polynomials_ansred} (see Remark \ref{rem:interpretation_of_the_distance_upper_bound}), and the notion of distance between strategies (Definition \ref{defn:strict_distance_strategies}), we can deduce that $\strategy$ is $\Es{\mttv\sim \mu}[\delta^\mttv]$-close to $\strategy'$. Now, as $\delta_{\ld}$ from \eqref{eq:defn_delta_ld} is monotonously increasing in all its  inputs (in particular, in its third input), and is concave with respect to its fourth input, we can deduce that 
	\[
	\Es{\mttv\sim \mu}[\delta^\mttv]\leq \delta_\ld(m,9,\heartsuit(n),2\cdot\underbrace{\Es{\mttv\sim \mu}[\eps^\mttv]}_{=\eps},q^{-1})=:\delta\ .
	\]
	Therefore, by Claim \ref{claim:close_strat_implies_close_correlations}, the value of $\strategy'$ is at least 
	\[
	1-\eps-10\sqrt\delta\geq 1-11\sqrt{\delta}\ .
	\]
	\\
	
	\emph{Second perturbation}:
	
	For every vertex $\mttv=(\mathtt{Player},\mttw)$ of  $\frak{Oracle}(\frak{DoubleCover}(\verifier_n))$, let $\cal{G}^\mttv$  be a PVM of dimension $N$ with outcomes in $k$-tuples of degree at most $9$ polynomials, where $k=2$ in case $\mathtt{Player}=\mttA$ or $\mttB$, and $k=\heartsuit(n)$ in case $\mathtt{Player}=\oracle$. Assume the $N$-dimensional quantum strategy $\strategy=\{\cal{P}\}$ for $\game$ defined by $\cal{P}^{\mttv,\mttSpace^\rho}=\cal{G}^\mttv_{[\eval_\rho(\cdot)]}$ has value $1-\eps$.
	
	An edge $e$ that is sampled according to $\mu^\mttv$ has a probability of at least $\frac{3}{10}$ to have the $\mttSpace$ associated with $\mttv$ be $\ALine$, namely for one of its endpoints to be $\mttv,\ALine^{\mathscr{L}}$. Hence, the strategy $\cal{P}$ loses against an edge with endpoint $\mttv,\ALine^{\mathscr{L}}$ with  probability of at most $\frac{10}{3}\eps^\mttv\leq  4\eps^\mttv$, given that such an edge was sampled. In such a case, $\game$ checks whether  \cref{clause3:Indifference_checks_in_ans_red} is satisfied.
	
	Let $f$ be an individual degree at most $9$ polynomial which is not indifferent to  the $i^{\rm th}$ variable of $S$. Then, $f=\sum_{j=0}^9 \alpha_j \sX_i^j$, where $\sX_i$ is the $i^{\rm th}$ variable of $S$, each $\alpha_j$ is a polynomial of individual degree at most $9$ in the other $m-1$ variables from $S$, and at least one of the $\alpha_j$'s for $j\geq 1$ is not identically zero --- denote by $1\leq l\leq 9$ the index such that $\alpha_l$ is not the zero polynomial. 
	For $u\in \FF_q^m$, recall from Theorem \ref{thm:classical_soundness_individual_low_degree}  the notation $\hat u^i$ for the vector in $\FF_q^{m-1}$ recovered by removing from $u$ its $i^{\rm th}$ coefficient.  
	As $\alpha_l$ is an individual degree at most $9$ polynomial, it has total degree at most $9(m-1)$. By applying the Schwartz-Zippel Lemma~\ref{lem:Schwartz-Zippel}, we deduce that for at least $1-\frac{9(m-1)}{q}$ of the points $u\in \FF_q^m$, we have $\alpha_l(\hat u^i)\neq 0$, and thus the restriction of $f$ to the axis parallel line $\mathscr{L}(u,e_i)$ is {not} constant. 
	
	Recall that the probability the measurement outcome of $\cal{G}^\mttv$ is a specific tuple $f_1,...,f_k$ is $\tau(\cal{G}^\mttv_{f_1,...,f_k})$. A tuple $f_1,...,f_k$  is said to be \emph{bad}  if there is an index $s\in [k]$ and a direction $i\in [m]$ such that $f_s$ is not indifferent to the $i^{\rm th}$ variable of $S$, although it should according to the game checks. By the analysis from the previous paragraph, given a bad tuple $f_1,...,f_k$, and sampling an axis parallel line $\mathscr{L}$ uniformly at random, there is a probability of at least $\nicefrac{1}{m}$ that $\mathscr{L}$ is in direction $i$, and a probability of at least $1-\frac{9(m-1)}{q}$ that $f_s(\mathscr{L})$ is not constant, which makes $\game$ reject $\eval_{\mathscr{L}}(f_1,...,f_k)$ due to \cref{clause3:Indifference_checks_in_ans_red}. 
	Therefore, 
	\begin{equation}\label{eq:lower_bound_theta^v}
	\frac{1}{m}\cdot\left(1-\frac{9(m-1)}{q}\right)\cdot \tau\left(\sum_{ f_1,...,f_k\ {\rm is\ bad}}\cal{G}^\mttv_{f_1,...,f_k}\right)\leq 4\eps^\mttv\ .
	\end{equation}
	By choosing $\gamma^\mttv=\frac{4m}{1-\frac{9(m-1)}{q}}\cdot \eps^\mttv$, one deduces that $\tau\left(\sum_{ f_1,...,f_k\ {\rm is\ bad}}\cal{G}^{\mttv}_{f_1,...,f_k}\right)\leq \gamma^\mttv$.
	We use the PVM $\cal{G}^\mttv$ to define a new PVM  $\cal{H}^\mttv$ with outcomes in $k$-tuples of individual degree at most $9$ polynomials (as is $\cal{G}^\mttv$): for a bad tuple $f_1,...,f_k$,  we let $\cal{H}^\mttv_{f_1,...,f_k}=0$; for a good tuple  $f_1,...,f_k$ such that not all the polynomials are $0$, we let $\cal{H}^\mttv_{f_1,...,f_k}=\cal{G}^\mttv_{f_1,...,f_k}$; for the tuple of only zero polynomials, we let $\cal{H}^\mttv_{0,...,0}=\cal{G}^\mttv_{0,...,0}+\sum_{f_1,...,f_k\ {\rm is\ bad}}\cal{G}^\mttv_{f_1,...,f_k}$. In other words, if we define the map $\Xi$ from $k$-tuples of individual degree at most $9$ polynomials to itself that is the identity on good tuples, but sends all bad tuples to the all zero tuple, then $\cal{H}^\mttv$ is the $\Xi$-evaluated PVM (Defintion \ref{defn:Data_proccessed_PVM}), namely $\cal{H}^\mttv=\cal{G}^\mttv_{[\Xi(\cdot)]}$. Now, 
	\[
	\begin{split}
	\sum_{f_1,...,f_k}\|\cal{H}^\mttv_{f_1,...,f_k}-\cal{G}^\mttv_{f_1,...,f_k}\|_{hs}^2&=\sum_{f_1,...,f_k\ {\rm is\ bad}}\|\cal{G}^\mttv_{f_1,...,f_k}\|_{hs}^2+    \left\|\sum_{f_1,...,f_k\ {\rm is\ bad}}\cal{G}^\mttv_{f_1,...,f_k}\right\|_{hs}^2\\
	&=_{\cal{G}\ {\rm is\ a \ PVM}}2\cdot\sum_{f_1,...,f_k\ {\rm is\ bad}}\tau(\cal{G}^\mttv_{f_1,...,f_k})\\
	&\leq 2\gamma^\mttv\ ,
	\end{split}
	\]
	which means 
	\begin{equation}\label{eq:H_and_G_are_close}
	\cal{H}^\mttv\approx_{2\gamma^\mttv}\cal{G}^\mttv\ .
	\end{equation}
	As $\cal{G}^\mttv$ and $\cal{H}^\mttv$ are PVMs, by Claim \ref{claim:data_processing3}, \eqref{eq:H_and_G_are_close} implies $\cal{H}^\mttv_{[\eval_\rho(\cdot)]}\approx_{2\gamma^\mttv}\cal{G}_{[\eval_\rho(\cdot)]}^\mttv$. Hence, if we define a strategy $\strategy'=\{\cal{Q}\}$ for $\game$ by letting $\cal{Q}^{\mttv,\mttSpace^\rho}=\cal{H}^\mttv_{[\eval_\rho(\cdot)]}$, then $\strategy'$ passes checks \labelcref{clause1:low_degree_check_in_ans_red} and \labelcref{clause3:Indifference_checks_in_ans_red} perfectly,  
	and by \eqref{eq:H_and_G_are_close} $\strategy$ and $\strategy'$ are $2\Es{\mttv\sim \mu}[\gamma^\mttv]$-close. Letting $\gamma=2\Es{\mttv\sim \mu}[\gamma^\mttv]=\frac{8m}{1-\frac{9(m-1)}{q}}\cdot \eps$, and using Claim \ref{claim:close_strat_implies_close_correlations}, the value of $\strategy'$ is at least 
	\[
	1-\eps-10\sqrt\gamma\geq 1 -\frac{81m}{1-\frac{9(m-1)}{q}}\cdot \sqrt \eps\ .
	\]
	\\
	
	\emph{Translating the resulting strategy into a high value strategy for $\frak{DoubleCover}(\game)$}:
	
	For every vertex $\mttv=(\mathtt{Player},\mttw)$ of  $\frak{Oracle}(\frak{DoubleCover}(\verifier_n))$, let $\cal{H}^\mttv$  be a PVM of dimension $N$ with outcomes in good $k$-tuples of degree at most $9$ polynomials (on the variable set $S$), where $k=2$ in case $\mathtt{Player}=\mttA$ or $\mttB$, and $k=\heartsuit(n)$ in case $\mathtt{Player}=\oracle$. Assume the $N$-dimensional quantum strategy $\strategy=\{\cal{P}\}$ for $\game$ defined by $\cal{P}^{\mttv,\mttSpace^\rho}=\cal{H}^\mttv_{[\eval_\rho(\cdot)]}$ has value $1-\eps$. Note that $\strategy$ passes checks \labelcref{clause1:low_degree_check_in_ans_red} and \labelcref{clause3:Indifference_checks_in_ans_red} of $\game$ perfectly. In particular, at oracle player vertices, the $\heartsuit(n)$-tuple of polynomials measured by $\cal{H}^{(\oracle,z)}$ is a degree $9$ PCP over $\FF_q$ in the sense of Definition \ref{defn:PCP_of_V_n_satisfied} --- not only these are polynomials from $\FF_q^S$ to $\FF_q$, the consisting polynomials are indeed indifferent to the relevant directions, as check \labelcref{clause3:Indifference_checks_in_ans_red}  is always passed.
	
	For a fixed $z\in \FF_2^r$, an edge $e$ sampled according to  $\mu^{(\oracle,z)}$ has a probability of at least $\frac{2}{5}$ to have the $\mttSpace$ associated with $(\oracle,z)$ be $\Point$, namely for one of its endpoints to be $(\oracle,z),\Point^p$. Hence, the probability $\cal{P}$ loses against an edge with endpoint $(\oracle,z),\Point^p$ is at most $\frac{5}{2}\eps^{(\oracle,z)}\leq 3\eps^{(\oracle,z)}$ (given such an edge was sampled). In this case, $\game$ checks \cref{clause4:PCP_checks_in_ans_red}. 
	
	A PCP $\Pi$ is sad to be \emph{bad} (with respect to $z$) if it does not pass the checks \eqref{eq:PCP_condition_1}, \eqref{eq:PCP_condition_2}, 
	\eqref{eq:Tseitin_satisfiable_induced} and \eqref{eq:induced_system_is_satisfied} perfectly for all $p\in \FF_q^m$, which is the same as failing with positive probability check \labelcref{clause4:PCP_checks_in_ans_red} given $(\oracle,z),\Point^p$ is an endpoint of the sampled edge and $\Pi$ was the outcome of measuring ${\cal{H}}^{(\oracle,z)}$. 
	By  the soundness clause of Corollary \ref{cor:functional_viewpoint_final}, given that $\Pi$ is a bad PCP, it will fail versus at least $1-\frac{63m}{q}$ of the points of $\FF_q^m$. Combined with the observation from the previous  paragraph, we can deduce that 
	\[
	\left(1-\frac{63m}{q}\right)\cdot \tau\left(\sum_{\Pi\ {\rm is\ bad}}\cal{H}^{(\oracle,z)}_\Pi\right)\leq 3\eps^{(\oracle,z)}\ ;
	\]
	namely, when sampling $\Pi$ according to $\cal{H}^{(\oracle,z)}$, there is a probability of at most $\frac{3}{1-\frac{63m}{1}}\cdot \eps^{(\oracle,z)}$ that $\Pi$ is bad. Therefore, the soundness clause of Corollary \ref{cor:functional_viewpoint_final} again, when sampling $\Pi$ according to $\cal{H}^{(\oracle,z)}$ the restriction of the polynomials $g^\frR_1,g^\frL_1,g^\frR_2,g^\frL_2$ in $\Pi$ to the subcube pass the game $\verifier_n$ given $\frS^\mttA(z)\frS^\mttB(z)$ were asked. 
	
	Let ${\rm Restrict}_1$ (respectively  ${\rm Restrict}_2$ and ${\rm Restrict}_{12}$) be the function that take a PCP $\Pi$ as input, and outoputs only the polynomilals $g^\frR_1,g^\frL_1$ (respectively $g^\frR_2,g^\frL_2$ and $g^\frR_1,g^\frL_1,g^\frR_2,g^\frL_2$) from it.
	For a fixed $z\in \FF_2^r$, the probability $e\sim \mu^{(\oracle,z)}$ is of the form 
	\begin{equation}\label{eq:edge_same_space_ans_red}
	(\mttA,\mttx),\Point^p-(\oracle,z),\Point^p\ ,
	\end{equation}
	where $\mttx=\frS^\mttA(z)$, is at least $\frac{1}{25}$ (and the same is true by replacing  for $(\mttA,\mttx)$ by $(\mttB,\mtty)$ for $\mtty=\frS^\mttB(z)$). In this case, check \labelcref{clause2:consistency_checks_in_ans_red} of $\game$ is checked. Therefore, the probability $\cal{P}$ loses against check \labelcref{clause2:consistency_checks_in_ans_red} given an edge of the form \eqref{eq:edge_same_space_ans_red} was sampled is at most $25\eps^{(\oracle,z)}$. Assume we mutually measured $(g^\frR_{\mttA,\mttx},g^\frL_{\mttA,\mttx}),(g^\frR_1,g^\frL_1)\sim (\cal{H}^{(\mttA,\mttx)},\cal{H}^{(\oracle,z)}_{[{\rm Restrict}_1(\cdot)]})$. Then, as all of these polynomials are of individual degree at most $9$, if $g^\frR_{\mttA,\mttx}\neq g^\frR_1$ or $g^\frL_{\mttA,\mttx}\neq g^\frL_1$, then by the Schwartz--Zippel  Lemma~\ref{lem:Schwartz-Zippel}, for at least $1-\frac{9m}{q}$ of the points in $\FF_q^m$ we have $g^\frR_{\mttA,\mttx}(p)\neq g^\frR_1(p)$ or $g^\frL_{\mttA,\mttx}(p)\neq g^\frL_1(p)$. Hence, by letting $\zeta^{(\oracle,z)}=\frac{25}{1-\frac{9m}{q}}\cdot \eps^{(\oracle,z)}$, and recalling the relation between distance and inconsistency of PVMs,  we deduce that 
	\[
	\cal{H}^{(\mttA,\mttx)}\approx_{2\zeta^{(\oracle,z)}} \cal{H}^{(\oracle,z)}_{[{\rm Restrict}_1(\cdot)]}\ .
	\]
	Similarly, we can deduce that
	\[
	\cal{H}^{(\mttB,\mtty)}\approx_{2\zeta^{(\oracle,z)}} \cal{H}^{(\oracle,z)}_{[{\rm Restrict}_2(\cdot)]}\ .
	\]
	By applying Claim \ref{claim:close_almost_prjective_POVMs_similar_joint} twice, we can deduce that the distribution on quadruples of degree at most $9$ polynomials induced by mutually measuring according to $(\cal{H}^{(\mttA,\mttx)},\cal{H}^{(\mttB,\mtty)})$ is at most $4\sqrt{2\zeta^{(\oracle,z)}}$ away in $L^1$-distance from the distribution induced by mutually measuring according to $(\cal{H}^{(\oracle,z)}_{[{\rm Restrict}_1(\cdot)]},\cal{H}^{(\oracle,z)}_{[{\rm Restrict}_2(\cdot)]})$, which in turn is equal to the distribution induced by measuring according to $\cal{H}^{(\oracle,z)}_{[{\rm Restrict}_{12}(\cdot)]}$.  As measuring according to $\cal{H}^{(\oracle,z)}$ produces a good PCP with probability of at least $1-\frac{3}{1-\frac{63m}{1}}\cdot \eps^{(\oracle,z)}$, by the soundness clause of Corollary \ref{cor:functional_viewpoint_final}, the restriction to the subcuce of the measurement outcome of $\cal{H}^{(\oracle,z)}_{[{\rm Restrict}_{12}(\cdot)]}$ passes $\verifier_n$ given $\frS^\mttA(z)\frS^\mttB(z)$ was asked with at least the same probability. Hence, the restriction to the subcuce of the mutual measurement according to $(\cal{H}^{(\mttA,\mttx)},\cal{H}^{(\mttB,\mtty)})$ passes $\verifier_n$ given $\frS^\mttA(z)\frS^\mttB(z)$ was asked with probability of at least $1-\frac{3}{1-\frac{63m}{1}}\cdot \eps^{(\oracle,z)}-4\sqrt{2\zeta^{(\oracle,z)}}$.
	
	Let $\strategy'=\{\cal{Q}\}$ be the strategy for $\frak{DoubleCover}(\verifier_n)$ defined by $\cal{Q}^{(\mathtt{Player},\mttw)}=\cal{H}^{(\mathtt{Player},\mttw)}_{[\Res(\cdot)]}$, where $\Res(\cdot)$ is the restriction to the subcube function from Definition \ref{defn:restriction_and_induction_polynomials}, not to be confused with ${\rm Restrict}_\circ$.
	According to the previous analysis, the value of $\strategy'$ versus  $\frak{DoubleCover}(\verifier_n)$ is at least 
	\[
	1-\Es{z\in \FF_2^r}\left[\frac{3}{1-\frac{63m}{1}}\cdot \eps^{(\oracle,z)}-4\sqrt{2\zeta^{(\oracle,z)}}\right]\ ,
	\]
	where the expectation is over a uniformly random $z\in\FF_2^r$. By the definition of $\zeta^{(\oracle,z)}$,  the concavity of $\sqrt \cdot$, and the fact $x\leq x^2$ for $x\geq 1$, we have that 
	\[
	\val(\frak{DoubleCover}(\verifier_n),\strategy')\geq 1-\frac{3}{1-\frac{63m}{q}}\Es{z\in \FF_2^r}[\eps^{(\oracle,z)}]-\frac{200}{1-\frac{9m}{q}}\cdot \sqrt{\Es{z\in \FF_2^r}[\eps^{(\oracle,z)}]}\ .
	\]
	By inspecting the type graph of $\game$, and recalling the marginal distribution $\mu$ induces on the vertices $(\mathtt{Player},\mttw)$, we have $\mu((\oracle,z))=\frac{8}{25}\cdot \frac{1}{2^r}$. Therefore, 
	\[
	\begin{split}
	\Es{z\in \FF_2^r}[\eps^{(\oracle,z)}]&=\sum_{z\in \FF_2^r}\frac{1}{2^r}\cdot\eps^{(\oracle,z)}\\
	&\leq \frac{25}{8}\sum_{z\in \FF_2^r}\mu((\oracle,z))\eps^{(\oracle,z)}\\
	&\leq 4\cdot \sum_{\mttv}\mu(\mttv)\eps^{\mttv}\\
	&=4\eps\ .
	\end{split}
	\]
	All in all, we found a strategy for $\frak{DoubleCover}(\verifier_n)$ with value of at least 
	\[
	1-\frac{12\eps}{1-\frac{63m}{q}}-\frac{400\sqrt\eps}{1-\frac{9m}{q}}
	\geq 
	1-\frac{412}{1-\frac{63m}{q}}\cdot \sqrt\eps\ .
	\]
	\\
	
	\emph{Combining all of the above to deduce soundness}:
	\begin{itemize}
		\item Starting with an $N$-dimensional value $1-\eps$ strategy for $\game$, the first perturbation allows us to find an $N$-dimensional value at least $1-\eps'$ strategy for $\game$ that always passes check \labelcref{clause1:low_degree_check_in_ans_red}, where $\eps'=11\sqrt{\delta}$ for $\delta=\delta_{\ld}(m,9,\heartsuit(n),2\eps,q^{-1})$. 
		\item
		By applying the second perturbation on the resulting strategy, we find an $N$-dimensional strategy for $\game$ with value of at least $1-\eps''$ for $\game$ that always pass both check \labelcref{clause1:low_degree_check_in_ans_red} and \labelcref{clause3:Indifference_checks_in_ans_red} perfectly, where $\eps''= \frac{81m}{1-\frac{9(m-1)}{q}}\cdot \sqrt{\eps'}$.
		\item Finally, by restricting the recovered strategy to the isolated player vertices, we showed that the resulting $N$-dimensional strategy for $\frak{DoubleCover}(\verifier_n)$ has value of at least $1-\eps'''$, where $\eps'''=\frac{412}{1-\frac{63m}{q}}\cdot \sqrt{\eps''}$. 
		\item Now,
		\[
		\eps'''\leq \frac{11\cdot 81m\cdot 412}{\left(1-\frac{63m}{q}\right)\left(1-\frac{9(m-1)}{q}\right)}\cdot\delta^{\nicefrac{1}{8}}\leq \frac{10^6\cdot m}{1-\frac{72m}{q}}\cdot\delta^{\nicefrac{1}{8}}\ ,
		\]
		which finishes the proof. 
	\end{itemize}
	
\end{proof}

Though we tried to emphasize the combinatorial aspects of the answer reduced game $\frak{AnsRed}(\verifier,\Lambda,\Delta,D,{T},Q,n,t)$, it is already quite ``normal form verify''ish in nature. The following is thus quite straightforward.
\begin{claim}[Algorithmic (partial) Answer Reduction]\label{claim:algorithmic_partial_ans_red}
	There is a polynomial time TM $\PartialAnsRed$ that takes as input a  $h$-level TNFV $\verifier=(\sampler,\length,\linproc,\decider)$, a positive integer $D$ and five $1$-input TMs $\Lambda,\Delta,{T},Q,\cal{FE}$,\footnote{The role of $\cal{FE}$ is to be the ``field exponent'', namely to  calculate the appropriate (log of the) field size $q=2^t$ for the $n^{\rm th}$ game of the normal form verifier.} and outputs a typed $\max(h,3)$-level TNFV 
	\[
	\verifier'=(\sampler',\length',\linproc',\decider)=\PartialAnsRed(\verifier,\Lambda,\Delta,D,T,Q,\cal{FE})
	\]
	with the following properties:
	\begin{itemize}
		\item \emph{Combinatorial Answer Reduction}: Given that the inputs satisfy that --- 
		\begin{itemize}[noitemsep,topsep=0pt,parsep=0pt,partopsep=0pt]
			\item[--] the input TMs always halt;
			\item[--] $\verifier$  is  purified and $2^\Lambda$-padded;
			\item[--] $\Delta(n)\geq \TIME(\linproc;n,\cdot,\cdot,\cdot,\cdot)\cdot 2^{\Lambda(n)+1}$  for every $n$;
			\item [--] 
			$ {Q}(n)\geq \TIME(\sampler;{n},\cdot,\cdot,\cdot,\cdot,\cdot)$ for every $n$;
			\item [--] $
			\forall n\in \mathbb{N}\ \colon \ \ T(n)\geq c\cdot(\TIME(\Lambda;n)^c+\TIME(\Delta;n)^c+2^{c\cdot\Lambda(n)}+\Delta(n)^c+\TIME(\linproc;n,\cdot,\cdot,\cdot,\cdot)^c)\ ,
			$
			where $c\geq 6$ is the positive integer implied by the $\poly$ notation in \eqref{eq:time_bound_L*};
			\item [--]  
			$|\verifier|,|\Lambda|,|\Delta|,|T|,|Q|\leq D$;
		\end{itemize}
		then $\verifier'_n=\frak{AnsRed}(\verifier,\Lambda,\Delta,D,{T},Q,n,2\cal{FE}(n)+1)$, where $\frak{AnsRed}$ is from Definition \ref{defn:combi_ans_red}.
		\item \emph{Running time and description length}: The running time of $\length'$ is 
		\begin{align*}
		\poly(\TIME(\Lambda;n),\TIME(\Delta;n),\TIME({T};n),\TIME(Q;n),\TIME(\cal{FE};n),\Lambda(n),\log(\Delta(n)),\log({T}(n)) ,Q(n),D,\cal{FE}(n))\ ,
		\end{align*}
		while the running times  of $\sampler'$ and $\linproc'$ depend polynomially on the same parameters and  also on $\TIME(\sampler;n,\cdot,\cdot,\cdot,\cdot,\cdot)$.
		
		For description lengths, we have that 
		\[
		\begin{split}
		|\sampler'|&=\poly(|\Lambda|,|\Delta|,|Q|,|T|,|\cal{FE}|,|\sampler|)\ ,\\
		|\length'|&=\poly(|\Lambda|,|\Delta|,|Q|,|T|,|\cal{FE}|)\ ,\\
		|\linproc'|&=\poly(|\Lambda|,|\Delta|,|Q|,|T|,|\cal{FE}|,|\sampler|,|\linproc|)\ .\\
		\end{split}
		\]
		Note  that the running times of $\length$ and $\linproc$ \textbf{do not participate} in the above bounds, only the \textbf{log} of the upper bound on them, namely $\log({T}(n))$ --- which is the goal of this transformation.   
	\end{itemize}

\end{claim}

\begin{proof}
	In the operation of all of the TMs in $\verifier'$ there is a pre-processing phase in which the following values are calculated:
	\begin{itemize}
		\item $t=2\cal{FE}(n)+1$ --- this  requires $O(\TIME(\cal{FE};n))$ time;
		\item use Fact \ref{fact:basis_F_q_over_F_2} to retrieve a basis for $\FF_q$ over $\FF_2$, where $q=2^t$ --- requires $\poly(t)=\poly(\cal{FE}(n))$ time;
		\item compute $\Lambda(n)$, $\Delta(n)$, ${T}(n)$ and $Q(n)$ --- requires $\poly(\TIME(\Lambda;n),\TIME(\Delta;n),\TIME({T};n),\TIME(Q;n))$ time;
		\item compute $M(n)$ and $s(n)$ as defined in Proposition \ref{prop:explicit-padded-succinct-deciders} --- this requires \[
		\poly(\TIME(\Lambda;n),\TIME(\Delta;n),\TIME(T;n),\TIME(Q;n),n,\log(T(n)) ,Q(n),D)-\textrm{time}\; ,\]
		and also $M(n),s(n)$ are $ \poly(\log(T(n)) ,Q(n),D)$-sized;
		\item compute $\diamondsuit(n)$ from \eqref{eq:defn_of_diamondsuit(n)},  $m=|S|$ from \eqref{eq:size_of_m_PCPs}  and $\heartsuit(n)$ from \eqref{eq:def_heartsuit} --- requires 
		\[
		\polylog(\Lambda(n),\log(\Delta(n)),\log(T(n)) ,Q(n),D)-{\rm time}\ .
		\]
	\end{itemize}

	\textbf{The Sampler}: $\sampler'$ is the product of $\sampler$ and $\sampler^{\ld}$ (from Fact \ref{fact:algorithmic_low_degree}) with parameters $(9,m,2^t,\cdot)$, where $\cdot$ does not affect the sampler. Running such a sampler is just the sum of the running times of the component samplers $\TIME(\sampler;n,\cdot,\cdot,\cdot,\cdot,\cdot)$ and 
	\[ \TIME(\sampler^{\ld};9,m,2^t,\cdot,\cdot,\cdot,\cdot,\cdot,\cdot)=\poly(t,m)=\poly(\Lambda(n),\log(\Delta(n)),\log({T}(n)) ,Q(n),D,\cal{FE}(n))\ .
	\]

	\textbf{The Answer length calculator}: $\length'$ follows Table \ref{tab:Answer_Lengths_combinatorial_ans_red}. This requires outputting (the encoding of) at most $t\cdot (9m+1)\cdot\heartsuit(n)$-many $1$s, which takes less than 
	\[
	\poly(t,m,\heartsuit(n))=\poly(\Lambda(n),\log(\Delta(n)),\log({T}(n)) ,Q(n),D,\cal{FE}(n))\quad\textrm{time}\ .
	\]
	
	\textbf{The Linear constraints processor}: $\linproc'$ needs to check given its input which of \cref{clause1:low_degree_check_in_ans_red}, \cref{clause2:consistency_checks_in_ans_red}, \cref{clause3:Indifference_checks_in_ans_red} and \cref{clause4:PCP_checks_in_ans_red} are needed. Implementing \cref{clause1:low_degree_check_in_ans_red} requires manipulating length $\poly(m)$ vectors of $\FF_q$ inputs, and doing arithmetic using them, which in turn takes $\poly(m,\log(q))=\poly(m,t)$-time. In any case, it  needs to output $t$-many constraints, writing which again takes at most $\poly(m,t)$-time.
	Implementing \cref{clause2:consistency_checks_in_ans_red} is just adding a fixed number of consistency equations, where this number does not surpass $2t(9m+1)=\poly(t,m)$-many equations; this again takes at most $\poly(m,t)$-time. Implementing \cref{clause3:Indifference_checks_in_ans_red} is adding at most $40tm^2$-many equations that force a certain variable to be zero, which again takes at most $\poly(m,t)$-time. Finally, implementing \cref{clause4:PCP_checks_in_ans_red} is slightly more involved: It requires us to calculate $\frS^\mttA(z)$ and $\frS^\mttB(z)$, which are calls to $\sampler$ and this takes at most $2\TIME(\sampler;n,\cdot,\cdot,\cdot,\cdot,\cdot)$-time. It requires us to calculate $\circuit$ which is the third output of $\paddedsuccinctdecider(\verifier,\Lambda,\Delta,D,{T},Q, n,\frS^\mttA(z),\frS^\mttB(z))$, which takes, according to Proposition \ref{prop:explicit-padded-succinct-deciders}  at most
	\[
	\poly(\TIME(\Lambda;n),\TIME(\Delta;n),\TIME(T;n),\TIME(Q;n),n,\log(T(n)) ,Q(n),D)\quad{\rm  time}\ .
	\]
	Finally, evaluating the Tseitin polynomial $T_\circuit$ at $p$ requires us to calculate the value at every non-input gate, which in turn is some arithmetic operation in $\FF_q$ which takes $\poly(t)$-time. As there are $s(n)$ non-input gates, this takes $\poly(t,s(n))=\poly(\cal{FE}(n),\log(T(n)) ,Q(n),D)$-time. Combining all of the above provides the running time bound. 
	\\
	
	\textbf{Description lengths}:   The desecription lengths of $\sampler',\length',\linproc'$ depend polynomially on the parameters that play a role in their operation, which is only $\Lambda,\Delta,D,T,Q,\cal{FE}$ for $\length'$, all of the above in addition to  $\sampler$  for $\sampler'$, and all of the above including both  $\sampler$ and $\linproc$ for $\linproc'$.
\end{proof}

\begin{remark}
	The reason Claim \ref{claim:algorithmic_partial_ans_red} is only \textbf{partial} answer reduction, is that it needs to be combined with the padding and purification transformations, as well as choosing  the TMs $\cal{FE},\Lambda,\Delta,\cal{T},Q$ depending only on $h,h'$ and $\lambda$. This is done in the following proof of Answer Reduction
\end{remark}
.

\subsection{Proving the main theorem of Answer Reduction: Theorem \ref{thm:main_ans_red}}\label{sec:proof_main_thm_AR}

The goal is to describe the TM $\AnsRed_{h,h'}$ and prove it possesses the desired properties according to the Theorem. To make it easier for the reader to follow, let us say where each TM that needs to be defined is located: $\Lambda$ and $Q$ are defined in \eqref{eq:def_Q_and_Lambda_proof_ans_red}; $\Delta$ is defined in \eqref{eq:def_Delta_proof_ans_red}; $T$ is defined in \eqref{eq:def_T_proof_ansred}; $D$ is defined in \eqref{eq:def_D_proof_ansred}; $\cal{FE}$ is defined in the paragraph preceding \eqref{eq:bounds_t_main_thm_AR}.

Throughout the proof, we need to consider many constants which control previous asymptotic notations --- behind every $g(n)=\poly(f(n))$ (respectively $g(n)=O(f(n))$) there is a universal constant $c$ such that $g(n)\leq cf(n)^c$ (respectively $g(n)\leq cf(n)$). 
We list all of the used constants in Table \ref{tab:constants_along_proof_of_ans_red}.
\begin{table}[httb!]
	\centering
	\captionsetup{singlelinecheck=off}
	\begin{tabular}{|c|c|}
		\hline
		Constant & Origin \\
		\hline
		\hline
		$c_\ld$ & Theorem \ref{thm:ldc-soundness} and \eqref{eq:delta_ld_bound_in_proof_of_AR}\\
		\hline
		$c_\qr(h')$ & Theorem \ref{thm:h_level_question_reduciton}\\
		\hline
		$c_1$ & \eqref{eq:def_Q_and_Lambda_proof_ans_red}\\
		\hline
		$c_2$ & \eqref{eq:runtime_bound_Q_and_Lambda}\\
		\hline
		$c_3$ &\eqref{eq:timebound_2^Lambda}\\
		\hline
		$c_4$ & Claim \ref{claim:Padding_NFV} and \eqref{eq:timebound_and_description_padded_TNFV}\\
		\hline
		$c_5$ & Claim \ref{claim:algorithmic_purification} and \eqref{eq:bounds_linproc2}\\
		\hline
		$c_6$ & \eqref{eq:def_Delta_proof_ans_red}\\
		\hline
		$c_7$ & \eqref{eq:def_T_proof_ansred}\\
		\hline
		$c_8$ & \eqref{eq:def_D_proof_ansred} \\
		\hline
		$c$ & \eqref{eq:soundness_bound_c_def_ansred_proof}\\
		\hline
	\end{tabular}
	\caption[]{Constants along the proof of Theorem \ref{thm:main_ans_red}.}
	\label{tab:constants_along_proof_of_ans_red}
\end{table}

Recall that $c_\qr(h')$ is the constant guaranteed in Theorem \ref{thm:h_level_question_reduciton}.
Let $c_1$ be a later to be fixed constant, such that $c_1\geq c_\qr(h')$, which in particular means it depends on $h'$. Define $\Lambda$ and ${Q}$ to be the TMs that take $n$ (in binary) as input and output $c_1(\lambda^{c_1}+ n^{c_1})$, namely
\begin{equation}\label{eq:def_Q_and_Lambda_proof_ans_red}
{Q}(n)=\Lambda(n)=c_1(\lambda^{c_1}+ n^{c_1})\ ,
\end{equation}
which have runtime and description length 
\begin{equation}\label{eq:runtime_bound_Q_and_Lambda}
\begin{split}
\TIME({Q};n)=\TIME(\Lambda;n)&= \polylog_{h'}(\lambda, n)\leq c_2(\log^{c_2}\lambda+\log^{c_2}n)\ ,    \\
|Q|=|\Lambda|&=\polylog_{h'}(\lambda)\leq c_2\log^{c_2}\lambda\ ,
\end{split}
\end{equation}
for some positive integer $c_2$ that depends on $c_1$ and thus on $h'$.
Let $2^{\Lambda}$ be the TM that outputs $2^{\Lambda(n)}$ many $1$s given $n$ as input, which has runtime and description length
\begin{equation}\label{eq:timebound_2^Lambda}
\begin{split}
\TIME(2^{\Lambda};n)&=\poly(\TIME(\Lambda;n),2^{\Lambda(n)})\leq c_3\cdot 2^{c_3(\lambda^{c_3}+n^{c_3})}\\        
|2^{\Lambda}|&=O(|\Lambda|)\leq c_3\log^{c_3}\lambda\ ,
\end{split}
\end{equation}
for some positive integer $c_3$ that depends on $c_1,c_2$ and thus on $h'$.

Let $\verifier^{(1)}=(\sampler,\length^{(1)},\linproc^{(1)},\decider)=\mathsf{Padding}(\verifier,2^\Lambda)$ --- the sampler stays the same in this transformation --- where $\mathsf{Padding}$ was defined in Claim \ref{claim:Padding_NFV}. By that claim,
\begin{equation}\label{eq:timebound_and_description_padded_TNFV}
\begin{split}
\TIME(\length^{(1)};n,\cdot,\cdot)&=O(\TIME(2^{\Lambda};n))\leq c_4\cdot 2^{c_4(\lambda^{c_4}+n^{c_4})} \ ,\\
|\length^{(1)}| &=O(|2^{\Lambda}|)\leq c_4\log^{c_4}\lambda\ ,\\
\TIME(\linproc^{(1)};n,\cdot,\cdot,\cdot,\cdot)&=\poly(\TIME(2^\Lambda;n),\TIME(\sampler;n,\cdot,\cdot,\cdot,\cdot,\cdot),\TIME(\length;n,\cdot,\cdot),\TIME(\linproc;n,\cdot,\cdot,\cdot,\cdot))\\
&\leq c_4(2^{c_4(\lambda^{c_4}+n^{c_4})}+\TIME(\sampler;n,\cdot,\cdot,\cdot,\cdot,\cdot)^{c_4}+\TIME(\length;n,\cdot,\cdot)^{c_4}+\TIME(\linproc;n,\cdot,\cdot,\cdot,\cdot)^{c_4})\ ,\\
|\linproc^{(1)}|&=\poly(|2^{\Lambda}|,|\sampler|,|\length|,|\linproc|)\leq c_4(\log^{c_4}\lambda+|\sampler|^{c_4}+|\length|^{c_4}+|\linproc|^{c_4})\ ,
\end{split}
\end{equation}
for some constant $c_4$ that depends on the asymptotic bounds from Claim \ref{claim:Padding_NFV} as well as $c_1,c_2,c_3$ (and thus $h'$).

Let $\verifier^{(2)}=(\sampler,\length^{(1)},\linproc^{(2)},\decider)=\mathsf{Purify}(\verifier^{(1)})$ --- the sampler and answer length calculator stay the same in this transformation ---  where $\mathsf{Purify}$ was defined in Claim \ref{claim:algorithmic_purification}. Then,  we have
\begin{equation}\label{eq:bounds_linproc2}
\begin{split}
\TIME(\linproc^{(2)};n,\cdot,\cdot,\cdot,\cdot)&=\poly(\TIME(\length^{(1)};n,\cdot,\cdot),\TIME(\linproc^{(1)};n,\cdot,\cdot,\cdot,\cdot))\\
&\leq c_5(2^{c_5(\lambda^{c_5}+n^{c_5})}+\TIME(\sampler;n,\cdot,\cdot,\cdot,\cdot,\cdot)^{c_5}+\TIME(\length;n,\cdot,\cdot)^{c_5}+\TIME(\linproc;n,\cdot,\cdot,\cdot,\cdot)^{c_5})  \ ,\\
|\linproc^{(2)}|&=\poly(|\linproc^{(1)}|,|\length^{(1)}|)\\
&\leq c_5(\log^{c_5}\lambda+|\sampler|^{c_5}+|\length|^{c_5}+|\linproc|^{c_5}) \ ,
\end{split}
\end{equation}
for some $c_5$ that depends on the asymptotic bounds from Claim \ref{claim:algorithmic_purification} as well as $c_1,c_2,c_3,c_4$.  Now, $\verifier^{(2)}$ is $2^{\Lambda}$-padded and purified.

We are left to define $\Delta,T,\cal{FE}$ and $D$.
Let $\Delta$ be the TM that calculates
\begin{equation}\label{eq:def_Delta_proof_ans_red}
\Delta(n)=2^{2\Lambda(n)+2}-2^{\Lambda(n)+1}.
\end{equation}
Its runtime bound and description length satisfy
\begin{equation}\label{eq:def_Delta_proof_ans_red}
\begin{split}
\TIME(\Delta;n)&=\poly(\TIME(\Lambda;n),\Lambda(n))\leq c_6(\lambda^{c_6}+ n^{c_6})\ ,\\
|\Lambda|&=O(|\Lambda|)\leq c_6\log^{c_6}\lambda\ ,
\end{split}
\end{equation}
for some positive integer $c_6$ that depends on $c_1,c_2$. This choice of $\Delta$ is made so that $\diamondsuit(n)$ is a nice expression $2\Lambda(n)+2$.

As $T$ needs to satisfy \eqref{eq:value_of_T(n)} given  that
\[
\TIME(\linproc;n,\cdot,\cdot,\cdot,\cdot)\leq 2^{c_\qr(h') (\lambda^{c_\qr(h')}+ n^{c_\qr(h')})}\ ,
\]
by the previous upper bounds on $\TIME(\Lambda;n)$, $\TIME(\Delta;n)$, $\Lambda(n)$ and $\Delta(n)$, we can choose a positive integer $c_7$, that depends on all the previous constants $c_1,...,c_6$ as well as the asymptotic notation from \eqref{eq:time_bound_L*} --- but not on anything else, and in particular not on $\linproc$ --- so that 
\begin{equation}\label{eq:def_T_proof_ansred}
T(n)=c_7\cdot2^{c_7(\lambda^{c_7}+n^{c_7})}
\end{equation}
indeed satisfies \eqref{eq:value_of_T(n)} under the upper bound assumption on $\TIME(\linproc;n,\cdot,\cdot,\cdot,\cdot)$.\footnote{Note that we do not assume at this point that $\linproc$ satisfies the runtime upper bound, only that $c_7$ can be calculated   to satisfy \eqref{eq:value_of_T(n)}  \textbf{in case} the upper bound holds. This is a crucial point, as the final sampler and answer length calculator depend on $T$, and are not allowed to depend on $\linproc$.} Moreover, 
\[
\TIME(T;n)=\poly_{h'}(\lambda,n)\quad{\rm and}\quad |T|=O_{h'}(\log\lambda)\ .
\]

As $D$ needs to bound $|\Lambda|$, $|\Lambda|$, $|T|$ and $|Q|$,  and also $|\verifier|$ but only under the assumption that $|\verifier|\leq 5c_\qr(h')\lambda^{c_\qr(h')}$, by the previous calculated bounds there is a large enough positive integer $c_8$  (that depends on all previous constants and thus on $h'$) so that 
\begin{equation}\label{eq:def_D_proof_ansred}
D=c_8 \lambda^{c_8}
\end{equation}
indeed upper bounds all of the above.

The choice of $\cal{FE}$ requires us to associate more constants with previously used asymptotic notation. Let  $c_\ld$ be the constant from the definition of $\delta_\ld$ \eqref{eq:defn_delta_ld}, namely 
\begin{equation}\label{eq:delta_ld_bound_in_proof_of_AR}
\delta_\ld(m,d,k,\eps,q^{-1})=c_\ld(m^{c_\ld}+d^{c_\ld}+k^{c_\ld})(\eps^{\nicefrac{1}{c_\ld}}+q^{-\nicefrac{1}{c_\ld}}+2^{-\nicefrac{md}{c_\ld}})\ .
\end{equation}
By running $\paddedsuccinctdecider$ from Proposition \ref{prop:explicit-padded-succinct-deciders} on $(\cdot,\cdot,\cdot,D,T,Q,n,\cdot,\cdot)$ --- which takes
\[
\poly(\TIME(\Lambda;n),\TIME(\Delta;n),\TIME(T;n),\TIME(Q;n),n,\log(T(n)) ,Q(n),D)=\poly_{h'}(\lambda,n)
\]
time --- we can retrieve $M(n)$ and $s(n)$ which are of size 
\[
\poly(\log(T(n)),Q(n),D)=\poly_{h'}(\lambda,n)\ ,
\]
and thus 
\[
\begin{split}
m=|S|=4\Lambda(n)+3\diamondsuit(n)+3M(n)+s(n)+12=\poly_{h'}(\lambda,n)\ .
\end{split}
\]
Note that 
\[
\heartsuit(n)= 12\Lambda(n)+12\diamondsuit(n)+6M(n)+s(n)+35\leq 4m\ ,
\]
and thus 
\[
m^{c_\ld}+9^{c_\ld}+\heartsuit(n)^{c_\ld}\leq (14m)^{c_\ld}\ .
\]
We now fix $c_1$ to be the smallest constant which is larger than $c_\qr(h')$ and makes $\Lambda(n)$ big enough so that $m$ satisfies $c_\ld (14m)^{c_\ld}\cdot 2^{-\nicefrac{9m}{c_\ld}}\leq m^{-16}$ for every $n\geq 2$. Now $\cal{FE}$ will be the TM that takes $n$ as input, outputs an integer $\frac{t-1}{2}$, where $t$ is the smallest positive odd integer  for which $q=2^t$ satisfies both
\begin{equation}\label{eq:bounds_t_main_thm_AR}
\frac{72m}{q}\leq \frac{1}{2}\qquad \textrm{and}\qquad c_\ld(14m)^{c_\ld}\cdot q^{\nicefrac{-1}{c_\ld}}\leq m^{-16}\;,
\end{equation}
for every $n\geq 2$.
Note that $t=\Theta(\log m)$, and as $\cal{FE}$ only needs to calculate $m$ and thus $M(n),s(n)$, it takes it at most $\poly_{h'}(\lambda,n)$ time. Also, its description is fixed up to appending the values $h'$ and $\lambda$, which means $|\cal{FE}|=\polylog_{h'}(\lambda)$.
By the above choices and equation \eqref{eq:delta_ld_bound_in_proof_of_AR}, we have
\[
\delta=\delta_\ld(m,9,\heartsuit(n),2\eps,q^{-1})\leq c_\ld(14m)^{c_\ld}\cdot (2\eps)^{\nicefrac{1}{c_\ld}}+2m^{-16}\ ;
\]
by the concavity of $(\cdot)^{\nicefrac{1}{8}}$  and the fact it only shrinks positive integers, we have
\[
\delta^{\nicefrac{1}{8}}\leq c_\ld(14m)^{c_\ld}\cdot (2\eps)^{\nicefrac{1}{8c_\ld}}+2m^{-2}\ ,
\]
and as $\frac{72m}{q}\leq \frac{1}{2}$ we have
\[
\frac{10^6\cdot m}{1-\frac{72m}{q}}\delta^{\nicefrac{1}{8}}\leq 2\cdot 10^6(c_\ld \cdot m\cdot(14m)^{c_\ld}\cdot (2\eps)^{\nicefrac{1}{8c_\ld}}+2m^{-1}).
\]
As  we have $m\geq \Lambda(n)\geq \lambda+n\geq \sqrt{\lambda n}$, for every positive integer $c\geq 2$ we have $m^{-1}\leq (\lambda n)^{-\nicefrac{1}{c}}$. All in all, there is a large enough positive integer $c$ (that depends on all previous constants and bounds, and thus on  $h'$) such that 
\begin{equation}\label{eq:soundness_bound_c_def_ansred_proof}
\frac{10^6\cdot m}{1-\frac{72m}{q}}\delta^{\nicefrac{1}{8}}\leq c((\lambda n)^c\eps^{\nicefrac{1}{c}}+(\lambda n)^{-\nicefrac{1}{c}})\ .    
\end{equation}

We can finally define $\verifier_\ar=(\sampler_\ar,\length_\ar,\linproc_\ar,\decider)=\PartialAnsRed(\verifier^{(2)},\Lambda,\Delta,{T},Q,\cal{FE})$, where $\PartialAnsRed$ is from Claim \ref{claim:algorithmic_partial_ans_red}. Let us prove that, indeed, $\verifier_\ar$ satisfies the requirements of Theorem \ref{thm:main_ans_red}:
\\

\textbf{Time bounds and description lengths}:

By Claim \ref{claim:algorithmic_partial_ans_red} and the above bonds,
we have
\[
\begin{split}
|\sampler_\ar|&=\poly(|\Lambda|,|\Delta|,|Q|,|T|,|\cal{FE}|,|\sampler|)=\poly_{h'}(\log\lambda,|\sampler|)\ ,\\
|\length_\ar|&=\poly(|\Lambda|,|\Delta|,|Q|,|T|,|\cal{FE}|)=\polylog_{h'}(\lambda)\ ,\\
|\linproc_\ar|&=\poly_{h'}(|\Lambda|,|\Delta|,|Q|,|T|,|\cal{FE}|,|\sampler|,|\linproc^{(2)}|)=\poly_{h'}(\log\lambda,|\verifier|)\ .\\
\end{split}
\]
For running times, by Claim \ref{claim:algorithmic_partial_ans_red}, we have that  $\TIME(\length_\ar;n,\cdot,\cdot)$ is bounded by 
\begin{align*}
\poly(&\TIME(\Lambda;n),\TIME(\Delta;n),\TIME({T};n),\TIME(Q;n),\TIME(\cal{FE};n),\\&\Lambda(n),\log(\Delta(n)),\log({T}(n)) ,Q(n),D,\cal{FE}(n))\\
&= \poly_{h'}(n,\lambda)\ ,
\end{align*}
while $\sampler_\ar$ and $\linproc_\ar$ running times also depend polynomially on $\TIME(\sampler;n,\cdot,\cdot,\cdot,\cdot,\cdot)$, i.e.,
\[
\TIME(\sampler_\ar;n,\cdot,\cdot,\cdot,\cdot,\cdot),\TIME(\linproc_\ar;n,\cdot,\cdot,\cdot,\cdot)=\poly_{h'}(n,\lambda,\TIME(\sampler;n,\cdot,\cdot,\cdot,\cdot,\cdot))\ ,
\]
as is needed  for the theorem to hold.
\\

\textbf{Completeness and Soundness}:

Assume 
\[
\begin{split}
&|\verifier|\leq 5\ c_\qr(h')\cdot\lambda^{c_\qr(h')}\ ,\\
\forall {n}\in \mathbb{N}\ \colon \ \ &\TIME(\sampler\ ;\ {n},\cdot,\cdot,\cdot,\cdot,\cdot)\leq c_\qr(h')(n^{c_\qr(h')}+\lambda^{c_\qr(h')})\\
\forall {n}\in \mathbb{N}\ \colon \ \ &\TIME(\length\ ;\ {n},\cdot,\cdot)\ ,\ \TIME(\linproc\ ;\ {n},\cdot,\cdot,\cdot,\cdot) \leq 2^{c_\qr(h')(n^{c_\qr(h')}+\lambda^{c_\qr(h')})}\ .
\end{split}
\]
As $c_1\geq c_\qr(h')$, the answer length of  $\verifier_n$ is bounded by $2^{\Lambda(n)}$, which, by Fact \ref{fact:padding_properties}, means $\verifier^{(1)}_n=\frak{Padding}(\verifier_n,2^{\Lambda(n)})$ has the same value as $\verifier_n$, and in particular has a perfect $\ZPC$ strategy if $\verifier_n$ has one. Now $\verifier^{(2)}_n=\frak{Pure}(\verifier^{(1)}_n)$, and again  this transformation preserves values (Fact \ref{fact:completeness_and_soundness_purification}). Let us check that $\verifier^{(2)}$ satisfies the conditions that ensure 
\[
(\verifier_\ar)_n=\frak{AnsRed}(\verifier^{(2)},\Lambda,\Delta,D,{T},Q,n,
2\cal{FE}(n)+1).
\]
If this is satisfied, then by Proposition \ref{prop:completeness_soundness_combi_ans_red}, the completeness is deduced from the completeness of $\verifier^{(2)}$ which we already settled, and the soundness is due to our choice of $t$ in \eqref{eq:bounds_t_main_thm_AR} and the bound from \eqref{eq:soundness_bound_c_def_ansred_proof}.
So, we are left to verify the conditions for the completeness and soundness of Proposition \ref{prop:completeness_soundness_combi_ans_red}:
\begin{itemize}[noitemsep,topsep=0pt,parsep=0pt,partopsep=0pt]
	\item[--] $\Lambda$   always halts;
	\item[--] $\verifier^{(2)}$ is purified  and $2^\Lambda$-padded;
	\item[--] $\Delta$ always halts; as $c_1\geq c_\qr(h')$, we have  $\TIME(\linproc;n,\cdot,\cdot,\cdot,\cdot)\leq 2^{\Lambda(n)}$, and thus
	\[\TIME(\linproc;n,\cdot,\cdot,\cdot,\cdot)\cdot 2^{\Lambda(n)+1}\leq 2^{2\Lambda(n)+1}\leq 2^{2\Lambda(n)+2}-2^{\Lambda(n)+1}=\Delta(n)\ ;\]
	\item [--] $Q$ always halts, and again as  $c_1\geq c_\qr(h')$ we have
	$\TIME(\sampler;{n},\cdot,\cdot,\cdot,\cdot,\cdot)\leq {Q}(n)$;
	\item [--] $T$ is always halting  and we chose $c_7$ exactly for \eqref{eq:time_bound_L*} to hold given the appropriate upper bound on $\TIME(\linproc;n,\cdot,\cdot,\cdot,\cdot)$;
	\item [--] $D$ a positive integer (in binary), and we chose $c_8$ so that, given the upper bound on $|\verifier|$, we will have  
	\[
	|\verifier|,|\Lambda|,|\Delta|,|T|,|Q|\leq D\ .
	\]
\end{itemize}

\section{Parallel repetition}\label{sec:parallel_rep}

\def\tvanch{\hat{\verifier}^\anch}
\def\tdanch{\hat{\decider}^\anch}
\def\tlanch{\hat{\linproc}^\anch}
\def\taanch{\hat{\length}^\anch}
\def\tsanch{\hat{\sampler}^\anch}
\def\vanch{\verifier^\anch}
\def\danch{\decider^\anch}
\def\lanch{\linproc^\anch}
\def\aanch{\length^\anch}
\def\sanch{\sampler^\anch}

\newcommand{\wtgame}{{\widetilde{\game}}}
\newcommand{\wtmttx}{{\widetilde{\mttx}}}
\newcommand{\wtmtty}{{\widetilde{\mtty}}}
\newcommand{\wtV}{\widetilde{V}}

In this section we define a gap amplification transformation on games. This transformation is based on taking repeated products (Definition \ref{def:product-game}) of a game with itself. It is designed so that, if a game has a $\ZPC$ strategy with value $1$, then this remains the case for the repeated game; while, if the game has value bounded away from $1$, then the value of the repeated game goes to zero (exponentially) with the number of repetitions. 

The {product} of two (tailored) games was introduced in Definition~\ref{def:product-game}. In Lemma~\ref{lem:sum-zpc} we showed that if $\game_1$ and $\game_2$ each have a perfect $\ZPC$ strategy, then so does $\game_1\otimes\game_2$. One might more generally expect that the value is multiplicative under game product --- however, this is \emph{not} the case (see e.g.~\cite[Example 4.3]{manvcinska2023products}). Fortunately, it is known that under mild assumptions on the structure of the game $\game$, the value of its $k$-fold repetition $\game^{\otimes k}$ does go to zero exponentially fast as $k$ goes to infinity (provided the value of $\game$ is strictly smaller than $1$).  Actually, under these mild assumptions, what is known is that if the \emph{non-synchronous} value of $\game$ is smaller than $1$, then the \emph{non-synchronous} value of the repeated game tends to zero exponentially in the number of repetitions (see Section \ref{sec:non-synch_setup} for the non-synchronous setup). To be able to translate this fact to our \emph{synchronous} setup (namely, the notions of value and entanglement lower bounds used throughout this paper), we need to use the results of \cite{vidick2022almost} that say, essentially, that if $\game$ has many self consistency checks, then its synchronous value is not much smaller than its non-synchronous one.

As usual, this transformation needs to be implemented on the level of tailored normal form verifiers. The following is the main theorem proved in this section. Recall  the asymptotic notation from Remark \ref{rem:asymptotic_notation}.

\begin{theorem}[Parallel Repetition. Proved in Section \ref{sec:main_par_rep}]
	\label{thm:repetition}
	There exists a $2$-input Turing machine
	$\ComputeParrepVerifier_h$, that takes as input a \textbf{typed} $h$-level TNFV $\verifier = (\sampler,\length,\linproc,\decider)$ with type set $\type$, and a $1$-input TM $\cal{K}$ (which induces, as usual, a partial  function $\cal{K}\colon \mathbb{N}\to \mathbb{N}$), 
	and  outputs an $(h+2)$-level \textbf{untyped} TNFV  
	\[ 
	\ComputeParrepVerifier_h(\verifier,\cal{K})=\verifier_\rep = (\sampler_\rep,\length_\rep,\linproc_\rep,
	\decider)
	\]
	with the following properties. There is a positive integer constant 
	\begin{equation}\label{eq:defn_c_rep}
	c=c_\rep(|\type|)
	\end{equation} 
	depending only on the number of types in $\verifier$, such that:

	\begin{itemize}
		\item \underline{Sampler properties}:  $\sampler_{\rep}$ depends only on $\cal{K}$ and the original sampler $\sampler$ (but not on $\length$ or $\linproc$), and  $\ComputeParrepVerifier_{h}$ can calculate its description in time $\poly(|\cal{K}|, |\sampler|)$ from them; in particular, $|S_\rep|\leq c\cdot (|\cal{K}|^c+|\sampler|^c)$. In addition, $\sampler_\rep$ runs in time
		\[
		\poly(n,\cal{K}(n),\TIME(\cal{K};n),\TIME(\sampler;n,\cdot,\cdot,\cdot,\cdot,\cdot))\;,
		\]
		namely
		\[
		\forall n\in \mathbb{N}\ \colon \ \ \TIME(\sampler_\rep;n,\cdot,\cdot,\cdot,\cdot,\cdot)\leq c\cdot(n^c+\cal{K}(n)^c+\TIME(\cal{K};n)^c+\TIME(\sampler;n,\cdot,\cdot,\cdot,\cdot,\cdot)^c)\ ,
		\]
		where $c$ is from \eqref{eq:defn_c_rep}.
		
		\item \underline{Answer length calculator properties}: $\length_\rep$  depends only on  $\cal{K},\sampler$ and $\length$,  and $\ComputeParrepVerifier_{h}$ can calculate its description in time $\poly(|\cal{K}|,|\sampler|,|\length|)$; in particular $|\length_\rep|\leq c\cdot (|\cal{K}|^c+|\sampler|^c+|\length|^c)$. 
		In addition, $\length_\ar$ runs in  time
		\[
		\poly(n,\cal{K}(n),\TIME(\cal{K};n),\TIME(\sampler;n,\cdot,\cdot,\cdot,\cdot,\cdot),\TIME(\length;n,\cdot,\cdot))\ ,
		\]
		namely 
		\[
		\forall n\in \mathbb{N}\ \colon \ \ \TIME(\length_\rep;n,\cdot,\cdot)\leq c\cdot(n^c+\cal{K}(n)^c+\TIME(\cal{K};n)^c+\TIME(\sampler;n,\cdot,\cdot,\cdot,\cdot,\cdot)^c+\TIME(\length;n,\cdot,\cdot)^c)\ .
		\]
		Finally,  if for every $\mttx\in \FF_2^{r(n)}$, where $r(n)=\sampler(n,{\rm Dimension},\cdot,\cdot,\cdot,\cdot)$,  and  $\kappa\in \{\frR,\frL\}$,  the output of $\length(n,\mttx,\kappa)$  never decodes (Definition \ref{defn:the_alphabet}) to an $\frak{error}$ sign, then for every $\mtty\in \FF_2^{\cal{K}(n)\cdot r(n)}$,  $\length_\rep(n,\mtty,\kappa)$  never decodes  to an $\frak{error}$ sign.
		
		\item \underline{Linear constraints process properties}: $\linproc_\rep$ depends on all inputs. Also,  $\ComputeParrepVerifier_{h}$ can calculate its description in time $\poly(|\cal{K}|, |\verifier|)$; in particular, $|\linproc_\rep|\leq c\cdot (|\cal{K}|^c+ |\verifier|^c)$. 
		In addition, $\linproc_\rep$ runs in   time
		\[
		\poly(n,\cal{K}(n),\TIME(\cal{K};n),\TIME(\sampler;n,\cdot,\cdot,\cdot,\cdot,\cdot),\TIME(\length;n,\cdot,\cdot),\TIME(\linproc;n,\cdot,\cdot,\cdot,\cdot))\ ,
		\]
		namely 
		\[
		\forall n\in \mathbb{N}\ \colon \ \ \TIME(\linproc_\rep;n,\cdot,\cdot,\cdot,\cdot)\leq c\cdot(n^c+\cal{K}(n)^c+\TIME(\cal{K};n)^c+\TIME(\sampler;n,\cdot,\cdot,\cdot,\cdot,\cdot)^c+\TIME(\length;n,\cdot,\cdot)^c+\TIME(\linproc;n,\cdot,\cdot,\cdot,\cdot)^c)\ .
		\]
		\item  \underline{Value properties}: For $c=c_\rep(|\type|)$ as in \eqref{eq:defn_c_rep} and 
		$\verifier_\rep$ the output of $\ComputeParrepVerifier$,  we have for all $n \geq 2$:
		\begin{enumerate}
			\item \textbf{Completeness}: If $\verifier_n$ has
			a value-$1$ $\ZPC$ strategy, then $(\verifier_{\rep})_n$ has a value-$1$ $\ZPC$
			strategy.
			\item \textbf{Soundness}: For all $\eps > 0$, let 
			\begin{equation}\label{eq:defn_p_par_rep}
			p=p(\eps,n) =  \frac{c}{\eps^c} \cdot 2^{-  \frac{\eps^{c}}{c} \cdot \cal{K}(n)\cdot \TIME(\length;n,\cdot,\cdot)^{-1}}\;. 
			\end{equation}
			Then, if ${(\verifier_\rep)}_n$ has a strategy with value of at least $p$,  then $\verifier_n$ has a strategy with value of at least $1-\eps$.
			\item \textbf{Entanglement bound}: For the same parameters as in the soundness property, 
			\[
			\Ent((\verifier_\rep)_n, p) \geq \Ent(\verifier_n, 1 - \eps)\;.
			\]
		\end{enumerate}
	\end{itemize}
\end{theorem}

The structure of this section is as follows: In Section~\ref{sec:anchoring-comb} we introduce a simple transformation that can be performed on any game, and is such that the resulting game, which is called \emph{anchored}, will behave well under repetition.  In Section~\ref{sec:parrep-comb} we introduce the  parallel repetition transformation, and analyze its completeness and soundness properties assuming the game was anchored beforehand. Finally, in Section~\ref{sec:main_par_rep}, we combine the two transformations (together with an in-between detyping) so as to prove the main theorem of this section (Theorem~\ref{thm:repetition}).

\subsection{The anchoring transformation} 
\label{sec:anchoring-comb}

The anchoring transformation is studied in~\cite{bavarian2017hardness} in the context of nonlocal games with quantum players sharing entanglement, and extended in~\cite{MIPRE} to normal form verifiers. 
Here we consider the anchoring transformation applied to a typed game $\game$.
Informally, anchoring a game consists in adding a single additional type $\Anchor$ such that, whenever a question of type $\Anchor$ is sampled, any answer is accepted. 

\begin{definition}[Combinatorial anchoring]\label{def:comb-anchoring}
	Let $\game$ be a  tailored nonlocal game with an $h$-level \textbf{typed} CL sampling scheme (Definition \ref{defn:typed_h-level_sampling_scheme}), and let $(\type,\cal{E})$ be the associated type graph. The anchoring $\frak{Anchor}(\game)$ of $\game$ is another typed tailored game $\game_\perp$ with an underlying $h$-level typed CL sampling scheme; its type set contains an additional element $\Anchor$, i.e. $\type_\perp =\type \sqcup \{\Anchor\}$. The anchored type graph contains the original type graph as the induced subgraph on the vertices $\type$, while attaching $\Anchor$ to every other type; namely  
	\[
	\cal{E}_\perp = \{tt'\in \type_\perp\times \type_\perp\mid  t=\Anchor\ \textrm{or}\  t'=\Anchor\ \textrm{or}\ tt'\in \cal{E}\}\ .
	\]
	
	The CLMs at each type $t\in \type$ are  kept as before, and we let $\frS^\Anchor$ be the zero function (which is $0$-level and thus does not change the level of the sampling scheme). The readable and unreadable answer lengths of the question $(\Anchor,\vec 0)$ are both zero. Finally, the decision function of the anchored game executes $D_{(t,\mttx)(t',\mtty)}$ whenever $t,t'\in \type$, and always accepts whenever $t$ or $t'$ equal $\Anchor$. 
\end{definition}

Recall that in a game with a typed CL sampling scheme (Definition \ref{defn:typed_h-level_sampling_scheme}) a uniform edge from the type graph is first sampled. Thus,  $\game_\perp$ runs $\game$ with probability $\frac{|\cal{E}|}{|\cal{E}|+2|\type|+1}$, and otherwise accepts. This translates to 
\begin{equation}\label{eq:anchoring-val}
\val^*(\frak{Anchor}(\game)) = \frac{2|\type|+1}{|\cal{E}|+2|\type|+1} + \frac{|\cal{E}|}{|\cal{E}|+2|\type|+1} \val^*(\game)\ .
\end{equation}
Actually, \textbf{every} strategy $\strategy$ for $\frak{Anchor}(\game)$ is also a strategy for $\game$ (as the anchor vertex is of length $0$), and the above relation is on that level, namely
\[
\val(\frak{Anchor}(\game),\strategy) = \frac{2|\type|+1}{|\cal{E}|+2|\type|+1} + \frac{|\cal{E}|}{|\cal{E}|+2|\type|+1} \val(\game,\strategy)\ .
\]
In particular, we have that $\val^*(\frak{Anchor}(\game)) = 1$ if and only if $\val^*(\game) = 1$. Moreover, if the latter is achieved with a $\ZPC$ strategy then so is the former; as is easily verified by using exactly the same strategy.

\begin{claim}[Algorithmic anchoring]\label{claim:alg_anchor}
	There is a polynomial time TM $\TMAnchoring$ that takes a typed $h$-level tailored normal form verifier   $\verifier=(\sampler,\length,\linproc,\decider)$   as  input, and outputs an $h$-level tailored normal form verifier  $\TMAnchoring(\verifier)=\verifier'=(\sampler',\length',\linproc',\decider)$, such that:
	\begin{enumerate}
		\item \label{enu:anch-completeness} \emph{Combinatorial anchoring}: If  $\verifier_n$ is well defined, then $\verifier'_n$ is well defined and $\verifier'_n=\frak{Anchor}(\verifier_n)$.
		\item \label{enu:anch-complexity} \emph{Complexity}: The
		time complexities of the TMs in $\verifier'$ satisfy, for every integer in binary $\overline{n}\in\{0,1\}^*$,
		\begin{align*}
		\TIME(\sampler';\overline{n},\cdot,\cdot,\cdot,\cdot,\cdot) & = O(\TIME(\sampler;\overline{n},\cdot,\cdot,\cdot,\cdot))\;,\\
		\TIME(\length';\overline{n},\cdot,\cdot) & = O(\TIME(\length;\overline{n},\cdot,\cdot))\;,\\
		\TIME(\linproc';\overline{n},\cdot,\cdot,\cdot,\cdot) & = O(\TIME(\linproc;\overline{n},\cdot,\cdot,\cdot,\cdot))\;.
		\end{align*}
		In addition,  the description   of $\sampler'$, $\length'$ and $\linproc'$
		can be computed in  time  which is linear in the descriptions of $\sampler$, $\length$ and $\linproc$. Moreover, the description of $\sampler'$ only depends on
		$\sampler$.
	\end{enumerate}
\end{claim}

\begin{proof}

	Let $\type$ be the type set of $\verifier$. Define the type set $\type_\perp = \type\sqcup \{\Anchor\}$. If the type graph of $\verifier$ is $(\type,\cal{E})$ then the type graph of $\verifier'$ is $(\type_\perp,\cal{E}_\perp)$ where according to Definition~\ref{def:comb-anchoring} we set
	\begin{equation}\label{eq:graph-anch}
	\cal{E}_\perp = \big\{tt'\in \type_\perp\times \type_\perp\mid  t=\anch\ \textrm{or}\  t'=\anch\ \textrm{or}\ tt'\in \cal{E}\big\}\;.
	\end{equation}
	The Turing machine $\sampler'$ is defined as follows:
	\begin{enumerate}
		\item On input $(\cdot,{\rm Graph},\cdot,\cdot,\cdot,\cdot)$ it executes $\sampler(\cdot,{\rm Graph},\cdot,\cdot,\cdot,\cdot)$ and modifies the output appropriately to match~\eqref{eq:graph-anch} (adding a type $\Anchor$ and a corresponding row and column of $1$'s to the adjacency matrix). 
		\item On input $(n,{\rm Dimension},\cdot,\cdot,\cdot,\cdot)$, it returns $\sampler(n,{\rm Dimension},\cdot,\cdot,\cdot,\cdot)$.
		\item On all other inputs
		\[
		(n,{\rm Action}, {\rm Type},j,\mttx,z)\ ,
		\] 
		if ${\rm Type}\neq \Anchor$, then it operates the same as $\sampler$. Otherwise, it encodes the zero map.
	\end{enumerate}
	Define the Turing machine $\linproc'$ that, on input
	$(n,(t,\mttx),(t',\mtty),a^\frR,b^\frR)$, if either
	$t$ or $t'$ is equal to $\Anchor$, then it accepts automatically (returns no constraints). 
	Otherwise, it returns the output of $\linproc(n,(t,\mttx),(t',\mtty),a^\frR,b^\frR)$. Finally, define the Turing machine $\length'$ that returns the same output as $\length$ when the type is not $\Anchor$, and returns $0$ when the type is $\Anchor$. 
	\\
	
	All the above TMs clearly run in time which is linear in their original counterparts, and their description is constant up to appending $\verifier$. Finally, $\sampler'$ indeed uses only $\sampler$ in its definition. 
\end{proof}

\subsection{The  parallel repetition transformation}
\label{sec:parrep-comb}

\newcommand{\tensk}{\otimes k}

Let $\game$ be a (untyped) tailored game with underlying graph $G=(V,E)$ and question distribution $\mu$ over edges. Recall that in Section~\ref{sec:composition_of_games} we define the tensor product of two games (Definition \ref{def:product-game}). Iterating the definition, we obtain the following ${k}$-fold tensored game $\game^{\tensk}$, which we make explicit for convenience: 
\begin{enumerate}
	\item The underlying graph is the $k$-fold tensor product of the graph $G$ with  itself, denoted $G^{\tensk}$, whose vertex set is $V^k$. The vertex $(\mttx_1,...,\mttx_k)$ is a neighbor of $(\mtty_1,...,\mtty_k)$ in $G^{\tensk}  $ if and only if $\mttx_i$ is a neighbor of $\mtty_i$ for every $i\in [k]$ in the original graph $G$.
	\item The question distribution is $\mu^k$; namely, $k$ edges $\mttx_i\mtty_i$ from $G$ are sampled independently according to $\mu$, and are combined to a single edge $(\mttx_1,...,\mttx_k)(\mtty_1,...,\mtty_k)$ in $G^{\tensk}$.
	\item The length functions satisfy  $\ell^{\tensk,\rvar}(\mttx)=\sum_i \ell^\rvar(\mttx_i)$ and $\ell^{\tensk,\lvar}(\mttx)=\sum_i \ell^\lvar(\mttx_i)$, and for each $\mttx=(\mttx_1,\ldots,\mttx_k)\in V^k$ the sets of formal generators are $S_\mttx^{\tensk,\rvar} = S_{\mttx_1}^\rvar\sqcup\cdots\sqcup S_{\mttx_k}^\rvar$ and  $S_\mttx^{\tensk,\lvar} = S_{\mttx_1}^\lvar\sqcup\cdots \sqcup S_{\mttx_k}^\lvar$; namely, the answer at $(\mttx_1,...,\mttx_k)$ is a $k$-tuple of answers to the respective $\mttx_i$'s. Note that there is a copy of $S_{\mttx_i}$ in $S_\mttx$, and thus linear constraints over $S_{\mttx_i}\cup S_{\mtty_i}$ can be naturally interpreted as  linear constraints over $S_{\mttx}\cup S_{\mtty}$.
	
	\item The function $L_{\mttx\mtty}^{\tensk}$ is obtained by returning the ``union'' of all $L_{\mttx_i\mtty_i}$. I.e.,  for  $\mttx=(\mttx_1,\ldots,\mttx_k)$ and  $\mtty=(\mtty_1,\ldots,\mtty_k)$, an answer to $\mttx$ is formatted as $a^\frR=(a_{1}^\frR,...,a_k^\frR),a^\frL=(a_{1}^\frL,...,a_k^\frL)$ and to $\mtty$ as $b^\frR=(b_{1}^\frR,...,b_k^\frR),b^\frL=(b_{1}^\frL,...,b_k^\frL)$. So, $L_{\mttx_i\mtty_i}(a_i^\frR,b_i^\frR)$ returns a sequence of linear constraints on variables in $S_{\mttx_i}\cup S_{\mtty_i}$ which we can interpret as linear constraints over $S_{\mttx}\cup S_{\mtty}$. Combining all of this we let
	\[ L_{\mttx\mtty}^{\tensk}\big(a^\frR ,b^\frR\big) \,=\, \bigsqcup_{i=1}^k L_{\mttx_i\mtty_i}\big(a^\frR_i,b^\frR_i \big)\;.\]
	Note that on the level of the decision predicate, this gives the following relation:
	\[
	D^{\tensk}_{\mttx\mtty}(a^\frR,a^\frL,b^\frR,b^\frL)=\prod_{i=1}^k D_{\mttx_i\mtty_i}(a_i^\frR,a_i^\frL,b_i^\frR,b_i^\frL)\ ,
	\]
	which is the standard view of parallel repetition. 
\end{enumerate}

The following theorem states the effect that the $k$-fold tensoring has on the value of a game, at least when it is applied to the detyping of a typed game that is anchored according to Definition~\ref{def:comb-anchoring}.

\begin{theorem}[Completeness and soundness of the parallel repeated game]\label{thm:parrep-comp-sound}
	Let $\game$ be a tailored nonlocal game with underlying typed $h$-level CL sampling scheme. Assume that the type graph of $\game$ has self-loops at each vertex, and let  $\Lambda$ be an upper bound on the answer length in $\game$. Let $\game'=\mathfrak{Anchor}(\game)$ (Definition \ref{def:comb-anchoring}) and $\wtgame=\frak{DeType}(\game')$ (Definition \ref{defn:combi_detyping}).
	Then, there is a positive integer constant $c_0$ depending only on $|\mathcal{T}|$, such that for any $k\geq 1$,
	the  game $\wtgame^{\tensk}$ has the following properties.
	\begin{enumerate}[label=\textcolor{black}{(\arabic*)}, ref= (\arabic*)]
		\item \label{clause:completeness_of_parrep}\emph{Perfect $\ZPC$ Completeness}: If $\game$ has a perfect $\ZPC$ strategy, then so does $\wtgame^{\tensk}$.
		\item \label{clause:soundness_of_parrep}\emph{Soundness}: 
		For any $\eps>0$, let  \begin{equation}\label{eq:p-bound}
		p \,=\, p(\eps)\,=\, \frac{c_0}{\eps^{c_0}} \cdot 2^{  - \frac{\eps^{c_0}}{c_0\cdot\Lambda}\cdot k }\;.
		\end{equation}
		If $\wtgame^{\tensk}$ has a strategy with value at least  $p$, then $\game$ has a strategy {of the same dimension} with value  at least $1-\eps$.   
		\item \label{clause:entanglement_of_parrep}\emph{Entanglement}: 
		For all $\eps > 0$ and $p$ as in~\eqref{eq:p-bound}, 
		\[
		\Ent(\wtgame^{\tensk},p)\geq \Ent(\game, 1 - \eps)\;.
		\]
	\end{enumerate}
\end{theorem}

\begin{proof}
	We first verify perfect completeness. Assume $\game$ has a perfect $\ZPC$ strategy. By the analysis immediately after Definition \ref{def:comb-anchoring},  the anchored game $\game'=\frak{Anchor}(\game)$ has a perfect $\ZPC$ strategy as well. Hence, by Corollary \ref{cor:comp_sound_combi_detyping}, the detyped game $\wtgame=\frak{DeType}(\game')$ has a perfect $\ZPC$ strategy. Finally, by Lemma \ref{lem:sum-zpc}, taking the sum (Definition \ref{def:sum-pvm}) of the perfect $\ZPC$ strategy for $\wtgame$ with itself $k$ times produces a perfect $\ZPC$ strategy for $\wtgame^{\tensk}$. This finishes the perfect completeness argument.


	Let us prove the soundness and entanglement lower bound. Recall the notions and notations of Section \ref{sec:non-synch_setup} on the non-synchronous setup.  Let $\delta>0$. If for the synchronous quantum value (Definition \ref{defn:value_of_a_game}) we have  $\val^*(\wtgame^{\tensk})\geq \delta$, then the same can be deduced for the non-synchronous value (Definition \ref{defn:non-synch_value}): $\valns(\wtgame^{\tensk})\geq \delta$, as every synchronous strategy is also a general strategy (Remark \ref{rem:synch_strategy_as_a_general_one}) --- note that the same holds for the entanglement lower bound. We now need to apply a deep theorem from the study of general quantum strategies.
	
	\begin{theorem}[Non-synchronous soundness of parallel repetition of anchored games, Theorem 6.1 of~\cite{bavarian2017hardness}]
		\label{thm:bvy}
		There exists a universal positive integer constant $c'$ such that for all two-player anchored games $\game$ and for all positive integers $k$, for all $0 < \eps \leq 1$, and for $\delta$ satisfying
		\begin{equation}\label{eq:p-bound-2}
		\delta = \frac{4}{\eps} \cdot 2^{  - \frac{\eps^{17}}{c'\cdot \Lambda}\cdot k }\;,
		\end{equation}
		we have that  every strategy for $\game^{\tensk}$ of dimension $n$ with value at least $\delta$ induces\footnote{Finding this induced strategy is not immediate at all, and is one of the main issues in the proof of this theorem.} a strategy of {the same dimension} $n$ for $\game$ with value $1-\eps$. Namely,
		\[
		\Entns(\game^{\tensk},\delta) \geq \Entns(\game,1-\eps)~.
		\]
	\end{theorem}
	\begin{remark}
		Here we do not repeat the precise definition of a game being ``anchored,'' referring to~\cite{bavarian2017hardness} for it. Suffice it to say that any game produced by the transformation $\frak{DeType}\circ \mathfrak{Anchor}(\cdot)$  is anchored. 
	\end{remark}
	As we fixed $\delta$ in the beginning of the proof, let $\eps_\delta$ be the value that satisfies  equation \eqref{eq:p-bound-2} with respect to it. Then, given a general strategy $\strategy$ for $\wtgame^{\tensk}$ with value $\delta$, there is a {general} strategy $\strategy'$ of the same dimension for $\wtgame$ with value of at least $1-\eps_\delta$. As, combinatorially, detyping (Definition \ref{defn:combi_detyping}) is just playing with some constant probability (that depends on the number of types $|\type|$) the double cover of the original game, and the double cover of the game has the {same} non-synchronous value as the game itself (Remark \ref{rem:genera_synch_as_double_cover}), we can deduce that $\game'$ has a general strategy with value of at least $1-C\eps_\delta$  and the same dimension ($C$ here depends only on $|\type|$). Furthermore, as anchoring is again just playing the original game with some positive probability (and accepting otherwise), the same general strategy has value of at least $1-C'C\eps_{\delta}$ against $\game$, where again $C'$ depends only on $|\type|$. Now, we assumed that the type graph of $\game$ contains all self loops, and thus for every vertex $\mttx$ we have 
	\[
	\frac{\mu(\mttx\mttx)}{\mu(\mttx)}\geq \frac{1}{2|\type|}\ .
	\]
	Therefore, we can apply Fact \ref{fact:non-synch_high_implies_synch_high}, and retrieve a {synchronous} strategy for $\game$ {of the same dimension} with value of at least $1-C''\cdot (4|\type|^2C'C)^{C''}\cdot \eps_{\delta}^{\nicefrac{1}{C''}}$, where this time  $C''$ is a universal positive integer constant. 
	
	Tracing back these equations, if we want to understand what is the $p$ for which $\val^*(\widetilde{\game}^{\tensk})\geq p$ implies $\val^*({\game})\geq 1-\eps$, we first need to choose $\eps_\delta$ so that 
	\begin{equation}\label{eq:condition_needed_on_eps_delta}
	\eps\geq C''\cdot (4|\type|^2C'C)^{C''}\cdot \eps_{\delta}^{\nicefrac{1}{C''}}\ .    
	\end{equation}
	In particular,  by letting   $c=(C''(4|\type|^2C'C)^{C''})^{C''}$ and $\eps_\delta=\frac{\eps^c}{c}$, \eqref{eq:condition_needed_on_eps_delta} is satisfied.
	We then plug this into equation \eqref{eq:p-bound-2} and deduce that for  
	\[
	p=\frac{4c}{\eps^c}\cdot 2^{  - \frac{\eps^{17c}}{c^{17}\cdot c'\cdot \Lambda}\cdot k }
	\]
	the conclusion is satisfied. As $c\geq 4$ and $c'\geq 1$, $c^{17}\cdot c'\geq 17c\geq 4c\geq c$, and thus, if we take $c_0=c^{17}\cdot c'$, the number 
	\[
	\frac{c_0}{\eps^{c_0}}\cdot 2^{  - \frac{\eps^{c_0}}{c_0\cdot \Lambda}\cdot k }
	\]
	is always larger than the aforementioned $p$, and we can conclude. Note that indeed $c_0$ depends only on $|\type|$ and nothing else.
	
\end{proof}

The following claim shows that, given a function $\cal{K}\colon\mathbb{N}\to \mathbb{N}$ and a normal form verifier $\verifier$, one can efficiently calculate a normal form verifier $\verifier'$ whose $n^{\rm th}$ game is  the $\cal{K}(n)$-fold tensor power of the $n^{\rm th}$ game  of $\verifier$, namely $\verifier'_n=\verifier_n^{\otimes \cal{K}(n)}$.

\begin{claim}[Algorithmic parallel repetition]\label{claim:alg_parallel_repetition}
	There is a $2$-input TM $\TMPR$ that takes as input an $h$-level {untyped} tailored normal form verifier $\verifier$ and a $1$-input always halting TM $\cal{K}$ (in particular, it induces a map $\cal{K}\colon \mathbb{N}\to\mathbb{N}$), and outputs a new $h$-level normal form verifier 
	\[
	\TMPR(\verifier,\cal{K})=\verifier'=(\sampler',\length',\linproc',\decider)
	\] 
	with the following properties.
	\begin{enumerate}
		\item \emph{Combinatorial parallel repetition}: If $\verifier_n$ is well defined, then $\verifier'_n$ is well defined and $\verifier'_n=\verifier_n^{\otimes \cal{K}(n)}$.
		\item \emph{Complexity}: The time bounds are 
		\begin{align}
		\TIME(\sampler';\overline{n},\cdot,\cdot,\cdot,\cdot,\cdot) & = \poly(\cal{K}(n),\TIME(\cal{K};n),\TIME(\sampler;\overline{n},\cdot,\cdot,\cdot,\cdot))\;,\notag\\
		\TIME(\length';\overline{n},\cdot,\cdot) & = \poly(\cal{K}(n),\TIME(\cal{K};n),\TIME(\sampler;\overline{n},\cdot,\cdot,\cdot,\cdot),\TIME(\length;\overline{n},\cdot,\cdot))\;,\label{eq:time_bounds_partial_par_rep}\\
		\TIME(\linproc';\overline{n},\cdot,\cdot,\cdot,\cdot) & = \poly(\cal{K}(n),\TIME(\cal{K};n),\TIME(\sampler;\overline{n},\cdot,\cdot,\cdot,\cdot),\TIME(\length;\overline{n},\cdot,\cdot),\TIME(\linproc;\overline{n},\cdot,\cdot,\cdot,\cdot))\;.\notag
		\end{align}
		In addition, the description lengths are linear in that of $\verifier$ and $\cal{K}$. Finally, the sampler $\sampler'$ depends only on $\cal{K}$ and $\sampler$, while the answer length calculator $\length'$ depends only on $\cal{K},\sampler$ and $\length$.
	\end{enumerate}
\end{claim}

\begin{proof}
	Let us define the TMs one by one.
	
	\textbf{The Sampler:}
	First, $\sampler'$ calculates $k=\cal{K}(n)$. If the input is $(n,{\rm Dimension},\cdot,\cdot,\cdot,\cdot)$, $\sampler'$ calculates  $\sampler(n,{\rm Dimension},\cdot,\cdot,\cdot,\cdot)=r$  and outputs $r\cdot k$. Otherwise, given an input of the form 
	\[
	(\on,{\rm  Action},{\rm Player},j,\mttx,z)\ ,
	\]
	as $\mttx$ and $z$ are length $r\cdot k$ bit-strings, they can be interpreted as $\mttx=(\mttx_1,...,\mttx_k)$ and $z=(z_1,...,z_k)$. So, $\sampler'$ does $k$ calls to $\sampler$, by evaluating it on $(\on,{\rm  Action},{\rm Player},j,\mttx_i,z_i)$ for every $i\in [k]$. It then concatenates the result so as to produce the appropriate output in $(\FF_2^{r})^k$. 
	\\
	
	This TM has a constant description up to appending $\cal{K}$ and $\sampler$. It runs in time linear in $\cal{K}(n)$ times $\TIME(\sampler;n,\cdot,\cdot,\cdot,\cdot,\cdot)$, as required, and depends only on $\sampler$ and $\cal{K}$.
	\\
	
	\textbf{The Answer length calculator}:
	Given  $(n,\mttx,\kappa)$, $\length'$ calls $\cal{K}(n)$ and $\sampler(n,{\rm Dimension},\cdot,\cdot,\cdot,\cdot)$ to retrieve $k$ and $r$ respectively. It then checks that $\mttx$ is indeed of length $k\cdot r$, and outputs $\frak{error}$ otherwise. Finally, if $\mttx$ is indeed of the appropriate length, it denotes it $\mttx=(\mttx_1,...,\mttx_k)\in (\FF_2^r)^k$, calculates $\length(n,\mttx_i,\kappa)$ for all $i\in [k]$, and concatenates the outputs to a single bit string.\footnote{As $\length'$ needs to output the (encoded) answer length in unary, this concatenation is exactly taking the sum of the appropriate lengths.}
	\\
	
	This TM has a constant description length up to appending $\cal{K}$ and $\sampler$ and $\length$. It runs in time linear in $\cal{K}(n)$,  $\TIME(\sampler;n,\cdot,\cdot,\cdot,\cdot,\cdot)$ and $\TIME(\length;n,\cdot,\cdot)$, as required, and depends only on $\sampler,\length$ and $\cal{K}$.
	\\
	
	\textbf{The Linear constraints processor}:
	The TM $\linproc'$ first  calls $\cal{K}(n)$ and $\sampler(n,{\rm Dimension},\cdot,\cdot,\cdot,\cdot)$ to retrieve $k$ and $r$ respectively. 
	Now, given  input
	$(n,{\mttx},{\mtty},{a^\frR},{b^\frR})$, it parses $\mttx,\mtty,a^\frR,b^\frR$ as $k$-tuples of questions and readable answers, respectively. It then calls $\length(n,\mttx_i,\kappa),\length(n,\mtty_i,\kappa)$ to retrieve $\ell^\kappa(\mttx_i)$ and $\ell^\kappa(\mtty_i)$  for every $i\in [k]$ and $\kappa\in \{\frR,\frL\}$. Then, it calls  $\linproc(n,\mttx_i,\mtty_i,a^\frR_i,b^\frR_i)$, for all $i \in [k]$. The output bit-strings represent constraints on some subset of the variables used in the parallel repeated game, i.e., the output of $\linproc(n,\mttx_i,\mtty_i,a^\frR_i,b^\frR_i)$ is (the encoding) of a sequence $(c^i_1,...,c^i_m)$, each of which is of length $\ell^\frR(\mttx_i)+\ell^\frL(\mttx_i)+\ell^\frR(\mtty_i)+\ell^\frL(\mtty_i)+1$,\footnote{As usual, if the outputs are not well formatted, $\linproc'$ should halt and output $\frak{error}$.} and we want to interpret it as a sequence of $m$ constraints of length $1+\sum_{j=1}^k\ell^\frR(\mttx_j)+\ell^\frL(\mttx_j)+\ell^\frR(\mtty_j)+\ell^\frL(\mtty_j)$. To do that, we need to pad them  with zeros appropriately so that $c^i_j$ would  be the same constraint as before, but on the variables at the $i^{\rm th}$ position. After doing it to the output of each call to $\linproc$, $\linproc'$ collects them all to a long list of constraints and outputs (the encoding) of it.
	\\
	
	This TM has a constant description up to appending $\cal{K}, \sampler, \length$ and $\linproc$. It runs in time linear in 
	\[
	\cal{K}(n)\ ,\   \TIME(\sampler;n,\cdot,\cdot,\cdot,\cdot,\cdot)\ ,\ \TIME(\length;n,\cdot,\cdot)
	\]
	and $\TIME(\linproc;n,\cdot,\cdot,\cdot,\cdot)$ as required.

\end{proof}

\subsection{Proving the main theorem of Parallel Repetition: Theorem \ref{thm:repetition}}\label{sec:main_par_rep}

\begin{theorem}[Anchored parallel repetition of tailored normal form verifiers]
	
	There exists a function 
	\begin{equation}
	c_\rep\colon \mathbb{N}\to \mathbb{N}
	\end{equation} 
	and a $2$-input polynomial-time Turing machine
	$\ComputeParrepVerifier_h$, that takes as input a typed $h$-level tailored normal form verifier $\verifier = (\sampler,\length,\linproc,\decider)$ and a $1$-input TM $\cal{K}$ (which induces, as usual, a partial  function $\cal{K}\colon \mathbb{N}\to \mathbb{N}$), 
	and  outputs an $(h+2)$-level \textbf{untyped} tailored normal form verifier  $\ComputeParrepVerifier_h(\verifier,\cal{K})=\verifier_\rep = (\sampler_\rep,\length_\rep,\linproc_\rep,
	\decider)$ 
	such that the following hold.
	\begin{enumerate}
		\item \emph{Combinatorial parallel repetition}: If $\verifier_n$ is well defined, then $(\verifier_{\rep})_n$ is well defined, and 
		\[
		(\verifier_{\rep})_n=(\frak{DeType}(\frak{Anchor}(\verifier_n)))^{\otimes \cal{K}(n)}\ ,
		\]
		where $\frak{Anchor}$ is the anchoring transformation (Definition \ref{def:comb-anchoring}), $\frak{DeType}$ is the detyping transformation (Definition \ref{defn:combi_detyping}), and $(\cdot)^{\otimes k}$ is the $k$-fold tensor power of a game (Definition \ref{def:product-game} and the beginning of Section \ref{sec:parrep-comb}).
		In particular, for $c=c_{\rep}(|\type|)$, where $|\type|$ is the number of types  in the type graph of $\verifier$ and $c_\rep$ is from \eqref{eq:defn_c_rep}, we have ---
		\begin{itemize}
			\item \label{enu:pr-completeness} \emph{Completeness}: If $\verifier_n$ has
			a value-$1$ $\ZPC$ strategy, then $(\verifier_{\rep})_n$ has a value-$1$ $\ZPC$
			strategy.
			
			\item \label{enu:pr-soundness} \emph{Soundness}: For all $\eps > 0$, letting $\Lambda(n)=\TIME(\length;n,\cdot,\cdot)$, if
			\begin{equation*}
			p \geq  \frac{c}{\eps^c} \cdot 2^{  -  \frac{\eps^{c}}{c\Lambda(n)} \cdot \cal{K}(n)}\;, 
			\end{equation*}
			then $\Ent((\verifier_\rep)_n, p) \geq \Ent(\verifier_n, 1 - \eps)$.
		\end{itemize}
		
		%
		
		\item \label{enu:pr-complexity} \emph{Complexity}: Letting $c=c_\rep(|\type|)$ as above, the  time complexities of the output verifier $(\verifier_{\rep})_n$ are
		\begin{align*}
		\TIME(\sampler_\rep;\overline{n},\cdot,\cdot,\cdot,\cdot) & \leq  c\big(\cal{K}(n)\cdot \TIME(\cal{K};n) \cdot \TIME(\sampler;\overline{n},\cdot,\cdot,\cdot,\cdot)\big)^c\;,\\
		\TIME(\length_\rep;\overline{n},\cdot,\cdot) & \leq  c\big(\cal{K}(n)\cdot \TIME(\cal{K};n) \cdot \TIME(\sampler;\overline{n},\cdot,\cdot,\cdot,\cdot)\cdot\TIME(\length;\overline{n},\cdot,\cdot)\big)^c\;,\\
		\ \TIME(\linproc_\rep;\overline{n},\cdot,\cdot,\cdot,\cdot) & \leq  c\big(\cal{K}(n)\cdot \TIME(\cal{K};n) \cdot \TIME(\sampler;\overline{n},\cdot,\cdot,\cdot,\cdot)\cdot\TIME(\length;\overline{n},\cdot,\cdot)\cdot \TIME(\linproc;\overline{n},\cdot,\cdot,\cdot,\cdot)\big)^c\;.
		\end{align*}  
		The repeated sampler $\sampler_\rep$ only depends on $\sampler$ and $\cal{K}$, the answer length calculator $\length_\rep$ only depends on $\sampler$, $\length$ and  $\cal{K}$. Finally, the description lengths of all output TMs are linear in the TMs they depend on. 
	\end{enumerate}
	
\end{theorem}

\begin{proof}
	Letting $\verifier_\rep=\TMPR(\mathsf{DeTyping}(\TMAnchoring(\verifier)),\cal{K})$, and using Claims \ref{claim:alg_parallel_repetition}, \ref{claim:DeTyping_NFV} and \ref{claim:alg_anchor}, we deduce that indeed 
	\[
	(\verifier_{\rep})_n=(\frak{DeType}(\frak{Anchor}(\verifier_n)))^{\otimes \cal{K}(n)}\ ,
	\]
	whenever $\verifier_n$  is well defined. Following all these claims as well as Theorem \ref{thm:parrep-comp-sound}, and taking $c_\rep(|\type|)$ large enough to bound all the constants appearing in them (in particular, bound $c_0$ from Theorem \ref{thm:parrep-comp-sound}), we deduce the soundness clause,  as well as the time complexities. Completeness is deduced by the same theorem and claims.
\end{proof}


\section{Proving the compression theorem}\label{sec:proof_of_compression}

Let us recall the version of the compression theorem, first phrased in Theorem \ref{thm:h_level_compression}, that we ought to prove. Recall also the asymptotic notation from Remark \ref{rem:asymptotic_notation}.

\begin{theorem}[Compression of tailored $h$-level  normal form verifiers] \label{thm:h_level_compression_recalled}
	
	For every positive integer $h$, there exist  two positive integers 
	\[
	c=c(h)\quad{\rm and}\quad C=C(h)
	\]
	that depend only on $h$,  and a $2$-input Turing machine $\Compress_h$, that takes as input a $h$-level
	TNFV $\verifier=(\sampler,\length,\linproc,\decider)$ and a positive integer $\lambda$ (in binary), and outputs  a \textbf{$\LevelConstant$-level} TNFV $\Compress_h(\verifier,\lambda)=\verifier'= (\sampler^\lambda, \length^\lambda, \linproc',\decider)$, such that:
	\begin{itemize}
		\item \underline{Sampler properties}:  The $5$-level CL sampler $\sampler^\lambda$ depends only on $\lambda$ and $h$, (but not the specific $\verifier$), and  $\Compress_h$ can calculate its description in time $\polylog_h (\lambda)$;\footnote{Recall our asymptotic notation from Remark \ref{rem:asymptotic_notation} to parse $\polylog_h$.} in particular, $|S^\lambda|\leq c\log^c\lambda$. In addition, $\sampler^\lambda$ runs in   $\poly_h(n,\lambda)$-time, namely
		\[
		\forall n\in \mathbb{N}\ \colon \ \ \TIME(\sampler^\lambda;n)\leq c\cdot (n^c+\lambda^c)\ .
		\]

		\item \underline{Answer length calculator properties}: $\length^\lambda$  depends only on $\lambda$ and $h$,  and $\Compress_h$ can calculate its description in time $\polylog_h (\lambda)$; in particular $|\length^\lambda|\leq c\log^c\lambda$. In addition, $\length^\lambda$ runs in   $\poly_h(n,\lambda)$-time, namely 
		\[
		\forall n\in \mathbb{N}\ \colon \ \ \TIME(\length^\lambda;n,\cdot,\cdot)\leq c\cdot (n^c+\lambda^c)\ .
		\]
		Finally,  given that $\mttx\in \FF_2^{r(n)}$, where $r(n)=\sampler^\lambda(n,{\rm Dimension},\cdot,\cdot,\cdot,\cdot)$,  and that $\kappa\in \{\frR,\frL\}$,  the output of $\length^\lambda(n,\mttx,\kappa)$  never decodes (Definition \ref{defn:the_alphabet}) to an $\frak{error}$ sign.
		
		\item \underline{Linear constraints processor properties}: $\linproc'$ depends on both $\lambda$ and $\verifier$, and $\Compress_h$ can calculate its description in time $\poly_h(\log \lambda, |\verifier|)$; in particular, $|\linproc'|\leq c\cdot (\log^c\lambda +|\verifier|^c)$. In addition, $\linproc'$ runs in $\poly_h(n,\lambda)$-time, namely 
		\[
		\forall {n}\in \mathbb{N}\ \colon \ \ \TIME(\linproc';{n},\cdot,\cdot,\cdot,\cdot) \ \leq\ 
		c\cdot (n^c+\lambda^c)\ .
		\]
		\item \underline{Decider properties}: The canonical decider $\decider$ (Definition \ref{def:canonical-decider}) is fixed  and runs in  time  which is linear in its input length. 
		
		\item  \underline{Value properties}: If $\verifier$ is $\lambda$-bounded, then 
		$\verifier'$, the output of $\Compress_h$,  satisfies: For all $n \geq C$,
		\begin{enumerate}
			\item \label{enu:h_level_compr-completeness} \textbf{Completeness}: If $\verifier_{2^n}$
			has a perfect $Z$-aligned permutation strategy that commutes along edges ($\ZPC$ strategy), then so does $\verifier'_n$.
			\item \label{enu:h_level_compr-soundness} \textbf{Soundness}:
			$\Ent(\verifier_{n}',\frac{1}{2}) \geq \max \left \{
			\Ent(\verifier_{2^n}, \frac{1}{2}), 2^{2^{\lambda n}-1}
			\right \}$.
		\end{enumerate}
	\end{itemize}
	
\end{theorem}

\begin{proof}
	Though our parameters are different, we use the same proof structure as \cite[Theorem 12.1]{MIPRE}. To that end, let $\cal{K}=\cal{K}_{\lambda,h}$ be a TM that takes an integer $n$ in binary as input, and outputs
	\begin{equation}\label{eq:defn_of_calK_proof_of_compression}
	\cal{K}(n)=c_0(n\lambda)^{c_0}\ 
	\end{equation}
	in binary, where $c_0$ is a integer parameter that will be fixed later, and which depends only on $h$. Note that $\TIME(\cal{K};n)=\polylog_h(n,\lambda)$ and $|\cal{K}|=\polylog_h(\lambda)$, which means there is a positive integer constant $C_0=C_0(h)$ such that $|\cal{K}|\leq C_0\log^{C_0}\lambda$ and $\TIME(\cal{K};n)\leq C_0(\log^{C_0} n+ \log ^{C_0}\lambda)$. Let us collect other constants that will be used along the proof: $c_1=c_\qr(h)$ defined in \eqref{eq:defn_c_qr} from  Theorem \ref{thm:h_level_question_reduciton}; $c_2=c_\ar(3,h)$ defined in  \eqref{eq:defn_c_ar} from Theorem \ref{thm:main_ans_red}; $c_3=c_\rep(9)$ defined in \eqref{eq:defn_c_rep} from Theorem \ref{thm:repetition}.

	Now, define three tailored normal form verifiers (TNFVs):
	\begin{enumerate}
		\item Let $\mathsf{QuestionReduction}_h(\verifier,\lambda)=\verifier^{(1)}=(\sampler^{(1)},\length^{(1)},\linproc^{(1)},\decider)$.
		\item Let $\AnsRed_{3,h}(\verifier^{(1)},\lambda)=\verifier^{(2)}=(\sampler^{(2)},\length^{(2)},\linproc^{(2)},\decider)$.
		\item Let $\ComputeParrepVerifier_3(\verifier^{(2)},\cal{K})=\verifier^{(3)}=(\sampler^{(3)},\length^{(3)},\linproc^{(3)},\decider)$.
		\item Finally, choose $\sampler^\lambda:=\sampler^{(3)}$, $\length^\lambda:=\length^{(3)}$ and $\linproc':=\linproc^{(3)}$.
	\end{enumerate}

	\textbf{Level:}

	By Theorem \ref{thm:h_level_question_reduciton}, the output of $\mathsf{QuestionReduction}_h$ is always a $3$-level TNFV, and thus $\verifier^{(1)}$ is such. By Theorem \ref{thm:main_ans_red}, given that its input was $3$-level, the output of $\AnsRed_{3,h}$ is always a $\max(3,3)=3$-level typed TNFV, and  thus $\verifier^{(2)}$ is such. By Theorem \ref{thm:repetition}$, \ComputeParrepVerifier_3$ outputs a $5$-level TNFV given that the input was a typed $3$-level TNFV, and thus $\verifier^{(3)}$ is a $5$-level TNFV, as needed.
	\\
	
	\textbf{Sampler properties:}
	
	By Theorem \ref{thm:h_level_question_reduciton}, the sampler $\sampler^{(1)}$ depends only on $\lambda$ and $h$, can be calculated in time $\polylog_h(\lambda)$, runs in time $c_1(n^{c_1}+\lambda^{c_1})$, and has description length bounded by $c_1\log^{c_1}\lambda$. 
	By Theorem \ref{thm:main_ans_red}, $\sampler^{(2)}$ depends on $\sampler^{(1)},\lambda,3$ and $h$, can be calculated in time 
	\[
	\poly_{3,h}(\log \lambda,\underbrace{|\sampler^{(1)}|}_{c_1\log^{c_1}\lambda})=\polylog_h(\lambda)\ ,
	\]
	has description length bounded by 
	\[
	|S^{(2)}|\leq c_2(\log^{c_2}\lambda+|\sampler^{(1)}|^{c_2})\leq 2c_1^{c_2}\cdot c_2\log^{c_1c_2}\lambda\ ,
	\]
	and runs in time at most 
	\[
	\TIME(\sampler^{(2)};n,\cdot,\cdot,\cdot,\cdot,\cdot)\leq c_2(n^{c_2}+\lambda^{c_2}+(\underbrace{\TIME(\sampler^{(1)};n,\cdot\cdot\cdot,\cdot,\cdot)}_{c_1(n^{c_1}+\lambda^{c_1})})^{c_2})\leq 2c_1^{c_2}\cdot c_2(n+\lambda)^{c_1c_2}\ .
	\]
	By Theorem \ref{thm:repetition}, 
	\[
	|\sampler^{(3)}|\leq c_3(|\cal{K}|^{c_3}+|\sampler^{(2)}|^{c_3})\leq c_3((C_0\log^{C_0} \lambda)^{c_3}+(c_2\log^{c_1c_2}\lambda)^{c_3})=\polylog_h(\lambda)\ ,
	\]
	and 
	\[
	\begin{split}
	\TIME(\sampler^{(3)};n,\cdot,\cdot,\cdot,\cdot,\cdot)&\leq c_3(n^{c_3}+\cal{K}(n)^{c_3}+\TIME(\cal{K};n)^{c_3}+\TIME(\sampler^{(2)};n,\cdot,\cdot,\cdot,\cdot,\cdot)^{c_3})\\
	&\leq c_3(n^{c_3}+(c_0n^{c_0}\lambda^{c_0})^{c_3}+(C_0\log^{C_0}n +C_0\log^{C_0}\lambda)^{c_3}+(2c_1^{c_2}\cdot c_2(n+\lambda)^{c_1c_2})^{c_3})\\
	&=\poly_h(n,\lambda)\ .
	\end{split}
	\]
	This proves the sampler properties of Theorem \ref{thm:h_level_compression_recalled}.
	\\
	
	\textbf{Answer length calculator properties:}
	
	By Theorem \ref{thm:main_ans_red}, $\length^{(2)}$ depends only on $3,h$ and $\lambda$, can be calculated from them in $\polylog_h(\lambda)$ time, and in particular $|\length^{(2)}|\leq c_2\log^{c_2}\lambda$. In addition, 
	\[
	\TIME(\length^{(2)};n,\cdot,\cdot)\leq c_2(n^{c_2}+\lambda^{c_2})\ .
	\]
	Also, whenever $\mttx\in \FF_2^{r_2(n)}$ for $r_2(n)=\sampler^{(2)}(n,{\rm Dimension},\cdot,\cdot,\cdot,\cdot)$, and $\kappa\in \{\frR,\frL\}$, the output of $\length^{(2)}(n,\mttx,\kappa)$ does not decode to $\frak{error}$.
	By Theorem \ref{thm:repetition}, $\length^{(3)}$ depends on $\cal{K},\sampler^{(2)}$ and $\length^{(2)}$, and can be calculated from their description in polynomial time. As we showed in the sampler properties $|\sampler^{(2)}|=\polylog_h(\lambda)$, and by the analysis above  $|\length^{(2)}|,|\cal{K}|=\polylog_h(\lambda)$. All in all, $|\length^{(3)}|=\poly(|\sampler^{(2)}|,|\length^{(2)}|,|\cal{K}|)=\polylog_h(\lambda)$. Furthermore, 
	\[
	\begin{split}
	\TIME(\length^{(3)};n,\cdot,\cdot)&\leq c_3\cdot(n^{c_3}+\cal{K}(n)^{c_3}+\TIME(\cal{K};n)^{c_3}+\TIME(\sampler^{(2)};n,\cdot,\cdot,\cdot,\cdot,\cdot)^{c_3}+\TIME(\length^{(2)};n,\cdot,\cdot)^{c_3}) \\
	&\leq c_3\cdot(n^{c_3}+(c_0n\lambda )^{c_0c_3}+(C_0\log^{C_0}n+C_0\log^{C_0}\lambda)^{c_3}\\
	&\quad +(2c_1^{c_2}\cdot c_2(n+\lambda)^{c_1c_2})^{c_3}+(c_2(n^{c_2}+\lambda^{c_2}))^{c_3})\\
	&=\poly_{h}(n,\lambda)\ .
	\end{split}
	\]
	Now, as $r_3(n)=\sampler^{(3)}(n,{\rm Dimension},\cdot,\cdot,\cdot,\cdot)=\cal{K}(n)\cdot r_2(n)$, and by the answer length calculator properties guaranteed by Theorem \ref{thm:repetition}, we can deduce that whenever $\mttx\in \FF_2^{r_3(n)}$ and $\kappa\in \{\frR,\frL\}$, the decoding of the output of $\length^{(3)}(n,\mttx,\kappa)$ is not $\frak{error}$. This finishes the proof of the answer length calculator properties of Theorem \ref{thm:h_level_compression_recalled}.
	\\
	
	\textbf{Linear constraints processor properties:}
	
	By Theorem \ref{thm:h_level_question_reduciton}, $\linproc^{(1)}$ depends on $\lambda,h$ and $\verifier$, and can be calculated in $\poly_h(\log \lambda,|\verifier|)$ time. In particular, $|\linproc^{(1)}|\leq c_1(\log^{c_1}\lambda +|\verifier|^{c_1})$. Moreover, $\TIME(\linproc^{(1)};n,\cdot,\cdot,\cdot,\cdot)\leq 2^{c_1(n^{c_1}+\lambda^{c_1})}$. From Theorem \ref{thm:main_ans_red}, we can deduce that $\linproc^{(2)}$ depends on $3,h,\lambda$ and $\verifier^{(1)}$, and can be calculated from them in time 
	$
	\poly_h(\log \lambda,{|\verifier^{(1)}|}).
	$
	By Theorem \ref{thm:h_level_question_reduciton}, 
	\begin{equation}\label{eq:verifir(1)_upper_bound_desecription}
	|\verifier^{(1)}|=\underbrace{|\sampler^{(1)}|}_{\leq c_1\log^{c_1}\lambda}+\underbrace{|\length^{(1)}|}_{\leq c_1\log^{c_1}\lambda}+\underbrace{|\linproc^{(1)}|}_{\leq c_1(\log^{c_1}\lambda+|\verifier|^{c_1})}+|\decider|\leq c_1(3\log^{c_1}\lambda+|\verifier|^{c_1})+\underbrace{|\decider|}_{O(1)}=\poly_h(\log \lambda,|\verifier|)\ ,
	\end{equation}
	and thus $|\verifier^{(2)}|=\poly_h(\log \lambda,|\verifier|)$. In addition, by Theorem \ref{thm:main_ans_red}, 
	\[
	\TIME(\linproc^{(2)};n,\cdot,\cdot,\cdot,\cdot)\leq c_2(n^{c_2}+\lambda^{c_2}+\TIME(\sampler^{(1)};n,\cdot,\cdot,\cdot,\cdot,\cdot)^{c_2})\ . 
	\]
	Now, by Theorem \ref{thm:repetition}, 
	\[
	\begin{split}
	\TIME(\linproc^{(3)};n,\cdot,\cdot,\cdot,\cdot)&\leq c_3\cdot(n^{c_3}+\cal{K}(n)^{c_3}+\TIME(\cal{K};n)^{c_3}+\TIME(\sampler^{(2)};n,\cdot,\cdot,\cdot,\cdot,\cdot)^{c_3}+\TIME(\length^{(2)};n,\cdot,\cdot)^{c_3}+\TIME(\linproc^{(2)};n,\cdot,\cdot,\cdot,\cdot)^{c_3})\\
	&\leq c_3\cdot(n^{c_3}+(c_0n\lambda )^{c_0c_3}+(C_0\log^{C_0}n+C_0\log^{C_0}\lambda)^{c_3}+(2c_1^{c_2}\cdot c_2(n+\lambda)^{c_1c_2})^{c_3}\\
	&\quad +(c_2(n^{c_2}+\lambda^{c_2}))^{c_3}+ (c_2(n^{c_2}+\lambda^{c_2})+(c_1(n^{c_1}+\lambda^{c_1}))^{c_2})^{c_3})\\
	&=\poly_h(n,\lambda)\ .
	\end{split}
	\]
	Also, the description of $\linproc^{(3)}$ can be calculated in time 
	\[
	\poly(\underbrace{|\cal{K}|}_{\polylog_h(\lambda)},\underbrace{|\verifier^{(2)}|}_{\poly_h(\log\lambda,|\verifier|)})=\poly_h(\log\lambda,|\verifier|)\ .
	\] This finishes the proof of the linear constraints processor properties of Theorem \ref{thm:h_level_compression_recalled}.
	\\
	
	\textbf{Value properties:}
	
	In this part, we may assume $\verifier$ is $\lambda$-bounded. Namely, 
	\[
	|\verifier|\leq \lambda\quad ,\quad \TIME(\sampler;n,\cdot,\cdot,\cdot,\cdot,\cdot),\TIME(\length;n,\cdot,\cdot),\TIME(\linproc;n,\cdot,\cdot,\cdot,\cdot) \leq n^\lambda\ .
	\]
	As $|\verifier|\leq \lambda$, and assuming $c_1$ is larger than $|\decider|$ regardless of $h$,  we have by \eqref{eq:verifir(1)_upper_bound_desecription} that 
	\begin{equation}\label{eq:bound_on_description_V1}
	|\verifier^{(1)}|\leq c_1(3\log^{c_1}\lambda+\lambda^{c_1}+1)\leq 5c_1\lambda^{c_1}\ .
	\end{equation}
	By Theorem \ref{thm:h_level_question_reduciton}, 
	\begin{equation}\label{eq:time_bounds_V1}
	\TIME(\sampler^{(1)};n,\cdot,\cdot,\cdot,\cdot,\cdot) \leq c_1 (n^{c_1}+\lambda^{c_1})\ ,\ \TIME(\length^{(1)};n,\cdot,\cdot),\TIME(\linproc^{(1)};n,\cdot,\cdot,\cdot,\cdot)\leq 2^{c_1(n^{c_1}+\lambda^{c_2})}\ .
	\end{equation}

	Let us assume $\verifier_{2^n}$ has a perfect $\ZPC$ strategy. By Theorem \ref{thm:h_level_question_reduciton}, given that $\verifier$ is $\lambda$-bounded, $\verifier^{(1)}_n$ has a perfect $\ZPC$ strategy. 
	Now, by \eqref{eq:bound_on_description_V1} and \eqref{eq:time_bounds_V1}, $\verifier^{(1)}$ satisfies the conditions of the value properties in Theorem \ref{thm:main_ans_red}, and thus  $\verifier^{(1)}_n$ having a perfect $\ZPC$ strategy implies $\verifier^{(2)}_n$ has a perfect $\ZPC$ strategy.
	As Theorem \ref{thm:repetition} has no conditions for the value properties, $\verifier^{(2)}_n$ having a perfect $\ZPC$ strategy implies $\verifier^{(3)}_n$ has a perfect $\ZPC$ strategy, proving the perfect completeness condition of Theorem \ref{thm:h_level_compression_recalled}.
	
	Let $\eps_1=\left(\frac{1}{2c_1}\right)^{16}$, which means $1-c_1\eps_1^{\nicefrac{1}{16}}=\nicefrac{1}{2}$ and $1-c_1\eps_1>\nicefrac{1}{2}$. Then, as $\verifier^{(1)}$ is $\lambda$-bounded, the entanglement lower bound from the value properties of Theorem \ref{thm:h_level_question_reduciton} holds, which means 
	\begin{equation}\label{eq:entanglement_ineq_2}
	\Ent(\verifier^{(1)}_n,1-\eps_1)\geq \underbrace{(1-c_1\eps_1)}_{\geq \nicefrac{1}{2}}\cdot 2^{2^{\lambda n}}\cdot \Ent (\verifier_{2^n},\underbrace{1-c_1\eps_1^{\nicefrac{1}{16}}}_{=\nicefrac{1}{2}})\geq 2^{2^{\lambda n}-1}\cdot \Ent (\verifier_{2^n},\nicefrac{1}{2})\ .
	\end{equation}
	As the last step in $\QueRed_h$ (see the proof of Theorem \ref{thm:h_level_question_reduciton} in Section \ref{sec:proof_of_QR}) is to apply $\mathsf{DeType}_1$ (Claim \ref{claim:DeTyping_NFV}), the game $\verifier^{(1)}_n$ is $\frak{DeType}(\game)$ (Definition \ref{defn:combi_detyping}) for some other game $\game$. Hence,  excluding the anchor vertices in $\verifier^{(1)}_n$ (whose lengths are anyway $0$ and whenever sampled against the game accepts), the underlying graph of $\verifier^{(1)}_n$ is bipartite. Therefore, by Remark \ref{rem:completeness_soundness_of_double_cover_bipartite_case}, the value and entanglement lower bounds of $\verifier^{(1)}_n$ and $\frak{DoubleCover}(\verifier^{(1)}_n)$ (recall the double cover game from Definition \ref{defn:double_cover}) are the same. In particular, 
	\[
	\Ent(\frak{DoubleCover}(\verifier^{(1)}_n),1-\eps_1)=\Ent(\verifier^{(1)}_n,1-\eps_1)\ .
	\]
	Let $C=((2c_1)^{16}c_2)^{c_2}$,\footnote{Note that as $c_1$ and $c_2$ are constants that depend only on $h$, the same is true for $C$.}   which implies in particular (as $\lambda$ is a positive integer) that $c_2(C\lambda)^{-\nicefrac{1}{c_2}}\leq \frac{\eps_1}{2} $, and let $\eps_2(n)=\left(\frac{\eps_1}{2c_2(n\lambda)^{c_2}}\right)^{c_2}$, which implies 
	\[
	c_2(n\lambda)^{c_2}(\eps_2(n))^{\nicefrac{1}{c_2}}=\frac{\eps_1}{2}\ .
	\]
	By \eqref{eq:bound_on_description_V1} and \eqref{eq:time_bounds_V1}, $\verifier^{(1)}$ satisfies the conditions of the value properties in Theorem \ref{thm:main_ans_red}, and thus
	\[
	\Ent(\verifier^{(2)}_n,1-\eps_2(n))\geq \Ent(\frak{DoubleCover}(\verifier^{(1)}_n),1-c_2((n\lambda)^{c_2}(\eps_2(n))^{\nicefrac{1}{c_2}}+(n\lambda)^{-\nicefrac{1}{c_2}}))\ .
	\]
	Assuming $n\geq C$, we have $c_2(n\lambda)^{-\nicefrac{1}{c_2}}\leq c_2(C\lambda)^{-\nicefrac{1}{c_2}}\leq \frac{\eps_1}{2}$ and combined with our choice of $\eps_2(n)$ we have 
	\[
	\Ent(\frak{DoubleCover}(\verifier^{(1)}_n),1-\underbrace{c_2(n\lambda)^{c_2}(\eps_2(n))^{\nicefrac{1}{c_2}}}_{=\nicefrac{\eps_1}{2}}-\underbrace{c_2(n\lambda)^{-\nicefrac{1}{c_2}}}_{\leq \nicefrac{\eps_1}{2}})\geq \Ent(\frak{DoubleCover}(\verifier^{(1)}_n),1-\eps_1)\ .
	\]
	All in all, for every $n\geq C$, we have 
	\begin{equation}\label{eq:entanglement_ineq_3}
	\Ent(\verifier^{(2)}_n,1-\eps_2(n))\geq \Ent(\verifier^{(1)}_n,1-\eps_1)\ .
	\end{equation}
	Recall the parameter from \eqref{eq:defn_p_par_rep} in Theorem \ref{thm:repetition}:
	\[
	p(\eps_2(n),n) =  \frac{c_3}{(\eps_2(n))^{c_3}} \cdot 2^{-  \frac{(\eps_2(n))^{c_3}}{c_3} \cdot \cal{K}(n)\cdot \TIME(\length^{(2)};n,\cdot,\cdot)^{-1}}=2^{\log \frac{c_3}{(\eps_2(n))^{c_3}}-\frac{(\eps_2(n))^{c_3}}{c_3} \cdot \cal{K}(n)\cdot \TIME(\length^{(2)};n,\cdot,\cdot)^{-1}}\ .
	\]
	By the value properties of Theorem \ref{thm:repetition}, 
	\begin{equation}\label{eq:entanglement_ineq_4}
	\Ent(\verifier^{(3)}_n,p(\eps_2(n),n))\geq \Ent(\verifier^{(2)},1-\eps_2(n))\ .
	\end{equation}
	If we choose $\cal{K}$ from \eqref{eq:defn_of_calK_proof_of_compression} so that 
	\begin{equation}\label{eq:condition_calK_needs_to_satisfy}
	\log \frac{c_3}{(\eps_2(n))^{c_3}}-\frac{(\eps_2(n))^{c_3}}{c_3} \cdot \cal{K}(n)\cdot \TIME(\length^{(2)};n,\cdot,\cdot)^{-1}\leq -1\ ,     
	\end{equation}
	then we have $p(\eps_2(n),n)\leq \nicefrac{1}{2}$ and thus
	\begin{equation}\label{eq:entanglement_ineq_5}
	\Ent(\verifier^{(3)}_n,\nicefrac{1}{2})\geq \Ent(\verifier^{(3)}_n,p(\eps_2(n),n))\ .
	\end{equation}
	Combining \eqref{eq:entanglement_ineq_2}, \eqref{eq:entanglement_ineq_3}, \eqref{eq:entanglement_ineq_4} and \eqref{eq:entanglement_ineq_5}, we deduce that for every $n\geq C$ (assuming \eqref{eq:condition_calK_needs_to_satisfy} is satisfied), 
	\[
	\Ent(\verifier^{(3)}_n,\nicefrac{1}{2})\geq 2^{2^{\lambda n}-1}\cdot \Ent (\verifier_{2^n},\nicefrac{1}{2})\geq \max\{2^{2^{\lambda n}-1}, \Ent (\verifier_{2^n},\nicefrac{1}{2})\}\ ,
	\]
	finishing the proof of the soundness condition from Theorem \ref{thm:h_level_compression_recalled}.
	So, let us finally choose $c_0$ and thus the TM $\cal{K}$ from \eqref{eq:defn_of_calK_proof_of_compression} so that \eqref{eq:condition_calK_needs_to_satisfy} is satisfied. Manipulating \eqref{eq:condition_calK_needs_to_satisfy}, we need to have
	\[
	\begin{split}
	\cal{K}(n)\geq \left(1+\log\frac{c_3}{(\eps_2(n))^{c_3}}\right)\cdot \frac{c_3}{(\eps_2(n))^{c_3}}\cdot \TIME(\length^{(2)};n,\cdot,\cdot)\ .
	\end{split}
	\]
	Recalling our choices of $\eps_1$ and $\eps_2(n)$, we have
	\[
	\frac{c_3}{(\eps_2(n))^{c_3}}=c_3(2c_2(n\lambda)^{c_2}(2c_1)^{16})^{c_2}\leq (4c_1c_2c_3)^{16c_2}(n\lambda)^{c_2^2}\ .
	\]
	Also, using the fact that for positive integers $1+\log_2x\leq 2x$, we have that 
	\[
	1+\log\frac{c_3}{(\eps_2(n))^{c_3}}\leq \frac{2c_3}{(\eps_2(n))^{c_3}}\leq 2(4c_1c_2c_3)^{16c_2}(n\lambda)^{c_2^2}\ .
	\]
	Let us also recall the upper bound on $\TIME(\length^{(2)};n,\cdot,\cdot)$ previously calculated,
	\[
	\TIME(\length^{(2)};n,\cdot,\cdot)\leq c_2(n^{c_2}+\lambda^{c_2})\leq 2c_2(n\lambda)^{c_2}\ .
	\]
	So, if we choose $c_0$ so that 
	\[
	c_0(n\lambda)^{c_0}=\cal{K}(n)\geq 4c_2(4c_1c_2c_3)^{32c_2}(n\lambda)^{c_2^2+c_2}\ ,
	\]
	we are done. As $4c_2(4c_1c_2c_3)^{32c_2}\geq  c_2^2+c_2$, choose $c_0=4c_2(4c_1c_2c_3)^{32c_2}$. Since each of $c_1,c_2,c_3$ depends only on $h$ (in the case of $c_3$, it is actually a universal constant), $c_0$ is only a function of $h$, as needed. This finishes the proof of Theorem \ref{thm:h_level_compression_recalled}.
\end{proof}

\bibliography{Bibliography}

\end{document}